\documentclass[12pt,brazil,english,spanish,american,oneside]{book}
\pdfoutput=1 
\usepackage[T1]{fontenc}
\usepackage[utf8]{inputenc}
\usepackage[a4paper]{geometry}
\usepackage{babel}
\addto\shorthandsspanish{\spanishdeactivate{~<>}}
\usepackage{pdfpages}
\usepackage{array}
\usepackage{float}
\usepackage{units}
\usepackage{mathtools}
\usepackage{tensor}
\usepackage{amsthm}
\usepackage{amsmath}
\usepackage{amssymb}
\usepackage{amsfonts}
\usepackage{graphicx}
\usepackage{setspace}
\usepackage{wasysym}
\usepackage{esint}
\usepackage[authoryear]{natbib}
\usepackage[unicode=true,pdfusetitle,
bookmarks=true,bookmarksnumbered=false,bookmarksopen=false,
breaklinks=false,pdfborder={0 0 0},backref=false,colorlinks=false,citecolor=blue]
{hyperref}
\usepackage[printonlyused]{acronym}

\newcommand{\bibfolder}{Bibliography}

\makeatletter

\floatstyle{ruled}
\newfloat{algorithm}{tbp}{loa}[chapter]
\providecommand{\algorithmname}{Algorithm}
\floatname{algorithm}{\protect\algorithmname}

\newenvironment{lyxlist}[1]
{\begin{list}{}
{\settowidth{\labelwidth}{#1}
 \setlength{\leftmargin}{\labelwidth}
 \addtolength{\leftmargin}{\labelsep}
 }}
{\end{list}}
  \theoremstyle{definition}
  \ifx\thechapter\undefined
    \newtheorem{defn}{\protect\definitionname}
  \else
    
  \fi
  \theoremstyle{definition}
  \ifx\thechapter\undefined
    \newtheorem{example}{\protect\examplename}
  \else
    \newtheorem{example}{\protect\examplename}[chapter]
  \fi
  \theoremstyle{plain}
  \ifx\thechapter\undefined
    \newtheorem{fact}{\protect\factname}
  \else
    
  \fi
  \theoremstyle{plain}
  \ifx\thechapter\undefined
    \newtheorem{cor}{\protect\corollaryname}
  \else
    \newtheorem{cor}{\protect\corollaryname}[chapter]
  \fi

\newtheorem{theorem}{Theorem}[chapter]
\newtheorem{lemma}{Lemma}[chapter]
\newtheorem{remark}{Remark}[chapter]
\newtheorem{proposition}{\bf{Proposition}}[chapter]
\newtheorem{definition}{\bf{Definition}}[chapter]
\newtheorem{property}{\bf{Property}}[chapter]
\newtheorem{method}{\bf{Method}}[chapter]
\newtheorem{corollary}{\bf Corollary}[chapter]
\newtheorem{assumption}{\bf Assumption}[chapter]

\makeatletter
\newcommand{\emptypage}[1]{%
	\cleardoublepage
	\begingroup
	\let\ps@plain\ps@empty
	\pagestyle{empty}
	#1
	\cleardoublepage\endgroup}
\makeatletter

\usepackage{microtype}

\usepackage{graphicx}

\usepackage{enumerate}

\usepackage{import}
\usepackage{xcolor}
\usepackage{amsmath}
\usepackage{bm}
\usepackage{dcolumn}
\usepackage{capt-of}
\usepackage{epigraph}
\usepackage{diagbox}
\usepackage{algorithm,algpseudocode}
\usepackage{multirow}

\usepackage{caption}
\usepackage{ae}
\usepackage{amssymb}
\usepackage[nottoc]{tocbibind}

\usepackage{lmodern}
\usepackage[lmodern]{quotchap}
\usepackage{dsfont}

\usepackage{afterpage}

\DeclareMathAlphabet{\mathpzc}{OT1}{pzc}{m}{it}

\DeclareMathSizes{12}{11}{6}{6}
\DeclareMathSizes{11}{11}{6}{6}
\DeclareMathSizes{10}{11}{6}{6}
\DeclareMathSizes{9}{11}{6}{6}
\@ifundefined{showcaptionsetup}{}{%
 \PassOptionsToPackage{caption=false}{subfig}}
\usepackage{subfig}
\makeatother

  \addto\captionsamerican{\renewcommand{\definitionname}{Definition}}
  \addto\captionsamerican{\renewcommand{\examplename}{Example}}
  \addto\captionsamerican{\renewcommand{\factname}{Fact}}
  \addto\captionsbrazil{\renewcommand{\definitionname}{Definição}}
  \addto\captionsbrazil{\renewcommand{\examplename}{Exemplo}}
  \addto\captionsbrazil{\renewcommand{\factname}{Fato}}
  \addto\captionsenglish{\renewcommand{\definitionname}{Definition}}
  \addto\captionsenglish{\renewcommand{\examplename}{Example}}
  \addto\captionsenglish{\renewcommand{\factname}{Fact}}
  \addto\captionsspanish{\renewcommand{\definitionname}{Definición}}
  \addto\captionsspanish{\renewcommand{\examplename}{Ejemplo}}
  \addto\captionsspanish{\renewcommand{\factname}{Hecho}}
  \providecommand{\definitionname}{Definition}
  \providecommand{\examplename}{Example}
  \providecommand{\factname}{Fact}
\addto\captionsamerican{\renewcommand{\corollaryname}{Corollary}}
\addto\captionsbrazil{\renewcommand{\algorithmname}{Algoritmo}}
\addto\captionsbrazil{\renewcommand{\corollaryname}{Corolário}}
\addto\captionsenglish{\renewcommand{\algorithmname}{Algorithm}}
\addto\captionsenglish{\renewcommand{\corollaryname}{Corollary}}
\addto\captionsspanish{\renewcommand{\algorithmname}{Algoritmo}}
\addto\captionsspanish{\renewcommand{\corollaryname}{Corolario}}
\providecommand{\corollaryname}{Corollary}

\newcommand{\mbf}[1]{\ensuremath{{\mathbf{#1}}}}

\newcommand{\ten}[1]{\ensuremath{{\cdot}10^{#1}}}

\newcommand{\frI}{\ensuremath{\mathcal{I}}}
\newcommand{\frL}{\ensuremath{\mathcal{L}}}
\newcommand{\frB}{\ensuremath{\mathcal{B}}}
\newcommand{\frC}[1]{\ensuremath{{\mathcal{C}_{#1}}}}
\newcommand{\frA}[1]{\ensuremath{{\mathcal{A}_{#1}}}}
\newcommand{\RIB}{\ensuremath{\mbf{R}^\mathcal{I}_\mathcal{B}}}
\newcommand{\RIL}{\ensuremath{\mbf{R}^\mathcal{I}_\mathcal{L}}}
\newcommand{\RLA}[1]{\ensuremath{\mbf{R}^\mathcal{L}_{\mathcal{A}_{#1}}}}
\newcommand{\RAB}[1]{\ensuremath{\mbf{R}^{\mathcal{A}_{#1}}_\mathcal{B}}}
\newcommand{\RLB}{\ensuremath{\mbf{R}^\mathcal{L}_\mathcal{B}}}
\newcommand{\RBC}[1]{\ensuremath{\mbf{R}^\mathcal{B}_{\mathcal{C}_{#1}}}}
\newcommand{\RBA}[1]{\ensuremath{\mbf{R}^\mathcal{B}_{\mathcal{A}_{#1}}}}

\newcommand{\RAC}[2]{\ensuremath{\mbf{R}^{\mathcal{A}_{#1}}_{\mathcal{C}_{#2}}}}

\newcommand{\dBC}[1]{\ensuremath{\mbf{d}^\mathcal{B}_{\mathcal{C}_{#1}}}}
\newcommand{\dLA}[1]{\ensuremath{\mbf{d}^\mathcal{L}_{\mathcal{A}_{#1}}}}
\newcommand{\dBA}[1]{\ensuremath{\mbf{d}^\mathcal{B}_{\mathcal{A}_{#1}}}}
\newcommand{\dAB}[1]{\ensuremath{\mbf{d}^{\mathcal{A}_{#1}}_{\mathcal{B}}}}

\newcommand{\dAC}[2]{\ensuremath{\mbf{d}^{\mathcal{A}_{#1}}_{\mathcal{C}_{#2}}}}

\newcommand{\deta}{\ensuremath{\dot{\bm{\eta}}}}
\newcommand{\dgamma}{\ensuremath{\dot{\bm{\gamma}}}}
\newcommand{\gone}{\ensuremath{{\gamma_1}}}

\newcommand{\gtwo}{\ensuremath{{\gamma_2}}}

\newcommand{\aR}{\ensuremath{{\alpha_\text{R}}}}
\newcommand{\daR}{\ensuremath{\dot{\alpha}_\text{R}}}
\newcommand{\aL}{\ensuremath{{\alpha_\text{L}}}}
\newcommand{\daL}{\ensuremath{\dot{\alpha}_\text{L}}}

\newcommand{\wLIL}{\ensuremath{\bm{\omega}^\mathcal{L}_{\mathcal{I}\mathcal{L}}}}

\newcommand{\wALA}[1]{\ensuremath{\bm{\omega}^{\mathcal{A}_{#1}}_{\mathcal{L} \mathcal{A}_{#1}}}}
\newcommand{\wBAB}[1]{\ensuremath{\bm{\omega}^\mathcal{B}_{\mathcal{A}_{#1}\mathcal{B}}}}
\newcommand{\wBLB}{\ensuremath{\bm{\omega}^\mathcal{B}_{\mathcal{L}\mathcal{B}}}}
\newcommand{\wBIB}{\ensuremath{\bm{\omega}^\mathcal{B}_{\mathcal{I}\mathcal{B}}}}
\newcommand{\wABA}[1]{\ensuremath{\bm{\omega}^{\mathcal{A}_{#1}}_{\mathcal{B} \mathcal{A}_{#1}}}}
\newcommand{\wiAi}[1]{\ensuremath{\bm{\omega}^{\mathcal{C}_{#1}}_{\mathcal{A}_{#1} \mathcal{C}_{#1}}}}
\newcommand{\wiBi}[1]{\ensuremath{\bm{\omega}^{\mathcal{C}_{#1}}_{\mathcal{B} \mathcal{C}_{#1}}}}

\newcommand{\mL}{\ensuremath{m_\mathcal{L}}}

\newcommand{\IL}{\ensuremath{\mbf{I}_\mathcal{L}}}
\newcommand{\Ii}[1]{\ensuremath{\mbf{I}_{#1}}}

\newcommand{\half}{\ensuremath{\frac{1}{2}}}

\newcommand{\Weta}{\ensuremath{\mbf{W}_{\bm{\eta}}}}
\newcommand{\ay}{\ensuremath{\mbf{a}_y}}

\newcommand{\eye}[1]{\ensuremath{\mbf{I}_{#1}}}
\newcommand{\eyenoarg}{\ensuremath{\mbf{I}}}
\newcommand{\zeros}[2]{\ensuremath{\bm{0}_{#1\times#2}}}
\newcommand{\ones}[2]{\ensuremath{\bm{1}_{#1\times#2}}}

\newcommand{\ktaub}{\ensuremath{\frac{k_\tau}{b}}}
\newcommand{\lambdaR}{\ensuremath{\lambda}_\text{R}}
\newcommand{\lambdaL}{\ensuremath{\lambda}_\text{L}}

\newcommand{\gforce}{\ensuremath{\bm{\vartheta}}}

\newcommand{\fR}{\ensuremath{\mbf{f}_\text{R}}}
\newcommand{\fL}{\ensuremath{\mbf{f}_\text{L}}}
\newcommand{\tauaR}{\ensuremath{\bm{\tau}_{\aR}}}
\newcommand{\tauaL}{\ensuremath{\bm{\tau}_{\aL}}}

\newcommand{\infR}{\ensuremath{f_\text{R}}}
\newcommand{\infL}{\ensuremath{f_\text{L}}}
\newcommand{\intauaR}{\ensuremath{\tau_{\aR}}}
\newcommand{\intauaL}{\ensuremath{\tau_{\aL}}}

\newcommand{\setreal}{\ensuremath{\mathbb{R}}}
\newcommand{\setrealmat}[2]{\ensuremath{\mathbb{R}^{#1\times#2}}}

\newcommand{\zdomain}{\ensuremath{\varsigma}}
\newcommand{\trace}[1]{\ensuremath{\text{trace}\left({#1}\right)}}

\newcommand{\Achi}{\ensuremath{\mbf{A}_{\bm{\chi}}}}
\newcommand{\Bchi}{\ensuremath{\mbf{B}_{\bm{\chi}}}}
\newcommand{\Fchi}{\ensuremath{\mbf{F}_{\bm{\chi}}}}

\newcommand{\iextension}[1]{\ensuremath{\square \left( #1 \right) }}

\newcommand{\zetaB}{\ensuremath{{\bm{\zeta}_\frB}}}
\newcommand{\phiB}{\ensuremath{{\phi_\frB}}}
\newcommand{\thetaB}{\ensuremath{{\theta_\frB}}}
\newcommand{\psiB}{\ensuremath{{\psi_\frB}}}
\newcommand{\etaB}{\ensuremath{{\bm{\eta}_\frB}}}

\newcommand{\Anu}{\ensuremath{\mbf{A}_{\bm{\nu}}}}
\newcommand{\Bnu}{\ensuremath{\mbf{B}_{\bm{\nu}}}}

\newcommand{\Hnu}{\ensuremath{\mbf{H}_{\bm{\nu}}}}

\newcommand{\setI}{\ensuremath{\mathbb{I}}}

\newcommand{\deps}{\ensuremath{\varepsilon}}
\newcommand{\real}[1]{\ensuremath{\text{Re}(#1)}}
\newcommand{\imag}[1]{\ensuremath{\text{Im}(#1)}}

\newcommand{\realset}{\ensuremath{\mathbb{R}}}
\newcommand{\realsetmat}[2]{\ensuremath{\mathbb{R}^{#1\times#2}}}
\newcommand{\intvalset}{\ensuremath{\mathbb{I}\mathbb{R}}}
\newcommand{\intvalsetmat}[2]{\ensuremath{\mathbb{I}\mathbb{R}^{#1\times#2}}}
\newcommand{\naturalset}{\ensuremath{\mathbb{N}}}

\newcommand{\modelset}{\ensuremath{\mathcal{M}}}

\newcommand{\standardcz}{\ensuremath{Z = \left\{ \mbf{G}, \mbf{c}, \mbf{A}, \mbf{b} \right\}}}
\newcommand{\standardib}{\ensuremath{\mathcal{B} = \cup_{j=1}^{n_b} B_{(j)}}}
\newcommand{\standardref}{\ensuremath{\cup_{j=1}^{n_b} \square(\bar{\mbf{f}}(B_{(j)}))}}
\newcommand{\particularref}{\ensuremath{\cup_{j=1}^{n_b} \square(\bar{\mbf{f}}(B_{(j)},\mbf{u}_{k-1}))}}

\newcommand{\lbound}{\ensuremath{\text{L}}}
\newcommand{\ubound}{\ensuremath{\text{U}}}
\newcommand{\midpoint}[1]{\ensuremath{\text{mid}({#1})}}

\newcommand{\diam}[1]{\ensuremath{\text{diam}({#1})}}
\newcommand{\rad}[1]{\ensuremath{\text{rad}({#1})}}

\newcommand{\ginter}[1]{\ensuremath{\cap_{\mbf{#1}}}}

\newcommand{\gzinclusion}{\ensuremath{\triangleleft}}

\newcommand{\ninf}[1]{\ensuremath{\|{#1}\|_\infty}}
\newcommand{\none}[1]{\ensuremath{\|{#1}\|_1}}

\newcommand{\conball}{\ensuremath{B_\infty(\mathbf{A},\mathbf{b})}}

\newcommand{\xim}{\ensuremath{\bm{\xi}_\text{m}}}
\newcommand{\xir}{\ensuremath{\bm{\xi}_\text{r}}}
\newcommand{\xil}{\ensuremath{\bm{\xi}^\text{L}}}
\newcommand{\xiu}{\ensuremath{\bm{\xi}^\text{U}}}

\newcommand{\arrowmatrix}[3]{\ensuremath{\underset{\xrightarrow[\scriptstyle{#2}]{\hphantom{#1}}}{\left[#1\right]}\left.\!\!\vphantom{#1}\right\downarrow\!{\scriptstyle{#3}}}}

\newcommand{\downarrowmatrix}[2]{\ensuremath{{\left[#1\right]}\left.\!\!\vphantom{#1}\right\downarrow\!{\scriptstyle{#2}}}}

\newcommand{\alamobravo}{ZMV}
\newcommand{\combastelbravo}{ZFO}

\newcommand{\dsum}[2]{\ensuremath{{\displaystyle\sum_{#1}^{#2}}}}
\newcommand{\dprod}[2]{\ensuremath{{\displaystyle\prod_{#1}^{#2}}}}

\newcommand{\seq}[1]{\ensuremath{{\protect\overset{\to}{#1}}}} 

\newcommand{\zerospace}{\ensuremath{{\,\!}}}
\newcommand{\nspace}{\ensuremath{\zerospace}}
\newcommand{\zspace}{\ensuremath{\zerospace}}
\newcommand{\inv}[1]{\ensuremath{{#1}\zerospace^{-1}}}

\newcommand{\noarg}{\ensuremath{\_\,}}
\newcommand{\emptyarg}{\noarg}

\allowdisplaybreaks

\begin{document}

\newcommand{\mytitlename}{Set-based state estimation and fault diagnosis using constrained zonotopes and applications}

\newcommand{\myauthorname}{Brenner Santana Rego}

\newcommand{\myadvisorname}{Guilherme Vianna Raffo}

\newcommand{\mycoadvisorname}{Davide Martino Raimondo}

\newcommand{\mydedication}{

To Maria José Rego

and Osvaldo Santana Ferreira

}

\newcommand{\myacknowledgments}{

To my parents Maria José Rego and Osvaldo Santana Ferreira, and also to my closest brother Felipe Santana Rego, for supporting me and cheering me up throughout this path.

To my advisor Prof. Guilherme Vianna Raffo, for accepting me as his student, guiding me and also supporting me. Also to my advisor Prof. Davide Martino Raimondo for the essential support and collaboration up to this moment. In addition, to Prof. Joseph K. Scott, who played a very important role in most of the contributions presented in this thesis.

To my colleagues and ex-colleagues from MACRO, ProVANT, MACSIN, and other laboratories at the Federal University of Minas Gerais, for their friendship and also for supporting me when going through hard times. A special thanks to my friends from the Identification and Control of Dynamic Systems Laboratory at the University of Pavia, for their cordiality and support during my stay in Italy for the doctoral sandwich program. I prefer to not list any names here, so that I can thank everyone in an equal manner.

To the funding agencies CAPES, CNPq, FAPEMIG, and the InSAC, for the financial support, which was essential for the ellaboration of this thesis.

To the Graduate Program in Electrical Engineering for providing the infrastructure required for the ellaboration of this thesis.
}

\newcommand{\portugueseabstract}{

\selectlanguage{brazil}%
Esta tese de doutorado desenvolve novos métodos para estimação de estados baseada em conjuntos e diagnóstico ativo de falhas de (i) sistemas não lineares em tempo discreto com incertezas limitadas, (ii) sistemas não lineares em tempo discreto cujas trajetórias satisfazem restrições de igualdade não linear (chamadas de invariantes), (iii) estimação de estados baseada em conjuntos e diagnóstico de falha ativa de sistemas descritores lineares, e (iv) estimação conjunta de estados e parâmetros de sistemas descritores não lineares. Estimação baseada em conjuntos tem como objetivo obter envoltórios precisos para os possíveis estados do sistema em cada instante de tempo, sujeito a incertezas desconhecidas-porém-limitadas. A maioria dos métodos existentes emprega uma estrutura predição-atualização padrão, com etapas de predição e atualização baseadas em diferentes técnicas e representações de conjunto. No entanto, obter envoltórios precisos para sistemas não lineares continua sendo um desafio significativo. Quando esses envoltórios são representados por conjuntos simples, como intervalos, elipsóides, paralelotopos e zonotopos, certas operações de conjunto podem ser muito conservadoras. Ainda, a utilização de politopos convexos generalizados é muito mais exigente do ponto de vista computacional. Para resolver este problema, esta tese de doutorado propõe novos métodos para propagar, com eficiência, zonotopos restritos (CZs) por meio de mapeamentos não lineares. Estes estendem métodos existentes baseados em zonotopos de maneira consistente. Além disso, esta tese de doutorado melhora a estrutura de predição-atualização padrão para sistemas com invariantes, adicionando uma etapa de consistência. Esta nova etapa utiliza invariantes para reduzir o conservadorismo do envoltório e é habilitada por novos algoritmos para refinar zonotopos restritos com base em restrições não lineares. Novos algoritmos de atualização são também desenvolvidos permitindo o uso de equações de medição não linear pela primeira fez em estimação de estados baseada em CZ. Além disso, esta tese de doutorado introduz uma nova abordagem para o diagnóstico ativo de falhas baseado em conjuntos para uma classe de sistemas não lineares em tempo discreto. Os novos envoltórios baseados em zonotopos restritos são utilizados para detecção passiva de falhas de uma classe de sistemas não lineares em tempo discreto. Ainda, uma parametrização afim dos conjuntos alcançáveis também é obtida, na direção do projeto de uma entrada ótima para a realização de diagnóstico ativo de falhas baseado em conjuntos para sistemas não lineares. Além disso, esta tese apresenta novos métodos baseados em zonotopos restritos para estimação de estados e diagnóstico ativo de falhas de sistemas descritores lineares. Em contraste com outras representações de conjuntos, restrições estáticas lineares nas variáveis de estado, típicas de sistemas descritores, podem ser incorporadas diretamente na descrição matemática de zonotopos restritos. Graças a esse recurso, os métodos baseados em CZs são capazes de fornecer envoltórios menos conservadores para as trajetórias de sistemas descritores lineares. Além disso, esta tese propõe uma nova representação para conjuntos ilimitados baseada em zonotopos, que permite desenvolver métodos de estimação de estados e diagnóstico ativo de falhas de sistemas descritores lineares sem assumir a existência de um conjunto admissível que envolva todas as trajetórias possíveis do sistema, e permitindo lidar com sistemas instáveis. 
Esta tese também desenvolve um novo método para o estimação conjunta de estados e parâmetros para sistemas descritores não lineares, que estende os métodos de estimação de estados não linear usando zonotopos restritos para incluir a estimação de parâmetros em uma estrutura unificada e, portanto, mantendo as dependências existentes entre os estados, variáveis algébricas e parâmetros desconhecidos, o que resulta em tanto a estimação de estados, quanto de parâmetros, sendo significativamente melhoradas. Por último, este texto apresenta a aplicação dos métodos propostos de estimação de estados e diagnóstico de falhas baseados em zonotopos restritos para veículos aéreos não tripulados, redes de distribuição de água e uma célula de íons de lítio. 

\textit{Palavras-chave:} Estimação de estados baseada em conjuntos, Diagnóstico de falhas, Zonotopos restritos.
\selectlanguage{american}%
}

\newcommand{\englishabstract}{
This doctoral thesis develops new methods for set-based state estimation and active fault diagnosis of (i) nonlinear discrete-time systems with bounded uncertainties, (ii) discrete-time nonlinear systems whose trajectories are known to satisfy nonlinear equality constraints (called invariants), (iii) set-based state estimation and active fault diagnosis of linear descriptor systems, and (iv) joint state and parameter estimation of nonlinear descriptor systems. Set-based estimation aims to compute tight enclosures of the possible system states in each time step subject to unknown-but-bounded uncertainties. Most existing methods employ a standard prediction-update framework with set-based prediction and update steps based on various set representations and techniques. However, achieving accurate enclosures for nonlinear systems remains a significant challenge. When these enclosures are represented by simple sets such as intervals, ellipsoids, parallelotopes, and zonotopes, certain set operations can be very conservative. Yet, using general convex polytopes is much more computationally demanding. To address this issue, the present doctoral thesis proposes new methods for efficiently propagating constrained zonotopes (CZs) through nonlinear mappings. These extend existing methods for zonotopes in a consistent way. Besides, this thesis improves the standard prediction-update framework for systems with invariants by adding a consistency step. This new step uses invariants to reduce conservatism and is enabled by new algorithms for refining constrained zonotopes based on nonlinear constraints. New update algorithms are also developed allowing nonlinear measurement equations in CZ-based state estimation for the first time. In addition, this doctoral thesis introduces a new approach for set-based active fault diagnosis of a class of nonlinear discrete-time systems. The new enclosures based on constrained zonotopes are employed for passive fault detection of a class of nonlinear discrete-time systems. Also, an affine parametrization of the reachable sets is obtained, in the direction of the design of an optimal input for set-based active fault diagnosis of nonlinear systems. In addition, this thesis presents new methods based on constrained zonotopes for set-valued state estimation and active fault diagnosis of linear descriptor systems. In contrast to other sets representations, linear static constraints on the state variables, typical of descriptor systems, can be directly incorporated in the mathematical description of constrained zonotopes. Thanks to this feature, set-based methods based on CZs can provide less conservative enclosures for the trajectories of linear descriptor systems. Moreover, this thesis proposes a new representation for unbounded sets based on zonotopes, which allows to develop methods for state estimation and active fault diagnosis of linear descriptor systems without assuming the knowledge of an admissible set that encloses all the possible trajectories of the system, and allowing to deal with unstable systems. This thesis also develops a new method for set-based joint state and parameter estimation of nonlinear discrete-time systems, which extends the nonlinear state estimation methods using constrained zonotopes to include parameter estimation in a unified framework, and therefore maintaining existing dependencies between states, algebraic variables, and unknown parameters, which results in the accuracy of both state and parameter estimation being significantly improved. Lastly, this manuscript presents the application of the proposed set-based state estimation and fault diagnosis methods using constrained zonotopes to unmanned aerial vehicles, water distribution networks, and a lithium-ion cell.	

\textit{Keywords:} Set-based state estimation, Fault diagnosis, Constrained zonotopes.
}

\thispagestyle{empty}
\pagenumbering{arabic}
\setcounter{page}{1}

\includegraphics[scale=0.145]{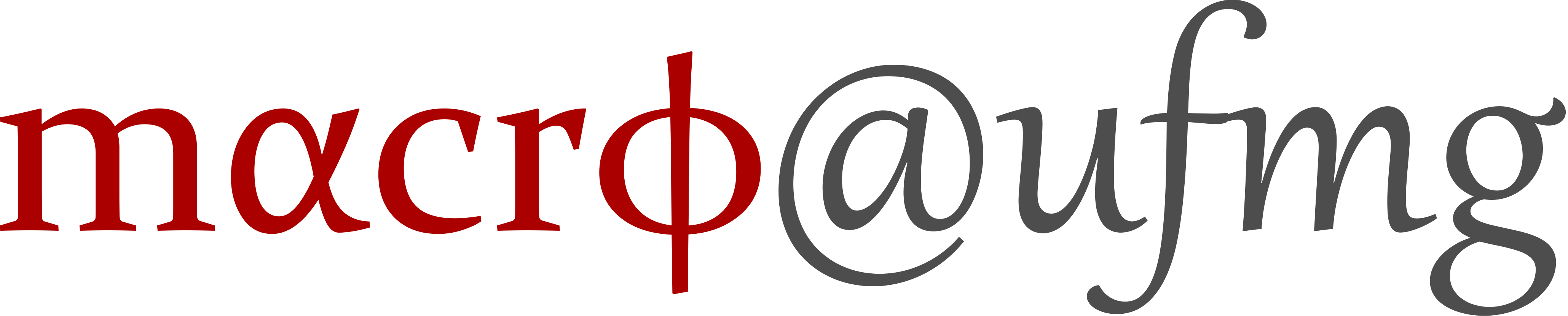}

{\small{}Universidade Federal de Minas Gerais}{\small \par}

{\small{}Programa de Pós-Graduação em Engenharia Elétrica}{\small \par}

{\small{}Research group MACRO - Mechatronics, Control and Robotics}{\small \par}

\vspace*{6.5cm}

\begin{center}
\bf{\Large{}\expandafter\MakeUppercase\expandafter{\mytitlename}}\vfill{}
	
\par\end{center}

\begin{center}
\textbf{\myauthorname}\\
Belo Horizonte, Brazil\\
2021
\par\end{center}

\pagebreak{}

\pagestyle{plain}

\pagestyle{empty}

\begin{center}
\textbf{\myauthorname}
\par\end{center}

\vfill{}

\begin{center}
\bf{\Large{}\expandafter\MakeUppercase\expandafter{\mytitlename}}
\par\end{center}{\Large \par}

\vfill{}

\begin{flushright}
\parbox[t]{0.6\columnwidth}{%
Thesis submitted to the Graduate Program in Electrical Engineering
of Escola de Engenharia at the Universidade Federal de Minas Gerais,
in partial fulfillment of the requirements for the degree of Doctor
in Electrical Engineering.

\vspace{1cm}



\ifdefined\mycoadvisorname%
\begin{tabular}{l l}
\textbf{Advisors:} & \myadvisorname \\%
& \mycoadvisorname
\end{tabular}
\else
\textbf{Advisor:} \myadvisorname%
\fi
}
\par\end{flushright}

\vspace*{\fill}

\begin{center}
Belo Horizonte, Brazil\\
2021
\par\end{center}


\includepdf{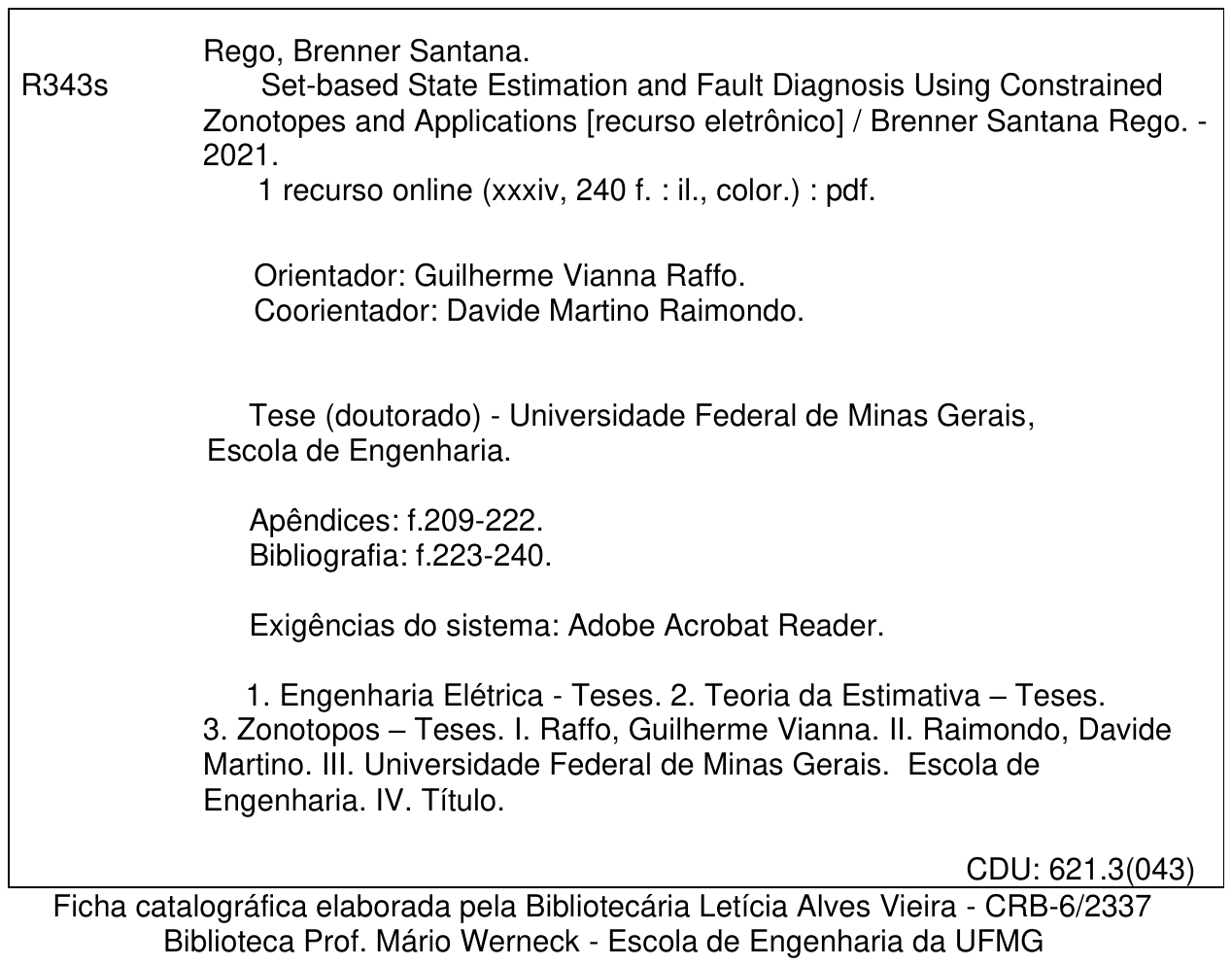}
\includepdf{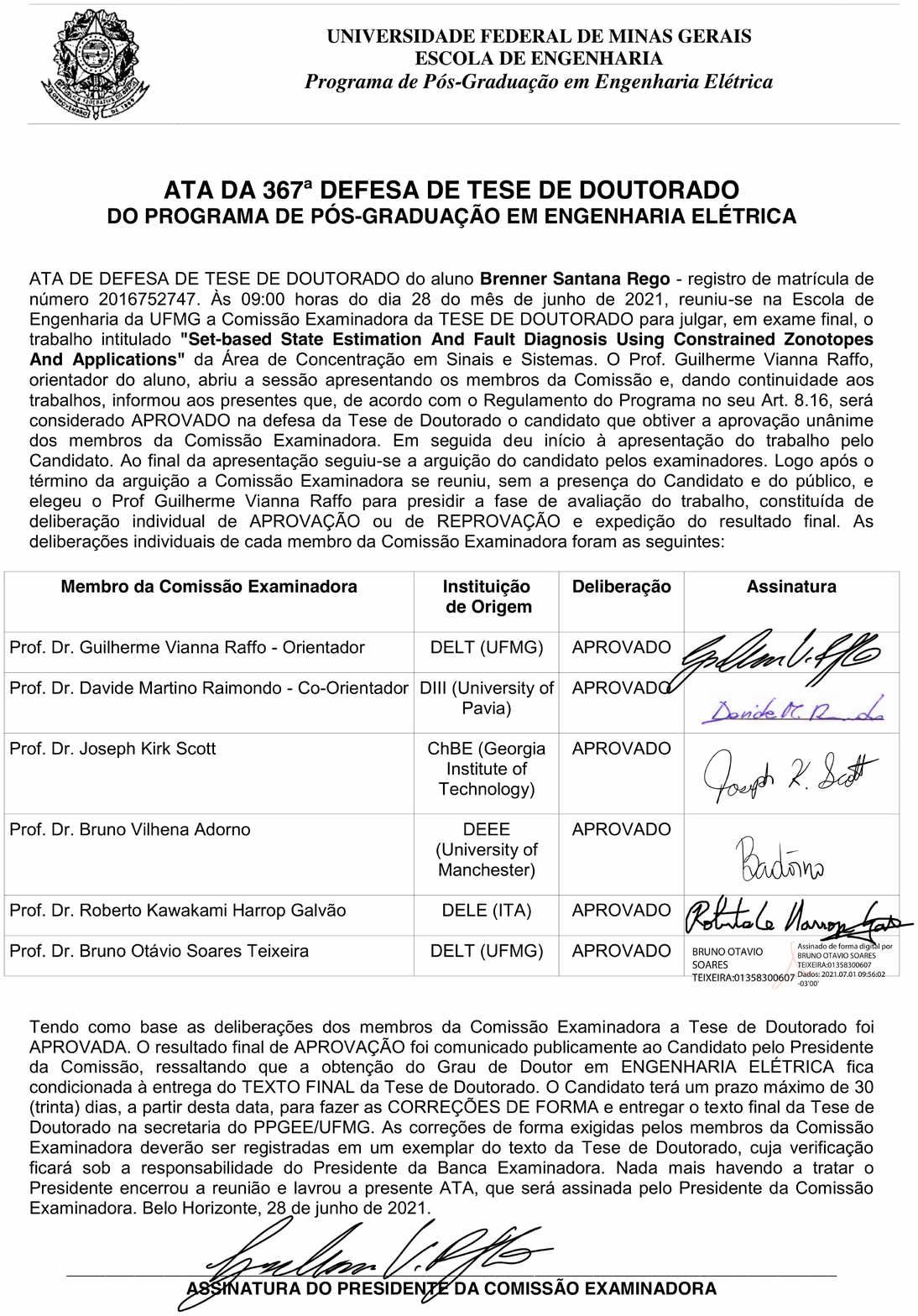}
%
\begin{singlespace}
	\pagebreak{}
\end{singlespace}

\selectlanguage{spanish}%
\textbf{\thispagestyle{empty}}

\vfill{}

\begin{flushright}
\begin{minipage}[t]{0.4\columnwidth}%
\selectlanguage{american}%
\begin{singlespace}
\begin{flushright}
\emph{\mydedication}
\par\end{flushright}\end{singlespace}
\selectlanguage{american}%
\end{minipage}
\par\end{flushright}

\vspace{3cm}

\selectlanguage{american}%
\begin{singlespace}
\pagebreak{}
\end{singlespace}

\chapter*{Acknowledgements}\thispagestyle{empty}

\myacknowledgments

\begin{singlespace}
\pagebreak{}
\end{singlespace}


\chapter*{Resumo}\thispagestyle{empty}

\portugueseabstract

\begin{onehalfspace}
	\pagebreak{}
\end{onehalfspace}

\selectlanguage{english}%

\chapter*{Abstract}\thispagestyle{empty}

\selectlanguage{american}%
\englishabstract

\pagebreak{}


%
%
%
%
%
%

\emptypage{\listoffigures}

\emptypage{\listoftables}

\selectlanguage{spanish}%

\chapter*{{\normalsize{}\addcontentsline{toc}{chapter}{Acronyms}}\foreignlanguage{american}{Acronyms}}\thispagestyle{empty}

\selectlanguage{american}%
\begin{lyxlist}{00.00.0000}
\item[{AFD}] Active Fault Diagnosis
\item[{ARR}] Average Radius Ratio
\item[{BMS}] Battery Management System
\item[{CAD}] Computer Aided Design
\item[{CG-rep}] Constrained Generator representation
\item[{CZ}] Constrained Zonotope
\item[{CZFO}] First-Order extension using Constrained Zonotopes
\item[{CZFO+C}] First-Order extension using Constrained Zonotopes with Consistency step
\item[{CZFO+F}] First-Order extension using Constrained Zonotopes considering Feasible set
\item[{CZFO+FC}] First-Order extension using Constrained Zonotopes considering Feasible set and with Consistency step
\item[{CZFO-J}] First-Order extension using Constrained Zonotopes, Joint approach
\item[{CZFO-N}] First-Order extension using Constrained Zonotopes, Naive approach
\item[{CZIB}] Constrained Zonotope and Interval Bundle
\item[{CZMV}] Mean Value extension using Constrained Zonotopes
\item[{CZMV+C}] Mean Value extension using Constrained Zonotopes with Consistency step
\item[{CZMV+F}] Mean Value extension using Constrained Zonotopes considering Feasible set
\item[{CZMV+FC}] Mean Value extension using Constrained Zonotopes considering Feasible set and with Consistency step
\item[{CZMV-J}] Mean Value extension using Constrained Zonotopes, Joint approach
\item[{CZMV-N}] Mean Value extension using Constrained Zonotopes, Naive approach
\item[{DC}] Difference of Convex functions
\item[{DFN}] Doyle-Fuller-Newman
\item[{DMA}] District Metered Area
\item[{ECM}] Electrical Circuit Model
\item[{EM}] Electrochemical Model
\item[{FBCP}] Forward-Backward Constraint Propagation
\item[{FD}] Fault Diagnosis
\item[{FIM}] Fisher Information Matrix
\item[{FTC}] Fault-Tolerant Control
\item[{FVM}] Finite Volume Method
\item[{G-rep}] Generator representation
\item[{GPS}] Global Positioning System
\item[{GPU}] Graphics Processing Unit
\item[{H-rep}] Half-space representation
\item[{IHISE}] Iterative Hydraulic Interval State Estimation
\item[{IHISE-CZ}] Iterative Hydraulic Interval State Estimation with Constrained Zonotope step
\item[{IMU}] Inertial Measurement Unit
\item[{ISAR}] Iterative Slope Approximation and Rescaling
\item[{ISAR-CZ}] Iterative Slope Approximation and Rescaling with Constrained Zonotope step
\item[{LDS}] Linear Descriptor Systems
\item[{LMI}] Linear Matrix Inequality
\item[{LP}] Linear Program
\item[{LPV}] Linear Parameter Varying
\item[{MCS}] Monte-Carlo Simulation
\item[{MIQP}] Mixed-Integer Quadratic Programming
\item[{MG-rep}] Mixed Generator representation
\item[{MPC}] Model Predictive Control
\item[{MPT}] Multi-Parametric Toolbox
\item[{MZ}] Mixed Zonotope
\item[{OCP}] Open Circuit Potential
\item[{ODE}] Ordinary Differential Equation
\item[{PDE}] Partial Differential Equation
\item[{SOC}] State Of Charge
\item[{SPMe}] Single Particle Model with electrolyte dynamics
\item[{SVD}] Singular Value Decomposition
\item[{UAV}] Unmanned Aerial Vehicle
\item[{UKF}] Unscented Kalman Filter
\item[{V-rep}] Vertex Representation
\item[{WDN}] Water Distribution Network
\item[{ZFO}] First-Order extension using Zonotopes
\item[{ZFO+F}] First-Order extension using Zonotopes with Feasible set
\item[{ZMV}] Mean Value extension using Zonotopes
\item[{ZMV+F}] Mean Value extension using Zonotopes with Feasible set

\end{lyxlist}

\pagebreak{}

\selectlanguage{spanish}%

\chapter*{{\normalsize{}\addcontentsline{toc}{chapter}{Notation}}\foreignlanguage{american}{Notation}}\thispagestyle{empty}

\selectlanguage{american}%
\begin{lyxlist}{1000000.00.0000}
\item[\bf General notation]
\item[$a$] Italic lower case letters denote scalars
\item[$\mbf{a}$] Boldface lower case letters denote vectors
\item[$\mbf{A}$] Boldface upper case letters denote matrices
\item[\bf General symbols and operators]
\item[$\naturalset$] Set of natural numbers
\item[$\naturalset^+$] Set of natural numbers with zero excluded
\item[$\realset$] Set of real numbers
\item[$\intvalset$] Set of real compact intervals
\item[$\zeros{n}{m}$] Matrix of zeros with $n$ rows and $m$ columns
\item[$\ones{n}{m}$] Matrix of ones with $n$ rows and $m$ columns
\item[$\eye{n}$] Identity matrix with dimension $n$
\item[{$\dot{\mbf{A}}$}] Time derivative of $\mbf{A}$
\item[{$\mbf{A}^T$}] Transpose of $\mbf{A}$
\item[{$\mbf{A}^{-1}$}] Inverse of $\mbf{A}$
\item[{$\mbf{A}^+$}] Pseudo-inverse of $\mbf{A}$
\item[{$\mbf{a}_i$}] $i$-th component of $\mbf{a}$
\item[{$\mbf{A}_{i,:}$}] $i$-th row of $\mbf{A}$
\item[{$\mbf{A}_{:,j}$}] $j$-th column of $\mbf{A}$
\item[{$\trace{\mbf{A}}$}] Trace of $\bm{A}$
\item[{$\text{diag}(\mbf{a})$}] Diagonal matrix whose diagonal is defined by $\mbf{a}$
\item[{$\text{blkdiag}(\mbf{A},\ldots)$}] Block diagonal matrix with blocks $\mbf{A},\ldots$.
\item[$\seq{\mbf{a}}$] Sequence of $\mbf{a}$ defined over a given time interval
\item[{$Z^+$}] Lifted zonotope for $Z$
\item[$\midpoint{\mbf{A}}$] Midpoint of $\mbf{A}$
\item[$\diam{\mbf{A}}$] Diameter of $\mbf{A}$
\item[$\text{rad}(\mbf{A})$] Radius of $\mbf{A}$
\item[$\nabla_x \mbf{f}$] Gradient of $\mbf{f}$ with respect to $\mbf{x}$
\item[$\mbf{H}_x \mbf{f}$] Upper triangular matrix denoting half the Hessian of $\mbf{f}$ with respect to $\mbf{x}$
\item[$\square(\mbf{f}(\cdot))$] Inclusion function of $\mbf{f}$
\item[$\square X$] Interval hull of $X$
\item[$\oplus$] Minkowski sum
\item[$\ginter{\mbf{A}}$] Generalized intersection with linear mapping $\mbf{A}$
\item[$\ginter{\mbf{A}_1,\mbf{A}_2}$] Two-side generalized intersection with linear mappings $\mbf{A}_1$ and $\mbf{A}_2$
\item[$\gzinclusion(\cdot,\cdot)$] CZ-inclusion operator
\item[$\{\mbf{G},\mbf{c}\}$] Zonotope with center $\mbf{c}$ and generators $\mbf{G}$
\item[$\{\mbf{G},\mbf{c},\mbf{A},\mbf{b}\}$] Constrained zonotope with center $\mbf{c}$, generators $\mbf{G}$, and constraints $(\mbf{A},\mbf{b})$
\item[$\{\mbf{M},\mbf{G},\mbf{c},\mbf{S},\mbf{A},\mbf{b}\}$] \hspace{-1mm}Mixed zonotope with center $\mbf{c}$, bounded generators $\mbf{G}$, unbounded generators $\mbf{M}$, and constraints $(\mbf{S},\mbf{A},\mbf{b})$
\selectlanguage{american}%
\end{lyxlist}

\pagebreak{}

\emptypage{\tableofcontents{}}

\pagestyle{headings}



\chapter{Introduction}\thispagestyle{headings} \label{cha:introduction}

\section{Motivation}

In view of the required reliability in many applications, the development of set-based methods has gained attention in the last decades in a wide range of applications \citep{Jaulin2001,Blanchini2015}. These are usually characterized by their robustness and reliability, since the obtained results are often guaranteed in a formal manner. These methods are based on computation with sets, either under set-based concepts such as positive invariance, set-membership methodology, or through numerical tools such as interval arithmetic \citep{Moore2009}. Recent works address the problems of robust control \citep{Le2011,Polyakov2013,Ping2015}, robust state estimation \citep{Bars2012,Jaulin2012,Jaulin2016,Mazenc2013,Chabane2014b,Chabane2014c,Chabane2014a}, parameter identification \citep{Herrero2016}, robot navigation \citep{Jaulin2013}, robot localization \citep{Neuland2014,Yu2016}, and collision avoidance \citep{Zhou2015}, for instance. 

\subsection{State estimation}

In recent decades, the importance of state estimation has gained attention in many fields of research \citep{Simon2006}. This includes a wide range of applications such as state-feedback control \citep{Jaulin2009,Goodarzi2017,Rego2019}, fault diagnosis \citep{Zhang2008,Combastel2015,Raimondo2016}, and robot localization \citep{Saeedi2016}. In contrast to Bayesian strategies such as Kalman filtering \citep{Teixeira2009,Simon2010}, set-valued state estimation methods aim to provide guaranteed enclosures of the system trajectories in applications affected by unknown-but-bounded uncertainties, without assuming knowledge of their stochastic properties \citep{Schweppe1968,Chisci1996}. %

The set-based state estimation problem has been extensively studied for linear discrete-time systems. The pioneering methods make use of ellipsoids to bound the trajectories of the system \citep{Schweppe1968}. Other classical methods propose recursive state estimation algorithms based on parallelotopes \citep{Chisci1996} and also zonotopes \citep{Combastel2003}. More recent works combine different classes of sets for set-valued estimation, such as ellipsoids and zonotopes \citep{Chabane2014}, and even combinations of zonotopic methods with Kalman filtering can be found \citep{Combastel2015}. However, since these sets are not closed in every set operation that arises in the state estimation problem, recently, \cite{Scott2016} use a generalization of zonotopes, called \emph{constrained zonotopes}, to overcome most of the difficulties encountered by the former algorithms. %
Whereas the state estimation problem is already consolidated for linear systems, nonlinear state estimation is still an open field, both for stochastic or set-based approaches. The exact characterization of sets containing the evolution of the system states is very difficult in the nonlinear case, if not intractable \citep{Kieffer1998,Kuhn1998,Platzer2007}. Therefore, in the set-membership framework the objective is to enclose such sets as tightly as possible by guaranteed outer bounds on the possible trajectories of the system states. Such outer bounds must be consistent with the previous estimate, known inputs, the current measurement, the bounds on the disturbances and uncertainties, and also bounds on the initial states. 

However, achieving accurate enclosures for nonlinear systems remains a significant challenge. When these enclosures are represented by simple sets such as intervals, ellipsoids, parallelotopes, and zonotopes, certain set operations can be very conservative. Yet, using general convex polytopes is much more computationally demanding. 
A common tool for computing these enclosures for nonlinear systems is \emph{interval analysis}. With a wide range of applications, such as global optimization, parameter estimation and robust control \citep{Jaulin2001}, interval analysis can be used to generate guaranteed bounds on the range of real valued functions, through interval extensions \citep{Moore2009}. However, severe overestimation often occurs due to interval dependency and the wrapping effect \citep{Kuhn1998}. The latter is  associated to the inability of intervals to capture dependencies between variables resulting from multivariate mappings, while the former arises from multiple occurrences of an interval variable in the same algebraic expression. Such problems are mitigated by computing unions of interval extensions over a partitioned domain (\emph{refinements}). Nonetheless, set inversion algorithms often are required to take measurements into account in interval-based observers, which can be very expensive \citep{Jaulin2009}.

\subsection{Fault diagnosis}

The requirements on reliability and safety of control systems under actuator and sensor malfunctions motivated the development of Fault-Tolerant Control (FTC) strategies since the 70's \citep{Montgomery1976a,Montgomery1976b,Chizeck1978}. Several techniques were developed to cope with faults in a great variety of systems and fault scenarios since then \citep{Zhang2008}. The purpose of such strategies is to accommodate component malfunctions by maintaining the overall system stable with acceptable closed-loop performance in fault scenarios, changing control objectives if necessary to cope with physical limitations imposed by the faults, while performing the original control task for the nominal plant. In particular, these control strategies have been of utmost importance in safety critical systems, such as nuclear power plants and aerospace systems, in which minor faults in components may lead to catastrophic consequences if not dealt with appropriately \citep{Hatami2016,Castaldi2014}.

Existing FTC strategies can be classified into two main categories: passive fault-tolerant control and active fault-tolerant control \citep{Zhang2008,Fekih2014}. Passive strategies are based on fixed robust controllers, designed to ensure that the closed-loop system remains insensitive to a predetermined repertory of faults. These strategies are characterized by their reduced complexity and are independent of Fault Diagnosis (FD) techniques. However, these methods are very conservative and capable of accommodating only a greatly limited number of fault scenarios. Examples of passive FTC strategies are based on linear quadratic control \citep{Hsieh2002}, sliding mode control \citep{Alwi2008,Hu2010,Shen2015}, linear $\mathcal{H}_\infty$ control \citep{Tao2015}, and also fuzzy $\mathcal{H}_\infty$ control \citep{Li2012}.

In view of the required reliability in FTC systems, the development of set-based fault diagnosis (FD) methods has gained attention in the last decades \citep{Xu2014,Raimondo2016}. Fault diagnosis aims to determine exactly which fault a process is subject to. Set-based FD methods are usually characterized by robustness and reliability, as diagnosis is often guaranteed. These methods are based on computation with sets, either under concepts such as positive invariance, consistency tests and set separation. Passive fault detection methods are quite common in the literature \citep{Guerra2008,Blesa2012,Xu2014,Vento2015}; however, passive fault detection alone cannot provide information on which fault occurred, in order to mitigate any harmful effects resulting from the fault. On the other hand, active fault diagnosis (AFD) methods allow to determine which fault has occurred, usually by the injection of an optimal input sequence in order to identify the fault \citep{Scott2014}. Set-based fault diagnosis often relies on the separation of reachable sets for models describing all the possible faulty scenarios. Therefore, high conservatism in the computation of these enclosures results in poor performance of the fault diagnosis procedure. Reduction of computational burden is another investigated issue in the literature, for instance by using offline computations based on multi-parametric programming \citep{Marseglia2017b}.

\subsection{Descriptor systems}

Another object of application of set-based state estimation and fault diagnosis involves many physical processes such as battery packs, robotic systems with holonomic and nonholonomic constraints, and socioeconomic systems \citep{JAN11,Yang2019}, which exhibit static relations between their internal variables. These processes are known as descriptor systems (or implicit systems), which have generalized dynamic and static behaviors described through differential and algebraic equations, respectively~\citep{Puig2018}. Descriptor systems appear in several contexts, such as linear control \citep{Laub1987}, fault-tolerant control \citep{Shi2014}, and fault diagnosis \citep{Wang2019b}. However, few strategies can deal effectively with state estimation and fault diagnosis of descriptor systems when uncertainties with unknown probability distribution are present \citep{Hamdi2012}.

Set-based strategies for state estimation of discrete-time descriptor systems with unknown-but-bounded uncertainties is a recent subject, often addressed using intervals, zonotopes and ellipsoids \citep{Efimov2015,Puig2018,Wang2018,Merhy2019}. Interval methods are used in \cite{Efimov2015} to design a state estimator for time-delay descriptor systems based on linear matrix inequalities and the Luenberger structure. However, interval arithmetics can lead to conservative enclosures due to the wrapping effect. A few methods are proposed in \cite{Puig2018} for set-based state estimation of linear descriptor systems by enclosing the intersection of two consistent sets with a zonotope bound. Nevertheless, since zonotopes cannot effectively capture the algebraic constraints (which are typical of descriptor systems), and also the intersection cannot be computed exactly, the resulting bound is conservative.

\section{Justification}

As commented in the previous section, set-based methods have gained attention in the last decades in a wide range of applications due to their robustness and reliability, since the obtained results are often guaranteed. Nevertheless, while set-based state estimation and fault diagnosis are relatively consolidated for linear systems, set-based state estimation and fault diagnosis of nonlinear discrete-time systems is still an open field. 

Set-based estimation aims to compute tight enclosures of the possible system states in each time step subject to unknown-but-bounded uncertainties. Most existing methods employ a standard prediction-update framework with set-based prediction and update steps based on various set representations and techniques. However, achieving accurate enclosures for nonlinear systems remains a significant challenge. When these enclosures are represented by simple sets such as intervals, ellipsoids, parallelotopes, and zonotopes, certain set operations can be very conservative. Yet, using general convex polytopes is much more computationally demanding. In addition, the set-based fault diagnosis considered in this doctoral thesis relies on the separation of reachable sets for all the possible models describing faulty scenarios. Therefore, high conservatism in the computation of these enclosures result in poor performance of the fault diagnosis procedure.

In view of the few available methods in the literature, and considering that both set-based state estimation and fault diagnosis are still open fields for several classes of discrete-time dynamical systems, this doctoral thesis focuses on the investigation of new methods for set-based state estimation and fault diagnosis. These explore the benefits of using constrained zonotopes as main set representation, which already presented advantages in set-based linear state estimation in previous works in comparison to other simple set representations.

\section{Objectives}

The main objective of this doctoral thesis is to investigate the development of new methods for set-based state estimation and fault diagnosis for a few classes of discrete-time dynamical systems, exploring the benefits of using constrained zonotopes as main set representation. 

\subsection{Specific Objectives}
Specifically, it is intended to investigate the following issues:
\begin{enumerate}
	\item Development of set-based state estimation methods for a class of discrete-time systems with nonlinear dynamics using constrained zonotopes. 
	\item Development of set-based state estimation methods for a class of discrete-time systems with nonlinear dynamics, nonlinear measurement, and nonlinear invariants, using constrained zonotopes. 	
	\item Development of set-based fault diagnosis methods for a class of discrete-time systems with nonlinear dynamics using constrained zonotopes. 		
	\item Development of set-based state estimation and fault diagnosis methods for linear descriptor systems using constrained zonotopes.
	\item Development of a new set representation to describe unbounded sets, aiming the design of new methods for set-based state estimation and fault diagnosis for linear descriptor systems.
	\item Development of set-based joint state and parameter estimation methods using constrained zonotopes for a class of nonlinear systems with nonlinear dynamics, nonlinear measurement, and nonlinear algebraic constraints.
	\item Application of the developed strategies in practical scenarios. In this thesis, unmanned aerial vehicles, water distribution networks, and a Lithium-ion cell, were considered.
\end{enumerate}

\section{Structure of the text}

\begin{figure}[!htb]
	\begin{footnotesize}
		\centering{
			\includegraphics[width=\textwidth]{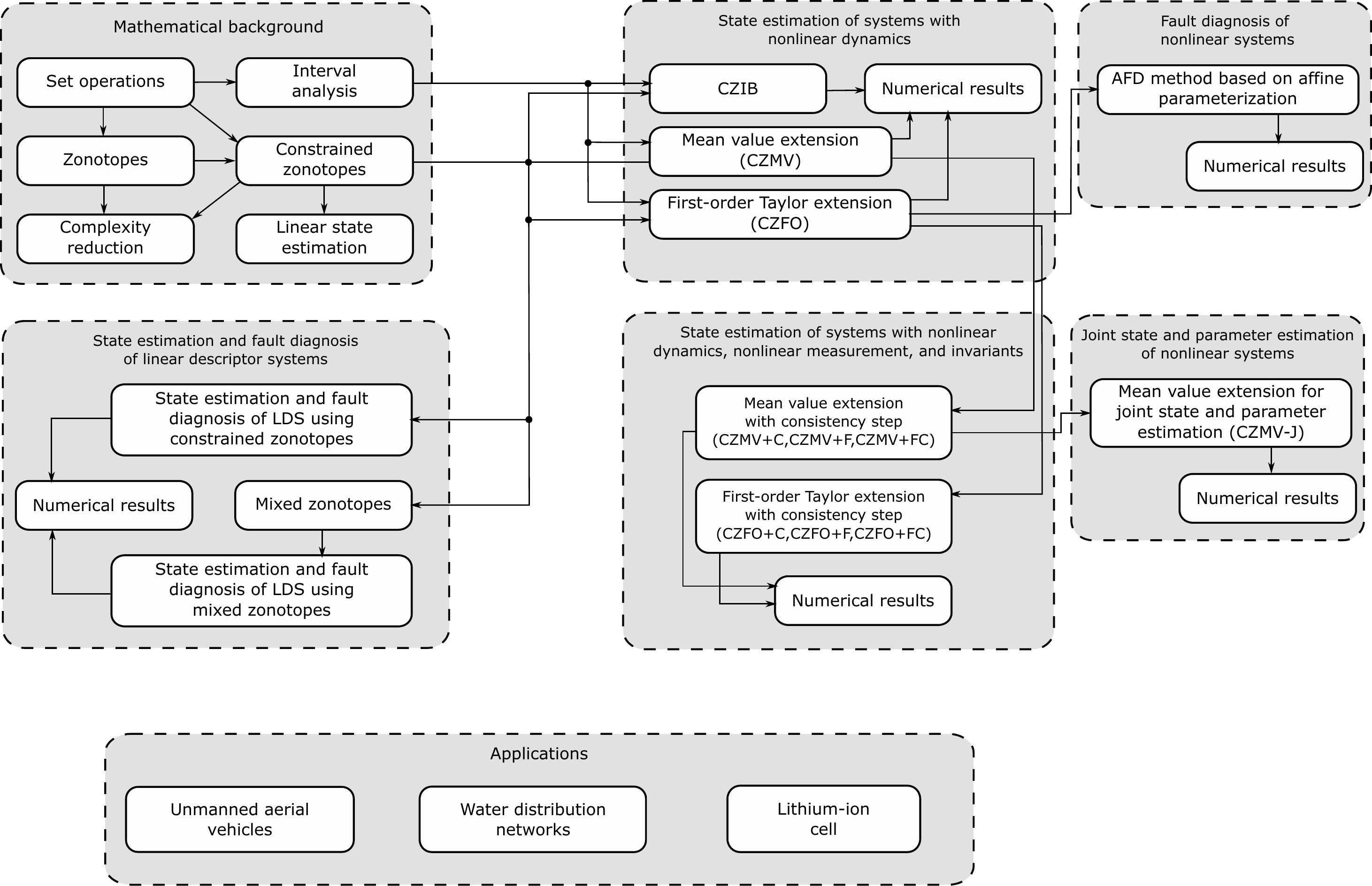}
			\caption{Flow chart of this doctoral thesis.}\label{fig:introduction_flowchart}}
	\end{footnotesize}
\end{figure}

Figure \ref{fig:introduction_flowchart} shows a flow chart used in the development of this doctoral thesis. This doctoral thesis is organized as follows:

\begin{itemize}
	\item \textbf{Chapter 2} presents a literature review highlighting the state of the art in set-based state estimation and fault diagnosis of nonlinear discrete-time systems and descriptor systems.
	\item \textbf{Chapter 3} presents the necessary mathematical background, concepts, and definitions used in this doctoral thesis. The following topics are introduced: (i) basic set operations; (ii) zonotopes; (iii) constrained zonotopes; (iv) complexity reduction methods; (v) interval analysis; and (iv) linear state estimation using constrained zonotopes.
	\item \textbf{Chapter 4} introduces new state estimation methods based on constrained zonotopes for discrete-time systems with nonlinear dynamics and linear measurement: (i) the constrained zonotope and interval bundle method; (ii) the mean value extension and (iii) the first-order Taylor extension using constrained zonotopes, which are generalizations of two existing zonotopic methods. 
	\item \textbf{Chapter 5} introduces new state estimation methods based on constrained zonotopes for discrete-time systems with nonlinear dynamics, nonlinear measurements, and nonlinear invariants. The proposed methods improve the standard prediction-update framework for systems with invariants by adding a consistency step (which uses invariants to reduce conservatism), besides presenting significant improvements to prediction and update steps proposed in Chapter \ref{cha:nonlineardynamics}, and allowing nonlinear measurement equations for the first time.
	\item \textbf{Chapter 6} develops a new approach for set-based active fault diagnosis for discrete-time systems with nonlinear dynamics using zonotopes and constrained zonotopes. Passive fault detection is proposed based on the new enclosures obtained in Chapter \ref{cha:nonlineardynamics}. An adaptation of the first-order Taylor extension is used for the reachable set parameterization, which leads to the development of an optimal method for input design for guaranteed set-based active fault diagnosis.
	\item \textbf{Chapter 7} presents new methods based on constrained zonotopes and unbounded sets for set-valued state estimation and active fault diagnosis of linear descriptor systems: (i) linear static constraints on the state variables are directly incorporated in the mathematical description of constrained zonotopes, leading to less conservative enclosures; and (ii) new methods for state estimation and active fault diagnosis of linear descriptor systems are developed based on a new representation for unbounded sets, without assuming the knowledge of an enclosure of all the possible trajectories of the system or the initial states.
	\item \textbf{Chapter 8} develops a new method for set-based joint state and parameter estimation of discrete-time systems with nonlinear dynamics, nonlinear measurement, and nonlinear algebraic constraints. The state estimation method proposed in Chapter \ref{cha:nonlinearmeasinv} is extended to include parameter estimation in a unified framework, and therefore maintaining existing dependencies between states, algebraic variables, and unknown parameters.
	\item \textbf{Chapter 9} addresses the application to unmanned aerial vehicles (UAVs): (i) a set-based linear state estimation using constrained zonotopes considering different sampling times, applied in state-feedback control for trajectory tracking of a suspended load using a tilt-rotor UAV; application of the (ii) state estimation and (iii) fault diagnosis methods developed in Chapters 4 and 6 to a quadrotor UAV, respectively.
	\item \textbf{Chapter 10} proposes two new methods for set-based state estimation and fault detection in water distribution networks (WDNs): (i) a new interval method is proposed based on iterative computation of tight enclosures of the nonlinear head-loss functions, and bounding the solution of the algebraic equations using rescaling; (ii) since intervals are not capable of capturing the dependencies between state variables, constrained zonotopes are used as an additional step, leading to two new algorithms capable of capturing the dependencies between hydraulic states.
	\item \textbf{Chapter 11} develops a discrete-time interval state estimator for a Lithium-ion cell, based on the forward-backward constraint propagation. The contributions include also the identification of parameter uncertainties of the electrical chemical model based on the Fisher Information Matrix. In practice, only cell current and voltage are usually measurable, and therefore, the model states need to be estimated. The proposed interval estimator provides a tight enclosure of the cell states, which is essential for fault detection and model-based control of a Lithium-ion cell.
	\item \textbf{Chapter 12} summarizes the main contributions developed in this doctoral thesis, and presents proposals for future work.	
	\item \textbf{Appendix A} details the computational complexities of some of the state estimation methods developed in this thesis.
\end{itemize}

\section{List of publications}

The following articles were published, accepted for publication, or submitted during the elaboration of this doctoral thesis:

\paragraph{Journal papers:}

\begin{enumerate}
	\item \citep{Santos2017} Santos, M. A., Rego, B. S., Raffo, G. V., \& Ferramosca, A. (2017b). Suspended load path tracking control strategy using a tilt-rotor UAV. \emph{Journal of Advanced Transportation}, 2017, 1--22.
	\item \citep{Rego2019} Rego, B. S. \& Raffo, G. V. (2019). Suspended load path tracking control using a tilt-rotor	UAV based on zonotopic state estimation. \emph{Journal of the Franklin Institute}, 356(4), 1695--1729.
	\item \citep{Rego2020} Rego, B. S., Raffo, G. V., Scott, J. K., \& Raimondo, D. M. (2020a). Guaranteed methods based on constrained zonotopes for set-valued state estimation of nonlinear discrete-time systems. \emph{Automatica}, 111, 108614.	
    \item \citep{Rego2020c} Rego, B. S., Scott, J. K., Raimondo, D. M., \& Raffo, G. V. (2021a). Set-valued state estimation of nonlinear discrete-time systems with nonlinear invariants based on constrained zonotopes. Automatica, 129, 109638.	
	\item \citep{Rego2020d} Rego, B. S., Vrachimis, S. F., Polycarpou, M. M., Raffo, G. V., \& Raimondo, D. M. (2021b). State estimation and leakage detection in water distribution networks using constrained zonotopes. \emph{IEEE Transactions on Control Systems Technology}. Minor revision after first round.	
	\item \citep{Rego2020e} Rego, B. S., Raimondo, D. M., \& Raffo, G. V.  (2021c). Set-based state estimation and active fault diagnosis of linear descriptor systems using mixed zonotopes. To be submitted.
	\item \citep{Rego2021d} Rego, B. S., Locatelli, D., Raimondo, D. M., \& Raffo, G. V. (2021a). Set-based joint state	and parameter estimation of discrete-time systems using constrained zonotopes. To be submitted.
	\item \citep{Rego2021e} Rego, B. S., Raimondo, D. M., \& Raffo, G. V. (2021b). Set-based joint state and parameter estimation of nonlinear systems with nonlinear algebraic constraints using constrained zonotopes. To be submitted.
\end{enumerate}

\paragraph{Conference papers:}

\begin{enumerate}
	\item \citep{Lara2017} Lara, A. V., Rego, B. S., Raffo, G. V., \& Arias-Garcia, J. (2017). Desenvolvimento de um ambiente de simulação de VANTs tilt-rotor para testes de estratégias de controle. In \emph{Anais do XIII Simpósio Brasileiro de Automação Inteligente} (pp. 2135--2141).
	\item \citep{Santos2017b} Santos, M. A., Cardoso, D. N., Rego, B. S., Raffo, G. V., \& Esteban, S. (2017a). A discrete robust adaptive control of a tilt-rotor UAV for an enlarged flight envelope. In \emph{Proc. of the 56th IEEE Conference on Decision and Control} (pp. 5208--5214).
	\item \citep{Rego2018b} Rego, B. S., Raimondo, D. M., \& Raffo, G. V. (2018a). Path tracking control with state estimation based on constrained zonotopes for aerial load transportation. In \emph{Proc. of the 57th IEEE Conference on Decision and Control} (pp. 1979--1984).
	\item \citep{Rego2018} Rego, B. S., Raimondo, D. M., \& Raffo, G. V. (2018b). Set-based state estimation of nonlinear systems using constrained zonotopes and interval arithmetic. \emph{In Proc. of the 2018 European Control Conference} (pp. 1584--1589).
	\item \citep{Rego2020b} Rego, B. S., Raimondo, D. M., \& Raffo, G. V. (2020b). Set-based state estimation and fault diagnosis of linear discrete-time descriptor systems using constrained zonotopes. In \emph{Proc. of the 1st IFAC-V World Congress} (pp. 4291-4296).
	\item \citep{Locatelli2021} Locatelli, D., Rego, B. S., Raffo, G. V., \& Raimondo, D. M. (2021). Interval state estimation based on constraint propagation for a lithium-ion cell using an equivalent circuit model. In \emph{Proc. of the 11th IFAC International Symposium on Advanced Control of Chemical Processes}. Accepted.	
\end{enumerate}

\chapter{Literature review}\thispagestyle{headings} \label{cha:litrev}

This chapter presents a detailed literature review on the main topics of interest in this doctoral thesis, namely set-based state estimation and set-based fault diagnosis of nonlinear systems. Brief literature reviews on set-based state estimation and fault diagnosis of descriptor systems, and joint state and parameter estimation, are also depicted.

\section{Reachability analysis and set-based state estimation of nonlinear systems}
\sectionmark{Reachability analysis and set-based state estimation}

Reachability analysis aim to construct compact sets that are guaranteed to enclose all possible trajectories of the system subject to unknown-but-bounded uncertainties. Interval analysis, proposed in \cite{Moore1966}, is a classic tool for computing these enclosures for nonlinear systems, with a wide range of other applications, such as global optimization, parameter estimation, and robust control \citep{Jaulin2001}. Nevertheless, severe conservatism may can occur in the computation of enclosures using intervals due to interval dependency and the wrapping effect. 

\cite{Kuhn1998} proposes methods to enclose the trajectories of discrete-time nonlinear systems using zonotopes, which are shown to be capable of notably mitigating the wrapping effect. \cite{Girard2006} use zonotopes to compute the exact reachable sets of linear discrete-time systems efficiently. Moreover, zonotopes are further used in \cite{Althoff2008} to bound the trajectories of nonlinear continuous-time systems by conservative linearization; however, in contrast to the linear discrete-time case, the method still results in substantial overestimation for systems with highly nonlinear behavior.

Seeking the reduction of conservatism in the nonlinear case, non-convex sets denoted by polynomial zonotopes are developed in \cite{Althoff2013} to bound the trajectories of nonlinear systems through conservative polynomialization. Nevertheless, the complexity of the representation described by polynomial zonotopes grows polynomially each time step, which is mitigated by the development of a sparse description in \cite{Kochdumper2020b} (called sparse polynomial zonotopes). Later, \cite{Kochdumper2020arxiv} extend the latter to constrained polynomial zonotopes, which allow the efficient computation of several set operations exactly for the first time (e.g., set union) using a single set.

Alternative techniques for reachability analysis include ellipsoidal methods \citep{Kurzhanski2000}; however, despite the efficiency of such methods, due to the limited complexity of ellipsoids, the resulting enclosure is often conservative even in the linear case. In addition, other techniques for reachability analysis consist of using Taylor models \citep{Chen2012}, which are based on approximating the nonlinear function by high-order Taylor expansions, allowing to combine different set representations such as intervals and zonotopes. Support functions are used in \cite{Girard2008} for reachability analysis of linear discrete-time systems, and in \cite{Frehse2015} for linear continuous-time systems; however, in practice, these methods are limited to approximation of the true set by half-spaces. Other frameworks include the use of Bernstein expansions \citep{Dang2012} to outer-approximate the reachable sets of polynomial systems by convex sets, while Hamilton-Jacobi methods have been proposed in \citep{Mitchell2005} to enclose the trajectories of nonlinear continuous-time systems through differential game formulation.

Set-based state estimation, using the standard recursive approach, involves first bounding the image of the current enclosure under the dynamics (prediction), and then enclosing the intersection of this set with the set of states consistent with a bounded-error measurement (update). For linear discrete-time systems, exact enclosures can be computed as convex polytopes. However, even for linear dynamics, polytope propagation requires demanding computations (e.g., polytope projection, Minkowski addition, or conversion between vertex and half-space representations) \citep{Walter1989,Shamma1999}. Thus, enclosures are often described by simpler sets including ellipsoids \citep{Schweppe1968,Durieu2001,Polyak2004}, parallelotopes \citep{Chisci1996,Vicino1996}, zonotopes \citep{Le2013,Combastel2003,Wang2018}, or combinations of these \citep{Chabane2014}. However, the mathematical limitations of these sets require certain operations to be over-approximated, sometimes quite significantly. Notably, this includes set intersection, which is critical for the update step in set-based state estimation \citep{Chisci1996,Le2013,Durieu2001}. \cite{Althoff2011} propose the use of zonotope bundles to describe intersections of zonotopes without explicit computation. However, the Minkowski sum and linear image are over-approximated. Efficient algorithms for linear set-based state estimation are proposed in \cite{Scott2016}, in which constrained zonotopes are introduced to overcome many of the limitations of zonotopes. These sets are closed under intersection, Minkowski sum, and linear image, and are capable of describing arbitrary convex polytopes if the complexity of the set description is not limited. 

In contrast to the linear case, effective set-based state estimation for nonlinear systems is still an open challenge \citep{Jaulin2009,Alamo2005a,Combastel2005,Wan2018}. Early approaches in this field used inclusion functions based on interval analysis \citep{Moore2009} to propagate bounds through the nonlinear dynamics, and used interval-based set inversion techniques to enclose the set of states consistent with the current measurement \citep{Jaulin2009b,Jaulin2009,Kieffer1998}. Improved accuracy is achieved using refinements (i.e., unions of intervals) as well as more advanced interval methods such as contractor and separator algebras \citep{Jaulin2009b,Jaulin2016}, and constraint propagation using tube contractors \citep{Rohou2018}. Unfortunately, even these methods often provide conservative bounds without extensive refinement, which is only tractable for systems with relatively few states \citep{Jaulin2009}.

A few alternatives for nonlinear set-based state estimation can be found in the literature, mostly based on zonotopes. \cite{Alamo2005a} and \cite{Combastel2005} both propose approximation procedures for propagating a zonotope through a nonlinear function. The first is based on the Mean Value Theorem and is referred to as the \emph{mean value extension}, while the second uses a first-order Taylor expansion with a rigorous remainder bound, and is referred to as the \emph{first-order Taylor extension}. Using these extensions for the prediction step, updates are then achieved by methods for over-approximating the intersection of a zonotope with a strip (i.e., a linear measurement with bounded error). An alternative zonotope-based prediction step using Difference of Convex functions (DC) programming is proposed in \cite{Alamo2008}, but with the same update as in \cite{Alamo2005a}. Even for linear measurements, the symmetry of zonotopes is known to cause significant errors in the update step \citep{Scott2016}. General convex polytopes in half-space representation are used in \cite{Wan2018} to enable an exact update. Prediction is then done by representing the polytope as an intersection of zonotopes and applying the mean value extension. Unfortunately, conversion between these representations is computationally demanding, and the increasing complexity of the zonotope bundle with time is not addressed. 

Still in the context of state estimation, several methods addresses the case of nonlinear systems whose solutions satisfy a set of potentially nonlinear equality constraints, referred to as \emph{invariants}. The trajectories of such systems evolve on a lower-dimensional manifold embedded in the state space. This is true for many systems of practical interest, including models of (bio)chemical reaction networks \citep{Shen2017}, attitude estimation in aircraft systems \citep{Goodarzi2017}, and the pose of the body frame in humanoids \citep{Rotella2014}.
In the stochastic state estimation framework, invariants have previously been used to force the estimated states to lie on the embedded manifold  \citep{Julier2007,Yang2009b,Teixeira2009,Simon2010,Eras2019}. In the set-based estimation framework, the aim is to use invariants to reduce the conservatism of the enclosure computed in each time step by eliminating enclosed regions that can be proven to violate the invariants, and hence cannot contain real trajectories. Such refinement is known to be very effective at reducing conservatism in interval-based nonlinear reachability calculations \citep{Scott2013a,Shen2017,Yang2020}. To the best of the author's knowledge, \cite{Yang2009} and \cite{Yang2018} are the only prior studies that have used invariants in set-based state estimation. \cite{Yang2009} propose a set-valued state estimator using ellipsoids. A linear matrix inequality approach is used to design the estimator taking into account the nonlinear state equality constraints. However, the method only applies to linear dynamics, and the nonlinear state constraints must be conservatively linearized. Moreover, an effective procedure for computing rigorous and accurate linearization error bounds is not provided. \cite{Yang2018} propose an effective method for using invariants to reduce the conservatism of a set-based state estimation method based on differential inequalities and interval analysis. However, the method is limited to systems that have been discretized by Euler approximation with a sufficiently small step size, in which the step size can also be difficult to compute and/or restrictive in some cases. Moreover, although the theory is general, the provided algorithm only applies to linear invariants and linear measurement equations.

\section{Set-based fault diagnosis of nonlinear systems}

A passive fault detection method is proposed in \cite{Guerra2008} for uncertain LPV systems based on state estimation by zonotopes. This method makes use of the zonotope methods proposed in \cite{Alamo2005a}, while the fault detection is performed by verifying if the intersection of the predicted zonotope with the measurement strips is empty. Another passive fault detection method is proposed in \cite{Blesa2012} for uncertain LPV systems based on consistency tests using polytopes and strips. The fault detection test is performed by verifying the intersection of a convex polytope containing all the feasible parameters and strips containing all parameters consistent with the current measurement. If this set is empty, then a fault has occurred. Later in \cite{Vento2015}, a set-based FD strategy is proposed for uncertain LPV hybrid systems. FD is done using the parity space and residual evaluation with computation by zonotopes, resulting in a hybrid diagnosis automaton. Lastly, a passive fault detection method is proposed in \cite{Xu2016} combining the set-theoretic methods based on zonotope computations and robust unknown input observers for linear discrete-time systems.

For better detectability of faults, a robust active FD strategy is proposed in \cite{Scott2013b} for linear discrete-time systems based on input design and later improved in \cite{Scott2014}. The occurrence of faults is modeled as abrupt model switching. This method is based on the computation of an optimal input sequence such that the output reachable sets (described by zonotopes) associated to each model are not overlapping by the injection of the sequence into the system. The design of such an input sequence is made by solving a mixed-integer quadratic program. Therefore, fault diagnosis is guaranteed along the respective time horizon by verifying   which output reachable set encloses the observed output sequence. A receding horizon approach is developed in \cite{Raimondo2013a} as an extension of the method in \cite{Scott2013b} for closed-loop active FD of linear discrete-time systems. After the introduction of the constrained zonotopes in \cite{Scott2016}, in which an effective passive fault detection method is also developed for linear systems, the open-loop active FD method in \cite{Scott2014} and the receding horizon approach in \cite{Raimondo2013a} were both improved in \cite{Raimondo2016}. These new linear active FD methods are based on constrained zonotopes, which allowed reduced conservativeness in comparison with zonotopes in the computation of the optimal input sequence to be injected into the system, with fault isolation guaranteed in a reduced time interval. Finally, to reduce the computational burden of the active FD methods proposed in \cite{Raimondo2016}, several operations are moved to prior offline computation in the active FD approach developed in \cite{Marseglia2017b} based on multi-parametric programming. Another receding horizon approach for active FD is proposed in \cite{Tabatabaeipour2015} based on optimal input design for linear discrete-time systems using convex polytopes. Nevertheless, the complexity of this active FD method scales exponentially with the system dimension due to the use of set-based state estimators based on convex polytopes.

Considering the design of reference signals instead of input sequences, extensions of the open-loop and receding horizon active FD approaches in \cite{Raimondo2016} are developed in \cite{Marseglia2017} for linear discrete-time systems considering state-feedback control. Using constrained zonotopes, the sequences of optimal reference inputs provided to the controller are obtained through the solution of a mixed-integer quadratic program. Since the injection of additional input sequences into the system is not required for fault isolation, this active FD method is more suitable for active FD of closed-loop systems. Fault isolation is performed similarly to \cite{Raimondo2016}. On the other hand, a nonlinear active FD method is proposed in \cite{Paulson2014} based on input design. This method is formulated for uncertain polynomial or rational discrete-time systems, with faults modeled by abrupt model switching. Robust fault isolation is achieved based on feasibility checking of inputs and measurement for the nonlinear dynamics of each model. After convex relaxations of the problem, the design of the optimal input sequence is performed through the solution of semidefinite or linear programs.

Lastly, \cite{Blanchini2018} proposes an active FD method based on separation of invariant sets for linear systems. This method is formulated for open-loop stable linear systems and is based on input design. Set computations are obtained using implicit representation of convex polytopes with later explicit characterization.

\section{Set-based state estimation and fault diagnosis of descriptor systems}

Set-based strategies for state estimation of discrete-time descriptor systems with unknown-but-bounded uncertainties is a recent subject, often addressed using intervals, zonotopes and ellipsoids \citep{Efimov2015,Puig2018,Wang2018,Merhy2019}. \cite{Efimov2015} use interval methods to design a state estimator for time-delay descriptor systems based on linear matrix inequalities and the Luenberger structure. However, interval arithmetics can lead to conservative enclosures due to the wrapping effect. A few methods are proposed in \cite{Puig2018} for set-based state estimation of linear descriptor systems by enclosing the intersection of two consistent sets with a zonotope bound. Nevertheless, since the intersection cannot be computed exactly, the resulting bound can be conservative. These strategies are extended in \cite{Wang2018} for linear parameter-varying descriptor systems, but conservative enclosures are still present since the intersection method is maintained. \cite{Merhy2019} use ellipsoids for state estimation of linear descriptor systems based on Luenberger type observers. Despite being able to provide stable bounds, the ellipsoidal estimation can be conservative since the complexity of the set is fixed and static relations are not directly incorporated. In addition, as in \cite{Puig2018}, restrictive assumptions on the rank of the system matrices are required to be able to design the proposed estimator. Moreover, due to the static constraints, the reachable sets of models of a descriptor system may be asymmetric even if the initial set is symmetric, and therefore methods based on the sets above can provide conservative enclosures. 

Fault diagnosis aims to determine exactly which fault a process is subject to. For the case of descriptor systems, this problem has been considered in recent works using zonotopes \citep{Yang2019,Wang2019b}. While \cite{Wang2019b} explore the use of unknown input observers for robust passive fault diagnosis limited to additive faults, \cite{Yang2019} propose a zonotope-based method for active fault diagnosis (AFD) of descriptor systems. The latter is based on the design of an input sequence for separation of the reachable sets. Unfortunately, the generator representation of zonotopes cannot incorporate exactly static relations between the state variables in general linear descriptor systems. Consequently, in practice, this may lead to a more difficult diagnosis.

\section{Set-based joint state and parameter estimation}

Set-based estimation has been widely used also in the parameter identification field, as an alternative to stochastic methods such as least squares or maximum likelihood, since it is able to provide guaranteed enclosures of the model parameters, when the uncertain model parameters have unknown stochastic properties but known bounds.

Zonotopes have been used to approximate the parametric set (i.e., the set of parameters consistent with the model and noisy measurements) for discrete-time systems with additive uncertainties in \cite{Bravo2006}, which was later extended to allow multiplicative uncertainties in \cite{Wang2017}. However, both methods are applied only to systems described by regression models, and rely on conservative intersections with strips to refine the parametric set. Intervals have been used in the context of optimal design of experiments in \cite{Vidal2019} and \cite{Mukkula2017}, to minimize the conservatism of the parametric enclosure. Moreover, a bisection-based interval algorithm has been used in \cite{Rumschinski2010} to deal with non-convex parameter sets using collections of intervals. Nevertheless, intervals are not able to capture dependencies between parameters, which may result in conservative enclosures due to  wrapping effect.  

Few state estimation strategies in the literature refine online the model parameter uncertainties in order to improve the accuracy of state estimation. Such methodology is referred to as \emph{joint state and parameter estimation}. 
Stochastic approaches have been used in order to take into account state and parameters dependencies (covariances for Gaussian filters) in an augmented state space.
A Kalman filtering strategy, based on multi-innovation recursive extended least squares algorithm has been proposed in \cite{Cui2020} to enhance the parameters estimation, while an iterative ensemble Kalman smoother is designed in \cite{Bocquet2013}. 
However, as in state estimation, bias issues introduced by Kalman filtering make such approaches unreliable in case the assumptions on the stochastic properties of the uncertainties are violated. 
Deterministic approaches include Luenberger-based observers \citep{Zhang2020} and set-based interval estimation \citep{raissi2004set}. \cite{raissi2004set} propose a prediction-update state and parameter estimator suitable for nonlinear continuous-time systems. However, besides not being able to capture the dependencies between states and parameters, the method can lead to high computational complexity. Moreover, \cite{chen1995set} propose an estimation method based on mixed-integer programming to obtain the smallest interval enclosing the states and parameters consistent with measurements. However, this requires the solution to an optimization problem, leading to high computational burden.

\section{Final remarks}

This chapter presented a literature review on the main topics of interest in this doctoral thesis, namely set-based state estimation and set-based fault diagnosis of nonlinear systems, set-based state estimation and fault diagnosis of descriptor systems, and joint state and parameter estimation.

As commented in Chapter \ref{cha:introduction}, and based on the literature review presented in this chapter, achieving accurate enclosures for the states of nonlinear systems remains a significant challenge. When these enclosures are represented by simple sets such as intervals, ellipsoids, parallelotopes, and zonotopes, certain set operations can be very conservative. Yet, using general convex polytopes is much more computationally demanding. Also, several set-based state estimation methods in the literature assume linear measurement, which is not true in most practical scenarios. Besides, in the presence of invariants, the existing methods are not able to incorporate this information efficiently due to the simple set representations used.

In addition, in set-based fault diagnosis, most methods in the literature are limited to linear discrete-time systems. It remains a challenge to develop fault diagnosis methods for nonlinear discrete-time systems, which make use of conservative reachable sets if the available methods for reachability analysis are employed.

Moreover, the techniques available for set-based state estimation and fault diagnosis of linear descriptor systems make use of the simple sets mentioned above. However, those set representations are not able to effectively incorporate equality constraints, typical of descriptor systems, in the set description, resulting in significant conservatism.

Lastly, the methods available for set-based joint state and parameter estimation in the literature are not able to effectively take into account the dependencies between states and parameters. This disadvantage is a result from the limitations of the employed set representations (mostly intervals), which suffer from severe wrapping effect.

\chapter{Mathematical background}\thispagestyle{headings} \label{cha:preliminaries}

This chapter presents basic concepts and definitions required in the rest of this doctoral thesis. The following topics are introduced: (i) basic set operations; (ii) zonotopes; (iii) constrained zonotopes; (iv) complexity reduction methods; (v) interval analysis; and (iv) linear state estimation using constrained zonotopes.

\section{Basic set operations and notation} \label{sec:pre_basicoperations}

This section describes some common set operations which are useful for computing enclosures required by set-based state estimation \citep{Le2013,Scott2016}. Consider sets $Z, W \subset \realset^{n}$, and $Y \subset \realset^{m}$, and a real matrix $\mbf{R} \in \realset^{m \times n}$. Define the linear mapping, Minkowski sum, and generalized intersection, as
\begin{align}
\mbf{R}Z & \triangleq \{ \mbf{R} \mbf{z} : \mbf{z} \in Z\}, \label{eq:pre_limage}\\
Z \oplus W & \triangleq \{ \mbf{z} + \mbf{w} : \mbf{z} \in Z,\, \mbf{w} \in W\}, \label{eq:pre_msum}\\
Z \cap_{\mbf{R}} Y & \triangleq \{ \mbf{z} \in Z : \mbf{R} \mbf{z} \in Y\}, \label{eq:pre_intersection}
\end{align}
respectively. Using ellipsoids or parallelotopes, the linear mapping \eqref{eq:pre_limage} can be computed efficiently and accurately, but the Minkoswki sum \eqref{eq:pre_msum} and intersection \eqref{eq:pre_intersection} must be conservatively outer-approximated \citep{Chisci1996,Schweppe1968}. On the other hand, for intervals, the Minkowski sum \eqref{eq:pre_msum} is efficient and exact, but \eqref{eq:pre_limage} and \eqref{eq:pre_intersection} are conservative due to the presence of the wrapping effect\footnote{The generalized intersection in \eqref{eq:pre_intersection} is not conservative when $\mbf{R} = \eyenoarg$, which corresponds to the standard intersection $\cap$.}. This well-known effect in interval analysis is associated to the inability of an interval variable to capture existing dependencies between its components \citep{Moore2009}. In contrast, convex polytopes are closed under \eqref{eq:pre_limage}--\eqref{eq:pre_intersection}. The operations \eqref{eq:pre_limage} and \eqref{eq:pre_msum} can be computed efficiently in vertex representation (V-rep), while \eqref{eq:pre_intersection} can be computed efficiently in half-space representation (H-rep). However, if multiple computations of \eqref{eq:pre_limage}--\eqref{eq:pre_intersection} are done using convex polytopes, this requires conversions between H-rep and V-rep which can be extremely expensive and numerically unstable in high dimensions \citep{Hagemann2015}. 

In this doctoral thesis, functions with set-valued arguments will be consistently used to denote exact image of the set under the function; e.g., $\bm{\mu}(X,W)\triangleq\{\bm{\mu}(\mathbf{x},\mathbf{w}): \mathbf{x}\in X, \ \mathbf{w}\in W\}$. In addition, the following notations are defined to be used in the proofs. Let $\bm{\kappa}$ be a function of class $\mathcal{C}^2$, and let $\mbf{z}$ denote its argument. Then, $\kappa_q$ denotes the $q$-th component of $\bm{\kappa}$, $\nabla \bm{\kappa}$ denotes the gradient of $\bm{\kappa}$, and $\mathbf{H} \kappa_q$ is an upper triangular matrix describing half of the Hessian of $\kappa_q$. Specifically, $H_{ii} \kappa_q = (1/2) \partial^2 \kappa_q/\partial z_i^2$, $H_{ij} \kappa_q = \partial^2 \kappa_q/\partial z_i \partial z_j$ for $i<j$, and $H_{ij} \kappa_q = 0$ for $i>j$.

\section{Zonotopes and constrained zonotopes} \label{sec:pre_zoncz}

\emph{Zonotopes} are centrally symmetric sets \citep{Kuhn1998} which allow both \eqref{eq:pre_limage} and \eqref{eq:pre_msum} to be computed exactly and with low computational burden. However, the intersection \eqref{eq:pre_intersection} does not result in a zonotope in general, can be difficult to compute, and must be outer-approximated \citep{Le2013,Bravo2006}.

\begin{definition} \rm \label{def:zonotopes}
	A set $Z \subset \realset^n$ is a \emph{zonotope} if there exists $(\mbf{G}_z,\mbf{c}_z) \in \realsetmat{n}{n_g} \times \realset^n$ such that
	\begin{equation} \label{eq:pre_grep}
	Z = \left\{ \mbf{c}_z + \mbf{G}_z \bm{\xi} : \| \bm{\xi} \|_\infty \leq 1 \right\}.
	\end{equation}	
\end{definition}

Equation \eqref{eq:pre_grep} is called \emph{generator representation} (G-rep). Each column of $\mbf{G}_z$ is a \emph{generator}, and $\mbf{c}_z$ is the \emph{center}. In this doctoral thesis, we refer to $\bm{\xi}$ as the \emph{generator variables}. By defining the $n_g$-dimensional unitary hypercube $B_\infty^{n_g} \triangleq \{\bm{\xi} \in \realset^{n_g} : \ninf{\bm{\xi}} \leq 1 \}$ (with $B_\infty \triangleq B_\infty^1$), a zonotope $Z$ can be alternatively interpreted as an affine transformation of $B_\infty^{n_g}$, given by $Z = \mbf{c}_z \oplus \mbf{G}_z B_\infty^{n_g}$. Since $B_\infty^{n_g}$ is symmetric, then the G-rep \eqref{eq:pre_grep} is limited to centrally symmetric convex polytopes. Another interpretation of a zonotope $Z$ is the Minkoswki sum of line segments, given by $Z = \bigoplus_{j=1}^{n_g} \mbf{g}_{z,j} B_\infty \oplus \mbf{c}_z$, where $\mbf{g}_{z,j}$ is the $j$-th column of $\mbf{G}_z$. In this doctoral thesis, we use the shorthand notation $Z = \{\mbf{G}_z, \mbf{c}_z\}$ for zonotopes.

Zonotopes have a few computational benefits in comparison to other set representations. Let $Z = \{\mbf{G}_z, \mbf{c}_z\} \subset \realset^n$, $W = \{\mbf{G}_w, \mbf{c}_w\} \subset \realset^n$, and $\mbf{R} \in \realsetmat{m}{n}$. The set operations \eqref{eq:pre_limage}--\eqref{eq:pre_msum} are computed efficiently in G-rep as 
\begin{align}
\mbf{R}Z & = \left\{ \mbf{R} \mbf{G}_z, \mbf{R} \mbf{c}_z \right\}, \label{eq:pre_zlimage}\\
Z \oplus W & = \left\{ \begin{bmatrix} \mbf{G}_z \,\; \mbf{G}_w \end{bmatrix}, \mbf{c}_z + \mbf{c}_w \right\}. \label{eq:pre_zmsum}
\end{align}
The operations \eqref{eq:pre_zlimage}--\eqref{eq:pre_zmsum} cause only a mild increase in the complexity of the G-rep \eqref{eq:pre_grep}. Unlike polytopes, this increased complexity can be addressed efficiently using order reduction algorithms (Section \ref{sec:complexityreduction}), to enclose a zonotope with another one with fewer number of generators, allowing one to balance accuracy and computational efficiency.

The term \emph{constrained zonotope} was first used in \cite{Ponsini2012} to refer to a new class of sets introduced in \cite{Ghorbal2010}. The latter is an extension of zonotopes defined by the logical product of a zonotope and polyhedral sets. In this doctoral thesis, we are interested in the constrained zonotopes (also an extension of zonotopes) defined as in \cite{Scott2016}. These are capable of describing also asymmetric convex polytopes, while maintaining many of the well-known computational benefits of zonotopes.

\begin{definition} \rm \label{def:pre_czonotopes}
	A set $Z \subset \realset^n$ is a \emph{constrained zonotope} if there exists $(\mbf{G}_z,\mbf{c}_z,\mbf{A}_z,\mbf{b}_z) \in \realsetmat{n}{n_g} \times \realset^n \times \realsetmat{n_c}{n_g} \times \realset^{n_c}$ such that
	\begin{equation} \label{eq:pre_cgrep}
	Z = \left\{ \mbf{c}_z + \mbf{G}_z \bm{\xi} : \| \bm{\xi} \|_\infty \leq 1, \mbf{A}_z \bm{\xi} = \mbf{b}_z \right\}.
	\end{equation}	
\end{definition}

We refer to \eqref{eq:pre_cgrep} as the \emph{constrained generator representation} (CG-rep). Each column of $\mbf{G}_z$ is a \emph{generator}, $\mbf{c}_z$ is the \emph{center}, $\mbf{A}_z \bm{\xi} = \mbf{b}_z$ are the \emph{constraints}, and $\bm{\xi}$ are the \emph{generator variables}. By defining the constrained unitary hypercube\footnote{We drop the use of the superscript $n_g$ for $B_\infty(\mbf{A}_z,\mbf{b}_z)$ since this dimension can be inferred from the number of columns of $\mbf{A}_z$.} $B_\infty(\mbf{A}_z,\mbf{b}_z) \triangleq \{\bm{\xi} \in \realset^{n_g} : \ninf{\bm{\xi}} \leq 1,\,  \mbf{A}_z \bm{\xi} = \mbf{b}_z \}$, a constrained zonotope $Z$ can be alternatively interpreted as an affine transformation of $B_\infty(\mbf{A}_z,\mbf{b}_z)$, given by $Z = \mbf{c} \oplus \mbf{G}_z B_\infty(\mbf{A}_z,\mbf{b}_z)$. Differently from zonotopes, the linear equality constraints in \eqref{eq:pre_cgrep} allow constrained zonotopes to represent any convex polytope provided that the complexity of the CG-rep \eqref{eq:pre_cgrep} (i.e., the number of generators and constraints) is not limited. In fact, $Z$ is a constrained zonotope iff it is a convex polytope (Property \ref{prop:pre_czhreptocgrep}). We use the compact notation $Z = \{\mbf{G}_z, \mbf{c}_z,\mbf{A}_z,\mbf{b}_z \}$ for constrained zonotopes.

In addition to \eqref{eq:pre_limage} and \eqref{eq:pre_msum}, the intersection \eqref{eq:pre_intersection} can also be computed exactly with constrained zonotopes. Let $Z = \{\mbf{G}_z, \mbf{c}_z, \mbf{A}_z, \mbf{b}_z\} \subset \realset^n$, $W = \{\mbf{G}_w, \mbf{c}_w, \mbf{A}_w, \mbf{b}_w\} \subset \realset^n$, $Y = \{\mbf{G}_y, \mbf{c}_y, \mbf{A}_y, \mbf{b}_y\} \subset \realset^m$, and $\mbf{R} \in \realsetmat{m}{n}$. The set operations \eqref{eq:pre_limage}--\eqref{eq:pre_intersection} are computed trivially in CG-rep (see \cite{Scott2016} for a proof) as
\begin{align}
\mbf{R}Z & = \left\{ \mbf{R} \mbf{G}_z, \mbf{R} \mbf{c}_z, \mbf{A}_z, \mbf{b}_z \right\}, \label{eq:pre_czlimage}\\
Z \oplus W & = \left\{ \begin{bmatrix} \mbf{G}_z \,\; \mbf{G}_w \end{bmatrix}, \mbf{c}_z + \mbf{c}_w, \begin{bmatrix} \mbf{A}_z & \bm{0} \\ \bm{0} & \mbf{A}_w \end{bmatrix}, \begin{bmatrix} \mbf{b}_z \\ \mbf{b}_w \end{bmatrix} \right\}\!, \label{eq:pre_czmsum}\\
Z \cap_{\mbf{R}} Y & = \left\{ \begin{bmatrix} \mbf{G}_z \,\; \bm{0} \end{bmatrix}, \mbf{c}_z, \begin{bmatrix} \mbf{A}_z & \bm{0} \\ \bm{0} & \mbf{A}_y \\ \mbf{R} \mbf{G}_z & -\mbf{G}_y \end{bmatrix}, \begin{bmatrix} \mbf{b}_z \\ \mbf{b}_y \\ \mbf{c}_y - \mbf{R} \mbf{c}_z \end{bmatrix} \right\}. \label{eq:pre_czintersection}
\end{align}

These operations can be performed efficiently and cause only a moderate increase in the complexity of the CG-rep. As with zonotopes, efficient methods for complexity reduction of constrained zonotopes, i.e. to enclose a constrained zonotope by another one with a fewer number of generators and constraints, are available (Section \ref{sec:complexityreduction}). Other useful operations with constrained zonotopes are presented in the following. Property \ref{prop:pre_czisemptyinside} provides simple methods to verify if a constrained zonotope is empty and if a given point belongs to it \citep{Scott2016}. Property \ref{prop:pre_czhreptocgrep} is a method to compute the corresponding CG-rep of a given convex polytope. Property \ref{prope:pre_czihull} provides a simple method for computing the interval hull of a constrained zonotope by solving $2n$ linear programs (LPs). For simplicity, the subscripts of the variables in \eqref{eq:pre_cgrep} will be suppressed henceforth when not necessary.

\begin{property} \rm \citep{Scott2016} \label{prop:pre_czisemptyinside}
	For every $\standardcz \subset \realset^{n}$ and $\mbf{z} \in \realset^n$, 
	\begin{align*}
	& Z \neq \emptyset \iff \underset{\bm{\xi}}{\min} \left\{ \ninf{\bm{\xi}} : \mbf{A} \bm{\xi} = \mbf{b} \right\} \leq 1, \\
	& \mbf{z} \in Z \iff \underset{\bm{\xi}}{\min} \left\{ \ninf{\bm{\xi}} : \begin{bmatrix} \mbf{G} \\ \mbf{A} \end{bmatrix} \bm{\xi} = \begin{bmatrix} \mbf{z} - \mbf{c} \\ \mbf{b} \end{bmatrix} \right\} \leq 1.
	\end{align*}
\end{property}

\begin{property} \rm (From Theorem 1 in \cite{Scott2016}) \label{prop:pre_czhreptocgrep}
	Let $P = \{\mbf{z} : \mbf{H} \mbf{z} \leq \mbf{k}\} \subset \realset^{n}$ be a convex polytope in H-rep, and choose $Z = \{\mbf{G}, \mbf{c}\} \subset \realset^{n}$ and $\bm{\sigma} \in \realset^{n}$ such that $P \subseteq Z$ and $\mbf{H}\mbf{z} \in \left[\bm{\sigma},\,\mbf{k}\right]$, $\forall \mbf{z} \in P$. Then, the convex polytope $P$ can be written in CG-rep as
	\begin{equation} \label{eq:pre_hreptocg}
	P = \left\{ \left[ \mbf{G} \,\; \bm{0} \right], \mbf{c}, \left[ \mbf{H}\mbf{G} \,\;\, \text{diag}\!\left( \dfrac{\bm{\sigma} - \mbf{k}}{2} \right) \right] , \dfrac{\mbf{k} + \bm{\sigma}}{2} - \mbf{H}\mbf{c} \right\}.
	\end{equation}
\end{property}

\begin{property}\rm  \citep{Scott2016,Rego2018} \label{prope:pre_czihull}
	Let $Z = \{\mbf{G}, \mbf{c}, \mbf{A}, \mbf{b} \} \subset \realset^{n}$, and let $\mbf{G}_{j,:}$ denote the $j$-th row of $\mbf{G}$. The \emph{interval hull} $[\bm{\zeta}^\text{L},\,\bm{\zeta}^\text{U}] \supseteq Z$ is obtained by solving the following linear programs for each $j = 1,2,\dots,n$:
	\begin{align*}
	\zeta_j^\text{L} & = \underset{\bm{\xi}}{\min} \left\{c_j + \mbf{G}_{j,:} \bm{\xi} : \ninf{\bm{\xi}} \leq 1,\, \mbf{A} \bm{\xi} = \mbf{b} \right\}, \\
	\zeta_j^\text{U} & = \underset{\bm{\xi}}{\max} \left\{c_j + \mbf{G}_{j,:} \bm{\xi} : \ninf{\bm{\xi}} \leq 1,\, \mbf{A} \bm{\xi} = \mbf{b} \right\}.
	\end{align*}
\end{property}

\section{Complexity reduction and rescaling} \label{sec:complexityreduction}

This section describes the methods used in this doctoral thesis for complexity reduction of zonotopes and constrained zonotopes. Other useful algorithms required by the techniques proposed in this thesis are also presented.

Method \ref{meth:genredA} is the pioneering general-purpose method for order reduction of zonotopes. It allows the efficient computation of a zonotope with a desired number of generators enclosing a more complex zonotope. Nevertheless, the resulting enclosure can be very conservative \citep{Scott2018}.

\begin{method} \rm (Generator reduction A) \citep{Combastel2003,Alamo2005a} \label{meth:genredA}
	Let $Z = \{\mbf{G},\mbf{c}\} \subset \realset^n$ be a zonotope, with $\mbf{G} \in \realsetmat{n}{n_g}$, $\mbf{c} \in \realset^n$, and consider $s \in \naturalset$ such that $n \leq s < n_g$. Denote by $\hat{\mbf{G}}$ the matrix resulting from reordering the columns of $\mbf{G}$ in decreasing 2-norm. Then, $$Z \subseteq \bar{Z} = \{[\mbf{H} \,\; \mbf{Q}], \mbf{c}\},$$ where $\bar{Z}$ has $s$ generators, $\mbf{H}$ is a matrix composed of the first $s-n$ columns of matrix $\hat{\mbf{G}}$, and $\mbf{Q}$ is a diagonal matrix given by
	\begin{equation*}
	Q_{ii} = \sum_{j=s-n+1}^{n_g} |\hat{G}_{ij}|,
	\end{equation*}	
	for $i = 1,2,\ldots,n$.
\end{method}

A weighted version of Method \ref{meth:genredA} is proposed in \cite{Combastel2015b}. However, the available criteria for choosing the weighting matrix depends on the structure of the state estimation algorithm, and are proposed only for few specific linear cases. A more accurate order reduction method has been proposed in \cite{Althoff2010}. Nevertheless, this exhibits a very high computational cost in higher dimensions. On the other hand, Method \ref{meth:genredB} was proposed in \cite{Scott2016}, with improved effectiveness and accuracy with respect to other available methods, as highlighted in \cite{Scott2018}.

\begin{method} \rm (Generator reduction B) \citep{Scott2016,Scott2018} \label{meth:genredB}
	Consider the zonotope $Z = \{\mbf{G},\mbf{c}\} \subset \realset^n$, with $\mbf{G} \in \realsetmat{n}{n_g}$, $\mbf{c} \in \realset^{n}$. A less conservative method for generator reduction of zonotopes is given by: 
	\begin{itemize}
		\item[\emph{1)}] Reorder the columns of $\mbf{G}$ as $[\mbf{T} \,\; \mbf{V}]$, where $\mbf{T} \in \realsetmat{n}{n}$ is an invertible matrix.
		\item[\emph{2)}] Choose a column $\mbf{v}$ of $\mbf{V} \in \realsetmat{n}{(n_g-n)}$ and rewrite $Z$ as $Z = X \oplus Y \triangleq \{[\mbf{T} \,\; \mbf{v}], \mbf{c}\} \oplus \{\mbf{V}_{-}, \bm{0}\}$, where $\mbf{V}_{-} \in \realsetmat{n}{(n_g-n-1)}$ is the matrix obtained by removing $\mbf{v}$ from $\mbf{V}$.
		\item[\emph{3)}] Reduce $X$ to an optimal parallelotope $\tilde{X}$ satisfying $X \subseteq \tilde{X}$.
		\item[\emph{4)}] Assign $\tilde{Z} = \tilde{X} + Y$, where $\tilde{Z}$ is a zonotope with $n_g-1$ generators.
	\end{itemize}
\end{method}

Step 1 in Method \ref{meth:genredB} is performed using Gauss-Jordan elimination, in which the sequence of column pivots yield the desired ordering $[\mbf{T} \,\; \mbf{V}]$. Moreover, Step 3 is done using the optimal reduction method proposed in \cite{Chisci1996}, according to which the minimum volume parallelotope containing $X$ is $\tilde{X} \triangleq \{\mbf{T}(\eyenoarg + \text{diag}(|\mbf{r}|)), \mbf{c}\}$, where $\mbf{r} = \mbf{T}^{-1} \mbf{v}$. It can be shown that $\ninf{\mbf{r}} \leq 1$. Lastly, in Step 2, the column $\mbf{v}$ is chosen by minimizing the volume error between $\tilde{X}$ and $X$. This error is denoted by $v(\tilde{X}) - v(X)$, which is computed using volume equation presented in \cite{Alamo2005a}, resulting in
\begin{align*}
	v(\tilde{X}) - v(X) = 2^n |\text{det} (\mbf{T})| \left( \prod_{i=1}^n (1+|r_i|) - \left( 1 + \sum_{i=1}^n |r_i| \right) \right).
\end{align*}
Implementation details of Method \ref{meth:genredB} are found in \cite{Scott2016}.

Method \ref{meth:rescaling} corresponds to the \emph{rescaling} procedure of constrained zonotopes proposed in the work of \cite{Scott2016}, which transfers part of the information present in the equality constraints to its generators. 

\begin{method} \rm (Rescaling) \citep{Scott2016} \label{meth:rescaling}
	Consider a constrained zonotope $Z = \{\mbf{G}, \mbf{c}, \mbf{A}, \mbf{b}\}$, with $\mbf{G} \in \realsetmat{n}{n_g}$, $\mbf{c} \in \realset^{n}$, $\mbf{A} \in \realsetmat{n_c}{n_g}$, and $\mbf{b} \in \realset^{n_c}$. If $\xil, \xiu \in \realset^{n_g}$ satisfy $B_\infty(\mbf{A},\mbf{b}) \subset [\xil, \xiu] \subset [-\bm{1},\bm{1}]$, then an equivalent CG-rep is
	\begin{equation} \label{eq:pre_rescalingdefinition}
	Z = \{\mbf{G}\text{diag}(\xir), \mbf{c} + \mbf{G} \xim, \mbf{A} \text{diag}(\xir), \mbf{b} - \mbf{A} \xim\},
	\end{equation}
	where $\xim = \half(\xil + \xiu)$ and $\xir = \half(\xiu - \xil)$. The \emph{rescaling} procedure corresponds to the process of computing the interval $[\xil, \xiu]$ and replacing $Z = \{\mbf{G}, \mbf{c}, \mbf{A}, \mbf{b}\}$ by \eqref{eq:pre_rescalingdefinition}. 
\end{method}

An efficient procedure based on interval arithmetic to obtain a tight interval $[\xil, \xiu]$, satisfying $B_\infty(\mbf{A},\mbf{b}) \subset [\xil, \xiu] \subset [-\bm{1},\bm{1}]$ in Method \ref{meth:rescaling}, is described in Algorithm \ref{alg:pre_rescaling}.
\begin{algorithm}[!htb]
	\caption{Rescaling using interval arithmetic \citep{Scott2016}} 
	\label{alg:pre_rescaling}
	\begin{algorithmic}[1]
		\State Assign $E \gets [ - \bm{1}, \bm{1}] \subset \realset^{n_g}, R \gets (-\bm{\infty}, +\bm{\infty}) = \realset^{n_g}, i = j = 1$.
		\State If $A_{ij} \neq 0$, assign
		\begin{align*}
		R_j & \gets R_j \cap \Big(A_{ij}^{-1} b_i - \sum_{k \neq j} A_{ij}^{-1} A_{ik} E_k \Big), \\
		E_j & \gets E_j \cap R_j.
		\end{align*}
		\State If $j < n_g$, assign $j \gets j + 1$ and go to Step 2. Otherwise, if $j = n_g$ and $i < n_c$, assign $(i,j) \gets (i+1,1)$ and go to Step 2. If $(i,j) = (n_c,n_g)$, assign $[\xil,\xiu] \gets E$ and terminate.
	\end{algorithmic}
	\normalsize
\end{algorithm}

Method \ref{meth:czconelim} was proposed in \cite{Scott2016} to compute a constrained zonotope enclosing another one having one less constraint. This is done by eliminating one constraint-generator pair from the latter with reduced conservativeness. Algorithm \ref{alg:pre_celim} summarizes the method.

\begin{method} \rm (Constraint elimination) \citep{Scott2016} \label{meth:czconelim} Consider the constrained zonotope $Z = \{\mbf{G}, \mbf{c}, \mbf{A}, \mbf{b}\}$, with $\mbf{G} \in \realsetmat{n}{n_g}$, $\mbf{c} \in \realset^{n}$, $\mbf{A} \in \realsetmat{n_c}{n_g}$, and $\mbf{b} \in \realset^{n_c}$. Let $\bm{\Lambda}_\text{G} \in \realsetmat{n}{n_c}$ and $\bm{\Lambda}_\text{A} \in \realsetmat{n_c}{n_c}$. Then, the following holds (Proposition 5 in \cite{Scott2016}):
	\begin{equation} \label{eq:constrelimination}
	Z \subseteq \tilde{Z} = \{ \mbf{G} - \bm{\Lambda}_\text{G} \mbf{A}, \mbf{c} + \bm{\Lambda}_\text{G} \mbf{b}, \mbf{A} - \bm{\Lambda}_\text{A} \mbf{A}, \mbf{b} - \bm{\Lambda}_\text{A} \mbf{b} \}.
	\end{equation}	
	Therefore, one pair of constraint-generator can be eliminated from $Z$ as follows. Let the $i$-th row of $\bm{\Lambda}_\text{A}$ be the $i$-th unit vector, then $\tilde{Z}$ has the trivial $i$-th constraint $\bm{0}^T\bm{\xi} = \bm{0}$, which can be removed. Choose any $i \in \{1,2,\ldots,n_c\}$ and $j \in \{1,2,\ldots,n_g\}$ such that $A_{ij} \neq 0$. From the equality constraints $\mbf{A} \bm{\xi} = \mbf{b}$, it holds that
	\begin{equation} \label{eq:partialsolve}
	\xi_j = A_{ij}^{-1}b_j - A_{ij}^{-1} \sum_{k \neq j} A_{ik} \xi_k.
	\end{equation}
	Thus, the elimination of the $i$-th constraint in $Z$ is performed by computing \eqref{eq:constrelimination} with $\bm{\Lambda}_\text{G} = \mbf{G} \mbf{E}_{ji} A_{ij}^{-1}$ and $\bm{\Lambda}_\text{A} = \mbf{A} \mbf{E}_{ji} A_{ij}^{-1}$, where $\mbf{E}_{ji}$ is a matrix of zeros with exception of an element one in the $j$-th row and $i$-th column. The resulting constrained zonotope has $n_g - 1$ generators and $n_c - 1$ constraints.
	
	To choose the $j$-th generator to eliminate with the least conservativeness, the constraints $\mbf{A} \bm{\xi} = \mbf{b}$ are first preconditioned using Gauss-Jordan elimination with full pivoting. Then, consider the variables $\xim, \xir$ from Method \ref{meth:rescaling}, and let $[\bm{\rho}^\text{L}, \bm{\rho}^\text{U}]$ denote the interval obtained in Algorithm \ref{alg:pre_rescaling}, rescaled as
	\begin{equation*}
	[\rho^\text{L}_j, \rho^\text{U}_j] \gets \left[\frac{\rho^\text{L}_j - \xi_{\text{m},j}}{\xi_{\text{r},j}}, \frac{\rho^\text{U}_j - \xi_{\text{m},j}}{\xi_{\text{r},j}} \right], ~\forall j \in \{1,2,\ldots,n_g\}.
	\end{equation*}
	The choice of $j \in \{1,2,\ldots,n_g\}$ is made such that the Hausdorff error between $Z$ and $\tilde{Z}$ in \eqref{eq:constrelimination} is minimized \citep{Scott2016}. Let $\mbf{r} \in \realset^{n_g}$ such that $r_j = \max\{0,\max\{|\rho_j^\text{L}|,|\rho_j^\text{U}|\}- 1\}$. 
	For computational reasons, for each $j$, the Hausdorff error is then approximated by $$\hat{H}_j = \underset{\mbf{d}}{\min}\{\|\mbf{G} \mbf{d}\|_2^2 + \|\mbf{d}\|_2^2 : \mbf{A} \mbf{d} = \bm{0}, d_j = r_j\}.$$ 
	The optimal $\mbf{d}^*$ is obtained by solving \citep{Scott2016}
	\begin{equation}\label{eq:pre_hausdorff}
	\begin{bmatrix} \eye{n_g+n_c} & \mbf{Q}^{-1} \mbf{e}_j \\ \mbf{e}_j^T & 0 \end{bmatrix} \begin{bmatrix} \mbf{d}^* \\ \bm{\lambda}^* \end{bmatrix} = \begin{bmatrix} \zeros{(n_g + n_c)}{1} \\ r_j \end{bmatrix},
	\end{equation}
	with
	\begin{equation*}
		\mbf{Q} = \begin{bmatrix} \mbf{G}^T \mbf{G} + \eye{n_g} & \mbf{A}^T \\ \mbf{A} & \zeros{n_c}{n_c} \end{bmatrix},
	\end{equation*}
	where $\mbf{e}_j \in \realset^{n_g + n_c}$ is the $j$-th standard unit vector augmented with $n_c$ zeros. Then, $\hat{H}_j$ is computed for each $j$, from which $j^* = \text{arg}~\underset{j}{\min} ~ \hat{H}_j$. Implementation details can be found in \cite{Scott2016}. 
	\begin{algorithm}[!htb]
		\caption{Constraint elimination \citep{Scott2016}}
		\label{alg:pre_celim}
		\small
		\begin{algorithmic}[1]
			\State Assign to $(\mbf{A},\mbf{b})$ the result obtained from Gauss-Jordan elimination applied to $\mbf{A} \bm{\xi} = \mbf{b}$ and reorder the columns of $\mbf{G}$ accordingly.
			\State Rescale $Z$ using Method \ref{meth:rescaling} and Algorithm \ref{alg:pre_rescaling}.
			\State Compute every $\hat{H}_j$ by solving \eqref{eq:pre_hausdorff}, assigning $\hat{H}_j = \|\mbf{G} \mbf{d}^*\|_2^2 + \|\mbf{d}^*\|_2^2$.
			\State Choose which generator to eliminate based on $j^* = \text{arg}~\underset{j}{\min} ~ \hat{H}_j$.
			\State Eliminate one constraint from $Z$ using \eqref{eq:constrelimination}, with $\bm{\Lambda}_\text{G} = \mbf{G} \mbf{E}_{j^*i} A_{ij^*}^{-1}$, and $\bm{\Lambda}_\text{A} = \mbf{A} \mbf{E}_{j^*i} A_{ij^*}^{-1}$.
		\end{algorithmic}
		\normalsize
	\end{algorithm}	
\end{method}

Method \ref{meth:czgenred} has been proposed in \cite{Scott2016} as a required intermediate step to apply generator reduction methods to constrained zonotopes. It is used after constraint elimination (Method \ref{meth:czconelim}), and allows zonotope-based order reduction methods, such as Methods \ref{meth:genredA} and \ref{meth:genredB}, to be extended to constrained zonotopes as well.

\begin{method} \rm (Lift-then-reduce) \citep{Scott2016} \label{meth:czgenred} Consider a constrained zonotope $Z = \{\mbf{G}, \mbf{c}, \mbf{A}, \mbf{b}\}$, with $\mbf{G} \in \realsetmat{n}{n_g}$, $\mbf{c} \in \realset^{n}$, $\mbf{A} \in \realsetmat{n_c}{n_g}$, and $\mbf{b} \in \realset^{n_c}$. Let $Z^+$ denote its \emph{lifted zonotope}, defined by
	\begin{equation*}
	Z^+ \triangleq \left\{ \begin{bmatrix} \mbf{G} \\ \mbf{A} \end{bmatrix}, \begin{bmatrix} \mbf{c} \\ \mbf{-b} \end{bmatrix} \right\} \subset \realset^{n+n_c}.
	\end{equation*}
	Then, it holds that $\mbf{z} \in Z \iff (\mbf{z},\mbf{0}) \in Z^+$ (Proposition 3 in \cite{Scott2016}). Therefore, generator reduction of constrained zonotopes is done as follows. Let $s \in \naturalset$ satisfy $(n+n_c) \leq s < n_g$, and let $\tilde{Z}^+$ be the zonotope with $s$ generators obtained by applying a zonotope order reduction method to $Z^+$. Then, $Z^+ \subseteq \tilde{Z}^+$, and $Z \subseteq \tilde{Z}$, where $\tilde{Z}$ is the constrained zonotope associated to $\tilde{Z}^+$, having $s < n_g$ generators and $n_c$ constraints.
\end{method}

\begin{remark} \rm \label{rem:liftthenreduce}
	Method \ref{meth:czgenred} is used as an intermediate step for generator reduction of constrained zonotopes in all the experiments in this thesis. Therefore, explicit references to this method will be omitted.
\end{remark}

\section{Interval analysis} \label{sec:pre_iarithmetic}

The methods developed in this doctoral thesis require some concepts from interval analysis, which are briefly introduced in this section. 

Interval analysis is a traditional tool for set-based state estimation of nonlinear systems \citep{Jaulin2001}. Let the set of compact intervals in $\realset$ be denoted by $\intvalset$. An \emph{interval} $X \in \intvalset$ is a real compact set defined by $X \triangleq \{ x \in \realset : x^\lbound \leq a \leq x^\ubound \}$, with shorthand notation $X = [x^\lbound,x^\ubound]$, where $x^\lbound, x^\ubound \in \realset$ are called \emph{endpoints}. The midpoint and radius are defined by $\midpoint{X} = \half(x^\ubound + x^\lbound)$ and $\rad{X} = \half(x^\ubound - x^\lbound)$. The diameter is defined by $\diam{X} = 2\rad{X}$. The four basic interval arithmetic operations are defined by $[x^\lbound,x^\ubound] \odot [y^\lbound,y^\ubound] = \{ x \odot y : x \in [x^\lbound,x^\ubound], y \in [y^\lbound,y^\ubound] \}$, where `$\odot$' denotes interval addition, subtraction, multiplication or division (the latter works provided that $0 \notin [y^\lbound,y^\ubound]$). Moreover, for any bounded set $X \subset \realset^n$, let $\square X$ refer to the interval hull of $X$.

\begin{definition} \rm \label{def:intvalvector}
	Given $(\mbf{x}^\text{L},\mbf{x}^\text{U}) \in \realset^{n} \times \realset^{n}$ satisfying $x_j^\text{L} \leq x_j^\text{U}$ for all $j=1,2,\ldots,n$, an \emph{interval vector} $X \in \intvalset^n$ is defined by $X \triangleq \{ \mbf{x} \in \realset^{n} : x_j^\text{L} \leq x_j \leq x_j^\text{U},\, j=1,2,\ldots,n \}$.
\end{definition}

An interval vector $X \in \intvalset^n$ can be written in G-rep as $\text{mid}(X) \oplus \text{diag}(\text{rad}(X)) B_\infty^n = \{\text{diag}(\text{rad}(X)), \text{mid}(X)\}$. Similarly to Definition \ref{def:intvalvector}, interval matrices are defined by $\{ \mbf{A} \in \realset^{n \times m}: A_{ij}^\lbound \leq A_{ij} \leq A_{ij}^\ubound,~i = 1,2, \ldots n,~j=1,2,\ldots,m\} \in \intvalsetmat{n}{m}$. Interval matrix operations are performed according to the rules of real matrix algebra. The midpoint and radius are defined componentwise for interval vectors and matrices \citep{Moore2009}. Moreover, an interval scalar $X = [x^\lbound, x^\ubound] \in \intvalset$ is said to be \emph{degenerate} iff $x^\text{L} = x^\text{U}$.  

\begin{definition} \rm \label{def:degenerate}
	An interval vector $[\mbf{x}^\lbound, \mbf{x}^\ubound] \in \intvalset^n$ is \emph{degenerate} iff $x_j^\text{L} = x_j^\text{U}$ for at least one $j \in \{1,2,\ldots,n\}$. Otherwise, it is said to be \emph{non-degenerate}. 
\end{definition}

Inclusion functions perform a central role in interval analysis. Let $\iextension{\mbf{f}(X)}$ denote an interval enclosure of a real valued function $\mbf{f}$ over an interval $X \subset \realset^n$, where $X$ is an interval enclosure of the domain of interest\footnote{The notation $\iextension{\mbf{f}(X)}$ is used through the doctoral thesis even when $X$ is not an interval. In this case, it is assumed that an interval hull of $X$ is employed in the operation.}. Then, $\mbf{f}(X) \subseteq \iextension{\mbf{f}(X)}$. Inclusion functions of elementary operations, such as $\{\sin, \cos, \ln, \exp\}$, are defined by their images over the respective interval arguments. For instance, the natural inclusion function of a real function is obtained by recursively computing inclusion functions for all arithmetic and elementary operations that compose it \citep{Moore2009}.

\begin{remark} \rm  \label{rem:intervaloverestimation}
	Well-known sources of conservatism in inclusion functions are the \emph{interval dependency} and the \emph{wrapping effect}. The latter results from the inability of an interval vector (or matrix) to capture any existing dependency between its components. On the other side, the interval dependency results from the multiple occurrences of an interval variable in the same algebraic expression \citep{Moore2009}.
\end{remark}

\subsection{Partitioning}

A usual method to reduce the conservatism of inclusion functions is the use of refinements \citep{Moore2009}. In this doctoral thesis, we recall the definition of \emph{interval bundles}, and define \emph{$n_d$-partitions}.

\begin{definition} \rm \label{def:pre_partitions}
	Given $n_b$ interval vectors $B_{(j)} \subset \realset^{n}$, with $j = 1,2,\ldots,n_b$, an \emph{interval bundle} is defined by $\mathcal{B} \triangleq \bigcup_{j=1}^{n_b} B_{(j)}$.
\end{definition}

Clearly an interval bundle is not a interval vector in the general case. However, it can be regarded as a list of interval vectors without computing the union explicitly. Moreover, the concept of an interval bundle is closely related to subpavings.

\begin{remark} \rm \label{rem:subpavings}
	A subpaving is defined as the union of non-overlapping intervals, and it is often associated with a tree structure. Subpavings are usually employed in interval branch-and-bound procedures and set inversion methods \citep{Jaulin2001}.
\end{remark} 

\begin{definition} \rm
	Let $B \in \intvalset^{n}$ be a non-degenerate interval vector. The \emph{$n_d$-partition} of $B$ is an interval bundle $\mathcal{B} = \cup_{j=1}^{n_b} B_{(j)}$ such that $\mathcal{B} = B$, $\text{diam}(B_{(j)}) = (1/n_d)\text{diam}(B)$, and the intersection of two intervals $B_{(j)}$ and $B_{(i)}$ produces a degenerate interval vector in $\realset^{n}$ for every $i\neq j$, with $i,j = 1,2,\dots,n_b$, and $n_d \in \naturalset^+$.
\end{definition}

The $n_d$-partition of an interval vector can be interpreted as a ``multi-dimensional grid'' with $n_d$ divisions per dimension. By definition, the diameter of each interval vector composing the partition is inversely proportional to the number of divisions, and proportional to the diameter of the original interval vector. Note that $n_b = n_d$ for the unidimensional case.

For a real valued function $f(\cdot)$ and a $n_d$-partition $\mathcal{B} = \cup_{j=1}^{n_b} B_{(j)}$ of an interval vector $B$, a \emph{refinement} of an inclusion function $\iextension{f(\cdot)}$ over $\mathcal{B}$ is defined by $\cup_{j=1}^{n_b} \iextension{f(B_{(j)})}$. The conservatism of a refinement is substantially reduced if compared with single inclusion functions over $B$ (see Chapter 6 in \cite{Moore2009} for details), and since each $\iextension{f(B_{(j)})}$ results in an interval vector, a refinement is also an interval bundle.

\section{Linear state estimation} \label{sec:pre_linearestimation}

To better introduce and contextualize the advantages of using constrained zonotopes, this section describes the existing method based on a prediction-update structure, proposed in \cite{Scott2016}, for set-based state estimation of linear discrete-time systems.

Consider a linear discrete-time system with time $k$, state $\mbf{x}_k \in \realset^{n}$, input $\mbf{u}_{k} \in \realset^{n_u}$, process uncertainty $\mbf{w}_k \in \realset^{n_w}$, measured output $\mbf{y}_k \in \realset^{n_y}$, and measurement uncertainty $\mbf{v}_k \in \realset^{n_v}$. In each time interval $[k-1,k]$, $k=1,2,\ldots$, the system evolves according to the model
\begin{equation}
\begin{aligned} \label{eq:pre_linearsystem}
\mbf{x}_k & = \mbf{A} \mbf{x}_{k-1} + \mbf{B} \mbf{u}_{k-1} + \mbf{B}_w \mbf{w}_{k-1}, \\
\mbf{y}_k & = \mbf{C} \mbf{x}_k + \mbf{D} \mbf{u}_{k} + \mbf{D}_v \mbf{v}_{k},
\end{aligned}
\end{equation}
with $\mbf{A} \in \realsetmat{n}{n}$, $\mbf{B} \in \realsetmat{n}{n_u}$, $\mbf{B}_w \in \realsetmat{n}{n_w}$, $\mbf{C} \in \realsetmat{n_y}{n}$, $\mbf{D} \in \realsetmat{n_y}{n_u}$, and $\mbf{D}_v \in \realsetmat{n_y}{n_v}$. The disturbances and uncertainties are assumed to be bounded, i.e., $\mbf{w}_{k} \in W$ and $\mbf{v}_k \in V$, where $W$ and $V$ are known convex polytopic sets.

Given an initial condition $\mbf{x}_0 \in \hat{X}_0$, the prediction-update algorithm consists in computing sets $\bar{X}_k$ and $\hat{X}_k$ such that
\begin{align}
\bar{X}_k & \supseteq \{ \mbf{A} \mbf{x}_{k-1} + \mbf{B} \mbf{u}_{k-1} + \mbf{B}_w \mbf{w}_{k-1} : \mbf{x}_{k-1} \in \hat{X}_{k-1}, \, \mbf{w}_{k-1} \in W \}, \label{eq:pre_linearprediction0}\\
\hat{X}_k & \supseteq \{ \mbf{x}_k \in \bar{X}_k :  \mbf{C} \mbf{x}_k + \mbf{D} \mbf{u}_{k} + \mbf{D}_v \mbf{v}_{k} = \mbf{y}_k, \, \mbf{v}_k \in V \}, \label{eq:pre_linearupdate0}
\end{align}
in which \eqref{eq:pre_linearprediction0} is referred to as the \emph{prediction step}, and \eqref{eq:pre_linearupdate0} as the \emph{update step}. 
In the method proposed in \cite{Scott2016}, both the prediction and update steps are computed using constrained zonotopes, according to
\begin{align} 
	\bar{X}_k & = \mbf{A} \hat{X}_{k-1} \oplus \mbf{B}_u \mbf{u}_{k-1} \oplus \mbf{B}_w W_{k-1}, \label{eq:pre_linearprediction}\\
	\hat{X}_k & = \bar{X}_k \cap_{\mbf{C}} ((\mbf{y}_k - \mbf{D}_u \mbf{u}_k) \oplus (-\mbf{D}_v V)). \label{eq:pre_linearupdate}
\end{align}
Unlike all the other sets mentioned in this chapter, the set operations \eqref{eq:pre_limage}--\eqref{eq:pre_intersection} present in \eqref{eq:pre_linearprediction} and \eqref{eq:pre_linearupdate} are all computed trivially (they are identities) and exactly using constrained zonotopes by \eqref{eq:pre_czlimage}--\eqref{eq:pre_czintersection}, and hence the equalities hold. Therefore, the enclosures in \eqref{eq:pre_linearprediction0} and \eqref{eq:pre_linearupdate0} are obtained accurately and efficiently. In addition, due to the linear increase in the number of generators and constraints, the only source of conservatism comes from the necessity of using complexity reduction methods (Section \ref{sec:complexityreduction}). Nevertheless, the results obtained using constrained zonotopes are compared in \cite{Scott2016} with methods based on polytopes \citep{Shamma1999}, paralellotopes \citep{Chisci1996}, and zonotopes \citep{Bravo2006}, in which constrained zonotopes provided more accurate enclosures for \eqref{eq:pre_linearprediction0}--\eqref{eq:pre_linearupdate0} while having a comparable computational burden.

\section{Final remarks}

This chapter introduced most of the required concepts, notation, and definitions used in this doctoral thesis. The following topic were presented: (i) basic set operations; (ii) zonotopes and their properties; (iii) constrained zonotopes and their properties; (iv) methods for reducing the complexity of zonotopes and constrained zonotopes, (v) interval analysis; and (vi) linear state estimation using constrained zonotopes.

The next chapter develops state estimation methods based on constrained zonotopes for discrete-time systems with nonlinear dynamics for the first time, considering   linear measurement: (i) the constrained zonotope and interval bundle approach; and the generalization of two existing nonlinear zonotopic methods, namely the (ii) mean value extension and the (iii) first-order Taylor extension using constrained zonotopes.

\chapter{State estimation of systems with nonlinear dynamics}\thispagestyle{headings} \label{cha:nonlineardynamics}
\chaptermark{Nonlinear dynamics}

This chapter introduces new nonlinear state estimation methods based on constrained zonotopes. Three novel nonlinear set-based state estimation methods are developed considering linear measurement, namely the constrained zonotope and interval bundle approach, and the consistent generalization of two existing zonotopic methods for nonlinear systems. The latter are referred to as the mean value extension and the first-order Taylor extension using constrained zonotopes.

\section{Problem formulation} \label{sec:ndyn_problemformulation}

Consider a class of discrete-time systems with nonlinear dynamics and linear measurements, described by
\begin{equation}
\begin{aligned} \label{eq:ndyn_system}
\mbf{x}_k & = \mbf{f}(\mbf{x}_{k-1}, \mbf{u}_{k-1}, \mbf{w}_{k-1}), \\
\mbf{y}_k & = \mbf{C} \mbf{x}_k + \mbf{D}_u \mbf{u}_k +\mbf{D}_v \mbf{v}_k,	
\end{aligned}
\end{equation}
for $k \geq 1$, with $\mbf{y}_0 = \mbf{C} \mbf{x}_0 + \mbf{D}_u \mbf{u}_0 +\mbf{D}_v \mbf{v}_0$, where $\mbf{x}_k \in \realset^{n}$ denotes the system state, $\mbf{u}_{k} \in \realset^{n_u}$ is a known input, $\mbf{w}_k \in \realset^{n_w}$ is the process disturbance, $\mbf{y}_k \in \realset^{n_y}$ is the measured output, and $\mbf{v}_k \in \realset^{n_v}$ is the measurement uncertainty, with $\mbf{x}_0$ the initial state. The nonlinear mapping $\mbf{f}$ is assumed to be of class $\mathcal{C}^2$, and the disturbances and uncertainties are assumed to be bounded, i.e., $\mbf{w}_{k} \in W$ and $\mbf{v}_k \in V$, where $W$ and $V$ are known convex polytopic sets.

This chapter proposes new methods to perform set-valued state estimation for nonlinear systems as in \eqref{eq:ndyn_system}. The exact characterization of sets $X_k$ containing the evolution of the system states is very difficult in the nonlinear case, if not intractable \citep{Kieffer1998,Kuhn1998,Platzer2007}. Therefore, in the set-membership framework the objective is to enclose such sets as tightly as possible by guaranteed outer bounds $\hat{X}_k$ on the possible trajectories of the system states $\mbf{x}_k$. Such outer bounds must be consistent with the previous estimate $\hat{X}_{k-1}$, known inputs $\mbf{u}_{k-1}$, the current measurement $\mbf{y}_k$, and also with the bounds on the disturbances and uncertainties $W$, $V$. Given an initial condition $\mbf{x}_0 \in \hat{X}_0$, a common approach is to proceed through the well-known prediction-update algorithm (a generalization of the linear case presented in Section \ref{sec:pre_linearestimation}), which consists in computing compact sets $\bar{X}_k$ and $\hat{X}_k$ such that
\begin{align}
\bar{X}_k & \supseteq \{ \mbf{f}(\mbf{x}_{k-1}, \mbf{u}_{k-1}, \mbf{w}_{k-1}) : \mbf{x}_{k-1} \in \hat{X}_{k-1}, \, \mbf{w}_{k-1} \in W \}, \label{eq:ndyn_prediction0}\\
\hat{X}_k & \supseteq \{ \mbf{x}_k \in \bar{X}_k : \mbf{C} \mbf{x}_k + \mbf{D}_u \mbf{u}_k + \mbf{D}_v \mbf{v}_k = \mbf{y}_k, \, \mbf{v}_k \in V \}, \label{eq:ndyn_update0}
\end{align}
in which \eqref{eq:ndyn_prediction0} is referred to as the \emph{prediction step}, and \eqref{eq:ndyn_update0} as the \emph{update step}. 

The goal is to obtain accurate outer bounds $\bar{X}_k$ and $\hat{X}_k$ according to \eqref{eq:ndyn_prediction0} and \eqref{eq:ndyn_update0}, respectively. Following the definitions of these enclosures as in \eqref{eq:ndyn_prediction0}--\eqref{eq:ndyn_update0}, and considering the initial condition $\mbf{x}_0 \in \hat{X}_0$, the property $\mbf{x}_k \in \hat{X}_k$ is guaranteed by construction for all $k \geq 1$ \citep{Chisci1996,Le2013,Alamo2008}.

\section{Prediction step} \label{sec:ndyn_nonlinearestimation}

This section presents three new different approaches for computing the prediction step \eqref{eq:ndyn_prediction0} based on constrained zonotopes. The first method (Section \ref{sec:ndyn_czib}) combines properties of interval arithmetic and constrained zonotopes, and yields a highly tunable and accurate state estimation algorithm. The other methods (Sections \ref{sec:ndyn_meanvalue} and \ref{sec:ndyn_firstorder}) expand the tools proposed in \cite{Scott2016} to the class of nonlinear discrete-time systems \eqref{eq:ndyn_system}. These methods extend, in a consistent way, two existing approaches for propagating zonotopes through nonlinear mappings. The key advantage of these new extensions is that they allow the entire state estimation procedure to be done using constrained zonotopes in CG-rep. Therefore, the update step can be done by exact intersection (with linear measurements), which is known to generate highly asymmetrical sets that cannot be accurately enclosed by ellipsoids, intervals, parallelotopes, and zonotopes. Using the methods developed in this section, such sets can be directly propagated to the next time step without prior simplification to a symmetric set. This overcomes a major source of conservatism in existing methods based on the aforementioned enclosures, while largely retaining the efficiency of computations with zonotopes. The content regarding the first method (Section \ref{sec:ndyn_czib}) was published in \cite{Rego2018}. The content concerning the two new extensions (Sections \ref{sec:ndyn_meanvalue} and \ref{sec:ndyn_firstorder}) was published in \cite{Rego2020}.

\subsection{CZIB approach} \label{sec:ndyn_czib}

This section presents the first proposed approach for set-based state estimation of the class of nonlinear systems described in Section \ref{sec:ndyn_problemformulation}. The method is referred to as the constrained zonotope and interval bundle (CZIB) approach, and combines important properties from interval arithmetic and constrained zonotopes. A highly tunable and accurate nonlinear state estimation algorithm is developed, capable of providing tight enclosures for the set-based state estimation problem described by \eqref{eq:ndyn_prediction0}--\eqref{eq:ndyn_update0}.

\subsubsection{Outline and assumptions} \label{sec:ndyn_cziboutline}

We first present the underlying ideas of the proposed CZIB approach. An additional assumption is made with respect to the set-based state estimation problem described in Section \ref{sec:ndyn_problemformulation}.
\begin{assumption} \label{ass:ndyn_czib}
	The nonlinear function $\mbf{f}$ is affine in $\mbf{w}$, i.e., $\mbf{f}(\mbf{x},\mbf{u},\mbf{w}) \triangleq \bar{\mbf{f}}(\mbf{x},\mbf{u}) + \mbf{D}_w \mbf{w}$, where $\bar{\mbf{f}} : \realset^n \times \realset^{n_u} \to \realset^n$, and $\mbf{D}_w \in \realsetmat{n}{n_w}$.
\end{assumption}

Assumption \ref{ass:ndyn_czib} is clearly restrictive and leads to a particular reformulation of the prediction step \eqref{eq:ndyn_prediction0}, given by
\begin{equation} \label{eq:ndyn_czibprediction}
	\bar{X}_k \supseteq \{ \bar{\mbf{f}}(\mbf{x}_{k-1}, \mbf{u}_{k-1}) + \mbf{D}_w \mbf{w}_{k-1} : \mbf{x}_{k-1} \in \hat{X}_{k-1}, \, \mbf{w}_{k-1} \in W \}.
\end{equation}
Nevertheless, note that even if $\hat{X}_0$ and $W$ are constrained zonotopes, the reachable sets of the nonlinear system \eqref{eq:ndyn_system} computed over the domains $\hat{X}_{k-1}$ and $W$ are not convex in general \citep{Althoff2013}. Thus, the main challenge is to compute the prediction step \eqref{eq:ndyn_czibprediction} such that the obtained enclosure $\bar{X}_k$ is a constrained zonotope, which is the main contribution of the CZIB approach as the first proposed solution to this problem.

To compute an enclosure $\bar{X}_k$ in CG-rep satisfying \eqref{eq:ndyn_czibprediction}, therefore bounding the possible trajectories of the nonlinear system \eqref{eq:ndyn_system} using constrained zonotopes, we subdivide the prediction step into three main intermediate steps:

\begin{enumerate}
	\item[\emph{\textbf{IS1)}}]\emph{Compute an interval bundle $\standardib$ satisfying $\hat{X}_{k-1} \subseteq \mathcal{B}$;}
	\item[\emph{\textbf{IS2)}}]\emph{Compute a refinement $\particularref$, where $\square(\bar{\mbf{f}}(\cdot))$ is an inclusion function of $\bar{\mbf{f}}(\cdot)$;}
	\item[\emph{\textbf{IS3)}}]\emph{Compute a constrained zonotope $\bar{X}_{k}$ satisfying $\standardref \oplus \mbf{D}_w W \subseteq \bar{X}_k$}.
\end{enumerate}

The intermediate step \emph{IS2} can be performed straightforwardly through interval arithmetic. Therefore, in this section we focus on developing methods to implement the intermediate steps \emph{IS1} and \emph{IS3}. These methods are presented in the next subsection, along with the complete CZIB approach. In addition, a method for computing a constrained zonotope satisfying the update step \eqref{eq:ndyn_update0} is discussed in Section \ref{sec:ndyn_linearupdate}.

\subsubsection{Intermediate steps} \label{sec:ndyn_czibsteps}

In order to obtain a tight enclosure satisfying \emph{IS1} described by an interval bundle, we require the ability to compute the interval hull of a constrained zonotope (Proposition \ref{prope:pre_czihull}). Our goal is to obtain a \emph{tight interval bundle} containing $\hat{X}_{k-1}$. The following proposition shows how to generate the tight interval bundle containing $\hat{X}_{k-1}$ from a given $n_d$-partition of the interval hull of $\hat{X}_{k-1}$ (recall Definition \ref{def:pre_partitions}).

\begin{definition} \rm \label{def:ndyn_tightibundle}
	Let $Z = \{\mbf{G}, \mbf{c}, \mbf{A}, \mbf{b} \} \subset \realset^{n}$. For a given $n_d \in \naturalset^+$, the \emph{tight interval bundle} $\mathcal{B} = \cup_{j=1}^{n_b} B_{(j)}$ with respect to $Z$ is the union of all intervals $B_{(j)}$ composing the $n_d$-partition of the interval hull of $Z$, such that
	\begin{equation} \label{eq:ndyn_tightibundle}
	Z \cap_{\eyenoarg} B_{(j)} \neq \emptyset,
	\end{equation}
	where $\eyenoarg$ denotes the identity matrix. 
\end{definition}

\begin{proposition} \rm \label{prop:ndyn_tightibundle}
	Let $Z = \{\mbf{G}, \mbf{c}, \mbf{A}, \mbf{b} \} \subset \realset^{n}$. Moreover, let $\mathcal{B} = \cup_{j=1}^{n_b} B_{(j)}$ be the tight interval bundle with respect to $Z$ for a given $n_d \in \naturalset^+$, and let $\tilde{\mathcal{B}} = \cup_{j=1}^{\tilde{n}_b} \tilde{B}_{(j)}$ and $\bar{\mathcal{B}} = \cup_{j=1}^{\bar{n}_b} \bar{B}_{(j)} \supset Z$ be interval bundles originated from the $n_d$-partition of the interval hull of $Z$ satisfying $\tilde{n}_b < n_b < \bar{n}_b$. Then, $\mathcal{B} \subset \bar{\mathcal{B}}$, and $Z \nsubseteq \tilde{\mathcal{B}}$.
\end{proposition}
\proof Since $\bar{\mathcal{B}}$ and $\mathcal{B}$ are associated to the same $n_d$-partition, and both $Z \subset \bar{\mathcal{B}}$ and $Z \subset \mathcal{B}$ hold, then at least $n_b$ intervals in $\bar{\mathcal{B}}$ must be equivalent to the intervals in $\mathcal{B}$. Therefore, $\cup_{j=1}^{\bar{n}_b} \bar{B}_{(j)} = (\cup_{j=1}^{n_b} B_{(j)}) \cup (\cup_{j=n_b+1}^{\bar{n}_b} \bar{B}_{(j)})$. From the properties of the set union operation, we have that $(\cup_{j=1}^{n_b} B_{(j)}) \subset (\cup_{j=1}^{n_b} B_{(j)}) \cup (\cup_{j=n_b+1}^{\bar{n}_b} \bar{B}_{(j)})$. Consequently, $\mathcal{B} \subset \bar{\mathcal{B}}$. On the other hand, by definition of $\mathcal{B}$, for every $B_j$ in $\mathcal{B}$, $Z \cap_{\eyenoarg} B_{(j)} \neq \emptyset$ must hold. Thus, $\mbf{z} \in Z \implies \mbf{z} \in B_{(j)}$ for at least one $j$. Let $\tilde{\mathcal{B}}$ be obtained by removing any $B_{(j)}$ from $\mathcal{B}$. Then, $\exists \mbf{z} \in Z : \mbf{z} \notin \tilde{B}_{(j)}$ for all $\tilde{B}_{(j)}$ in $\tilde{\mathcal{B}}$. Consequently, $\tilde{\mathcal{B}} \nsupseteq Z$. \qed

\begin{remark} \rm
	Since every interval vector is a constrained zonotope, the relation \eqref{eq:ndyn_tightibundle} can be easily verified using \eqref{eq:pre_czintersection} and Property \ref{prop:pre_czisemptyinside} for each $B_{(j)}$, from which the tight interval bundle with respect to $\hat{X}_{k-1}$ is effectively obtained. Further reductions in conservativeness can be achieved by increasing the number of divisions $n_d$ when computing the tight interval bundle.
\end{remark}

\begin{figure}[!tb]
	\begin{scriptsize}
		\centering{
			\def\svgwidth{0.7\columnwidth}
			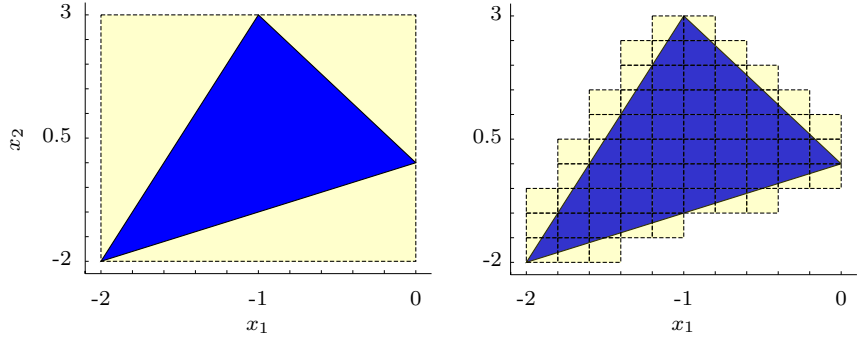
			\caption{Solid lines (blue) depicts a constrained zonotope $Z = \{\mbf{G},\mbf{c},\mbf{A},\mbf{b}\}$ defined by $\mbf{G} = \protect\begin{bmatrix} 1 & 0 & 1 \protect\\ 1 & 2 & -1 \protect\end{bmatrix}$, $\mbf{c} = \protect\begin{bmatrix} 0 \protect\\ 0 \protect\end{bmatrix}$, $\mbf{A} = \protect[ -2 \;\, 1 \;\, {-}1 \protect]$, and $\mbf{b} = 2$. Dashed lines denote the interval hull of $Z$ (left), and the tight interval bundle containing $Z$ considering a 10-partition (right).}\label{fig:ndyn_ihull_tibundle}}
	\end{scriptsize}
\end{figure}

Figure \ref{fig:ndyn_ihull_tibundle} illustrates the underlying ideas regarding Proposition \ref{prop:ndyn_tightibundle}. We are then able to state Algorithm \ref{alg:ndyn_CZIB_IS1}, which performs the intermediate step \emph{IS1}.

\begin{algorithm}[!htb]
	\caption{Intermediate step \emph{IS1}}
	\label{alg:ndyn_CZIB_IS1}
	\small
	\begin{algorithmic}[1]
		\State Compute the interval hull of $\hat{X}_{k-1}$ using Property \ref{prope:pre_czihull}
		\State Compute the $n_d$-partition of the interval hull according to Definition \ref{def:pre_partitions}
		\State Compute the tight interval bundle $\standardib$ containing $\hat{X}_{k-1}$ through Definition \ref{def:ndyn_tightibundle}
	\end{algorithmic}
	\normalsize
\end{algorithm}

Intermediate step \emph{IS2} can be done straightforwardly using interval arithmetic. The associated procedure is presented in Algorithm \ref{alg:ndyn_CZIB_IS2}, which essentially generates a refinement of $\bar{\mbf{f}}(\mbf{x}, \mbf{u}_{k-1})$ over $\standardib$. This procedure can be performed using different inclusion functions\footnote{E.g., natural inclusion function, mean value extension, and others (see \cite{Moore2009} for definitions).}, which may be tested offline for choosing which leads to less conservatism, according to the nonlinear function $\bar{\mbf{f}}$. 

\begin{algorithm}[!htb]
	\caption{Intermediate step \emph{IS2}}
	\label{alg:ndyn_CZIB_IS2}
	\small
	\begin{algorithmic}[1]
		\For {$j = 1,...,n_b$}
		\State Compute an inclusion function of $\bar{\mbf{f}}(\mbf{x}_{k-1}, \mbf{u}_{k-1})$ for $\mbf{x}_{k-1} \in B_{(j)}$, from $\standardib$ obtained through Algorithm \ref{alg:ndyn_CZIB_IS1}
		\EndFor
	\end{algorithmic}
	\normalsize
\end{algorithm}

\begin{remark} \rm \label{rem:ndyn_czibparallelization}
	As in the case of algorithms involving zonotope bundles \citep{Althoff2011}, the computational burden of Algorithm \ref{alg:ndyn_CZIB_IS2} can be reduced using code parallelization techniques.
\end{remark}

In order to propose an algorithm for the intermediate step \emph{IS3}, we first describe a method to compute the interval hull of an interval bundle. Then, we define the concept of \emph{endpoint intervals}, which, in practice, are the intervals that intersect the boundaries of the interval hull of the bundle.

\begin{proposition} \rm \label{prop:ndyn_ibundleihull}
	Let $\mathcal{B} = \cup_{j=1}^{n_b} B_{(j)}$ be an interval bundle in $\realset^{n}$, with $B^{(j)} \triangleq [ \bm{\beta}_{(j)}^{\text{L}}, \bm{\beta}_{(j)}^\text{U}]$. The \emph{interval hull} of $\mathcal{B}$, denoted by $[\bm{\zeta}^\text{L}, \bm{\zeta}^\text{U}]$, is obtained component-wise by searching for $\zeta_i^\text{L} = {\min}\{ \beta_{(j),i}^\text{L} \}$, $\zeta_i^\text{U} = {\max}\{ \beta_{(j),i}^\text{U} \}$
	for each $i = 1,2,\dots,n$, where $\beta_{(j),i}^\text{L}$ and $\beta_{(j),i}^\text{U}$ denote the $i$-th component of $\bm{\beta}_{(j)}^\text{L}$ and $\bm{\beta}_{(j)}^\text{U}$, respectively.
\end{proposition}
\proof The proof follows from Definition \ref{def:intvalvector}. \qed 

\begin{definition} \rm \label{def:ndyn_endpointintval}
	Let $\mathcal{B} = \cup_{j=1}^{n_b} B_{(j)}$ be an interval bundle in $\realset^{n}$. The $j$-th interval vector $B_{(j)}$ is an \emph{endpoint interval} of $\mathcal{B}$ iff at least one component of any of the endpoints of $B_{(j)}$ coincides with at least one component of any endpoint of the interval hull of $\mathcal{B}$.
\end{definition}

To obtain an accurate enclosure, the convex hull of $\particularref$ must be obtained to solve the intermediate step \emph{IS3} with reduced conservatism. However, due to the complexity of this operation, our solution consists in subdividing \emph{IS3} into three other steps:
\begin{itemize}
	\item[\emph{\textbf{IS3a)}}] Build a convex polytope using the midpoints of the endpoint intervals of the bundle $\cup_{j=1}^{n_b} \square(\bar{\mbf{f}}(B_{(j)},\mbf{u}_{k-1}))$ as vertices. Denote it by $X_k^{\alpha}$.
	\item[\emph{\textbf{IS3b)}}] Move the hyperplanes of $X_k^{\alpha}$ to obtain a polytope $X_k^{\beta}$ effectively enclosing the bundle $\particularref$, and convert the resulting set to CG-rep.
\end{itemize}

Algorithm \ref{alg:ndyn_CZIB_IS3a} clarifies the steps necessary to perform \emph{IS3a}.

\begin{algorithm}[!htb]
	\caption{Step \emph{IS3a}}
	\label{alg:ndyn_CZIB_IS3a}
	\begin{algorithmic}[1]
		\State Compute the endpoint intervals of $\particularref$ according to Definition \ref{def:ndyn_endpointintval}
		\State Build a convex polytope in V-rep using the midpoints of the endpoint intervals
		\State Convert the resulting set from V-rep to H-rep
	\end{algorithmic}
\end{algorithm}

\begin{figure}[!htb]
	\begin{scriptsize}
		\centering{
			\def\svgwidth{0.7\columnwidth}
			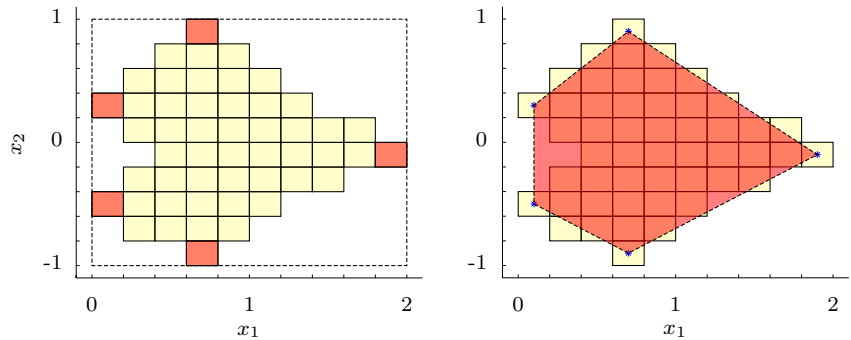
			\caption{Solid lines depicts an interval bundle $\mathcal{B}$. Left: the interval hull (dashed) and the endpoint intervals (red) of $\mathcal{B}$. Right: the auxiliary convex polytope, formed by the midpoints of the endpoint intervals (red, dashed).}\label{fig:ndyn_ivertices_auxcz}}
	\end{scriptsize}
\end{figure}

Figure \ref{fig:ndyn_ivertices_auxcz} demonstrates the essential stages to obtain the auxiliary convex polytope. Algorithm \ref{alg:ndyn_CZIB_IS3a} requires the conversion from V-rep to H-rep, which can be expensive. However, in the proposed state estimation algorithm such conversion is performed only once per iteration. In addition, \emph{IS3b} is solved by means of Algorithm \ref{alg:ndyn_CZIB_IS3b}, which requires the computation of $n_h$ LPs at each time step.

\begin{remark} \rm \label{rem:CZIB_notsimplex}
In some cases, the refinement $\particularref$ may not have enough endpoint intervals to form a simplex in $\realset^{n_x}$ using the corresponding midpoints. In such situation, to generate additional midpoints before performing the conversion to H-rep, a ``small'' zero-centered interval vector can be added to the existing midpoints in V-rep through Minkowski sum.
\end{remark}

\begin{algorithm}[!htb]
	\caption{Step \emph{IS3b}}
	\label{alg:ndyn_CZIB_IS3b}
	\begin{algorithmic}[1]
		\Statex Given $X^{\alpha}_k$ in the H-rep $\{\mbf{x} \in \realset^n : \mbf{H}\mbf{x} \leq\mbf{k}\}, \mbf{k} \in \realset^{n_h}$
		\For{$i = 1,2,\dots n_h$}
		\State ${k}'_i = \underset{k_i}{\min} \{k_i : \mbf{H}_i \mbf{x}^{\{j,l\}} \leq k_i ,\, \forall j,l \}$,
		\Statex $\quad \,\,\,$where $\mbf{x}^{\{j,l\}}$ denotes the $l$-th vertex of the $j$-th interval vector in $\particularref$
		\EndFor
		\State Convert the resulting $\{\mbf{H},\mbf{k}'\}$ to CG-rep using \eqref{eq:pre_hreptocg}
	\end{algorithmic}
\end{algorithm}

We now conclude our solution to \emph{IS3}. Given the constrained zonotope $$X^{\beta}_k \supset \cup_{j=1}^{n_b} \square(\bar{\mbf{f}}(B_{(j)},\mbf{u}_{k-1})),$$ obtained using Algorithms \ref{alg:ndyn_CZIB_IS3a} and \ref{alg:ndyn_CZIB_IS3b}, a constrained zonotope $\bar{X}_k$ bounding the trajectories of the system can be obtained through
\begin{equation} \label{eq:ndyn_CZIBpredictiondone}
\bar{X}_k = (X^{\beta}_k \cap H) \oplus \mbf{D}_w W,
\end{equation}
where $H$ denotes the interval hull of $\particularref$. The intersection with the interval hull $H$ is introduced to obtain tighter bounds, since the vertices of the constrained zonotope $X^{\beta}_k$ may extrapolate $H$. This intersection is computed using \eqref{eq:pre_czintersection}. The linear image and the Minkowski sum in \eqref{eq:ndyn_CZIBpredictiondone} are computed exactly using \eqref{eq:pre_czlimage} and \eqref{eq:pre_czmsum}, respectively.

By solving the intermediate steps \emph{IS1}, \emph{IS2} and \emph{IS3}, the constrained zonotope $\bar{X}_k$ satisfying the prediction step \eqref{eq:ndyn_czibprediction} is finally obtained, and is given by \eqref{eq:ndyn_CZIBpredictiondone}. The complete CZIB approach for the prediction step is summarized in Algorithm \ref{alg:ndyn_czib}.

\begin{algorithm}
	\caption{CZIB approach for the prediction step \eqref{eq:ndyn_prediction0}}
	\label{alg:ndyn_czib}
	\small
	\begin{algorithmic}[1]
		\Statex \emph{Prediction step}
		\State $\quad$ Compute $\standardib \supset \hat{X}_{k-1}$ using Algorithm \ref{alg:ndyn_CZIB_IS1}
		\State $\quad$ Compute $\particularref$ using Algorithm \ref{alg:ndyn_CZIB_IS2}
		\State $\quad$ Compute $\bar{X}_k$ through Algorithms \ref{alg:ndyn_CZIB_IS3a}, \ref{alg:ndyn_CZIB_IS3b} and \eqref{eq:ndyn_CZIBpredictiondone}
	\end{algorithmic}
	\normalsize
\end{algorithm}

The compromise between efficiency and accuracy in the CZIB approach is given by the number of divisions chosen in Algorithm \ref{alg:ndyn_CZIB_IS1}, and the inclusion function chosen to be used in Algorithm \ref{alg:ndyn_CZIB_IS2}. Moreover, due to the proposed transformations performed in the prediction step (from CG-rep to interval bundle and vice-versa), the increase in the complexity of the constrained zonotope $\hat{X}_{k}$ is not associated with the time $k$, but rather with the number of existing endpoint intervals (see Figure \ref{fig:ndyn_ivertices_auxcz}). Therefore, the complexity of the enclosure $\bar{X}_k$ can be limited by choosing a desired number of endpoint intervals among the existing ones according to a chosen criterion (e.g, minimum allowed distance between vertices). 

\begin{remark} \rm \label{rem:ndyn_CZIBbundlesonly}
It is possible to perform both the prediction step \eqref{eq:ndyn_czibprediction} and the update step discussed in Section \ref{sec:ndyn_linearupdate}, using only interval bundles, i.e., without performing transformations to CG-rep and vice-versa. However, in this case the polytopic sets $W$ and $V$ must be outer-approximated by interval bundles, resulting in undesired conservatism. Besides, the linear mapping in \eqref{eq:ndyn_czibprediction} would be subject to the wrapping effect due to the use of pure interval arithmetic. Moreover, in order to perform the update step using interval bundles, a set of states consistent with the measurement $\mbf{y}_k$ must be obtained explicitly \citep{Alamo2005a}. If the latter is also an interval bundle, the intersection of the two bundles requires the computation of intersections of every possible pair-wise combination of intervals contained in the different bundles, therefore leading to a worst-case factorial increase in the number of intervals composing the resulting set.
\end{remark}

\subsection{Mean value extension} \label{sec:ndyn_meanvalue}

This section presents the second new method for enclosing the range of a nonlinear function $\bm{\mu}$ over a set of inputs described by constrained zonotopes. This method is referred to as the \emph{mean value extension} of $\bm{\mu}$ (because it relies on the Mean Value Theorem), and is a consistent generalization of the zonotope-based method proposed in \cite{Alamo2005a}. Due to significant differences with respect to its zonotopic counterpart, a new theorem (Theorem \ref{thm:ndyn_meanvalue}) together with a detailed proof is provided for the new method.

The method in \cite{Alamo2005a} relies on a \emph{zonotope inclusion} operator that computes a zonotopic enclosure of the product of an interval matrix with a unitary box. We first generalize this operator to the product of an interval matrix with a constrained zonotope.

\begin{theorem} \rm \label{thm:ndyn_czinclusion}
	Let $Z = \{\mbf{G},\mbf{c},\mbf{A},\mbf{b}\} \subset \realset^{n_z}$ be a constrained zonotope with $n_g$ generators and $n_c$ constraints, let $\mbf{J} \in \intvalsetmat{n_s}{n_z}$ be an interval matrix, and consider the set $S = \mbf{J} Z \triangleq \{\hat{\mbf{J}} \mbf{z} : \hat{\mbf{J}} \in \mbf{J}, \mbf{z} \in Z\} ~\subset \realset^{n_s}$.
	Let $\bar{Z} = \{\bar{\mbf{G}},\bar{\mbf{c}}\} \subset \realset^{n_z}$ be a zonotope with $\bar{n}_g$ generators satisfying $Z \subseteq \bar{Z}$, let $\mbf{m} \in \intvalset^{n_s}$ be an interval vector such that $\mbf{m} \supseteq (\mbf{J} - \midpoint{\mbf{J}}) \bar{\mbf{c}}$ and $\midpoint{\mbf{m}} = \bm{0}$, and let $\mbf{P} \in \realsetmat{n}{n}$ be a diagonal matrix defined by
	\begin{equation} \label{eq:ndyn_czinclusionP}
	P_{ii} = \half \diam{m_i} +  \half \sum_{j=1}^{\bar{n}_g} \sum_{k=1}^{n_z} \diam{J_{ik}} |\bar{G}_{kj}|,
	\end{equation}
	for all $i=1,2,\dots,n_s$. Then, $S$ is contained in the \emph{CZ-inclusion} operator
	\begin{equation} \label{eq:ndyn_czinclusion}
	S \subseteq \gzinclusion(\mbf{J},Z) \triangleq \midpoint{\mbf{J}}Z \oplus \mbf{P}B_\infty^{n_s}.
	\end{equation}
\end{theorem}

\proof 
Choose any $\mbf{s} \in S$. It will be shown that $\mbf{s} \in \gzinclusion(\mbf{J},Z)$. By the definition of $S$, there must exist $\mbf{z} \in Z$ and $\hat{\mbf{J}} \in \mbf{J}$ such that $\mbf{s} = \hat{\mbf{J}} \mbf{z}$. Adding and subtracting $\midpoint{\mbf{J}} \mbf{z}$,
\begin{equation*}
\mbf{s} = \midpoint{\mbf{J}} \mbf{z} + (\hat{\mbf{J}} - \midpoint{\mbf{J}}) \mbf{z}.	
\end{equation*}
Since $\mbf{z}\in Z\subseteq \bar{Z}$, there exists $\bm{\delta} \in B_\infty^{\bar{n}_g}$ such that $\mbf{z} = \bar{\mbf{c}} + \bar{\mbf{G}} \bm{\delta}$. Therefore, $\mbf{s} = \midpoint{\mbf{J}}\mbf{z} + (\hat{\mbf{J}} - \midpoint{\mbf{J}}) (\bar{\mbf{c}} + \bar{\mbf{G}} \bm{\delta}).$ By the choice of $\mbf{m}$, there must exist $\hat{\mbf{m}}\in\mbf{m}$ such that
\begin{align}
\label{Eq: CZ Inc Proof - z equality}
\mbf{s} &= \midpoint{\mbf{J}}\mbf{x} + \hat{\mbf{m}} + (\hat{\mbf{J}} - \midpoint{\mbf{J}}) \bar{\mbf{G}} \bm{\delta}.
\end{align}
Let $\bm{\eta}= \hat{\mbf{m}}+(\hat{\mbf{J}} - \midpoint{\mbf{J}}) \bar{\mbf{G}} \bm{\delta}$. Then 
\begin{align*}
\eta_{i} &= \hat{m}_i+\sum_{j=1}^{\bar{n}_g} ((\hat{\mbf{J}} - \midpoint{\mbf{J}}) \bar{\mbf{G}})_{ij} \delta_j, \\
&= \hat{m}_i+\sum_{j=1}^{\bar{n}_g} \left(\sum_{k=1}^{n_z} (\hat{J}_{ik}-\midpoint{J_{ik}}) \bar{G}_{kj}\right) \delta_j.
\end{align*}
By the triangle inequality and the fact that $|\delta_j|\leq 1$,
\begin{align*}
|\eta_{i}| &\leq |\hat{m}_i|+\sum_{j=1}^{\bar{n}_g} \left(\sum_{k=1}^{n_z} |(\hat{J}_{ik}-\midpoint{J_{ik}})|| \bar{M}_{kj}|\right)|\delta_j|, \\
&\leq \half \diam{m_i} +  \half \sum_{j=1}^{\bar{n}_g} \sum_{k=1}^{n_z} \diam{J_{ik}} |\bar{M}_{kj}|.
\end{align*}
Therefore, $\bm{\eta}\in \mbf{P}B_\infty^{n_s}$. From \eqref{Eq: CZ Inc Proof - z equality}, this implies that
\begin{equation*}
\mbf{s} = \midpoint{\mbf{J}}\mbf{z} + \bm{\eta}
\in \midpoint{\mbf{J}} Z \oplus \mbf{P} B_\infty^{n_s} = \gzinclusion(\mbf{J},Z).
\end{equation*}
Thus $S \subseteq \gzinclusion(\mbf{J},Z)$.
\qed

\begin{remark} \rm  \label{rem:ndyn_czinclusion}
	In Theorem \ref{thm:ndyn_czinclusion}, a zonotope $\bar{Z}$ satisfying $Z \subseteq \bar{Z}$ can be easily obtained by eliminating all the $n_c$ constraints from $Z$ using the method in \cite{Scott2016}. Moreover, $\mbf{m}$ can be obtained by simply evaluating $(\mbf{J} - \midpoint{\mbf{J}}) \bar{\mbf{c}}$ with interval arithmetic. These methods are used in this doctoral thesis. Finally, the enclosure \eqref{eq:ndyn_czinclusion} has $n_g + n_s$ generators and $n_c$ constraints.
\end{remark}

The following theorem provides the mean value extension for constrained zonotopes. 

\begin{theorem} \rm \label{thm:ndyn_meanvalue}
	Let $\bm{\mu} : \realset^n \times \realset^{n_w} \to \realset^n$ be continuously differentiable and $\nabla_x \bm{\mu}$ denote the gradient of $\bm{\mu}$ with respect to its first argument. Let $X\subset \realset^n$ and $W \subset \realset^{n_w}$ be constrained zonotopes and choose any $\bm{\gamma}_x \in X$. If $Z_w$ is a constrained zonotope such that $\bm{\mu}(\bm{\gamma}_x,W) \subseteq Z_w$, and $\mbf{J}_x \in \intvalsetmat{n}{n}$ is an interval matrix satisfying $\nabla^T_x \bm{\mu}(X, W)\subseteq \mbf{J}_x$, then $$\bm{\mu}(X,W) \subseteq Z_w \oplus \gzinclusion\left(\mbf{J}_x,  X - \bm{\gamma}_x \right).$$
\end{theorem}

\proof Choose any $(\mbf{x},\mbf{w})\in X\times W$. It will be shown that $\bm{\mu}(\mbf{x},\mbf{w}) \in Z \oplus \gzinclusion\left(\mbf{J}_x,  X - \bm{\gamma}_x \right)$. For any $i=1,2,\ldots,n$, the Mean Value Theorem ensures that $\exists \bm{\delta}^{[i]}\in X$ such that
\begin{align*}
\mu_i(\mbf{x},\mbf{w}) &= \mu_i(\bm{\gamma}_x,\mbf{w}) + \nabla^T_x \mu_i(\bm{\delta}^{[i]},\mbf{w}) (\mbf{x} - \bm{\gamma}_x).
\end{align*}
But the vector $\nabla^T_x \mu_i(\bm{\delta}^{[i]},\mbf{w})$ is contained in the $i$-th row of $\mbf{J}_x$ by hypothesis, and since this is true for all $i=1,2,\ldots,n$, there exists a real matrix $\hat{\mbf{J}}_x \in \mbf{J}_x$ such that $\bm{\mu}(\mbf{x},\mbf{w}) = \bm{\mu}(\bm{\gamma}_x,\mbf{w}) + \hat{\mbf{J}}_x (\mbf{x} - \bm{\gamma}_x).$ 

By Theorem \ref{thm:ndyn_czinclusion} and the choice of $Z$, it follows that $\bm{\mu}(\mbf{x},\mbf{w}) \in Z_w \oplus \gzinclusion\left(\mbf{J}, X - \bm{\gamma}_x \right)$, as desired. \qed

\begin{remark} \rm \label{rem:ndyn_affine}
	The interval matrix $\mbf{J}_x$ required by Theorem \ref{thm:ndyn_meanvalue} can be obtained by computing the interval hulls of $X$ and $W$ as in Property \ref{prope:pre_czihull} and, then, bounding $\nabla^T_x \bm{\mu}(X, W)$ using interval arithmetic. Similarly, the constrained zonotope $Z_w \supseteq \bm{\mu}(\bm{\gamma}_x,W)$ can be obtained by bounding $\bm{\mu}(\bm{\gamma}_x,W)$ with interval arithmetic. Alternatively, another mean value extension can be applied around some $\bm{\gamma}_w \in W$ to obtain $\bm{\mu}(\bm{\gamma}_x,W) \subseteq Z_w \triangleq \bm{\mu}(\bm{\gamma}_x,\bm{\gamma}_w) \oplus \gzinclusion\left(\mbf{J}_w,W - \bm{\gamma}_w\right)$, where $\mbf{J}_w$ is an interval enclosure of $\nabla^T_w\bm{\mu}(\bm{\gamma}_x, W)$. Finally, if $\bm{\mu}$ is affine in $\mbf{w}$, i.e, $\bm{\mu}(\mbf{x},\mbf{w}) \triangleq \bm{\beta}_x(\mbf{x}) + \mbf{B}_w(\mbf{x}) \mbf{w}$, then an exact enclosure of $\bm{\mu}(\bm{\gamma}_x,W)$ is $Z_w = \bm{\beta}_x(\bm{\gamma}_x) \oplus \mbf{B}_w(\bm{\gamma}_x) W$, which is a particular case of the mean value extension.
\end{remark}

Since the CG-rep \eqref{eq:pre_cgrep} is an alternative representation for convex polytopes \citep{Scott2016}, the mean value extension developed in Theorem \ref{thm:ndyn_meanvalue} provides a new method for propagating convex polytopes implicitly through nonlinear mappings. A related approach can be found in \cite{Wan2018}, where convex polytopes are represented by intersections of zonotopes (i.e., zonotope bundles \citep{Althoff2011}). However, while effective complexity reduction algorithms are available for constrained zonotopes \citep{Scott2016}, efficient methods for complexity control of zonotope bundles have not yet been proposed. 

\begin{remark} \rm \label{rem:ndyn_meanvaluecomplexity}
	The enclosure obtained in Theorem \ref{thm:ndyn_meanvalue} has at most $n_g + n_{g_w} + 2n$ generators and $n_c + n_{c_w}$ constraints (considering $Z_w$ computed as in the alternatives presented in Remark \ref{rem:ndyn_affine}), with $n_g$ and $n_{g_w}$ denoting the number of generators of $X$ and $W$, and $n_c$ and $n_{c_w}$ the number of constraints, respectively. Thus, the complexity of the resulting set increases linearly with respect to the number of constraints and generators.
\end{remark}

\subsubsection{Selection of approximation point} \label{sec:ndyn_choosehmeanvalue}

The method proposed in this section requires a choice of $\bm{\gamma}_x \in X = \{\mbf{G}_x,\mbf{c}_x,\mbf{A}_x,\mbf{b}_x\}$ in order to compute a constrained zonotope enclosure for the prediction step \eqref{eq:ndyn_prediction0}. As shown in Section \ref{sec:ndyn_example2}, this choice may drastically affect the accuracy of the obtained enclosure. 

In the mean value extension for intervals and zonotopes, a usual choice of $\bm{\gamma}_x \in X$ is the center of $X$ \citep{Alamo2005a,Moore2009}. However, with constrained zonotopes, since the center of the CG-rep may not belong to $X$\footnote{From $X = \mbf{c}_x \oplus \mbf{G}_x B_\infty(\mbf{A}_x,\mbf{b}_x)$, $\mbf{c}_x \notin X$ as long as $\nexists \bm{\xi} \in B_\infty(\mbf{A}_x,\mbf{b}_x)$ satisfying $\mbf{G}_x \bm{\xi} = \bm{0}$.}, a different point $\bm{\gamma}_x \in X$ must be chosen. A simple and inexpensive choice is the center of the interval hull of $X$. Unfortunately, even this point may not belong to $X$ in some cases\footnote{An example is the polytope with vertices $(0,0,0)$, $(1,1,0)$, $(0,1,0)$, $(0,1,1)$.}. Nevertheless, this choice can be applied rigorously by simply checking if $\bm{\gamma}_x \in X$ beforehand by solving an LP \citep{Scott2016}. 

In the following, we analyze alternative choices of $\bm{\gamma}_x \in X$ valid for the mean value extension (Theorem \ref{thm:ndyn_meanvalue}). This extension relies on the CZ-inclusion operator (Theorem \ref{thm:ndyn_czinclusion}) and, therefore, requires the computation of a zonotope enclosing $X - \bm{\gamma}_x$. In this work, we assume that this zonotope is computed through constraint elimination (see Remark \ref{rem:ndyn_czinclusion}). Let $\{\mbf{G}^{(\ell)}, \mbf{c}^{(\ell)}, \mbf{A}^{(\ell)}, \mbf{b}^{(\ell)}\}$ denote the constrained zonotope obtained by reducing to $\ell$ the number of remaining constraints in $X - \bm{\gamma}_x$. Following the constraint elimination algorithm in \cite{Scott2016}, for each $\ell=n_c,n_c-1,\dots,1$, the remaining constraints $\mbf{A}^{(\ell)} \bm{\xi} = \mbf{b}^{(\ell)}$ are first preconditioned through Gauss-Jordan elimination with full pivoting and then subjected to a rescaling procedure before the next constraint is eliminated. The entire procedure can be represented by the following recursive equations (see Proposition 5 and the Appendix in \cite{Scott2016} for details), where $\bar{(\cdot)}$ denotes variables after preconditioning, $\tilde{(\cdot)}$ denotes variables after rescaling, and $\bm{\Lambda}_\text{G}$, $\bm{\Lambda}_\text{A}$, $\bm{\xi}_\text{m}$, and $\bm{\xi}_\text{r}$ are defined as in \cite{Scott2016}:
\begin{equation} \label{eq:ndyn_rescaling}
\begin{aligned}
\tilde{\mbf{c}}^{(\ell)} & = \mbf{c}^{(\ell)} + \bar{\mbf{G}}^{(\ell)} \bm{\xi}_\text{m}^{(\ell)}, & \mbf{c}^{(\ell-1)} & = \tilde{\mbf{c}}^{(\ell)} + \bm{\Lambda}_\text{G}^{(\ell)} \tilde{\mbf{b}}^{(\ell)}, \\ \tilde{\mbf{G}}^{(\ell)} & = \bar{\mbf{G}}^{(\ell)} \text{diag}(\bm{\xi}_\text{r}^{(\ell)}), & \mbf{G}^{(\ell-1)} & = \tilde{\mbf{G}}^{(\ell)} - \bm{\Lambda}_\text{G}^{(\ell)} \tilde{\mbf{A}}^{(\ell)},\\
\tilde{\mbf{A}}^{(\ell)} & = \bar{\mbf{A}}^{(\ell)} \text{diag}(\bm{\xi}_\text{r}^{(\ell)}), & \mbf{A}^{(\ell-1)} & = \tilde{\mbf{A}}^{(\ell)} - \bm{\Lambda}_\text{A}^{(\ell)} \tilde{\mbf{A}}^{(\ell)}, \\
\tilde{\mbf{b}}^{(\ell)} & = \bar{\mbf{b}}^{(\ell)} - \bar{\mbf{A}}^{(\ell)} \bm{\xi}_\text{m}^{(\ell)}, & \mbf{b}^{(\ell-1)} & = \tilde{\mbf{b}}^{(\ell)} - \bm{\Lambda}_\text{A}^{(\ell)} \tilde{\mbf{b}}^{(\ell)}.
\end{aligned}
\end{equation}

Careful examination of the algorithm in \cite{Scott2016} reveals that the actions taken during preconditioning, rescaling, and constraint elimination are all independent of the center of the original constrained zonotope, which in this case is $\mbf{c}^{(n_c)} = \mbf{c}_x - \bm{\gamma}_x$. Therefore, with exception of the center, the variables $({\cdot})^{(\ell)}$ can be obtained by eliminating the constraints of $X$ prior to choosing $\bm{\gamma}_x$. Considering procedure \eqref{eq:ndyn_rescaling}, the following corollary provides a choice of $\bm{\gamma}_x$ that leads to a tight enclosure by reducing the conservativeness of the CZ-inclusion operator $\gzinclusion~(\mbf{J},  X - \bm{\gamma}_x)$.

\begin{corollary} \rm \label{col:ndyn_C2}
	Let $X = \{\mbf{G}_x,\mbf{c}_x,\mbf{A}_x,\mbf{b}_x\} \subset \realset^n$, and consider $\bm{\mu}$, $W$, and $\mbf{J}$ as defined in Theorem \ref{thm:ndyn_meanvalue}. Assume that $\bar{\mbf{G}}^{(\ell)}$, $ \bm{\xi}_\text{m}^{(\ell)}$,$\bm{\Lambda}_\text{G}^{(\ell)}$, and $\tilde{\mbf{b}}^{(\ell)}$ are obtained by eliminating all $n_c$ constraints from $X$ according to \eqref{eq:ndyn_rescaling}, and set
	\begin{equation} \label{eq:ndyn_choice3}
	\bm{\gamma}_x = \mbf{c}_x + \sum_{\ell=1}^{n_c} \left(\bar{\mbf{G}}^{(\ell)} \bm{\xi}_\text{m}^{(\ell)} + \bm{\Lambda}_\text{G}^{(\ell)} \tilde{\mbf{b}}^{(\ell)}\right).
	\end{equation}
	Let $\bar{X} = \{\mbf{G}^{(0)},\mbf{c}^{(0)}\}$ be obtained by eliminating all $n_c$ constraints from $X - \bm{\gamma}_x$ according to \eqref{eq:ndyn_rescaling}, let $\mbf{m} \supseteq (\mbf{J}-\midpoint{\mbf{J}})\mbf{c}^{(0)}$ be computed by standard interval arithmetic, and suppose that $\gzinclusion\left(\mbf{J},  X - \bm{\gamma}_x \right)$ is computed as in Theorem \ref{thm:ndyn_czinclusion} with this choice of $\bar{X}$ and $\mbf{m}$. Finally, let $Z \supseteq \bm{\mu}(\bm{\gamma}_x,W)$. If $\bm{\gamma}_x \in X$, then $\bm{\mu}(X,W) \subseteq  Z \oplus \gzinclusion\left(\mbf{J},  X - \bm{\gamma}_x \right)$. Moreover, $\gzinclusion\left(\mbf{J},  X - \bm{\gamma}_x \right) \subseteq \gzinclusion~(\mbf{J},  X - \hat{\bm{\gamma}}_x)$ for any $\hat{\bm{\gamma}}_x \in X$, $\hat{\bm{\gamma}}_x \neq \bm{\gamma}_x$.
\end{corollary}
\proof 
For $\bm{\gamma}_x \in X$, $\bm{\mu}(X,W) \subseteq  Z \oplus \gzinclusion\left(\mbf{J},  X - \bm{\gamma}_x \right)$ follows directly from Theorem \ref{thm:ndyn_meanvalue}. Now, let us show that $\gzinclusion\left(\mbf{J},  X - \bm{\gamma}_x \right) \subseteq \gzinclusion~(\mbf{J},  X - \hat{\bm{\gamma}}_x)$ holds for any $\hat{\bm{\gamma}}_x \in X$, $\hat{\bm{\gamma}}_x \neq \bm{\gamma}_x$. Recursive computation of \eqref{eq:ndyn_rescaling} leads to
\begin{equation} \label{eq:ndyn_pbar}
\mbf{c}^{(0)} = \mbf{c}^{(n_c)} + \sum_{\ell=1}^{n_c} \left( \bar{\mbf{G}}^{(\ell)} \bm{\xi}_\text{m}^{(\ell)} + \bm{\Lambda}_\text{G}^{(\ell)} \tilde{\mbf{b}}^{(\ell)} \right),
\end{equation}
where $\mbf{c}^{(n_c)} = \mbf{c}_x - \bm{\gamma}_x$. Therefore, $\mbf{c}^{(0)} = \bm{0}$ iff $\bm{\gamma}_x$ is given by \eqref{eq:ndyn_choice3}, and then $\mbf{m} = \bm{0}$, and $\text{diam}(\mbf{m}) = \mbf{0}$. Note that in \eqref{eq:ndyn_czinclusionP}, $\bar{\mbf{M}} \triangleq \mbf{G}^{(0)}$ is invariant with respect to $\bm{\gamma}_x$, and since $\mbf{J} \supseteq \nabla_x^T\bm{\mu}(X,W)$, then $\mbf{J}$ is also invariant with respect to $\bm{\gamma}_x$. Consequently, the second term in \eqref{eq:ndyn_czinclusionP} is not a function of $\bm{\gamma}_x$. Therefore, $\mbf{P} B_\infty^n \subseteq \hat{\mbf{P}}B_\infty^n = (1/2)\text{diag}(\diam{\hat{\mbf{m}}})B_\infty^n \oplus \mbf{P} B_\infty^n$, with $\hat{\mbf{m}}$ computed using $\hat{\bm{\gamma}}_x \in X$. The result then follows from \eqref{eq:ndyn_czinclusion}.
\qed

By Corollary \ref{col:ndyn_C2}, the enclosure obtained in Theorem \ref{thm:ndyn_meanvalue} is tightened by choosing $\bm{\gamma}_x$ such that $\mbf{c}^{(0)}$ is equal to zero. Unfortunately, the $\bm{\gamma}_x$ given by \eqref{eq:ndyn_choice3} may not belong to $X$. Then, an alternative to obtain tight bounds is to reduce the size of the box $\mbf{m}$ by solving 
\begin{equation} \label{eq:ndyn_choice3optimal}
\underset{\bm{\gamma}_x}{\min}~\{\|\diam{\mbf{m}}\|_1 : \bm{\gamma}_x \in X\},
\end{equation}
with $\mbf{m} \supseteq (\mbf{J}-\midpoint{\mbf{J}})\bar{\mbf{p}}$ computed using interval arithmetic, where $\bar{\mbf{p}} \triangleq \mbf{c}^{(0)}$. Recall that $\mbf{c}^{(0)}$ is the center of the zonotope obtained by eliminating all the constraints of $X-\bm{\gamma}_x$.

\begin{lemma} \rm \label{lem:ndyn_choosemindiam}
	Let $X = \{\mbf{G}_x,\mbf{c}_x,\mbf{A}_x,\mbf{b}_x\} \subset \realset^n$, $\mbf{J} \in \intvalsetmat{n}{n}$. Assume that $\bar{\mbf{G}}^{(\ell)}$,  $\bm{\xi}_\text{m}^{(\ell)}$, $\bm{\Lambda}_\text{G}^{(\ell)}$, and $\tilde{\mbf{b}}^{(\ell)}$ are obtained by eliminating all $n_c$ constraints of $X$ according to \eqref{eq:ndyn_rescaling}. Then, $\bm{\gamma}_x = \mbf{c}_x + \mbf{G}_x \bm{\xi}^*$ is the solution to \eqref{eq:ndyn_choice3optimal} iff $\bm{\xi}^*$ is the solution to the linear program
	\begin{equation} \label{eq:ndyn_choosehmindiam}
	\underset{\bm{\xi}}{\min}~ \|\bm{\Theta} \bar{\mbf{p}}\|_1, \quad
	\text{s.t.} \quad \mbf{A}_x \bm{\xi} = \mbf{b}_x, \quad \ninf{\bm{\xi}} \leq 1,
	\end{equation}
	with $\bar{\mbf{p}} = -\mbf{G}_x \bm{\xi} + \sum_{\ell=1}^{n_c} \left( \bar{\mbf{G}}^{(\ell)} \bm{\xi}_\text{m}^{(\ell)} + \bm{\Lambda}_\text{G}^{(\ell)} \tilde{\mbf{b}}^{(\ell)} \right)$, $\Theta_{jj} = \sum_{i=1}^n \diam{J_{ij}}$, and $\Theta_{ij} = 0$ for $i\neq j$.
\end{lemma}
\proof Each element of $(\mbf{J} - \midpoint{\mbf{J}})\in \intvalsetmat{n}{n}$ is a symmetric interval satisfying $(J_{ij} - \midpoint{J_{ij}}) = (1/2) \diam{J_{ij}}[-1,1]$, and for every $a \in \realset$, $a [-1,1] = |a| [-1,1]$ holds. Therefore $m_i = \sum_{j=1}^{n} (1/2) \diam{J_{ij}} |\bar{p}_j| [-1,1]$. Consequently, $\diam{m_i} = \sum_{j=1}^n \diam{J_{ij}} |\bar{p}_j|$, and
\begin{align*}
\|\diam{\mbf{m}}\|_1 & = \sum_{i=1}^n \sum_{j=1}^n \diam{J_{ij}} |\bar{p}_j| = \sum_{j=1}^n \left( \sum_{i=1}^n \diam{J_{ij}} \right) |\bar{p}_j| \\
& = \sum_{j=1}^n \Theta_{jj} |\bar{p}_j| =  \|\bm{\Theta} \bar{\mbf{p}}\|_1.
\end{align*}
The equality $\bar{\mbf{p}} = -\mbf{G}_x \bm{\xi} + \sum_{\ell=1}^{n_c} \left( \bar{\mbf{G}}^{(\ell)} \bm{\xi}_\text{m}^{(\ell)} + \bm{\Lambda}_\text{G}^{(\ell)} \tilde{\mbf{b}}^{(\ell)} \right)$ and the constraints in \eqref{eq:ndyn_choosehmindiam} follow directly from \eqref{eq:ndyn_pbar} and $\bm{\gamma}_x \in X$. \qed

Lemma \ref{lem:ndyn_choosemindiam} yields an optimal choice of $\bm{\gamma}_x \in X$ that can be used in Theorem \ref{thm:ndyn_meanvalue} to reduce conservatism in the CZ-inclusion operator and requires only the solution of an LP. Note that formulating \eqref{eq:ndyn_choosehmindiam} requires the knowledge of $\bar{\mbf{G}}^{(\ell)}$, $\bm{\xi}_\text{m}^{(\ell)}$, $\bm{\Lambda}_\text{G}^{(\ell)}$, and $\tilde{\mbf{b}}^{(\ell)}$, which are obtained from the iterated constraint elimination process. As stated before, constraint elimination can be performed over $X$ to obtain the required data prior to the solution of \eqref{eq:ndyn_choosehmindiam}. Once the optimal $\bm{\gamma}_x$ is obtained, constraint elimination can be repeated, or equivalently, the zonotope obtained using $\bm{\gamma}_x = \bm{0}$ can simply be translated by $-\bm{\gamma}_x$.

\begin{remark} \rm
	Note that if the $\bm{\gamma}_x$ given by Corollary \ref{col:ndyn_C2} belongs to $X$, then this coincides with the solution provided by Lemma \ref{lem:ndyn_choosemindiam}.
\end{remark}

We summarize the proposed choices of $\bm{\gamma}_x \in X$ for use in Theorem \ref{thm:ndyn_meanvalue} as follows:
\begin{itemize}
	\item[\textbf{\textit{C1)}}] $\bm{\gamma}_x$ is given by the center of the interval hull of $X$ if it satisfies $\bm{\gamma}_x \in X$;
	\item[\textbf{\textit{C2)}}] $\bm{\gamma}_x$ is obtained by solving \eqref{eq:ndyn_choosehmindiam}.
\end{itemize}

A comparison between the different choices of $\bm{\gamma}_x \in X$ is illustrated in Section \ref{sec:ndyn_example2}.

\subsection{First-order Taylor extension} \label{sec:ndyn_firstorder}

This section presents the third new method for enclosing the range of a nonlinear function $\bm{\eta}$ over a set of inputs described by constrained zonotopes. This method is referred to as the \emph{first-order Taylor extension} of $\bm{\eta}$ because it relies on a first-order Taylor expansion with a rigorous remainder bound, and is a consistent generalization of the zonotope-based method proposed in \cite{Combastel2005}. In contrast to Theorem \ref{thm:ndyn_meanvalue}, for the sake of simplicity of the proof, this function has only one argument. Even so, it is possible to consider both states and process uncertainties by concatenating them into a single vector. Due to substantial changes with respect to the zonotopic method, the new approach comes with a new theorem (Theorem \ref{thm:ndyn_firstorder}) and a detailed proof. In the main result below, $(\cdot)_{i,:}$ denotes the $i$-th row of a matrix, and $(\cdot)_{ij}$ denotes the element from its $i$-th row and $j$-th column.

\begin{theorem} \label{thm:ndyn_firstorder} \rm
	Let $\bm{\eta}: \realset^{n_z} \to \realset^{n}$ be of class $\mathcal{C}^2$ and $\mbf{z} \in \realset^{n_z}$ denote its argument. Let $Z = \{\mbf{G}, \mbf{c}, \mbf{A}, \mbf{b}\} \subset \realset^{n_z}$ be a constrained zonotope with $m_g$ generators and $m_c$ constraints. For each $q = 1,2,\dots,n$, let $\mbf{Q}^{[q]}\in\mathbb{IR}^{n_g\times n_g}$ and $\tilde{\mbf{Q}}^{[q]}\in\mathbb{IR}^{m_g\times m_g}$ be interval matrices satisfying $\mbf{Q}^{[q]} \supseteq \mbf{H} \eta_q (Z)$ and $\tilde{\mbf{Q}}^{[q]} \supseteq \mbf{G}^T \mbf{Q}^{[q]} \mbf{G}$. Moreover, define        
	\begin{align*}
	& \tilde{c}_q = \trace{\midpoint{\tilde{\mbf{Q}}^{[q]}}}/2, \quad \tilde{\mbf{G}}_{q,:} = \big[ \cdots \,\; \underbrace{\midpoint{\tilde{Q}^{[q]}_{ii}}/2}_{\forall i} \,\; \cdots \,\; \underbrace{\left(\midpoint{\tilde{Q}^{[q]}_{ij}} + \midpoint{\tilde{Q}^{[q]}_{ji}}\right)}_{\forall i<j} \,\; \cdots \big],\\
	& \tilde{\mbf{G}}_{d} = \text{diag}(\mbf{d}), \quad d_q = \sum_{i,j} \left| \rad{\tilde{Q}^{[q]}_{ij}} \right|, \quad \tilde{\mbf{A}} = \left[ \tilde{\mbf{A}}_{\bm{\zeta}} \,\; \tilde{\mbf{A}}_{\bm{\xi}} \,\; \zeros{\frac{m_c}{2}(1+m_c)}{n} \right],\\
	& \tilde{\mbf{A}}_{\bm{\zeta}} = \arrowmatrix{\begin{matrix} & \vdots & \\ \cdots & \half A_{ri} A_{si} & \cdots \\ & \vdots & \end{matrix}}{\forall i}{\forall r \leq s}, \quad \tilde{\mbf{A}}_{\bm{\xi}} = \arrowmatrix{\begin{matrix} & \vdots & \\ \cdots & A_{ri} A_{sj} + A_{rj} A_{si} & \cdots \\ & \vdots & \end{matrix}}{\forall i < j}{\forall r \leq s}, \\
	& \tilde{\mbf{b}} = \downarrowmatrix{\begin{matrix} \vdots \\ b_{r} b_{s} - \half \sum_i A_{ri} A_{si} \\ \vdots\end{matrix}}{\forall r \leq s},
	\end{align*}
	with indices $i,j = 1,2,\dots,m_g$ and $r,s = 1,2,\dots,m_c$. Finally, choose any $\bm{\gamma}_z \in Z$ and let $\mbf{L}\in\mathbb{IR}^{n\times m}$ be an interval matrix satisfying $\mbf{L}_{q,:} \supseteq (\mbf{c} - \bm{\gamma}_z)^T \mbf{Q}^{[q]}$ for all $q = 1,\dots,n$. Then,
	\begin{equation} \label{eq:ndyn_firstorderextension}
	\bm{\eta}(Z) \subseteq \bm{\eta}(\bm{\gamma}_z) \oplus \nabla^T \bm{\eta}(\bm{\gamma}_z)(Z - \bm{\gamma}_z) \oplus R,
	\end{equation}
	where $R = \tilde{\mbf{c}} \oplus [\tilde{\mbf{G}} \,\; \tilde{\mbf{G}}_{d}] B_\infty(\tilde{\mbf{A}}, \tilde{\mbf{b}}) \oplus \gzinclusion (\mbf{L}, (\mbf{c} - \bm{\gamma}_z) \oplus 2\mbf{G} B_\infty(\mbf{A},\mbf{b}) )$.
\end{theorem}

\proof
Choose any $\mbf{z}\in Z$ and $q\in\{1,\ldots,n\}$. By Taylor's theorem applied to $\eta_q$ with reference point $\bm{\gamma}_z$, there must exist $\bm{\Gamma}^{[q]} \in \mbf{H} \eta_q (Z) \subseteq \mbf{Q}^{[q]}$ such that\footnote{Let $\bm{\Upsilon}^{[q]}$ belong to the standard Hessian matrix of $\eta_q (Z)$. Then, $(1/2)(\mbf{z} - \bm{\gamma}_z)^T \bm{\Upsilon}^{[q]} (\mbf{z} - \bm{\gamma}_z) =  (\mbf{z} - \bm{\gamma}_z)^T \bm{\Gamma}^{[q]} (\mbf{z} - \bm{\gamma}_z)$ holds. See \citep{Combastel2005} for a motivation on this approach.}
\begin{equation*}
\eta_q(\mbf{z}) = \eta_q (\bm{\gamma}_z) + \nabla^T \eta_q (\bm{\gamma}_z)(\mbf{z} - \bm{\gamma}_z) + (\mbf{z} - \bm{\gamma}_z)^T \bm{\Gamma}^{[q]} (\mbf{z} - \bm{\gamma}_z).
\end{equation*}
Since $\mbf{z}\in Z$, there must exist $\bm{\xi} \in B_\infty(\mbf{A},\mbf{b})$ such that $\mbf{z}=\mbf{c} + \mbf{G}\bm{\xi}$. Thus, defining $\mbf{p} = \mbf{c} - \bm{\gamma}_z$ for brevity,
\begin{equation*}
\eta_q(\mbf{z}) = \eta_q (\bm{\gamma}_z) + \nabla^T \eta_q (\bm{\gamma}_z)(\mbf{p} + \mbf{G}\bm{\xi}) + (\mbf{p} + \mbf{G}\bm{\xi})^T \bm{\Gamma}^{[q]}  (\mbf{p} + \mbf{G}\bm{\xi}).
\end{equation*}
Expanding the product $(\mbf{p} + \mbf{G}\bm{\xi})^T \bm{\Gamma}^{[q]}  (\mbf{p} + \mbf{G}\bm{\xi})$ yields $\mbf{p}^T \bm{\Gamma}^{[q]} (\mbf{p} + 2 \mbf{G} \bm{\xi}) + \bm{\xi}^T \tilde{\bm{\Gamma}}^{[q]} \bm{\xi}$, 
with $\tilde{\bm{\Gamma}}^{[q]} = \mbf{G}^T \bm{\Gamma}^{[q]} \mbf{G} \in \tilde{\mbf{Q}}^{[q]}$.
Since $\tilde{\bm{\Gamma}}^{[q]} \in \tilde{\mbf{Q}}^{[q]}$, it follows that $\tilde{\Gamma}^{[q]}_{ij} = \midpoint{\tilde{Q}^{[q]}_{ij}} + \rad{\tilde{Q}^{[q]}_{ij}} \Lambda^{[q]}_{ij}$ for some $\Lambda^{[q]}_{ij} \in B_\infty^1$. Additionally, $\xi_i \in [-1,1]$ implies that $\xi_i^2 \in [0, 1]$, and hence $\xi_i^2 = \half + \half \zeta_i$ for some $\zeta_i \in [-1,1]$. Considering these two facts,
\begin{align*}
\bm{\xi}^T \tilde{\bm{\Gamma}}^{[q]} \bm{\xi} & = \half \sum_i \midpoint{\tilde{Q}^{[q]}_{ii}} + \half \sum_i \midpoint{\tilde{Q}^{[q]}_{ii}} \zeta_i \\
& \quad + \sum_{i<j}(\midpoint{\tilde{Q}^{[q]}_{ij}} + \midpoint{\tilde{Q}^{[q]}_{ji}}) \xi_i \xi_j + \sum_{i,j} \rad{\tilde{Q}^{[q]}_{ij}} \xi_i \xi_j \Lambda^{[q]}_{ij},
\end{align*}
where the third summation results from the fact that $\xi_i \xi_j = \xi_j \xi_i$. Thus, by defining the new generator variables
\begin{equation*} 
\bar{\bm{\xi}} = \big[ \,\; \cdots \,\; \underbrace{\zeta_i}_{\forall i} \,\; \cdots \,\; \underbrace{\xi_i \xi_j}_{\forall i<j} \,\; \cdots \,\; \underbrace{\xi_i \xi_j \Lambda^{[q]}_{ij}}_{\forall i,j,q} \,\; \cdots \,\; \big]^T,
\end{equation*}
with $i,j = 1,2,\dots,m_g$, $q = 1,2,\dots,n$, we have that $\bm{\xi}^T \tilde{\bm{\Gamma}}^{[q]} \bm{\xi} = \tilde{c}_q + [\tilde{\mbf{G}} \,\; \bar{\mbf{G}}_{d}]_{q,:} \bar{\bm{\xi}}$, where
$\bar{\mbf{G}}_{d} = \text{blkdiag}(\mbf{N}^{[1]}, \mbf{N}^{[2]}, \dots, \mbf{N}^{[n]})$\footnote{In this work $\text{blkdiag}(\mbf{A},\mbf{B},\dots)$ denotes a block diagonal matrix with blocks $\mbf{A},\mbf{B},\dots$.}, 
\begin{equation*}
\mbf{N}^{[q]} = \big[ \,\; \cdots \,\; \underbrace{\rad{\tilde{Q}^{[q]}_{ij}}}_{\forall i,j} \,\; \cdots \,\; \big] \in \realsetmat{1}{m_g^2}.
\end{equation*}

Therefore, we have established that $\eta_q(\mbf{z}) = \eta_q (\bm{\gamma}_z) + \nabla^T \eta_q (\bm{\gamma}_z)(\mbf{z} - \bm{\gamma}_z) + \mbf{p}^T \bm{\Gamma}^{[q]} (\mbf{p} + 2 \mbf{G} \bm{\xi}) + \tilde{c}_q + [\tilde{\mbf{G}} \,\; \bar{\mbf{G}}_{d}]_{q,:} \bar{\bm{\xi}}.$ This holds for every $q = 1,2,\dots,n$. Moreover, $\mbf{L}$ satisfies $\mbf{L}_{q,:} \supseteq \mbf{p}^T \mbf{Q}^{[q]}$ for all $q = 1,2,\dots,n$ by definition, so there must exist $\hat{\mbf{L}}\in\mbf{L}$ such that $\hat{\mbf{L}}_{q,:}=\mbf{p}^T \bm{\Gamma}^{[q]}$ for all $q = 1,2,\dots,n$. Therefore,
\begin{equation}
\label{eq:ndyn_eta vector expansion}
\bm{\eta}(\mbf{z}) = \bm{\eta}(\bm{\gamma}_z) + \nabla^T \bm{\eta} (\bm{\gamma}_z)(\mbf{z} - \bm{\gamma}_z)+ \hat{\mbf{L}}(\mbf{p} + 2 \mbf{G} \bm{\xi}) + \tilde{\mbf{c}} + [\tilde{\mbf{G}} \,\; \bar{\mbf{G}}_{d}] \bar{\bm{\xi}}.
\end{equation}

Furthermore, the equality constraints $\mbf{A} \bm{\xi} = \mbf{b}$ imply that $\mbf{A} \bm{\xi} \bm{\xi}^T \mbf{A}^T = \mbf{b} \mbf{b}^T$. Thus, considering $\xi_i^2 = \half + \half \zeta_i$, the $r$-th row and $s$-th column of this matrix equality yields
\begin{equation*}
\half \sum_i A_{ri} A_{si} \zeta_i + \sum_{i<j} (A_{ri} A_{sj} + A_{rj} A_{si}) \xi_i \xi_j = b_r b_s - \half \sum_i A_{ri} A_{si},
\end{equation*}
with $r,s = 1,2,\dots,m_c$.
Such constraints are linear in $\bar{\bm{\xi}}$ and non-repeating for $r \leq s$, therefore $\bar{\mbf{A}} \bar{\bm{\xi}} = \tilde{\mbf{b}}$ holds, where $\bar{\mbf{A}} = [\tilde{\mbf{A}}_{\bm{\zeta}} \,\; \tilde{\mbf{A}}_{\bm{\xi}} \,\; \bm{0}_{\tilde{m}_c \times n m_g^2}]$, with $\tilde{m}_c = \frac{m_c}{2}(1+m_c)$. Hence, $\bm{\xi} \in B_\infty(\mbf{A},\mbf{b}) \implies \bar{\bm{\xi}} \in B_\infty(\bar{\mbf{A}},\tilde{\mbf{b}})$. Combining this with \eqref{eq:ndyn_eta vector expansion}, we have proven the enclosure $\bm{\eta}(Z) \subseteq \bm{\eta} (\bm{\gamma}_z) \oplus \nabla^T \bm{\eta} (\bm{\gamma}_z)(Z - \bm{\gamma}_z) \oplus \mbf{L} (\mbf{p} \oplus 2 \mbf{G} B_\infty(\mbf{A},\mbf{b})) \oplus \tilde{\mbf{c}} \oplus [\tilde{\mbf{G}} \,\; \bar{\mbf{G}}_{d}] B_\infty (\bar{\mbf{A}}, \tilde{\mbf{b}})$.

In fact, this enclosure can be greatly simplified by noting that the columns of $\bar{\mbf{A}}$ corresponding to the variables $[\,\; \cdots \,\; \xi_i \xi_j \Lambda^{[q]}_{ij}\,\; \cdots \,\;]$ are all zero, and hence 
\begin{equation*}
[\tilde{\mbf{G}} \,\; \bar{\mbf{G}}_{d}] B_\infty (\bar{\mbf{A}}, \tilde{\mbf{b}}) = \tilde{\mbf{G}} B_\infty ([\tilde{\mbf{A}}_{\bm{\zeta}} \,\; \tilde{\mbf{A}}_{\bm{\xi}}], \tilde{\mbf{b}}) \oplus \bar{\mbf{G}}_{d} B_\infty^{nm_g^2}.
\end{equation*}
Since $\bar{\mbf{G}}_{d}$ is block diagonal and each $\mbf{N}^{[q]}$ is a row vector, $\bar{\mbf{G}}_{d} B_\infty^{nm_g^2}$ is an interval and is equivalent to $\tilde{\mbf{G}}_{d} B_\infty^{n}$, with $\tilde{\mbf{G}}_{d}$ defined as in the statement of the theorem. Thus,
\begin{align*}
[\tilde{\mbf{G}} \,\; \bar{\mbf{G}}_{d}] B_\infty (\bar{\mbf{A}}, \tilde{\mbf{b}}) &= \tilde{\mbf{G}} B_\infty ([\tilde{\mbf{A}}_{\bm{\zeta}} \,\; \tilde{\mbf{A}}_{\bm{\xi}}], \tilde{\mbf{b}}) \oplus \tilde{\mbf{G}}_{d} B_\infty^{n}, \\
&= [\tilde{\mbf{G}} \,\; \tilde{\mbf{G}}_{d}] B_\infty ([\tilde{\mbf{A}}_{\bm{\zeta}} \,\; \tilde{\mbf{A}}_{\bm{\xi}} \,\; \mbf{0}_{\tilde{m}_c\times n}], \tilde{\mbf{b}}) = [\tilde{\mbf{G}} \,\; \tilde{\mbf{G}}_{d}] B_\infty (\tilde{\mbf{A}}, \tilde{\mbf{b}}).
\end{align*}
Therefore, $\bm{\eta}(Z) \subseteq \bm{\eta} (\bm{\gamma}_z) \oplus \nabla^T \bm{\eta} (\bm{\gamma}_z)(Z - \bm{\gamma}_z)
\oplus \mbf{L} (\mbf{p} \oplus 2 \mbf{G} B_\infty(\mbf{A},\mbf{b})) \oplus \tilde{\mbf{c}} \oplus [\tilde{\mbf{G}} \,\; \tilde{\mbf{G}}_{d}] B_\infty (\tilde{\mbf{A}}, \tilde{\mbf{b}}),$ 
and \eqref{eq:ndyn_firstorderextension} follows immediately from the definition of $R$. \qed

\begin{remark} \rm
	Regarding the definitions of $\tilde{\mbf{G}}$, $\tilde{\mbf{A}}_{\bm{\zeta}}$, $\tilde{\mbf{A}}_{\bm{\xi}}$, and $\tilde{\mbf{b}}$ in Theorem \ref{thm:ndyn_firstorder}, the ordering of the indices $i<j$ and $r\leq s$ is irrelevant, as long as it is the same for all variables.
\end{remark}

\begin{remark} \rm
	The interval matrices $\mbf{Q}^{[q]}$ required by Theorem \ref{thm:ndyn_firstorder} can be obtained by computing the interval hull of $Z$ (Property \ref{prope:pre_czihull}) and then bounding $\mbf{H} \eta_q (Z)$ using interval arithmetic. Moreover, $\tilde{\mbf{Q}}^{[q]}$ and $\mbf{L}$ can be obtained by evaluating $\mbf{G}^T \mbf{Q}^{[q]} \mbf{G}$ and $(\mbf{c} - \bm{\gamma}_z)^T \mbf{Q}^{[q]}$ using interval arithmetic.
\end{remark}

As stated before, process disturbances can be taken into account in \eqref{eq:ndyn_firstorderextension} by considering the augmented vector $\mbf{z} = (\mbf{x},\mbf{w})$ with $Z \triangleq X \times W \subset \realset^{n+n_w}$ and $\bm{\gamma}_z \triangleq (\bm{\gamma}_x,\bm{\gamma}_w) \in Z$. With $X = \{\mbf{G}_x, \mbf{c}_x, \mbf{A}_x, \mbf{b}_x\}$ and $W = \{\mbf{G}_w, \mbf{c}_w, \mbf{A}_w, \mbf{b}_w\}$, the Cartesian product $Z$ is easily computed by
\begin{equation*}
	X \times W = \left\{ \begin{bmatrix} \mbf{G}_x & \bm{0} \\ \bm{0} & \mbf{G}_w \end{bmatrix}, \begin{bmatrix} \mbf{c}_x \\ \mbf{c}_w \end{bmatrix}, \begin{bmatrix} \mbf{A}_x & \bm{0} \\ \bm{0} & \mbf{A}_w \end{bmatrix}, \begin{bmatrix} \mbf{b}_x \\ \mbf{b}_w \end{bmatrix} \right\}.
\end{equation*}

\begin{remark} \rm \label{rem:ndyn_firstordercomplexity}
	In Theorem \ref{thm:ndyn_firstorder}, $\tilde{\mbf{G}}$ has $\sum_{j=1}^{m_g}j = \frac{1}{2}m_g(m_g+1)$ columns, $\tilde{\mbf{G}}_{\mbf{d}} \in \realsetmat{n}{n}$, $\tilde{\mbf{A}}$ has $\sum_{s=1}^{m_c}s = \frac{1}{2}m_c(m_c+1)$ rows, and $\gzinclusion (\mbf{L}, (\mbf{c} - \bm{\gamma}_z) \oplus 2\mbf{G} B_\infty(\mbf{A},\mbf{b}) )$ has $m_g+n$ generators and $m_c$ constraints (Remark \ref{rem:ndyn_czinclusion}). Therefore, the resulting enclosure in \eqref{eq:ndyn_firstorderextension} has $\frac{1}{2}m_g^2 + \frac{5}{2}m_g + 2n$ generators and $\frac{1}{2}m_c^2 + \frac{5}{2}m_c$ constraints. If $Z \triangleq X \times W$, then the enclosure has $\frac{1}{2}(n_g + n_{g_w})^2 + \frac{5}{2}(n_g + n_{g_w}) + 2n$ generators and $\frac{1}{2}(n_c + n_{c_w})^2 + \frac{5}{2}(n_c + n_{c_w})$ constraints, which is a polynomial increase in complexity in terms of both generators and constraints.
\end{remark}

\subsubsection{Selection of approximation point} \label{sec:ndyn_choosehfirstorder}

In this section, we focus on proposing suitable choices of $\bm{\gamma}_z \triangleq (\bm{\gamma}_x,\bm{\gamma}_w) \in X \times W$ valid for the first-order Taylor extension (Theorem \ref{thm:ndyn_firstorder}). As with the mean value extension, the usual choice of $\bm{\gamma}_z \in X \times W$ in first-order Taylor extensions for intervals and zonotopes is the center of $X$ \citep{Moore2009,Combastel2005}. The next corollary shows that this choice leads to a tight enclosure if it in fact belongs to $X \times W$.

\begin{corollary} \rm \label{col:ndyn_C3}
	Let $Z = \{\mbf{G},\mbf{c},\mbf{A},\mbf{b}\} = X \times W \subset \realset^{n + n_w}$, and consider $\bm{\eta}$, $\tilde{\mbf{c}}$, $\tilde{\mbf{G}}$, $\tilde{\mbf{G}}_\mbf{d}$, $\tilde{\mbf{A}}$, $\tilde{\mbf{b}}$, $\mbf{L}$, and $\mbf{Q}^{[q]}$ as defined in Theorem \ref{thm:ndyn_firstorder}, with $q \in \{1,2,\ldots,n\}$. If $\bm{\gamma}_z = \mbf{c} \in Z$, then $\bm{\eta}(Z) \subseteq \bm{\eta}(\bm{\gamma}_z) \oplus \nabla^T \bm{\eta}(\bm{\gamma}_z)(Z - \bm{\gamma}_z) \oplus \tilde{\mbf{c}} \oplus [ \tilde{\mbf{G}} \,\; \tilde{\mbf{G}}_{\mbf{d}} ] B_\infty(\tilde{\mbf{A}}, \tilde{\mbf{b}})$.
\end{corollary}
\proof 
For $\bm{\gamma}_z = \mbf{c}$, $\mbf{L}_{q,:} \supseteq (\mbf{c} - \bm{\gamma}_z)^T \mbf{Q}^{[q]} = \bm{0} $, $q = 1,2,\ldots,n$. Therefore $\mbf{L} = \mbf{0}$ holds, and $\gzinclusion (\mbf{L}, (\mbf{c} - \bm{\gamma}_z) \oplus 2\mbf{G} B_\infty(\mbf{A},\mbf{b}) ) = \{\mbf{0}\}$. The result then follows from \eqref{eq:ndyn_firstorderextension}. 
\qed

By inspecting Corollary \ref{col:ndyn_C3}, it is clear that the enclosure in \eqref{eq:ndyn_firstorderextension} is tightened since $\gzinclusion (\mbf{L}, (\mbf{c} - \bm{\gamma}_z) \oplus 2\mbf{G} B_\infty(\mbf{A},\mbf{b}) ) = \{\mbf{0}\}$. However, the presence of the equality constraints in $X \times W$ may imply that its center is not in $X \times W$ (i.e., when $\nexists \bm{\xi} \in  B_\infty(\mbf{A},\mbf{b})$ such that $\mbf{G}\bm{\xi} = \mbf{0}$). Therefore, a good alternative may be to consider the closest point in $X \times W$ to its center, obtained by means of Proposition \ref{propo:ndyn_closest}. By the definition of $\mbf{L}$, this heuristic leads to smaller values of $\text{diam}(\mbf{L})$, and therefore reduces the size of $\gzinclusion~(\mbf{L}, (\mbf{c} - \bm{\gamma}_z) \oplus 2\mbf{G} B_\infty(\mbf{A},\mbf{b}) )$ (see \eqref{eq:ndyn_czinclusionP}). 

\begin{proposition} \rm \label{propo:ndyn_closest} 
	Let $Z = \{\mbf{G}, \mbf{c}, \mbf{A}, \mbf{b} \} \subset \realset^{n}$ and $\mbf{s} \in \realset^{n}$. A point $\mbf{z} \in Z$ that minimizes $\none{\mbf{z} - \mbf{s}}$ is given by $\mbf{z}^* = \mbf{c} + \mbf{G} \bm{\xi}^*$, where $\bm{\xi}^*$ is a solution to the linear program
	\begin{equation*}
	\underset{\bm{\xi}}{\min}~\| \mbf{c} - \mbf{s} + \mbf{G} \bm{\xi} \|_1 \quad \text{\rm s.t.} \quad \mbf{A}\bm{\xi} = \mbf{b}, \quad \ninf{\bm{\xi}} \leq 1.
	\end{equation*}
\end{proposition}
\proof By definition,
\begin{align*}
\none{\mbf{z}^* - \mbf{s}} &= \none{\mbf{c} - \mbf{s} + \mbf{G} \bm{\xi}^*} \\
&\leq \none{\mbf{c} - \mbf{s} + \mbf{G} \bm{\xi}}, \quad \forall \bm{\xi}\in B_{\infty}(\mbf{A},\mbf{b}).
\end{align*}
But, for any $\mbf{z}\in Z$, there exists $\bm{\xi}\in B_{\infty}(\mbf{A},\mbf{b})$ such that $\mbf{z} = \mbf{c} + \mbf{G} \bm{\xi}$. Therefore, $\none{\mbf{z}^* - \mbf{s}} \leq \none{\mbf{z} - \mbf{s}}$, $\forall \mbf{z}\in Z$.
\qed

On the other hand, a third option is to use Proposition \ref{propo:ndyn_centerchange} to obtain an alternative CG-rep of $X\times W$ with any desired center. In this case, the new center is chosen as some point in $X$, $\bar{\bm{\gamma}}_z \in X \times W$, and then $\bm{\gamma}_z$ is chosen as $\bm{\gamma}_z = \bar{\bm{\gamma}}_z$. 

\begin{proposition}[Move center] \rm \label{propo:ndyn_centerchange}
	Let $Z = \{\mbf{G},\mbf{c},\mbf{A},\mbf{b}\} \subset \realset^n$, and let $\tilde{\bm{\xi}}^\text{L}, \tilde{\bm{\xi}}^\text{U} \in \realset^{n_g}$ satisfy $\conball \subseteq [\tilde{\bm{\xi}}^\text{L},\,\tilde{\bm{\xi}}^\text{U}]$. Choose any desired center $\hat{\mbf{c}} \in \realset^n$ lying in the range of $\mbf{G}$ and let $\xil, \xiu \in \realset^{n_g}$ be solutions to the linear program
	\begin{align*}
	\underset{\xil,\xiu}{\min}~&\left\| \half (\xiu - \xil) \right\|_1 \\
	\text{s.t.} \quad & \mbf{c} + \half \mbf{G} (\xil + \xiu) = \hat{\mbf{c}}, \quad \xil \leq \tilde{\bm{\xi}}^\text{L}, \quad \xiu \geq \tilde{\bm{\xi}}^\text{U}.
	\end{align*}
	Letting $\xim = \half (\xil + \xiu)$ and $\mbf{E}_r = \half\text{diag}(\xiu - \xil)$, an equivalent CG-rep of $Z$ with center $\hat{\mbf{c}}$ is given by
	\begin{align} 
	Z = \left\{\begin{bmatrix} \mbf{G} \mbf{E}_r & \bm{0} \end{bmatrix}, \hat{\mbf{c}}, \begin{bmatrix} \mbf{A} \mbf{E}_r & \bm{0} \\ \bm{0} & \mbf{A} \\ \mbf{G} \mbf{E}_r & -\mbf{G} \end{bmatrix} , \begin{bmatrix}\mbf{b} - \mbf{A} \xim  \\ \mbf{b} \\ \mbf{c} - \hat{\mbf{c}} \end{bmatrix}  \right\}. \label{eq:ndyn_centerchange}
	\end{align}
\end{proposition}

\proof It is first shown that $Z$ is contained in the set 
\begin{equation*}
\bar{Z} = \{ \mbf{GE}_r,\,\mbf{c} + \mbf{G} \xim,\, \mbf{AE}_r,\, \mbf{b} - \mbf{A} \xim \}.
\end{equation*}
Choose any $\mbf{z} \in Z$. There must exist $\bm{\xi} \in B_\infty (\mbf{A},\mbf{b})$ such that $\mbf{z} = \mbf{c} + \mbf{G} \bm{\xi}$. Since $\bm{\xi} \in [\xil,\,\xiu]$, there must exist $\bm{\delta} \in B_\infty^{n_g}$ such that $ \bm{\xi} = \xim + \mbf{E}_r\bm{\delta}$. Thus,
\begin{align*}
\mbf{z} \in Z & \implies \exists \bm{\delta} \in B_\infty^{n_g} : \mbf{z} = \mbf{c} + \mbf{G} (\xim + \mbf{E}_r \bm{\delta}), \\ & \qquad \qquad \qquad \quad \; \mbf{A}(\xim + \mbf{E}_r \bm{\delta})= \mbf{b} \implies \mbf{z} \in \bar{Z}.
\end{align*}
Therefore, $Z \subseteq \bar{Z}$ and it is true that $Z=\bar{Z} \cap Z$. Since $\xil$ and $\xiu$ satisfy $\mbf{c} + \half \mbf{G} (\xil + \xiu) = \mbf{c} + \mbf{G} \xim = \hat{\mbf{c}}$, then representing $\bar{Z} \cap Z$ as in \eqref{eq:pre_czintersection} gives \eqref{eq:ndyn_centerchange}. \qed

\begin{remark} \rm \label{rem:ndyn_centerchange}
	The linear program in Proposition \ref{propo:ndyn_centerchange} does not require $[\xil, \xiu] \subseteq B_\infty^{n_g}$. Therefore, the midpoint $\half (\xil + \xiu)$ can assume any desired value, and it is always possible to satisfy $\mbf{c} + \half \mbf{G} (\xil + \xiu) = \hat{\mbf{c}}$ if $\hat{\mbf{c}}$ is in the range of $\mbf{G}$. Thus, the linear program is always feasible.
\end{remark}

\begin{remark} \rm The optimization problems in this doctoral thesis can be readily rewritten as standard form programs by using additional decision variables and constraints \citep{Boyd2004}.
\end{remark}

Therefore, based on Corollary \ref{col:ndyn_C3}, the proposed alternatives for use in Theorem \ref{thm:ndyn_firstorder} are summarized as follows:
\begin{itemize}
	\item[\textbf{\textit{C3)}}] $\bm{\gamma}_z = (\bm{\gamma}_x, \bm{\gamma}_w)$ is given by the closest point in $X \times W$ to the center of $X \times W$, computed through Proposition \ref{propo:ndyn_closest};	
	\item[\textbf{\textit{C4)}}] $\bm{\gamma}_z = (\bm{\gamma}_x, \bm{\gamma}_w)$ is the center of $X \times W$, if it satisfies $\bm{\gamma}_z \in X \times W$. Otherwise, $\bar{\bm{\gamma}}_z \in X \times W$ is chosen as the center of the interval hull of $X \times W$, the center of $X \times W$ is moved to $\bar{\bm{\gamma}}_z$ using Proposition \ref{propo:ndyn_centerchange}, and $\bm{\gamma}_z$ is given by $\bm{\gamma}_z \triangleq \bar{\bm{\gamma}}_z$.\footnote{Note that this choice may lead to the same value of $\bm{\gamma}_z$ provided by \emph{C1}, but $X \times W$ is described by a different CG-rep with center $\bm{\gamma}_z$. If this point is not in $X \times W$, $\bar{\bm{\gamma}}_z$ can be chosen as the point obtained from Proposition \ref{propo:ndyn_closest} instead.}
\end{itemize}

\begin{remark} \rm \label{rem:ndyn_firstorderchoices}
	Using heuristic \textit{C4}, the new center is guaranteed to belong to $X \times W$. Therefore, the latter can be chosen as approximation point, leading to a less conservative enclosure. However, since Proposition \ref{propo:ndyn_centerchange} provides a substantially more complex enclosure than the original set, given by \eqref{eq:ndyn_centerchange}, \textit{C4} results in an increased computational burden with respect to \textit{C3}.
\end{remark}

\section{Update step} \label{sec:ndyn_linearupdate}

As described in the previous section, an enclosure for the prediction step \eqref{eq:ndyn_prediction0} can be obtained in CG-rep using either the CZIB approach (Algorithm \ref{alg:ndyn_czib}), the mean value extension (Theorem \ref{thm:ndyn_meanvalue}), or the first-order Taylor extension (Theorem \ref{thm:ndyn_firstorder}). Consequently, taking into account the linearity of the measurement in \eqref{eq:ndyn_system}, an exact bound for the update step \eqref{eq:ndyn_update0} can be obtained efficiently by computing the generalized intersection of two constrained zonotopes, as explained below. 

Given the prediction set $\bar{X}_k \subset \realset^n $, the constrained zonotope $V \subset \realset^{n_v}$ describing bounds on measurement errors, the current input $\mbf{u}_k \in \realset^{n_u}$, and the measurement $\mbf{y}_k \in \realset^{n_y}$, an exact enclosure for the update step is obtained using the definition \eqref{eq:pre_intersection}, given by 
\begin{equation} \label{eq:ndyn_updatelinear}
\hat{X}_k = \bar{X}_k \cap_{\mbf{C}} ((\mbf{y}_k - \mbf{D}_u \mbf{u}_k) \oplus (-\mbf{D}_v V)).
\end{equation}

It is well known that the intersection in \eqref{eq:ndyn_updatelinear} can not be computed exactly using zonotopes, and must be over-approximated \citep{Le2013,Alamo2005a}. As a consequence, the enclosures of the system states obtained after many iterations of prediction and update may be quite conservative using zonotopes. However, with constrained zonotopes all operations in \eqref{eq:ndyn_updatelinear} are easily computed through \eqref{eq:pre_czlimage}--\eqref{eq:pre_czintersection}. These lead to an enclosure with $n_g + n_{g_v}$ generators, and $n_c + n_{c_v} + n_y$ constraints, where $n_g$ and $n_c$ are the number of generators and constraints of $\bar{X}_k$, respectively.

\begin{remark} \rm
	Unlike the CZIB approach, iterated computations of the extensions proposed in Theorems \ref{thm:ndyn_meanvalue} and \ref{thm:ndyn_firstorder}, and the update step \eqref{eq:ndyn_updatelinear} result in at most a linear and quadratic increase in the complexity of the CG-rep \eqref{eq:pre_cgrep}, respectively. As with zonotopes, this complexity increase can be efficiently addressed using the complexity reduction methods discussed in Section \ref{sec:complexityreduction}, that outer-approximate a constrained zonotope by another with lower complexity with reasonable conservativeness. 	
\end{remark}

\section{Complexity analysis} \label{sec:ndyn_complexity}

Table \ref{tab:ndyn_complexitybasic} shows the computational complexity\footnote{We use the standard $O({\cdot})$ notation defined in \cite{Cormen2009}.} of the basic operations used in the constrained zonotope methods. The complexities of basic operations using zonotopes are also presented for comparison. These complexities assume generic inputs with dimensions $\mbf{R} \in \realsetmat{n_r}{n}$ in \eqref{eq:pre_limage}; $Y =\{\mbf{G}_y, \mbf{c}_y, \mbf{A}_y, \mbf{b}_y\}$ in \eqref{eq:pre_intersection}, with $\mbf{G}_y \in \realsetmat{n_r}{n_{g_r}}$, $\mbf{c}_y \in \realset^{n_r}$, $\mbf{A}_y \in \realsetmat{n_{c_r}}{n_{g_r}}$, and $\mbf{b}_y \in \realset^{n_{c_r}}$; $\mbf{J} \in \intvalsetmat{n_s}{n}$ in Theorem \ref{thm:ndyn_czinclusion}; $X = \{\mbf{G}_x,\mbf{c}_x\}$ or $X = \{\mbf{G}_x,\mbf{c}_x,\mbf{A}_x,\mbf{b}_x\}$, with $\mbf{G}_x \in \realsetmat{n}{n_g}$, $\mbf{c}_x \in \realset^n$, $\mbf{A}_x \in \realsetmat{n_c}{n_g}$, and $\mbf{b}_x \in \realset^{n_c}$; $k_g$ and $k_c$ are the number of generators and constraints removed in the order reduction process, respectively. `Set inclusion' refers to the zonotope inclusion operator in \cite{Alamo2005a} for zonotopes and the CZ-inclusion operator (Theorem \ref{thm:ndyn_czinclusion}) for constrained zonotopes. `Closest point' and `Change center' correspond to Propositions \ref{propo:ndyn_closest} and \ref{propo:ndyn_centerchange}, respectively, which are LPs. For the latter, the bounds $\tilde{\bm{\xi}}^\text{L}, \tilde{\bm{\xi}}^\text{U}$ are obtained using Algorithm 1 in \cite{Scott2016}. Note that the interval hull of zonotopes does not require the solution of LPs (see Remark 3 in \cite{Kuhn1998}). In addition, we consider that each LP is solved at least with the performance of the simplex method presented in \cite{Kelner2006}, which is $O(N_dN_c^3)$ with $N_d$ and $N_c$ the number of decision variables and constraints, respectively. Note that these numbers can be inferred for each respective LP directly from Table \ref{tab:ndyn_complexitybasic}. 

\begin{table}[!htb]
	\scriptsize
	\centering
	\caption{Computational complexity $O(\cdot)$ of basic operations.}
	\begin{tabular}{c c c} \hline
		Operation & Zonotopes & Constrained zonotopes\\ \hline
		Linear mapping & $nn_gn_r$ & $nn_gn_r$\\
		Minkowski sum & $n$ & $n$ \\		
		Generalized intersection & -- & $nn_gn_r + n_rn_{g_r}$\\
		Interval hull & $nn_g$ & $nn_g(n_g+n_c)^3$ \\
		Set inclusion & $nn_g$ & $nn_sn_g+n_c(n_g+n_c)^3+nn_g^2n_c$ \\
		Closest point & -- & $(n+n_g)(n+n_g+n_c)^3$ \\
		Change center& -- & $n_g(n+n_g)^3 + n_g^2n_c$ \\
		Generator reduction & $n^2n_g+k_gnn_g$ & $(n+n_c)^2n_g+k_g(n+n_c)n_g$ \\
		Constraint elimination & -- & $k_c(n_g+n_c)^3 + k_cnn_g^2$ \\
		\hline		
	\end{tabular} \normalsize
	\label{tab:ndyn_complexitybasic}
\end{table}

Table \ref{tab:ndyn_complexity} shows the computational complexity of the mean value extension and first-order Taylor extension, for the prediction and update steps, as well as complexity reduction to the same number of generators and constraints of the set prior to prediction. Specifically, we use the mean value extension (Theorem \ref{thm:ndyn_meanvalue}) and the first-order Taylor extension (Theorem \ref{thm:ndyn_firstorder}) for the prediction steps, while the update steps are both given by the generalized intersection \eqref{eq:ndyn_updatelinear}. These methods are denoted by CZMV and CZFO, respectively. The computational complexities of their zonotope counterparts are also presented for comparison, denoted analogously by ZMV and ZFO, which use the mean value approach in \cite{Alamo2005a} and the first-order Taylor approach in \cite{Combastel2005}, respectively, for the prediction step. The update algorithm proposed in \cite{Bravo2006} is used for both ZMV and ZFO because it provided the best trade-off between accuracy and efficiency in our numerical experiments with zonotopes. Complexity reduction is applied after the update step in all four methods using the reduction methods in \cite{Scott2016} for constrained zonotopes and Method 4 in \cite{Scott2018} for zonotopes. For constrained zonotopes, constraint elimination is performed prior to generator reduction. The complexities in Table \ref{tab:ndyn_complexity} take into account the growth of the number of generators and constraints after each step (see Remarks \ref{rem:ndyn_meanvaluecomplexity} and \ref{rem:ndyn_firstordercomplexity}). The dimensions in Table \ref{tab:ndyn_complexity} are specified by the definitions $\hat{X}_{k-1} = \{\mbf{G}_x,\mbf{c}_x\}$, $W = \{\mbf{G}_w, \mbf{c}_w\}$, and $V = \{\mbf{G}_v, \mbf{c}_v\}$ or $\hat{X}_{k-1} = \{\mbf{G}_x,\mbf{c}_x,\mbf{A}_x,\mbf{b}_x\}$, $W = \{\mbf{G}_w, \mbf{c}_w, \mbf{A}_w, \mbf{b}_w\}$, and $V = \{\mbf{G}_v, \mbf{c}_v, \mbf{A}_v, \mbf{b}_v\}$ with $\mbf{G}_x \in \realsetmat{n}{n_g}$, $\mbf{c}_x \in \realset^n$, $\mbf{A}_x \in \realsetmat{n_c}{n_g}$, $\mbf{b}_x \in \realset^{n_c}$, $\mbf{G}_w \in \realsetmat{n_w}{n_{g_w}}$, $\mbf{c}_w \in \realset^{n_w}$, $\mbf{A}_w \in \realsetmat{n_{c_w}}{n_{g_w}}$, $\mbf{b}_w \in \realset^{n_{c_w}}$, $\mbf{G}_v \in \realsetmat{n_v}{n_{g_v}}$, $\mbf{c}_v \in \realset^{n_v}$, $\mbf{A}_v \in \realsetmat{n_{c_v}}{n_{g_v}}$, $\mbf{b}_v \in \realset^{n_{c_v}}$, $\mbf{u}_k \in \realset^{n_u}$, and $\mbf{y}_k \in \realset^{n_y}$. For simplicity, we define $m = n + n_w$, $m_g = n_g + n_{g_w}$, $m_c = n_c + n_{c_w}$, $\delta_n = n - n_y$, $\delta_w = n_{g_w} - n_{c_w}$, $\delta_v = n_{g_v} - n_{c_v}$, and $\tilde{\delta} = m_g^2 - m_c^2$. Moreover, we consider that scalar real function and scalar inclusion function evaluations have complexity $O(1)$. These correspond to evaluations of the nonlinear dynamics in \eqref{eq:ndyn_system} and its derivatives using real and interval arithmetic, respectively. For a detailed derivation of the computational complexities in Tables \ref{tab:ndyn_complexitybasic} and \ref{tab:ndyn_complexity}, see Appendix \ref{app:computationalcomplexity}.

The dominant terms in the prediction step of CZMV come from the computation of the interval hulls of $X$ and $W$ and the CZ-inclusion operators $\gzinclusion\left(\mbf{J}_x,  X - \bm{\gamma}_x \right)$ and $\gzinclusion\left(\mbf{J}_w,  W - \bm{\gamma}_w \right)$ in Theorem \ref{thm:ndyn_meanvalue} and Remark \ref{rem:ndyn_affine}. In the case of CZFO, the dominant terms come from the computation of the interval matrices $\tilde{\mbf{Q}}^{[q]}$, the interval hull of $Z = X \times W$, and the CZ-inclusion operator $\gzinclusion (\mbf{L}, (\mbf{c} - \bm{\gamma}_z) \oplus 2\mbf{G} B_\infty(\mbf{A},\mbf{b}) )$ in Theorem \ref{thm:ndyn_firstorder}. Note that the worst-case complexities of the prediction steps of the proposed algorithms are higher than the zonotope methods, while the update steps are cheaper due to the generalized intersection \eqref{eq:ndyn_updatelinear}. Even so, the complexity of the proposed methods are still polynomial. For a simplified analysis, assuming that all of the variables in Table \ref{tab:ndyn_complexity} increase linearly with $n$, the total complexities for ZMV, ZFO, CZMV and CZFO are $O(n^4)$, $O(n^5)$, $O(n^5)$, and $O(n^8)$, respectively. On the other hand, even basic polytope operations are known to be exponential \citep{Hagemann2015}. Besides, despite the higher complexities of CZMV and CZFO in comparison to the zonotope methods, they provide more accurate enclosures as shown in the next section.

\begin{table*}[!tb]
	\scriptsize
	\centering
	\caption{Computational complexity $O(\cdot)$ of the state estimators.}
	\begin{tabular}{c c} \hline
		Step & ZMV \\ \hline
		Prediction & $n^2n_g + nn_wn_{g_w}$ \\
		Update & $n_y(n^3(m_g+n)+n^2(m_g+n)^2+n_u+n_vn_{g_v})$ \\
		Reduction & $n^2(m_g+n) + n(n_{g_w}+n)(m_g+n)$ \\
		\hline
		Step & ZFO \\ \hline
		Prediction & $n(m^2m_g + mm_g^2)$ \\
		Update & $n_y(n^3(m_g^2+n)+n^2(m_g^2+n)^2+n_u+n_vn_{g_v})$  \\
		Reduction & $n^2(m_g^2+n) + n(m_g^2+n)^2$ \\
		\hline		
		Step & CZMV \\ \hline
		\multirow{1}{*}{Prediction} & $n(m^2m_g+mm_g^2) + (mm_g+m_c)(m_g+m_c)^3$ \\
		Update & $n_yn(m_g^2+n) + n_yn_u + n_yn_vn_{g_v}$ \\
		\multirow{2}{*}{Reduction} & $(n_{c_w}+n_{c_v}+n_y)(m_g+m_c+n_{g_v}+n_{c_v}+n+n_y)^3$ \\
		& $+ (n+n_c)^2(n_g+\delta_n+\delta_w+\delta_v)+(n+n_c)(\delta_n+\delta_w+\delta_v)(n_g+\delta_n+\delta_w+\delta_v)$ \\
		\hline		
		Step & CZFO \\ \hline
		\multirow{1}{*}{Prediction} & $n(m^2m_g+mm_g^2) + (mm_g+m_c)(m_g+m_c)^3$\\
		Update & $n_yn(m_g^2+n) + n_yn_u + n_yn_vn_{g_v}$ \\
		\multirow{2}{*}{Reduction} & $(m_c^2+n_{c_v}+n_y)(m_g^2+m_c^2+n_{g_v}+n_{c_v}+n+n_y)^3 $ \\
		& $+ (n+n_c)^2(\tilde{\delta}+\delta_n + \delta_v)+(n+n_c)(\tilde{\delta}+\delta_n+\delta_v)^2$ \\
		\hline				
	\end{tabular} \normalsize
	\label{tab:ndyn_complexity}
\end{table*}

\section{Numerical examples}

\subsection{Example 1} \label{sec:ndyn_example1}

This section presents a numerical example that demonstrates the performance of the CZIB approach. 

We address the second example presented in \cite{Alamo2005a}. Consider a isothermal gas-phase reactor, charged with an initial amount of species $A$ and $B$. The species are allowed to react according to the reversible reaction $2A \rightleftharpoons B$. The system states are defined as the partial pressure of each species in the reactor. The objective is to estimate all the system states while measuring the total pressure of the vessel as the reaction proceeds. The dynamic equations of the system are given by (discretized through Euler's method\footnote{All the simulations in this doctoral thesis are executed using discrete-time models. Therefore, the discretization errors are not present in the simulations and are out of the scope of the thesis.})
\begin{align*}
x_{1,k} & = x_{1,k-1} + T_s \left( -2k_1 x_{1,k-1}^2  + 2k_2 x_{2,k-1}\right) + k_3 w_{1,k-1}, \\
x_{2,k} & = x_{2,k-1} + T_s \left( k_1 x_{1,k-1}^2  - k_2 x_{2,k-1}\right) + k_3 w_{2,k-1},
\end{align*}
where $k_1 = 0.16/60$ s$^{-1}$ atm$^{-1}$, $k_2 = 0.0064/60$ s$^{-1}$, $k_3 = 0.0001$ and $T_s = 6$ s \citep{Alamo2005a}. The measured output is given by $y_k = [ 1 \,\; 1 ] \mbf{x}_k + v_k$. The existing disturbances are bounded by $\ninf{\mbf{w}_k} \leq 1$, 
and $\ninf{v_k} \leq 0.3$. The initial states are assumed to belong to the constrained zonotope
\begin{equation*}
\bar{X}_0 \triangleq \left\{ \begin{bmatrix} 2.5 & 0 & 1.0 \\ 0 & 0.5 & 1.0 \end{bmatrix}, \begin{bmatrix} 2.5 \\ 1.0 \end{bmatrix}, \begin{bmatrix} 1 & 0 & 1 \end{bmatrix}, 1 \right\}.
\end{equation*}

In the simulation, we use 12-partitions in Algorithm \ref{alg:ndyn_CZIB_IS1} and the natural inclusion function \citep{Moore2009} in Algorithm \ref{alg:ndyn_CZIB_IS2}. A simulation using 3-partitions is also conducted to evaluate the resulting conservatism. The inclusion functions were computed using INTLAB \citep{Rump1999}. The conversion from V-rep to H-rep in Algorithm \ref{alg:ndyn_CZIB_IS3a} was performed using the MPT toolbox \citep{MPT3}. Moreover, we perform an initial correction step $\hat{X}_0 = \bar{X}_0 \ginter{C} Y_0$ to obtain tighter initial bounds, where $Y_0 = \mbf{y}_0 \oplus (- \mbf{D}_v V_0)$. For comparison, a third simulation is conducted using the algorithm proposed in \cite{Combastel2005}, denoted by ZComb, which is based on first-order Taylor approximation using zonotopes for the prediction step, and update step based on singular-value decomposition (SVD), with complexity limited to 20 generators using the reduction algorithm described in Method \ref{meth:genredA} in Chapter \ref{cha:preliminaries}. The initial state is $\mbf{x}_0 = [ 4.9 \,\; 1.5 ]^T$.

\begin{figure}[!tb]
	\begin{footnotesize}
		\centering{
			\def\svgwidth{0.8\columnwidth}			
			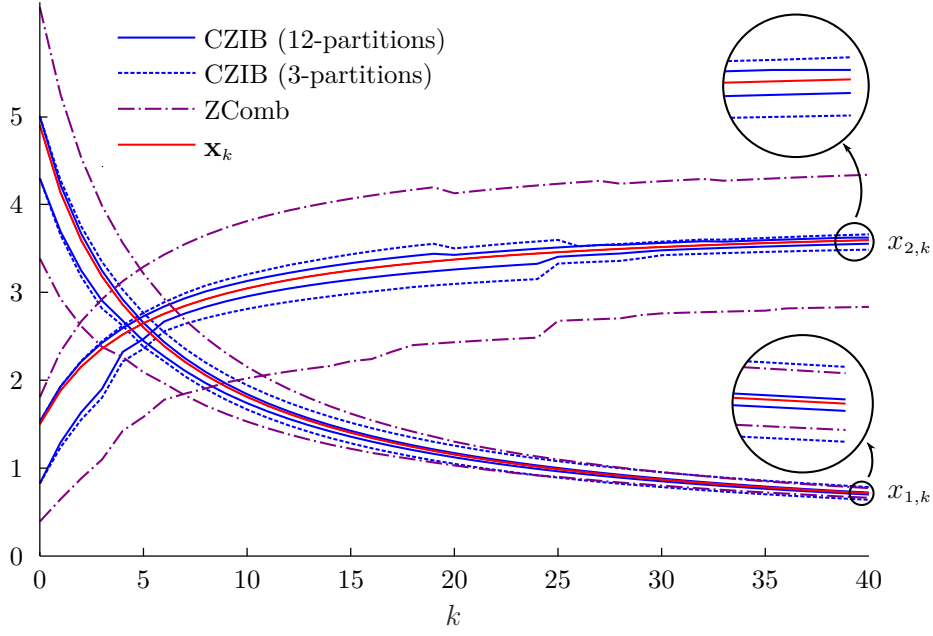\vspace{-2mm}
			\caption{Time evolution of the estimated bounds and the real states.}\label{fig:ndyn_czibresults}}
	\end{footnotesize}
\end{figure}

Figure \ref{fig:ndyn_czibresults} shows the time evolution of the system states, and the estimated bounds provided by the CZIB approach\footnote{The plotted curves are the bounds of the interval hull of $\hat{X}_k$.} and ZComb. Although reasonable for $x_1$, the bounds obtained using the latter are substantially conservative for $x_2$. Besides, the bounds obtained using the 12-partition CZIB are significantly less conservative in all cases, demonstrating the attained reduction in overestimation using the proposed method. In order to overcome the ``simplex problem'' (see Remark \ref{rem:CZIB_notsimplex}), a Minkowski sum was performed using a zero-centered interval vector with $10\%$ of the total diameter of the interval bundle, at every time step it occurred. In addition, the CG-rep of the constrained zonotope $\hat{X}_{40}$ in the 12-partition simulation was given by 13 generators and 9 constraints, without requiring the use of complexity reduction algorithms. The average execution times for the 3-partition and 12-partition simulations were $0.0855$ and $0.6013$ seconds per iteration, respectively, in MATLAB 9.1 and CPLEX 12.8, with 8GB RAM and Intel Core i7 4510U 3.1 GHz processor, using only sequential computation.

\subsection{Example 2} \label{sec:ndyn_example2}

This section presents numerical results for the two new set-based state estimation methods enabled by the new extensions proposed in Sections \ref{sec:ndyn_meanvalue} and \ref{sec:ndyn_firstorder}. Specifically, we use the mean value extension (Theorem \ref{thm:ndyn_meanvalue}) and the first-order Taylor extension (Theorem \ref{thm:ndyn_firstorder}) for the prediction steps, while the update step is given by the exact intersection \eqref{eq:ndyn_updatelinear} for both methods. These methods are denoted by CZMV and CZFO, respectively. The results are compared against two zonotopic methods denoted analogously by ZMV and ZFO, which use the mean value approach in \cite{Alamo2005a} and the first-order Taylor approach in \cite{Combastel2005}, respectively, for the prediction step. The update algorithm proposed in \cite{Bravo2006} is used for both ZMV and ZFO because it provided the best trade-off between accuracy and efficiency in our numerical experiments with zonotopes. Complexity reduction is applied after the update step in all four methods using Methods \ref{meth:czgenred} and \ref{meth:czconelim} for constrained zonotopes and Method \ref{meth:genredB} for zonotopes. The imposed complexity limits are described separately for each example below.

To demonstrate the effect of the different choices of the approximation point $\bm{\gamma}_x$, we first analyze one iteration of the prediction step for the nonlinear system \citep{Raimondo2012}
\begin{equation} \label{eq:ndyn_example2}
\begin{aligned}
x_{1,k} & = 3 x_{1,k-1} - \frac{x_{1,k-1}^2}{7} - \frac{4 x_{1,k-1} x_{2,k-1}}{4 + x_{1,k-1}} + w_{1,k-1}, \\
x_{2,k} & = -2 x_{2,k-1} + \frac{3 x_{1,k-1} x_{2,k-1}}{4 + x_{1,k-1}} + w_{2,k-1},
\end{aligned}
\end{equation}
with
\begin{equation} \label{eq:ndyn_example2_choicesofh_X0}
X_0 = \left\{ \begin{bmatrix} 0.2 & 0.4 & 0.2 \\ 0.2 & 0 & -0.2 \end{bmatrix}, \begin{bmatrix} -1 \\ 1 \end{bmatrix}, \begin{bmatrix} 2 & 2 & 2  \end{bmatrix}, -3 \right\},
\end{equation}
where $\mbf{w}_k \in \realset^2$ denotes process uncertainties, which are zero in this first scenario.

\begin{figure}[!tb]
	\centering{
		\def\svgwidth{0.7\columnwidth}
		{\scriptsize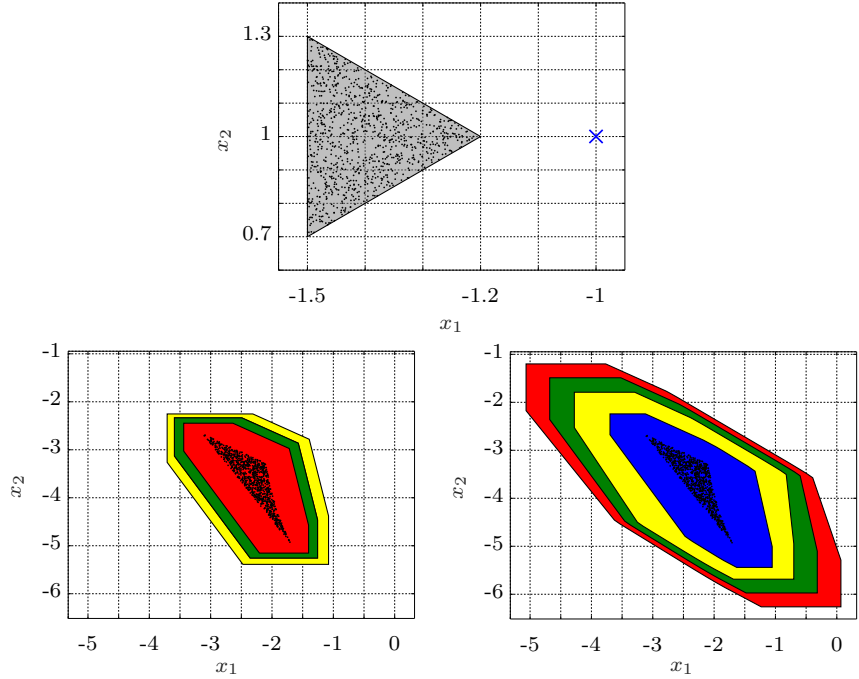}
		\caption{Top: The constrained zonotope $X_0$ with `$\times$' denoting its center. Left: Enclosures obtained by applying Theorem \ref{thm:ndyn_meanvalue} to \eqref{eq:ndyn_example2}. Right: Enclosures obtained by applying Theorem \ref{thm:ndyn_firstorder} to \eqref{eq:ndyn_example2}. The real vector $\bm{\gamma}_x$ is determined by \emph{C1} (green), \emph{C2} (red), \emph{C3} (yellow), and \emph{C4} (blue). For the mean value extension (left), \emph{C4} is overlapped with \emph{C1}. Black dots denote uniform samples from $X_0$ propagated through \eqref{eq:ndyn_example2}.}\label{fig:ndyn_example2_choicesofh}}
\end{figure}

Figure \ref{fig:ndyn_example2_choicesofh} shows the constrained zonotope $X_0$ and the enclosures of the one-step reachable set obtained by Theorem \ref{thm:ndyn_meanvalue} using \emph{C1}--\emph{C4}. Since the complexity of the enclosure for \emph{C4} is higher than for the other methods (see Proposition \ref{propo:ndyn_centerchange}), the reduction methods in \cite{Scott2016} were used to reduce the number of generators and constraints in this enclosure to match the other methods before comparison. In this example, the choice of $\bm{\gamma}_x$ has a moderate impact in the enclosure computed by Theorem \ref{thm:ndyn_meanvalue}, with \emph{C2} providing the least conservative result, as expected. Therefore, \emph{C2} is employed in Theorem \ref{thm:ndyn_meanvalue} henceforth. 

Figure \ref{fig:ndyn_example2_choicesofh} also shows the enclosures of the one-step reachable set obtained by Theorem \ref{thm:ndyn_firstorder} with \emph{C1}--\emph{C4}. Clearly, the enclosures produced by Theorem \ref{thm:ndyn_firstorder} are strongly affected by the choice of $\bm{\gamma}_x$, with \emph{C4} providing the least conservative result. In addition, note that the enclosures provided by the first-order Taylor extension are more conservative than those obtained by the mean value extension. However, experience with zonotopes and intervals (see \cite{Raimondo2012} for detailed examples) suggests that the relative merits of these two methods will depend on the dynamics of the system, as well as the shape and size of the set $X_0$, and the maximum allowed number of generators and constraints. This is corroborated by the next results.

We consider now the linear measurement equation
\begin{equation} \label{eq:ndyn_example2_meas}
\begin{bmatrix} y_{1,k} \\ y_{2,k} \end{bmatrix} = \begin{bmatrix} 1 & 0 \\ -1 & 1 \end{bmatrix} \begin{bmatrix} x_{1,k} \\ x_{2,k} \end{bmatrix} + \begin{bmatrix} v_{1,k} \\ v_{2,k} \end{bmatrix},
\end{equation}
with bounds $\ninf{\mbf{w}_k} \leq 0.4$ and $\ninf{\mbf{v}_k} \leq 0.4$, where $\mbf{v}_k \in \realset^2$ denotes measurement uncertainties. The initial states $\mbf{x}_0$ are bounded by the zonotope\footnote{Note that $X_0$, $W$ and $V$ are expressed as zonotopes for a fair comparison with the zonotope methods.}
\begin{equation} \label{eq:ndyn_example2_X0}
X_0 = \left\{ \begin{bmatrix} 0.1 & 0.2 & -0.1 \\ 0.1 & 0.1 & 0 \end{bmatrix}, \begin{bmatrix} 0.5 \\ 0.5 \end{bmatrix}\right\}.
\end{equation}

\begin{figure}[!tb]
	\centering{
		\def\svgwidth{0.6\columnwidth}
		{\scriptsize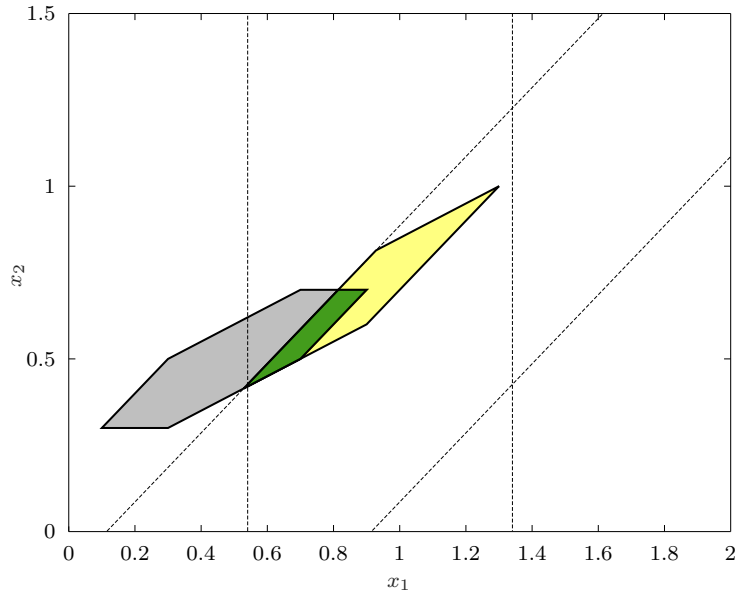}
		\caption{The initial set $X_0$ (gray), the initial bounded uncertain measurements (dashed lines), the intersection computed as in \cite{Bravo2006} (yellow), and the constrained zonotope (green) computed by \eqref{eq:ndyn_updatelinear}.}\label{fig:ndyn_example2_update}}
\end{figure}

To generate process measurements, system \eqref{eq:ndyn_example2} was simulated with $\mbf{x}_0 = (0.8,0.65) \in X_0$ and process and measurement uncertainties drawn from uniform random distributions. The number of generators and constraints of the constrained zonotopes was limited to 20 and 5, respectively, while the number of generators of the zonotopes was limited to 20. Figure \ref{fig:ndyn_example2_update} shows the results of the initial update step using the intersection algorithm in \cite{Bravo2006} and the generalized intersection \eqref{eq:ndyn_updatelinear} computed using \eqref{eq:pre_czintersection}, which yields a constrained zonotope. Clearly, since the generalized intersection is not a symmetric set, it cannot be described by a zonotope. In contrast, the resulting constrained zonotope corresponds to the exact intersection, providing far less conservative bounds in the first update step.

\begin{figure}[!tb]
	\centering{
		\def\svgwidth{0.7\columnwidth}
		{\scriptsize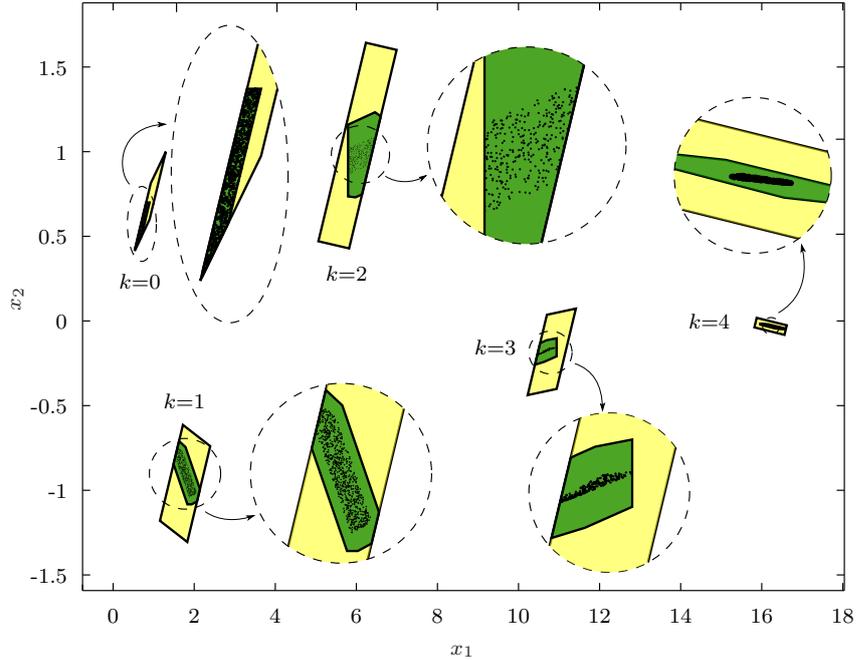}
		\caption{Results from the first four time steps of set-valued state estimation (after update) using the constrained zonotopic method CZMV (green) and the zonotopic method {\alamobravo} (yellow). Black dots denote uniform samples from $X_0$ propagated through \eqref{eq:ndyn_example2} that are consistent with the current measurement.} \label{fig:ndyn_example2_meanvalue_4iter}}
\end{figure}

Figure \ref{fig:ndyn_example2_meanvalue_4iter} shows the first four time steps of CZMV with $\bm{\gamma}_x$ given by \emph{C2} in a scenario without process uncertainties ($\mbf{w}_k = \bm{0}$). For comparison purposes, the zonotopes computed using ZMV are also depicted. CZMV provides much less conservative enclosures than {\alamobravo} for this example, demonstrating the effectiveness of the proposed nonlinear state estimation strategy.

\begin{figure}[!tb]
	\centering{
		\def\svgwidth{0.8\columnwidth}
		{\scriptsize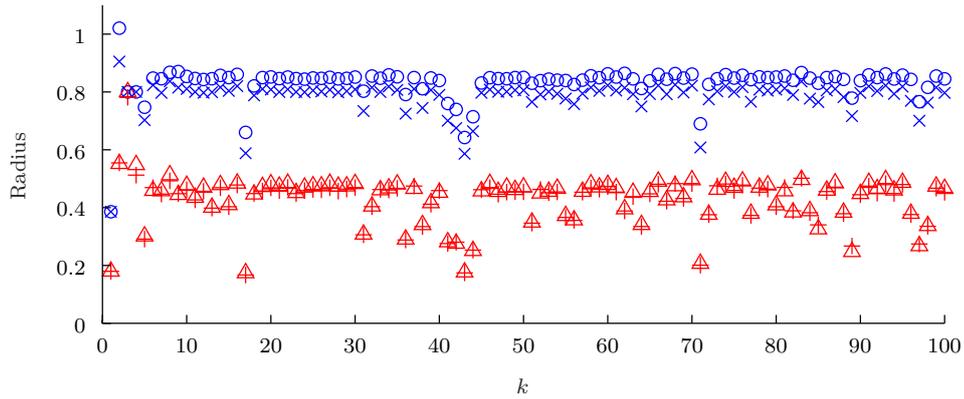}
		\caption{Radii of the update sets obtained by applying {\alamobravo} ($\circ$), ZFO ($\times$), CZMV ($\triangle$), and CZFO ($+$) to \eqref{eq:ndyn_example2} with process disturbances.}\label{fig:ndyn_example2_comparison_radiusw}}
\end{figure}

Figure \ref{fig:ndyn_example2_comparison_radiusw} shows the radii (half the length of the longest edge of the interval hull) of the sets provided by CZMV using \emph{C2} and {\alamobravo} over 100 time steps considering process disturbances, and also compares the radii of the update sets computed by {\combastelbravo} and CZFO with $(\bm{\gamma}_x,\bm{\gamma}_w)$ given by \emph{C4}. Since \eqref{eq:ndyn_example2} is affine in $\mbf{w}_k$, the enclosure $Z \supseteq \bm{\mu}(\bm{\gamma}_x,W)$ in Theorem \ref{thm:ndyn_meanvalue} was computed as described at the end of Remark \ref{rem:ndyn_affine}. CZMV provided less conservative bounds than the zonotopes computed by {\alamobravo}, with a CZMV-to-{\alamobravo} average radius ratio (ARR, i.e., the ratio of the radius of the CZMV set at $k$ over the ZMV set at $k$ averaged over all time steps $k$) of only $51.4\%$. 
In addition, as in the previous case, CZFO provides less conservative bounds than {\combastelbravo}, with the CZFO-to-ZFO ARR being only $53.66\%$. The size of the sets provided by CZMV and CZFO were quite similar, with CZFO being less conservative (the CZFO-to-CZMV ARR was $98.75\%$). 
The ARR for different numbers of constraints are shown in Table \ref{tab:ndyn_example2ARR}, with the average computed considering in addition simulations with different numbers of generators. Execution times are shown in Table \ref{tab:ndyn_example2times}. These were obtained using MATLAB 9.1 with CPLEX 12.8 and INTLAB 9, in a laptop with 8GB RAM and an Intel Core i7 4510U 3.1 GHz processor.

The use of constrained zonotopes in CZMV and CZFO results in sets that are slightly more complex than those generated by ZMV and ZFO (specifically, the set description involves five equality constraints that are not present in the zonotopes from ZMV and ZFO). However, this example shows that this increase in complexity is compensated by greatly improved accuracy.

\begin{table}[!htb]
	\footnotesize
	\centering
	\caption{Average radius ratio of the estimators with varying numbers of constraints. Each average is taken over 3 separate simulations using $n_g \in \{ 8,12,20\}$.}
	\begin{tabular}{c c c c} \hline
		$n_c$ & CZMV/ZMV & CZFO/ZFO & CZFO/CZMV \\ \hline
		$1$ & $54.1\%$ & $59.8\%$ & $104.5\%$ \\	    
		$3$ & $51.6\%$ & $54.0\%$ & $99.0\%$ \\		
		$5$ & $51.6\%$ & $53.7\%$ & $98.5\%$ \\
		\hline
	\end{tabular} \normalsize
	\label{tab:ndyn_example2ARR}
\end{table}

\begin{table}[!htb]
	\footnotesize
	\centering
	\caption{Average total times per iteration of the estimators with varying numbers of constraints. Each average is taken over 30 separate simulations using $n_g \in \{8,12,20\}$. Times spent only on complexity reduction are shown in parenthesis.}
	\begin{tabular}{c c c c c} \hline
		$n_c$ & ZMV & ZFO & CZMV & CZFO \\ \hline
		$0$ & $5.79~(0.24)$ ms & $14.03~(2.93)$ ms & -- & -- \\
		$1$ & -- & -- & $19.2~(1.6)$ ms & $76.4~(57.1)$ ms \\
		$3$ & -- & -- & $20.7~(1.8)$ ms & $4.28~(4.26)$ s \\		
		$5$ & -- & -- & $22.5~(1.9)$ ms & $8.93~(8.91)$ s \\
		\hline
	\end{tabular} \normalsize
	\label{tab:ndyn_example2times}
\end{table}

A third example, consisting of a realistic scenario of nonlinear set-based state estimation of a quadrotor UAV using the mean value and first-order Taylor extensions  proposed in this chapter, can be found in Chapter \ref{cha:appuav}. 

\section{Final remarks}

This chapter developed new set-based state estimation methods based on constrained zonotopes. Three new nonlinear set-based state estimation methods were developed, specifically the CZIB approach, the mean value extension, and the first-order Taylor extension using constrained zonotopes. The measurements were effectively incorporated in the state estimation methods by means of generalized intersection of constrained zonotopes considering linear measurement. 

The advantages of the new nonlinear state estimation methods were corroborated by numerical experiments. The first example demonstrated the reduced conservativeness of the CZIB approach in comparison to the zonotope-based method by \cite{Combastel2005}. The second example demonstrated the accuracy of the proposed extensions for nonlinear set-based state estimation using constrained zonotopes. In each experiment, the new methods were compared with existing zonotope-based approaches.

The next chapter develops new methods for set-based state estimation of systems with nonlinear measurement and invariants. These are a generalization of the mean value and first-order Taylor extensions introduced in this chapter.

\chapter{State estimation of systems with nonlinear measurement and invariants}\thispagestyle{headings} \label{cha:nonlinearmeasinv}
\chaptermark{Nonlinear measurement and invariants}

This chapter presents new methods for set-valued state estimation of discrete-time nonlinear systems whose trajectories are known to satisfy nonlinear equality constraints, called \emph{invariants} (e.g., conservation laws). The new methods based on constrained zonotopes improve the standard prediction-update framework for systems with invariants by adding a \emph{consistency step}. This new step uses invariants to reduce conservatism and is enabled by new algorithms for refining constrained zonotopes based on nonlinear constraints. This chapter also presents significant improvements to prediction and update steps proposed in Chapter \ref{cha:nonlineardynamics}. Specifically, new update algorithms are developed that allow nonlinear measurement equations for the first time, and existing prediction methods based on conservative approximation techniques are modified to allow a more flexible choice of the approximation point, which can lead to tighter enclosures. Numerical results demonstrate that the resulting methods can provide significantly tighter enclosures than existing zonotope-based methods while maintaining comparable efficiency. The content presented in this chapter was published in Automatica \citep{Rego2020c}.

This chapter is organized as follows. The set-based state estimation problem and the class of nonlinear systems considered are described in Section \ref{sec:nmeas_problemformulation}. The main results are given in Section \ref{sec:nmeas_estimation}, including the new consistency and update algorithms and the improvements to the prediction algorithms developed in Chapter \ref{cha:nonlineardynamics}. Numerical examples are presented in Section \ref{sec:nmeas_examples}, and Section \ref{sec:nmeas_conclusions} concludes the chapter.

\section{Problem formulation} \label{sec:nmeas_problemformulation}

Consider a class of nonlinear discrete-time systems described by
\begin{equation}
\begin{aligned} \label{eq:nmeas_system}
\mbf{x}_k & = \mbf{f}(\mbf{x}_{k-1}, \mbf{u}_{k-1}, \mbf{w}_{k-1}), \\
\mbf{y}_k & = \mbf{g}(\mbf{x}_k, \mbf{u}_{k}, \mbf{v}_{k}),	
\end{aligned}
\end{equation}
for $k \geq 1$, with $\mbf{y}_0 = \mbf{g}(\mbf{x}_0, \mbf{u}_0, \mbf{v}_0)$, where $\mbf{x}_k \in \realset^{n}$ denotes the system state, $\mbf{u}_{k} \in \realset^{n_u}$ is the known input, $\mbf{w}_k \in \realset^{n_w}$ is the process uncertainty, $\mbf{y}_k \in \realset^{n_y}$ is the measured output, and $\mbf{v}_k \in \realset^{n_v}$ is the measurement uncertainty. The nonlinear mappings  $\mbf{f}:\realset^{n}\times\realset^{n_u}\times\realset^{n_w}\to\realset^{n}$ and $\mbf{g}:\realset^{n}\times\realset^{n_u}\times\realset^{n_v}\to\realset^{n_y}$ are assumed to be of class $\mathcal{C}^2$. The initial condition and uncertainties are assumed to be unknown-but-bounded, i.e., $\mbf{x}_0 \in \bar{X}_0$, $\mbf{w}_{k} \in W$, and $\mbf{v}_k \in V$, where $\bar{X}_0$, $W$, and $V$ are known convex polytopic sets.

This chapter presents an improved set-valued state estimation method for systems satisfying known invariants, as defined in the following assumption.

\begin{assumption} \rm
	\label{ass:nmeas_invariant}
	There exists a $\mathcal{C}^2$ function $\mbf{h} : \realset^n \to \realset^{n_h}$ such that, for every solution of \eqref{eq:nmeas_system} with $\mbf{x}_0\in \bar{X}_0$, $\mbf{w}_{k} \in W$, and $\mbf{v}_k \in V$,
	\begin{alignat}{1}
	\label{eq:nmeas_invariantdef}
	\mbf{h}(\mbf{x}_0) = \bm{0} \quad\implies\quad \mbf{h}(\mbf{x}_k) = \bm{0}, \ \ \forall k \geq 0.
	\end{alignat}
	We refer to the components of $\mbf{h}$ as \emph{invariants}.
\end{assumption}

\begin{remark} \rm \label{rem:nmeas_invariantcondition}
	A sufficient condition for \eqref{eq:nmeas_invariantdef} is that $\mbf{h}(\mbf{f}(\mbf{x}_k, \mbf{u}_{k}, \mbf{w}_k)) = \bm{0}$ for all $\mbf{x}_k$ such that $\mbf{h}(\mbf{x}_k) = \bm{0}$, for all $\mbf{w}_k \in W$ and $\mbf{u}_k$ with $k \geq 0$.
\end{remark}

Many systems of practical interest obey invariants describing, e.g., material conservation laws in chemical systems, conservation of energy or momentum in mechanical systems, or the isometry inherent to orientation dynamics in aerospace and robotic systems \citep{Shen2017,Goodarzi2017,Rotella2014}. Prior work on nonlinear reachability analysis has shown that, if used properly, even simple physical information in the form of invariants can dramatically improve the accuracy of reachability bounds computed by interval methods \citep{Scott2013a,Shen2017,Yang2020}. Similarly, our aim here is to develop new algorithms for effectively using invariants to improve the accuracy of the state-of-the-art state estimation algorithms based on constrained zonotopes proposed in Chapter \ref{cha:nonlineardynamics}.

For any $k\geq 0$, let $X_k$ denote the set of all states $\mbf{x}_k$ that are consistent with (i) the nonlinear model \eqref{eq:nmeas_system}, (ii) the measured output sequence up to time $k$, $(\mbf{y}_0,\ldots,\mbf{y}_k)$, and (iii) the unknown-but-bounded uncertainties $\mbf{x}_0\in\{\mbf{x}\in \bar{X}_0:\mbf{h}(\mbf{x})=\mbf{0}\}$, $\mbf{w}_k\in W$, and $\mbf{v}_k\in W$. %
Since exact characterization of $X_k$ is generally intractable \citep{Kuhn1998,Platzer2007}, the objective of set-valued state estimation is to approximate $X_k$ as accurately as possible by a guaranteed enclosure $\tilde{X}_k \supseteq X_k$. We accomplish this here by extending the standard prediction-update estimation framework with a new consistency step for tightening the enclosures using invariants. The general scheme is given by the following recursion:
\begin{align}
\bar{X}_k & \supseteq \{ \mbf{f}(\mbf{x}_{k-1}, \mbf{u}_{k-1}, \mbf{w}_{k-1}) : \mbf{x}_{k-1} \in \tilde{X}_{k-1}, \, \mbf{w}_{k-1} \in W \}, \label{eq:nmeas_prediction0}\\
\hat{X}_k & \supseteq \{ \mbf{x}_{k} \in \bar{X}_k : \mbf{g}(\mbf{x}_{k}, \mbf{u}_{k}, \mbf{v}_{k}) = \mbf{y}_k , \, \mbf{v}_{k} \in V \}, \label{eq:nmeas_update0}\\
\tilde{X}_k & \supseteq \{ \mbf{x}_{k} \in \hat{X}_k : \mbf{h}(\mbf{x}_{k}) = \bm{0} \},
\label{eq:nmeas_consistency0}
\end{align}
where \eqref{eq:nmeas_prediction0} is the \emph{prediction step}, \eqref{eq:nmeas_update0} is the \emph{update step}, \eqref{eq:nmeas_consistency0} is the \emph{consistency step}, and the scheme is initialized with $\bar{X}_0$ in the update step.
According to the definition of $X_k$, we have that $X_0=\{\mbf{x}_{0}\in \bar{X}_0: \mbf{h}(\mbf{x}_{0})=\mbf{0}, \ \mbf{g}(\mbf{x}_{0}, \mbf{u}_0, \mbf{v}_{0}) = \mbf{y}_0 , \, \mbf{v}_{0} \in V \}$. This immediately implies that $\tilde{X}_0\supseteq X_0$. If $\tilde{X}_{k-1}$ is a valid enclosure of $X_{k-1}$ for some $k\geq 1$, then standard results in set-valued state estimation show that $\hat{X}_{k}\supseteq X_{k}$ \citep{Chisci1996,Le2013}. Since any $\mbf{x}_{k-1}\in X_{k-1}$ emanates from some $\mbf{x}_0\in \bar{X}_0$ satisfying $\mbf{h}(\mbf{x}_0)=\mbf{0}$ by definition, Assumption \ref{ass:nmeas_invariant} implies that $\mbf{h}(\mbf{x}_k)=\mbf{0}$, and it follows that $\tilde{X}_{k}\supseteq X_{k}$ as well. By induction, we conclude that $\tilde{X}_{k}\supseteq X_{k}$ for all $k\geq 0$ as desired.

In the remainder of the chapter, our goal is to develop methods for computing accurate enclosures for each of the three steps \eqref{eq:nmeas_prediction0}--\eqref{eq:nmeas_consistency0}. Building on the results presented in Chapter \ref{cha:nonlineardynamics}, the main results of this chapter include generalizations of the prediction methods proposed in Sections \ref{sec:ndyn_meanvalue} and \ref{sec:ndyn_firstorder} with improved accuracy, new update methods that are applicable to nonlinear measurement equations, and methods for the new consistency step to make effective use of invariants.

\section{Recursive algorithm} \label{sec:nmeas_estimation}
This section presents new methods for computing enclosures for each step in the extended prediction-update-consistency algorithm \eqref{eq:nmeas_prediction0}--\eqref{eq:nmeas_consistency0} using constrained zonotopes. The proposed recursive scheme is summarized in Algorithm \ref{alg:nmeas_estimation}. In this algorithm, complexity reduction methods can be used after each step to limit the set complexity increase. We begin with two core lemmas required for all three steps. 

\begin{algorithm}[!htb]
	\caption{Recursive state estimation algorithm}
	\label{alg:nmeas_estimation}
	\begin{algorithmic}[1]	
		\State (Prediction step) Given the constrained zonotopes $\tilde{X}_{k-1} \times W  \subset \realset^n \times \realset^{n_w}$, and input $\mbf{u}_{k-1} \in \realset^{n_u}$, compute the predicted constrained zonotope $\bar{X}_k$ satisfying \eqref{eq:nmeas_prediction0}.
		\State (Update step) Given the constrained zonotopes $\bar{X}_{k} \times V \subset \realset^n \times \realset^{n_v}$, input $\mbf{u}_{k} \in \realset^{n_u}$, and measurement $\mbf{y}_k \in \realset^{n_y}$, compute a refined constrained zonotope $\hat{X}_k$ satisfying \eqref{eq:nmeas_update0}.
		\State (Consistency step) Given the constrained zonotope $\hat{X}_{k} \subset \realset^n$, compute a refined constrained zonotope $\tilde{X}_k$ satisfying \eqref{eq:nmeas_consistency0}.
	\end{algorithmic}
	\normalsize
\end{algorithm}

\begin{lemma} \rm \label{lem:nmeas_mve}
	Let $\bm{\alpha} : \realset^n \times \realset^{n_w} \to \realset^{n_\alpha}$ be of class $\mathcal{C}^1$, and let $\nabla_x \bm{\alpha}$ denote the gradient of $\bm{\alpha}$ with respect to its first argument. Let $X\subset \realset^n$ and $W \subset \realset^{n_w}$ be constrained zonotopes, and let $\mbf{J}_x \in \intvalsetmat{n_\alpha}{n}$ be an interval matrix satisfying 
	\begin{equation}
	\label{eq:nmeas_MV Lemma J contains the image}
	\nabla^T_x \bm{\alpha}(\square X, W)\triangleq\{\nabla^T_x \bm{\alpha}(\mbf{x},\mbf{w}): \mbf{x}\in\square X, \ \mbf{w}\in W\}\subseteq \mbf{J}_x.
	\end{equation}
	For every $\mbf{x} \in X$, $\mbf{w} \in W$, and $\bm{\gamma}_x \in \square X$, there exists $\hat{\mbf{J}}_x \in \mbf{J}_x$ such that 
	\begin{equation*}
	\bm{\alpha}(\mbf{x},\mbf{w}) = \bm{\alpha}(\bm{\gamma}_x,\mbf{w}) + \hat{\mbf{J}}_x (\mbf{x} - \bm{\gamma}_x).
	\end{equation*}
\end{lemma}

\proof Choose any $(\mbf{x},\mbf{w})\in X\times W$. Since $\mbf{x} \in X \subseteq \square X$ and $\bm{\gamma}_x \in \square X$, the Mean Value Theorem ensures that, for any $i=1,2,\ldots,n$, $\exists \bm{\delta}^{[i]}\in \square X$ such that $\alpha_i(\mbf{x},\mbf{w}) = \alpha_i(\bm{\gamma}_x,\mbf{w}) + \nabla^T_x \alpha_i(\bm{\delta}^{[i]},\mbf{w}) (\mbf{x} - \bm{\gamma}_x)$. But $\nabla^T_x \alpha_i(\bm{\delta}^{[i]},\mbf{w})$ is contained in the $i$-th row of $\mbf{J}_x$ by hypothesis, and since this is true for all $i=1,2,\ldots,n$, $\exists \hat{\mbf{J}}_x\in \mbf{J}_x$ such that $\bm{\alpha}(\mbf{x},\mbf{w}) = \bm{\alpha}(\bm{\gamma}_x,\mbf{w}) + \hat{\mbf{J}}_x (\mbf{x} - \bm{\gamma}_x).$ \qed

Lemma \ref{lem:nmeas_mve} provides an exact linear representation of the nonlinear function $\bm{\alpha}$ between two points based on the Mean Value Theorem, which is useful for computations with constrained zonotopes. This lemma is very similar to Theorem \ref{thm:ndyn_meanvalue} in Chapter \ref{cha:nonlineardynamics}. The only difference is that Theorem \ref{thm:ndyn_meanvalue} requires the approximation point $\bm{\gamma}_x$ to lie in $X$, while Lemma \ref{lem:nmeas_mve} allows $\bm{\gamma}_x$ to be chosen from the larger set $\square X$. This is important because obtaining a point in $X$ (or testing a given point for membership) requires solving a linear program (Property \ref{prop:pre_czisemptyinside}), whereas obtaining a point in $\square X$ is trivial. The proof of Lemma \ref{lem:nmeas_mve} is given above for completeness, but it follows easily from the proof of Theorem \ref{thm:ndyn_meanvalue} by replacing the condition $\bm{\gamma}_x \in X$ with $\bm{\gamma}_x \in \square X$ throughout.

The next lemma provides an alternative method for obtaining an exact linear representation of a nonlinear function between two points based on Taylor's Theorem. This lemma is similar to Theorem \ref{thm:ndyn_firstorder} in Chapter \ref{cha:nonlineardynamics}, with the difference again that the approximation point is chosen from $\square Z$ rather than $Z$. 
\begin{lemma} \rm \label{lem:nmeas_foe}
	Let $\bm{\beta}: \realset^{m} \to \realset^{n}$ be of class $\mathcal{C}^2$ and let $\mbf{z} \in \realset^{m}$ denote its argument. Let $Z = \{\mbf{G}, \mbf{c}, \mbf{A}, \mbf{b}\} \subset \realset^{m}$ be a constrained zonotope with $m_g$ generators and $m_c$ constraints. For each $q = 1,2,\dots,n$, let $\mbf{Q}^{[q]}\in\mathbb{IR}^{m\times m}$ and $\tilde{\mbf{Q}}^{[q]}\in\mathbb{IR}^{m_g\times m_g}$ be interval matrices satisfying $\mbf{Q}^{[q]} \supseteq \mbf{H} \beta_q (\square Z)$ and $\tilde{\mbf{Q}}^{[q]} \supseteq \mbf{G}^T \mbf{Q}^{[q]} \mbf{G}$. Moreover, define  
	\begin{align*}
	& \tilde{c}_q = \trace{\midpoint{\tilde{\mbf{Q}}^{[q]}}}/2, \quad \tilde{\mbf{G}}_{q,:} = \big[ ~\cdots \; \underbrace{\midpoint{\tilde{Q}^{[q]}_{ii}}/2}_{\forall i} \; \cdots \; \underbrace{\left(\midpoint{\tilde{Q}^{[q]}_{ij}} + \midpoint{\tilde{Q}^{[q]}_{ji}}\right)}_{\forall i<j} \; \cdots~ \big],\\
	& \tilde{\mbf{G}}_{\mbf{d}} = \text{diag}(\mbf{d}), \quad d_q = \sum_{i,j} \left| \rad{\tilde{Q}^{[q]}_{ij}} \right|, \quad \tilde{\mbf{A}} = \left[ \tilde{\mbf{A}}_{\bm{\zeta}} \,\; \tilde{\mbf{A}}_{\bm{\xi}} \,\; \zeros{\frac{m_c}{2}(1+m_c)}{n} \right], \\
	& \tilde{\mbf{A}}_{\bm{\zeta}} = \arrowmatrix{\begin{matrix} & \vdots & \\ ~\cdots &  \half A_{ri} A_{si} & \cdots~ \\ & \vdots & \end{matrix}}{\forall i}{\forall r \leq s}, \quad \tilde{\mbf{b}} = \downarrowmatrix{\begin{matrix} \vdots \\ b_{r} b_{s} - \half \sum_i A_{ri} A_{si} \\ \vdots\end{matrix}}{\forall r \leq s}, \\
	& \tilde{\mbf{A}}_{\bm{\xi}} = 
	\arrowmatrix{\begin{matrix} & \vdots & \\ ~\cdots & A_{ri} A_{sj} + A_{rj} A_{si} & \cdots~ \\ & \vdots & \end{matrix}}{\forall i < j}{\forall r \leq s}, 
	\end{align*}
	with indices $i,j = 1,2,\dots,m_g$ and $r,s = 1,2,\dots,m_c$. Finally, choose any $\bm{\gamma}_z \in \square Z$ and let $\mbf{L}\in\mathbb{IR}^{n\times m}$ be an interval matrix satisfying $\mbf{L}_{q,:} \supseteq (\mbf{c} - \bm{\gamma}_z)^T \mbf{Q}^{[q]}$ for all $q = 1,\dots,n$. For every $\mbf{z} \in Z$,  there exist $\bm{\xi} \in B_\infty(\mbf{A},\mbf{b})$, $\tilde{\bm{\xi}} \in B_\infty(\tilde{\mbf{A}},\tilde{\mbf{b}})$, and $\hat{\mbf{L}} \in \mbf{L}$ such that
	\begin{equation*}
	\begin{aligned}
	\bm{\beta}(\mbf{z}) = \bm{\beta}(\bm{\gamma}_z) & + \nabla^T \bm{\beta}(\bm{\gamma}_z)(\mbf{z} - \bm{\gamma}_z) \\ & + \tilde{\mbf{c}} + [ \tilde{\mbf{G}} \,\; \tilde{\mbf{G}}_{\mbf{d}} ] \tilde{\bm{\xi}} + \hat{\mbf{L}} ((\mbf{c} - \bm{\gamma}_z) + 2\mbf{G} \bm{\xi}).
	\end{aligned}
	\end{equation*}
\end{lemma}

\proof This follows by replacing $\bm{\gamma}_z \in Z$ with $\bm{\gamma}_z \in \square Z$ in the proof of Theorem \ref{thm:ndyn_firstorder}. \qed 

\subsection{Prediction step}
This section presents two different approaches for the prediction step in Algorithm \ref{alg:nmeas_estimation}. These methods are improved versions of the mean value and first-order Taylor extensions developed in Chapter \ref{cha:nonlineardynamics}, which allow for a more flexible choice of the approximation point enabled by Lemmas \ref{lem:nmeas_mve} and \ref{lem:nmeas_foe} above. 

\begin{proposition}\rm \label{thm:nmeas_mvepred}
	Let $\mbf{f} : \realset^n \times \realset^{n_u} \times \realset^{n_w} \to \realset^n$ be of class $\mathcal{C}^1$, and let $\nabla_x \mbf{f}$ denote the gradient of $\mbf{f}$ with respect to its first argument. Let $\mbf{u} \in \realset^{n_u}$, and let $X\subset \realset^n$ and $W \subset \realset^{n_w}$ be constrained zonotopes. Choose any $\bm{\gamma}_x \in \square X$. If $Z_w$ is a constrained zonotope such that $\mbf{f}(\bm{\gamma}_x,\mbf{u},W) \subseteq Z_w$ and $\mbf{J}_x \in \intvalsetmat{n}{n}$ is an interval matrix satisfying $\nabla^T_x \mbf{f}(\square X,\mbf{u},W)\subseteq \mbf{J}_x$, then $\mbf{f}(X,\mbf{u},W) \subseteq Z_w \oplus \gzinclusion\left(\mbf{J}_x,  X - \bm{\gamma}_x \right)$.
\end{proposition}
\proof Choose any $(\mbf{x},\mbf{w})\in X\times W$. Lemma \ref{lem:nmeas_mve} ensures that there exists a real matrix $\hat{\mbf{J}}_x \in \mbf{J}_x$ such that $\mbf{f}(\mbf{x},\mbf{u},\mbf{w}) = \mbf{f}(\bm{\gamma}_x,\mbf{u},\mbf{w}) + \hat{\mbf{J}}_x (\mbf{x} - \bm{\gamma}_x).$ By Theorem \ref{thm:ndyn_czinclusion} and the choice of $Z_w$, it follows that $\mbf{f}(\mbf{x},\mbf{u},\mbf{w}) \in Z_w \oplus \gzinclusion\left(\mbf{J}_x, X - \bm{\gamma}_x\right)$, as desired. \qed

\begin{proposition} \label{thm:nmeas_foepred} \rm
	Let $\mbf{f}: \realset^{n} \times \realset^{n_u} \times \realset^{n_w} \to \realset^{n}$ be of class $\mathcal{C}^2$, let $\mbf{u} \in \realset^{n_u}$, and let $X = \{\mbf{G}_x, \mbf{c}_x, \mbf{A}_x, \mbf{b}_x\}$ and $W = \{\mbf{G}_w, \mbf{c}_w, \mbf{A}_w, \mbf{b}_w\}$ be constrained zonotopes with $n_g$ generators and $n_c$ constraints, and with $n_{g_w}$ generators and $n_{g_w}$ constraints, respectively. Denote $\mbf{z}=(\mbf{x},\mbf{w})$ and $Z = X \times W = \{\mbf{G}, \mbf{c}, \mbf{A}, \mbf{b}\} \subset \realset^{n+n_w}$. For each $q = 1,2,\dots,n$, let $\mbf{Q}^{[q]}\in\mathbb{IR}^{(n+n_w)\times (n+n_w)}$ and $\tilde{\mbf{Q}}^{[q]}\in\mathbb{IR}^{(n_g+n_{g_w})\times(n_g+n_{g_w})}$ be interval matrices satisfying $\mbf{Q}^{[q]} \supseteq \mbf{H}_z f_q (\square X,\mbf{u}, \square W)$ and $\tilde{\mbf{Q}}^{[q]} \supseteq \mbf{G}^T \mbf{Q}^{[q]} \mbf{G}$. Moreover, define $\tilde{\mbf{c}}$, $\tilde{\mbf{G}}$, $\tilde{\mbf{G}}_\mbf{d}$, $\tilde{\mbf{A}}$, and $\tilde{\mbf{b}}$, as in Lemma \ref{lem:nmeas_foe}.
	Finally, choose any $\bm{\gamma}_z=(\bm{\gamma}_x, \bm{\gamma}_w)\in \square Z$ and let $\mbf{L}\in\mathbb{IR}^{n\times m}$ be an interval matrix satisfying $\mbf{L}_{q,:} \supseteq (\mbf{c} - \bm{\gamma}_z)^T \mbf{Q}^{[q]}$ for all $q = 1,\dots,n$. Then,
	\begin{equation} \label{eq:nmeas_firstorderextension}
	\mbf{f}(X,\mbf{u},W) \subseteq \mbf{f}(\bm{\gamma}_x, \mbf{u}, \bm{\gamma}_w) \oplus \nabla^T_z \mbf{f}(\bm{\gamma}_x, \mbf{u}, \bm{\gamma}_w)(Z - \bm{\gamma}_z) \oplus R,
	\end{equation}
	where $R = \tilde{\mbf{c}} \oplus \left[ \tilde{\mbf{G}} \,\; \tilde{\mbf{G}}_{\mbf{d}} \right] B_\infty(\tilde{\mbf{A}}, \tilde{\mbf{b}}) \oplus \gzinclusion (\mbf{L}, (\mbf{c} - \bm{\gamma}_z) \oplus 2\mbf{G} B_\infty(\mbf{A},\mbf{b}) )$.
\end{proposition}

\proof Choose any $(\mbf{x},\mbf{w}) = \mbf{z} \in Z$. Lemma \ref{lem:nmeas_foe} ensures that there exist $\bm{\xi} \in B_\infty(\mbf{A},\mbf{b})$, $\tilde{\bm{\xi}} \in B_\infty(\tilde{\mbf{A}},\tilde{\mbf{b}})$, and $\hat{\mbf{L}} \in \mbf{L}$, such that
\begin{align*}
\mbf{f}(\mbf{x},\mbf{u},\mbf{w}) = \mbf{f}(\bm{\gamma}_x, \mbf{u}, \bm{\gamma}_w) & + \nabla^T \mbf{f}(\bm{\gamma}_x, \mbf{u}, \bm{\gamma}_w)(\mbf{z} - \bm{\gamma}_z) \\ & + \tilde{\mbf{c}} + [ \tilde{\mbf{G}} \,\; \tilde{\mbf{G}}_{\mbf{d}} ] \tilde{\bm{\xi}} + \hat{\mbf{L}} ((\mbf{c} - \bm{\gamma}_z) + 2\mbf{G} \bm{\xi}).
\end{align*}
Therefore, $\mbf{f}(\mbf{x}, \mbf{u}, \mbf{w}) \in \mbf{f} (\bm{\gamma}_x, \mbf{u}, \bm{\gamma}_w) \oplus \nabla^T \mbf{f} (\bm{\gamma}_x, \mbf{u}, \bm{\gamma}_w)(Z - \bm{\gamma}_z) \oplus \gzinclusion(\mbf{L}, (\mbf{c} - \bm{\gamma}_z) \oplus 2 \mbf{G} B_\infty(\mbf{A},\mbf{b})) \oplus \tilde{\mbf{c}} \oplus [\tilde{\mbf{G}} \,\; \bar{\mbf{G}}_{\mbf{d}}] B_\infty (\bar{\mbf{A}}, \tilde{\mbf{b}})$. Thus, \eqref{eq:nmeas_firstorderextension} follows immediately from the definition of $R$. \qed

\begin{remark}\rm \label{rem:nmeas_mvepred} 
	The constrained zonotope $Z_w$ in Proposition \ref{thm:nmeas_mvepred} can be obtained using the mean value extension $\mbf{f}(\bm{\gamma}_x,\mbf{u}, W) \subseteq Z_w = \mbf{f}(\bm{\gamma}_x,\mbf{u},\bm{\gamma}_w) \oplus \gzinclusion \left(\mbf{J}_w,  W - \bm{\gamma}_w \right)$ for a chosen point $\bm{\gamma}_w \in \square W$, with $\mbf{J}_w$ being an interval matrix satisfying $\mbf{J}_w \supseteq \nabla^T_w \mbf{f}(\bm{\gamma}_x,\mbf{u}, \square W)$. In this chapter, the interval matrices $\mbf{J}_x$, $\mbf{J}_w$ (Proposition \ref{thm:nmeas_mvepred}), $\mbf{Q}^{[q]}$, $\tilde{\mbf{Q}}^{[q]}$, $\mbf{L}$ (Proposition \ref{thm:nmeas_foepred}), and similar interval matrices in Propositions \ref{thm:nmeas_mveupdate} and \ref{thm:nmeas_foeupdate}, and Corollaries \ref{thm:nmeas_mveequality} and \ref{thm:nmeas_foeequality}, are all computed using interval arithmetic.
\end{remark}
\begin{remark}\rm \label{rem:nmeas_complexitypred}
	The complexity of the enclosures obtained by Propositions \ref{thm:nmeas_mvepred} and \ref{thm:nmeas_foepred} are similar to the methods in Chapter \ref{cha:nonlineardynamics}. Specifically, if $X$ and $W$ have $n_g$ and $n_{g_w}$ generators, and $n_c$ and $n_{c_w}$ constraints, respectively, then Proposition \ref{thm:nmeas_mvepred} gives $n_g+n_{g_w}+2n$ generators and $n_c + n_{c_w}$ constraints, and Proposition \ref{thm:nmeas_foepred} gives $0.5(n_g+n_{g_w})^2 + 2.5(n_g+n_{g_w}) + 2n$ generators and $0.5(n_c+n_{c_w})^2 + 2.5(n_c+n_{c_w})$ constraints.
\end{remark}

Propositions \ref{thm:nmeas_mvepred} and \ref{thm:nmeas_foepred} permit $\bm{\gamma}_x$ and $\bm{\gamma}_z$ to be chosen from $\square X$ and $\square Z$, respectively, whereas the corresponding results in Chapter \ref{cha:nonlineardynamics} required these points to be chosen from the smaller sets $X$ and $Z$. The following example illustrates the potential advantage of these extensions.

\begin{figure}[!tb]
	\centering{
		\def\svgwidth{0.7\columnwidth}
		{\scriptsize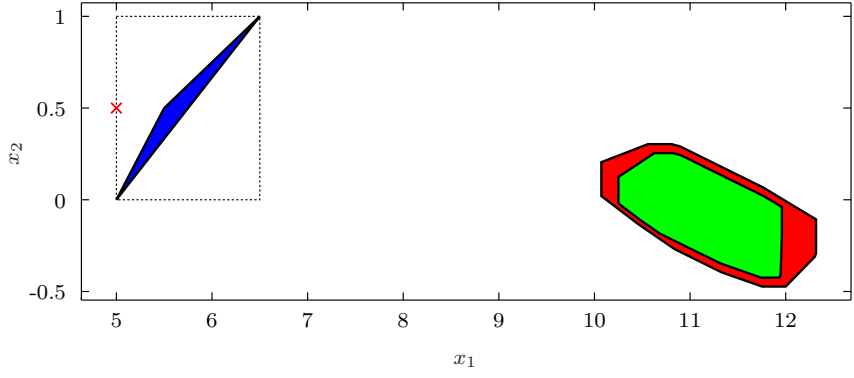}
		\caption{The sets $X$ (blue), $\square X$ (dashed lines), the center of $X$ ($\times$), the enclosures obtained using Proposition \ref{thm:nmeas_foepred} with $\bm{\gamma}_x$ as the center of $X$ (green), and using Theorem \ref{thm:ndyn_firstorder} with $\bm{\gamma}_x$ as the closest point in $X$ to its center (red).}\label{fig:nmeas_choiceofgamma}}
\end{figure}

\begin{example}
Consider the nonlinear mapping $\mbf{f} : \realset^2 \to \realset^2$ defined by
\begin{equation} \label{eq:nmeas_examplechoiceofgamma}
f_1 (\mbf{x})= 3 x_1 - \frac{x_1^2}{7} - \frac{4 x_1 x_2}{4 + x_1}, \quad f_2 (\mbf{x}) = -2 x_2 + \frac{3 x_1 x_2}{4 + x_1},
\end{equation}
and the constrained zonotope
\begin{equation*}
X = \left\{ \begin{bmatrix} 0.5 & 1 & -0.5 \\ 0.5 & 0.5 & 0 \end{bmatrix}, \begin{bmatrix} 5 \\ 0.5 \end{bmatrix}, \begin{bmatrix} -1 & 1 & -1 \end{bmatrix}, 2 \right\}.
\end{equation*}
As shown in Figure \ref{fig:nmeas_choiceofgamma}, the center $\mbf{c}$ in this CG-rep of $X$ does not actually lie in $X$, but does lie in $\square X$. Therefore, it is a valid choice of $\bm{\gamma}_x$ in Proposition \ref{thm:nmeas_foepred}, but not in Theorem \ref{thm:ndyn_firstorder} in Chapter \ref{cha:nonlineardynamics}. Figure \ref{fig:nmeas_choiceofgamma} shows the enclosures of $\mbf{f}(X)$ obtained using Proposition \ref{thm:nmeas_foepred} with this choice of $\bm{\gamma}_x$ and using Theorem \ref{thm:ndyn_firstorder} with $\bm{\gamma}_x$ chosen as the closest point in $X$ to $\mbf{c}$, which is the best heuristic proposed in Chapter \ref{cha:nonlineardynamics}. The enclosure obtained using Proposition \ref{thm:nmeas_foepred} is tighter. Thus, allowing $\bm{\gamma}_x$ to be chosen from the larger region $\square X$ can lead to less conservative results.
\end{example}

\subsection{Update step}
This section presents both mean-value and first-order Taylor methods for the update step in Algorithm \ref{alg:nmeas_estimation}, considering nonlinear measurement equations in contrast to the linear update step used in Chapter \ref{cha:nonlineardynamics}. Specifically, Lemmas \ref{lem:nmeas_mve} and \ref{lem:nmeas_foe} are used, respectively, to formulate the required enclosure in \eqref{eq:nmeas_update0} as the generalized intersection of two constrained zonotopes.

\begin{proposition} \rm \label{thm:nmeas_mveupdate}
	Let $\mbf{g}: \realset^n \times  \realset^{n_u} \times \realset^{n_v} \to \realset^{n_y}$ be of class $\mathcal{C}^1$, let $\mbf{u} \in \realset^{n_u}$, let $X \subset \realset^n$ and $V \subset \realset^{n_v}$ be constrained zonotopes, and choose any $\mbf{y} \in \realset^{n_y}$ such that $\mbf{y}=\mbf{g}(\mbf{x},\mbf{u},\mbf{v})$ for some $(\mbf{x},\mbf{v}) \in X \times V$. Choose any $\bm{\gamma}_x \in \square X$ and any $\tilde{\mbf{J}}_x \in \realsetmat{n_y}{n}$. If $Z_v$ is a constrained zonotope such that $-\mbf{g}(\bm{\gamma}_x,\mbf{u},V) \subseteq Z_v$ and $\mbf{J}_x \in \intvalsetmat{n_y}{n}$ is an interval matrix satisfying $\nabla^T_x \mbf{g}(\square X,\mbf{u},V) \subseteq \mbf{J}_x$, then $$\{ \mbf{x} \in X : \mbf{g}(\mbf{x}, \mbf{u}, \mbf{v}) = \mbf{y}, \, \mbf{v} \in V \} \subseteq X \cap_{\mbf{C}} Y,$$ where $\mbf{C} = \tilde{\mbf{J}}_x$, and $Y = (\mbf{y} + \tilde{\mbf{J}}_x\bm{\gamma}_x) \oplus Z_v \oplus \gzinclusion(\tilde{\mbf{J}}_x - \mbf{J}_x, X - \bm{\gamma}_x)$.
\end{proposition}

\proof
Choose any $(\mbf{x},\mbf{v})\in X \times V$ satisfying $\mbf{g}(\mbf{x},\mbf{u},\mbf{v}) = \mbf{y}$. Lemma \ref{lem:nmeas_mve} ensures that there exists a real matrix $\hat{\mbf{J}}_x\in \mbf{J}_x$ such that	$\mbf{g}(\mbf{x},\mbf{u},\mbf{v}) = \mbf{g}(\bm{\gamma}_x,\mbf{u},\mbf{v}) + \hat{\mbf{J}}_x (\mbf{x} - \bm{\gamma}_x)$. Since $\hat{\mbf{J}}_x = \tilde{\mbf{J}}_x + (\hat{\mbf{J}}_x - \tilde{\mbf{J}}_x)$ holds, then $\mbf{g}(\mbf{x},\mbf{u},\mbf{v}) = \mbf{g}(\bm{\gamma}_x,\mbf{u},\mbf{v}) + \tilde{\mbf{J}}_x (\mbf{x} - \bm{\gamma}_x) + (\hat{\mbf{J}}_x - \tilde{\mbf{J}}_x)(\mbf{x} - \bm{\gamma}_x)$. Consequently,
\begin{align*}
\tilde{\mbf{J}}_x \mbf{x} & = \mbf{g}(\mbf{x},\mbf{u},\mbf{v}) + \tilde{\mbf{J}}_x \bm{\gamma}_x - \mbf{g}(\bm{\gamma}_x,\mbf{u},\mbf{v}) + (\tilde{\mbf{J}}_x - \hat{\mbf{J}}_x) (\mbf{x} - \bm{\gamma}_x) \\
& = \mbf{y} + \tilde{\mbf{J}}_x \bm{\gamma}_x - \mbf{g}(\bm{\gamma}_x,\mbf{u},\mbf{v}) + (\tilde{\mbf{J}}_x - \hat{\mbf{J}}_x) (\mbf{x} - \bm{\gamma}_x) \\
& \in (\mbf{y} + \tilde{\mbf{J}}_x\bm{\gamma}_x) \oplus Z_v \oplus \gzinclusion(\tilde{\mbf{J}}_x - \mbf{J}_x, X - \bm{\gamma}_x) = Y.
\end{align*}
Then, we conclude that $\{\mbf{x} \in X : \mbf{g}(\mbf{x},\mbf{u},\mbf{v}) = \mbf{y}, \mbf{v} \in V\} \subseteq \{\mbf{x} \in X : \tilde{\mbf{J}}_x \mbf{x} \in Y\} = X \cap_{\mbf{C}} Y$. \qed

\begin{remark} \rm \label{rem:Jtildeupdate}
	The constrained zonotope $Z_v$ in Proposition \ref{thm:nmeas_mveupdate} can be obtained as $Z_v = -\mbf{g}(\bm{\gamma}_x,\mbf{u},\bm{\gamma}_v) \oplus \gzinclusion \left(-\mbf{J}_v,  V - \bm{\gamma}_v \right) \supseteq - \mbf{g}(\bm{\gamma}_x,\mbf{u},V)$ for some $\bm{\gamma}_v \in \square V$ and interval matrix $\mbf{J}_v \supseteq \nabla^T_v \mbf{g}(\bm{\gamma}_x,\mbf{u}, \square V)$. The matrix $\tilde{\mbf{J}}_x$ is a free parameter in Proposition \ref{thm:nmeas_mveupdate}. Choosing $\tilde{\mbf{J}}_x = \text{mid}(\mbf{J}_x)$ gives $\text{mid}(\tilde{\mbf{J}}_x - \mbf{J}_x) = \mbf{0}$, then $\gzinclusion(\tilde{\mbf{J}}_x - \mbf{J}_x, X - \bm{\gamma}_x) = \midpoint{\tilde{\mbf{J}}_x - \mbf{J}_x}(X - \bm{\gamma}_x) \oplus \mbf{P}B_\infty^{n_y} = \mbf{P}B_\infty^{n_y}$, with $\mbf{P}$ defined as in Theorem \ref{thm:ndyn_czinclusion}. This choice is adopted throughout this chapter.
\end{remark}

\begin{proposition} \rm \label{thm:nmeas_foeupdate}
	Let $\mbf{g}: \realset^n \times \realset^{n_u} \times \realset^{n_v} \to \realset^{n_y}$ be of class $\mathcal{C}^2$, let $\mbf{u} \in \realset^{n_u}$, let $X = \{\mbf{G}_x, \mbf{c}_x, \mbf{A}_x, \mbf{b}_x\}$ and $V = \{\mbf{G}_v, \mbf{c}_v, \mbf{A}_v, \mbf{b}_v\}$ be constrained zonotopes with $n_g$ generators and $n_c$ constraints, and with $n_{g_v}$ generators and $n_{g_c}$ constraints, respectively, and choose any $\mbf{y} \in \realset^{n_y}$ such that $\mbf{y}=\mbf{g}(\mbf{x},\mbf{u},\mbf{v})$ for some $(\mbf{x},\mbf{v}) \in X \times V$. Denote $\mbf{z}=(\mbf{x},\mbf{v})$ and $Z = X \times V = \{\mbf{G}, \mbf{c}, \mbf{A}, \mbf{b}\} \subset \realset^{n+n_v}$. For each $q = 1,2,\dots,n_y$, let $\mbf{Q}^{[q]}\in\mathbb{IR}^{(n+n_v)\times (n+n_v)}$ and $\tilde{\mbf{Q}}^{[q]}\in\mathbb{IR}^{(n_g+n_{g_v})\times (n_g+n_{g_v})}$ be interval matrices satisfying $\mbf{Q}^{[q]} \supseteq \mbf{H}_z g_q (\square X,\mbf{u}, \square V)$ and $\tilde{\mbf{Q}}^{[q]} \supseteq \mbf{G}^T \mbf{Q}^{[q]} \mbf{G}$. Moreover, define $\tilde{\mbf{c}}$, $\tilde{\mbf{G}}$, $\tilde{\mbf{G}}_\mbf{d}$, $\tilde{\mbf{A}}$, and $\tilde{\mbf{b}}$, as in Lemma \ref{lem:nmeas_foe}.    
	Finally, choose any $\bm{\gamma}_z = (\bm{\gamma}_x, \bm{\gamma}_v) \in \square Z$, and let $\mbf{L}\in\mathbb{IR}^{n_y\times (n+n_v)}$ be an interval matrix satisfying $\mbf{L}_{q,:} \supseteq (\mbf{c} - \bm{\gamma}_z)^T \mbf{Q}^{[q]}$ for all $q = 1,\dots,n_y$. Then, 
	\begin{equation*}
	\{ \mbf{x} \in X : \mbf{g}(\mbf{x}, \mbf{u}, \mbf{v}) = \mbf{y}, \, \mbf{v} \in V \} \subseteq X \cap_{\mbf{C}} Y,
	\end{equation*}
	where $\mbf{C} = \nabla^T_x \mbf{g}(\bm{\gamma}_x,\mbf{u},\bm{\gamma}_v)$, $Y = (\mbf{y} - \mbf{g}(\bm{\gamma}_x,\mbf{u},\bm{\gamma}_v) + \nabla^T_z \mbf{g}(\bm{\gamma}_x,\mbf{u},\bm{\gamma}_v)\bm{\gamma}_z) \oplus (- \nabla^T_v \mbf{g}(\bm{\gamma}_x,\mbf{u},\bm{\gamma}_v) V) \oplus (-R)$, and $R = \tilde{\mbf{c}} \oplus [ \tilde{\mbf{G}} \,\; \tilde{\mbf{G}}_{\mbf{v}} ] B_\infty(\tilde{\mbf{A}}, \tilde{\mbf{b}}) \oplus \gzinclusion (\mbf{L}, (\mbf{c} - \bm{\gamma}_z) \oplus 2\mbf{G} B_\infty(\mbf{A},\mbf{b}) )$.
\end{proposition}

\proof
Choose $(\mbf{x},\mbf{v}) = \mbf{z} \in Z$ such that $\mbf{g}(\mbf{x}, \mbf{u}, \mbf{v}) = \mbf{y}$. Lemma \ref{lem:nmeas_foe} ensures that there exist $\bm{\xi} \in B_\infty(\mbf{A},\mbf{b})$, $\tilde{\bm{\xi}} \in B_\infty(\tilde{\mbf{A}},\tilde{\mbf{b}})$, and $\hat{\mbf{L}} \in \mbf{L}$, such that 
\begin{align*}
	\mbf{g}(\mbf{x},\mbf{u},\mbf{v}) & = \mbf{g} (\bm{\gamma}_x,\mbf{u},\bm{\gamma}_v) + \nabla^T_x \mbf{g}(\bm{\gamma}_x,\mbf{u},\bm{\gamma}_v)(\mbf{x} - \bm{\gamma}_x) \\ & + \nabla^T_v \mbf{g}(\bm{\gamma}_x,\mbf{u},\bm{\gamma}_v)(\mbf{v} - \bm{\gamma}_v) + \hat{\mbf{L}}(\mbf{p} + 2 \mbf{G} \bm{\xi}) + \tilde{\mbf{c}} + [\tilde{\mbf{G}} \,\; \bar{\mbf{G}}_{\mbf{v}}] \bar{\bm{\xi}},
\end{align*}
where $\mbf{p} = \mbf{c} - \bm{\gamma}_z$. Since $\mbf{g}(\mbf{x}, \mbf{u}, \mbf{v}) = \mbf{y}$, 
\begin{equation*}
\begin{aligned}
	\nabla^T_x \mbf{g} (\bm{\gamma}_x,\mbf{u},\bm{\gamma}_v)\mbf{x} & = \mbf{y} - \mbf{g}(\bm{\gamma}_x,\mbf{u},\bm{\gamma}_v) + \nabla^T_z \mbf{g}(\bm{\gamma}_x,\mbf{u},\bm{\gamma}_v)\bm{\gamma}_z \\ & \quad- \nabla^T_v \mbf{g}(\bm{\gamma}_x,\mbf{u},\bm{\gamma}_v)\mbf{v} - \hat{\mbf{L}}(\mbf{p} + 2 \mbf{G} \bm{\xi}) - \tilde{\mbf{c}} - [\tilde{\mbf{G}} \,\; \bar{\mbf{G}}_{\mbf{v}}] \bar{\bm{\xi}} \\
	& \in (\mbf{y} - \mbf{g}(\bm{\gamma}_x,\mbf{u},\bm{\gamma}_v) + \nabla^T_z \mbf{g}(\bm{\gamma}_x,\mbf{u},\bm{\gamma}_v)\bm{\gamma}_z) \\ & \quad \oplus (- \nabla^T_v \mbf{g}(\bm{\gamma}_x,\mbf{u},\bm{\gamma}_v)V) \oplus (-R) = Y
\end{aligned}
\end{equation*}
Then, we conclude that $\{\mbf{x} \in X : \mbf{g}(\mbf{x},\mbf{u},\mbf{v}) = \mbf{y}, \mbf{v} \in V\} \subseteq \{\mbf{x} \in X : \nabla^T_x \mbf{g}(\bm{\gamma}_x,\mbf{u},\bm{\gamma}_v)\mbf{x} \in Y\} = X \cap_{\mbf{C}} Y$. \qed

\begin{remark} \rm \label{rem:nmeas_complexityupdate}  
	If $X$ and $V$ have $n_g$ and $n_{g_v}$ generators, and $n_c$ and $n_{c_v}$ constraints, respectively, then the enclosure obtained by Proposition \ref{thm:nmeas_mveupdate} has $2n_g+n_{g_v}+2n_y$ generators and $2n_c + n_{c_v} + n_y$ constraints, and the enclosure obtained by Proposition \ref{thm:nmeas_foeupdate} has $0.5(n_g+n_{g_v})^2 + 2.5(n_g+n_{g_v}) + 2n_y$ generators and $0.5(n_c+n_{c_v})^2 + 2.5(n_c+n_{c_v}) +n_y$ constraints.
\end{remark}

\begin{remark} \rm \label{rem:nmeas_affine}
	If $\mbf{f}$ and $\mbf{g}$ are affine in $\mbf{w}$ and $\mbf{v}$ (i.e., $\mbf{f}(\mbf{x},\mbf{u},\mbf{w}) = \bm{\phi}(\mbf{x},\mbf{u}) + \bm{\Phi}(\mbf{x},\mbf{u}) \mbf{w}$ and $\mbf{g}(\mbf{x},\mbf{u},\mbf{v}) = \bm{\psi}(\mbf{x},\mbf{u}) + \bm{\Psi}(\mbf{x},\mbf{u}) \mbf{v}$), then the constrained zonotopes $Z_w \supseteq \mbf{f} (\bm{\gamma}_x,\mbf{u},W)$ and $Z_v \supseteq -\mbf{g} (\bm{\gamma}_x,\mbf{u},V)$ in Propositions \ref{thm:nmeas_mvepred} and \ref{thm:nmeas_mveupdate} can be computed exactly by $Z_w = \bm{\phi}(\bm{\gamma}_x,\mbf{u}) \oplus \bm{\Phi}(\bm{\gamma}_x,\mbf{u}) W$ and $Z_v = - \bm{\psi}(\bm{\gamma}_x,\mbf{u}) \oplus (- \bm{\Psi}(\bm{\gamma}_x,\mbf{u}) V)$, respectively.
\end{remark}

\subsection{Consistency step}
This section presents both mean value and first-order Taylor methods for the consistency step in Algorithm \ref{alg:nmeas_estimation}. As in the previous section, the obtained enclosure is formulated as the generalized intersection of two constrained zonotopes. Since the proposed methods for the consistency step are direct consequences of Propositions \ref{thm:nmeas_mveupdate} and \ref{thm:nmeas_foeupdate}, these are presented as corollaries.

\begin{corollary} \rm \label{thm:nmeas_mveequality}
	Let $\mbf{h}: \realset^n \to \realset^{n_h}$ be of class $\mathcal{C}^1$, and let $X \subset \realset^n$ be a constrained zonotope. Choose any $\bm{\gamma}_x \in \square X$ and any $\tilde{\mbf{J}}_x \in \realsetmat{n_h}{n}$. If $\mbf{J}_x \in \intvalsetmat{n_h}{n}$ is an interval matrix satisfying $\nabla^T_x \mbf{h}(\square X) \subseteq \mbf{J}_x$, then $\{ \mbf{x} \in X : \mbf{h}(\mbf{x}) = \bm{0}\} \subseteq X \cap_{\mbf{D}} H$, where $\mbf{D} = \tilde{\mbf{J}}_x$, and $H = (\tilde{\mbf{J}}_x\bm{\gamma}_x - \mbf{h}(\bm{\gamma}_x)) \oplus \gzinclusion(\tilde{\mbf{J}}_x - \mbf{J}_x, X - \bm{\gamma}_x)$.
\end{corollary}

\proof 
Choose any $\mbf{x} \in X$ satisfying $\mbf{h}(\mbf{x})=\bm{0}$. Lemma \ref{lem:nmeas_mve} ensures that there exists a real matrix $\hat{\mbf{J}}_x\in \mbf{J}_x$ such that $\mbf{h}(\mbf{x}) = \mbf{h}(\bm{\gamma}_x) + \hat{\mbf{J}}_x (\mbf{x} - \bm{\gamma}_x)$. Since $\hat{\mbf{J}}_x = \tilde{\mbf{J}}_x + (\hat{\mbf{J}}_x - \tilde{\mbf{J}}_x)$ holds, then $\mbf{h}(\mbf{x}) = \mbf{h}(\bm{\gamma}_x) + \tilde{\mbf{J}}_x (\mbf{x} - \bm{\gamma}_x) + (\hat{\mbf{J}}_x - \tilde{\mbf{J}}_x)(\mbf{x} - \bm{\gamma}_x)$. Consequently,
\begin{align*}
\tilde{\mbf{J}}_x \mbf{x} & = \mbf{h}(\mbf{x}) + \tilde{\mbf{J}}_x \bm{\gamma}_x - \mbf{h}(\bm{\gamma}_x) + (\tilde{\mbf{J}}_x - \hat{\mbf{J}}_x) (\mbf{x} - \bm{\gamma}_x) \\
& = \bm{0} + \tilde{\mbf{J}}_x \bm{\gamma}_x - \mbf{h}(\bm{\gamma}_x) + (\tilde{\mbf{J}}_x - \hat{\mbf{J}}_x) (\mbf{x} - \bm{\gamma}_x) \\
& \in ( \tilde{\mbf{J}}_x\bm{\gamma}_x - \mbf{h}(\bm{\gamma}_x)) \oplus \gzinclusion(\tilde{\mbf{J}}_x - \mbf{J}_x, X - \bm{\gamma}_x) = H.
\end{align*}
Therefore, $\{\mbf{x} \in X : \mbf{h}(\mbf{x}) = \bm{0}\} \subseteq \{\mbf{x} \in X : \tilde{\mbf{J}}_x \mbf{x} \in H \} = X \cap_{\mbf{D}} H$.	
\qed

\begin{remark} \rm \label{rem:Jtildeequality}
	As in the update step, the matrix $\tilde{\mbf{J}}_x$ is a free parameter in Proposition \ref{thm:nmeas_mveequality}. If $\tilde{\mbf{J}}_x = \text{mid}(\mbf{J}_x)$, then $\text{mid}(\tilde{\mbf{J}}_x - \mbf{J}_x) = \mbf{0}$, and $\gzinclusion(\tilde{\mbf{J}}_x - \mbf{J}, X - \bm{\gamma}_x) = \midpoint{\tilde{\mbf{J}}_x - \mbf{J}_x}(X - \bm{\gamma}_x) \oplus \mbf{P}B_\infty^{n_y} = \mbf{P}B_\infty^{n_y}$. Therefore, this choice is adopted also for the consistency step. 
\end{remark}

\begin{corollary}
	\rm \label{thm:nmeas_foeequality}
	Let $\mbf{h}: \realset^n \to \realset^{n_h}$ be of class $\mathcal{C}^2$ and let $X = \{\mbf{G}, \mbf{c}, \mbf{A}, \mbf{b}\}$ be a constrained zonotope with $n_g$ generators and $n_c$ constraints. For each $q = 1,2,\dots,n_h$, let $\mbf{Q}^{[q]}\in\mathbb{IR}^{n\times n}$ and $\tilde{\mbf{Q}}^{[q]}\in\mathbb{IR}^{n_g\times n_g}$ be interval matrices satisfying $\mbf{Q}^{[q]} \supseteq \mbf{H}_x h_{q} (\square X)$ and $\tilde{\mbf{Q}}^{[q]} \supseteq \mbf{G}^T \mbf{Q}^{[q]} \mbf{G}$. Moreover, define $\tilde{\mbf{c}}$, $\tilde{\mbf{G}}$, $\tilde{\mbf{G}}_\mbf{d}$, $\tilde{\mbf{A}}$, and $\tilde{\mbf{b}}$, as in Lemma \ref{lem:nmeas_foe}.    
	Finally, choose any $\bm{\gamma}_x \in \square X$ and let $\mbf{L}\in\mathbb{IR}^{n_h \times n}$ be an interval matrix satisfying $\mbf{L}_{q,:} \supseteq (\mbf{c} - \bm{\gamma}_x)^T \mbf{Q}^{[q]}$ for all $q = 1,\dots,n_h$. Then, 
	\begin{equation*}
	\{ \mbf{x} \in X : \mbf{h}(\mbf{x}) = \bm{0} \} \subseteq X \cap_{\mbf{D}} H,
	\end{equation*}
	where $\mbf{D} = \nabla^T_x \mbf{h}(\bm{\gamma}_x)$, $H = (- \mbf{h}(\bm{\gamma}_x) + \nabla^T_x \mbf{h}(\bm{\gamma}_x)\bm{\gamma}_x) \oplus (-R)$, and $R = \tilde{\mbf{c}} \oplus [ \tilde{\mbf{G}} \,\; \tilde{\mbf{G}}_{\mbf{d}} ] B_\infty(\tilde{\mbf{A}}, \tilde{\mbf{b}}) \oplus \gzinclusion (\mbf{L}, (\mbf{c} - \bm{\gamma}_x) \oplus 2\mbf{G} B_\infty(\mbf{A},\mbf{b}) )$.
\end{corollary}

\proof
Choose $\mbf{x} \in X$ such that $\mbf{h}(\mbf{x}) = \bm{0}$. Lemma \ref{lem:nmeas_foe} ensures that there exist $\bm{\xi} \in B_\infty(\mbf{A},\mbf{b})$, $\tilde{\bm{\xi}} \in B_\infty(\tilde{\mbf{A}},\tilde{\mbf{b}})$, and $\hat{\mbf{L}} \in \mbf{L}$, such that
\begin{align*}
\mbf{h}(\mbf{x}) & = \mbf{h} (\bm{\gamma}_x) + \nabla^T_x \mbf{h}(\bm{\gamma}_x)(\mbf{x} - \bm{\gamma}_x) \\ & + \hat{\mbf{L}}(\mbf{p} + 2 \mbf{G} \bm{\xi}) + \tilde{\mbf{c}} + [\tilde{\mbf{G}} \,\; \bar{\mbf{G}}_{\mbf{v}}] \bar{\bm{\xi}}.
\end{align*}
with $\mbf{p} = \mbf{c} - \bm{\gamma}_x$. Since $\mbf{h}(\mbf{x}) = \bm{0}$, we have $\nabla^T_x \mbf{h} (\bm{\gamma})\mbf{x} = - \mbf{h}(\bm{\gamma}_x) + \nabla^T_x \mbf{h}(\bm{\gamma}_x)\bm{\gamma}_x - \hat{\mbf{L}}(\mbf{p} + 2 \mbf{G} \bm{\xi}) - \tilde{\mbf{c}} - [\tilde{\mbf{G}} \,\; \bar{\mbf{G}}_{\mbf{v}}] \bar{\bm{\xi}}$, and therefore
\begin{equation*}
\nabla^T_x \mbf{h} (\bm{\gamma}_x)\mbf{x} \in (- \mbf{h}(\bm{\gamma}_x) + \nabla^T_x \mbf{h}(\bm{\gamma}_x)\bm{\gamma}_x)  \oplus (-R) = H.
\end{equation*}
We conclude that $\{\mbf{x} \in X : \mbf{h}(\mbf{x}) = \bm{0}\} \subseteq \{\mbf{x} \in X : \nabla^T_x \mbf{h}(\bm{\gamma}_x)\mbf{x} \in H\} = X \cap_{\mbf{D}} H$.
\qed

\begin{remark} \rm \label{rem:consistencycomplexity}
	If $X$ has $n_g$ generators and $n_c$ constraints, then the enclosure obtained from Proposition \ref{thm:nmeas_mveequality} has $2n_g+n_y$ generators and $2n_c + n_y$ constraints, and the enclosure obtained from Proposition \ref{thm:nmeas_foeequality} has $0.5n_g^2 + 2.5n_g + 2n_y$ generators and $0.5n_c^2 + 2.5n_c +n_y$ constraints.
\end{remark}

\begin{remark} \rm \label{rem:nmeas_priorbounds}
	The enclosures in Propositions \ref{thm:nmeas_mveequality} and \ref{thm:nmeas_foeequality} can be tightened if an enclosure $X_\text{F}$ in CG-rep of the feasible state set $\{\mbf{x} \in \realset^n : \mbf{h}(\mbf{x}) = \mbf{0}\}$ is known \emph{a priori}. Such an enclosure can be obtained offline by using, for instance, the contractor programming methods in \cite{Chabert2009}. In both propositions, $\mbf{h}$ is conservatively approximated over $X$, and the size of the resulting constrained zonotope $Z$ is proportional to the size of $X$. If $X$ is large, significant improvement can result from setting $X\gets X\cap X_{\text{F}}$ prior to applying the proposition. This situation is likely in practice because, within the overall estimation framework \eqref{eq:nmeas_prediction0}--\eqref{eq:nmeas_consistency0}, the set $\hat{X}_k$ will play the role of $X$ in Propositions \ref{thm:nmeas_mveequality} and \ref{thm:nmeas_foeequality}, and $\hat{X}_k$ can be very conservative before accounting for the invariant $\mbf{h}(\mbf{x}_k) = \mbf{0}$ (see Section \ref{sec:nmeas_exinvariants}).
\end{remark}

\begin{remark} \rm
	In the literature, different methods has been proposed to solve a more general problem of computing the intersection of sets with nonlinear equations, but applied to differents problem such as reachability analysis \citep{Kochdumper2020}. Nevertheless, none of the existing methods can be applied to sets described by constrained zonotopes, and the obtained enclosure is generally not convex.
\end{remark}

\subsection{Selection of approximation point} \label{sec:nmeas_choiceofgamma}

The methods proposed in this chapter require heuristics to choose an approximation point $(\bm{\gamma}_x,\bm{\gamma}_w,\bm{\gamma}_v) \in \square X\times \square W \times \square V$, where $X$ stands for either $\tilde{X}_{k-1}$, $\bar{X}_{k}$, or $\hat{X}_{k}$ depending on the step in \eqref{eq:nmeas_prediction0}--\eqref{eq:nmeas_consistency0}. As discussed in Chapter \ref{cha:nonlineardynamics}, the center of the CG-rep of $X \times W \times V$ cannot be chosen in general because it may not belong to either $X \times W \times V$ or $\square X\times \square W \times \square V$. However, in contrast to Chapter \ref{cha:nonlineardynamics}, the center of the interval $\square X\times \square W \times \square V$ is a valid choice here. This first heuristic is summarized as follows:
\begin{itemize}
	\item[\bf\emph{C1)}] $(\bm{\gamma}_x,\bm{\gamma}_w,\bm{\gamma}_v)$ is given by the center of $\square X\times \square W \times \square V$.
\end{itemize}

Despite its efficiency, \emph{C1} is not optimal in any sense and can lead to conservative enclosures. Following Chapter \ref{cha:nonlineardynamics}, we next present an improved heuristic \emph{C2} specifically for use with the methods based on mean value extensions in Propositions \ref{thm:nmeas_mvepred} and \ref{thm:nmeas_mveupdate}, and Corollary \ref{thm:nmeas_mveequality} (the exact heuristic in Chapter \ref{cha:nonlineardynamics} is not optimal here because it restricts $(\bm{\gamma}_x,\bm{\gamma}_w,\bm{\gamma}_v)$ to $X\times W \times V$ rather than $\square X\times \square W \times \square V$). Propositions \ref{thm:nmeas_mvepred} and \ref{thm:nmeas_mveupdate}, and Corollary \ref{thm:nmeas_mveequality} all apply the CZ-inclusion operator defined in Theorem \ref{thm:ndyn_czinclusion} with the second argument taking the form $X - \bm{\gamma}_x$. The idea behind \emph{C2} is therefore to choose $\bm{\gamma}_x$ so as to minimize the conservatism of this CZ-inclusion operator.  

In this sense, consider the CZ-inclusion operator $\gzinclusion(\mbf{J},Z-\bm{\gamma})$ for arbitrary $Z = \{\mbf{G},\mbf{c},\mbf{A},\mbf{b}\} \subset \realset^{n_z}$, $\bm{\gamma} \in \realset^{n_z}$, and $\mbf{J} \in \intvalsetmat{m}{n_z}$. As per Theorem \ref{thm:ndyn_czinclusion} and Remark \ref{rem:ndyn_czinclusion}, computing $\gzinclusion(\mbf{J},Z-\bm{\gamma})$ requires a zonotope $\{\bar{\mbf{G}},\bar{\mbf{c}}\} \supseteq (Z-\bm{\gamma})$ that is computed by eliminating all constraints from $(Z-\bm{\gamma})$ using the constraint elimination algorithm given by Method \ref{meth:czconelim}. Based on that algorithm, Chapter \ref{cha:nonlineardynamics} derived a closed form expression for the resulting center $\bar{\mbf{c}}$ as a function of $(\mbf{G},\mbf{c},\mbf{A},\mbf{b})$ and $\bm{\gamma}$, obtained by displacing \eqref{eq:ndyn_pbar} by $-\bm{\gamma}$, which takes the form
\begin{equation*}
\bar{\mbf{c}} = \mbf{c}-\bm{\gamma} + \bm{\delta}(\mbf{G},\mbf{A},\mbf{b}).
\end{equation*}
The definition of $\bm{\delta}(\mbf{G},\mbf{A},\mbf{b})$ can be deduced from \eqref{eq:ndyn_pbar} and is omitted here for brevity. This $\bar{\mbf{c}}$ is then used to compute $\mbf{m}\supset(\mbf{J} - \midpoint{\mbf{J}}) \bar{\mbf{c}}$ using interval arithmetic, and the size of the final enclosure $\gzinclusion(\mbf{J},Z-\bm{\gamma})$ is proportional to $\text{rad}(\mbf{m}) = (1/2)\text{diam}(\mbf{m})$. Thus, the aim is to choose $\bm{\gamma}$ so as to minimize $\text{diam}(\mbf{m})$.

\begin{proposition} \rm \label{lem:nmeas_choosemindiam}
	Let $Z = \{\mbf{G},\mbf{c},\mbf{A},\mbf{b}\} \subset \realset^m$, $\mbf{J} \in \intvalsetmat{m}{n}$, and $[\mbf{z}^\text{L},\mbf{z}^\text{U}] = \square Z$. For any choice of $\bm{\gamma}\in\square Z$, let $\mbf{m}_{\bm{\gamma}} \supseteq (\mbf{J} - \midpoint{\mbf{J}}) \bar{\mbf{c}}_{\bm{\gamma}}$ be an interval vector computed using interval arithmetic, where $\bar{\mbf{c}}_{\bm{\gamma}} = \mbf{c} - \bm{\gamma} + \bm{\delta}(\mbf{G},\mbf{A},\mbf{b})$. Then, $\bm{\gamma}^* \in \square Z$ minimizes $\|\diam{\mbf{m}_{\bm{\gamma}}}\|_1$ iff it is the solution to the linear program (LP) 
	\begin{equation} \label{eq:nmeas_choosehmindiam}
	\underset{\bm{\gamma}}{\min}~ \|\bm{\Theta} \bar{\mbf{c}}_{\bm{\gamma}}\|_1, \quad
	\text{s.t.} \quad \mbf{z}^\text{L} \leq \bm{\gamma} \leq \mbf{z}^\text{U} ,
	\end{equation}
	where $\Theta_{jj} = \sum_{i=1}^m \diam{J_{ij}}$ and $\Theta_{ij} = 0$ for all $i\neq j$.
\end{proposition}

\proof Each component of $(\mbf{J} - \midpoint{\mbf{J}})\in \intvalsetmat{m}{n}$ is an interval satisfying $(J_{ij} - \midpoint{J_{ij}}) = (1/2) \diam{J_{ij}}[-1,1]$. Moreover, $a [-1,1] = |a| [-1,1]$ holds for every $a \in \realset$. Therefore $m_{\bm{\gamma},i} = \sum_{j=1}^{n} (1/2) \diam{J_{ij}} |\bar{c}_{\bm{\gamma},j}| [-1,1]$. Consequently, $\diam{m_{\bm{\gamma},i}}  \sum_{j=1}^n \diam{J_{ij}} |\bar{c}_{\bm{\gamma},j}|$, and
\begin{align*}
\|\diam{\mbf{m}_{\bm{\gamma}}}\|_1 & = \sum_{i=1}^m \sum_{j=1}^n \diam{J_{ij}} |\bar{c}_{\bm{\gamma},j}| = \sum_{j=1}^n \left( \sum_{i=1}^m \diam{J_{ij}} \right) |\bar{c}_{\bm{\gamma},j}| \\
& = \sum_{j=1}^n \Theta_{jj} |\bar{c}_{\bm{\gamma},j}| =  \|\bm{\Theta} \bar{\mbf{c}}_{\bm{\gamma}}\|_1.
\end{align*}
The constraints in \eqref{eq:nmeas_choosehmindiam} follow directly from the requirement that $\bm{\gamma} \in \square Z$. \qed

This heuristic is summarized as follows:
\begin{itemize}
	\item[\bf\emph{C2)}] $\bm{\gamma}_x$, $\bm{\gamma}_w$, and $\bm{\gamma}_v$ are given by the points obtained from Proposition \ref{lem:nmeas_choosemindiam} for $(\mbf{J},X - \bm{\gamma}_x)$ in Proposition \ref{thm:nmeas_mvepred}, $(\tilde{\mbf{J}} - \mbf{J},X - \bm{\gamma}_x)$ in Proposition \ref{thm:nmeas_mveupdate} and Corollary \ref{thm:nmeas_mveequality}, $(\mbf{J}_w,W - \bm{\gamma}_w)$, and $(\mbf{J}_v,V - \bm{\gamma}_v)$, respectively.
\end{itemize}

Next, we present a heuristic specifically for the methods based on first-order Taylor extensions in Propositions \ref{thm:nmeas_foepred} and \ref{thm:nmeas_foeupdate}, and Corollary \ref{thm:nmeas_foeequality}. The conservatism of these methods is directly related to the conservatism in the remainder $R$, which is mostly affected by the size of the interval matrices $\mbf{Q}^{[q]}$, $\tilde{\mbf{Q}}^{[q]}$, and $\mbf{L}$. The matrices $\mbf{Q}^{[q]}$ and $\tilde{\mbf{Q}}^{[q]}$ are unaffected by the choice of $(\bm{\gamma}_x,\bm{\gamma}_w,\bm{\gamma}_v)$. However, similarly to Theorem \ref{thm:ndyn_firstorder}, the radius of $\mbf{L}$ is proportional to the differences $\mbf{c}_x - \bm{\gamma}_x$, $\mbf{c}_w - \bm{\gamma}_w$, and $\mbf{c}_v - \bm{\gamma}_v$. Therefore, based on Corollary \ref{col:ndyn_C3} and Proposition \ref{propo:ndyn_closest}, we propose the following heuristic for the first-order Taylor extensions:
\begin{itemize}
	\item[\bf\emph{C3)}] $(\bm{\gamma}_x,\bm{\gamma}_w,\bm{\gamma}_v)$ is the closest point to the center of $X\times W \times V$ that belongs to $\square X\times \square W \times \square V$, obtained by solving the respective LPs ${\min}~ \{\| \bm{\gamma} - \mbf{c}_x \|_1 : \bm{\gamma}  \in \square X\}$, ${\min}~ \{\| \bm{\gamma} - \mbf{c}_w \|_1 : \bm{\gamma}  \in \square W\}$, and ${\min}~ \{\| \bm{\gamma} - \mbf{c}_v \|_1 : \bm{\gamma}  \in \square V\}$.
\end{itemize}

\subsection{Linear case}

This section investigates the results obtained by the prediction, update, and consistency steps for nonlinear systems developed in this chapter when applied to linear systems. In this case, the resulting enclosures are straightforward. Consider the linear discrete-time system
\begin{subequations} \label{eq:nmeas_systemlinear}
	\begin{align}
	\mbf{x}_k & = \mbf{A} \mbf{x}_{k-1} + \mbf{B}_u \mbf{u}_{k-1} + \mbf{B}_w \mbf{w}_{k-1}, \label{eq:nmeas_systemlineardynamics} \\
	\mbf{y}_k & = \mbf{C} \mbf{x}_{k} + \mbf{D}_u \mbf{u}_{k} + \mbf{D}_v \mbf{v}_{k}, \label{eq:nmeas_systemlinearmeasurement}
	\end{align}
\end{subequations}
where $\mbf{A} \in \realsetmat{n}{n}$, $\mbf{B}_u \in \realsetmat{n}{n_u}$, $\mbf{B}_w \in \realsetmat{n}{n_w}$, $\mbf{C} \in \realsetmat{n_y}{n}$, $\mbf{D}_u \in \realsetmat{n_y}{n_u}$, $\mbf{D}_v \in \realsetmat{n_y}{n_v}$, with known polytopic bounds $(\mbf{x}_0,\mbf{w}_k,\mbf{v}_k) \in X_0 \times W \times V$. Moreover, assume that the trajectories of \eqref{eq:nmeas_systemlinear} satisfy the linear invariants $\mbf{E}\mbf{x}_k = \mbf{d}$, with $\mbf{E} \in \realsetmat{n_\text{d}}{n}$, and $\mbf{d} \in \realset^{n_\text{d}}$. 
Given the previous set $\tilde{X}_{k-1}$, the prediction step \eqref{eq:nmeas_prediction0} and the update step \eqref{eq:nmeas_update0} are computed exactly for \eqref{eq:nmeas_systemlineardynamics}--\eqref{eq:nmeas_systemlinearmeasurement} as in Section \ref{sec:pre_linearestimation}:
\begin{align}
\bar{X}_k & = \mbf{A} \tilde{X}_{k-1} \oplus \mbf{B}_u \mbf{u}_{k-1} \oplus \mbf{B}_w W, \label{eq:nmeas_predictionlinear} \\
\hat{X}_k & = \bar{X}_{k} \cap_{\mbf{C}} ((\mbf{y}_k - \mbf{D}_u \mbf{u}) \oplus (-\mbf{D}_v V)). \label{eq:nmeas_updatelinear}
\end{align}
All the set operations in \eqref{eq:nmeas_predictionlinear}--\eqref{eq:nmeas_updatelinear} can be computed straightforwardly using \eqref{eq:pre_czlimage}--\eqref{eq:pre_czintersection}. To compute the consistency step \eqref{eq:nmeas_consistency0}, note that in this case this can be written as $\tilde{X}_k \supseteq \{ \mbf{x} \in \hat{X}_k : \mbf{E}\mbf{x}_k \in \{\mbf{d}\} \}$, where $\{\mbf{d}\}$ denotes a singleton that contains only the point $\mbf{d}$. Therefore, if $\hat{X}_{k} = \{\hat{\mbf{G}}_k, \hat{\mbf{c}}_k, \hat{\mbf{A}}_k, \hat{\mbf{b}}_k\}$, then $\tilde{X}_k$ is given by
\begin{equation} \label{eq:nmeas_consistencylinear}
\tilde{X}_k = \hat{X}_{k} \cap_{\mbf{E}} \{\mbf{d}\} = \left\{ \hat{\mbf{G}}_k, \hat{\mbf{c}}_k, \begin{bmatrix} \hat{\mbf{A}}_k \\ \mbf{E} \hat{\mbf{G}}_k \end{bmatrix}, \begin{bmatrix} \hat{\mbf{b}}_k \\  \mbf{d} - \mbf{E} \hat{\mbf{c}}_k \end{bmatrix} \right\}.
\end{equation}
Hence, the consistency step can be computed exactly as well. Therefore, similarly to the method described in Section \ref{sec:pre_linearestimation} and as a direct extension for linear invariants of the linear method proposed in \cite{Scott2016}, the only source of conservatism in the set-valued state estimation of \eqref{eq:nmeas_systemlinear} using constrained zonotopes through the steps \eqref{eq:nmeas_predictionlinear}--\eqref{eq:nmeas_consistencylinear} arises if the complexity of the sets are limited, which requires the use of complexity reduction methods (Section \ref{sec:complexityreduction}).

\begin{remark} \rm \label{rem:nmeas_linearexact}
	If complexity reduction methods are not employed, the enclosures obtained for both the prediction step \eqref{eq:nmeas_predictionlinear} and consistency step \eqref{eq:nmeas_consistencylinear} are exact using constrained zonotopes. Therefore, for linear systems, in the case the complexity of the sets is not limited, since the trajectories of the system satisfy $\mbf{E}\mbf{x}_k = \mbf{d}$ by assumption, it is sufficient to compute the consistency step \eqref{eq:nmeas_consistencylinear} only at $k = 0$.
\end{remark}

\section{Computational complexity}

Table \ref{tab:nmeas_complexity} shows the computational complexity of the methods proposed in this chapter for the prediction, update, and consistency steps, using the mean value extension (Propositions \ref{thm:nmeas_mvepred} and \ref{thm:nmeas_mveupdate}, and Corollary \ref{thm:nmeas_mveequality}) and the first-order Taylor extension (Propositions \ref{thm:nmeas_foepred} and \ref{thm:nmeas_foeupdate}, and Corollary \ref{thm:nmeas_foeequality}). To derive these complexities, we consider that the enclosures $(\tilde{X}_{k-1}, \bar{X}_k, \hat{X}_k)$ have ($\tilde{n}_g$, $\bar{n}_g$, $\hat{n}_g$) generators and ($\tilde{n}_c$, $\bar{n}_c$, $\hat{n}_c$) constraints, respectively. Moreover, $(W,V)$ have $(n_{g_w},n_{g_v})$ generators and $(n_{c_w},n_{c_v})$ constraints, and we define $(m_w,m_{g_w},m_{c_w}) \triangleq (n + n_w, \tilde{n}_g + n_{g_w}, \tilde{n}_c + n_{c_w})$, and $(m_v,m_{g_v},m_{c_v}) \triangleq (n + n_v, \bar{n}_g + n_{g_v}, \bar{n}_c + n_{c_v})$. As in Chapter \ref{cha:nonlineardynamics}, we assume that evaluations of nonlinear real functions and nonlinear inclusion functions have complexity $O(1)$, and that all the LPs (including the ones necessary to compute the interval hulls) are solved at least with the performance of the method proposed in \cite{Kelner2006}. This method has (simplified) polynomial complexity $O(N_dN_c^3)$, with $N_d$ and $N_c$ the number of decision variables and constraints, respectively.

The complexities of the prediction, update, and consistency steps, using the mean value and first-order Taylor extensions, are similar to the previous prediction methods proposed in Chapter \ref{cha:nonlineardynamics}. In all the complexities shown in Table \ref{tab:nmeas_complexity}, the higher order terms such as $(m_wm_{g_w}+m_{c_w})(m_{g_w}+m_{c_w})^3$ come from both the interval hull computations and the constraint elimination procedure used to obtain the zonotope enclosure required by Theorem 1. The other terms come from matrix products that appear in the proposed expressions to compute the respective CG-rep variables.

Table \ref{tab:nmeas_complexity} also shows a simplified complexity analysis of the proposed methods for each step. In this analysis, we consider that every variable is proportional to the space dimension $n$, and that $(\tilde{X}_{k-1}, \bar{X}_k, \hat{X}_k)$ have the same number of generators and constraints (this can be achieved by using generator reduction and constraint elimination methods after each step). Details on the complexities of the basic operations with constrained zonotopes are found in Appendix \ref{app:computationalcomplexity}.

\begin{table}[!htb] 
	\scriptsize
	\centering
	\caption{Computational complexity $O(\cdot)$ of the prediction, update, and consistency steps using constrained zonotopes.}
	\begin{tabular}{c c c} \hline
		Step & Mean value extension & Simplified \\ \hline
		Prediction & $nm_wm_{g_w} + (m_wm_{g_w}+m_{c_w})(m_{g_w}+m_{c_w})^3$ & $n^5$ \\
		Update & $n_y(m_vm_{g_v} + n_y) + (m_vm_{g_v}+m_{c_v})(m_{g_v}+m_{c_v})^3$ & $n^5$ \\
		Consistency & $n_h(n\hat{n}_g + n_h) + (n\hat{n}_g+\hat{n}_c)(\hat{n}_g+\hat{n}_c)^3$ & $n^5$ \\
		\hline
		Step & First order extension & Simplified \\ \hline
		Prediction & $n(m_w^2m_{g_w}+m_wm_{g_w}^2) + (m_wm_{g_w}+m_{c_w})(m_{g_w}+m_{c_w})^3$ & $n^5$\\
		Update & $n_y^2 + n_y(m_v^2m_{g_v}+m_vm_{g_v}^2) + (m_vm_{g_v}+m_{c_v})(m_{g_v}+m_{c_v})^3$ & $n^5$ \\
		Consistency & $n_h^2 + n_h(n^2\hat{n}_g+n\hat{n}_g^2) + (n\hat{n}_g+\hat{n}_c)(\hat{n}_g+\hat{n}_c)^3$ & $n^5$ \\
		\hline		
	\end{tabular} \normalsize
	\label{tab:nmeas_complexity}
\end{table}

\section{Numerical examples} \label{sec:nmeas_examples}

This section evaluates the accuracy of the set-valued state estimation methods proposed in Section \ref{sec:nmeas_estimation}. Let CZMV denote the method based entirely on the mean value extension, using Proposition \ref{thm:nmeas_mvepred} for the prediction step and Proposition \ref{thm:nmeas_mveupdate} for the update step, but with no consistency step. Moreover, let CZMV+C denote the method CZMV with the addition of the consistency step using Corollary \ref{thm:nmeas_mveequality}, let CZMV+F denote CZMV with the addition of an intersection with an enclosure of the feasible state set as described in Remark \ref{rem:nmeas_priorbounds}, and let CZMV+FC denote CZMV with the addition of both the intersection in Remark \ref{rem:nmeas_priorbounds} and then the consistency step using Corollary \ref{thm:nmeas_mveequality}. These are referred to collectively as CZMV-like methods. Analogously, let CZFO denote the method based entirely on first-order Taylor extensions, using Proposition \ref{thm:nmeas_foepred} for the prediction step and Proposition \ref{thm:nmeas_foeupdate} for the update step, and let CZFO+C, CZFO+F, and CZFO+FC denote CZFO with the addition of, respectively, the consistency step using Proposition \ref{thm:nmeas_foeequality}, the intersection described in Remark \ref{rem:nmeas_priorbounds}, and both the intersection and the consistency step. In CZMV-like methods, complexity reduction is applied after the consistency step using the complexity reduction methods described in Section \ref{sec:complexityreduction}, with constraint elimination performed prior to generator reduction. Due to the quadratic complexity increase of the enclosure in each intermediate step of CZFO-like methods (see Remarks \ref{rem:nmeas_complexitypred}, \ref{rem:nmeas_complexityupdate} and \ref{rem:Jtildeequality}), complexity reduction is applied after all three steps in these methods. Heuristic \emph{C2} is used for choosing $(\bm{\gamma}_x,\bm{\gamma}_w,\bm{\gamma}_v)$ in CZMV-like methods, and heuristic \emph{C3} in CZFO-like methods.

We also compare our results with two nonlinear zonotope methods with prediction steps based on the Mean Value Theorem proposed in \cite{Alamo2005a} and Taylor's Theorem proposed in \cite{Combastel2005}. These are denoted by ZMV and ZFO\footnote{In a simplified analysis, ZMV and ZFO have computational complexities $O(n^4)$ and $O(n^5)$, respectively. See Section \ref{sec:ndyn_complexity} for details.}, respectively. In both zonotope methods, the nonlinear update step is given by the intersection method presented in \cite{Bravo2006}, with strips computed as in \cite{Alamo2005a}. Generator reduction is applied after the update step using Method 4 in \cite{Scott2018}. In addition, we denote by ZMV+F and ZFO+F the methods ZMV and ZFO with the addition of the intersection discussed in Remark \ref{rem:nmeas_priorbounds}. Since zonotopes are not closed under intersection, this intersection enclosed by converting the \emph{a priori} enclosure $X_\text{F}$ from CG-rep to half-space representation as described in \cite{Scott2016}, representing $X_\text{F}$ as an intersection of strips, and then using the method for bounding the intersection of a zonotope with a set of strips from \cite{Bravo2006}. As with constrained zonotopes, these are referred to as ZMV-like and ZFO-like methods, respectively.

\subsection{A system with nonlinear measurement equations}
\label{sec:nmeas_exnonlinearoutput}

\begin{figure*}[!tb]
	\centering{
		\def\svgwidth{0.8\textwidth}
		{\scriptsize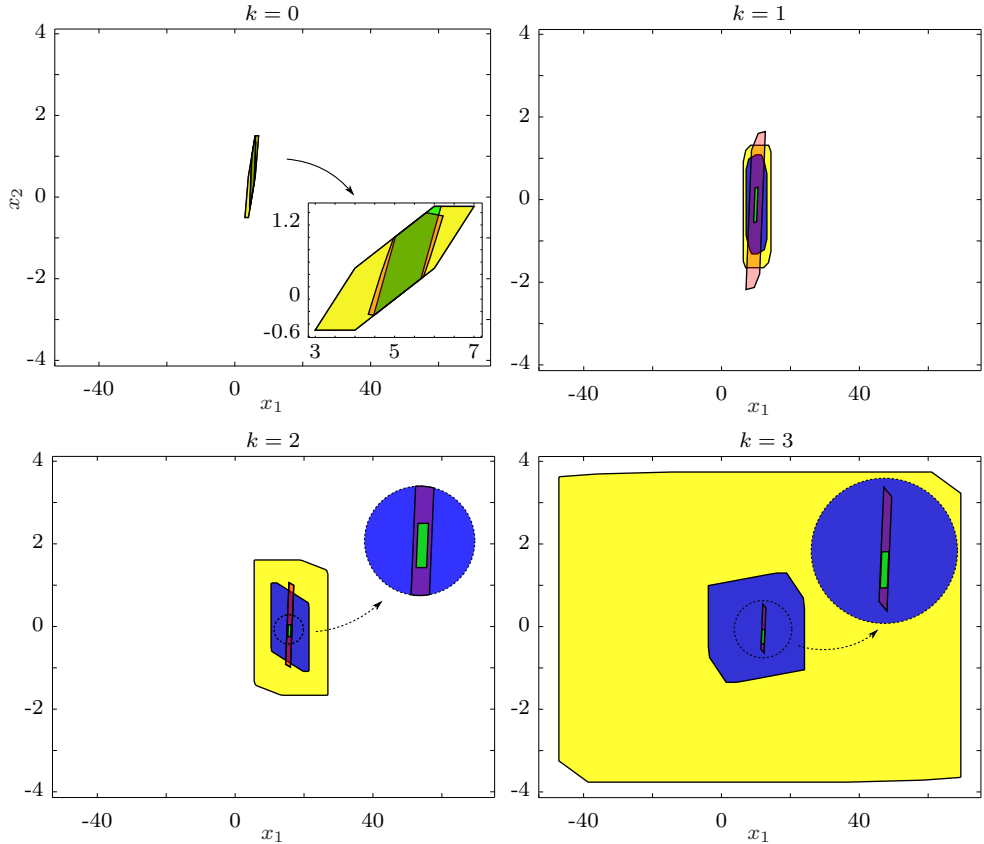}
		\caption{The enclosures $\tilde{X}_k$ from the first four time steps of set-valued state estimation in the example in Section \ref{sec:nmeas_exnonlinearoutput} using ZMV (yellow), ZFO (blue), CZMV (green), and CZFO (orange).}\label{fig:nmeas_example1sets}}
\end{figure*}

\begin{figure}[!tb]
	\centering{
		\def\svgwidth{0.8\columnwidth}
		{\scriptsize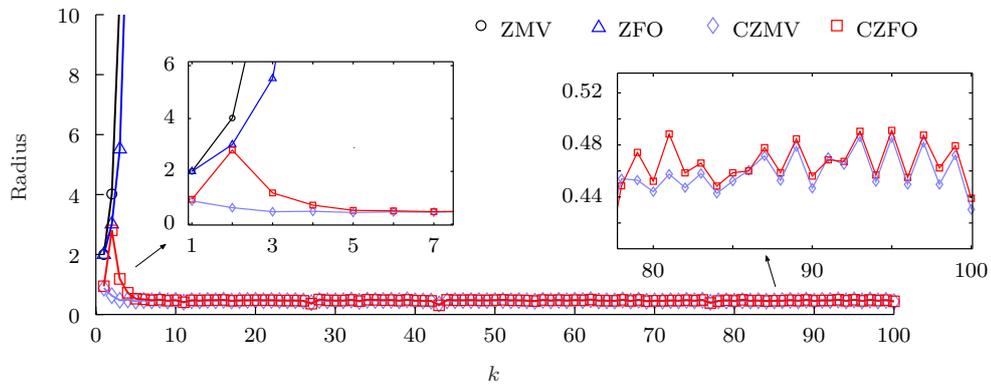}
		\caption{The radii of the estimated enclosures $\tilde{X}_k$ in the example in Section \ref{sec:nmeas_exnonlinearoutput} obtained using ZMV, ZFO, CZMV, and CZFO.}\label{fig:nmeas_example1radius}}
\end{figure}

To demonstrate the effectiveness of the methods proposed in this chapter for set-valued state estimation of systems with nonlinear outputs, we first consider the nonlinear discrete-time system $\mbf{x}_k = \mbf{f}(\mbf{x}_{k-1}) + \mbf{w}_{k-1}$, where $\mbf{f}$ is defined by \eqref{eq:nmeas_examplechoiceofgamma} and $\mbf{w}_k \in \realset^2$ denotes process uncertainties with $\|\mbf{w}_k\|_\infty \leq 0.4$. The measurements are given by
\begin{equation} \label{eq:nmeas_example1output}
\begin{aligned}
y_{1,k} & =  x_{1,k} - \sin\left(\frac{x_{2,k}}{2}\right) + v_{1,k}, \\
y_{2,k} & = -x_{1,k}x_{2,k} + x_{2,k} + v_{2,k},
\end{aligned}
\end{equation}
with $\|\mbf{v}_k\|_\infty \leq 0.4$. Finally, let
\begin{equation} \label{eq:nmeas_example1reachX0}
X_0 = \left\{ \begin{bmatrix} 0.5 & 1 & -0.5 \\ 0.5 & 0.5 & 0 \end{bmatrix}, \begin{bmatrix} 5 \\ 0.5 \end{bmatrix} \right\}.
\end{equation}

Figure \ref{fig:nmeas_example1sets} shows the estimated enclosures $\tilde{X}_k$ (since there are no invariants, these are $\tilde{X}_k = \hat{X}_k$) for $k = 0,1,2,3$, obtained using ZMV, ZFO, CZMV, and CZFO. In this case, $Z_w$ and $Z_v$ are computed as in Remark \ref{rem:nmeas_affine}. The number of generators and constraints is limited to 8 and 3, respectively. The simulations were run in MATLAB 9.1 with INTLAB 12 and CPLEX 12.8, on a laptop with 32 GB RAM and an Intel Core i7-9750H processor. The first set $\tilde{X}_0$ coincides with $X_0$ for both ZMV and ZFO, which demonstrates that the update step method using zonotopes can be very conservative with nonlinear measurements, making the first update ineffective in this example. On the other hand, the sets $\tilde{X}_0$ obtained by CZMV and CZFO have reduced volume relative to $X_0$, showing that $X_0$ was effectively tightened by the first measurement. In addition, in contrast to CZMV and CZFO, the size of the enclosures $\tilde{X}_k$ for both ZMV and ZFO increases substantially with time. This is corroborated by Figure \ref{fig:nmeas_example1radius}, which illustrates the radii of the sets $\tilde{X}_k$ (half of the length of the longest edge of the interval hull). Note that the radii of the sets obtained by ZMV and ZFO increase to infinity, while the radii of the sets obtained by CZMV and CZFO remain finite. This result corroborates the improved accuracy achieved by using constrained zonotopes for computing the update step with nonlinear measurement equations. Lastly, Table \ref{tab:nmeas_example1times} shows the average computational times per time step of each method, together with the computational times spent in complexity reduction of the enclosures. The latter is included to distinguish the computational burden of the proposed methods from the complexity reduction procedures. Note that CZMV and CZFO were able to provide accurate enclosures with an increase of 114\% and 52.3\% of the execution times with respect to ZMV and ZFO, respectively. Nevertheless, as mentioned above, the sets obtained by ZMV and ZFO increased to infinity in few steps, and therefore cannot provide any information about the state trajectories.

\begin{table}[!htb] 
	\scriptsize
	\centering
	\caption{Total and complexity reduction average execution times per time step of the state estimators for the first example.}
	\begin{tabular}{c c c c c} \hline
		& ZMV  & CZMV & ZFO & CZFO \\ \hline
		Total & $30.5$ ms & $65.3$ ms & $47.4$ ms & $72.2$ ms \\
		Red. & $4.4$ ms & $4.8$ ms & $6.6$ ms & $14.2$ ms \\
		\hline 
	\end{tabular} \normalsize
	\label{tab:nmeas_example1times}
\end{table}

\subsection{A system with nonlinear measurements and invariants}
\label{sec:nmeas_exinvariants}

The second example involves state estimation of the attitude of a flying robot. The robot is driven by angular velocity $\check{\mbf{u}}_k \in \realset^3$, with attitude expressed as a rotation quaternion $\mbf{x}_k \in \realset^4$ satisfying $\|\mbf{x}_k\|_2^2 = 1$, which defines the invariant $h(\mbf{x}_k)  = \|\mbf{x}_k\|_2^2 -1 = 0$ to be used in the consistency step \eqref{eq:nmeas_consistency0}. The known value $\mbf{u}_k$ of the physical input $\check{\mbf{u}}_k$ is measured by gyroscopes and therefore is considered to be corrupted by additive noise $\mbf{w}_k \in \realset^{3}$. Physically, the system is driven by the uncorrupted signal $\check{\mbf{u}}_k = \mbf{u}_k - \mbf{w}_k$. The attitude $\mbf{x}_k$ evolves in discrete time according to \citep{Teixeira2009,Lefferts1982} 
\begin{equation} \label{eq:nmeas_examplequaternion}
\mbf{x}_{k}  = \left(\cos(p(\mbf{u}_k,\mbf{w}_k)) \eye{4} - \frac{T_s}{2} \frac{\sin(p(\mbf{u}_k,\mbf{w}_k))}{p(\mbf{u}_k,\mbf{w}_k)} \bm{\Omega}(\mbf{u}_k,\mbf{w}_k) \right) \mbf{x}_{k-1},
\end{equation}
where $T_s$ is the sampling time and
\begin{equation*}
p(\mbf{u}_k,\mbf{w}_k) = \frac{T_s}{2}\| \check{\mbf{u}}_k\|_2, \; \bm{\Omega}(\mbf{u}_k,\mbf{w}_k) = \begin{bmatrix} 0 & \check{u}_{3,k} & -\check{u}_{2,k} & \check{u}_{1,k} \\ -\check{u}_{3,k} & 0 & \check{u}_{1,k} & \check{u}_{2,k} \\ \check{u}_{2,k} & -\check{u}_{1,k} & 0 & \check{u}_{3,k} \\ -\check{u}_{1,k} & -\check{u}_{2,k} & -\check{u}_{3,k} & 0 \end{bmatrix}.
\end{equation*}
The known value $\mbf{u}_k$ is
\begin{equation}
\mbf{u}_k = \check{\mbf{u}}_k + \mbf{w}_k = \begin{bmatrix} 0.3 \sin((2\pi/12)kT_s) \\ 0.3 \sin((2 \pi/12)kT_s - 6) \\ 0.3 \sin((2\pi/12)kT_s - 12) \end{bmatrix} + \mbf{w}_k,
\end{equation}
with $\|\mbf{w}_k\|_\infty \leq 3.0 {\times} 10^{-3}$. The measurement is given by $\mbf{y}_k = (\mbf{C}(\mbf{x}_k) \mbf{r}^{[1]}, \mbf{C}(\mbf{x}_k) \mbf{r}^{[2]}) + \mbf{v}_k$, with $\mbf{r}^{[1]} = [1 \; 0 \; 0]^T$, $\mbf{r}^{[2]} = [0 \; 1 \; 0]^T$, $\|\mbf{v}_k\|_\infty \leq 0.15$, and $\mbf{C}(\mbf{x}_k)$ is a rotation matrix defined by
\begin{align*}
\mbf{C}(\mbf{x}_k) & \triangleq \left[\begin{matrix} x_{1,k}^2 - x_{2,k}^2 - x_{3,k}^2 + x_{4,k}^2 &            2(x_{1,k}x_{2,k} + x_{3,k}x_{4,k}) \\
2(x_{1,k}x_{2,k} - x_{3,k}x_{4,k}) &  -x_{1,k}^2 + x_{2,k}^2 - x_{3,k}^2 + x_{4,k}^2  \\
2(x_{1,k}x_{3,k} + x_{2,k}x_{4,k}) &    2(-x_{1,k}x_{4,k} + x_{2,k}x_{3,k})  \end{matrix} \right.\\
& \quad\quad\quad\quad\quad\quad\quad\quad\quad\quad\quad\quad \left. \begin{matrix} 2(x_{1,k}x_{3,k} - x_{2,k}x_{4,k}) \\  2(x_{1,k}x_{4,k} + x_{2,k}x_{3,k})  \\ - x_{1,k}^2 - x_{2,k}^2 + x_{3,k}^2 + x_{4,k}^2 \end{matrix} \right].
\end{align*}
The sampling time is $T_s = 0.2$s, and the initial state belongs to the zonotope $X_0 = \{0.18\eye{4}, [0\; 1 \; 0 \; 0 ]^T\}$. For the purpose of generating trajectories of \eqref{eq:nmeas_examplequaternion}, the initial state is $\mbf{x}_0 = [0\; 1 \; 0 \; 0 ]^T$.

\begin{figure}[!tb]
	\centering{
		\def\svgwidth{0.7\columnwidth}
		{\scriptsize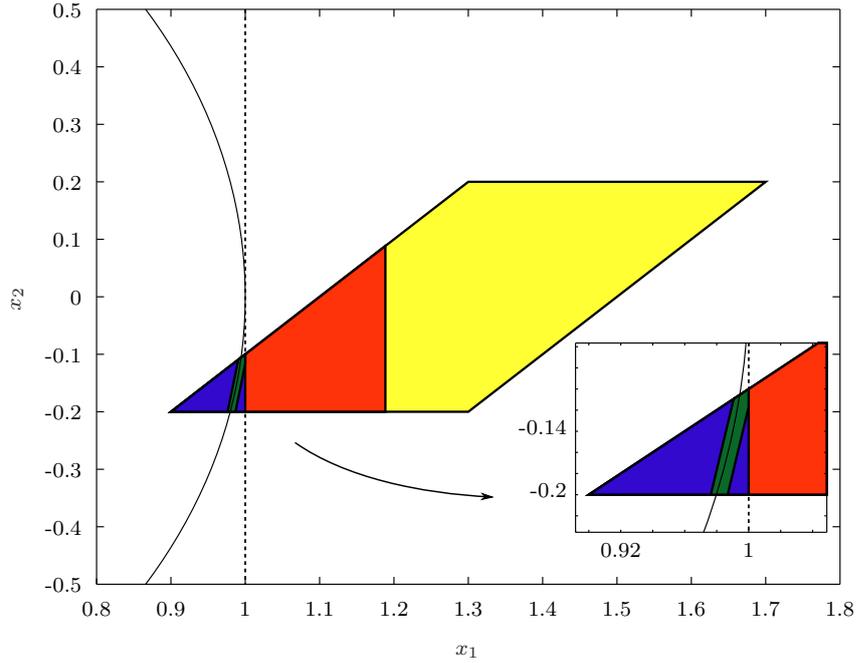}
		\caption{The zonotope $X_0$ (yellow), the enclosure $\tilde{X}_0$ obtained in the consistency step \eqref{eq:nmeas_consistency0} using Corollary \ref{thm:nmeas_mveequality} with $\hat{X}_0 = X_0$ (red), the set $X_0 \cap \{\eye{2},\mbf{0}\}$ (blue), and the enclosure $\tilde{X}_0$ obtained in the consistency step \eqref{eq:nmeas_consistency0} using Corollary \ref{thm:nmeas_mveequality} with $\hat{X}_0 = X_0 \cap \{\eye{2},\mbf{0}\}$ (green). The resulting enclosures contain each other according to the sequence above. The dashed line denotes the box $\{\eye{2},\mbf{0}\}$. The circle that describes the feasible state set of $\|\mbf{x}_0\|_2^2 = 1$ is also depicted. }\label{fig:nmeas_bigbox}}
\end{figure}

In the following, for the sake of clarity we first illustrate the observation described in Remark \ref{rem:nmeas_priorbounds} for the consistency step in a sub-example with $\mbf{x}_k \in \realset^2$, $\|\mbf{x}_k\|_2^2 = 1$, and $$X_0 = \left\{\begin{bmatrix} 0.2 & 0.2 \\ 0 & 0.2 \end{bmatrix}, \begin{bmatrix} 1.3 \\ 0 \end{bmatrix}\right\}.$$ Note that in this case the set $\{\eye{2},\mbf{0}\}$ is a valid enclosure for the feasible state set of the invariant $\|\mbf{x}_k\|_2^2 = 1$. %
Figure \ref{fig:nmeas_bigbox} shows the initial set $X_0$ and the enclosure $\tilde{X}_0$ obtained using Corollary \ref{thm:nmeas_mveequality} for the consistency step with $\hat{X}_0 = X_0$. Note that, although tightened, the resulting set is still very conservative. Figure \ref{fig:nmeas_bigbox} also shows the intersection $X_0 \cap \{\eye{2},\mbf{0}\}$, which is tighter than the previous result. Finally, we illustrate the enclosure $\tilde{X}_0$ obtained using Corollary \ref{thm:nmeas_mveequality} with $\hat{X}_0 = X_0 \cap \{\eye{2},\mbf{0}\}$, which is the least conservative result. This demonstrates the improved accuracy that can be achieved if an enclosure of the feasible state set is known \emph{a priori}.

\begin{figure}[!tb]
	\centering{
		\def\svgwidth{0.7\columnwidth}
		{\scriptsize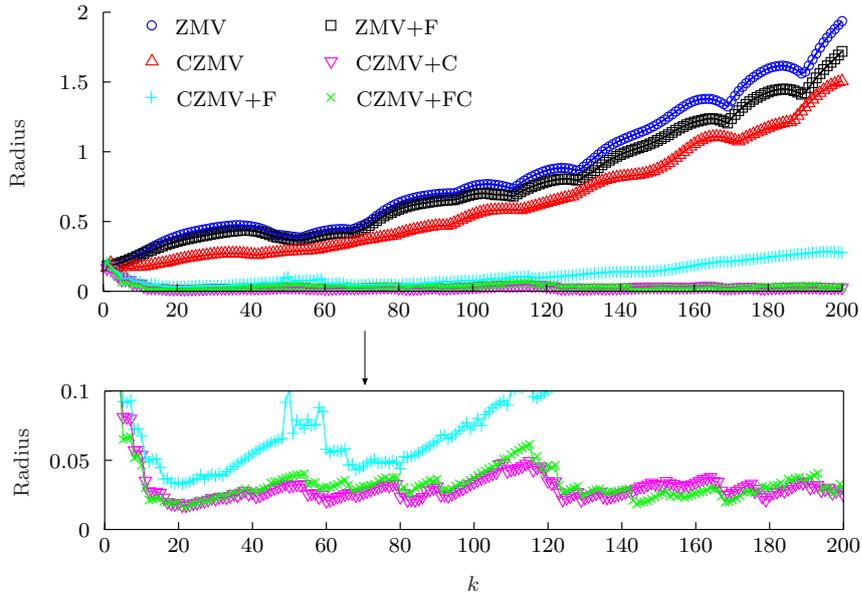}
		\caption{The radii of the estimated enclosures $\tilde{X}_k$ for \eqref{eq:nmeas_examplequaternion} obtained using ZMV-like and CZMV-like methods.}\label{fig:nmeas_attituderadiusmv}}
\end{figure}

Figure \ref{fig:nmeas_attituderadiusmv} illustrates the radii of the enclosures $\tilde{X}_k$ obtained for the trajectories of the system \eqref{eq:nmeas_examplequaternion} using ZMV-like and CZMV-like methods. We consider the enclosure $\{\eye{4},\mbf{0}\}$ of the feasible state set of the invariant $\|\mbf{x}_k\|_2^2 = 1$. In this case, $Z_w$ and $Z_v$ are computed as in Remarks \ref{rem:nmeas_mvepred} and \ref{rem:nmeas_affine}, respectively. The number of generators and constraints is limited to 12 and 5, respectively. Note that the zonotope methods were not able to provide useful enclosures for \eqref{eq:nmeas_examplequaternion}, i.e., the sizes of the enclosures increase with time and do not provide useful information, even when considering the intersection with $\{\eye{4},\mbf{0}\}$. Note that the enclosures provided by CZMV and CZMV+F also are not useful in this case, even though CZMV+F is much tighter than the others. On the other hand, CZMV+C and CZMV+FC both provided good enclosures with stable size, with the latter providing more accurate sets in the initial time steps, as expected. This demonstrates the advantage of including the consistency step \eqref{eq:nmeas_consistency0} in state estimation using the mean value extension to take into account the invariant $\|\mbf{x}_k\|_2^2 = 1$. In addition, note that the radii of the enclosures provided by CZMV+C and CZMV+FC are much smaller than the radius of $\{\eye{4},\mbf{0}\}$, showing that significant accuracy can be obtained by combining the state estimation procedure with the invariant, in comparison with using only the information available about the feasible state set.

\begin{figure}[!tb]
	\centering{
		\def\svgwidth{0.7\columnwidth}
		{\scriptsize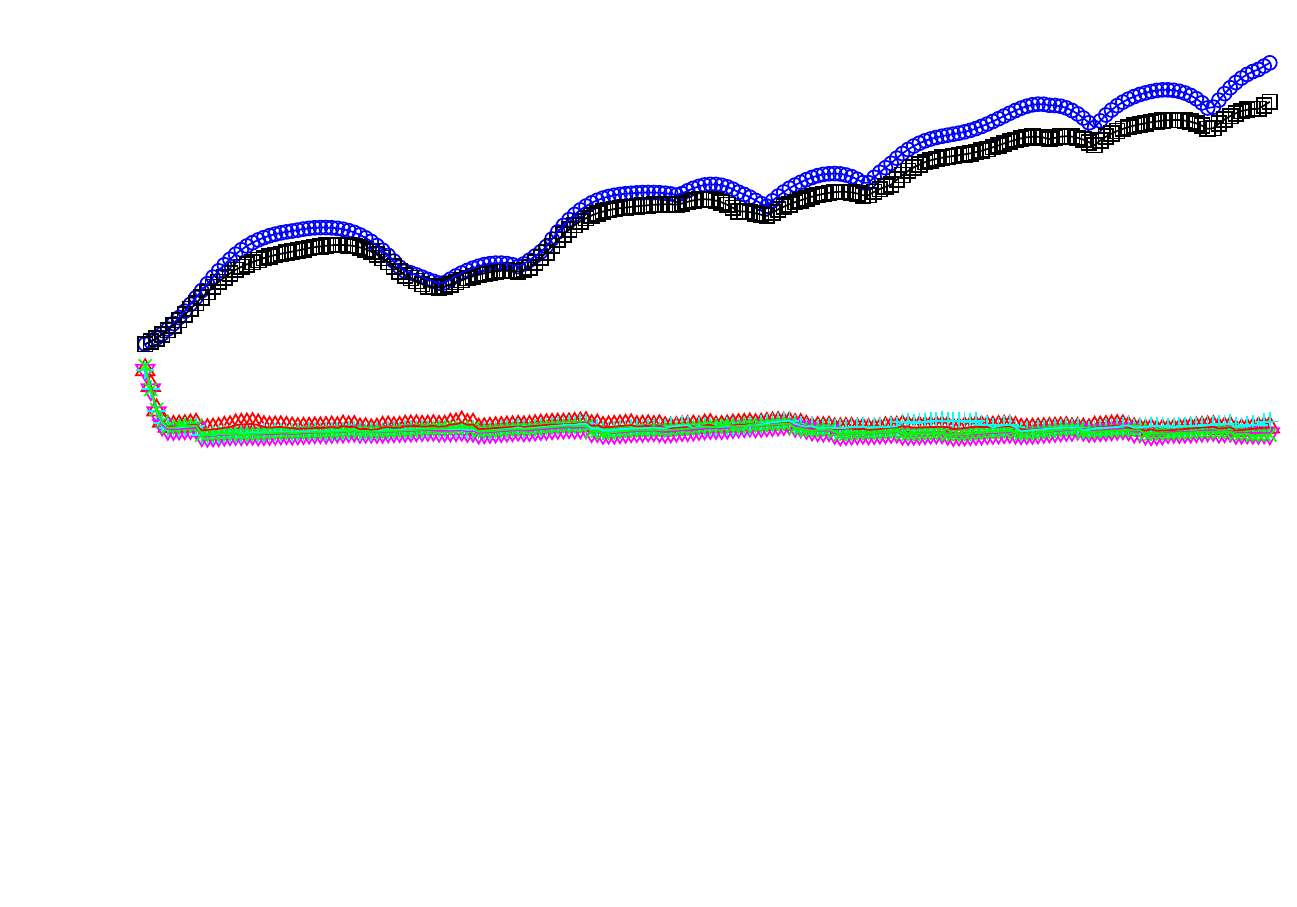}
		\caption{The radii of the estimated enclosures $\tilde{X}_k$ for \eqref{eq:nmeas_examplequaternion} obtained using ZFO-like and CZFO-like methods.}\label{fig:nmeas_attituderadiusfo}}
\end{figure}

Figure \ref{fig:nmeas_attituderadiusfo} shows the radii of the enclosures $\tilde{X}_k$ obtained for the trajectories of \eqref{eq:nmeas_examplequaternion} using ZFO-like and CZFO-like methods. Once again, the enclosures computed by zonotopes do not provide useful information since these increase with time, even when considering the intersection with $\{\eye{4},\mbf{0}\}$. On the other hand, even CZFO provides tight enclosures for this example. This demonstrates that the first-order Taylor extension is able to provide significantly less conservative bounds than the mean value extension in this case, since the nonlinear measurements are polynomials of second order, and therefore the interval matrices $\mbf{Q}^{[q]}$ in Proposition \ref{thm:nmeas_foeupdate} are singletons. Nevertheless, CZFO+C and CZFO+FC both provide still sharper enclosures, with comparable sizes due to the limited complexity of the sets. To provide a comprehensive comparison between all of the methods, Table \ref{tab:nmeas_example2ARR} shows the average radius ratio for this example (ARR, i.e., the ratio of the radius of the set provided by one method over the radius of the set provided by another method at $k$, averaged over all time steps), and Table \ref{tab:nmeas_example2times} shows the average computational times per time step of each method. Note that, in contrast to the analogous state estimation algorithms for linear measurements in Chapter \ref{cha:nonlineardynamics}, the computational times of CZMV-like methods were competitive with ZMV-like methods, and CZFO-like methods with ZFO-like methods as well. The increased times of the zonotope methods arise from the iterative computation of strips based on interval analysis in \cite{Alamo2005a} and the intersection with strips given in \cite{Bravo2006} to perform the update step. In this sense, using the mean value extension with constrained zonotopes, one can achieve about 93\% less conservative bounds (CZMV+C--to--ZMV+F ARR of only 6.7\%) in comparison to zonotopes, with a mild increase of 22.2\% in the average execution time. On the other hand, using the first-order Taylor extension, one can achieve about 95\% less conservative bounds (CZFO+C--to--ZFO+F ARR of 5.1\%), with an increase of 10.5\% in the average execution time. This demonstrates the joint accuracy and efficiency provided by the proposed methods based on constrained zonotopes. These ARR are highlighted in Table \ref{tab:nmeas_example2ARR}, and correspond to a comparison between the most accurate results obtained by ZMV-like and CZMV-like methods, and between ZFO-like and CZFO-like methods.

Lastly, note that the CZFO+C--to--CZFO ARR was of 69.64\%, showing again the improved accuracy obtained by taking into account the invariant through the consistency step. In addition, CZFO-like methods provided better enclosures than CZMV-like methods in this example. Nevertheless, this comes with an important increase in computational time as shown in Table \ref{tab:nmeas_example2times}. This demonstrates that the choice between CZMV-like methods and CZFO-like methods for state estimation can provide a trade-off between accuracy and efficiency, and therefore will depend on the current application.

\begin{table*}[!htb]
	\scriptsize
	\centering
	\caption{Average radius ratio of the enclosures obtained by the state estimators (column per row) for the system \eqref{eq:nmeas_examplequaternion}.}
	\begin{tabular}{c c c c c c c} \hline
		$\backslash$ & ZMV  & ZMV+F     & CZMV      & CZMV+C   & CZMV+F   & CZMV+FC  \\ \hline
		ZMV     & $1$       & $0.9187$  & $0.7193$  & $0.0636$ & $0.1491$ & $0.0669$ \\	    
		ZMV+F   & $1.0891$  & $1$       & $0.7834$  & \boxed{$0.0671$} & $0.1606$ & $0.0707$  \\		
		CZMV    & $1.4091$  & $1.2946$  & $1$       & $0.0857$ & $0.2024$ & $0.0901$ \\
		CZMV+C  & $29.2088$ & $26.5557$ & $21.4655$ & $1$      & $3.8897$ & $1.0665$ \\
		CZMV+F  & $8.0755$  & $7.4009$  & $5.6895$  & $0.3745$ &      $1$ & $0.4023$ \\
		CZMV+FC & $28.7622$ & $26.1185$ & $21.1725$ & $0.9697$ & $3.8402$ & $1$      \\ \hline
		ZFO     & $1.6829$  & $1.5384$  & $1.2325$  & $0.0849$ & $0.2420$ & $0.0898$ \\
		ZFO+F   & $1.8040$  & $1.6484$  & $1.3211$  & $0.0891$ & $0.2581$ & $0.0942$ \\
		CZFO    & $34.1796$ & $31.0484$ & $25.3451$ & $1.1983$ & $4.6785$ & $1.2562$ \\
		CZFO+C  & $50.4008$ & $45.7732$ & $37.1612$ & $1.7337$ & $6.8068$ & $1.8039$ \\
		CZFO+F  & $33.4713$ & $30.4709$ & $24.4434$ & $1.2814$ & $4.5068$ & $1.3433$ \\
		CZFO+FC & $50.2958$ & $45.6286$ & $37.2526$ & $1.6647$ & $6.8396$ & $1.7317$ \\
		\hline
		$\backslash$ & ZFO  & ZFO+F     & CZFO     & CZFO+C   & CZFO+F   & CZFO+FC \\ \hline
		ZMV     & $0.6447$  & $0.6057$  & $0.0489$ & $0.0358$ & $0.0436$ & $0.0385$ \\	    
		ZMV+F   & $0.6988$  & $0.6562$  & $0.0521$ & $0.0379$ & $0.0465$ & $0.0408$ \\		
		CZMV    & $0.9272$  & $0.8701$  & $0.0680$ & $0.0484$ & $0.0593$ & $0.0523$ \\
		CZMV+C  & $16.3264$ & $15.2013$ & $0.8974$ & $0.6056$ & $0.8472$ & $0.6363$ \\
		CZMV+F  & $5.2585$  & $4.9195$  & $0.3382$ & $0.2289$ & $0.2919$ & $0.2460$ \\
		CZMV+FC & $15.8969$ & $14.7785$ & $0.8571$ & $0.5744$ & $0.8144$ & $0.6018$ \\ \hline
		ZFO     & $1$       & $0.9367$  & $0.0671$ & $0.0487$ & $0.0621$ & $0.0519$ \\
		ZFO+F   & $1.0683$  & $1$       & $0.0709$ & \boxed{$0.0512$} & $0.0656$ & $0.0546$ \\
		CZFO    & $18.5748$ & $17.2934$ & $1$      & $0.6964$ & $0.9717$ & $0.7290$ \\
		CZFO+C  & $27.7376$ & $25.7968$ & $1.4945$ & $1$      & $1.4372$ & $1.0499$ \\
		CZFO+F  & $19.2573$ & $17.9445$ & $1.1000$ & $0.7504$ & $1$      & $0.7929$ \\
		CZFO+FC & $27.1388$ & $25.2164$ & $1.4323$ & $0.9617$ & $1.3947$ & $1$ \\
		\hline
	\end{tabular} \normalsize
	\label{tab:nmeas_example2ARR}
\end{table*}

\begin{table}[!htb]
	\scriptsize
	\centering
	\caption{Total and complexity reduction average execution times per time step of the state estimators for the system \eqref{eq:nmeas_examplequaternion}.}
	\begin{tabular}{c c c c c c c} \hline
		& ZMV  & ZMV+F     & CZMV      & CZMV+C   & CZMV+F   & CZMV+FC \\ \hline
		Total & $0.4552$ s & $0.4610$ s & $0.4478$ s & $0.5635$ s & $0.4519$ s & $0.5942$ s  \\
		Red. & $0.37$ ms & $0.31$ ms & $10.6$ ms & $71.8$ ms & $19.8$ ms & $96.5$ ms  \\
		\hline
		& ZFO       & ZFO+F     & CZFO     & CZFO+C   & CZFO+F   & CZFO+FC \\ \hline
		Total & $1.0403$ s & $1.0563$ s & $1.0907$ s & $1.1671$ s & $1.1009$ s & $1.3750$ s \\
		Red.  & $2.3$ ms & $2.2$ ms & $90.3$ ms & $0.1266$ s & $98.7$ ms & $0.3306$ s \\		
		\hline 
	\end{tabular} \normalsize
	\label{tab:nmeas_example2times}
\end{table}

\section{Final remarks} \label{sec:nmeas_conclusions}

This chapter developed new approaches for set-valued state estimation of nonlinear discrete-time systems with nonlinear measurements and nonlinear invariants. The state trajectories were enclosed using the standard prediction-update algorithm with the addition of a new consistency step accounting for the nonlinear invariants. New methods were proposed for the update and consistency steps using generalized intersections of constrained zonotopes. In addition, our previous methods for the prediction step were generalized to allow the approximation points for the mean value and first-order Taylor extensions to lie in a larger region. Numerical results demonstrate that the methods proposed in this chapter can provide significantly tighter enclosures compared to existing zonotope methods. The improved accuracy is achieved with a mild increase in computational cost. Nevertheless, future work will seek to reduce the execution times, since these can be a major issue in many practical applications.

The next chapter develops a new method for set-based fault diagnosis of nonlinear discrete-time systems. Enclosures described by constrained zonotopes are employed in the passive fault detection and active fault isolation processes, based on the nonlinear state estimation methods introduced in this doctoral thesis.

\chapter{Fault diagnosis of nonlinear systems}\thispagestyle{headings} \label{cha:faultdiagnosis}

This chapter introduces a new approach for set-based active fault diagnosis (AFD) of a class of nonlinear discrete-time systems. The new enclosures based on constrained zonotopes proposed in the previous chapters are employed for passive fault detection of a class of nonlinear discrete-time systems. Moreover, an adaptation of the first-order Taylor extension (Section \ref{sec:ndyn_firstorder}) is used to obtain an affine parametrization of the reachable sets, in the direction of the design of an optimal input for set-based active fault diagnosis of nonlinear systems.

\section{Problem formulation}

Consider a discrete-time system with nonlinear dynamics and linear measurement given by
\begin{equation}
\begin{aligned} \label{eq:fault_system}
\mbf{x}_k & = \mbf{f}^{[i]}(\mbf{x}_{k-1}, \mbf{u}_{k-1}, \mbf{w}_{k-1}),\\
\mbf{y}_k & = \mbf{s}^{[i]} + \mbf{C}^{[i]} \mbf{x}_k + \mbf{D}_v^{[i]} \mbf{v}_k,	
\end{aligned}
\end{equation}
where $\mbf{x}_k \in \realset^n$ is the system state, $\mbf{u}_k \in \realset^{n_u}$ is the system input, $\mbf{y}_k \in \realset^{n_y}$ is the measured output, $\mbf{w}_k \in \realset^{n_w}$ is the process noise, and $\mbf{v}_k \in \realset^{n_v}$ is the measurement noise. In each time interval $[k-1,k]$, the system evolves according to one of the possible models in \eqref{eq:fault_system} indexed by $i \in \modelset = \{1, \dots, n_i\}$, with $i = 1$ denoting the nominal model. Measurement bias is modeled by $\mbf{s}^{[i]}$. Uncertainties are assumed to be bounded by convex polytopic sets $\mbf{w}_k \in W$, $\mbf{v}_k \in V$, and the system input is bounded by an admissible set given by a convex polytope $\mbf{u}_k \in U$. Note that a convex polytope in H-rep can be converted to CG-rep using Proposition \ref{prop:pre_czhreptocgrep}.

The objective of AFD is to find which model describes the process behaviour. In this chapter, the dynamics of such system are assumed to not change during the diagnosis procedure, i.e. the AFD is fast enough to avoid the switching between models. In this sense, an input sequence $(\mbf{u}_k, \mbf{u}_{k+1}, \ldots, \mbf{u}_{k+N-1})$ of minimal length $N$ is designed such that the output sequence $(\mbf{y}_{k+1}^{[i]}, \mbf{y}_{k+2}^{[i]}, \ldots, \mbf{y}_{k+N}^{[i]})$ is consistent with only one $i \in \modelset$. If feasible, this problem may admit multiple solutions. For this reason, we introduce a cost function and select among the feasible input sequences the optimal one.

\section{Passive fault detection} \label{sec:fault_passivefaultdetection}

The set-based state estimation methods proposed in Section \ref{sec:ndyn_nonlinearestimation} can be effectively employed to perform guaranteed passive fault detection of nonlinear discrete-time systems as in \eqref{eq:fault_system}. Let $\hat{X}_{k-1}^{[i]}$ denote the previous update set described by a constrained zonotope satisfying $\mbf{x}_{k-1} \in \hat{X}_{k-1}^{[i]}$, and let $\bar{X}_k^{[i]}$ denote the prediction set described by a constrained zonotope satisfying
\begin{equation} \label{eq:fault_predictionfaultdetection}
	\bar{X}_k^{[i]} \supseteq \{ \mbf{f}^{[i]}(\mbf{x}_{k-1}, \mbf{u}_{k-1}, \mbf{w}_{k-1}) : \mbf{x}_{k-1} \in \hat{X}_{k-1}^{[i]}, \, \mbf{w}_{k-1} \in W \}.
\end{equation}
Therefore, define the estimated measurement set $\hat{Y}_k^{[i]}$ by
\begin{equation} \label{eq:fault_estimatedmeasurementset}
	\hat{Y}_k^{[i]} = \mbf{s}^{[i]} \oplus \mbf{C}^{[i]} \bar{X}_k^{[i]} \oplus \mbf{D}_v^{[i]} V.
\end{equation}

Assuming the $i$-th model to be active at $k = 0$, and $\mbf{x}_0 \in \hat{X}_0^{[i]}$, a constrained zonotope satisfying \eqref{eq:fault_predictionfaultdetection} can be obtained by using any of the methods proposed in Section \ref{sec:ndyn_nonlinearestimation}, including the CZIB approach if Assumption \ref{ass:ndyn_czib} in Section \ref{sec:ndyn_cziboutline} is satisfied. The estimated measurement set in \eqref{eq:fault_estimatedmeasurementset} can be computed through \eqref{eq:pre_czlimage}--\eqref{eq:pre_czmsum}. Given the current output $\mbf{y}_k$, if the inclusion
\begin{equation} \label{eq:fault_detectioninclusion}
	\mbf{y}_k \in \hat{Y}_k^{[i]}
\end{equation}
does not hold, then a fault has occurred at a given time $k_f \leq k$, i.e., the $i$-th model was not active in the time interval $[k_f,\; k]$. Therefore, passive fault detection can be effectively performed by checking if the inclusion \eqref{eq:fault_detectioninclusion} holds. This inclusion can be verified by solving an LP (Property \ref{prop:pre_czisemptyinside}).

\section{Reachable sets} \label{sec:fault_reachablesets}

This section focuses on the computation of reachable sets for the nonlinear discrete-time system \eqref{eq:fault_system}, aiming the design of an optimal input sequence for guaranteed active fault diagnosis.

\subsection{Guaranteed enclosure}

The following theorem is based on the first-order Taylor extension proposed in Chapter \ref{cha:nonlineardynamics} (Section \ref{sec:ndyn_firstorder}) and provides a simplified enclosure of the image of a nonlinear function of class $\mathcal{C}^2$ with domain given by constrained zonotopes. The obtained enclosure is also a constrained zonotope.

\begin{theorem} \label{thm:fault_firstorder} \rm
	Let $\bm{\eta}: \realset^n \times \realset^{n_u} \times \realset^{n_w} \to \realset^n$ be of class $\mathcal{C}^2$, and let $\mbf{x} \in \realset^n$, $\mbf{u} \in \realset^{n_u}$ and $\mbf{w} \in \realset^{n_w}$ denote its arguments. Let $X \subset \realset^n$, $U \subset \realset^{n_u}$, and $W \subset \realset^{n_w}$ be constrained zonotopes such that $\mbf{x} \in X$, $\mbf{u} \in U$, $\mbf{w} \in W$, and choose any $\bm{\gamma}_x \in X$, $\bm{\gamma}_u \in U$, $\bm{\gamma}_w \in W$. For each $q = 1,2,\dots,n$, define an interval matrix $\mbf{Q}^{[q]} \in \intvalsetmat{n+n_u+n_w}{n+n_u+n_w}$ satisfying $\mbf{Q}^{[q]} \supseteq \mbf{H} \eta_q (X,U,W)$, and define an interval vector $\bm{\varphi} \in \intvalset^{n}$ whose $q$-th component satisfies
	\begin{equation} \label{eq:fault_ellgeneral}
	\varphi_q \supseteq \begin{bmatrix} X - \bm{\gamma}_x \\ \mbf{u} - \bm{\gamma}_u \\ W - \bm{\gamma}_w \end{bmatrix}^T \mbf{Q}^{[q]} \begin{bmatrix} X - \bm{\gamma}_x \\ \mbf{u} - \bm{\gamma}_u \\ W - \bm{\gamma}_w \end{bmatrix}.
	\end{equation}
	In addition, let $R \triangleq \left\{\text{diag}(\rad{\bm{\varphi}}), \midpoint{\bm{\varphi}} \right\}$. Then, for any $\mbf{u} \in U$,
	\begin{equation} \label{eq:fault_simplifiedfirstorder}
	\begin{aligned}
	\bm{\eta}(X,\mbf{u},W) \subseteq \bm{\eta}(\bm{\gamma}_x,\bm{\gamma}_u,\bm{\gamma}_w) \oplus \nabla^T_x \bm{\eta}(\bm{\gamma}_x,\bm{\gamma}_u,\bm{\gamma}_w)(X - \bm{\gamma}_x) & \oplus \nabla^T_u \bm{\eta}(\bm{\gamma}_x,\bm{\gamma}_u,\bm{\gamma}_w)(\mbf{u} - \bm{\gamma}_u) \\ & \oplus \nabla^T_w \bm{\eta}(\bm{\gamma}_x,\bm{\gamma}_u,\bm{\gamma}_w)(W - \bm{\gamma}_w) \oplus R.
	\end{aligned}
	\end{equation}
\end{theorem}

\proof

Let $(\mbf{x}, \mbf{u}, \mbf{w}) \in X \times U \times W$, $\mbf{z} \triangleq (\mbf{x}, \mbf{u}, \mbf{w})$, and $\bm{\gamma}_z \triangleq (\bm{\gamma}_x,\bm{\gamma}_u,\bm{\gamma}_w)$. From the Taylor's theorem around $\bar{\mbf{z}}$, the $q$-th component of $\bm{\eta}(\mbf{z})$ is given by \citep{Combastel2005}
\begin{align*}
\eta_q(\mbf{z}) & = \eta_q (\bar{\mbf{z}}) + \nabla^T_z \eta_q (\bar{\mbf{z}})(\mbf{z} - \bar{\mbf{z}})+ (\mbf{z} - \bar{\mbf{z}})^T \bm{\Gamma}^{[q]} (\mbf{z} - \bar{\mbf{z}}),
\end{align*}
for one $\bm{\Gamma}^{[q]} \in \mbf{Q}^{[q]}$. Noting that $(\mbf{z} - \bar{\mbf{z}})^T \bm{\Gamma}^{[q]} (\mbf{z} - \bar{\mbf{z}}) \in \varphi_q$ for every $(\mbf{x},\mbf{u},\mbf{w}) \in X \times U \times W$, $\bm{\Gamma}^{[q]} \in \mbf{Q}^{[q]}$, and also $\nabla^T_z \eta_q (\bar{\mbf{z}})(\mbf{z} - \bar{\mbf{z}}) = \nabla^T_x \eta_q (\bm{\gamma}_x,\bm{\gamma}_u,\bm{\gamma}_w)(\mbf{x} - \bm{\gamma}_x) + \nabla^T_u \eta_q (\bm{\gamma}_x,\bm{\gamma}_u,\bm{\gamma}_w)(\mbf{u} - \bm{\gamma}_u) + \nabla^T_w \eta_q (\bm{\gamma}_x,\bm{\gamma}_u,\bm{\gamma}_w)(\mbf{w} - \bm{\gamma}_w)$, it follows that
\begin{align*}
\eta_q(\mbf{x},\mbf{u},\mbf{w}) \in \eta_q (\bm{\gamma}_x,\bm{\gamma}_u,\bm{\gamma}_w) \oplus \nabla^T_x \eta_q (\bm{\gamma}_x,\bm{\gamma}_u,\bm{\gamma}_w)(X - \bm{\gamma}_x) & \oplus \nabla^T_u \eta_q (\bm{\gamma}_x,\bm{\gamma}_u,\bm{\gamma}_w)(\mbf{u} - \bm{\gamma}_u) \\ & \oplus \nabla^T_w \eta_q (\bm{\gamma}_x,\bm{\gamma}_u,\bm{\gamma}_w)(W - \bm{\gamma}_w) \oplus \varphi_q, 
\end{align*}
for every $\mbf{x} \in X$, $\mbf{w} \in W$. Therefore, $\bm{\gamma} = \midpoint{\bm{\gamma}} \oplus \half \text{diag}(\diam{\bm{\gamma}}) B_\infty^n$ implies that
\begin{equation*}
\begin{aligned}
\bm{\eta}(X,\mbf{u},W) \subseteq \bm{\eta}(\bm{\gamma}_x,\bm{\gamma}_u,\bm{\gamma}_w) \oplus \nabla^T_x \bm{\eta}(\bm{\gamma}_x,\bm{\gamma}_u,\bm{\gamma}_w)(X - \bm{\gamma}_x) & \oplus \nabla^T_u \bm{\eta}(\bm{\gamma}_x,\bm{\gamma}_u,\bm{\gamma}_w)(\mbf{u} - \bm{\gamma}_u) \\ & \oplus \nabla^T_w \bm{\eta}(\bm{\gamma}_x,\bm{\gamma}_u,\bm{\gamma}_w)(W - \bm{\gamma}_w) \oplus R.
\end{aligned}
\end{equation*}
\qed

Theorem \ref{thm:fault_firstorder} provides a simplified constrained zonotope enclosing the trajectories of a nonlinear system. Nevertheless, the computation of the intervals $\varphi_q$ is not trivial in the general case, and particular solutions are proposed. To obtain an affine parameterization in $\mbf{u}_k$ of the reachable sets of \eqref{eq:fault_system}, aiming the input design for active fault diagnosis, since the input $\mbf{u}_k$ is unknown \emph{a priori}, the interval matrices $\mbf{Q}^{[q]} = \iextension{\mbf{H} \eta_q (X,U,W)}$ are evaluated conservatively for the entire admissible set $U$ in the reachable set computations. In the following, a new approach is proposed to compute reachable state and output sets for \eqref{eq:fault_system}, yielding an affine parameterization in the input $\mbf{u}_k$.

\subsection{Affine parameterization of the reachable set}

Consider constrained zonotopes $X_k \subset \realset^n$, $W \subset \realset^{n_w}$, $V \subset \realset^{n_v}$. It is assumed that $\mbf{x}_k \in X_k$, and $(\mbf{w}_{k+m}, \mbf{v}_{k+m}) \in W \times V$ for $m \in \{0, 1, \ldots, N\}$. For each time $k$, choose $\bm{\gamma}_{x,k} \in X_k$, $\bm{\gamma}_{u,k} \in U$, $\bm{\gamma}_{w,k} \in W$. Then, we consider intervals $\varphi_q^{[i]}$ computed using interval arithmetic, satisfying
\begin{equation} \label{eq:ellaffine}
\varphi_q^{[i]} \supseteq \begin{bmatrix} \square X_k - \bm{\gamma}_{x,k} \\ \square U - \bm{\gamma}_{u,k} \\ \square W - \bm{\gamma}_{w,k} \end{bmatrix}^T \mbf{Q}^{[q,i]} \begin{bmatrix} \square X_k - \bm{\gamma}_{x,k} \\ \square U - \bm{\gamma}_{u,k} \\ \square W - \bm{\gamma}_{w,k} \end{bmatrix},
\end{equation}
where $\mbf{Q}^{[q,i]} = \square \mbf{H} f^{[i]}_q (\square X,\square U,\square W)$.
Note that $\varphi_q^{[i]}$ computed as in \eqref{eq:ellaffine} is not a function of the input $\mbf{u}_k$. Therefore, Theorem \ref{thm:fault_firstorder} immediately provides an enclosure $\Phi_{k+1}^{[i]} \supseteq \mbf{f}^{[i]}(X_k,\mbf{u}_{k},W)$, given by
\begin{equation} \label{eq:fault_affinemap0}
\begin{aligned}
\Phi_{k+1}^{[i]} & = \mbf{f}^{[i]}(\bm{\gamma}_{x,k},\bm{\gamma}_{u,k},\bm{\gamma}_{w,k}) \oplus \nabla^T_x \mbf{f}^{[i]}(\bm{\gamma}_{x,k},\bm{\gamma}_{u,k},\bm{\gamma}_{w,k})(X_k - \bm{\gamma}_{x,k}) \oplus \nabla^T_u \mbf{f}^{[i]}(\bm{\gamma}_{x,k},\bm{\gamma}_{u,k},\bm{\gamma}_{w,k})(\mbf{u}_k - \bm{\gamma}_{u,k}) \\
& \oplus \nabla^T_w \mbf{f}^{[i]}(\bm{\gamma}_{x,k},\bm{\gamma}_{u,k},\bm{\gamma}_{w,k})(W - \bm{\gamma}_{w,k}) \oplus R^{[i]}(\bm{\gamma}_{x,k},\bm{\gamma}_{u,k},\bm{\gamma}_{w,k},X_{k},U,W).
\end{aligned}
\end{equation}

Let $\mbf{A}_k^{[i]} \triangleq \nabla^T_x \mbf{f}^{[i]}(\bm{\gamma}_{x,k},\bm{\gamma}_{u,k},\bm{\gamma}_{w,k})$, $\mbf{B}_k^{[i]} \triangleq \nabla^T_u \mbf{f}^{[i]}(\bm{\gamma}_{x,k},\bm{\gamma}_{u,k},\bm{\gamma}_{w,k})$, $\mbf{F}_k^{[i]} \triangleq \nabla^T_w \mbf{f}^{[i]}(\bm{\gamma}_{x,k},\bm{\gamma}_{u,k},\bm{\gamma}_{w,k})$, and $\mbf{r}_k^{[i]} \triangleq \mbf{f}^{[i]}(\bm{\gamma}_{x,k},\bm{\gamma}_{u,k},\bm{\gamma}_{w,k}) - \mbf{A}_k^{[i]}\bm{\gamma}_{x,k}$, where dependences on $(\bm{\gamma}_{x,k},\bm{\gamma}_{u,k},\bm{\gamma}_{w,k})$ were omitted for brevity. For clarity, we rewrite \eqref{eq:fault_affinemap0} as follows:
\begin{equation} \label{eq:affinemap}
\Phi_{k+1}^{[i]}(\mbf{u}_k) = \mbf{r}_k^{[i]} \oplus \mbf{A}_k^{[i]} X_k \oplus \mbf{B}_k^{[i]} (\mbf{u}_k - \bm{\gamma}_{u,k}) \oplus \mbf{F}_k^{[i]} (W - \bm{\gamma}_{w,k}) \oplus R_k^{[i]}(X_k,U,W),
\end{equation}
which is affine in $\mbf{u}_k$ (the dependency on the inputs is depicted henceforth). 

Similarly to the approach in \cite{Scott2014}, for each model $i \in \modelset$ we define the \emph{reachable state and output sets} of \eqref{eq:fault_system} on the interval $[k+1,k+N]$ as
\begin{align}
\seq{\Phi}\nspace_{k+1:k+N}^{[i]}(\seq{\mbf{u}}) & = \seq{\mbf{r}}\nspace^{[i]} \oplus \seq{\mbf{A}}\nspace^{[i]} X_k \oplus \seq{\mbf{B}}\nspace^{[i]} (\seq{\mbf{u}} - \seq{\bm{\gamma}}_u) \oplus \seq{\mbf{F}}\nspace^{[i]} (\seq{W} - \seq{\bm{\gamma}}_w) \oplus \seq{R}\nspace^{[i]}(X_k,\seq{U},\seq{W}), \label{eq:reachablestate}\\
\seq{\Psi}\nspace_{k+1:k+N}^{[i]}(\seq{\mbf{u}}) & = \seq{\mbf{s}}\nspace^{[i]} \oplus \seq{\mbf{C}}\nspace^{[i]} \seq{\Phi}\nspace_{k+1:k+N}^{[i]}(\seq{\mbf{u}}) \oplus \seq{\mbf{D}}\nspace_v^{[i]} \seq{V}, \label{eq:fault_reachableoutput}
\end{align}
where $\seq{U} = U \times U \times \ldots \times U$, $\seq{W} = W \times W \times \ldots \times W$, $\seq{V} = V \times V \times \ldots \times V$, with $N-1$ Cartesian products, $\seq{\bm{\gamma}}_u$ and $\seq{\bm{\gamma}}_w$ are sequences satisfying $\seq{\bm{\gamma}}_u \in \seq{U}$, $\seq{\bm{\gamma}}_w \in \seq{W}$, respectively, and $\seq{\mbf{u}} = (\mbf{u}_k, \mbf{u}_{k+1}, \ldots, \mbf{u}_{k+N-1})$. The variables $\seq{\mbf{r}}\nspace^{[i]}$, $\seq{\mbf{A}}\nspace^{[i]}$, $\seq{\mbf{B}}\nspace^{[i]}$, $\seq{\mbf{F}}\nspace^{[i]}$, $\seq{R}\nspace^{[i]}$, $\seq{\mbf{s}}\nspace^{[i]}$, $\seq{\mbf{C}}\nspace^{[i]}$, $\seq{\mbf{D}}\nspace_v^{[i]}$, are defined according to 
\begin{align}
\seq{\mbf{r}}\nspace^{[i]} & = \begin{bmatrix} \mbf{r}_k^{[i]} \\ \mbf{A}_{k+1}^{[i]} \mbf{r}_k^{[i]} + \mbf{r}_{k+1}^{[i]} \\ \vdots \\ \dsum{n=0}{N-2} \left(\dprod{j=n+1}{N-1} \mbf{A}_{k+j}^{[i]}\right) \mbf{r}_{k+n}^{[i]} + \mbf{r}_{k+N-1}^{[i]} \end{bmatrix}, ~
\seq{\mbf{A}}\nspace^{[i]} = \begin{bmatrix} \mbf{A}_k^{[i]} \\ \mbf{A}_{k+1}^{[i]} \mbf{A}_k^{[i]} \\ \vdots \\ \dprod{j=0}{N-1} \mbf{A}_{k+j}^{[i]}  \end{bmatrix}, \\
\seq{\mbf{B}}\nspace^{[i]} & = \begin{bmatrix} \mbf{B}_k^{[i]} & \bm{0} & \cdots & \bm{0} \\ \mbf{A}_{k+1}^{[i]} \mbf{B}_k^{[i]} & \mbf{B}_{k+1}^{[i]} & \cdots & \bm{0} \\ \vdots & \vdots & \ddots & \vdots \\ \left(\dprod{j=1}{N-1} \mbf{A}_{k+j}^{[i]} \right) \mbf{B}_k^{[i]} & \left(\dprod{j=2}{N-1} \mbf{A}_{k+j}^{[i]} \right) \mbf{B}_{k+1}^{[i]} & \cdots &  \mbf{B}_{k+N-1}^{[i]}\end{bmatrix}, \\
\seq{\mbf{F}}\nspace^{[i]} & = \begin{bmatrix} \mbf{F}_k^{[i]} & \bm{0} & \cdots & \bm{0} \\ \mbf{A}_{k+1}^{[i]} \mbf{F}_k^{[i]} & \mbf{F}_{k+1}^{[i]} & \cdots & \bm{0} \\ \vdots & \vdots & \ddots & \vdots \\ \left(\dprod{j=1}{N-1} \mbf{A}_{k+j}^{[i]} \right) \mbf{F}_k^{[i]} & \left(\dprod{j=2}{N-1} \mbf{A}_{k+j}^{[i]} \right) \mbf{F}_{k+1}^{[i]} & \cdots &  \mbf{F}_{k+N-1}^{[i]}\end{bmatrix}, \\	
\seq{R}\nspace^{[i]} & = R^{[i]}_k(X_k,U,W) \times R^{[i]}_{k+1}(\Phi_{k+1}^{[i]}(\seq{\mbf{u}}),U,W) \times \ldots \times R_{k+N-1}^{[i]}(\Phi_{k+N-1}^{[i]}(\seq{\mbf{u}}),U,W), \label{eq:Rtilde} \\
\seq{\mbf{s}}\nspace^{[i]} & = \begin{bmatrix} \mbf{s}_k^{[i]} \\ \mbf{s}_{k+1}^{[i]} \\ \vdots \\ \mbf{s}_{k+N-1}^{[i]} \end{bmatrix}, ~
\seq{\mbf{C}}\nspace^{[i]} = \begin{bmatrix} \mbf{C}^{[i]} & \bm{0} & \cdots & \bm{0} \\ \bm{0} & \mbf{C}^{[i]} & \cdots & \bm{0} \\ \vdots & \vdots & \ddots & \vdots \\ \bm{0} &  \bm{0} & \cdots &  \mbf{C}^{[i]}\end{bmatrix}, ~
\seq{\mbf{D}}\nspace_v^{[i]} = \begin{bmatrix} \mbf{D}_v^{[i]} & \bm{0} & \cdots & \bm{0} \\ \bm{0} & \mbf{D}_v^{[i]} & \cdots & \bm{0} \\ \vdots & \vdots & \ddots & \vdots \\ \bm{0} &  \bm{0} & \cdots & \mbf{D}_v^{[i]} \end{bmatrix}, 
\end{align}
where the matrix products are defined from the left, i.e. $\prod_{j=1}^{3} \mbf{A}_{k+j}^{[i]} \triangleq \mbf{A}_{k+3}^{[i]} \mbf{A}_{k+2}^{[i]} \mbf{A}_{k+1}^{[i]}$. 

Note that the set $\seq{R}\nspace^{[i]}$ is a function of $\Phi_{k+j}^{[i]}(\seq{\mbf{u}})$, $j=1,2,\ldots,N-1$, and therefore also of the input sequence $\seq{\mbf{u}}$ (see \eqref{eq:affinemap}). To circumvent this dependency of $\seq{R}\nspace^{[i]}$ on the input sequence, we propose the use of conservative enclosures $\underline{\Phi}\nspace_{k+j}^{[i]}$ satisfying $\underline{\Phi}\nspace_{k+j}^{[i]} \supseteq \Phi_{k+j}^{[i]}(\seq{\mbf{u}})$ in \eqref{eq:Rtilde}. These are computed recursively by
\begin{equation} \label{eq:fault_conservativereachable}
\underline{\Phi}\nspace_{k+j}^{[i]} = \mbf{r}_{k+j-1}^{[i]} \oplus \mbf{A}_{k+j-1}^{[i]} \underline{\Phi}_{k+j-1}^{[i]} \oplus \mbf{B}_{k+j-1}^{[i]} (U - \bm{\gamma}_{u,k+j-1}) \oplus \mbf{F}_{k+j-1}^{[i]} (W - \bm{\gamma}_{w,k+j-1}) \oplus R_{k+j-1}^{[i]}(\underline{\Phi}_{k+j-1}^{[i]},U,W),
\end{equation}
where $\underline{\Phi}_{k}^{[i]} \triangleq X_k$. Therefore, we define the respective conservative reachable state and output sets satisfying $\underline{\seq{\Phi}}\nspace_{k+1:k+N}^{[i]}(\seq{\mbf{u}}) \supseteq \seq{\Phi}\nspace_{k+1:k+N}^{[i]}(\seq{\mbf{u}})$, and $\underline{\seq{\Psi}}\nspace_{k+1:k+N}^{[i]}(\seq{\mbf{u}}) \supseteq \seq{\Psi}\nspace_{k+1:k+N}^{[i]}(\seq{\mbf{u}})$. These are computed by
\begin{align}
\underline{\seq{\Phi}}\nspace_{k+1:k+N}^{[i]}(\seq{\mbf{u}}) & = \seq{\mbf{r}}\nspace^{[i]} \oplus \seq{\mbf{A}}\nspace^{[i]} X_k \oplus \seq{\mbf{B}}\nspace^{[i]} (\seq{\mbf{u}} - \seq{\bm{\gamma}}_u) \oplus \seq{\mbf{F}}\nspace^{[i]} (\seq{W} - \seq{\bm{\gamma}}_w) \oplus \underline{\seq{R}}\nspace^{[i]}(X_k,\seq{U},\seq{W}), \label{eq:fault_conservativereachablestate}\\
\underline{\seq{\Psi}}\nspace_{k+1:k+N}^{[i]}(\seq{\mbf{u}}) & = \seq{\mbf{s}}\nspace^{[i]} \oplus \seq{\mbf{C}}\nspace^{[i]} \underline{\seq{\Phi}}\nspace_{k+1:k+N}^{[i]}(\seq{\mbf{u}}) \oplus \seq{\mbf{D}}\nspace_v^{[i]} \seq{V}, \label{eq:fault_conservativereachableoutput}
\end{align}
where
\begin{equation*}
\underline{\seq{R}}\nspace^{[i]} \triangleq R^{[i]}_k(X_k,U,W) \times R^{[i]}_{k+1}(\underline{\Phi}_{k+1}^{[i]},U,W) \times \ldots \times R_{k+N-1}^{[i]}(\underline{\Phi}_{k+N-1}^{[i]},U,W).
\end{equation*}
Note that the proposed conservative enclosure \eqref{eq:fault_conservativereachable} is not a function of the input $\mbf{u}_{k+j-1}$, $j = 1,2,\ldots,N$. Consequently, \eqref{eq:fault_conservativereachablestate}--\eqref{eq:fault_conservativereachableoutput} are all affine in the input sequence $\seq{\mbf{u}}$ as desired.

\section{Input design for active fault diagnosis} \label{sec:fault_inputdesign}

This section proposes a new method for input design aiming at active fault diagnosis of the nonlinear discrete-time system \eqref{eq:fault_system}. The method is based on the open-loop fault diagnosis approach proposed in \cite{Scott2014} for linear systems using zonotopes. This method is then extended using constrained zonotopes based on the approach proposed in \cite{Raimondo2016}.

\subsection{Zonotope method}

Let $X_k \subset \realset^n$, $W \subset \realset^{n_w}$, and $V \subset \realset^{n_v}$ be zonotopes. In this subsection we assume that $U$ is described by a zonotope, but with its H-rep also being available. We first present the required lemma below.
\begin{lemma} \rm \citep{Dobkin1993,Scott2014} \label{lem:intersection}
	Let $Z = \{\mbf{G}_z, \mbf{a}_z + \mbf{b}_z\}$ and $Y = \{\mbf{G}_y, \mbf{a}_y + \mbf{b}_y\}$. Then $Z \cap Y = \emptyset \iff \mbf{a}_y - \mbf{a}_z \notin \{ \mbf{G}_z, \mbf{b}_z\} \oplus \{ \mbf{G}_y, - \mbf{b}_y\}$.
\end{lemma}

The following definition is similar to Definition 1 in \cite{Scott2014}, but it differs by the fact that the conservative reachable output sets formulated by \eqref{eq:fault_conservativereachableoutput} are used in place of the exact enclosures.

\begin{definition} \rm \label{def:fault_separatinginput}
	An input sequence $\seq{\mbf{u}}$ is said to be a \emph{separating input} on the time interval $[k+1, k+N]$ if
	\begin{equation} \label{eq:fault_separatinginputdef}
	\underline{\seq{\Psi}}\nspace_{k+1:k+N}^{[i]}(\seq{\mbf{u}}) \cap \underline{\seq{\Psi}}\nspace_{k+1:k+N}^{[j]}(\seq{\mbf{u}}) = \emptyset,
	\end{equation}
	for every $i,j \in \modelset$, $i \neq j$.
\end{definition}

Let $\seq{\mbf{u}} \in \seq{U}$ be an input sequence defined in the time interval $[k,\, k+N-1]$, and consider the output sequence $\seq{\mbf{y}} = (\mbf{y}_{k+1}, \mbf{y}_{k+2}, \ldots, \mbf{y}_{k+N})$ observed by the injection of $\seq{\mbf{u}}$ into the nonlinear system \eqref{eq:fault_system}\footnote{The input and output sequences are displaced by one time step since system \eqref{eq:fault_system} is causal.}. If $\seq{\mbf{u}}$ is a separating input as in Definition \ref{def:fault_separatinginput}, then the inclusion
\begin{equation} \label{eq:fault_diagnosisinclusion}
	\seq{\mbf{y}} \in \underline{\seq{\Psi}}\nspace_{k+1:k+N}^{[i]}(\seq{\mbf{u}})
\end{equation}
is satisfied only for the model $i \in \modelset$ that is active in $[k+1,\, k+N]$. Therefore, the injection of a separating input in \eqref{eq:fault_system} allows guaranteed active fault diagnosis in the time interval $[k+1,\, k+N]$ by verifying the inclusion \eqref{eq:fault_diagnosisinclusion} for every $i \in \modelset$. Since $\underline{\seq{\Psi}}\nspace_{k+1:k+N}^{[i]}(\seq{\mbf{u}})$ is a zonotope, this inclusion can be easily verified by solving an LP (similarly to Property \ref{prop:pre_czisemptyinside}). The following theorem is an extension of Theorem 3 in \cite{Scott2014} to the approach presented in this section.

\begin{theorem} \rm \label{thm:fault_separatinginput}
	Let $\underline{\seq{\Psi}}\nspace_{k+1:k+N}^{[i]}(\seq{\mbf{u}}) = \{\mbf{G}^\Psi_{k+1:k+N}(i), \mbf{c}^\Psi_{k+1:k+N}(\seq{\mbf{u}},i) \}$. An input sequence $\seq{\mbf{u}}$ is a separating input iff
	\begin{equation} \label{eq:separationcondition}
		\seq{\mbf{N}}(i,j) \seq{\mbf{u}} \notin \seq{Z}(i,j),
	\end{equation}
	for all $i,j \in \modelset$, $i \neq j$, where $\seq{\mbf{N}}(i,j) = \seq{\mbf{C}}\nspace^{[j]} \seq{\mbf{B}}\nspace^{[j]} - \seq{\mbf{C}}\nspace^{[i]}\seq{\mbf{B}}\nspace^{[i]}$, and $\seq{Z}(i,j) = \big\{ [ \mbf{G}^\Psi_{k+1:k+N}(i) \,\; \mbf{G}^\Psi_{k+1:k+N}(j)]$, $ \mbf{c}^\Psi_{k+1:k+N} (\seq{\bm{\gamma}}_u,i) - \mbf{c}^\Psi_{k+1:k+N} (\seq{\bm{\gamma}}_u,j) + \seq{\mbf{N}}(i,j) \seq{\bm{\gamma}}_u \big\}$.
\end{theorem}

\proof The relation below follows immediately from \eqref{eq:fault_conservativereachablestate}:
\begin{equation*}
	\underline{\seq{\Phi}}\nspace_{k+1:k+N}^{[i]}(\seq{\mbf{u}}) = \underline{\seq{\Phi}}\nspace_{k+1:k+N}^{[i]}(\seq{\bm{\gamma}}_u) \oplus \seq{\mbf{B}}_u^{[i]} (\seq{\mbf{u}} - \seq{\bm{\gamma}}_u),
\end{equation*}
which means that the center of $\underline{\seq{\Phi}}\nspace_{k+1:k+N}^{[i]}(\seq{\mbf{u}})$ is displaced from the center of $\underline{\seq{\Phi}}\nspace_{k+1:k+N}^{[i]}(\seq{\bm{\gamma}}_u)$ by $\seq{\mbf{B}}\nspace^{[i]} (\seq{\mbf{u}} - \seq{\bm{\gamma}}_u)$, and the respective generator matrix is unaffected. Consequently, it follows from \eqref{eq:fault_conservativereachableoutput} that 
\begin{equation*}
	\underline{\seq{\Psi}}\nspace_{k+1:k+N}^{[i]}(\seq{\mbf{u}}) = \underline{\seq{\Psi}}\nspace_{k+1:k+N}^{[i]}(\seq{\bm{\gamma}}_u) \oplus \seq{\mbf{C}}\nspace^{[i]} \seq{\mbf{B}}\nspace^{[i]} (\seq{\mbf{u}} - \seq{\bm{\gamma}}_u).
\end{equation*}
Therefore, $\mbf{c}^\Psi_{k+1:k+N}(\seq{\mbf{u}},i) = \mbf{c}^\Psi_{k+1:k+N}(\seq{\bm{\gamma}}_u,i) + \seq{\mbf{C}}\nspace^{[i]} \seq{\mbf{B}}\nspace^{[i]}(\seq{\mbf{u}} - \seq{\bm{\gamma}}_u)$. Finally, if $\seq{\mbf{N}}(i,j)$ and $\seq{Z}(i,j)$ are defined as in the statement of the theorem, then Lemma \ref{lem:intersection} implies that \eqref{eq:fault_separatinginputdef} holds for every $i,j \in \modelset$, $i \neq j$, iff \eqref{eq:separationcondition} holds for every $i,j \in \modelset$, $i \neq j$. \qed

Let $n_q$ denote the number of all possible combinations between $i,j \in \modelset$, $i \neq j$, and define $\mbf{N}^{[q]} \triangleq \mbf{N}(i,j)$, $\seq{Z}\nspace^{[q]} \triangleq \seq{Z}(i,j)$, $q \in \{ 1,2,\ldots,n_q\}$. Lemma \ref{lem:fault_verifyseparating} is proposed in \cite{Scott2014} and can be used to verify if a given input sequence $\seq{\mbf{u}}$ is a separating input according to Theorem \ref{thm:fault_separatinginput} by means of the solution of $n_q$ LPs.

\begin{lemma} \rm \citep{Scott2014} \label{lem:fault_verifyseparating}
	Let $\seq{Z}\nspace^{[q]} = \{\mbf{G}^{[q]}, \mbf{c}^{[q]}\}$. For each $\seq{\mbf{u}} \in \seq{U}$ and $q \in \{1,2,\ldots,n_q\}$, define
	\begin{align*}
		\hat{\delta}^{[q]}(\seq{\mbf{u}}) & = \underset{\hat{\delta}^{[q]}, \bm{\xi}^{[q]}}{\min} \delta^{[q]} \\
		\text{s.t.} & \quad \seq{\mbf{N}}\nspace^{[q]} \seq{\mbf{u}} = \mbf{G}^{[q]} \bm{\xi}^{[q]} + \mbf{c}^{[q]} \\
		& \quad \ninf{\bm{\xi}} \leq 1 + \delta^{[q]}.
	\end{align*}
	Then $\seq{\mbf{N}}\nspace^{[q]} \seq{\mbf{u}} \notin \seq{Z}\nspace^{[q]} \iff \hat{\delta}^{[q]}(\seq{\mbf{u}}) > 0$.
\end{lemma}

\proof See Lemma 4 in \cite{Scott2014}. \qed

Based on Theorem \ref{thm:fault_separatinginput}, we therefore consider the design of an optimal separating input by solving the optimization problem
\begin{align}
	\underset{\seq{\mbf{u}}}{\min} & ~ J(\seq{\mbf{u}}) \label{eq:fault_optimalseparating}\\
	\text{s.t.}	& ~ \seq{\mbf{u}} \in \seq{U}, \nonumber\\
	& ~ \seq{\mbf{N}}\nspace^{[q]} \seq{\mbf{u}} \notin \seq{Z}\nspace^{[q]}, ~ \forall q  = 1,2,\ldots,n_q, \nonumber	
\end{align}
where $J(\seq{\mbf{u}})$ is a cost function chosen to minimize any harmful effects caused to the system when injecting the separating input. For simplicity we define $J(\seq{\mbf{u}}) = \|\seq{\mbf{u}}\|_{\seq{\mbf{R}}}^2 = \sum_{j=0}^{N-1} \mbf{u}_{k+j}^T \mbf{R} \mbf{u}_{k+j}$, with $\seq{\mbf{R}} \triangleq \text{blkdiag}(\mbf{R},\mbf{R},\ldots,\mbf{R})$.

As in \cite{Scott2014}, the optimization problem \eqref{eq:fault_optimalseparating} is a bilevel program and can be rewritten as a mixed-integer quadratic program (MIQP) as follows. Define the \emph{minimum separation threshold} $\varepsilon > 0$, then the optimization problem \eqref{eq:fault_optimalseparating} can be written as
\begin{align}
	\underset{\seq{\mbf{u}}}{\min} & ~ J(\seq{\mbf{u}}) \label{eq:optimalseparating2}\\
	\text{s.t.}	& ~ \seq{\mbf{u}} \in \seq{U}, \nonumber\\
	 & ~ \varepsilon \leq \hat{\delta}^{[q]} \leq \hat{\delta}_\text{m}^{[q]} , ~ \forall q  = 1,2,\ldots,n_q, \nonumber
\end{align}
with $\hat{\delta}^{[q]}$ defined as in Lemma \ref{lem:fault_verifyseparating}, and $\hat{\delta}^{[q]}_\text{m}$ is given by $\hat{\delta}^{[q]}_\text{m} = \underset{\seq{\mbf{u}} \in \seq{U}}{\max} ~ \hat{\delta}^{[q]} (\seq{\mbf{u}})$ (see Section 5.1 in \cite{Scott2016} for details in the computation of $\hat{\delta}^{[q]}_\text{m}$). By replacing the inner optimization programs in \eqref{eq:optimalseparating2} by their necessary and sufficient conditions for optimality (Proposition 3.4.1 in \cite{Bertsekas1999}), and by introducing binary variables to avoid non-convex constraints (see Section 5 in \cite{Scott2016}), the optimization problem \eqref{eq:optimalseparating2} can be rewritten as the single level program
\begin{align}
	& \underset{\seq{\mbf{u}}, \delta^{[q]}, \bm{\xi}^{[q]}, \mbf{\lambda}^{[q]}, \bm{\mu}_1^{[q]}, \bm{\mu}_2^{[q]}, \mbf{p}_1^{[q]}, \mbf{p}_2^{[q]}}{\min} ~ J(\seq{\mbf{u}}) \label{eq:fault_optimalseparatingfinal} \\
	\text{s.t.}	& ~ \seq{\mbf{u}} \in \seq{U}, \nonumber\\
	& \left .\begin{aligned}
	& \varepsilon \leq \hat{\delta}^{[q]} \leq \hat{\delta}_\text{m}^{[q]}, \\
	& \seq{\mbf{N}}\nspace^{[q]} \seq{\mbf{u}} = \mbf{G}^{[q]} \bm{\xi}^{[q]} + \mbf{c}^{[q]}, \\
	& \ninf{\bm{\xi}^{[q]}} \leq 1 + \delta^{[q]}, \\
	& (\mbf{G}^{[q]})^T \bm{\lambda}^{[q]} = \bm{\mu}_1^{[q]} - \bm{\mu}_2^{[q]}, \\
	& 1 = (\bm{\mu}_1^{[q]} + \bm{\mu}_2^{[q]})^T \bm{1}, \\
	& \bm{0} \leq \bm{\mu}_1^{[q]}, \bm{\mu}_2^{[q]}, \\
	& \mbf{p}_1^{[q]}, \mbf{p}_2^{[q]} \in \{0,1\}^{n_g^{[q]}}, \\
	& \bm{\mu}_1^{[q]} \leq \mbf{p}_1^{[q]}, \\
	& \bm{\mu}_2^{[q]} \leq \mbf{p}_2^{[q]}, \\
	& \xi_j^{[q]} - 1 - \delta^{[q]} \in [-2(1 + \hat{\delta}_\text{m}^{[q]})(1 - p_{1,j}^{[q]}),\, 0], \\
	& \xi_j^{[q]} + 1 + \delta^{[q]} \in [0,\, 2(1 + \hat{\delta}_\text{m}^{[q]})(1 - p_{2,j}^{[q]})],
	\end{aligned}\right\} ~ \forall q  = 1,2,\ldots,n_q, \nonumber
\end{align}
which is a MIQP, with $j = 1,2,\ldots,n_g^{[q]}$. The constraints $\seq{\mbf{N}}\nspace^{[q]} \seq{\mbf{u}} = \mbf{G}^{[q]} \bm{\xi}^{[q]} + \mbf{c}^{[q]}$ and $\ninf{\bm{\xi}^{[q]}} \leq 1 + \delta^{[q]}$ come from Lemma \ref{lem:fault_verifyseparating}; $\varepsilon \leq \hat{\delta}^{[q]} \leq \hat{\delta}_\text{m}^{[q]}$ guarantees the boundedness of \eqref{eq:fault_optimalseparatingfinal}; $(\mbf{G}^{[q]})^T \bm{\lambda}^{[q]} = \bm{\mu}_1^{[q]} - \bm{\mu}_2^{[q]}$, $1 = (\bm{\mu}_1^{[q]} + \bm{\mu}_2^{[q]})^T \bm{1}$, and $\bm{0} \leq \bm{\mu}_1^{[q]}, \bm{\mu}_2^{[q]}$ are a result from Karush-Kuhn-Tucker conditions; and $\mbf{p}_1^{[q]}, \mbf{p}_2^{[q]} \in \{0,1\}^{n_g^{[q]}}$, $\bm{\mu}_1^{[q]} \leq \mbf{p}_1^{[q]}$, $\bm{\mu}_2^{[q]} \leq \mbf{p}_2^{[q]}$, $\xi_j^{[q]} - 1 - \delta^{[q]} \in [-2(1 + \hat{\delta}_\text{m}^{[q]})(1 - p_{1,j}^{[q]}),\, 0]$, $\xi_j^{[q]} + 1 + \delta^{[q]} \in [0,\, 2(1 + \hat{\delta}_\text{m}^{[q]})(1 - p_{2,j}^{[q]})]$ are mixed-integer reformulations of non-convex complementarity constraints \citep{Scott2014}.

\subsection{Constrained zonotope method} \label{sec:fault_czmethod}

Let $X_k \subset \realset^n$, $W \subset \realset^{n_w}$, and $V \subset \realset^{n_v}$ now be constrained zonotopes. Similarly to the open-loop method in \cite{Raimondo2016}, the approach presented in the previous subsection can be extended to constrained zonotopes as follows.

\begin{theorem} \rm \label{thm:fault_separatinginputcz}
	Let $\underline{\seq{\Psi}}\nspace_{k+1:k+N}^{[i]}(\seq{\mbf{u}}) = \{\mbf{G}^\Psi_{k+1:k+N}(i), \mbf{c}^\Psi_{k+1:k+N}(\seq{\mbf{u}},i), \mbf{A}^\Psi_{k+1:k+N}(i), \mbf{b}^\Psi_{k+1:k+N}(i) \}$. An input sequence $\seq{\mbf{u}}$ is a separating input iff
	\begin{equation} \label{eq:fault_separationconditioncz}
	\seq{\mbf{N}}(i,j) \seq{\mbf{u}} \notin \seq{Z}(i,j),
	\end{equation}
	for all $i,j \in \modelset$, $i \neq j$, where $\seq{\mbf{N}}(i,j) = \seq{\mbf{C}}\nspace^{[j]} \seq{\mbf{B}}\nspace^{[j]} - \seq{\mbf{C}}\nspace^{[i]}\seq{\mbf{B}}\nspace^{[i]}$, and 
	\begin{align*}
	\seq{Z}(i,j) =  & \left\{ [ \mbf{G}^\Psi_{k+1:k+N}(i) \,\; -\mbf{G}^\Psi_{k+1:k+N}(j)],\, \mbf{c}^\Psi_{k+1:k+N} (\seq{\bm{\gamma}}_u,i) - \mbf{c}^\Psi_{k+1:k+N} (\seq{\bm{\gamma}}_u,j) + \seq{\mbf{N}}(i,j) \seq{\bm{\gamma}}_u, \right.
	 \\& \left. \begin{bmatrix} \mbf{A}^\Psi_{k+1:k+N}(i) & \bm{0} \\ \bm{0} & \mbf{A}^\Psi_{k+1:k+N}(j) \end{bmatrix}, \begin{bmatrix} \mbf{b}^\Psi_{k+1:k+N}(i) \\ \mbf{b}^\Psi_{k+1:k+N}(j) \end{bmatrix} \right\}.
	\end{align*}
\end{theorem}

\proof The following is similar to the proof of Theorem 1 in \cite{Raimondo2016}. From the generalized intersection \eqref{eq:pre_czintersection} and recalling that $\mbf{c}^\Psi_{k+1:k+N}(\seq{\mbf{u}},i) = \mbf{c}^\Psi_{k+1:k+N}(\seq{\bm{\gamma}}_u,i) + \mbf{C}^{[i]} \mbf{B}^{[i]} (\seq{\mbf{u}} - \seq{\bm{\gamma}}_u)$, it holds that 
\begin{align*}
	& \underline{\seq{\Psi}}\nspace_{k+1:k+N}^{[i]}(\seq{\mbf{u}}) \cap \underline{\seq{\Psi}}\nspace_{k+1:k+N}^{[j]}(\seq{\mbf{u}}) = \emptyset \iff \nexists \bm{\xi} : \ninf{\bm{\xi}} \leq 1, \\
	& \begin{bmatrix} \mbf{A}^\Psi_{k+1:k+N}(i) & \bm{0} \\ \bm{0} & \mbf{A}^\Psi_{k+1:k+N}(j) \\ \mbf{G}^\Psi_{k+1:k+N}(i) & -\mbf{G}^\Psi_{k+1:k+N}(j) \end{bmatrix} \bm{\xi} =  \begin{bmatrix} \mbf{b}^\Psi_{k+1:k+N}(i) \\ \mbf{b}^\Psi_{k+1:k+N}(j) \\ \mbf{c}^\Psi_{k+1:k+N}(\seq{\bm{\gamma}}_u,j) - \mbf{c}^\Psi_{k+1:k+N}(\seq{\bm{\gamma}}_u,i) + \seq{\mbf{N}}(i,j)(\seq{\mbf{u}} - \seq{\bm{\gamma}}_u) \end{bmatrix}.
\end{align*}
The relation above is equivalent to
\begin{align*}
   & \underline{\seq{\Psi}}\nspace_{k+1:k+N}^{[i]}(\seq{\mbf{u}}) \cap \underline{\seq{\Psi}}\nspace_{k+1:k+N}^{[j]}(\seq{\mbf{u}}) = \emptyset \iff \\ & \seq{\mbf{N}}(i,j) \seq{\mbf{u}} \notin \left\{ [\mbf{G}^\Psi_{k+1:k+N}(i) \,\; -\mbf{G}^\Psi_{k+1:k+N}(j)],\, \mbf{c}^\Psi_{k+1:k+N}(\seq{\bm{\gamma}}_u,i) - \mbf{c}^\Psi_{k+1:k+N}(\seq{\bm{\gamma}}_u,j) + \seq{\mbf{N}}(i,j)\seq{\bm{\gamma}}_u, \right. \\
	& \qquad \qquad \quad \left. \begin{bmatrix} \mbf{A}^\Psi_{k+1:k+N}(i) & \bm{0} \\ \bm{0} & \mbf{A}^\Psi_{k+1:k+N}(j) \end{bmatrix}, \begin{bmatrix} \mbf{b}^\Psi_{k+1:k+N}(i) \\ \mbf{b}^\Psi_{k+1:k+N}(j) \end{bmatrix} \right\},
\end{align*}
which right hand side is \eqref{eq:fault_separationconditioncz}. \qed

We now reformulate Theorem \ref{thm:fault_separatinginputcz} in terms of lifted zonotopes.

\begin{cor} \rm \label{cor:separatinginputcz}
	Let $\seq{Z}(i,j) = \{\mbf{G}^{Z}(i,j), \mbf{c}^{Z}(i,j), \mbf{A}^{Z}(i,j), \mbf{b}^{Z}(i,j) \}$ be a constrained zonotope as defined in Theorem \ref{thm:fault_separatinginputcz}. An input sequence $\seq{\mbf{u}}$ is a separating input iff
	\begin{equation} \label{eq:separationconditioncz2}
	\begin{bmatrix} \seq{\mbf{N}}(i,j) \\ \bm{0} \end{bmatrix} \seq{\mbf{u}} \notin \seq{Z}\nspace^+(i,j) = \left\{ \begin{bmatrix} \mbf{G}^{Z}(i,j) \\ \mbf{A}^{Z}(i,j) \end{bmatrix}, \begin{bmatrix} \mbf{c}^{Z}(i,j) \\ -\mbf{b}^{Z}(i,j) \end{bmatrix} \right\},
	\end{equation}
	for all $i,j \in \modelset$, $i \neq j$.
\end{cor}

\proof See Corollary 2 in \cite{Raimondo2016}. \qed

Let $\seq{\mbf{M}}\nspace^{[q]} = [\seq{\mbf{N}}\nspace^T(i,j) \,\; \bm{0}]^T$,  $\seq{Z}\nspace^{+[q]} = \{\mbf{G}^{+[q]}, \mbf{c}^{+[q]}\} = \seq{Z}\nspace^+(i,j)$, $q \in \{1,2,\ldots,n_q\}$. Since $\seq{Z}\nspace^{+[q]}$ is a zonotope, the relation \eqref{eq:separationconditioncz2} can be verified by solving an LP similar to Lemma \ref{lem:fault_verifyseparating}.  We then consider the design of the optimal separating input by solving
\begin{align}
\underset{\seq{\mbf{u}}}{\min} & ~ J(\seq{\mbf{u}}) \label{eq:fault_optimalseparatingcz}\\
\text{s.t.}	& ~ \seq{\mbf{u}} \in \seq{U}, \nonumber\\
& ~ \seq{\mbf{M}}\nspace^{[q]} \seq{\mbf{u}} \notin \seq{Z}\nspace^{+[q]}, ~ \forall q  = 1,2,\ldots,n_q, \nonumber	
\end{align}
where for simplicity $J(\seq{\mbf{u}}) = \|\seq{\mbf{u}}\|_{\seq{\mbf{R}}}^2 = \sum_{j=0}^{N-1} \mbf{u}_{k+j}^T \mbf{R} \mbf{u}_{k+j}$, $\seq{\mbf{R}} = \text{blkdiag}(\mbf{R},\mbf{R},\ldots,\mbf{R})$. Analogously to the previous subsection, the optimization problem \eqref{eq:fault_optimalseparatingcz} is a bilevel program, but can be reformulated as an MIQP given by
\begin{align}
	& \underset{\seq{\mbf{u}}, \delta^{[q]}, \bm{\xi}^{[q]}, \bm{\lambda}^{[q]}, \bm{\mu}_1^{[q]}, \bm{\mu}_2^{[q]}, \mbf{p}_1^{[q]}, \mbf{p}_2^{[q]}}{\min} ~ J(\seq{\mbf{u}}) \label{eq:fault_optimalseparatingczfinal} \\
	\text{s.t.}	& ~ \seq{\mbf{u}} \in \seq{U}, \nonumber\\
	& \left .\begin{aligned}
	& \varepsilon \leq \hat{\delta}^{[q]} \leq \hat{\delta}_\text{m}^{[q]}, \\
	& \seq{\mbf{M}}\nspace^{[q]} \seq{\mbf{u}} = \mbf{G}^{+[q]} \bm{\xi}^{[q]} + \mbf{c}^{+[q]}, \\
	& \ninf{\bm{\xi}^{[q]}} \leq 1 + \delta^{[q]}, \\
	& (\mbf{G}^{+[q]})^T \bm{\lambda}^{[q]} = \bm{\mu}_1^{[q]} - \bm{\mu}_2^{[q]}, \\
	& 1 = (\bm{\mu}_1^{[q]} + \bm{\mu}_2^{[q]})^T \bm{1}, \\
	& \bm{0} \leq \bm{\mu}_1^{[q]}, \bm{\mu}_2^{[q]}, \\
	& \mbf{p}_1^{[q]}, \mbf{p}_2^{[q]} \in \{0,1\}^{n_g^{[q]}}, \\
	& \bm{\mu}_1^{[q]} \leq \mbf{p}_1^{[q]}, \\
	& \bm{\mu}_2^{[q]} \leq \mbf{p}_2^{[q]}, \\
	& \xi_j^{[q]} - 1 - \delta^{[q]} \in [-2(1 + \hat{\delta}_\text{m}^{[q]})(1 - p_{1,j}^{[q]}),\, 0], \\
	& \xi_j^{[q]} + 1 + \delta^{[q]} \in [0,\, 2(1 + \hat{\delta}_\text{m}^{[q]})(1 - p_{2,j}^{[q]})],
	\end{aligned}\right\} ~ \forall q  = 1,2,\ldots,n_q, \nonumber
\end{align}
with $j = 1,2,\ldots,n_g^{[q]}$. The meaning of each constraint in \eqref{eq:fault_optimalseparatingczfinal} is similar to the ones from the zonotope method presented in the previous section.

\begin{remark} \rm \label{rem:fault_miqpmultiplesolutions}
	The MIQPs \eqref{eq:fault_optimalseparatingfinal} and \eqref{eq:fault_optimalseparatingczfinal} may admit multiple optimal solutions due to the non-convex nature of mixed-integer problems. In such cases, any of the resulting optimal input sequences is considered valid for active fault diagnosis.
\end{remark}

\section{Numerical example} \label{sec:fault_examples}

\begin{figure}[htb]
	\centering{
		\includegraphics[width=0.4\columnwidth]{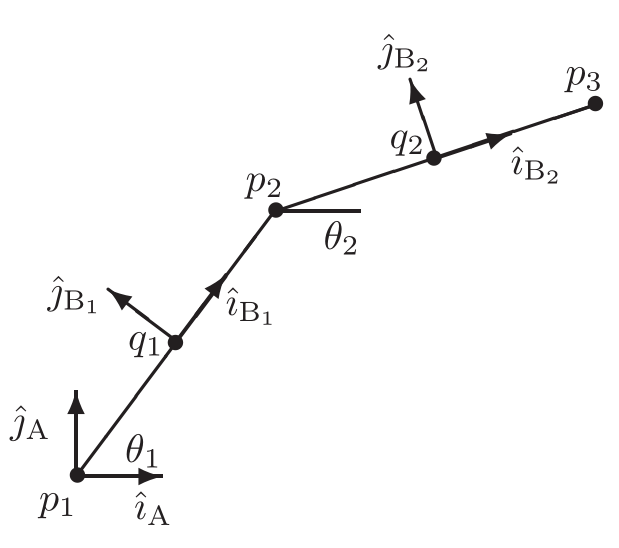}
		\caption{The planar robotic manipulator \citep{Damato2011}.}
		\label{fig:fault_roboticmanipulator}}
\end{figure}

This section demonstrates the performance of the proposed active-fault diagnosis approach for nonlinear discrete-time systems described by \eqref{eq:fault_system} in a numerical example. The experiment consists of fault diagnosis of a planar robotic manipulator composed of two revolute joints (Figure \ref{fig:fault_roboticmanipulator}) which is subject to actuator faults. 

Let $\theta_1$ and $\theta_2$ denote the angles of the first and second revolute joint, respectively, and let $u$ denote the torque applied to the first joint. The second joint is passive (i.e., it is not actuated). The equations of motion that describe the dynamics of the robotic manipulator are given by \citep{Damato2011}
\begin{equation} \label{eq:fault_roboticmanipulator0}
\begin{aligned}
\kappa^{[i]} u & = \left( \frac{1}{3}m_1 l_1^2 + m_2 l_1^2 \right) \ddot{\theta}_1 + \frac{1}{2} m_2 l_1 l_2 \sin(\theta_1 - \theta_2) \dot{\theta}_2^2 + \frac{1}{2} m_2 l_1 l_2 \cos(\theta_1 - \theta_2) \ddot{\theta}_2 \\
& \quad + (k_1 + k_2) \theta_1 - k_2 \theta_2 + (c_1 + c_2)\dot{\theta}_1 - c_2 \dot{\theta}_2 \text{,}\\
0 & = \left( \frac{1}{3} m_2 l_2^2 \right) \ddot{\theta}_2 - \frac{1}{2} m_2 l_1 l_2 \sin(\theta_1 - \theta_2)\dot{\theta}_1^2 + \frac{1}{2} m_2 l_1 l_2 \cos(\theta_1 - \theta_2) \ddot{\theta}_1 \\
& \quad - k_2 \theta_1 + k_2 \theta_2 - c_2 \dot{\theta}_1 + c_2 \dot{\theta}_2 \text{,}
\end{aligned}
\end{equation}
where $m_1 = 2$ kg, $m_2 = 1$ kg, $l_1 = 3$ m, $l_2 = 2$ m, $k_1 = 7$ N${\cdot}$m/rad, $k_2 = 5$ N${\cdot}$m/rad, $c_1 = 10$ kg${\cdot}$m$^2$/rad, $c_2 = 1$ kg${\cdot}$m$^2$/rad. The actuator faults are modeled by the parameter $\kappa^{[i]}$. Two scenarios are considered, the nominal scenario in which $\kappa^{[1]} = 1$, and the faulty one given by $\kappa^{[2]} = 0$. By defining the variables
\begin{equation*}
\begin{aligned}
& \mbf{q} \triangleq \begin{bmatrix} \theta_1 \\ \theta_2 \end{bmatrix},\\
& \mbf{R}(\mbf{q}) \triangleq \begin{bmatrix} \frac{1}{3}m_1 l_1^2 + m_2 l_1^2 & \frac{1}{2} m_2 l_1 l_2 \cos(\theta_1 - \theta_2) \\ \frac{1}{2} m_2 l_1 l_2 \cos(\theta_1 - \theta_2) & \frac{1}{3} m_2 l_2^2 \end{bmatrix}, \quad \mbf{P}^{[i]} \triangleq \begin{bmatrix} \kappa^{[i]} \\ 0 \end{bmatrix} \text{,} \\
& \mbf{S}(\mbf{q},\dot{\mbf{q}}) \triangleq \begin{bmatrix} \frac{1}{2} m_2 l_1 l_2 \sin(\theta_1 - \theta_2) \dot{\theta}_2^2 + (k_1 + k_2) \theta_1 - k_2 \theta_2 + (c_1 + c_2) \dot{\theta}_1 - c_2 \dot{\theta}_2 \\ -\frac{1}{2} m_2 l_1 l_2 \sin(\theta_1 - \theta_2) \dot{\theta}_1^2 - k_2 \theta_1 + k_2 \theta_2 - c_2 \dot{\theta}_1 + c_2 \dot{\theta}_2 \end{bmatrix},
\end{aligned}
\end{equation*}
the dynamic equations \eqref{eq:fault_roboticmanipulator0} can be written in compact form as $\mbf{R}(\mbf{q}) \ddot{\mbf{q}} + \mbf{S}(\mbf{q},\dot{\mbf{q}}) = \mbf{P}^{[i]} u$. In addition, by defining the system state $\mbf{x} = [\theta_1 \,\; \theta_2 \,\; \dot{\theta}_1 \,\; \dot{\theta}_2]^T$, it leads to the following state-space representation of the robotic manipulator dynamics:
\begin{equation} \label{eq:fault_robmanipulatorstatespace}
\dot{\mbf{x}} = \left[\begin{array}{c} x_3 \\ x_4 \\ \mbf{R} (x_1,x_2)^{-1} ( - \mbf{S}(x_1,x_2,x_3,x_4) + \mbf{P}^{[i]}u )\end{array}\right] \text{.}
\end{equation}

We consider that $\theta_1$ and $\theta_2$ are measured by sensors. Then, the corresponding measurement equation is given by
\begin{equation} \label{eq:fault_robmanipulatormeasurement}
\begin{bmatrix} y_{1,k} \\ y_{2,k}  \end{bmatrix} = \begin{bmatrix} x_{1,k} \\ x_{2,k} \end{bmatrix} + \begin{bmatrix} v_{1,k} \\ v_{2,k} \end{bmatrix},
\end{equation}
with $\ninf{\mbf{v}_k} \leq 0.1$. In this example, the nonlinear equations \eqref{eq:fault_robmanipulatorstatespace} are discretized through Euler approximation with sampling time 0.01 s. Moreover, the initial state $\mbf{x}_0$ is bounded by the zonotope
\begin{equation*}
	X_0 = \left\{ 0.01 {\cdot} \eye{4}, [ {\pi}/{2} \,\; {\pi}/{2} \,\; 0 \,\; 0 ]^T \right\}.
\end{equation*}

For fault diagnosis of the system \eqref{eq:fault_robmanipulatorstatespace}--\eqref{eq:fault_robmanipulatormeasurement}, we consider the design of an input sequence with length $N = 5$, and choose $\seq{\bm{\gamma}}_u = \bm{0}$, $\seq{\bm{\gamma}}_w = \bm{0}$. The cost function in \eqref{eq:fault_optimalseparatingfinal} is chosen as $J(\seq{\mbf{u}}) = \sum_{j=0}^{N-1}  r u_j^2$, with $r = 1$. The admissible inputs must satisfy $|u_k| \leq 100$ N$\cdot$m for $k \in [0,N-1]$. Moreover, $\bm{\gamma}_{x,0} = [ \pi/2 \,\; \pi/2 \,\; 0 \,\; 0]^T$, and $\bm{\gamma}_{x,k}$ is obtained by propagating $(\bm{\gamma}_{x,k-1},\bm{\gamma}_{u,k-1},\bm{\gamma}_{w,k-1})$ through the discretized version of the nonlinear dynamics \eqref{eq:fault_robmanipulatorstatespace} for each model, with $k=1,2,\ldots,N-1$. The number of generators of the zonotopes $\seq{Z}\nspace^{[q]}$ is limited to 20 using the generator reduction algorithm described by Method \ref{meth:genredB} (Section \ref{sec:complexityreduction}). The minimum separation threshold was chosen as $\varepsilon = 1.0 \ten{-8}$.

\begin{figure}[!tb]
	\centering{
		\def\svgwidth{0.7\columnwidth}
		{\scriptsize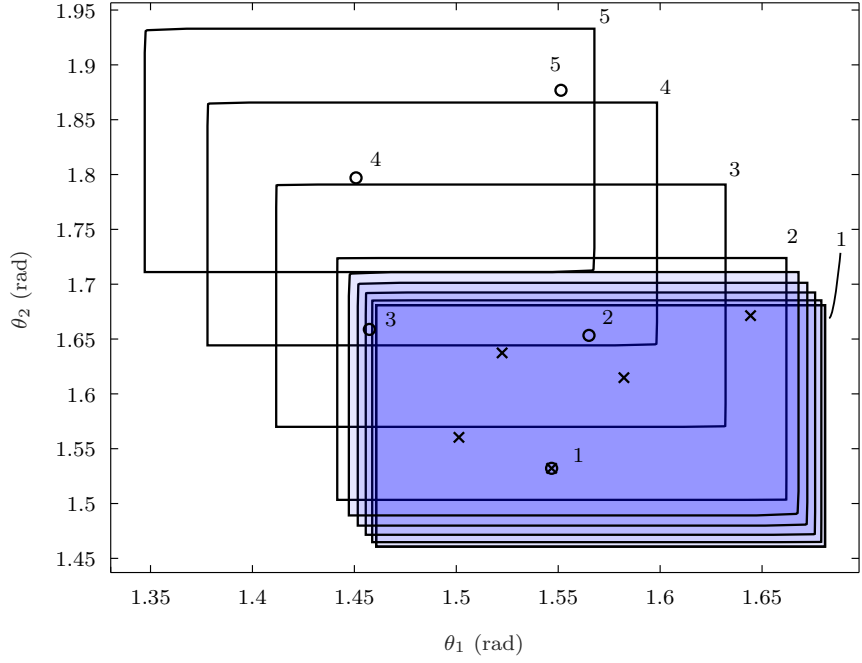}
		\caption{The reachable output sets for models 1 (colorless sets) and 2 (blue sets) along with the output $\seq{\mbf{y}}$ observed by the injection of $\seq{\mbf{u}}$ into models 1 ($\circ$) and 2 ($\times$) for $k=1,2,\ldots,5$. For clarity, numbers the note the corresponding time step $k$ for the output sequence $\seq{\mbf{y}}\nspace^{[1]}$ observed from model 1 and its output reachable sets.}\label{fig:fault_robmanipulator_reachable}}
\end{figure}

The MIQP \eqref{eq:fault_optimalseparatingfinal} was solved using MATLAB 9.1 and GUROBI 8.0.1\footnote{\url{http://www.gurobi.com}}, which resulted in the optimal separating input
\begin{equation} \label{eq:fault_robmanipulatorseparatinginput}
\seq{\mbf{u}} = ({-}100, {-}91.52, {-}73.88, {-}45.12, 0) ~\text{N}{\cdot}\text{m}.
\end{equation}
This input sequence was obtained with execution time\footnote{Laptop with 8GB RAM and Intel Core i7 4510U 3.1 GHz processor.} $0.035$ s. Figure \ref{fig:fault_robmanipulator_reachable} shows the output reachable sets for models 1 and 2 (nominal and faulty, respectively), along with the respective output sequences observed by the injection of \eqref{eq:fault_robmanipulatorseparatinginput} into \eqref{eq:fault_robmanipulatorstatespace} for $k=1,2,\ldots,5$, which are denoted here by $\seq{\mbf{y}}\nspace^{[1]}$ and $\seq{\mbf{y}}\nspace^{[2]}$. Note that, as expected the following relations were satisfied:
\begin{equation*}
	\seq{\mbf{y}}\nspace^{[1]} \in \underline{\seq{\Psi}}\nspace_{1:5}^{[1]}(\seq{\mbf{u}}), ~ \seq{\mbf{y}}\nspace^{[2]} \in \underline{\seq{\Psi}}\nspace_{1:5}^{[2]}(\seq{\mbf{u}}), ~\seq{\mbf{y}}\nspace^{[1]} \notin \underline{\seq{\Psi}}\nspace_{1:5}^{[2]}(\seq{\mbf{u}}), ~ \seq{\mbf{y}}\nspace^{[2]} \notin \underline{\seq{\Psi}}\nspace_{1:5}^{[1]}(\seq{\mbf{u}}).
\end{equation*}
Therefore, the active model in \eqref{eq:fault_robmanipulatorstatespace} was guaranteed to be identified by the injection of the separating input \eqref{eq:fault_robmanipulatorseparatinginput}, allowing robust fault diagnosis of the robotic manipulator subject to actuator faults using the proposed active fault diagnosis method.

A second experiment, consisting of fault diagnosis of quadrotor UAV which is subject to actuator and sensor faults, can be found in Chapter \ref{cha:appuav}. In this experiment, the method developed in Section \ref{sec:fault_czmethod} based on constrained zonotopes is also demonstrated.

\section{Final remarks}

This chapter introduced a new approach for set-based active fault diagnosis of a class of nonlinear discrete-time systems with nonlinear dynamics and linear measurement. A simple and effective method for passive fault detection was developed based on the state estimation methods proposed in Chapter \ref{cha:nonlineardynamics}. A modified version of the first-order Taylor extension (Theorem \ref{thm:fault_firstorder}) was proposed to compute affine parametrizations of the reachable state and output sets of the nonlinear discrete-time systems as a function of the system input. By being affine in the system input, this allowed the development of a nonlinear set-based active fault diagnosis method based on the design of optimal separating inputs, by extending the linear open-loop approach in \cite{Scott2014} using zonotopes, and the method in \cite{Raimondo2016} using constrained zonotopes. The resulting problem consisted in solving an MIQP to find the optimal input sequence that allows guaranteed fault diagnosis in a given time horizon.

The performance of the proposed active fault diagnosis method was demonstrated in a numerical experiment with a robotic manipulator subject to actuator faults. The diagnosis of the current possible faults was guaranteed in the experiment by injecting the designed optimal input sequence into the system.

The next chapter introduces new methods based on constrained zonotopes for set-based state estimation and active fault diagnosis of discrete-time linear descriptor systems.

\chapter{Descriptor systems}\thispagestyle{headings} \label{cha:descriptor}

This chapter presents new methods based on constrained zonotopes for set-valued state estimation and active fault diagnosis of linear descriptor systems (LDS). In contrast to the other sets representations mentioned in Chapter \ref{cha:preliminaries}, linear static constraints on the state variables, typical of descriptor systems, can be directly incorporated in the mathematical description of constrained zonotopes. Thanks to this feature, set-based methods based on CZs can provide less conservative enclosures. In addition, this chapter proposes a new representation for unbounded sets based on zonotopes, which allows to develop methods for state estimation and AFD of linear descriptor systems without assuming the knowledge of an admissible set that encloses all the possible trajectories of the system. Therefore, the proposed methods lead to more accurate results in state estimation in comparison to existing methods based on zonotopes, without requiring rank assumptions on the structure of the descriptor system and with a fair trade-off between accuracy and efficiency. The superiority of the proposed approaches with respect to zonotope-based methods is highlighted in numerical examples. The content concerning the new methods based on constrained zonotopes proposed in this chapter was published in \cite{Rego2020b}.

\section{Problem formulation} \label{sec:desc_problemformulation}

\subsection{State estimation}

Consider a linear discrete-time descriptor system with time $k$, state $\mbf{x}_k \in \realset^{n}$, input $\mbf{u}_{k} \in \realset^{n_u}$, process uncertainty $\mbf{w}_k \in \realset^{n_w}$, measured output $\mbf{y}_k \in \realset^{n_y}$, and measurement uncertainty $\mbf{v}_k \in \realset^{n_v}$. In each time interval $[k-1,k]$, $k=1,2,\ldots$, the system evolves according to the model
\normalsize
\begin{equation}
\begin{aligned} \label{eq:desc_system}
\mbf{E} \mbf{x}_k & = \mbf{A} \mbf{x}_{k-1} + \mbf{B} \mbf{u}_{k-1} + \mbf{B}_w \mbf{w}_{k-1}, \\
\mbf{y}_k & = \mbf{C} \mbf{x}_k + \mbf{D} \mbf{u}_{k} + \mbf{D}_v \mbf{v}_{k},
\end{aligned}
\end{equation}
\normalsize
with $\mbf{E} \in \realsetmat{n}{n}$, $\mbf{A} \in \realsetmat{n}{n}$, $\mbf{B} \in \realsetmat{n}{n_u}$, $\mbf{B}_w \in \realsetmat{n}{n_ w}$, $\mbf{C} \in \realsetmat{n_y}{n}$, $\mbf{D} \in \realsetmat{n_y}{n_u}$, and $\mbf{D}_v \in \realsetmat{n_y}{n_v}$. In the following, it is assumed that $\text{rank}(\mbf{E}) \leq n$. Note that when $\mbf{E}$ is singular, one has $n - \text{rank}(\mbf{E})$ purely static constraints. It is assumed that the initial state $\mbf{x}_0 \in X_0$ and $(\mbf{w}_{k},\mbf{v}_{k}) \in W \times V$ for all $k \geq 0$, where $X_0$, $W$ and $V$ are known convex polytopic sets. Moreover, the initial condition $(\mbf{x}_0,\mbf{u}_0,\mbf{w}_0,\mbf{v}_{0})$ is assumed to be feasible, i.e., consistent with the static relations in \eqref{eq:desc_system}, and the output $\mbf{y}_0$ is computed as $\mbf{y}_0 = \mbf{C} \mbf{x}_0 + \mbf{D} \mbf{u}_{0} + \mbf{D}_v \mbf{v}_{0}$. For any $k\geq 0$, the objective of state estimation is to approximate the feasible trajectories of \eqref{eq:desc_system} as accurately as possible by a guaranteed enclosure $\tilde{X}_k$.

\subsection{Fault diagnosis}

Consider a linear discrete-time descriptor system whose dynamics obeys one of possible $n_m$ known models
\begin{equation}
\begin{aligned} \label{eq:desc_systemfaulty}
\mbf{E}^{[i]} \mbf{x}_k^{[i]} & = \mbf{A}^{[i]}  \mbf{x}_{k-1}^{[i]} + \mbf{B}^{[i]}  \mbf{u}_{k-1} + \mbf{B}_w^{[i]}  \mbf{w}_{k-1}, \\
\mbf{y}_k^{[i]} & = \mbf{C}^{[i]}  \mbf{x}_k^{[i]} + \mbf{D}^{[i]} \mbf{u}_{k} + \mbf{D}_v^{[i]} \mbf{v}_{k},
\end{aligned}
\end{equation}
for $k \geq 1$, with $\mbf{E}^{[i]} \in \realsetmat{n}{n}$, $\mbf{A}^{[i]} \in \realsetmat{n}{n}$, $\mbf{B}^{[i]} \in \realsetmat{n}{n_u}$, $\mbf{B}_w^{[i]} \in \realsetmat{n}{n_w}$, $\mbf{C}^{[i]} \in \realsetmat{n_y}{n}$, $\mbf{D}^{[i]} \in \realsetmat{n_y}{n_u}$, and $\mbf{D}_v^{[i]} \in \realsetmat{n_y}{n_v}$, $i \in \modelset \triangleq \{1,2,\ldots,n_m\}$. Also, $\text{rank}(\mbf{E}^{[i]}) \leq n$, and let $\mbf{x}_0^{[i]} \in X_0$, $(\mbf{w}_{k}, \mbf{v}_k) \in W \times V$, and $\mbf{u}_k \in U$, with $X_0$, $W$, $V$ and $U$ being known convex polytopic sets. %

Similar to chapter \ref{cha:faultdiagnosis}, in this chapter the goal of AFD is to find which model describes the process behaviour. The dynamics of the system are assumed to not change during the diagnosis procedure, i.e. the AFD is fast enough to avoid the switching between models. In this sense, a sequence $(\mbf{u}_0, \mbf{u}_1, ..., \mbf{u}_N)$ of minimal length $N$ is designed such that any output $\mbf{y}_N^{[i]}$ is consistent with only one $i \in \modelset$. If feasible, this problem may admit multiple solutions. For this reason, we introduce a cost function and select among the feasible input sequences the optimal one. 

\subsection{Motivational example} \label{sec:desc_motivational}

\begin{figure}[!tb]
	\centering{s
		\def\svgwidth{0.6\columnwidth}
		{\scriptsize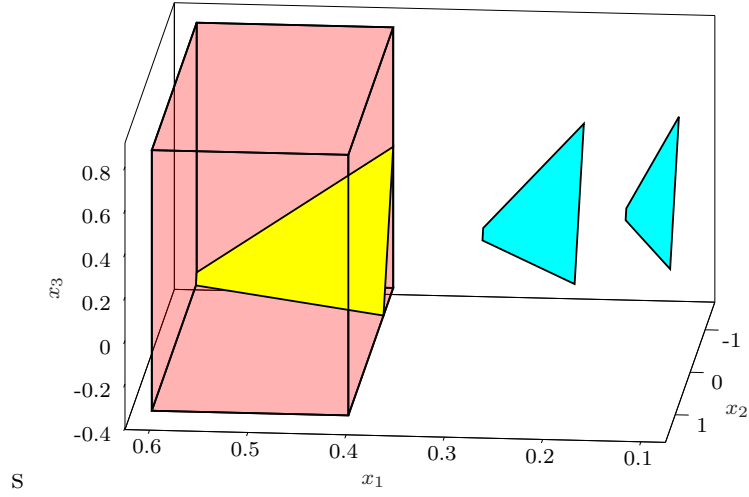}
		\caption{The feasible sets for the system \eqref{eq:desc_systemmotivational}: the initial set $X_0$ (red), the set $S_0$ (yellow), and $S_k$ (cyan) for $k = 1,2$.}\label{fig:desc_motivational}}
\end{figure}

In this subsection, we present a motivational example illustrating that, in the general case, the feasible sets of the linear discrete-time descriptor system \eqref{eq:desc_system} cannot be described exactly by zonotopes even when $X_0$, $W$, and $V$ are all zonotopes. For the sake of simplicity, consider
\normalsize
\begin{equation}
\begin{aligned} \label{eq:desc_systemmotivational}
\mbf{E} \mbf{x}_k & = \mbf{A} \mbf{x}_{k-1}, \\
\end{aligned}
\end{equation}
\normalsize
with 
\normalsize
\begin{equation*}
\begin{aligned}
& \mbf{E} = \begin{bmatrix} 1 & 0 & 0 \\ 0 & 1 & 0 \\ 0 & 0 & 0 \end{bmatrix}, \; \mbf{A} = \begin{bmatrix} 0.5 & 0 & 0 \\ 0.8 & 0.95 & 0 \\ 0.3 & 0.1 & 0.1 \end{bmatrix}.
\end{aligned}
\end{equation*}
\normalsize

Let $\mbf{x}_0 \in X_0$, where $X_0$ is a zonotope given by
\begin{equation} \label{eq:desc_motivationalX0}
X_0 = \left\{\begin{bmatrix} 0.1 & 0 & 0 \\ 0 & 1.5 & 0 \\ 0 & 0 & 0.6 \end{bmatrix}, \begin{bmatrix} 0.5 \\ 0 \\ 0.25 \end{bmatrix} \right\}.
\end{equation}
Since the last row of $\mbf{E}$ is zero, the dynamics \eqref{eq:desc_systemmotivational} further constrains $\mbf{x}_{k-1}$ for $k \geq 1$. In particular, $\mbf{x}_0$ should be consistent with both \eqref{eq:desc_motivationalX0} and the constraints in \eqref{eq:desc_systemmotivational}. Therefore, $\mbf{x}_0 \in S_0$ with $S_0 = \{\mbf{x} \in X_0: \eqref{eq:desc_systemmotivational} \text{ holds for }k = 1\}$. Similarly, we define $S_{k} = \{\mbf{x} \in \realset^n : \eqref{eq:desc_systemmotivational} \text{ holds}, \mbf{E} \mbf{x} = \mbf{A} \mbf{x}', \mbf{x}' \in S_{k-1}\}$, with $k \geq 1$. Figure \ref{fig:desc_motivational} shows the zonotope $X_0$ and the convex polytopes $S_k$ for $k = 0,1,2$. These latter are computed exactly using the MPT toolbox \citep{MPT3}. As it can be noticed, the sets $S_k$ are not symmetric, and therefore cannot be described accurately by zonotopes. Consequently, in the general case the use of zonotopes for reachability analysis, set-based estimation, and fault diagnosis of descriptor systems could result very conservative. The next sections present methods to address these problems using constrained zonotopes, as well as a new proposal for representing unbounded sets.

\section{Set-based state estimation of LDS using constrained zonotopes} \label{sec:desc_estimationCZ}
\sectionmark{State estimation of LDS using constrained zonotopes}

This section presents a new method for set-based state estimation of system \eqref{eq:desc_system}. %
Consider an initial condition $X_0$ and an input $\mbf{u}_k$ with $k \geq 0$, let
\begin{equation} \label{eq:desc_initialset}
\hat{X}_0 = \{ \mbf{x} \in X_0 : \mbf{C} \mbf{x} + \mbf{D} \mbf{u}_{0} + \mbf{D}_v \mbf{v} = \mbf{y}_0, \, \mbf{v} \in V \}.
\end{equation}
The linear prediction step \eqref{eq:pre_linearprediction0} must be reformulated when $\mbf{E}$ is not invertible. We first consider the following assumption.
\begin{assumption} \rm \label{ass:desc_admissible}
	There exists a known CZ $X_\text{a} = \{\mbf{G}_\text{a}, \mbf{c}_\text{a}$, $\mbf{A}_\text{a}$, $\mbf{b}_\text{a}\}  \subset \realset^n$ such that $\mbf{x}_k \in X_\text{a}$ for all $k \geq 0$.
\end{assumption}

\begin{remark} \rm Even though $X_\text{a}$ can be arbitrarily large, Assumption \ref{ass:desc_admissible} is restrictive in the sense that the descriptor system \eqref{eq:desc_system} must be stable. Unstable systems are out of the scope of this section and are considered later in Section \ref{sec:desc_estimationMZ}.
\end{remark}

The set estimation in Section \ref{sec:pre_linearestimation} relied on \eqref{eq:pre_czlimage}--\eqref{eq:pre_czintersection} to compute the steps \eqref{eq:pre_linearprediction0}--\eqref{eq:pre_linearupdate0}. In order to extend this method to the case of descriptor systems, when computing $\bar{X}_k$ it is necessary to take into account the static constraints arising from the possible singularity of $\mbf{E}$. The proposed method is based on singular value decomposition (SVD). Let $\mbf{E} = \mbf{U} \mbf{\Sigma} \mbf{V}^T$, where $\mbf{U}$ and $\mbf{V}$ are invertible by construction. Since $\mbf{E}$ is square, then $\mbf{\Sigma}$ is also square. Without loss of generality, let $\mbf{\Sigma}$ be arranged as $\mbf{\Sigma} = \text{blkdiag}(\tilde{\mbf{\Sigma}}, \mbf{0})$,
where $\tilde{\mbf{\Sigma}} \in \realsetmat{n_z}{n_z}$ is diagonal with all the $n_z = \text{rank}(\mbf{E})$ nonzero singular values of $\mbf{E}$. Moreover, let $\mbf{z}_k = (\tilde{\mbf{z}}_k,\check{\mbf{z}}_k) = \inv{\mbf{T}} \mbf{x}_k, \; \tilde{\mbf{z}}_k \in \realset^{n_z}, \check{\mbf{z}}_k \in \realset^{n-n_z}$,
\begin{equation}\label{eq:desc_SVDmatrices} 
\begin{aligned}
\begin{bmatrix} \tilde{\mbf{A}} \\ \check{\mbf{A}} \end{bmatrix} & = \begin{bmatrix} \tilde{\mbf{\Sigma}}^{-1} & \mbf{0} \\ \mbf{0} & \eyenoarg \end{bmatrix} \inv{\mbf{U}} \mbf{A} \mbf{T}, \\
\begin{bmatrix} \tilde{\mbf{B}} \\ \check{\mbf{B}} \end{bmatrix} & = \begin{bmatrix} \tilde{\mbf{\Sigma}}^{-1} & \mbf{0} \\ \mbf{0} & \eyenoarg \end{bmatrix} \inv{\mbf{U}} \mbf{B}, \begin{bmatrix} \tilde{\mbf{B}}_w \\ \check{\mbf{B}}_w \end{bmatrix} = \begin{bmatrix} \tilde{\mbf{\Sigma}}^{-1} & \mbf{0} \\ \mbf{0} & \eyenoarg \end{bmatrix} \inv{\mbf{U}} \mbf{B}_w,  \\
\end{aligned}
\end{equation}
with $\mbf{T} = \inv{(\mbf{V}^T)}$, $\tilde{\mbf{A}} \in \realsetmat{n_z}{n}$, $\tilde{\mbf{B}} \in \realsetmat{n_z}{n_u}$, and $\tilde{\mbf{B}}_w \in \realsetmat{n_z}{n_w}$. Then, system \eqref{eq:desc_system} can be rewritten with decoupled dynamics given by \citep{Jonckheere1988}
\begin{subequations} \label{eq:desc_systemSVD}
	\begin{align} 
	\tilde{\mbf{z}}_k & = \tilde{\mbf{A}} \mbf{z}_{k-1} + \tilde{\mbf{B}} \mbf{u}_{k-1} + \tilde{\mbf{B}}_w \mbf{w}_{k-1},  \label{eq:desc_systemSVDdynamics} \\
	\mbf{0} & = \check{\mbf{A}} \mbf{z}_{k-1} + \check{\mbf{B}} \mbf{u}_{k-1} + \check{\mbf{B}}_w \mbf{w}_{k-1}, \label{eq:desc_systemSVDconstraints} \\
	\mbf{y}_k & = \mbf{C} \mbf{T} \mbf{z}_k + \mbf{D} \mbf{u}_{k} + \mbf{D}_v \mbf{v}_{k}. \label{eq:desc_systemSVDoutput}
	\end{align}
\end{subequations}
Consider $W = \{\mbf{G}_w, \mbf{c}_w, \mbf{A}_w, \mbf{b}_w\}$, and $\hat{X}_{0}$ given by \eqref{eq:desc_initialset}. From \eqref{eq:desc_systemSVDconstraints}, the state $\mbf{z}_0$ must satisfy $\check{\mbf{A}} \mbf{z}_0 + \check{\mbf{B}} \mbf{u}_0 + \check{\mbf{B}}_w \mbf{w}_0 = \mbf{0}$. This constraint can be incorporated in the CG-rep of the initial set $\hat{Z}_0$ as follows. Let $\inv{\mbf{T}} \hat{X}_{0} \triangleq \{\mbf{G}_0,\mbf{c}_0,\mbf{A}_0,\mbf{b}_0\}$. Then, with a slight abuse of notation\footnote{This is a abuse of notation since $\hat{Z}_0 \neq \inv{\mbf{T}} \hat{X}_{0}$. The constraints $\check{\mbf{A}} \mbf{z}_0 + \check{\mbf{B}} \mbf{u}_0 + \check{\mbf{B}}_w \mbf{w}_0 = \mbf{0}$ are incorporated in $\hat{Z}_0$, while not in $\inv{\mbf{T}} \hat{X}_0$.}, define $\hat{Z}_0 \triangleq \{\hat{\mbf{G}}_0, \hat{\mbf{c}}_0,\hat{\mbf{A}}_0,\hat{\mbf{b}}_0\}$, with $\hat{\mbf{G}}_0 = [\mbf{G}_0 \,\; \mbf{0}]$, $\hat{\mbf{c}}_0 = \mbf{c}_0$,
\begin{equation*}
\begin{aligned}
\hat{\mbf{A}}_0 = \begin{bmatrix} \mbf{A}_0 & \mbf{0} \\ \check{\mbf{A}} \mbf{G}_0 & \check{\mbf{B}}_w \mbf{G}_w \end{bmatrix}, \; \hat{\mbf{b}}_0 = \begin{bmatrix}  \mbf{b}_0 \\ -\check{\mbf{A}} \mbf{c}_0 - \check{\mbf{B}}_w \mbf{c}_w - \check{\mbf{B}} \mbf{u}_{0} \end{bmatrix}.
\end{aligned}
\end{equation*}
Note that the extra columns in $\hat{\mbf{G}}_0$ and $\hat{\mbf{A}}_0$ come from $\mbf{w}_0 \in W$. Having defined $\hat{Z}_0$, for state estimation purposes, the static relation \eqref{eq:desc_systemSVDconstraints} can be shifted forward to time $k$ without loss of information. By doing so, from \eqref{eq:desc_systemSVD}, the variables $\tilde{\mbf{z}}_k$ are fully determined by \eqref{eq:desc_systemSVDdynamics}, while $\check{\mbf{z}}_k$ are obtained a posteriori by $\check{\mbf{A}} \mbf{z}_{k} + \check{\mbf{B}} \mbf{u}_{k} + \check{\mbf{B}}_w \mbf{w}_{k} = \mbf{0}$. 

Consider the set $Z_\text{a} = \inv{\mbf{T}} X_\text{a} = \{\inv{\mbf{T}}\mbf{G}_\text{a},\inv{\mbf{T}}\mbf{c}_\text{a},\mbf{A}_\text{a},\mbf{b}_\text{a}\}$, and let $\inv{\mbf{T}} \mbf{c}_\text{a} = [ \tilde{\mbf{c}}_\text{a}^T \,\; \check{\mbf{c}}_\text{a}^T ]^T$, $\inv{\mbf{T}} \mbf{G}_\text{a} = [ \tilde{\mbf{G}}_\text{a}^T \,\; \check{\mbf{G}}_\text{a}^T ]^T$. An effective enclosure of the prediction step for the descriptor system \eqref{eq:desc_systemSVD} can be obtained in CG-rep as follows. 

\begin{lemma} \rm \label{lem:desc_predictioncz}
	For $k \geq 1$, consider $\mbf{z}_{k-1} \in \hat{Z}_{k-1} = \{\hat{\mbf{G}}_{k-1}, \hat{\mbf{c}}_{k-1}$, $\hat{\mbf{A}}_{k-1}, \hat{\mbf{b}}_{k-1}\}$, $\mbf{w}_{k-1}, \mbf{w}_k \in W = \{\mbf{G}_w, \mbf{c}_w, \mbf{A}_w, \mbf{b}_w\}$, and system \eqref{eq:desc_systemSVD}. If Assumption \ref{ass:desc_admissible} holds, then $\mbf{z}_k \in \bar{Z}_k = \{\bar{\mbf{G}}_k, \bar{\mbf{c}}_k, \bar{\mbf{A}}_k, \bar{\mbf{b}}_k\}$, with
	\begin{align*}
	\bar{\mbf{G}}_k & =\begin{bmatrix} \tilde{\mbf{A}} \hat{\mbf{G}}_{k-1} & \tilde{\mbf{B}}_w \mbf{G}_w & \mbf{0} & \mbf{0}\\
	\mbf{0} & \mbf{0} & \check{\mbf{G}}_\text{a} & \mbf{0}\end{bmatrix},
	\bar{\mbf{c}}_k = \begin{bmatrix} \tilde{\mbf{A}} \hat{\mbf{c}}_{k-1} + \tilde{\mbf{B}} \mbf{u}_{k-1} + \tilde{\mbf{B}}_w \mbf{c}_w \\ \check{\mbf{c}}_\text{a}\end{bmatrix}, \\  
	\bar{\mbf{A}}_k & = \begin{bmatrix} \multicolumn{4}{c}{\text{blkdiag}(\hat{\mbf{A}}_{k-1}, \mbf{A}_w, \mbf{A}_\text{a}, \mbf{A}_w)} \\
	\check{\mbf{A}} \begin{bmatrix} \tilde{\mbf{A}} \hat{\mbf{G}}_{k-1}  \\ \mbf{0} \end{bmatrix} & \check{\mbf{A}} \begin{bmatrix} \tilde{\mbf{B}}_w \mbf{G}_w \\ \mbf{0} \end{bmatrix} & \check{\mbf{A}} \begin{bmatrix} \mbf{0} \\ \check{\mbf{G}}_\text{a} \end{bmatrix} & \check{\mbf{B}}_w \mbf{G}_w \end{bmatrix}, \\
	\bar{\mbf{b}}_k & = \begin{bmatrix} [\hat{\mbf{b}}_{k-1}^T \,\; \mbf{b}_w^T \,\; \mbf{b}_\text{a}^T \,\; \mbf{b}_w^T]^T \\ -\check{\mbf{A}} \begin{bmatrix} \tilde{\mbf{A}} \hat{\mbf{c}}_{k-1} + \tilde{\mbf{B}} \mbf{u}_{k-1} + \tilde{\mbf{B}}_w \mbf{c}_w \\ \check{\mbf{c}}_\text{a}\end{bmatrix} - \check{\mbf{B}} \mbf{u}_k - \check{\mbf{B}}_w \mbf{c}_w \end{bmatrix}.                            
	\end{align*}
\end{lemma}
\proof
Since by assumption $(\mbf{z}_{k-1},\mbf{w}_{k-1},\mbf{w}_k) \in \hat{Z}_{k-1} \times W \times W$, there exists $(\bm{\xi}_{k-1}, \bm{\delta}_{k-1}, \bm{\delta}_{k}) \in B_\infty(\hat{\mbf{A}}_{k-1}, \hat{\mbf{b}}_{k-1}) \times B_\infty(\mbf{A}_{w}, \mbf{b}_{w}) \times B_\infty(\mbf{A}_w, \mbf{b}_w)$ such that $\mbf{z}_{k-1} = \hat{\mbf{c}}_{k-1} + \hat{\mbf{G}}_{k-1} \bm{\xi}_{k-1}$, $\mbf{w}_{k-1} = \mbf{c}_w + \mbf{G}_w \bm{\delta}_{k-1}$, and $\mbf{w}_{k} = \mbf{c}_w + \mbf{G}_w \bm{\delta}_{k}$. Besides, Assumption \ref{ass:desc_admissible} implies that $\mbf{z}_k \in Z_\text{a}$. Thus, there must exist $\bm{\xi}_\text{a} \in B_\infty (\mbf{A}_\text{a}, \mbf{b}_\text{a})$ such that $\check{\mbf{z}}_{k} = \check{\mbf{c}}_\text{a} + \check{\mbf{G}}_\text{a}\bm{\xi}_\text{a}$. Therefore, substituting these equalities in \eqref{eq:desc_systemSVDdynamics} leads to
\begin{equation} \label{eq:desc_lema1proof1} 
\begin{aligned}
(\tilde{\mbf{z}}_{k}, \check{\mbf{z}}_{k}) = (\tilde{\mbf{A}} \hat{\mbf{c}}_{k-1} + \tilde{\mbf{B}} \mbf{u}_{k-1} + \tilde{\mbf{B}}_w \mbf{c}_{w} + \tilde{\mbf{A}} \hat{\mbf{G}}_{k-1} \bm{\xi}_{k-1} + \tilde{\mbf{B}}_w \mbf{G}_{w} \bm{\delta}_{k-1}, \check{\mbf{c}}_\text{a} + \check{\mbf{G}}_\text{a} \bm{\xi}_\text{a}).
\end{aligned}
\end{equation}
From the constraint \eqref{eq:desc_systemSVDconstraints} shifted to time $k$, we have that
\begin{equation} \label{eq:desc_lema1proof2}
\begin{aligned}
& \check{\mbf{B}}_w \mbf{c}_w +  \check{\mbf{B}}_w \mbf{G}_w \bm{\delta}_k + \check{\mbf{A}} \begin{bmatrix} \tilde{\mbf{A}} \hat{\mbf{c}}_{k-1} + \tilde{\mbf{B}} \mbf{u}_{k-1} + \tilde{\mbf{B}}_w \mbf{c}_{w}  \\  \check{\mbf{c}}_\text{a} \end{bmatrix} \\ & + \check{\mbf{A}} \begin{bmatrix} \tilde{\mbf{A}} \hat{\mbf{G}}_{k-1} & \tilde{\mbf{B}}_w \mbf{G}_{w} & \mbf{0} \\ \mbf{0} & \mbf{0} & \check{\mbf{G}}_\text{a} \end{bmatrix} \begin{bmatrix} \bm{\xi}_{k-1} \\ \bm{\delta}_{k-1} \\ \bm{\xi}_\text{a}\end{bmatrix} + \check{\mbf{B}} \mbf{u}_{k} = \mbf{0}.
\end{aligned}		
\end{equation}
Rearranging \eqref{eq:desc_lema1proof1} and \eqref{eq:desc_lema1proof2}, grouping $(\bm{\xi}_{k-1}, \bm{\delta}_{k-1}, \bm{\xi}_\text{a},$ $ \bm{\delta}_{k})$, and writing in the CG-rep \eqref{eq:pre_cgrep} proves the lemma. \qed

Lemma \ref{lem:desc_predictioncz} provides a predicted enclosure of the state $\mbf{z}_k$ in which the equality constraints \eqref{eq:desc_systemSVDconstraints}, shifted to time $k$, are directly taken into account. This is possible thanks to the fact that CZs incorporate equality constraints (see \eqref{eq:pre_cgrep}). Finally, the prediction-update algorithm proposed for descriptor systems consists in the computation of CZs $\bar{Z}_k$, $\hat{Z}_k$, and $\hat{X}_k$, such that
\begin{align}
\bar{Z}_k & = \{\bar{\mbf{G}}_k, \bar{\mbf{c}}_k, \bar{\mbf{A}}_k, \bar{\mbf{b}}_k\}, \label{eq:desc_predictionSVDcz} \\
\hat{Z}_k & = \bar{Z}_k \cap_{\mbf{C}\mbf{T}} ((\mbf{y}_k - \mbf{D}_u \mbf{u}_k) \oplus (-\mbf{D}_v V_k)), \label{eq:desc_updateSVDcz}  \\
\hat{X}_k & = \mbf{T} \hat{Z}_k. \label{eq:desc_finalSVDcz}
\end{align}
For this algorithm, the initial set is $\hat{Z}_0$. The algorithm \eqref{eq:desc_predictionSVDcz}--\eqref{eq:desc_finalSVDcz} operates recursively with $\bar{Z}_k$ and $\hat{Z}_k$ for $k \geq 1$ in the transformed state-space \eqref{eq:desc_systemSVD}, while the estimated enclosure in the original state-space \eqref{eq:desc_system} is given by $\hat{X}_k$.

\begin{remark} \rm \label{rem:desc_admissible}
	The set $Z_\text{a}$ is used only to predict an enclosure for the components $\check{\mbf{z}}_k$. This way, the static relation \eqref{eq:desc_systemSVDconstraints} is incorporated as constraints to the variables $\bm{\xi}_\text{a}$ in $Z_\text{a}$.%
\end{remark}

\begin{remark} \rm \label{rem:desc_convervativeness}
	By construction, the CG-rep \eqref{eq:desc_predictionSVDcz} corresponds to the exact feasible state set of \eqref{eq:desc_systemSVD} at $k$ for the known state and uncertainty bounds. In addition, \eqref{eq:desc_updateSVDcz}--\eqref{eq:desc_finalSVDcz} can be computed exactly. However, in practice, in order to limit the complexity of the resulting sets these are outer approximated by using order reduction algorithms. In this case, equalities \eqref{eq:desc_predictionSVDcz}-\eqref{eq:desc_finalSVDcz} are replaced by the relation $\supset$. 
\end{remark}

\section{Active fault diagnosis of LDS using constrained zonotopes}  \label{sec:desc_AFDCZ}
\sectionmark{Fault diagnosis of LDS using constrained zonotopes}

The previous section presented a method to address the problem of set-based estimation of descriptor systems using constrained zonotopes. In the following, this tool is used in the design of a CZ-based AFD method accounting for a finite number of possible abrupt faults.

\subsection{State and output reachable sets} \label{sec:desc_AFDCZ_reachable}

Let $\seq{\mbf{u}} = (\mbf{u}_0, ..., \mbf{u}_N)\in \mathbb{R}^{(N+1)n_u}$, $\seq{\mbf{w}} = (\mbf{w}_0, \ldots$, $\mbf{w}_N)\in \mathbb{R}^{(N+1)n_w}$, and $\seq{W} = W \times \ldots \times W$. Consider a variable transformation similar to the one used in the previous section. With a slight abuse of notation, let $\mbf{z}_k = (\tilde{\mbf{z}}_k,\check{\mbf{z}}_k) = (\inv{\mbf{T}} \mbf{x}_k, \mbf{w}_k),$ $\tilde{\mbf{z}}_k \in \realset^{n_z}$, $\check{\mbf{z}}_k \in \realset^{n+n_w-n_z}$,
with $\mbf{T}^{[i]} = \inv{((\mbf{V}^{[i]})^T)}$, $\mbf{V}^{[i]}$ being obtained from the SVD $\mbf{E}^{[i]} = \mbf{U}^{[i]} \mbf{\Sigma}^{[i]} (\mbf{V}^{[i]})^T$. Then, \eqref{eq:desc_systemfaulty} can be rewritten as
\begin{subequations} \label{eq:desc_systemSVDfault}
	\begin{align} 
	\tilde{\mbf{z}}_k^{[i]} & = \tilde{\mbf{A}}_z^{[i]} \mbf{z}_{k-1}^{[i]} + \tilde{\mbf{B}}^{[i]} \mbf{u}_{k-1}, \label{eq:systemSVDfaultdynamics} \\
	\mbf{0} & = \check{\mbf{A}}_z^{[i]} \mbf{z}_{k}^{[i]} + \check{\mbf{B}}^{[i]} \mbf{u}_{k}, \label{eq:desc_systemSVDfaultconstraints} \\
	\mbf{y}_k^{[i]} & = \mbf{F}^{[i]} \mbf{z}_k^{[i]} + \mbf{D}^{[i]} \mbf{u}_{k} + \mbf{D}_v^{[i]} \mbf{v}_{k}, \label{eq:systemSVDfaultoutput}
	\end{align}
\end{subequations}
with $\mbf{F}^{[i]} = \mbf{C}^{[i]} \mbf{T}^{[i]} \mbf{P}$, where $\mbf{P} = [ \eye{n} \,\; \zeros{n}{n_w}]$, $\tilde{\mbf{A}}_z^{[i]} = [\tilde{\mbf{A}}^{[i]} \,\; \tilde{\mbf{B}}_w^{[i]}]$, and $\check{\mbf{A}}_z^{[i]} = [\check{\mbf{A}}^{[i]} \,\; \check{\mbf{B}}_w^{[i]}]$. Note that the $\tilde{(\cdot)}$ and $\check{(\cdot)}$ variables are defined according to \eqref{eq:desc_SVDmatrices} for each $i$, and equation \eqref{eq:desc_systemSVDfaultconstraints} has been already shifted to time $k$. 

For each model $i$, consider the CZ $Z_\text{a}^{[i]} = \inv{(\mbf{T}^{[i]})} X_\text{a} \times W = \{ \mbf{G}_\text{a}^{[i]}, \mbf{c}_\text{a}^{[i]}, \mbf{A}_\text{a}^{[i]}, \mbf{b}_\text{a}^{[i]}\}$, where $X_\text{a}$ satisfies Assumption 1, the set $ \{\mbf{G}_z^{[i]},\mbf{c}_z^{[i]},\mbf{A}_z^{[i]},\mbf{b}_z^{[i]}\} = \inv{(\mbf{T}^{[i]})} X_0 \times W $, and define the initial feasible set $Z_0^{[i]} (\mbf{u}_0) = \{\mbf{z} \in \inv{(\mbf{T}^{[i]})} X_0 \times W : \eqref{eq:desc_systemSVDfaultconstraints} \text{ holds for }k=0\}$. This set is given by $Z_0^{[i]}(\mbf{u}_0) = \{\mbf{G}_0^{[i]},\mbf{c}_0^{[i]},\mbf{A}_0^{[i]},$ $\mbf{b}_0^{[i]}(\mbf{u}_0)\}$, where $\mbf{G}_0^{[i]} = \mbf{G}_z^{[i]}$, $\mbf{c}_0^{[i]} = \mbf{c}_z^{[i]}$,
\begin{equation} \label{eq:initialAb}
\mbf{A}_0^{[i]} = \begin{bmatrix} \mbf{A}_z^{[i]} \\ \check{\mbf{A}}_z^{[i]} \mbf{G}_{0}^{[i]} \end{bmatrix}, \; \mbf{b}_0^{[i]}(\mbf{u}_0) = \begin{bmatrix} \mbf{b}_z^{[i]} \\ -\check{\mbf{A}}_z^{[i]} \mbf{c}_0^{[i]} - \check{\mbf{B}}^{[i]} \mbf{u}_{0} \end{bmatrix}.
\end{equation}
In addition, define the solution mappings $(\bm{\phi}_k^{[i]},\bm{\psi}_k^{[i]}) : \realset^{(k+1)n_u} \times \realset^n \times \realset^{(k+1)n_w} \times \realset^{n_v} \to \realset^{n+n_w} \times \realset^{n_y}$ such that $\bm{\phi}_k^{[i]}(\seq{\mbf{u}},\mbf{z}_0, \seq{\mbf{w}})$ and $\bm{\psi}_k^{[i]}(\seq{\mbf{u}},\mbf{z}_0, \seq{\mbf{w}}, \mbf{v}_k)$ are the state and output of \eqref{eq:desc_systemSVDfault} at $k$, respectively. Then, for each $i \in \modelset$, define \emph{state and output reachable sets} at time $k$ as
\begin{equation*}
\begin{aligned}
Z_k^{[i]}(\seq{\mbf{u}}) = & \{ \bm{\phi}_k^{[i]}(\seq{\mbf{u}},\mbf{z}_0, \seq{\mbf{w}}) : (\mbf{z}_0^{[i]}, \seq{\mbf{w}}) \in Z_0^{[i]}(\mbf{u}_0) \times \seq{W} \},\\
Y_k^{[i]}(\seq{\mbf{u}}) = & \{ \bm{\psi}_k^{[i]}(\seq{\mbf{u}},\mbf{z}_0, \seq{\mbf{w}}, \mbf{v}_k) : (\mbf{z}_0,\seq{\mbf{w}},\mbf{v}_k) \in Z_0^{[i]}(\mbf{u}_0) \times \seq{W} \times V\}.
\end{aligned}
\end{equation*}
Using \eqref{eq:pre_czlimage}--\eqref{eq:pre_czmsum}, and taking note that by assumption $\mbf{z}_k^{[i]} \in Z_\text{a}^{[i]} \subset \realset^n \times W$ for every $k \geq 0$, the set $Z_N^{[i]}(\seq{\mbf{u}})$ is given by the CZ $\{ \mbf{G}_N^{[i]}, \mbf{c}_N^{[i]}(\seq{\mbf{u}}), \mbf{A}_N^{[i]}, \mbf{b}_N^{[i]}(\seq{\mbf{u}})\},$ 
where $\mbf{G}_N^{[i]}$, $\mbf{c}_N^{[i]}(\seq{\mbf{u}})$, $\mbf{A}_N^{[i]}$, and $\mbf{b}_N^{[i]}(\seq{\mbf{u}})$ are obtained by the recursive relations
\normalsize
\begin{align*}
& \mbf{c}_k^{[i]}(\seq{\mbf{u}}) = \!\begin{bmatrix} \tilde{\mbf{A}}_z^{[i]} \mbf{c}_{k-1}^{[i]}(\seq{\mbf{u}}) + \tilde{\mbf{B}}^{[i]} \mbf{u}_{k-1} \\ \check{\mbf{c}}_\text{a}^{[i]} \end{bmatrix}\!, \mbf{G}_k^{[i]} = \begin{bmatrix} \tilde{\mbf{A}}_z^{[i]} \mbf{G}_{k-1}^{[i]}\!\! & \mbf{0} \\ \mbf{0} & \check{\mbf{G}}_\text{a}^{[i]}\end{bmatrix}\!, \\
& \mbf{A}_k^{[i]} = \begin{bmatrix} \mbf{A}_{k-1}^{[i]} & \mbf{0} \\ \mbf{0} & \mbf{A}_\text{a}^{[i]} \\ \multicolumn{2}{c}{\check{\mbf{A}}_z^{[i]} \mbf{G}_{k}^{[i]}} \end{bmatrix}, \mbf{b}_k^{[i]}(\seq{\mbf{u}}) = \begin{bmatrix} \mbf{b}_{k-1}^{[i]}(\seq{\mbf{u}}) \\  \mbf{b}_\text{a}^{[i]} \\ -\check{\mbf{A}}_z^{[i]} \mbf{c}_k^{[i]}(\seq{\mbf{u}}) - \check{\mbf{B}}^{[i]} \mbf{u}_{k} \end{bmatrix},  
\end{align*}
\normalsize
for $k = 1,2,\ldots,N$.
Note that the third constraint in $(\mbf{A}_k^{[i]}$, $\mbf{b}_k^{[i]}(\seq{\mbf{u}}))$ comes from the fact that \eqref{eq:desc_systemSVDfaultconstraints} must hold. Using the initial values \eqref{eq:initialAb}, the variables $\mbf{c}_N^{[i]}(\seq{\mbf{u}})$ and $\mbf{b}_N^{[i]}(\seq{\mbf{u}})$ can be written as explicit functions of the input sequence $\seq{\mbf{u}}$ as
\begin{align}
\mbf{c}_N^{[i]}(\seq{\mbf{u}}) & = \begin{bmatrix} \tilde{\mbf{A}}_z^{[i]} \\ \mbf{0} \end{bmatrix}^{N} \mbf{c}_z^{[i]} + \sum_{m=1}^{N} \left( \begin{bmatrix} \tilde{\mbf{A}}_z^{[i]} \\ \mbf{0} \end{bmatrix}^{m-1} \begin{bmatrix} \mbf{0} \\ \check{\mbf{c}}_\text{a}^{[i]} \end{bmatrix} \right) + \mbf{H}_N^{[i]} \seq{\mbf{u}} \label{eq:desc_stateckcz},\\
\mbf{b}_N^{[i]}(\seq{\mbf{u}}) & = \bm{\alpha}_N^{[i]} + \bm{\Lambda}_N^{[i]} \mbf{c}_z^{[i]} + \bm{\Omega}_N^{[i]} \seq{\mbf{u}}, \label{eq:statebk} 
\end{align}
where $\bm{\alpha}_N^{[i]} = \bm{\beta}_N^{[i]} + \bm{\Upsilon}_N^{[i]} \mbf{p}_N^{[i]}$, $\bm{\Lambda}_N^{[i]} = \bm{\Upsilon}_N^{[i]} \mbf{Q}_N^{[i]}$, $\bm{\Omega}_N^{[i]} = \bm{\Gamma}_N^{[i]}+ \bm{\Upsilon}_N^{[i]} \seq{\mbf{H}}\zerospace^{[i]}$, %
\begin{align*}
& \bm{\beta}_N^{[i]} = \begin{bmatrix} [(\mbf{b}_z^{[i]})^T \; \mbf{0}] & [(\mbf{b}_\text{a}^{[i]})^T \; \mbf{0}] & \cdots & [ (\mbf{b}_\text{a}^{[i]})^T \; \mbf{0} ] \end{bmatrix}^T, \\
& \mbf{p}_N^{[i]} = \left[ \begin{matrix} \begin{bmatrix} \mbf{0} \\ \mbf{0} \end{bmatrix}^T & \begin{bmatrix} \mbf{0} \\ \check{\mbf{c}}_\text{a}^{[i]} \end{bmatrix}^T & \cdots & \sum_{m=1}^{N} \left( \begin{bmatrix} \tilde{\mbf{A}}_z^{[i]} \\ \mbf{0} \end{bmatrix}^{m-1} \begin{bmatrix} \mbf{0} \\ \check{\mbf{c}}_\text{a}^{[i]} \end{bmatrix} \right)^T \end{matrix} \right]^T, \\
& \bm{\Upsilon}_N^{[i]} = \text{blkdiag} \big( \big[\mbf{0} \,\; (-\check{\mbf{A}}_z^{[i]})^T\big]^T , \ldots, \big[\mbf{0} \,\; (-\check{\mbf{A}}_z^{[i]})^T\big]^T \big), \\
& \bm{\Gamma}_N^{[i]} = \text{blkdiag} \big(\big[\mbf{0} \,\; (-\check{\mbf{B}}^{[i]})^T\big]^T, \ldots, \big[\mbf{0} \,\; (-\check{\mbf{B}}^{[i]})^T\big]^T \big), \\
& \mbf{Q}_N^{[i]} = \left[\begin{matrix} \cdots & \left(\begin{bmatrix} \tilde{\mbf{A}}_z^{[i]} \\ \mbf{0} \end{bmatrix}^\ell\right)^T & \cdots \end{matrix}\right]^T\!\!, \; \seq{\mbf{H}}\zerospace^{[i]} = \big[\begin{matrix} \cdots & (\mbf{H}_\ell^{[i]})^T & \cdots \end{matrix}\big]^T\!\!, \\
& \mbf{H}_h^{[i]} = \left[\begin{matrix} \cdots & \underbrace{\begin{bmatrix} \tilde{\mbf{A}}_z^{[i]} \\ \mbf{0} \end{bmatrix}^{h-m} \begin{bmatrix} \tilde{\mbf{B}}^{[i]} \\ \mbf{0} \end{bmatrix}}_{m = 1,2,\ldots,h} & \cdots & \underbrace{\begin{bmatrix} \mbf{0} \\ \mbf{0} \end{bmatrix}}_{N-h+1 \text{ terms}} & \cdots \end{matrix} \right], 
\end{align*}
with $\ell = 0,\ldots,N$. The variables $\bm{\beta}_N$, $\mbf{p}_N$, $\bm{\Upsilon}_N$, and $\bm{\Gamma}_N$ have $N+1$ block matrices. In addition, the expression $\mbf{H}_h^{[i]}$ holds for $h = 1,\ldots,N$, while $\mbf{H}_0^{[i]} = \zeros{n}{(N{+}1)n_u}$. 

Since $Z_N^{[i]}(\seq{\mbf{u}})$ is a CZ, the output reachable set $Y_N^{[i]}(\seq{\mbf{u}})$ is then a CZ obtained in accordance with \eqref{eq:systemSVDfaultoutput} as $Y_N^{[i]}(\seq{\mbf{u}}) = \mbf{F}^{[i]}Z_N^{[i]} (\seq{\mbf{u}}) \oplus \mbf{D}^{[i]}\mbf{u}_N \oplus \mbf{D}_v^{[i]} V.$ Using properties \eqref{eq:pre_czlimage} and \eqref{eq:pre_czmsum}, and letting $V = \{ \mbf{G}_v, \mbf{c}_v, \mbf{A}_v,$ $ \mbf{b}_v\}$, this set is $Y_N^{[i]}(\seq{\mbf{u}}) = \{ \mbf{G}_N^{Y[i]}, \mbf{c}_N^{Y[i]}(\seq{\mbf{u}})$, $\mbf{A}_N^{Y[i]},$ $ \mbf{b}_N^{Y[i]}(\seq{\mbf{u}})\}$, with
\begin{subequations} \label{eq:outputreachable}
	\begin{align}
	& \mbf{c}_N^{Y[i]}(\seq{\mbf{u}}) = \mbf{F}^{[i]} \mbf{c}_N^{[i]}(\seq{\mbf{u}}) + \mbf{D}^{[i]}\mbf{u}_N + \mbf{D}_v^{[i]} \mbf{c}_v, \label{eq:desc_outputckcz}\\
	& \mbf{G}_N^{Y[i]} = [ \mbf{F}^{[i]} \mbf{G}_N^{[i]} \,\; \mbf{G}_v ],\; \mbf{A}_N^{Y[i]} = \text{blkdiag}(\mbf{A}_N^{[i]},\mbf{A}_v), \label{eq:outputGkAk}\\
	& \mbf{b}_N^{Y[i]}(\seq{\mbf{u}}) = [(\mbf{b}_N^{[i]}(\seq{\mbf{u}}))^T \,\; \mbf{b}_v^T ]^T \label{eq:desc_outputbkcz}. 
	\end{align}
\end{subequations}

\subsection{Input design}

Consider an input sequence $\seq{\mbf{u}} \in \seq{U}$ to be injected into the set of models \eqref{eq:desc_systemSVDfault}, and let $\mbf{y}^{[i]}_N$ denote the observed output. We are interested in the design of an input sequence such that the relation $\mbf{y}_N^{[i]} \in Y_N^{[i]}(\seq{\mbf{u}})$ is valid for only one $i \in \modelset$.
\begin{definition} \rm \label{def:separatinginput}
	An input sequence $\seq{\mbf{u}}$ is said to be a \emph{separating input} on $k \in [0, N]$ if, for every $i,j \in \modelset$, $i \neq j$,
	\begin{equation} \label{eq:separatinginputdef}
	Y_N^{[i]}(\seq{\mbf{u}}) \cap Y_N^{[j]}(\seq{\mbf{u}}) = \emptyset.
	\end{equation}
\end{definition}
Clearly, if \eqref{eq:separatinginputdef} holds for all $i,j \in \modelset$, $i \neq j$, then $\mbf{y}_N^{[i]} \in Y_N^{[i]}(\seq{\mbf{u}})$ must hold only for one $i$. In the case that this is not valid for any $i \in \modelset$, one concludes that the real dynamics does not belong to the set of models \eqref{eq:desc_systemSVDfault}. The following theorem is based on the computation of $Y_N^{[i]}(\seq{\mbf{u}})$ expressed by \eqref{eq:outputreachable} and the results in \cite{Raimondo2016}.

\begin{theorem} \rm \label{thm:separatinginput}
	An input $\seq{\mbf{u}} \in \seq{U}$ is a separating input iff
	\begin{equation} \label{eq:desc_separationconditionczlast}
	\begin{bmatrix} \mbf{N}(i,j) \\ \bm{\Omega}(i,j) \end{bmatrix} \seq{\mbf{u}} \!\notin\! \mathcal{Y}(i,j) \!=\! \left\{ \begin{bmatrix} \mbf{G}^Y_N(i,j) \\ \mbf{A}^Y_N(i,j) \end{bmatrix} , \begin{bmatrix} \mbf{c}^Y_N(i,j) \\ - \mbf{b}^Y_N(i,j) \end{bmatrix} \right\},
	\end{equation}
	$\forall i,j \in \modelset$, $i \neq j$, where $\mbf{N}(i,j) = \mbf{F}^{[j]} \seq{\mbf{H}}\zerospace^{[j]} - \mbf{F}^{[i]} \seq{\mbf{H}}\zerospace^{[i]} + [\mbf{0} \,\; (\mbf{D}^{[j]}-\mbf{D}^{[i]})]$, $\bm{\Omega}(i,j) = [(\bm{\Omega}_N^{[i]})^T \; \mbf{0} \; (\bm{\Omega}_N^{[j]})^T \; \mbf{0}]^T$, and 
	\begin{align*}
	& \mbf{G}^Y_N(i,j) = [ \mbf{G}^{Y[i]}_N \; {-}\mbf{G}^{Y[j]}_N ],\, \mbf{c}^Y_N(i,j) = \mbf{c}^{Y[i]}_N(\seq{\mbf{0}}) - \mbf{c}^{Y[j]}_N(\seq{\mbf{0}}),\\
	& \mbf{A}^Y_N(i,j) = \begin{bmatrix} \mbf{A}_N^{Y[i]} & \mbf{0} \\ \mbf{0}  & \mbf{A}_N^{Y[j]}  \end{bmatrix},\; \mbf{b}^Y_N(i,j) = \begin{bmatrix} \mbf{b}_N^{Y[i]}(\seq{\mbf{0}}) \\ \mbf{b}_N^{Y[j]}(\seq{\mbf{0}}) \end{bmatrix},
	\end{align*}
	where $\seq{\mbf{0}}$ denotes the zero input sequence.
\end{theorem}

\proof
The relations below follow from \eqref{eq:desc_stateckcz}--\eqref{eq:desc_outputckcz} and \eqref{eq:desc_outputbkcz}:
\begin{align}
\mbf{c}^{Y[i]}_N(\seq{\mbf{u}}) & = \mbf{c}^{Y[i]}_N(\seq{\mbf{0}}) + \mbf{F}^{[i]} \mbf{H}_N^{[i]} \seq{\mbf{u}} + \mbf{D}^{[i]} \mbf{u}_N, \label{eq:outputck0}\\
\mbf{b}^{Y[i]}_N(\seq{\mbf{u}}) & = \mbf{b}^{Y[i]}_N(\seq{\mbf{0}}) + [ (\bm{\Omega}_N^{[i]})^T \; \mbf{0}]^T \seq{\mbf{u}}.\label{eq:outputbk0}
\end{align}
From \eqref{eq:pre_czintersection} with $\mbf{R} = \mbf{I}$, \eqref{eq:separatinginputdef} is true iff $\nexists \bm{\xi} \in B_\infty$ such that
\begin{equation*}
\begin{bmatrix} \mbf{A}_N^{Y[i]} & \mbf{0} \\ \mbf{0} & \mbf{A}_N^{Y[j]} \\ \mbf{G}_N^{Y[i]} & - \mbf{G}_N^{Y[j]} \end{bmatrix} \bm{\xi} = \begin{bmatrix} \mbf{b}_N^{Y[i]}(\seq{\mbf{u}}) \\ \mbf{b}_N^{Y[j]}(\seq{\mbf{u}}) \\ \mbf{c}_N^{Y[j]}(\seq{\mbf{u}}) - \mbf{c}_N^{Y[i]}(\seq{\mbf{u}}) \end{bmatrix}.
\end{equation*}
According to \eqref{eq:outputck0}--\eqref{eq:outputbk0}, one has $\mbf{c}_N^{Y[j]}(\seq{\mbf{u}}) - \mbf{c}_N^{Y[i]}(\seq{\mbf{u}}) = - \mbf{c}^Y_N(i,j) + \mbf{N}(i,j) \seq{\mbf{u}}$, $[(\mbf{b}_N^{Y[i]}(\seq{\mbf{u}}))^T  \;  (\mbf{b}_N^{Y[j]}(\seq{\mbf{u}}))^T]^T $ $ \!=\! \mbf{b}^Y_N(i,j)$ $+ \bm{\Omega}(i,j) \seq{\mbf{u}}$,
with $\mbf{c}^Y_N(i,j)$, $\mbf{b}^Y_N(i,j)$, $\mbf{N}(i,j)$, and $\bm{\Omega}(i,j)$ defined as in the statement of the theorem. Then, \eqref{eq:separatinginputdef} holds iff $\nexists \bm{\xi} \in B_\infty$ such that $\mbf{G}^Y_N(i,j) \bm{\xi} = - \mbf{c}^Y_N(i,j) + \mbf{N}(i,j) \seq{\mbf{u}}$, and $\mbf{A}^Y_N(i,j) \bm{\xi} = \mbf{b}^Y_N(i,j) +  \bm{\Omega}(i,j) \seq{\mbf{u}}$. This is equivalent to
\begin{align*}
\nexists \bm{\xi} \in B_\infty: \begin{bmatrix} \mbf{G}^Y_N(i,j) \\ \mbf{A}^Y_N(i,j) \end{bmatrix} \bm{\xi} + \begin{bmatrix} \mbf{c}^Y_N(i,j) \\ - \mbf{b}^Y_N(i,j) \end{bmatrix} = \begin{bmatrix} \mbf{N}(i,j) \\ \bm{\Omega}(i,j) \end{bmatrix} \seq{\mbf{u}}, 
\end{align*}
which in turn holds iff \eqref{eq:desc_separationconditionczlast} is satisfied with $\mbf{G}^Y_N(i,j)$ and $\mbf{A}^Y_N(i,j)$ defined as in the statement of the theorem. \qed

Let $n_q$ denote the number of all possible combinations of $i,j \in \modelset$, $i \neq j$, and define $\mbf{N}^{\mathcal{Y}[q]} = [\mbf{N}^T(i,j) \; \bm{\Omega}^T(i,j)]^T$, $\mathcal{Y}^{[q]} = \mathcal{Y}(i,j) = \{\mbf{G}^{\mathcal{Y}[q]}, \mbf{c}^{\mathcal{Y}[q]}\}$, for each $q \in \{ 1,2,\ldots,n_q\}$, with $\{\mbf{G}^{\mathcal{Y}[q]}$, $\mbf{c}^{\mathcal{Y}[q]}\}$ being the right hand side of \eqref{eq:desc_separationconditionczlast}. As it can be noticed, $\mathcal{Y}^{[q]}$ is a zonotope. Then, the relation $\mbf{N}^{\mathcal{Y}[q]} \seq{\mbf{u}} \notin \mathcal{Y}^{[q]}$ can be verified by solving a linear program (LP) similar to what was proposed in \cite{Scott2014}. In this sense, the following lemma provides an effective way to verify if a given input sequence is a separating input according to Theorem \ref{thm:separatinginput}, consequently satisfying \eqref{eq:separatinginputdef}.

\begin{lemma} \rm \label{lem:desc_verifyseparating_lastcz}
	Let $\mathcal{Y}^{[q]} = \{\mbf{G}^{\mathcal{Y}[q]}, \mbf{c}^{\mathcal{Y}[q]}\}$. For each $\seq{\mbf{u}} \in \seq{U}$ and $q \in \{1,2,\ldots,n_q\}$, define $\hat{\delta}^{[q]}(\seq{\mbf{u}}) = \underset{\delta^{[q]}, \bm{\xi}^{[q]}}{\min} \delta^{[q]}$, subject to
	\begin{equation*}
	\mbf{N}^{\mathcal{Y}[q]} \seq{\mbf{u}} = \mbf{G}^{\mathcal{Y}[q]} \bm{\xi}^{[q]} + \mbf{c}^{\mathcal{Y}[q]}, \quad \ninf{\bm{\xi}} \leq 1 + \delta^{[q]}.
	\end{equation*}
	Then $\mbf{N}^{\mathcal{Y}[q]} \seq{\mbf{u}} \notin \mathcal{Y}^{[q]} \iff \hat{\delta}^{[q]}(\seq{\mbf{u}}) > 0$.
\end{lemma}
\proof See Lemma 4 in \cite{Scott2014}. \qed

For the AFD of the $n_m$ models in \eqref{eq:desc_systemSVDfault} of the descriptor system, we consider the design of a separating input of minimum length according to the optimization problem
\begin{equation} \label{eq:desc_optimalseparatingcz}
\underset{\seq{\mbf{u}} \in \seq{U}}{\min} ~ \{J(\seq{\mbf{u}})
: \mbf{N}^{\mathcal{Y}[q]} \seq{\mbf{u}} \notin \mathcal{Y}^{[q]}, ~ \forall q  = 1,2,\ldots,n_q\},   
\end{equation}
with $J(\seq{\mbf{u}})$ chosen to minimize any harmful effects caused by injecting $\seq{\mbf{u}}$ into \eqref{eq:desc_systemSVDfault}. For simplicity, we may choose $J(\seq{\mbf{u}}) = \sum_{j=0}^{N} \mbf{u}_{j}^T \mbf{R} \mbf{u}_{j}$, where $\mbf{R}$ is a weighting matrix. As in \cite{Scott2014}, this is a bilevel optimization problem and can be rewritten as a mixed-integer quadratic program by defining a \emph{minimum separation threshold} $\varepsilon > 0$ such that $\varepsilon \leq \hat{\delta}^{[q]}(\seq{\mbf{u}})$, for all $q  = 1,2,\ldots,n_q$, similar to the one presented in Chapter \ref{cha:faultdiagnosis}.

\section{Mixed zonotopes} \label{sec:desc_mixedzonotopes}

In this section, we propose a new set representation referred to as \emph{mixed zonotope}, which is an extension of constrained zonotopes that describes a larger class of sets. Mixed zonotopes inherit most of the properties of constrained zonotopes, while having the additional ability to describe unbounded sets.

\begin{definition} \rm \label{def:desc_mzonotopes}
	A set $Z \subseteq \realset^n$ is a \emph{mixed zonotope} if there exists $(\mbf{M}_z,\mbf{G}_z,\mbf{c}_z,\mbf{S}_z,\mbf{A}_z,\mbf{b}_z) \in \realsetmat{n}{n_\delta} \times \realsetmat{n}{n_g} \times \realset^n \times \realsetmat{n_c}{n_\delta} \times \realsetmat{n_c}{n_g} \times \realset^{n_c}$ such that
	\begin{equation} \label{eq:desc_mgrep}
	\begin{aligned}
	Z = \left\{ \mbf{c}_z + \mbf{M}_z \bm{\delta} + \mbf{G}_z \bm{\xi}: \right. & \left. \bm{\delta} \in \realset^{n_\delta}, \| \bm{\xi} \|_\infty \leq 1, \right. \left. \mbf{S}_z \bm{\delta} + \mbf{A}_z \bm{\xi} = \mbf{b}_z \right\}.
	\end{aligned}
	\end{equation}	
\end{definition}

We refer to \eqref{eq:desc_mgrep} as \emph{mixed generator representation} (MG-rep). Each column of $\mbf{M}_z$ is an \emph{unbounded generator}, each column of $\mbf{G}_z$ is a \emph{bounded generator}, $\mbf{c}_z$ is the \emph{center}, and $\mbf{S}_z \bm{\delta} + \mbf{A}_z \bm{\xi} = \mbf{b}_z$ are the \emph{constraints}. Note that, by the definition of the MG-rep \eqref{eq:desc_mgrep}, if $\mbf{S}_z$ is a matrix of zeros, then the mixed zonotope $Z$ is \emph{unbounded} in the directions given by the columns of $\mbf{M}_z$. This allows the MG-rep \eqref{eq:desc_mgrep} to describe symmetrically unbounded sets, such as strips (Proposition \ref{thm:desc_stripmgrep}) and the entire space $\realset^n$ (Proposition \ref{thm:desc_realspacemgrep}). We use the shorthand notation $Z = \{\mbf{M}_z,\mbf{G}_z,\mbf{c}_z,\mbf{S}_z,\mbf{A}_z,\mbf{b}_z\}$ for mixed zonotopes. Moreover, we denote by `$\noarg\!$' an empty argument to this shorthand notation. For instance, zonotopes are expressed in MG-rep as $\{\mbf{G}_z,\mbf{c}_z\} = \{\noarg,\mbf{G}_z,\mbf{c}_z,\noarg,\noarg,\noarg\}$, and constrained zonotopes as $\{\mbf{G}_z,\mbf{c}_z,\mbf{A}_z,\mbf{b}_z\} = \{\noarg,\mbf{G}_z,\mbf{c}_z,\noarg,\mbf{A}_z,\mbf{b}_z\}$.

Consider sets $Z, W \subseteq \realset^{n}$, $Y \subseteq \realset^{m}$, and a real matrix $\mbf{R} \in \realset^{m \times n}$. Similarly to constrained zonotopes, if $Z$, $W$, $Y$ are in MG-rep, i.e., $Z \triangleq \{\mbf{M}_z,\mbf{G}_z,\mbf{c}_z,\mbf{S}_z,\mbf{A}_z,\mbf{b}_z\}$, $W \triangleq \{\mbf{M}_w,\mbf{G}_w,\mbf{c}_w,\mbf{S}_w,$ $\mbf{A}_w,\mbf{b}_w\}$, and $Y \triangleq \{\mbf{M}_y,\mbf{G}_y,\mbf{c}_y,\mbf{S}_y,\mbf{A}_y,\mbf{b}_y\}$, then the elementary set operations \eqref{eq:pre_limage}--\eqref{eq:pre_intersection} are computed trivially as
\begin{align}
\mbf{R}Z & = \left\{ \mbf{R} \mbf{M}_z, \mbf{R} \mbf{G}_z, \mbf{R} \mbf{c}_z, \mbf{S}_z, \mbf{A}_z, \mbf{b}_z \right\}, \label{eq:desc_mzlimage}\\
Z \oplus W & =\left\{ \begin{bmatrix} \mbf{M}_z \,\; \mbf{M}_w \end{bmatrix}, \begin{bmatrix} \mbf{G}_z \,\; \mbf{G}_w \end{bmatrix},  \mbf{c}_z + \mbf{c}_w, \begin{bmatrix} \mbf{S}_z & \bm{0} \\ \bm{0} & \mbf{S}_w \end{bmatrix}, \begin{bmatrix} \mbf{A}_z & \bm{0} \\ \bm{0} & \mbf{A}_w \end{bmatrix}, \begin{bmatrix} \mbf{b}_z \\ \mbf{b}_w \end{bmatrix} \right\}\!, \label{eq:desc_mzmsum}\\
Z \cap_{\mbf{R}} Y & = \left\{ \begin{bmatrix} \mbf{M}_z \,\; \bm{0} \end{bmatrix}, \begin{bmatrix} \mbf{G}_z \,\; \bm{0} \end{bmatrix}, \mbf{c}_z, \begin{bmatrix} \mbf{S}_z & \bm{0} \\ \bm{0} & \mbf{S}_y \\ \mbf{R} \mbf{M}_z & -\mbf{M}_y \end{bmatrix}, \begin{bmatrix} \mbf{A}_z & \bm{0} \\ \bm{0} & \mbf{A}_y \\ \mbf{R} \mbf{G}_z & -\mbf{G}_y \end{bmatrix}, \begin{bmatrix} \mbf{b}_z \\ \mbf{b}_y \\ \mbf{c}_y - \mbf{R} \mbf{c}_z \end{bmatrix} \right\}. \label{eq:desc_mzintersection}
\end{align}

The demonstration of \eqref{eq:desc_mzlimage}--\eqref{eq:desc_mzintersection} is analogous to \eqref{eq:pre_czlimage}--\eqref{eq:pre_czintersection}, which is presented in \cite{Scott2016}. Moreover, note that every zonotope is a mixed zonotope, and also every constrained zonotope is a mixed zonotope. In the following, we demonstrate that other classes of sets are also mixed zonotopes.

\begin{proposition} \rm \label{thm:desc_realspacemgrep}
	The Euclidean space $\realset^n$ is a mixed zonotope.
\end{proposition}
\proof Let $\mbf{M}_\text{R} \in \realsetmat{n}{n_\delta}$ be a full row rank matrix, and let $\mbf{c}_\text{R} \in \realset^n$. Since $\text{rank}(\mbf{M}_\text{R}) = n$, for every $\mbf{z} \in \realset^n$ there exist at least one $\bm{\delta} \in \realset^{n_\delta}$ such that $\mbf{z} = \mbf{c}_\text{R} + \mbf{M}_\text{R} \bm{\delta}$. Therefore, $\realset^n \subseteq \left\{ \mbf{M}_\text{R}, \noarg, \mbf{c}_\text{R}, \noarg, \noarg, \noarg \right\}$. On the other side, choose one $\bm{\delta} \in \realset^{n_\delta}$, and define $\mbf{r} \triangleq \mbf{c}_\text{R} + \mbf{M}_\text{R} \bm{\delta}$. Since $\text{rank}(\mbf{M}_\text{R}) = n$, there must exist $\mbf{z} \in \realset^n$ such that $\mbf{z} = \mbf{r}$. Therefore, $\left\{ \mbf{M}_\text{R}, \noarg, \mbf{c}_\text{R}, \noarg, \noarg, \noarg \right\} \subseteq \realset^n$, which implies that
\begin{equation} \label{eq:disc_Rnmgrep}
\realset^n = \left\{ \mbf{M}_\text{R}, \noarg, \mbf{c}_\text{R}, \noarg, \noarg, \noarg \right\}.
\end{equation}
\qed

\begin{proposition} \rm \label{thm:desc_stripmgrep}
	Every strip is a mixed zonotope.
\end{proposition}
\proof Consider a strip $S = \{ \mbf{x} \in \realset^n : |\bm{\rho}_s^T \mbf{x} - d_s| \leq \sigma_s\}$, with $\bm{\rho}_s \in \realset^n$, $d_s,\sigma_s \in \realset$, $\sigma_s \geq 0$. This strip can be described in the MG-rep \eqref{eq:desc_mgrep} with one bounded generator and $n$ unbounded generators as follows. Note that an equivalent definition of the strip $S$ is $\{ \mbf{x} \in \realset^n : \bm{\rho}_s^T \mbf{x} \in [-\sigma_s + d_s, \sigma_s + d_s]\}$. Therefore, by writing the interval $[-\sigma_s + d_s, \sigma_s + d_s]$ in G-rep as $\{\sigma_s, d_s\}$, or equivalently in MG-rep as $[-\sigma_s + d_s, \sigma_s + d_s] = \{\emptyarg,\sigma_s,d_s,\emptyarg,\emptyarg,\emptyarg\}$, and noting that the set of real vectors $\realset^n$ can be described in MG-rep as $\realset^n = \{\eye{n},\emptyarg,\bm{0},\emptyarg,\emptyarg,\emptyarg\}$, therefore we have that, by definition of $S$ and the generalized intersection \eqref{eq:desc_mzintersection}, the strip $S$ can be described in MG-rep as
\begin{equation} \label{eq:desc_stripmgrep}
S = \realset^n \cap_{\bm{\rho}^T} \{\sigma_s, d_s\} = \left\{ \eye{n}, \zeros{n}{1}, \zeros{n}{1}, \bm{\rho}^T, -\sigma_s, d_s \right\}.
\end{equation}
\qed

\begin{proposition} \rm \label{thm:desc_hyperplanemgrep}
	Every hyperplane is a mixed zonotope.
\end{proposition}
\proof Consider a hyperplane $H \triangleq \{ \mbf{x} \in \realset^n : \bm{\rho}_s^T \mbf{x} = d_s\}$, with $\bm{\rho}_s \in \realset^n$, $d_s \in \realset$. Note that $H$ is a particular case of a strip given by $\{ \mbf{x} \in \realset^n : |\bm{\rho}_s^T \mbf{x} - d_s| \leq 0\}$ (i.e., a degenerated strip). Therefore, by Proposition \ref{thm:desc_stripmgrep}, $H$ can be described in MG-rep as
\begin{equation} \label{eq:desc_hyperplanemgrep}
H = \realset^n \cap_{\bm{\rho}^T} \{0, d_s\} = \left\{ \eye{n}, \zeros{n}{1}, \zeros{n}{1}, \bm{\rho}^T, 0, d_s \right\} = \left\{ \eye{n}, \noarg, \zeros{n}{1}, \bm{\rho}^T, \noarg, d_s \right\}.
\end{equation}
\qed

\begin{figure}[!htb]
	\begin{scriptsize}
		\centering{
			\def\svgwidth{1\textwidth}			
			{\scriptsize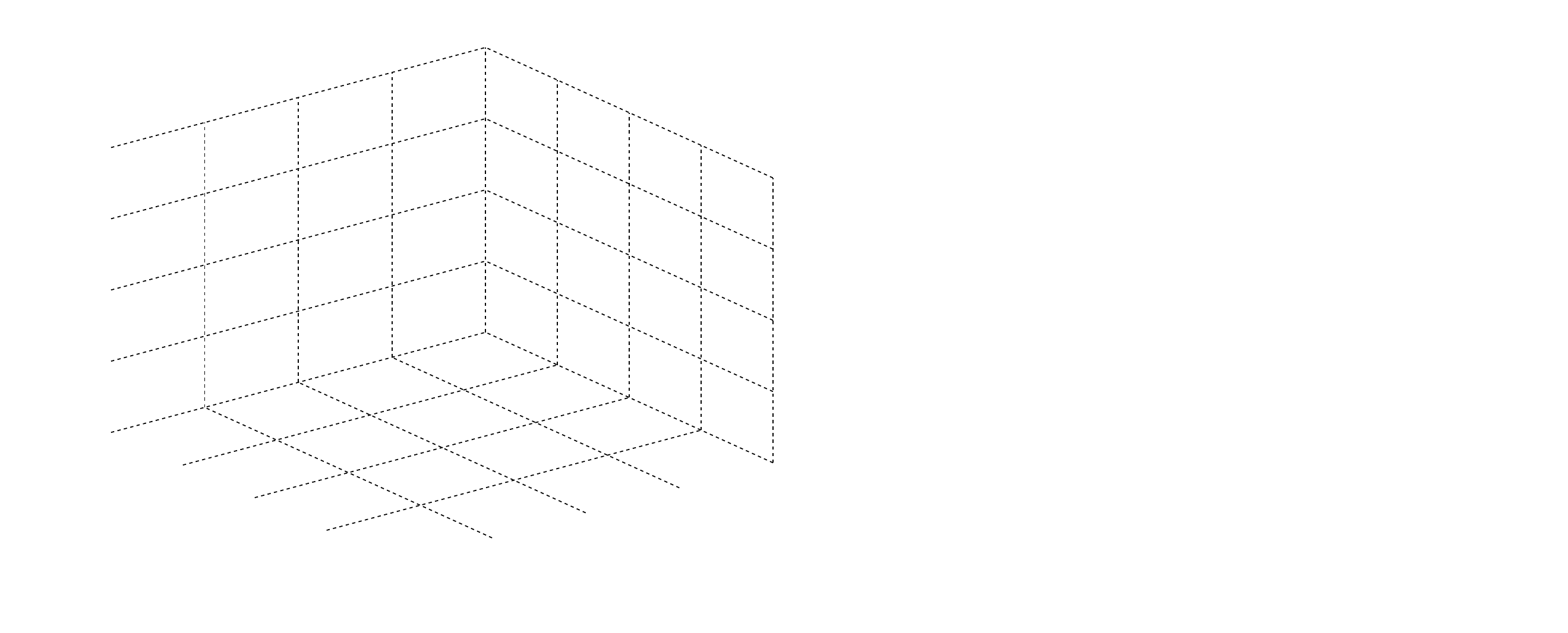}
			\caption{Examples of mixed zonotopes: the Euclidean space $\realset^n = \left\{ \mbf{M}_\text{R}, \noarg, \mbf{c}_\text{R}, \noarg, \noarg, \noarg \right\}$ (left), where $\mbf{M}_\text{R}$ and $\mbf{c}_\text{R}$ can be chosen as $\eye{n}$ and $\zeros{n}{1}$, respectively, and the strip $S = \{ \mbf{x} \in \realset^2 : |[-1\,\; 1] \mbf{x} - 1| \leq 0.5\} = \left\{ \eye{2}, \zeros{2}{1}, \zeros{2}{1}, [-1\,\; 1], -0.5, 1 \right\}$ (right).}\label{fig:desc_mzonotopeexamples}}			
	\end{scriptsize}
\end{figure}

Figure \ref{fig:desc_mzonotopeexamples} shows examples of unbounded sets described by mixed zonotopes. The ability of describing unbounded sets allows the set-based state estimation of the linear descriptor system \eqref{eq:desc_system} to be performed using MG-rep  (see Section \ref{sec:desc_estimationMZ}): (i) without the necessity of knowing a bounded initial set satisfying $\mbf{x}_0 \in X_0$; and (ii) without requiring the existence of an admissible set $X_\text{a}$ satisfying Assumption \ref{ass:desc_admissible}. A few other applications of mixed zonotopes are discussed in Section \ref{sec:desc_mzexamples}. In the following, we propose methods for complexity reduction of the MG-rep \eqref{eq:desc_mgrep}.

\subsection{Complexity reduction of mixed zonotopes} \label{sec:desc_mzcomplexityreduction}

Similarly to zonotopes and constrained zonotopes, the set operations \eqref{eq:desc_mzlimage}--\eqref{eq:desc_mzintersection} result in a linear increase of number of bounded and unbounded generators, and constraints of the MG-rep \eqref{eq:desc_mgrep}. If these operations are performed iteratively, which is the case in set-based state estimation, the complexity of the set may increase indefinitely. This section presents methods for complexity reduction of mixed zonotopes, which outer-approximate a mixed zonotope by another one with fewer bounded and unbounded generators, and constraints.

\subsubsection{Elimination of unbounded generators}

The proposed procedure to eliminate unbounded generators is similar to the constraint elimination of constrained zonotopes described by Method \ref{meth:czconelim}, as explained below. 

Let $Z = \{\mbf{M}_z,\mbf{G}_z,\mbf{c}_z,\mbf{S}_z,\mbf{A}_z,\mbf{b}_z\} \subseteq \realset^n$ be a mixed zonotope with $n_g$ bounded generators, $n_\delta$ unbounded generators, and $n_c$ constraints. It will be shown that a number of unbounded generators equal to the row rank of $\mbf{S}_z$ can always be eliminated, and additionally, this procedure does not introduce conservativeness, i.e., the resulting set is equivalent to the previous one. The procedure is similar to Method \ref{meth:czconelim}, with the difference that the chosen generator variable to be eliminated is an unbounded one. This is done by solving the constraints $\mbf{S}_z \bm{\delta} + \mbf{A}_z \bm{\xi} = \mbf{b}_z$ in the variables $\bm{\delta}$, and substituting these in $\mbf{c}_z + \mbf{M}_z \bm{\delta} + \mbf{G}_z \bm{\xi}$ in \eqref{eq:desc_mgrep}. Note that this simultaneously eliminates the same number of constraints from $Z$. However, in contrast to eliminating a bounded generator, since $\bm{\delta}$ is unbounded, hence no information is lost in the process. Moreover, note that in the case that all the unbounded generators of $Z$ can be eliminated through this procedure, then $Z$ is a constrained zonotope. Algorithm \ref{alg:desc_mzugennelim} summarizes the proposed procedure.

\begin{algorithm}[!htb]
	\caption{Elimination of an unbounded generator from a mixed zonotope}
	\label{alg:desc_mzugennelim}
	\small
	\begin{algorithmic}[1]
		\State Let $Z \triangleq \{\mbf{M},\mbf{G},\mbf{c},\mbf{S},\mbf{A},\mbf{b}\}$. Choose one unbounded generator $\delta_j$ to eliminate from $Z$ and one $i \in \{1,\ldots,n_c\}$, such that $S_{i,j} \neq 0$.
		\State Solve the constraint $\mbf{S}_{i,:} \bm{\delta} + \mbf{A}_{i,:} \bm{\xi} = \mbf{b}_{i,:}$ for $\delta_j$.
		\State Replace the result above in $\mbf{S} \bm{\delta} + \mbf{A} \bm{\xi} = \mbf{b}$ and $\mbf{c} + \mbf{M} \bm{\delta} + \mbf{G} \bm{\xi}$.
	\end{algorithmic}
	\normalsize
\end{algorithm}	

\subsubsection{Constraint elimination}

Let $Z = \{\mbf{M}_z,\mbf{G}_z,\mbf{c}_z,\mbf{S}_z,\mbf{A}_z,\mbf{b}_z\} \subseteq \realset^n$. Constraint elimination of mixed zonotopes can be done exactly as with constrained zonotopes by using Method \ref{meth:czconelim}. However, since the removal of an unbounded generator is not conservative, if $\text{rank}(\mbf{S}_z) > 0$ then $k_\delta = \text{rank}(\mbf{S}_z)$ unbounded generators chosen to be eliminated together with $k_\delta$ constraints prior to removing any bounded generator from $Z$. Constraint elimination is done in this order for reduced conservativeness.

\subsubsection{Elimination of bounded generators}

To eliminate a bounded generator from $Z = \{\mbf{M}_z,\mbf{G}_z,\mbf{c}_z,\mbf{S}_z,\mbf{A}_z,\mbf{b}_z\} \subseteq \realset^n$, we consider the case in which $\mbf{S}_z$ can be reduced to a matrix of zeros through the unbounded generator elimination procedure. In other words, let $Z^- = \{\mbf{M}_z^-,\mbf{G}_z^-,\mbf{c}_z^-,\mbf{S}_z^-,\mbf{A}_z^-,\mbf{b}_z^-\} \subseteq \realset^n$ be the equivalent mixed zonotope obtained after removing $k_\delta = \text{rank}(\mbf{S}_z)$ constraints and unbounded generators from $Z$. We assume that $\mbf{S}_z^- = \bm{0}$. Then, the resulting mixed zonotope $Z^-$ can be decoupled as $Z^- = \mbf{c}_z^- \oplus \mbf{M}_z^- \realset^{n_\delta^-} \oplus \mbf{G}_z^- B_\infty(\mbf{A}_z^-,\mbf{b}_z^-)$. 

Therefore, bounded generator elimination can be done exactly as in Method \ref{meth:czgenred}, i.e., by first lifting $\mbf{G}_z^- B_\infty(\mbf{A}_z^-,\mbf{b}_z^-)$ and then using a zonotope order reduction method. The resulting mixed zonotope is given by $\bar{Z} = \mbf{c}_z^- \oplus \mbf{M}_z^- \realset^{n_\delta} \oplus \bar{\mbf{G}}_z B_\infty(\bar{\mbf{A}}_z,\mbf{b}_z^-) \supseteq Z^-$, with $\bar{\mbf{G}}_z$ and $\bar{\mbf{A}}_z$ being the resulting matrices provided by the zonotope generator reduction method.

\subsection{Concept examples} \label{sec:desc_mzexamples}

\subsubsection{Exact intersection of a zonotope and a strip}

Since both zonotope and strip are mixed zonotopes, then the intersection of a zonotope and a strip can be computed trivially and efficiently in MG-rep, using the generalized intersection \eqref{eq:pre_czintersection} as illustrated below.

Consider the zonotope \citep{Bravo2006}
\begin{equation*}
Z = \left\{\begin{bmatrix} 0.2812 & 0.1968 & 0.4235 \\ 0.0186 & -0.2063 & -0.2267 \end{bmatrix}, \begin{bmatrix} 0 \\ 0 \end{bmatrix} \right\},
\end{equation*}
and the strip $S = \{ \mbf{x} \in \realset^2 : |\bm{\rho}_s^T \mbf{x} - d_s| \leq \sigma_s\}$, with $\bm{\rho}_s = [1 \; -1]^T$, $d_s = 0$, and $\sigma_s = 0.1$. The intersection $Z \cap S$ is computed by writing $S$ in the MG-rep \eqref{eq:desc_stripmgrep} and then using \eqref{eq:pre_czintersection}. The resulting set is
\begin{equation*}
\begin{aligned}
Z \cap S = & \left\{\begin{bmatrix} 0 & 0 \\ 0 & 0 \end{bmatrix}, \begin{bmatrix} 0.2812 & 0.1968 & 0.4235 & 0 \\ 0.0186 & -0.2063 & -0.2267 & 0 \end{bmatrix}, \begin{bmatrix} 0 \\ 0 \end{bmatrix}, \right. \\ & \left. \begin{bmatrix} 1 & -1 \\ -1 & 1 \\ 0 & -1 \end{bmatrix}, \begin{bmatrix} 0 & 0 & 0 & -0.1 \\ 0.2812 & 0.1968 & 0.4235 & 0 \\ 0.0186 & -0.2063 & -0.2267 & 0 \end{bmatrix}, \begin{bmatrix} 0 \\ 0 \\ 0 \end{bmatrix} \right\}.
\end{aligned}
\end{equation*} 

\begin{figure*}[!htb]
	\centering{
		\def\svgwidth{0.7\columnwidth}
		{\scriptsize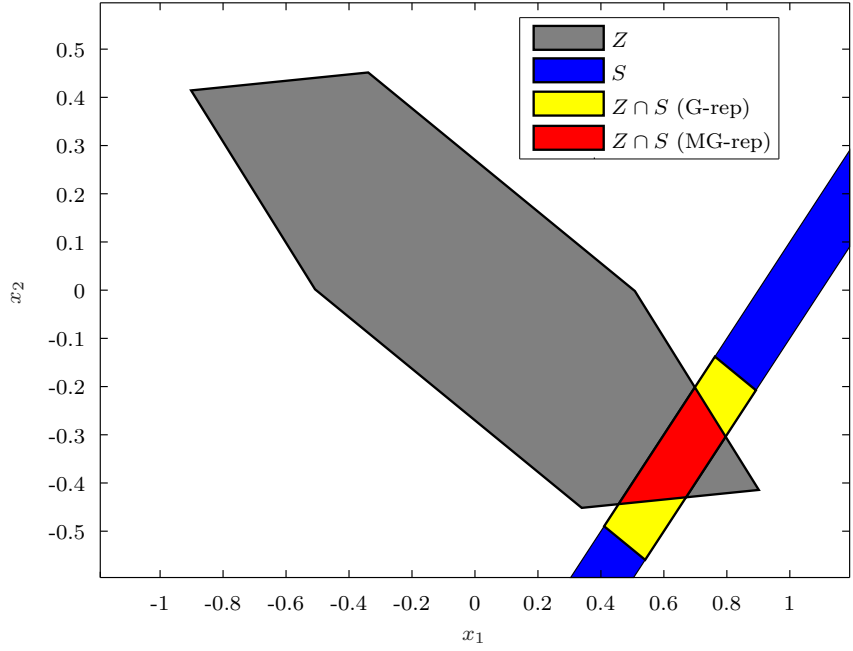}
		\caption{The zonotope $Z$ (gray), the strip $S$ (blue), the zonotope obtained using the intersection method \citep{Bravo2006} (yellow), and the mixed zonotope $Z \cap S$ computed in MG-rep (red)}\label{fig:desc_intersectionzonstrip}}
\end{figure*}

Figure \ref{fig:desc_intersectionzonstrip} shows the zonotope $Z$ (gray) and strip $S$ (blue), as well as the zonotope obtained using the intersection method \citep{Bravo2006} (yellow) and the mixed zonotope $Z \cap S$ computed in MG-rep (red). The latter is obtained using Proposition \ref{thm:desc_stripmgrep} and \eqref{eq:desc_mzintersection}, and as it can be noticed, it corresponds to the exact intersection of the zonotope $Z$ and the strip $S$, .

\subsubsection{Unbounded intersection between strips}

Let $S_1 \subset \realset^n$ and $S_2 \subset \realset^n$ be strips. The intersection $S_1 \cap S_2$ for $n>2$ is not bounded in general case, and therefore cannot be computed using zonotopes or constrained zonotopes. On the other side, since $S_1$ and $S_2$ can be written in MG-rep using Proposition \ref{thm:desc_stripmgrep}, the set $S_1 \cap S_2$ is then a mixed zonotope and can be computed using \eqref{eq:desc_mzintersection}.

\begin{figure*}[!htb]
	\centering{
		\def\svgwidth{0.7\columnwidth}
		{\scriptsize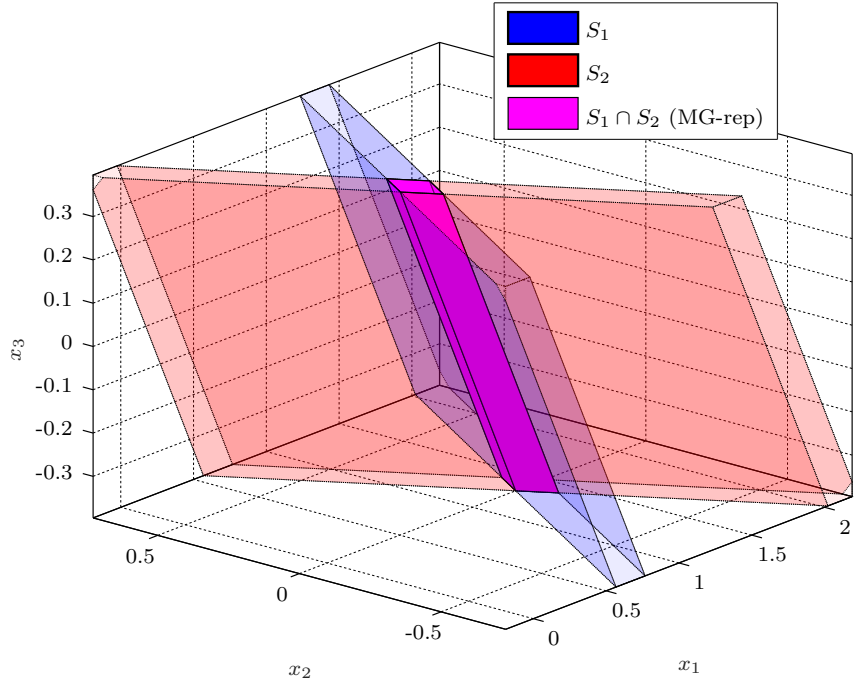}
		\caption{The strips $S_1$ (blue) and $S_2$ (red), and the intersection $S_1 \cap S_2$ computed computed in MG-rep (magenta).}\label{fig:desc_intersectionstrips}}
\end{figure*}

Let $S_1 \triangleq \{ \mbf{x} \in \realset^3 : |\bm{\rho}_1^T \mbf{x} - d_1| \leq \sigma_1\}$ and $S_2 \triangleq \{ \mbf{x} \in \realset^3 : |\bm{\rho}_2^T \mbf{x} - d_2| \leq \sigma_2\}$, with $\bm{\rho}_1 = [1 \,\; -1 \,\; 1]^T$, $\bm{\rho}_2 = [1 \,\; 1 \,\; 1]^T$, $d_1 = 1$, $d_2 = 1$, $\sigma_1 = 0.1$, and $\sigma_2 = 0.1$. Figure \ref{fig:desc_intersectionstrips} shows the strips $S_1$ (blue) and $S_2$ (red), as well as the intersection $S_1 \cap S_2$ computed in MG-rep (magenta). Note that the intersection is unbounded, and therefore cannot be expressed as a constrained zonotope or a strip, while it can be described exactly using the MG-rep \eqref{eq:desc_mgrep}.

\subsubsection{Two-side generalized intersection}

Consider sets $Z \subset \realset^{n_z}$, $Y \subset \realset^{n_y}$ and real matrices $\mbf{R}_z \in \realsetmat{n_z}{n}$, $\mbf{R}_y \in \realsetmat{n_y}{n}$. Define the two-side generalized intersection between $Z$ and $Y$ as
\begin{equation} \label{eq:desc_twointersection}
Z \cap_{\mbf{R}_z, \mbf{R}_y} Y \triangleq \{ \mbf{x} \in \realset^n: \mbf{R}_z \mbf{x} \in Z,~ \mbf{R}_y \mbf{x} \in Y\}.
\end{equation}

Note that the generalized intersection \eqref{eq:pre_intersection} is a particular case of \eqref{eq:desc_twointersection} in which $\mbf{R}_z = \eyenoarg$ and $\mbf{R}_y = \mbf{R}$. When $Z$ and $Y$ are in MG-rep, the two-side generalized intersection \eqref{eq:desc_twointersection} can be computed as follows. Let $Z \cap_{\mbf{R}_z, \mbf{R}_y} Y$ be rewritten as
\begin{equation*}
Z \cap_{\mbf{R}_z, \mbf{R}_y} Y = (\realset^n \cap_{\mbf{R}_z} Z) \cap_{\mbf{R}_y} Y.
\end{equation*} 

Let $Z = \{\mbf{M}_z, \mbf{G}_z, \mbf{c}_z, \mbf{S}_z, \mbf{A}_z, \mbf{b}_z\}$, $Y = \{\mbf{M}_y, \mbf{G}_y, \mbf{c}_y, \mbf{S}_y, \mbf{A}_y, \mbf{b}_y\}$, and let $\realset^n$ be given in MG-rep as $\realset^n = \{\eye{n}, \emptyarg, \mbf{0}, \emptyarg, \emptyarg, \emptyarg\}$. Then, using \eqref{eq:pre_intersection} twice leads to
\begin{equation*}
Z \cap_{\mbf{R}_z, \mbf{R}_y} Y = \left\{ [\eye{n} \,\; \mbf{0} \,\; \mbf{0}], [\mbf{0} \,\; \mbf{0}], \mbf{0}, \begin{bmatrix} \mbf{0} & \mbf{S}_z & \mbf{0} \\ \mbf{0} & \bm{0} & \mbf{S}_y \\ \mbf{R}_z & -\mbf{M}_z & \mbf{0} \\ \mbf{R}_y & \mbf{0} & -\mbf{M}_y \end{bmatrix}, \begin{bmatrix} \mbf{A}_z & \bm{0} \\ \bm{0} & \mbf{A}_y \\ - \mbf{G}_z & \bm{0} \\ \bm{0} & -\mbf{G}_y \end{bmatrix}, \begin{bmatrix} \mbf{b}_z \\ \mbf{b}_y \\ \mbf{c}_z \\  \mbf{c}_y \end{bmatrix} \right\}.
\end{equation*}

\section{Set-based state estimation of LDS using mixed zonotopes}  \label{sec:desc_estimationMZ}
\sectionmark{State estimation of LDS using mixed zonotopes}

This section presents a new method for set-based state estimation of system \eqref{eq:desc_system} using mixed zonotopes. Given an initial condition $X_0$ and an input $\mbf{u}_k$ with $k \geq 0$, let
\begin{equation} \label{eq:desc_initialsetmz}
\hat{X}_0 = \{ \mbf{x} \in X_0 : \mbf{C} \mbf{x} + \mbf{D} \mbf{u}_{0} + \mbf{D}_v \mbf{v} = \mbf{y}_0, \, \mbf{v} \in V \}.
\end{equation}

The method proposed in this section is also based on singular value decomposition. Let $\mbf{E} = \mbf{U} \mbf{\Sigma} \mbf{V}^T$, where $\mbf{U}$ and $\mbf{V}$ are invertible by construction. Since $\mbf{E}$ is square, then $\mbf{\Sigma}$ is also square. Without loss of generality, let $\mbf{\Sigma}$ be arranged as $\mbf{\Sigma} = \text{blkdiag}(\tilde{\mbf{\Sigma}}, \mbf{0})$, where $\tilde{\mbf{\Sigma}} \in \realsetmat{n_z}{n_z}$ is diagonal with all the $n_z = \text{rank}(\mbf{E})$ nonzero singular values of $\mbf{E}$. Moreover, let $\mbf{z}_k = (\tilde{\mbf{z}}_k,\check{\mbf{z}}_k) = \inv{\mbf{T}} \mbf{x}_k, \; \tilde{\mbf{z}}_k \in \realset^{n_z}, \check{\mbf{z}}_k \in \realset^{n-n_z}$,
\begin{equation}\label{eq:desc_SVDmatricesmz} 
\begin{aligned}
\begin{bmatrix} \tilde{\mbf{A}} \\ \check{\mbf{A}} \end{bmatrix} & = \begin{bmatrix} \tilde{\mbf{\Sigma}}^{-1} & \mbf{0} \\ \mbf{0} & \eyenoarg \end{bmatrix} \inv{\mbf{U}} \mbf{A} \mbf{T}, \\
\begin{bmatrix} \tilde{\mbf{B}} \\ \check{\mbf{B}} \end{bmatrix} & = \begin{bmatrix} \tilde{\mbf{\Sigma}}^{-1} & \mbf{0} \\ \mbf{0} & \eyenoarg \end{bmatrix} \inv{\mbf{U}} \mbf{B}, \begin{bmatrix} \tilde{\mbf{B}}_w \\ \check{\mbf{B}}_w \end{bmatrix} = \begin{bmatrix} \tilde{\mbf{\Sigma}}^{-1} & \mbf{0} \\ \mbf{0} & \eyenoarg \end{bmatrix} \inv{\mbf{U}} \mbf{B}_w,  \\
\end{aligned}
\end{equation}
with $\mbf{T} = \inv{(\mbf{V}^T)}$, $\tilde{\mbf{A}} \in \realsetmat{n_z}{n}$, $\tilde{\mbf{B}} \in \realsetmat{n_z}{n_u}$, $\tilde{\mbf{B}}_w \in \realsetmat{n_z}{n_w}$. Then, as in Section \ref{sec:desc_estimationCZ}, system \eqref{eq:desc_system} can be rewritten as
\begin{subequations} \label{eq:desc_systemSVDmz}
	\begin{align} 
	\tilde{\mbf{z}}_k & = \tilde{\mbf{A}} \mbf{z}_{k-1} + \tilde{\mbf{B}} \mbf{u}_{k-1} + \tilde{\mbf{B}}_w \mbf{w}_{k-1},  \label{eq:desc_systemSVDdynamicsmz} \\
	\mbf{0} & = \check{\mbf{A}} \mbf{z}_{k} + \check{\mbf{B}} \mbf{u}_{k} + \check{\mbf{B}}_w \mbf{w}_{k}, \label{eq:desc_systemSVDconstraintsmz} \\
	\mbf{y}_k & = \mbf{C} \mbf{T} \mbf{z}_k + \mbf{D} \mbf{u}_{k} + \mbf{D}_v \mbf{v}_{k}. \label{eq:desc_systemSVDoutputmz}
	\end{align}
\end{subequations}
Let $W = \{\mbf{M}_w, \mbf{G}_w, \mbf{c}_w, \mbf{S}_w, \mbf{A}_w, \mbf{b}_w\}$, and $\hat{X}_{0}$ given by \eqref{eq:desc_initialsetmz}. From \eqref{eq:desc_systemSVDconstraintsmz}, the state $\mbf{z}_0$ must satisfy $\check{\mbf{A}} \mbf{z}_0 + \check{\mbf{B}} \mbf{u}_0 + \check{\mbf{B}}_w \mbf{w}_0 = \mbf{0}$. This constraint can be incorporated in the MG-rep of the initial set $\hat{Z}_0$ as follows. Let $\inv{\mbf{T}} \hat{X}_{0} \triangleq \{\mbf{M}_0, \mbf{G}_0,\mbf{c}_0,\mbf{S}_0,\mbf{A}_0,\mbf{b}_0\}$. Then, $\hat{Z}_0 \triangleq \{\hat{\mbf{M}}_0,\hat{\mbf{G}}_0,\hat{\mbf{G}}_0, \hat{\mbf{c}}_0,\hat{\mbf{S}}_0,\hat{\mbf{A}}_0,\hat{\mbf{b}}_0\}$, with $\hat{\mbf{M}}_0 = [\mbf{M}_0 \,\; \mbf{0}] ,\hat{\mbf{G}}_0 = [\mbf{G}_0 \,\; \mbf{0}]$, $\hat{\mbf{c}}_0 = \mbf{c}_0$,
\begin{equation*}
\begin{aligned}
\hat{\mbf{S}}_0 = \begin{bmatrix} \mbf{S}_0 & \mbf{0} \\  \mbf{0} & \mbf{S}_w \\ \check{\mbf{A}} \mbf{M}_0 & \check{\mbf{B}}_w \mbf{M}_w \end{bmatrix},\; \hat{\mbf{A}}_0 = \begin{bmatrix} \mbf{A}_0 & \mbf{0} \\ \mbf{0} & \mbf{A}_w \\ \check{\mbf{A}} \mbf{G}_0 & \check{\mbf{B}}_w \mbf{G}_w \end{bmatrix}, \; \hat{\mbf{b}}_0 = \begin{bmatrix}  \mbf{b}_0 \\  \mbf{b}_w \\ -\check{\mbf{A}} \mbf{c}_0 - \check{\mbf{B}}_w \mbf{c}_w - \check{\mbf{B}} \mbf{u}_{0} \end{bmatrix}.
\end{aligned}
\end{equation*}
Note that the extra columns in $\hat{\mbf{G}}_0$ and $\hat{\mbf{A}}_0$ come from $\mbf{w}_0 \in W$. Moreover, from \eqref{eq:desc_systemSVDmz}, the variables $\tilde{\mbf{z}}_k$ are fully determined by \eqref{eq:desc_systemSVDdynamicsmz}, while $\check{\mbf{z}}_k$ are obtained a posteriori by $\check{\mbf{A}} \mbf{z}_{k} + \check{\mbf{B}} \mbf{u}_{k} + \check{\mbf{B}}_w \mbf{w}_{k} = \mbf{0}$. 

An effective enclosure of the prediction step for the descriptor system \eqref{eq:desc_systemSVDmz} can be obtained in MG-rep as follows. 

\begin{lemma} \label{lem:desc_predictionmz} \rm
	Consider the transformed state-space \eqref{eq:desc_systemSVDmz}. Let $\mbf{z}_{k-1} \in \hat{Z}_{k-1} \triangleq \{\hat{\mbf{M}}_{k-1}, \hat{\mbf{G}}_{k-1},$ $ \hat{\mbf{c}}_{k-1}$, $\hat{\mbf{S}}_{k-1}$, $\hat{\mbf{A}}_{k-1}, \hat{\mbf{b}}_{k-1}\}$, and $\mbf{w}_{k-1}, \mbf{w}_k \in W \triangleq \{\mbf{M}_w, \mbf{G}_w, \mbf{c}_w, \mbf{S}_w, \mbf{A}_w, \mbf{b}_w\}$, for all $k \geq 1$. Moreover, let $\mbf{c}_\text{R} \in \realset^{n-n_z}$, and let $\mbf{M}_\text{R} \in \realsetmat{n-n_z}{n_{\delta_\text{R}}}$ be a full row rank matrix with $n_\delta \geq n-n_z$. Then $\mbf{z}_k \in \bar{Z}_k \triangleq \{\bar{\mbf{M}}_{k}, \bar{\mbf{G}}_{k}, \bar{\mbf{c}}_{k}$, $\bar{\mbf{S}}_{k}$, $\bar{\mbf{A}}_{k}, \bar{\mbf{b}}_{k}\}$, with
	\begin{align*}
	\bar{\mbf{M}}_k & = \begin{bmatrix} \tilde{\mbf{A}} \hat{\mbf{M}}_{k-1} & \tilde{\mbf{B}}_w \mbf{M}_{w} & \bm{0} & \bm{0} \\ \mbf{0} & \mbf{0} & \bm{0} & \mbf{M}_\text{R} \end{bmatrix}\!, \bar{\mbf{G}}_k {=} \begin{bmatrix} \tilde{\mbf{A}} \hat{\mbf{G}}_{k-1} & \tilde{\mbf{B}}_w \mbf{G}_w & \mbf{0} \\
	\mbf{0} & \mbf{0} & \mbf{0}\end{bmatrix}\!, \bar{\mbf{c}}_k {=} \begin{bmatrix} \tilde{\mbf{A}} \hat{\mbf{c}}_{k-1} + \tilde{\mbf{B}} \mbf{u}_{k-1} + \tilde{\mbf{B}}_w \mbf{c}_w \\ \mbf{c}_\text{R}\end{bmatrix}\!, \\  
	\bar{\mbf{S}}_k & = \begin{bmatrix} \hat{\mbf{S}}_{k-1} & \mbf{0} & \mbf{0} & \bm{0} \\ \mbf{0} & \mbf{S}_w & \mbf{0} & \bm{0} \\ \mbf{0} & \mbf{0} & \mbf{S}_w  & \bm{0} \\  \check{\mbf{A}} \begin{bmatrix} \tilde{\mbf{A}} \hat{\mbf{M}}_{k-1} \\ \bm{0} \end{bmatrix} & \check{\mbf{A}} \begin{bmatrix} \tilde{\mbf{B}}_w \mbf{M}_w \\ \bm{0} \end{bmatrix} & \mbf{0} & \check{\mbf{A}} \begin{bmatrix} \bm{0} \\ \mbf{M}_\text{R} \end{bmatrix} \end{bmatrix}\!, \bar{\mbf{A}}_k {=} \begin{bmatrix} \hat{\mbf{A}}_{k-1} & \mbf{0} & \mbf{0} \\ 
	\mbf{0} & \mbf{A}_w & \mbf{0} \\
	\mbf{0} & \mbf{0} & \mbf{A}_w \\
	\check{\mbf{A}} \begin{bmatrix} \tilde{\mbf{A}} \hat{\mbf{G}}_{k-1}  \\ \mbf{0} \end{bmatrix} & \check{\mbf{A}} \begin{bmatrix} \tilde{\mbf{B}}_w \mbf{G}_w \\ \mbf{0} \end{bmatrix} & \check{\mbf{B}}_w \mbf{G}_w \end{bmatrix}\!, \\
	\bar{\mbf{b}}_k & = \begin{bmatrix} \hat{\mbf{b}}_{k-1} \\ \mbf{b}_w \\ \mbf{b}_w \\ -\check{\mbf{A}} \begin{bmatrix} \tilde{\mbf{A}} \hat{\mbf{c}}_{k-1} + \tilde{\mbf{B}} \mbf{u}_{k-1} + \tilde{\mbf{B}}_w \mbf{c}_w \\ \mbf{c}_\text{R}\end{bmatrix} - \check{\mbf{B}} \mbf{u}_k - \check{\mbf{B}}_w \mbf{c}_w \end{bmatrix}.                                
	\end{align*}
\end{lemma}
\proof
Since by assumption $(\mbf{z}_{k-1},\mbf{w}_{k-1},\mbf{w}_k) \in \hat{Z}_{k-1} \times W \times W$, there exists $((\bm{\xi}_{k-1},$ $ \bm{\delta}_{k-1}),$ $(\bm{\varphi}_{k-1},\bm{\vartheta}_{k-1}), $ $(\bm{\varphi}_{k},\bm{\vartheta}_{k})) \in \{(\bm{\xi}, \bm{\delta}) \in B_\infty^{n_g} \times \realset^{n_\delta} : \hat{\mbf{S}}_{k-1} \bm{\delta} + \hat{\mbf{A}}_{k-1} \bm{\xi} = \hat{\mbf{b}}_{k-1}\} \times \{(\bm{\varphi}, \bm{\vartheta}) \in B_\infty^{n_{g_w}} \times \realset^{n_{\delta_w}} : \mbf{S}_w \bm{\vartheta} + \mbf{A}_w \bm{\varphi} = \mbf{b}_w\} \times \{(\bm{\varphi}, \bm{\vartheta}) \in B_\infty^{n_{g_w}} \times \realset^{n_{\delta_w}} : \mbf{S}_w \bm{\vartheta} + \mbf{A}_w \bm{\varphi} = \mbf{b}_w\}$ such that $\mbf{z}_{k-1} = \hat{\mbf{c}}_{k-1} + \hat{\mbf{G}}_{k-1} \bm{\xi}_{k-1} + \hat{\mbf{M}}_{k-1} \bm{\delta}_{k-1}$, $\mbf{w}_{k-1} = \mbf{c}_w + \mbf{S}_w \bm{\vartheta}_{k-1} + \mbf{G}_w \bm{\varphi}_{k-1}$, and $\mbf{w}_{k} = \mbf{c}_w + \mbf{S}_w \bm{\vartheta}_k + \mbf{G}_w \bm{\varphi}_{k}$. Moreover, since $\check{\mbf{z}}_k \in \realset^{n - n_z}$, then there must exist $\bm{\delta}_\text{R} \in \realset^{n_{\delta_\text{R}}}$ such that $\check{\mbf{z}}_{k} = \mbf{c}_\text{R} + \mbf{M}_\text{R} \bm{\delta}_\text{R}$ for any $\mbf{c}_\text{R} \in \realset^{n - n_z}$, and any full row rank $\mbf{M}_\text{R} \in \realsetmat{n - n_z}{n_{\delta_\text{R}}}$ with $n_{\delta_\text{R}} \geq n - n_z$. Then, substituting these equalities in \eqref{eq:desc_systemSVDdynamicsmz} leads to
\begin{equation} \label{eq:desc_mzlema1proof1}
\begin{aligned}
\begin{bmatrix} \tilde{\mbf{z}}_{k} \\ \check{\mbf{z}}_{k} \end{bmatrix} = & \begin{bmatrix} \tilde{\mbf{A}} \hat{\mbf{c}}_{k-1} + \tilde{\mbf{B}} \mbf{u}_{k-1} + \tilde{\mbf{B}}_w \mbf{c}_{w} \\ \mbf{c}_\text{R} \end{bmatrix} \\ & + \begin{bmatrix} \tilde{\mbf{A}} \hat{\mbf{G}}_{k-1} & \tilde{\mbf{B}}_w \mbf{G}_{w} \\ \mbf{0} & \mbf{0} \end{bmatrix} \begin{bmatrix} \bm{\xi}_{k-1} \\ \bm{\varphi}_{k-1} \end{bmatrix} + \begin{bmatrix} \tilde{\mbf{A}} \hat{\mbf{M}}_{k-1} & \tilde{\mbf{B}}_w \mbf{M}_{w} & \bm{0} \\ \mbf{0} & \mbf{0} & \mbf{M}_\text{R} \end{bmatrix} \begin{bmatrix} \bm{\delta}_{k-1} \\ \bm{\vartheta}_{k-1} \\ \bm{\delta}_\text{R} \end{bmatrix}.
\end{aligned}		
\end{equation}
From the constraint \eqref{eq:desc_systemSVDconstraintsmz}, we have that%
\begin{equation} \label{eq:desc_mzlema1proof2}
\begin{aligned}
& \check{\mbf{B}}_w \mbf{c}_w +  \check{\mbf{B}}_w \mbf{G}_w \bm{\varphi}_k + \check{\mbf{B}}_w \mbf{M}_w \bm{\vartheta}_k + \check{\mbf{B}} \mbf{u}_{k} + \check{\mbf{A}} \begin{bmatrix} \tilde{\mbf{A}} \hat{\mbf{c}}_{k-1} + \tilde{\mbf{B}} \mbf{u}_{k-1} + \tilde{\mbf{B}}_w \mbf{c}_{w}  \\  \mbf{c}_\text{R} \end{bmatrix} \\ & + \check{\mbf{A}} \begin{bmatrix} \tilde{\mbf{A}} \hat{\mbf{G}}_{k-1} & \tilde{\mbf{B}}_w \mbf{G}_{w} \\ \mbf{0} & \mbf{0} \end{bmatrix} \begin{bmatrix} \bm{\xi}_{k-1} \\ \bm{\varphi}_{k-1} \end{bmatrix} + \check{\mbf{A}} \begin{bmatrix} \tilde{\mbf{A}} \hat{\mbf{M}}_{k-1} & \tilde{\mbf{B}}_w \mbf{M}_{w} & \bm{0} \\ \mbf{0} & \mbf{0} & \mbf{M}_\text{R} \end{bmatrix} \begin{bmatrix} \bm{\delta}_{k-1} \\ \bm{\vartheta}_{k-1} \\ \bm{\delta}_\text{R} \end{bmatrix} = \mbf{0}.
\end{aligned}		
\end{equation}
Rearranging \eqref{eq:desc_mzlema1proof1} and \eqref{eq:desc_mzlema1proof2}, grouping the variables $(\bm{\xi}_{k-1}, \bm{\varphi}_{k-1}, \bm{\varphi}_{k})$ and $(\bm{\delta}_{k-1}, \bm{\vartheta}_{k-1}, \bm{\vartheta}_k, \bm{\delta}_\text{R})$, and writing in the MG-rep \eqref{eq:desc_mgrep}, proves the lemma. \qed

\begin{remark} \rm The enclosure $\bar{Z}_k$ provided by Lemma \ref{lem:desc_predictionmz} has $n_\delta + 2 n_{\delta_w} + n_{\delta_\text{R}}$ unbounded generators, $n_g + 2 n_{g_w}$ bounded generators, and $n_c + 2 n_{c_w} + n - n_z$ constraints.
\end{remark}

Lemma \ref{lem:desc_predictionmz} provides a predicted enclosure of the state $\mbf{z}_k$ in which the equality constraints \eqref{eq:desc_systemSVDconstraints} are directly incorporated. As in the case of CZs, this is possible thanks to the fact that MZs incorporate equality constraints. The prediction-update algorithm proposed for descriptor systems consists in the computation of MZs $\bar{Z}_k$, $\hat{Z}_k$, and $\hat{X}_k$, such that
\begin{align}
\bar{Z}_k & = \{\bar{\mbf{M}}_k, \bar{\mbf{G}}_k, \bar{\mbf{c}}_k, \bar{\mbf{S}}_k, \bar{\mbf{A}}_k, \bar{\mbf{b}}_k\}, \label{eq:desc_predictionSVDmz} \\
\hat{Z}_k & = \bar{Z}_k \cap_{\mbf{C}\mbf{T}} ((\mbf{y}_k - \mbf{D}_u \mbf{u}_k) \oplus (-\mbf{D}_v V_k)), \label{eq:desc_updateSVDmz}  \\
\hat{X}_k & = \mbf{T} \hat{Z}_k. \label{eq:desc_finalSVDmz}
\end{align}
For this algorithm, the initial set is $\hat{Z}_0$. The algorithm \eqref{eq:desc_predictionSVDmz}--\eqref{eq:desc_finalSVDmz} operates recursively with $\bar{Z}_k$ and $\hat{Z}_k$ for $k \geq 1$ in the transformed state-space \eqref{eq:desc_systemSVDmz}, while the estimated enclosure in the original state-space \eqref{eq:desc_system} is given by $\hat{X}_k$.

\begin{remark} \rm \label{rem:desc_mzestconvervativeness}
	By construction, the MG-rep \eqref{eq:desc_predictionSVDmz} corresponds to the exact feasible state set of \eqref{eq:desc_systemSVDmz} at $k$ for the known state and uncertainty bounds. In addition, \eqref{eq:desc_updateSVDmz}--\eqref{eq:desc_finalSVDmz} can be computed exactly. However, in practice, in order to limit the complexity of the resulting sets these are outer approximated by using order reduction algorithms. In this case, equalities \eqref{eq:desc_predictionSVDmz}-\eqref{eq:desc_finalSVDmz} are replaced by the relation $\supset$. 
\end{remark}

\begin{remark} \rm In contrast to constrained zonotopes, state estimation of the descriptor system \eqref{eq:desc_system} using mixed zonotopes does not require knowledge of a bounded set enclosing all the states (Assumption \ref{ass:desc_admissible}). In other words, this means that the method proposed in this section can be applied also to unstable descriptor systems.
\end{remark}

\begin{remark} \rm \label{rem:desc_unboundedX0}
	Since $X_0$ is described in MG-rep \eqref{eq:desc_mgrep}, the method proposed in this section then does not require the knowledge of a bounded initial set. The enclosures $\hat{X}_k$ will be bounded depending on observability properties of the system.
\end{remark}

\section{Active fault diagnosis of LDS using mixed zonotopes}  \label{sec:desc_AFDMZ}

The previous section presented a method to address the problem of the set-based estimation of descriptor systems using mixed zonotopes. In the following, this tool is used in the design of an MZ-based AFD method accounting for a finite number of possible abrupt faults.

Let $\seq{\mbf{u}} = (\mbf{u}_0, ..., \mbf{u}_N)\in \mathbb{R}^{(N+1)n_u}$, $\seq{\mbf{w}} = (\mbf{w}_0, \ldots$, $\mbf{w}_N)\in \mathbb{R}^{(N+1)n_w}$, and $\seq{W} = W \times \ldots \times W$. Let $\mbf{z}_k = (\tilde{\mbf{z}}_k,\check{\mbf{z}}_k) = (\inv{\mbf{T}} \mbf{x}_k, \mbf{w}_k),$ $\tilde{\mbf{z}}_k \in \realset^{n_z}$, $\check{\mbf{z}}_k \in \realset^{n+n_w-n_z}$,
with $\mbf{T}^{[i]} = \inv{((\mbf{V}^{[i]})^T)}$, $\mbf{V}^{[i]}$ being obtained from the SVD $\mbf{E}^{[i]} = \mbf{U}^{[i]} \mbf{\Sigma}^{[i]} (\mbf{V}^{[i]})^T$. Then, \eqref{eq:desc_systemfaulty} can be rewritten as
\begin{subequations} \label{eq:desc_systemSVDfaultmz}
	\begin{align} 
	\tilde{\mbf{z}}_k^{[i]} & = \tilde{\mbf{A}}_z^{[i]} \mbf{z}_{k-1}^{[i]} + \tilde{\mbf{B}}^{[i]} \mbf{u}_{k-1}, \label{eq:systemSVDfaultdynamicsmz} \\
	\mbf{0} & = \check{\mbf{A}}_z^{[i]} \mbf{z}_{k}^{[i]} + \check{\mbf{B}}^{[i]} \mbf{u}_{k}, \label{eq:desc_systemSVDfaultconstraintsmz} \\
	\mbf{y}_k^{[i]} & = \mbf{F}^{[i]} \mbf{z}_k^{[i]} + \mbf{D}^{[i]} \mbf{u}_{k} + \mbf{D}_v^{[i]} \mbf{v}_{k}, \label{eq:systemSVDfaultoutputmz}
	\end{align}
\end{subequations}
with $\mbf{F}^{[i]} = \mbf{C}^{[i]} \mbf{T}^{[i]} \mbf{L}$, where $\mbf{L} = [ \eye{n} \,\; \zeros{n}{n_w}]$, $\tilde{\mbf{A}}_z^{[i]} = [\tilde{\mbf{A}}^{[i]} \,\; \tilde{\mbf{B}}_w^{[i]}]$, and $\check{\mbf{A}}_z^{[i]} = [\check{\mbf{A}}^{[i]} \,\; \check{\mbf{B}}_w^{[i]}]$. Note that the $\tilde{(\cdot)}$ and $\check{(\cdot)}$ variables are defined according to \eqref{eq:desc_SVDmatricesmz} for each $i \in \modelset$. 

For each model $i \in \modelset$, consider the set $\{\mbf{M}_z^{[i]}, \mbf{G}_z^{[i]},\mbf{c}_z^{[i]}, \mbf{S}_z^{[i]}, \mbf{A}_z^{[i]},\mbf{b}_z^{[i]}\} = \inv{(\mbf{T}^{[i]})} X_0 \times W $, and define the initial feasible set $Z_0^{[i]} (\mbf{u}_0) = \{\mbf{z} \in \inv{(\mbf{T}^{[i]})} X_0 \times W : \eqref{eq:desc_systemSVDfaultconstraintsmz} \text{ holds for }k=0\}$. This set is given by $Z_0^{[i]}(\mbf{u}_0) = \{\mbf{M}_0^{[i]}, \mbf{G}_0^{[i]},\mbf{c}_0^{[i]},\mbf{S}_0^{[i]},\mbf{A}_0^{[i]},$ $\mbf{b}_0^{[i]}(\mbf{u}_0)\}$, where $\mbf{M}_0^{[i]} = \mbf{M}_z^{[i]}$, $\mbf{G}_0^{[i]} = \mbf{G}_z^{[i]}$, $\mbf{c}_0^{[i]} = \mbf{c}_z^{[i]}$,
\begin{equation} \label{eq:initialAbmz}
\mbf{S}_0^{[i]} = \begin{bmatrix} \mbf{S}_z^{[i]} \\ \check{\mbf{A}}_z^{[i]} \mbf{M}_{0}^{[i]} \end{bmatrix}, \; \mbf{A}_0^{[i]} = \begin{bmatrix} \mbf{A}_z^{[i]} \\ \check{\mbf{A}}_z^{[i]} \mbf{G}_{0}^{[i]} \end{bmatrix}, \; \mbf{b}_0^{[i]}(\mbf{u}_0) = \begin{bmatrix} \mbf{b}_z^{[i]} \\ -\check{\mbf{A}}_z^{[i]} \mbf{c}_0^{[i]} - \check{\mbf{B}}^{[i]} \mbf{u}_{0} \end{bmatrix}.
\end{equation}

Let $\realset^{n-n_z} = \{\mbf{M}_\text{R}, \noarg, \mbf{c}_\text{R}, \noarg, \noarg, \noarg\}$, with $\mbf{M}_\text{R}$ full row rank, let $W = \{\mbf{M}_w, \mbf{G}_w, \mbf{c}_w, \mbf{S}_w,$ $\mbf{A}_w, \mbf{b}_w\}$, and $Z_\text{A} = \{\mbf{M}_\text{A}, \mbf{G}_\text{A}, \mbf{c}_\text{A}, \mbf{S}_\text{A}, \mbf{A}_\text{A}, \mbf{b}_\text{A}\} \triangleq \realset^{n-n_z} \times W$.
In addition, define the solution mappings $(\bm{\phi}_k^{[i]},\bm{\psi}_k^{[i]}) : \realset^{(k+1)n_u} \times \realset^n \times \realset^{(k+1)n_w} \times \realset^{n_v} \to \realset^{n+n_w} \times \realset^{n_y}$ such that $\bm{\phi}_k^{[i]}(\seq{\mbf{u}},\mbf{z}_0, \seq{\mbf{w}})$ and $\bm{\psi}_k^{[i]}(\seq{\mbf{u}},\mbf{z}_0, \seq{\mbf{w}}, \mbf{v}_k)$ are the state and output of \eqref{eq:desc_systemSVDfault} at $k$, respectively. Then, for each $i \in \modelset$, define state and output reachable sets at time $k$ as
\begin{equation*}
\begin{aligned}
Z_k^{[i]}(\seq{\mbf{u}}) \triangleq & \{ \bm{\phi}_k^{[i]}(\seq{\mbf{u}},\mbf{z}_0, \seq{\mbf{w}}) : (\mbf{z}_0^{[i]}, \seq{\mbf{w}}) \in Z_0^{[i]}(\mbf{u}_0) \times \seq{W} \},\\
Y_k^{[i]}(\seq{\mbf{u}}) \triangleq & \{ \bm{\psi}_k^{[i]}(\seq{\mbf{u}},\mbf{z}_0, \seq{\mbf{w}}, \mbf{v}_k) : (\mbf{z}_0,\seq{\mbf{w}},\mbf{v}_k) \in Z_0^{[i]}(\mbf{u}_0) \times \seq{W} \times V\}.
\end{aligned}
\end{equation*}

Let $N \in \naturalset$, $N \geq 0$, and define $\seq{\bm{\phi}}\zspace^{[i]}(\seq{\mbf{u}}, \mbf{z}_0, \seq{\mbf{w}}) \triangleq (\bm{\phi}_0^{[i]},\bm{\phi}_1^{[i]},\ldots,\bm{\phi}_N^{[i]})(\seq{\mbf{u}},\mbf{z}_0, \seq{\mbf{w}})$, and $\seq{\bm{\psi}}\zspace^{[i]}(\seq{\mbf{u}},\mbf{z}_0,$ $\seq{\mbf{w}},\seq{\mbf{v}}) \triangleq (\bm{\psi}_0^{[i]},\bm{\psi}_1^{[i]},\ldots,\bm{\psi}_N^{[i]})(\seq{\mbf{u}},\mbf{z}_0, \seq{\mbf{w}}, \seq{\mbf{v}})$. Given the time horizon $k \in [0,N]$, we define the \emph{state and output reachable tubes} as
\begin{equation*}
\begin{aligned}
\seq{Z}\zspace^{[i]}(\seq{\mbf{u}}) \triangleq & \{ \seq{\bm{\phi}}\zspace^{[i]}(\seq{\mbf{u}},\mbf{z}_0, \seq{\mbf{w}}) : (\mbf{z}_0^{[i]}, \seq{\mbf{w}}) \in Z_0^{[i]}(\mbf{u}_0) \times \seq{W} \},\\
\seq{Y}\zspace^{[i]}(\seq{\mbf{u}}) \triangleq & \{ \seq{\bm{\psi}}\zspace^{[i]}(\seq{\mbf{u}},\mbf{z}_0, \seq{\mbf{w}}, \seq{\mbf{v}}) : (\mbf{z}_0,\seq{\mbf{w}},\seq{\mbf{v}}) \in Z_0^{[i]}(\mbf{u}_0) \times \seq{W} \times \seq{V}\}.
\end{aligned}
\end{equation*}

\subsection{State reachable tube}

Let $Z_z = \{\mbf{M}_z, \mbf{G}_z, \mbf{c}_z, \mbf{S}_z, \mbf{A}_z, \mbf{b}_z \} \triangleq (\inv{(\mbf{T}^{[i]})} X_0) \times W$, and $Z_0 \triangleq \{\mbf{z} \in Z_z : \mbf{0} = \check{\mbf{A}}_z \mbf{z} + \check{\mbf{B}} \mbf{u}_0\}$. Moreover, let $\realset^{n-n_z} = \{\mbf{M}_\text{R}, \noarg, \mbf{c}_\text{R}, \noarg, \noarg, \noarg\}$, with $\mbf{M}_\text{R}$ full row rank, let $W = \{\mbf{M}_w, \mbf{G}_w, \mbf{c}_w, \mbf{S}_w, \mbf{A}_w, \mbf{b}_w\}$, and $Z_\text{A} = \{\mbf{M}_\text{A}, \mbf{G}_\text{A}, \mbf{c}_\text{A}, \mbf{S}_\text{A}, \mbf{A}_\text{A}, \mbf{b}_\text{A}\} \triangleq \realset^{n-n_z} \times W$. 

Using \eqref{eq:desc_mzlimage}--\eqref{eq:desc_mzmsum}, the set $Z_k^{[i]}(\seq{\mbf{u}})$ is given by the MG-rep $\{ \mbf{M}_k^{[i]}, \mbf{G}_k^{[i]}, \mbf{c}_k^{[i]}(\seq{\mbf{u}}), \mbf{S}_k^{[i]}, \mbf{A}_k^{[i]},$ $\mbf{b}_k^{[i]}(\seq{\mbf{u}})\},$ 
where $\mbf{M}_k^{[i]}$, $\mbf{G}_k^{[i]}$, $\mbf{c}_k^{[i]}(\seq{\mbf{u}})$, $\mbf{S}_k^{[i]}$, $\mbf{A}_k^{[i]}$, and $\mbf{b}_k^{[i]}(\seq{\mbf{u}})$ are given by the recursive relations
\normalsize
\begin{align*}
& \mbf{c}_k^{[i]}(\seq{\mbf{u}}) = \begin{bmatrix} \tilde{\mbf{A}}_z^{[i]} \mbf{c}_{k-1}^{[i]}(\seq{\mbf{u}}) + \tilde{\mbf{B}}^{[i]} \mbf{u}_{k-1} \\ \mbf{c}_\text{A} \end{bmatrix}, \; \mbf{M}_k^{[i]} = \begin{bmatrix} \tilde{\mbf{A}}_z^{[i]} \mbf{M}_{k-1}^{[i]} & \mbf{0} \\ \mbf{0} & \mbf{M}_\text{A} \end{bmatrix}, \; \mbf{G}_k^{[i]} = \begin{bmatrix} \tilde{\mbf{A}}_z^{[i]} \mbf{G}_{k-1}^{[i]} & \mbf{0} \\ \mbf{0} & \mbf{G}_\text{A} \end{bmatrix}, \\
& \mbf{S}_k^{[i]} = \begin{bmatrix} \mbf{S}_{k-1}^{[i]} & \mbf{0} \\ \mbf{0} & \mbf{S}_\text{A} \\ \multicolumn{2}{c}{\check{\mbf{A}}_z^{[i]} \mbf{M}_{k}^{[i]}} \end{bmatrix}, \; \mbf{A}_k^{[i]} = \begin{bmatrix} \mbf{A}_{k-1}^{[i]} & \mbf{0} \\ \mbf{0} & \mbf{A}_\text{A} \\ \multicolumn{2}{c}{\check{\mbf{A}}_z^{[i]} \mbf{G}_{k}^{[i]}} \end{bmatrix}, \; \mbf{b}_k^{[i]}(\seq{\mbf{u}}) = \begin{bmatrix} \mbf{b}_{k-1}^{[i]}(\seq{\mbf{u}}) \\  \mbf{b}_\text{A} \\ -\check{\mbf{A}}_z^{[i]} \mbf{c}_k^{[i]}(\seq{\mbf{u}}) - \check{\mbf{B}}^{[i]} \mbf{u}_{k} \end{bmatrix},  
\end{align*}
\normalsize
for $k = 1,2,\ldots,N$.
Note that the third constraint in $(\mbf{S}_k^{[i]}, \mbf{A}_k^{[i]}, \mbf{b}_k^{[i]}(\seq{\mbf{u}}))$, comes from the fact that \eqref{eq:desc_systemSVDfaultconstraintsmz} must hold. Using the initial values \eqref{eq:initialAbmz}, the state reachable tube $\seq{Z}\zerospace^{[i]}(\seq{\mbf{u}})$ can be written as an explicit function of the input sequence $\seq{\mbf{u}}$ as $\seq{Z}\zerospace^{[i]}(\seq{\mbf{u}}) = \{\seq{\mbf{M}}\zerospace^{[i]}, \seq{\mbf{G}}\zerospace^{[i]}, \seq{\mbf{c}}\zerospace^{[i]}(\seq{\mbf{u}}), \seq{\mbf{S}}\zerospace^{[i]}, \seq{\mbf{A}}\zerospace^{[i]}, \seq{\mbf{b}}\zerospace^{[i]}(\seq{\mbf{u}}) \}$, where
\begin{equation} \label{eq:desc_stateckmz}
\seq{\mbf{c}}\zerospace^{[i]}(\seq{\mbf{u}}) = \mbf{Q}_N^{[i]} \mbf{c}_z + \mbf{p}_N^{[i]} + \seq{\mbf{H}}\zerospace^{[i]} \seq{\mbf{u}},\quad \seq{\mbf{M}}\zerospace^{[i]} = [ \mbf{Q}_N^{[i]} \mbf{M}_z^{[i]} \,\; \seq{\mbf{P}}\zerospace^{[i]}_\text{M}], \quad \seq{\mbf{G}}\zerospace^{[i]} = [ \mbf{Q}_N^{[i]} \mbf{G}_z^{[i]} \,\; \seq{\mbf{P}}\zerospace^{[i]}_\text{G}],
\end{equation}
\begin{equation}
\seq{\mbf{A}}\zerospace^{[i]} = \begin{bmatrix} \begin{bmatrix} \mbf{A}_z \\ \check{\mbf{A}}_z^{[i]} \mbf{G}_z^{[i]} \end{bmatrix} & \begin{bmatrix} \mbf{0} \\ \mbf{0} \end{bmatrix} & \cdots & \begin{bmatrix} \mbf{0} \\ \mbf{0} \end{bmatrix} \\
\check{\mbf{A}}_z^{[i]} \begin{bmatrix} \tilde{\mbf{A}}_z^{[i]} \\ \mbf{0} \end{bmatrix} \mbf{G}_z^{[i]} & \check{\mbf{A}}_z^{[i]} \begin{bmatrix} \mbf{0} \\ \mbf{G}_\text{A} \end{bmatrix} & \cdots & \begin{bmatrix} \mbf{0} \\ \mbf{0} \end{bmatrix} \\
\check{\mbf{A}}_z^{[i]} \begin{bmatrix} \tilde{\mbf{A}}_z^{[i]} \\ \mbf{0} \end{bmatrix}^2 \mbf{G}_z^{[i]} & \check{\mbf{A}}_z^{[i]} \begin{bmatrix} \tilde{\mbf{A}}_z^{[i]} \\ \mbf{0} \end{bmatrix} \begin{bmatrix} \mbf{0} \\ \mbf{G}_\text{A} \end{bmatrix}  & \cdots & \begin{bmatrix} \mbf{0} \\ \mbf{0} \end{bmatrix} \\
\vdots & \vdots & \ddots & \vdots \\
\check{\mbf{A}}_z^{[i]} \begin{bmatrix} \tilde{\mbf{A}}_z^{[i]} \\ \mbf{0} \end{bmatrix}^N \mbf{G}_z^{[i]} & \check{\mbf{A}}_z^{[i]} \begin{bmatrix} \tilde{\mbf{A}}_z^{[i]} \\ \mbf{0} \end{bmatrix}^{N-1} \begin{bmatrix} \mbf{0} \\ \mbf{G}_\text{A} \end{bmatrix}  & \cdots & \check{\mbf{A}}_z^{[i]} \begin{bmatrix} \mbf{0} \\ \mbf{G}_\text{A} \end{bmatrix} \\
\mbf{0} & \mbf{A}_\text{A} & \cdots & \mbf{0} \\
\mbf{0} & \mbf{0} & \cdots & \mbf{0} \\
\vdots & \vdots & \ddots & \vdots \\
\mbf{0} & \mbf{0} & \cdots & \mbf{A}_\text{A}
\end{bmatrix},
\end{equation}
\begin{equation}
\seq{\mbf{S}}\zerospace^{[i]} = \begin{bmatrix} \begin{bmatrix} \mbf{S}_z \\ \check{\mbf{A}}_z^{[i]} \mbf{M}_z^{[i]} \end{bmatrix} & \begin{bmatrix} \mbf{0} \\ \mbf{0} \end{bmatrix} & \cdots & \begin{bmatrix} \mbf{0} \\ \mbf{0} \end{bmatrix} \\
\check{\mbf{A}}_z^{[i]} \begin{bmatrix} \tilde{\mbf{A}}_z^{[i]} \\ \mbf{0} \end{bmatrix} \mbf{M}_z^{[i]} & \check{\mbf{A}}_z^{[i]} \begin{bmatrix} \mbf{0} \\ \mbf{M}_\text{A} \end{bmatrix} & \cdots & \begin{bmatrix} \mbf{0} \\ \mbf{0} \end{bmatrix} \\
\check{\mbf{A}}_z^{[i]} \begin{bmatrix} \tilde{\mbf{A}}_z^{[i]} \\ \mbf{0} \end{bmatrix}^2 \mbf{M}_z^{[i]} & \check{\mbf{A}}_z^{[i]} \begin{bmatrix} \tilde{\mbf{A}}_z^{[i]} \\ \mbf{0} \end{bmatrix} \begin{bmatrix} \mbf{0} \\ \mbf{M}_\text{A} \end{bmatrix}  & \cdots & \begin{bmatrix} \mbf{0} \\ \mbf{0} \end{bmatrix} \\
\vdots & \vdots & \ddots & \vdots \\
\check{\mbf{A}}_z^{[i]} \begin{bmatrix} \tilde{\mbf{A}}_z^{[i]} \\ \mbf{0} \end{bmatrix}^N \mbf{M}_z^{[i]} & \check{\mbf{A}}_z^{[i]} \begin{bmatrix} \tilde{\mbf{A}}_z^{[i]} \\ \mbf{0} \end{bmatrix}^{N-1} \begin{bmatrix} \mbf{0} \\ \mbf{M}_\text{A} \end{bmatrix}  & \cdots & \check{\mbf{A}}_z^{[i]} \begin{bmatrix} \mbf{0} \\ \mbf{M}_\text{A} \end{bmatrix} \\
\mbf{0} & \mbf{S}_\text{A} & \cdots & \mbf{0} \\
\mbf{0} & \mbf{0} & \cdots & \mbf{0} \\
\vdots & \vdots & \ddots & \vdots \\
\mbf{0} & \mbf{0} & \cdots & \mbf{S}_\text{A},
\end{bmatrix} 
\end{equation}
\begin{equation} \label{eq:desc_statebkmz}
\seq{\mbf{b}}\zerospace^{[i]} = \seq{\bm{\alpha}}\zerospace^{[i]} + \seq{\bm{\Lambda}}\zerospace^{[i]} \mbf{c}_z + \seq{\bm{\Omega}}\zerospace^{[i]} \seq{\mbf{u}},
\end{equation}
with
\begin{equation}
\mbf{Q}_N^{[i]} = \begin{bmatrix} \begin{bmatrix} \tilde{\mbf{A}}_z^{[i]} \\ \mbf{0} \end{bmatrix}^0 \\ \begin{bmatrix} \tilde{\mbf{A}}_z^{[i]} \\ \mbf{0} \end{bmatrix}^1 \\ \begin{bmatrix} \tilde{\mbf{A}}_z^{[i]} \\ \mbf{0} \end{bmatrix}^2 \\ \vdots \\ \begin{bmatrix} \tilde{\mbf{A}}_z^{[i]} \\ \mbf{0} \end{bmatrix}^N \end{bmatrix}, \quad
\mbf{p}_N^{[i]} = \begin{bmatrix} \begin{bmatrix} \mbf{0} \\ \mbf{0} \end{bmatrix} \\ \begin{bmatrix} \mbf{0} \\ \mbf{c}_\text{A} \end{bmatrix} \\ \begin{bmatrix} \tilde{\mbf{A}}_z^{[i]} \\ \mbf{0} \end{bmatrix} \begin{bmatrix} \mbf{0} \\ \mbf{c}_\text{A} \end{bmatrix} + \begin{bmatrix} \mbf{0} \\ \mbf{c}_\text{A} \end{bmatrix} \\ \vdots \\ \sum_{m=1}^N \begin{bmatrix} \tilde{\mbf{A}}_z^{[i]} \\ \mbf{0} \end{bmatrix}^{m-1} \begin{bmatrix} \mbf{0} \\ \mbf{c}_\text{A} \end{bmatrix} \end{bmatrix},
\end{equation}
\begin{equation}
\seq{\mbf{H}}\zerospace^{[i]} = \begin{bmatrix} \mbf{0} \\ \mbf{H}_1^{[i]} \\ \mbf{H}_2^{[i]} \\ \vdots \\ \mbf{H}_N^{[i]} \end{bmatrix}, \quad \mbf{H}_h^{[i]} = \left[\begin{matrix} \cdots & \underbrace{\begin{bmatrix} \tilde{\mbf{A}}_z^{[i]} \\ \mbf{0} \end{bmatrix}^{h-m} \begin{bmatrix} \tilde{\mbf{B}}^{[i]} \\ \mbf{0} \end{bmatrix}}_{m = 1,2,\ldots,h} & \cdots & \underbrace{\begin{bmatrix} \mbf{0} \\ \mbf{0} \end{bmatrix}}_{N-h+1 \text{ terms}} & \cdots \end{matrix} \right],
\end{equation}
\begin{equation}
\seq{\mbf{P}}\zerospace^{[i]}_\text{G} = \begin{bmatrix} \begin{bmatrix} \mbf{0} \\ \mbf{0} \end{bmatrix} & \begin{bmatrix} \mbf{0} \\ \mbf{0} \end{bmatrix} & \cdots & \begin{bmatrix} \mbf{0} \\ \mbf{0} \end{bmatrix} \\ \begin{bmatrix} \mbf{0} \\ \mbf{G}_\text{A} \end{bmatrix} & \begin{bmatrix} \mbf{0} \\ \mbf{0} \end{bmatrix} & \cdots & \begin{bmatrix} \mbf{0} \\ \mbf{0} \end{bmatrix} \\
\begin{bmatrix} \tilde{\mbf{A}}_z^{[i]} \\ \mbf{0} \end{bmatrix} \begin{bmatrix} \mbf{0} \\ \mbf{G}_\text{A} \end{bmatrix} & \begin{bmatrix} \mbf{0} \\ \mbf{G}_\text{A} \end{bmatrix} & \cdots & \begin{bmatrix} \mbf{0} \\ \mbf{0} \end{bmatrix} \\
\vdots & \vdots & \ddots & \vdots \\
\begin{bmatrix} \tilde{\mbf{A}}_z^{[i]} \\ \mbf{0} \end{bmatrix}^{N-1}  \begin{bmatrix} \mbf{0} \\ \mbf{G}_\text{A} \end{bmatrix} & \begin{bmatrix} \tilde{\mbf{A}}_z^{[i]} \\ \mbf{0} \end{bmatrix}^{N-2} \begin{bmatrix} \mbf{0} \\ \mbf{G}_\text{A} \end{bmatrix}  & \cdots & \begin{bmatrix} \mbf{0} \\ \mbf{G}_\text{A} \end{bmatrix} \end{bmatrix},
\end{equation}
\begin{equation}
\seq{\mbf{P}}\zerospace^{[i]}_\text{M} = \begin{bmatrix} \begin{bmatrix} \mbf{0} \\ \mbf{0} \end{bmatrix} & \begin{bmatrix} \mbf{0} \\ \mbf{0} \end{bmatrix} & \cdots & \begin{bmatrix} \mbf{0} \\ \mbf{0} \end{bmatrix} \\ \begin{bmatrix} \mbf{0} \\ \mbf{M}_\text{A} \end{bmatrix} & \begin{bmatrix} \mbf{0} \\ \mbf{0} \end{bmatrix} & \cdots & \begin{bmatrix} \mbf{0} \\ \mbf{0} \end{bmatrix} \\
\begin{bmatrix} \tilde{\mbf{A}}_z^{[i]} \\ \mbf{0} \end{bmatrix} \begin{bmatrix} \mbf{0} \\ \mbf{M}_\text{A} \end{bmatrix} & \begin{bmatrix} \mbf{0} \\ \mbf{M}_\text{A} \end{bmatrix} & \cdots & \begin{bmatrix} \mbf{0} \\ \mbf{0} \end{bmatrix} \\
\vdots & \vdots & \ddots & \vdots \\
\begin{bmatrix} \tilde{\mbf{A}}_z^{[i]} \\ \mbf{0} \end{bmatrix}^{N-1}  \begin{bmatrix} \mbf{0} \\ \mbf{M}_\text{A} \end{bmatrix} & \begin{bmatrix} \tilde{\mbf{A}}_z^{[i]} \\ \mbf{0} \end{bmatrix}^{N-2} \begin{bmatrix} \mbf{0} \\ \mbf{M}_\text{A} \end{bmatrix}  & \cdots & \begin{bmatrix} \mbf{0} \\ \mbf{M}_\text{A} \end{bmatrix} \end{bmatrix},
\end{equation}
\begin{equation}
\seq{\bm{\alpha}}\zerospace^{[i]} = \seq{\bm{\beta}} + \seq{\bm{\Upsilon}}\zerospace^{[i]} \mbf{p}_N^{[i]}, \quad \seq{\bm{\Lambda}}\zerospace^{[i]} = \seq{\bm{\Upsilon}}\zerospace^{[i]} \mbf{Q}_N^{[i]}, \quad \seq{\bm{\Omega}}\zerospace^{[i]} = \seq{\bm{\Gamma}}\zerospace^{[i]} + \seq{\bm{\Upsilon}}\zerospace^{[i]} \seq{\mbf{H}}\zerospace^{[i]},
\end{equation}
\begin{equation}
\seq{\bm{\beta}} = \begin{bmatrix} \begin{bmatrix} \mbf{b}_z \\ \mbf{0} \end{bmatrix} \\ \mbf{0} \\ \mbf{0} \\ \vdots \\ \mbf{0} \\ \mbf{b}_\text{A} \\ \mbf{b}_\text{A} \\ \vdots \\ \mbf{b}_\text{A} \end{bmatrix}, \quad 
\seq{\bm{\Upsilon}}\zerospace^{[i]} = \begin{bmatrix} \mbf{0} & \mbf{0} & \mbf{0} & \cdots & \mbf{0} \\ -\check{\mbf{A}}_z^{[i]} & \mbf{0} & \mbf{0} & \cdots & \mbf{0} \\ \mbf{0} &  -\check{\mbf{A}}_z^{[i]} & \mbf{0} & \cdots & \mbf{0} \\ \mbf{0} & \mbf{0} & -\check{\mbf{A}}_z^{[i]} & \cdots & \mbf{0} \\ \vdots & \vdots & \vdots & \ddots & \vdots \\
\mbf{0} & \mbf{0} & \mbf{0} & \cdots & -\check{\mbf{A}}_z^{[i]} \\ \mbf{0} & \mbf{0} & \mbf{0} & \cdots & \mbf{0} \\  \mbf{0} & \mbf{0} & \mbf{0} & \cdots & \mbf{0} \\ \vdots & \vdots & \vdots & \ddots & \vdots  \\ \mbf{0} & \mbf{0} & \mbf{0} & \cdots & \mbf{0} \end{bmatrix},
\end{equation}
\begin{equation}
\seq{\bm{\Gamma}}\zerospace^{[i]} = \begin{bmatrix} \mbf{0} & \mbf{0} & \mbf{0} & \cdots & \mbf{0} \\ -\check{\mbf{B}}^{[i]} & \mbf{0} & \mbf{0} & \cdots & \mbf{0} \\ \mbf{0} & -\check{\mbf{B}}^{[i]} & \mbf{0} & \cdots & \mbf{0} \\ \mbf{0} & \mbf{0} & -\check{\mbf{B}}^{[i]} & \cdots & \mbf{0} \\ \vdots & \vdots & \vdots & \ddots & \vdots \\
\mbf{0} & \mbf{0} & \mbf{0} & \cdots & -\check{\mbf{B}}^{[i]} \\ \mbf{0} & \mbf{0} & \mbf{0} & \cdots & \mbf{0} \\ \mbf{0} & \mbf{0} & \mbf{0} & \cdots & \mbf{0} \\ \vdots & \vdots & \vdots & \ddots & \vdots  \\ \mbf{0} & \mbf{0} & \mbf{0} & \cdots & \mbf{0} \end{bmatrix},
\end{equation}
\begin{equation*}
\mbf{c}_\text{A} = (\mbf{c}_\text{R}, \mbf{c}_w), \quad \mbf{G}_\text{A} = \begin{bmatrix} \mbf{0} \\ \mbf{G}_w \end{bmatrix}, \quad \mbf{M}_\text{A} = \begin{bmatrix} \mbf{M}_\text{R} & \mbf{0} \\ \mbf{0} & \mbf{M}_w \end{bmatrix}, \quad \mbf{S}_\text{A} = [\mbf{0} \,\; \mbf{S}_w] \quad \mbf{A}_\text{A} = \mbf{A}_w, \quad \mbf{b}_\text{A} = \mbf{b}_w.
\end{equation*}

\subsection{Output reachable tube}

Since $\seq{Z}\zerospace^{[i]}(\seq{\mbf{u}})$ is a mixed zonotope, the output reachable tube $\seq{Y}\zspace^{[i]}(\seq{\mbf{u}})$ is then a mixed zonotope obtained in accordance with \eqref{eq:systemSVDfaultoutputmz} as 
\begin{equation} \label{eq:desc_outputreachablemz}
\seq{Y}\zerospace^{[i]}(\seq{\mbf{u}}) = \seq{\mbf{F}}\zerospace^{[i]} \seq{Z}\zerospace^{[i]}(\seq{\mbf{u}}) \oplus \seq{\mbf{D}}\zerospace^{[i]} \seq{\mbf{u}} \oplus \seq{\mbf{D}}\zerospace^{[i]}_v \seq{V},
\end{equation}
where
\begin{equation*}
\seq{\mbf{F}}\zerospace^{[i]} = 
\begin{bmatrix} \mbf{F}^{[i]} & \mbf{0} & \cdots & \mbf{0} \\ \mbf{0} & \mbf{F}^{[i]} & \cdots & \mbf{0} \\
\vdots & \vdots & \ddots & \vdots \\
\mbf{0} & \mbf{0} & \cdots & \mbf{F}^{[i]}
\end{bmatrix}, \;
\seq{\mbf{D}}\zerospace^{[i]} = 
\begin{bmatrix} \mbf{D}^{[i]} & \mbf{0} & \cdots & \mbf{0} \\ \mbf{0} & \mbf{D}^{[i]} & \cdots & \mbf{0} \\
\vdots & \vdots & \ddots & \vdots \\
\mbf{0} & \mbf{0} & \cdots & \mbf{D}^{[i]}
\end{bmatrix}, \;
\seq{\mbf{D}}\zerospace^{[i]}_v = 
\begin{bmatrix} \mbf{D}^{[i]}_v & \mbf{0} & \cdots & \mbf{0} \\ \mbf{0} & \mbf{D}^{[i]}_v & \cdots & \mbf{0} \\
\vdots & \vdots & \ddots & \vdots \\
\mbf{0} & \mbf{0} & \cdots & \mbf{D}^{[i]}_v
\end{bmatrix}.
\end{equation*}
Using properties \eqref{eq:desc_mzlimage} and \eqref{eq:desc_mzmsum}, and letting $V = \{ \mbf{M}_v, \mbf{G}_v, \mbf{c}_v, \mbf{S}_v, \mbf{A}_v, \mbf{b}_v\}$, the output reachable tube is given by the MG-rep $\seq{Y}\zerospace^{[i]}(\seq{\mbf{u}}) = \{\seq{\mbf{M}}\zerospace^{Y[i]}, \seq{\mbf{G}}\zerospace^{Y[i]}, \seq{\mbf{c}}\zerospace^{Y[i]}(\seq{\mbf{u}}), \seq{\mbf{S}}\zerospace^{Y[i]}, \seq{\mbf{A}}\zerospace^{Y[i]}, \seq{\mbf{b}}\zerospace^{Y[i]}(\seq{\mbf{u}}) \}$, in which
\begin{align}
& \seq{\mbf{c}}\zerospace^{Y[i]}(\seq{\mbf{u}}) = \seq{\mbf{F}}\zerospace^{[i]} \seq{\mbf{c}}\zerospace^{[i]}(\seq{\mbf{u}}) + \seq{\mbf{D}}\zerospace^{[i]} \seq{\mbf{u}} + \seq{\mbf{D}}\zerospace^{[i]}_v \seq{\mbf{c}}_v, \label{eq:desc_outputckmz} \\
& \seq{\mbf{M}}\zerospace^{Y[i]} = \big[\seq{\mbf{F}}\zerospace^{[i]} \seq{\mbf{M}}\zerospace^{[i]} \,\; \seq{\mbf{D}}\zerospace^{[i]}_v \seq{\mbf{M}}\zerospace^{[i]}_v \big],\quad \seq{\mbf{G}}\zerospace^{Y[i]} = \big[\seq{\mbf{F}}\zerospace^{[i]} \seq{\mbf{G}}\zerospace^{[i]} \,\; \seq{\mbf{D}}\zerospace^{[i]}_v \seq{\mbf{G}}\zerospace^{[i]}_v \big],\\
& \seq{\mbf{S}}\zerospace^{Y[i]} = \begin{bmatrix} \seq{\mbf{S}}\zerospace^{[i]} & \mbf{0} \\ \mbf{0} & \seq{\mbf{S}}\zerospace^{[i]}_v \end{bmatrix},\quad \seq{\mbf{A}}\zerospace^{Y[i]} = \begin{bmatrix} \seq{\mbf{A}}\zerospace^{[i]} & \mbf{0} \\ \mbf{0} & \seq{\mbf{A}}\zerospace^{[i]}_v \end{bmatrix},\quad \seq{\mbf{b}}\zerospace^{Y[i]}(\seq{\mbf{u}}) = \begin{bmatrix} \seq{\mbf{b}}\zerospace^{[i]}(\seq{\mbf{u}}) \\ \seq{\mbf{b}}\zerospace^{[i]}_v \end{bmatrix}.  \label{eq:desc_outputbkmz}
\end{align}

\subsection{Separation of output reachable tubes}

Consider an input sequence $\seq{\mbf{u}} \in \seq{U}$ to be injected into the set of models \eqref{eq:desc_systemSVDfaultmz}, and let $\seq{\mbf{y}}\zspace^{[i]} \triangleq (\mbf{y}_0^{[i]}, \mbf{y}_1^{[i]}, \ldots, \mbf{y}_N^{[i]})$ denote the observed output sequence of model $i \in \modelset$. We are interested in the design of an input sequence such that the relation $\seq{\mbf{y}}\zspace^{[i]} \in \seq{Y}\zspace^{[i]}(\seq{\mbf{u}})$ is valid for only one $i \in \modelset$.
\begin{definition} \rm \label{def:separatinginputtube}
	An input sequence $\seq{\mbf{u}}$ is said to be a \emph{separating input} on $k \in [0, N]$ if, for every $i,j \in \modelset$, $i \neq j$,
	\begin{equation} \label{eq:separatinginputdeftube}
	\seq{Y}\zspace^{[i]}(\seq{\mbf{u}}) \cap \seq{Y}\zspace^{[j]}(\seq{\mbf{u}}) = \emptyset.
	\end{equation}
\end{definition}
Clearly, if \eqref{eq:separatinginputdeftube} holds for all $i,j \in \modelset$, $i \neq j$, then $\seq{\mbf{y}}\zspace^{[i]} \in \seq{Y}\zspace^{[i]}(\seq{\mbf{u}})$ must hold only for one $i \in \modelset$. In the case that this is not valid for any $i \in \modelset$, one concludes that the real dynamics does not belong to the set of models \eqref{eq:desc_systemSVDfaultmz}. The following theorem is based on the computation of $\seq{Y}\zspace^{[i]}(\seq{\mbf{u}})$ expressed by \eqref{eq:desc_outputreachablemz} and the results in \cite{Raimondo2016}.

\begin{theorem} \rm \label{thm:desc_separatinginputmz}
	An input $\seq{\mbf{u}} \in \seq{U}$ is a separating input iff
	\begin{equation} \label{eq:desc_separationconditionmztube}
	 \begin{bmatrix} \seq{\mbf{N}}(i,j) \\ \seq{\bm{\Omega}}(i,j) \end{bmatrix} \seq{\mbf{u}} \notin \seq{\mathcal{Y}}(i,j) \triangleq \left\{ \begin{bmatrix} \seq{\mbf{M}}\zerospace^{Y}(i,j) \\ \seq{\mbf{S}}\zerospace^{Y}(i,j) \end{bmatrix}, \begin{bmatrix} \seq{\mbf{G}}\zerospace^Y(i,j) \\ \seq{\mbf{A}}\zerospace^Y(i,j) \end{bmatrix}, \; \begin{bmatrix} \seq{\mbf{c}}\zerospace^Y(i,j) \\ -\seq{\mbf{b}}\zerospace^Y(i,j) \end{bmatrix}, \noarg, \noarg, \noarg \right\}
	\end{equation}
	$\forall i,j \in \modelset$, $i \neq j$, where $\mbf{N}(i,j) = \seq{\mbf{F}}\zerospace^{[j]} \seq{\mbf{H}}\zerospace^{[j]} + \seq{\mbf{D}}\zerospace^{[j]} - (\seq{\mbf{F}}\zerospace^{[i]} \seq{\mbf{H}}\zerospace^{[i]} + \seq{\mbf{D}}\zerospace^{[i]})$, $\bm{\Omega}(i,j) = [(\bm{\Omega}_N^{[i]})^T \; \mbf{0} \; (\bm{\Omega}_N^{[j]})^T \; \mbf{0}]^T$, and 
	\begin{align*}
	& \seq{\mbf{M}}\zerospace^Y(i,j) \triangleq [\seq{\mbf{M}}\zerospace^{Y[i]} \,\; - \seq{\mbf{M}}\zerospace^{Y[j]}], \quad \seq{\mbf{G}}\zerospace^Y(i,j) \triangleq [\seq{\mbf{G}}\zerospace^{Y[i]} \,\; - \seq{\mbf{G}}\zerospace^{Y[j]}],\quad \seq{\mbf{c}}\zspace^Y(i,j) = \seq{\mbf{c}}\zspace^{Y[i]}(\seq{\mbf{0}}) - \seq{\mbf{c}}\zspace^{Y[j]}(\seq{\mbf{0}}),\\
	& \seq{\mbf{S}}\zspace^Y(i,j) \triangleq \begin{bmatrix} \seq{\mbf{S}}\zerospace^{Y[i]} & \mbf{0} \\ \mbf{0} & \seq{\mbf{S}}\zerospace^{Y[j]} \end{bmatrix}, \quad \seq{\mbf{A}}\zspace^Y(i,j) \triangleq \begin{bmatrix} \seq{\mbf{A}}\zerospace^{Y[i]} & \mbf{0} \\ \mbf{0} & \seq{\mbf{A}}\zerospace^{Y[j]} \end{bmatrix},\quad \seq{\mbf{b}}\zspace^Y(i,j) = \begin{bmatrix} \seq{\mbf{b}}\zspace^{Y[i]}(\seq{\mbf{0}}) \\ \seq{\mbf{b}}\zspace^{Y[j]}(\seq{\mbf{0}}) \end{bmatrix},
	\end{align*}
	where $\seq{\mbf{0}}$ denotes the zero input sequence.
\end{theorem}

\proof
The relations below follow from \eqref{eq:desc_stateckmz}, \eqref{eq:desc_statebkmz}, \eqref{eq:desc_outputckmz}, and \eqref{eq:desc_outputbkmz},%
\begin{align}
& \seq{\mbf{c}}\zerospace^{Y[i]}(\seq{\mbf{u}}) = \seq{\mbf{c}}\zerospace^{Y[i]}(\mbf{0}) + \seq{\mbf{F}}\zerospace^{[i]} \seq{\mbf{H}}\zerospace^{[i]} \seq{\mbf{u}} + \seq{\mbf{D}}\zerospace^{[i]} \seq{\mbf{u}} = \seq{\mbf{c}}\zerospace^{Y[i]}(\mbf{0}) + (\seq{\mbf{F}}\zerospace^{[i]} \seq{\mbf{H}}\zerospace^{[i]} + \seq{\mbf{D}}\zerospace^{[i]}) \seq{\mbf{u}}, \label{eq:desc_outputcseq0} \\
& \seq{\mbf{b}}\zerospace^{Y[i]}(\seq{\mbf{u}}) = \seq{\mbf{b}}\zerospace^{Y[i]}(\mbf{0}) + \begin{bmatrix} \seq{\bm{\Omega}}\zerospace^{[i]} \\ \mbf{0} \end{bmatrix} \seq{\mbf{u}}. \label{eq:desc_outputbseq0}
\end{align}
From \eqref{eq:desc_mzintersection} with $\mbf{R} = \mbf{I}$, \eqref{eq:separatinginputdeftube} is true iff $\nexists \bm{\xi} \in B_\infty, \bm{\delta} \in \realset$ such that
\begin{equation*}
\begin{bmatrix} \seq{\mbf{S}}\zerospace^{Y[i]} & \mbf{0} \\ \mbf{0} & \seq{\mbf{S}}\zerospace^{Y[j]} \\ \seq{\mbf{M}}\zerospace^{Y[i]} & -\seq{\mbf{M}}\zerospace^{Y[j]} \end{bmatrix} \bm{\delta} + \begin{bmatrix} \seq{\mbf{A}}\zerospace^{Y[i]} & \mbf{0} \\\mbf{0} & \seq{\mbf{A}}\zerospace^{Y[j]} \\ \seq{\mbf{G}}\zerospace^{Y[i]} & -\seq{\mbf{G}}\zerospace^{Y[j]} \end{bmatrix} \bm{\xi} = \begin{bmatrix} \seq{\mbf{b}}\zerospace^{Y[i]}(\seq{\mbf{u}}) \\ \seq{\mbf{b}}\zerospace^{Y[j]}(\seq{\mbf{u}}) \\  \seq{\mbf{c}}\zerospace^{Y[j]}(\seq{\mbf{u}}) - \seq{\mbf{c}}\zerospace^{Y[i]}(\seq{\mbf{u}}) \end{bmatrix}.
\end{equation*}
According to \eqref{eq:desc_outputcseq0}--\eqref{eq:desc_outputbseq0}, one has $\seq{\mbf{c}}\zerospace^{Y[i]}(\seq{\mbf{u}}) = \seq{\mbf{c}}\zerospace^{Y[i]}(\mbf{0}) + \seq{\mbf{F}}\zerospace^{[i]} \seq{\mbf{H}}\zerospace^{[i]} \seq{\mbf{u}} + \seq{\mbf{D}}\zerospace^{[i]} \seq{\mbf{u}} = \seq{\mbf{c}}\zerospace^{Y[i]}(\mbf{0}) + (\seq{\mbf{F}}\zerospace^{[i]} \seq{\mbf{H}}\zerospace^{[i]} + \seq{\mbf{D}}\zerospace^{[i]}) \seq{\mbf{u}} = - \seq{\mbf{c}}\zerospace^Y(i,j) + \seq{\mbf{N}}(i,j) \seq{\mbf{u}}$, $[(\mbf{b}_N^{Y[i]}(\seq{\mbf{u}}))^T  \;  (\mbf{b}_N^{Y[j]}(\seq{\mbf{u}}))^T]^T $ $ \!=\! \seq{\mbf{b}}\zspace^Y_N(i,j)$ $+ \bm{\Omega}(i,j) \seq{\mbf{u}}$,
with $\seq{\mbf{c}}\zspace^Y(i,j)$, $\seq{\mbf{b}}\zspace^Y(i,j)$, $\seq{\mbf{N}}(i,j)$, and $\seq{\bm{\Omega}}(i,j)$ defined as in the statement of the theorem. Then, \eqref{eq:separatinginputdeftube} holds iff $\nexists \bm{\xi} \in B_\infty$ and $\bm{\delta} \in \realset$ such that $\seq{\mbf{M}}\zerospace^{Y}(i,j) \bm{\delta} + \seq{\mbf{G}}\zerospace^{Y}(i,j) \bm{\xi} = - \seq{\mbf{c}}\zerospace^Y(i,j) + \seq{\mbf{N}}(i,j) \seq{\mbf{u}}$, and $\seq{\mbf{S}}\zerospace^{Y}(i,j) \bm{\delta} + \seq{\mbf{A}}\zerospace^{Y}(i,j) \bm{\xi} = \seq{\mbf{b}}\zerospace^{Y}(i,j) + \seq{\bm{\Omega}}(i,j) \seq{\mbf{u}}$. This is equivalent to
\begin{align*}
\nexists (\bm{\delta},\bm{\xi}) \in \realset \times B_\infty: \begin{bmatrix} \seq{\mbf{M}}\zerospace^{Y}(i,j) \\ \seq{\mbf{S}}\zerospace^{Y}(i,j) \end{bmatrix} \bm{\delta} + \begin{bmatrix} \seq{\mbf{G}}\zerospace^{Y}(i,j) \\ \seq{\mbf{A}}\zerospace^{Y}(i,j) \end{bmatrix} \bm{\xi} + \begin{bmatrix} \seq{\mbf{c}}\zerospace^{Y}(i,j) \\ - \seq{\mbf{b}}\zerospace^{Y}(i,j) \end{bmatrix} = \begin{bmatrix} \seq{\mbf{N}}(i,j) \\ \seq{\bm{\Omega}}(i,j) \end{bmatrix} \seq{\mbf{u}}, 
\end{align*}
which in turn holds iff \eqref{eq:desc_separationconditionmztube} is satisfied, with $\mbf{M}^Y_N(i,j)$, $\mbf{G}^Y_N(i,j)$, $\mbf{S}^Y_N(i,j)$, and $\mbf{A}^Y_N(i,j)$ defined as in the statement of the theorem. \qed

For simplicity, let $(\cdot)^{[q]} \triangleq (\cdot)(i,j)$, with $q \in \{1,2,\ldots,n_q\}$, where $n_q$ denotes the number of possible combinations of $i,j \in \modelset$, $i < j$.

\begin{lemma} \label{thm:desc_separatinginputmzequiv} \rm 
Let 
\begin{equation*}
\mbf{N}^{\dagger[q]} \triangleq \begin{bmatrix} \seq{\mbf{N}}\zerospace^{[q]} \\ \seq{\bm{\Omega}}\zerospace^{[q]} \end{bmatrix}, ~ \mbf{G}^{\dagger[q]} \triangleq \begin{bmatrix} \seq{\mbf{G}}\zerospace^{Y[q]} \\ \seq{\mbf{A}}\zerospace^{Y[q]} \end{bmatrix}, ~ \mbf{M}^{\dagger[q]} \triangleq \begin{bmatrix} \seq{\mbf{M}}\zerospace^{Y[q]} \\ \seq{\mbf{S}}\zerospace^{Y[q]} \end{bmatrix}, ~ \mbf{c}^{\dagger[q]} \triangleq \begin{bmatrix} \seq{\mbf{c}}\zerospace^{Y[q]} \\ - \seq{\mbf{b}}\zerospace^{Y[q]} \end{bmatrix}.
\end{equation*}
Moreover, let $\mbf{M}^{\dagger[q]}$ be full column rank. Then, there exist matrices $\mbf{M}^{-[q]}$, $\mbf{M}^{+[q]}$, $\mbf{G}^{-[q]}$, $\mbf{G}^{+[q]}$, $\mbf{N}^{-[q]}$, $\mbf{N}^{+[q]}$, and vectors $\mbf{c}^{-[q]}$, $\mbf{c}^{+[q]}$, such that
\begin{equation*}
\mbf{N}^{\dagger[q]} \seq{\mbf{u}} \notin \seq{\mathcal{Y}}\zerospace^{[q]} = \{\mbf{M}^{\dagger[q]}, \mbf{G}^{\dagger[q]}, \mbf{c}^{\dagger[q]}, \noarg, \noarg, \noarg\} \iff \mathring{\mbf{N}}^{[q]} \seq{\mbf{u}} \notin \mathring{\mathcal{Y}}^{[q]} \triangleq \{ \mathring{\mbf{G}}^{[q]}, \mathring{\mbf{c}}^{[q]} \},
\end{equation*}
with $\mathring{\mbf{G}}^{[q]} \triangleq \mbf{G}^{-[q]} - \mbf{M}^{-[q]}\inv{(\mbf{M}^{+[q]})} \mbf{G}^{+[q]}$, $\mathring{\mbf{c}}^{[q]} \triangleq \mbf{c}^{-[q]} - \mbf{M}^{-[q]}\inv{(\mbf{M}^{+[q]})} \mbf{c}^{+[q]}$, $\mathring{\mbf{N}}^{[q]} \triangleq \mbf{N}^{-[q]} - \mbf{M}^{-[q]}\inv{(\mbf{M}^{+[q]})} \mbf{N}^{+[q]}$.
\end{lemma}

\proof
Recall that $\mbf{N}^{\dagger[q]} \seq{\mbf{u}} \notin \seq{\mathcal{Y}}\zerospace^{[q]} \iff \nexists (\bm{\delta},\bm{\xi}) \in \realset^{n_\delta} \times B_\infty^{n_g} : \mbf{M}^{\dagger[q]} \bm{\delta} + \mbf{G}^{\dagger[q]} \bm{\xi} + \mbf{c}^{\dagger[q]} = \mbf{N}^{\dagger[q]} \seq{\mbf{u}}$. Since $\mbf{M}^{\dagger[q]}$ is full column rank, then by rearranging the rows of $\mbf{M}^{\dagger[q]} \bm{\delta} + \mbf{G}^{\dagger[q]} \bm{\xi} + \mbf{c}^{\dagger[q]} = \mbf{N}^{\dagger[q]} \seq{\mbf{u}}$, there exists an invertible matrix $\mbf{M}^{+[q]}$ satisfying
\begin{equation} \label{eq:lemmazonotopeequiv1}
\begin{bmatrix} \mbf{M}^{+[q]} \\ \mbf{M}^{-[q]} \end{bmatrix} \bm{\delta} + \begin{bmatrix} \mbf{G}^{+[q]} \\ \mbf{G}^{-[q]} \end{bmatrix} \bm{\xi} + \begin{bmatrix} \mbf{c}^{+[q]} \\ \mbf{c}^{-[q]} \end{bmatrix} = \begin{bmatrix} \mbf{N}^{+[q]} \\ \mbf{N}^{-[q]} \end{bmatrix} \seq{\mbf{u}}.
\end{equation}
Therefore, $\mbf{M}^{+[q]} \bm{\delta} + \mbf{G}^{+[q]} \bm{\xi} + \mbf{c}^{+[q]} = \mbf{N}^{+[q]} \seq{\mbf{u}}$ holds, implying that $\bm{\delta} = \inv{(\mbf{M}^{+[q]})}(- \mbf{G}^{+[q]} \bm{\xi} - \mbf{c}^{+[q]} + \mbf{N}^{+[q]} \seq{\mbf{u}})$. Substituting the latter in the lower part of \eqref{eq:lemmazonotopeequiv1} leads to
\begin{equation*}
(\mbf{G}^{-[q]} - \mbf{M}^{-[q]}\inv{(\mbf{M}^{+[q]})} \mbf{G}^{+[q]}) \bm{\xi} + (\mbf{I} - \mbf{M}^{-[q]}\inv{(\mbf{M}^{+[q]})}) \mbf{c}^{+[q]} = (\mbf{N}^{-[q]} - \mbf{M}^{-[q]}\inv{(\mbf{M}^{+[q]})} \mbf{N}^{+[q]}) \seq{\mbf{u}},
\end{equation*}
which proves the lemma. \qed

Let $\mathring{\mbf{N}}^{[q]}$ and $\mathring{\mathcal{Y}}^{[q]} = \{\mathring{\mbf{G}}^{[q]}, \mathring{\mbf{c}}^{[q]}\}$ be defined as in Lemma \ref{thm:desc_separatinginputmzequiv}, for each $q \in \{ 1,2,\ldots,n_q\}$. As in the constrained zonotope method, $\mathring{Y}^{[q]}$ is a zonotope. Then, the relation $\mathring{\mbf{N}}^{[q]} \seq{\mbf{u}} \notin \mathring{\mathcal{Y}}^{[q]}$ can be verified by solving an LP. In this sense, the following lemma provides an effective way to verify if a given input sequence is a separating input according to Theorem \ref{thm:desc_separatinginputmz}, consequently satisfying \eqref{eq:separatinginputdeftube}.

\begin{lemma} \rm \label{lem:desc_verifyseparating_tubemz}
	Let $\mathring{\mathcal{Y}}^{[q]} = \{\mathring{\mbf{G}}^{[q]}, \mathring{\mbf{c}}^{[q]}\}$. For each $\seq{\mbf{u}} \in \seq{U}$ and $q \in \{1,2,\ldots,n_q\}$, define $\hat{\delta}^{[q]}(\seq{\mbf{u}}) = \underset{\delta^{[q]}, \bm{\xi}^{[q]}}{\min} \delta^{[q]}$, subject to
	\begin{equation*}
	\mathring{\mbf{N}}^{[q]} \seq{\mbf{u}} = \mathring{\mbf{G}}^{[q]} \bm{\xi}^{[q]} + \mathring{\mbf{c}}^{[q]}, \quad \ninf{\bm{\xi}} \leq 1 + \delta^{[q]}.
	\end{equation*}
	Then $\mathring{\mbf{N}}^{[q]} \seq{\mbf{u}} \notin \mathring{\mathcal{Y}}^{[q]} \iff \hat{\delta}^{[q]}(\seq{\mbf{u}}) > 0$.
\end{lemma}
\proof See Lemma 4 in \cite{Scott2014}. \qed

For the AFD of the $n_m$ models in \eqref{eq:desc_systemSVDfaultmz} of the descriptor system, we consider the design of a separating input of minimum length according to the optimization problem
\begin{equation} \label{eq:desc_optimalseparatingmz}
\underset{\seq{\mbf{u}} \in \seq{U}}{\min} ~ \{J(\seq{\mbf{u}})
: \mathring{\mbf{N}}^{[q]} \seq{\mbf{u}} \notin \mathring{\mathcal{Y}}^{[q]}, ~ \forall q  = 1,2,\ldots,n_q\},   
\end{equation}
with $J(\seq{\mbf{u}})$ chosen to minimize any harmful effects caused by injecting $\seq{\mbf{u}}$ into \eqref{eq:desc_systemSVDfaultmz}. For simplicity we may choose $J(\seq{\mbf{u}}) = \sum_{j=0}^{N} \mbf{u}_{j}^T \mbf{R} \mbf{u}_{j}$, where $\mbf{R}$ is a weighting matrix. As in \cite{Scott2014}, this is a bilevel optimization problem and can be rewritten as a mixed-integer quadratic program by defining a minimum separation threshold $\varepsilon > 0$ such that $\varepsilon \leq \hat{\delta}^{[q]}(\seq{\mbf{u}})$, for all $q  = 1,2,\ldots,n_q$, similar to the MIQP presented in Chapter \ref{cha:faultdiagnosis}.

\section{Numerical examples} \label{sec:desc_examples}

\subsection{State estimation}

This section first evaluates the accuracy of the state estimation methods proposed in Sections \ref{sec:desc_estimationCZ} and \ref{sec:desc_estimationMZ} for descriptor systems using CZs and MZs, respectively (denoted by CZ and MZ). 
Consider system \eqref{eq:desc_system} with matrices $\mbf{E} = \text{diag}(1,1,0)$, $\mbf{B}_w = \text{diag}(0.1,1.5,0.6)$, $\mbf{D}_v = \text{diag}(0.5,1.5)$,
\begin{equation*}
\mbf{A} = \begin{bmatrix} 0.5 & 0 & 0 \\ 0.8 & 0.95 & 0 \\ -1 & 0.5 & 1 \end{bmatrix}, \; \mbf{B} = \begin{bmatrix} 1 & 0 \\ 0 & 1 \\ 0 & 0 \end{bmatrix}, \; \mbf{C} = \begin{bmatrix} 1 & 0 & 1 \\ 1 & -1 & 0 \end{bmatrix}, 
\end{equation*}
and $\mbf{D} = \mbf{0}$. The initial state $\mbf{x}_0$ is bounded by the zonotope
\begin{equation} \label{eq:desc_estimationx0cz}
X_0 = \left\{ \text{diag}(0.1, 1.5, 0.6),\, [ 0.5 \; 0.5 \; 0.25]^T \right\},
\end{equation}
and the uncertainties are random uniform noises bounded by $\|\mbf{w}_k\|_\infty \leq 1$, $\|\mbf{v}_k\|_\infty \leq 1$. The CZ in Assumption 1 is $X_\text{a} = \{50{\cdot}\eye{3},\mbf{0}\}$. On the other hand, for MZs the considered initial set is the entire state space $\realset^3$. The simulation is conducted for $\mbf{x}_0 = [ 0.5 \; 0.5 \; 0.25 ]^T$, and the complexity of the CZs and MZs is limited to 30 (bounded) generators and 5 constraints using the constraint elimination algorithm described by Method \ref{meth:czconelim} and the generator reduction described by Method \ref{meth:czgenred}. For MZs, unbounded generators are eliminated prior to eliminating constraints following the scheme proposed in Section \ref{sec:desc_mzcomplexityreduction}.
\begin{figure}[!tb]
	\centering{
		\def\svgwidth{0.7\columnwidth}
		{\scriptsize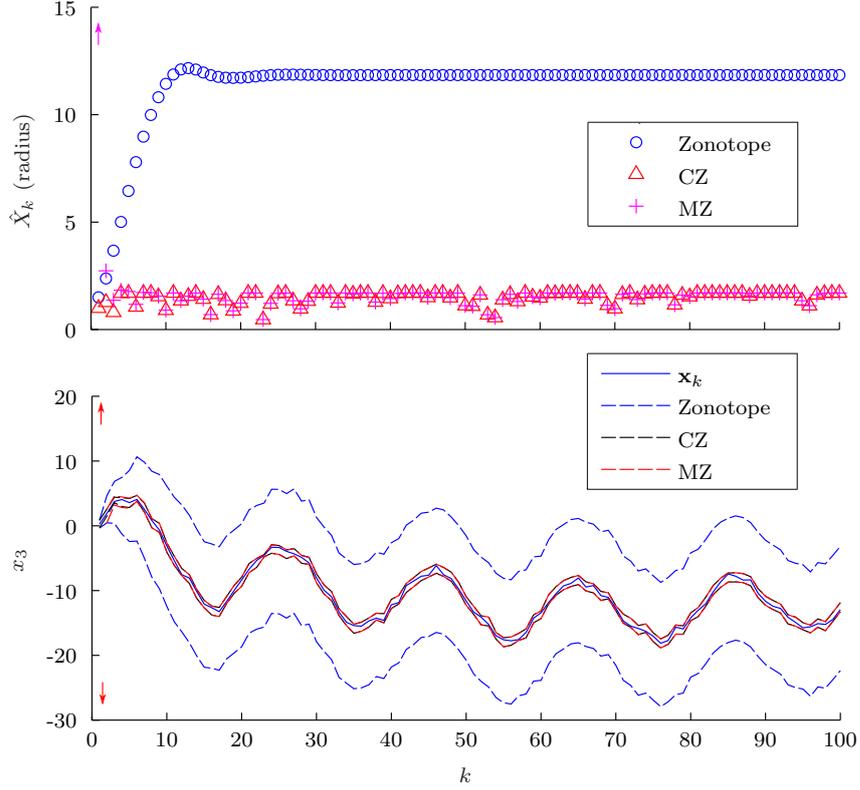}
		\caption{The radii of the enclosures $\hat{X}_k$ obtained using zonotope method proposed in \cite{Puig2018}, the constrained zonotope method proposed in Section \ref{sec:desc_estimationCZ}, and the mixed zonotope method proposed in Section \ref{sec:desc_estimationMZ} (top), as well as the projections of $\hat{X}_k$ onto $x_3$ (bottom).}\label{fig:desc_czestimationx3}}
\end{figure}

Figure \ref{fig:desc_czestimationx3} shows the radii of the enclosures $\hat{X}_k$ for $k \in [0,100]$ obtained using the state estimation algorithms proposed in Sections \ref{sec:desc_estimationCZ} and \ref{sec:desc_estimationMZ}. Results obtained using the zonotope method in \cite{Puig2018} are presented for comparison\footnote{Specifically, this is the set-membership approach proposed in \cite{Puig2018} with Kalman correction matrix.}. 
The complexity of the zonotopes is limited to 30 generators using Method \ref{meth:genredB}. Figure \ref{fig:desc_czestimationx3} shows also the projections of $\hat{X}_k$ onto $x_3$. As it can be noticed, both CZ and MZ provide substantially sharper bounds in comparison to zonotopes. This is possible since the enclosures proposed in Lemmas \ref{lem:desc_predictioncz} and \ref{lem:desc_predictionmz} take into account the static constraints explicitly, while zonotopes provide only a conservative bound of the corresponding feasible region. In addition, as it can be noticed, MZ is able to provide enclosures that are as accurate as the ones obtained using CZ, but without requiring the knowledge of bounded initial and admissible sets. As a side note, the enclosure $\hat{X}_k$ provided by MZ at $k=1$ is unbounded (denoted by the arrows towards minus and plus infinity), but this enclosure becomes bounded at $k=2$ since more measurements are added to the computed set. %

\subsection{Fault diagnosis using constrained zonotopes} \label{sec:desc_examplesAFDCZ}

\begin{figure*}[!tb]
	\centering{
		\def\svgwidth{\columnwidth}
		{\scriptsize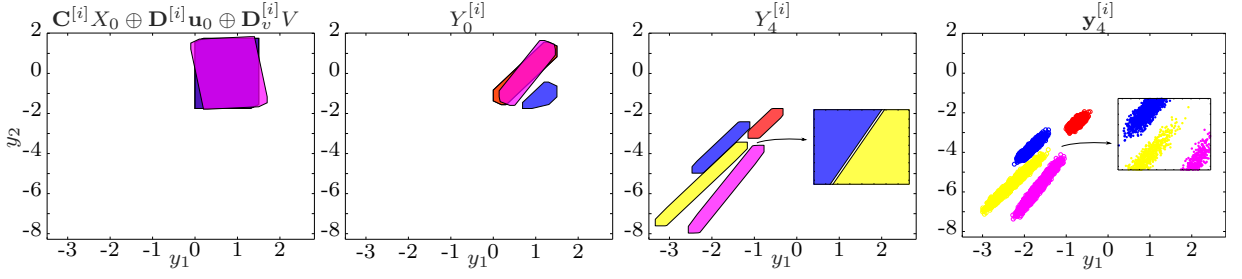}
		\vspace{-2mm}
		\caption{The sets $\mbf{C}^{[i]} X_0 \oplus \mbf{D}^{[i]} \mbf{u}_0 \oplus \mbf{D}_v^{[i]} V$, the output reachable sets $Y_0^{[i]}$, $Y_4^{[i]}$, and 2000 samples for $\mbf{y}^{[i]}_4$, with $i=1$ (yellow), $i=2$ (red), $i=3$ (blue), and $i=4$ (magenta), obtained by the injection of the input sequence $\seq{\mbf{u}}$.}\label{fig:desc_fdiczlast}}
\end{figure*}

We now evaluate the effectiveness of the AFD method proposed in Section \ref{sec:desc_AFDCZ}. Consider the set of models \eqref{eq:desc_systemfaulty} with model $i = 1$ described in the previous example, and
\begin{equation*}
\begin{aligned}
& \mbf{A}^{[2]} = \begin{bmatrix} 0.5 & 0 & 0 \\ 0.8 & 0.6 & 0 \\ -1 & 0.5 & 1 \end{bmatrix}, \; \mbf{B}^{[3]} = \begin{bmatrix} 1 & 0 \\ 0 & 0 \\ -1 & 0 \end{bmatrix}, \; \mbf{C}^{[4]} = \begin{bmatrix} 1 & 0.1 & 1 \\ 1 & -1 & 0.1 \end{bmatrix},
\end{aligned}
\end{equation*}
$\mbf{E}^{[i]} = \mbf{E}^{[1]}$, $\mbf{B}^{[2]} = \mbf{B}^{[4]} = \mbf{B}^{[1]}$, $\mbf{C}^{[2]} = \mbf{C}^{[3]} = \mbf{C}^{[1]}$, $\mbf{B}_w^{[i]} = \mbf{B}_w^{[1]}$, and $\mbf{D}_v^{[i]} = \mbf{D}_v^{[1]}$, $\mbf{D}^{[i]} = \mbf{D}^{[1]}$, $i \in \{2,3,4\}$. The initial state $\mbf{x}_0^{[i]}$ is bounded by \eqref{eq:desc_estimationx0cz}, 
the uncertainties are bounded by $\|\mbf{w}_k\|_\infty \leq 0.1$, $\|\mbf{v}_k\|_\infty \leq 0.1$, and the input is limited by $\|\mbf{u}_k\|_\infty \leq 1$. Let $X_\text{a} = \{50{\cdot}\eye{3},\mbf{0}\}$ and $\varepsilon = 0.01$. All the models $i \in \modelset$ are considered to be faulty and must be separated. The number of generators of $\mathcal{Y}(i,j)$ was limited to two times its dimension using Method \ref{meth:genredB}. The minimum length optimal input sequence that solves \eqref{eq:desc_optimalseparatingcz} was obtained using CPLEX 12.8 and MATLAB 9.1, with $J(\seq{\mbf{u}}) = \seq{\mbf{u}}\zerospace^T \seq{\mbf{u}}$, and is
\begin{equation*}
\seq{\mbf{u}} = \left(\begin{bmatrix} 1 \\ 1 \end{bmatrix},\begin{bmatrix} 0.73 \\ 1 \end{bmatrix}, \begin{bmatrix} 0 \\ 0.92 \end{bmatrix}, \begin{bmatrix} 0 \\ 0 \end{bmatrix}, \begin{bmatrix} -0.45 \\ 0 \end{bmatrix} \right).
\end{equation*}
Figure \ref{fig:desc_fdiczlast} shows the output reachable sets $Y_0^{[i]}$ and $Y_4^{[i]}$ for $i=1,2,3,4$, resulting from the injection of the designed $\seq{\mbf{u}}$. We also show the sets defined by $\mbf{C}^{[i]} X_0 \oplus \mbf{D}^{[i]} \mbf{u}_0 \oplus \mbf{D}_v^{[i]} V$, which do not take into account the equality constraint \eqref{eq:desc_systemSVDfaultconstraints}. Note that these sets are completely overlapped for $i \in \{1,2,3,4\}$, and $Y_0^{[i]}$ are completely overlapped for $i \in \{1,2,4\}$. On the other hand, $Y_4^{[i]}$ are disjoint for every $i \in \modelset$, showing that the injection of $\seq{\mbf{u}}$ guarantees fault diagnosis at $k = 4$. Figure \ref{fig:desc_fdiczlast} also shows clouds containing 2000 samples of the output $\mbf{y}^{[i]}_4$ for each model $i$, demonstrating that the outputs of the models \eqref{eq:desc_systemfaulty} are in fact separated.

\subsection{Fault diagnosis using mixed zonotopes}

\begin{figure*}[!tb]
	\centering{
		\def\svgwidth{\columnwidth}
		{\scriptsize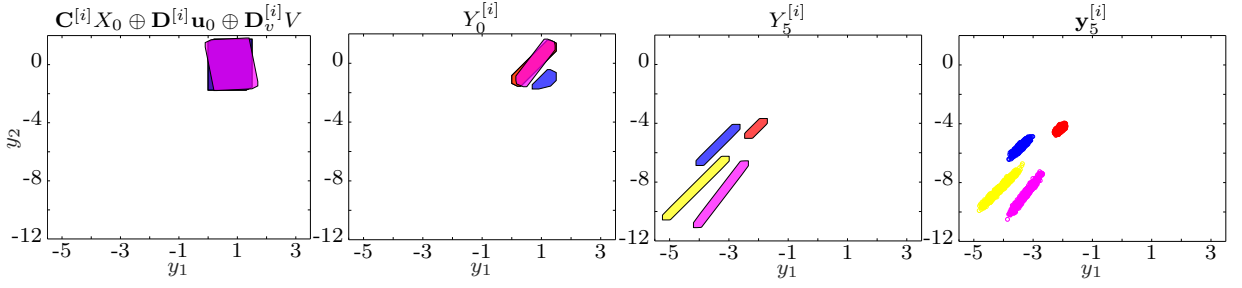}
		\vspace{-2mm}
		\caption{The sets $\mbf{C}^{[i]} X_0 \oplus \mbf{D}^{[i]} \mbf{u}_0 \oplus \mbf{D}_v^{[i]} V$, the output reachable sets $Y_0^{[i]}$, $Y_5^{[i]}$, and 1000 samples for $\mbf{y}^{[i]}_5$, with $i=1$ (yellow), $i=2$ (red), $i=3$ (blue), and $i=4$ (magenta), obtained by the injection of the input sequence $\seq{\mbf{u}}$.}\label{fig:desc_fdimztube}}
\end{figure*}

We now evaluate the effectiveness of the AFD method based on mixed zonotopes proposed in Section \ref{sec:desc_AFDMZ}. We consider the same set of models described in the previous example (Section \ref{sec:desc_examplesAFDCZ}), with also the same uncertainty and input bounds. The number of generators of the zonotope $\mathring{\mathcal{Y}}(i,j)$ (Lemma \ref{thm:desc_separatinginputmzequiv}) was limited to 1.6\footnote{The tuning parameters for this experiment differ from the previous one due to the higher dimension of the output reachable tubes in comparison to the output reachable sets.} times its dimension using Method \ref{meth:genredB}. The minimum length optimal input sequence that solves \eqref{eq:desc_optimalseparatingmz} was obtained using CPLEX 12.8 and MATLAB 9.1, with $J(\seq{\mbf{u}}) = \seq{\mbf{u}}\zerospace^T \seq{\mbf{u}}$, $\varepsilon = 1 \ten{-6}$, and $N=5$, and it is given by
\begin{equation*}
\seq{\mbf{u}} = \left(\begin{bmatrix} 1 \\ 0.7636 \end{bmatrix},\begin{bmatrix} 1 \\ 0.2664 \end{bmatrix}, \begin{bmatrix} 0.9892 \\ 1 \end{bmatrix}, \begin{bmatrix} -0.0879 \\ 1 \end{bmatrix}, \begin{bmatrix} -0.3478 \\ 1 \end{bmatrix}, \begin{bmatrix} -0.7202 \\ -1 \end{bmatrix} \right).
\end{equation*}
Figure \ref{fig:desc_fdimztube} shows the output reachable sets $Y_0^{[i]}$ and $Y_5^{[i]}$ for $i=1,2,3,4$, resulting from the injection of the designed $\seq{\mbf{u}}$. We also show the sets defined by $\mbf{C}^{[i]} X_0 \oplus \mbf{D}^{[i]} \mbf{u}_0 \oplus \mbf{D}_v^{[i]} V$, which do not take into account the equality constraint \eqref{eq:desc_systemSVDfaultconstraintsmz}. Note that these sets are completely overlapped for $i \in \{1,2,3,4\}$, and $Y_0^{[i]}$ are completely overlapped for $i \in \{1,2,4\}$. On the other hand, $Y_5^{[i]}$ are disjoint for every $i \in \modelset$, showing that the injection of $\seq{\mbf{u}}$ guarantees fault diagnosis at $k = 5$. Figure \ref{fig:desc_fdimztube} also shows clouds containing 1000 samples of the output $\mbf{y}^{[i]}_5$ for each model $i$, demonstrating that the outputs of the models \eqref{eq:desc_systemfaulty} are in fact separated. In comparison to the results presented in the previous section using constrained zonotopes, the output reachable tube method using mixed zonotopes did not require the knowledge of an admissible state set. However, a longer input sequence was required for fault diagnosis. The reasons for this disadvantage are a subject of future investigation.

\section{Final remarks} \label{sec:desc_finalremarks}

This chapter proposed novel algorithms for set-based state estimation and AFD of linear descriptor systems with unknown-but-bounded uncertainties. The methods use constrained zonotopes and mixed zonotopes. The latter is a generalization of zonotopes capable of describing unbounded sets. Both methods lead to significantly tighter results than zonotope approaches. In addition, AFD was enabled without assuming rank properties on the structure of the system. The effectiveness of the new methods was corroborated by numerical examples.

\chapter{Joint state and parameter estimation}\thispagestyle{headings} \label{cha:stateparameter}

This chapter presents a new method for set-based joint state and parameter estimation of nonlinear discrete-time systems. The new method is based on the algorithms proposed in Chapter \ref{cha:nonlinearmeasinv}. By extending the nonlinear state estimation methods using constrained zonotopes to include parameter estimation in a unified framework, and therefore, maintaining existing dependencies between states, algebraic variables, and unknown parameters, the accuracy of both state and parameter estimation is significantly improved. The advantages of the new approach are highlighted in two numerical examples.

The chapter is organized as follows. The class of nonlinear systems considered in this chapter is presented in Section \ref{sec:joint_problemformulation}, together with the proposed estimator framework. Section \ref{sec:joint_stateestimation} develops the proposed method for set-based joint state and parameters estimation using constrained zonotopes. The numerical examples are presented in Section \ref{sec:joint_example}, and Section \ref{sec:joint_remarks} concludes the chapter.

\section{Problem formulation} \label{sec:joint_problemformulation}

Consider a class of nonlinear discrete-time system with unknown parameters and algebraic variables, described by
\begin{subequations} \label{eq:joint_system}
	\begin{align}
	\mbf{x}_k & = \mbf{f}(\mbf{x}_{k-1}, \mbf{u}_{k-1}, \bm{\lambda}_{k-1}, \mbf{p}, \mbf{w}_{k-1}), \label{eq:joint_systemf} \\
	\mbf{0} &   = \mbf{h}(\mbf{x}_{k}, \mbf{u}_k, \bm{\lambda}_{k}, \mbf{p}, \mbf{d}_k), \label{eq:joint_systemh} \\
	\mbf{y}_k & = \mbf{g}(\mbf{x}_{k}, \mbf{u}_{k}, \bm{\lambda}_{k}, \mbf{p}, \mbf{v}_{k}), \label{eq:joint_systemg}
	\end{align}
\end{subequations}
where $\mbf{x}_k \in \realset^{n}$ are the system states, $\mbf{u}_{k} \in \realset^{n_u}$ are the known inputs, $\mbf{w}_k \in \realset^{n_w}$ and $\mbf{d}_k \in \realset^{n_d}$ are the process uncertainties, $\mbf{y}_k \in \realset^{n_y}$ are the measured outputs, and $\mbf{v}_k \in \realset^{n_v}$ are the measurement uncertainties. Moreover, $\bm{\lambda}_{k} \in \realset^{n_\lambda}$ are the algebraic variables (i.e., static variables satisfying the algebraic constraints \eqref{eq:joint_systemh}), and $\mbf{p} \in \realset^{n_p}$ are unknown but constant model parameters. The nonlinear functions $\mbf{f}: \realset^n \times \realset^{n_u} \times \realset^{n_\lambda} \times \realset^{n_p} \times \realset^{n_w} \to \realset^n$, $\mbf{g}: \realset^n  \times \realset^{n_u} \times \realset^{n_\lambda} \times \realset^{n_p} \times \realset^{n_v} \to \realset^{n_y}$, and $\mbf{h}: \realset^n  \times \realset^{n_u} \times \realset^{n_\lambda} \times \realset^{n_p} \times \realset^{n_d} \to \realset^{n_h}$, are assumed to be of class $\mathcal{C}^2$. The initial condition, algebraic variables, unknown model parameters, and uncertainties are assumed to be unknown-but-bounded, i.e., $\mbf{x}_0 \in X_0$, $\bm{\lambda}_k \in L_k$, $\mbf{p} \in P$, $\mbf{w}_{k} \in W$, $\mbf{d}_{k} \in D$, and $\mbf{v}_k \in V$, $\forall k \geq 0$, where $X_0$, $L_k$, $P$, $W$, $D$, and $V$ are known convex polytopic sets.

\begin{remark} \rm \label{rem:joint_algebraicdisturbances}
	The nonlinear algebraic constraints \eqref{eq:joint_systemh} may not hold exactly in practice. This condition is described by the presence of the uncertainty $\mbf{d}_k \in D$.
\end{remark}

\begin{remark} \rm \label{rem:joint_invariant}
	The nonlinear system \eqref{eq:joint_system} is equivalent to \eqref{eq:nmeas_system} in the particular case where $n_\lambda = n_p = n_d = 0$, and if Assumption \ref{ass:nmeas_invariant} holds.
\end{remark}

For any $k\geq 0$, the objective of this chapter is to approximate the solution set of \eqref{eq:joint_system} as accurately as possible by a guaranteed enclosure $\hat{Z}_k \subset \realset^{n+n_\lambda+n_p}$ satisfying $(\mbf{x}_k,\bm{\lambda}_k,\mbf{p}) \in \hat{Z}_k$. We accomplish this here by extending the prediction-update-consistency structure proposed in Chapter \ref{cha:nonlinearmeasinv} to a joint state and parameter estimation framework, considering the presence of both algebraic variables $\bm{\lambda}_k$ and unknown model parameters $\mbf{p}$. The generalized scheme is given by the following recursion:
\begin{align}
\bar{Z}_k & \supseteq \{ (\mbf{f}(\mbf{x}_{k-1}, \mbf{u}_{k-1}, \bm{\lambda}_{k-1}, \mbf{p}, \mbf{w}_{k-1}),\bm{\lambda}_{k}, \mbf{p}) : (\mbf{x}_{k-1}, \bm{\lambda}_{k-1}, \mbf{p}) \in \hat{Z}_{k-1}, \, \bm{\lambda}_k \in L_k ,\,\mbf{w}_{k-1} \in W \}, \label{eq:joint_prediction0}\\
\tilde{Z}_k & \supseteq \{ (\mbf{x}_{k},\bm{\lambda}_k,\mbf{p}) \in \bar{Z}_k : \mbf{h}(\mbf{x}_{k},\mbf{u}_k, \bm{\lambda}_{k}, \mbf{p}, \mbf{d}_k) = \bm{0}, \, \mbf{d}_k \in D \},\label{eq:joint_consistency0}\\
\hat{Z}_k & \supseteq \{ (\mbf{x}_{k},\bm{\lambda}_k,\mbf{p}) \in \tilde{Z}_k : \mbf{g}(\mbf{x_{k}}, \mbf{u}_{k}, \bm{\lambda}_{k}, \mbf{p}, \mbf{v}_{k}) = \mbf{y}_k , \, \mbf{v}_{k} \in V \}, \label{eq:joint_update0}
\end{align}
where \eqref{eq:joint_prediction0} is the \emph{joint prediction step}, \eqref{eq:joint_consistency0} is the \emph{joint consistency step}, \eqref{eq:joint_update0} is the \emph{joint update step}, and the scheme is initialized with $\bar{Z}_0 \triangleq \bar{X}_0 \times L_0 \times P$ in the consistency step. %
If $\hat{Z}_{k-1}$ is a valid enclosure of $(\mbf{x}_{k-1},\bm{\lambda}_{k-1},\mbf{p})$ for some $k \geq 1$, then standard results in set-valued state estimation show that $(\mbf{x}_k,\bm{\lambda}_k,\mbf{p}) \in \bar{Z}_k$ given by \eqref{eq:joint_prediction0}. By construction, this leads to $(\mbf{x}_k,\bm{\lambda}_k,\mbf{p}) \in \tilde{Z}_k$ and $(\mbf{x}_k,\bm{\lambda}_k,\mbf{p}) \in \hat{Z}_k$, from \eqref{eq:joint_consistency0} and \eqref{eq:joint_update0}, respectively.

\begin{remark} \rm \label{rem:joint_consistencyfirst}
	In contrast to the method proposed in Chapter \ref{cha:nonlinearmeasinv}, to improve the accuracy in the update step \eqref{eq:joint_update0}, in this chapter we consider computing the joint consistency step prior to the joint update step. This is because the known set $L_k$ enclosing the algebraic variables $\bm{\lambda}_k$ may be very conservative in practice, with $\bm{\lambda}_k$ being usually determined by the nonlinear algebraic constraints \eqref{eq:joint_systemh}. 
\end{remark}

In the remainder of the chapter, our goal is to develop methods for computing accurate enclosures for each of the three steps \eqref{eq:joint_prediction0}--\eqref{eq:joint_update0}. The main results of this chapter include generalizations of the prediction, update, and consistency methods proposed in Chapter \ref{cha:nonlinearmeasinv}, which consider state variables, algebraic variables, and unknown parameters, as well as nonlinear dynamics, nonlinear measurement, and nonlinear algebraic constraints.

\section{Proposed scheme based on constrained zonotopes}  \label{sec:joint_stateestimation}

In the following, we rewrite Lemma \ref{lem:nmeas_mve} to consider more general arguments than just states and process disturbances. This generalized result will be necessary for the new method developed in this chapter.

\begin{lemma} \rm \label{lem:joint_mve}
	Let $\bm{\alpha} : \realset^{n_r} \times \realset^{n_s} \to \realset^{n_\alpha}$ be of class $\mathcal{C}^1$, and let $\nabla_r \bm{\alpha}$ denote the gradient of $\bm{\alpha}$ with respect to its first argument. Let $R\subset \realset^{n_r}$ and $S \subset \realset^{n_s}$ be constrained zonotopes, and let $\mbf{J} \in \intvalsetmat{n_\alpha}{n_r}$ be an interval matrix satisfying 
	\begin{equation}
	\label{eq:joint_mvelemma_J}
	\nabla^T_r \bm{\alpha}(\square R, S)\triangleq\{\nabla^T_r \bm{\alpha}(\mbf{r},\mbf{s}): \mbf{r}\in\square R, \ \mbf{s} \in S\}\subseteq \mbf{J}.
	\end{equation}
	Then, for every $\mbf{r} \in R$, $\mbf{s} \in S$, and $\bm{\gamma}_r \in \square R$, there exists $\hat{\mbf{J}} \in \mbf{J}$ such that 
	\begin{equation} \label{eq:joint_mvelemma}
	\bm{\alpha}(\mbf{r},\mbf{s}) = \bm{\alpha}(\bm{\gamma}_r,\mbf{s}) + \hat{\mbf{J}} (\mbf{r} - \bm{\gamma}_r).
	\end{equation}
\end{lemma}

\proof Choose any $(\mbf{r},\mbf{s})\in R\times S$. Since $\mbf{r} \in R \subseteq \square R$ and $\gamma_r \in \square R$, the Mean Value Theorem ensures that, for any $i=1,2,\ldots,n_\alpha$, $\exists \bm{\delta}^{[i]}\in \square R$ such that $\alpha_i(\mbf{r},\mbf{s}) = \alpha_i(\bm{\gamma}_r,\mbf{s}) + \nabla^T_r \alpha_i(\bm{\delta}^{[i]},\mbf{s}) (\mbf{r} - \bm{\gamma}_r)$. But $\nabla^T_r \alpha_i(\bm{\delta}^{[i]},\mbf{s})$ is contained in the $i$-th row of $\mbf{J}$ by hypothesis, and since this is true for all $i=1,2,\ldots,n_\alpha$, $\exists \hat{\mbf{J}}\in \mbf{J}$ such that $\bm{\alpha}(\mbf{r},\mbf{s}) = \bm{\alpha}(\bm{\gamma}_r,\mbf{s}) + \hat{\mbf{J}} (\mbf{r} - \bm{\gamma}_r).$ \qed

The following proposition provides an enclosure for the state $\mbf{x}_k$ portion in the prediction step \eqref{eq:joint_prediction0}. A method to compute an enclosure $\bar{Z}_k$ satisfying the joint prediction step is derived based on the result from Proposition \ref{thm:joint_mvestatepred}, which is given by Corollary \ref{thm:joint_mvejointprediction}.

\begin{proposition} (State prediction) \rm \label{thm:joint_mvestatepred}
	Let $\mbf{f} : \realset^n \times \realset^{n_u} \times \realset^{n_\lambda} \times \realset^{n_p} \times \realset^{n_w} \to \realset^n$ be of class $\mathcal{C}^1$, and let $\nabla_x \mbf{f}$ denote the gradient of $\mbf{f}$ with respect to its first argument. Let $\mbf{u} \in \realset^{n_u}$, and let $X\subset \realset^n$, $L \subset \realset^{n_\lambda}$, $P \subset \realset^{n_p}$, and $W \subset \realset^{n_w}$ be constrained zonotopes. Let $Z = X \times L \times P$, and choose any $\bm{\gamma}_z = (\bm{\gamma}_x, \bm{\gamma}_\lambda, \bm{\gamma}_p) \in \square X \times \square L \times \square P$. If $Z_w$ is a constrained zonotope such that $\mbf{f}(\bm{\gamma}_x,\mbf{u},\bm{\gamma}_\lambda,\bm{\gamma}_p,W) \subseteq Z_w$ and $\mbf{J} \in \intvalsetmat{n}{n}$ is an interval matrix satisfying $\nabla^T_z \mbf{f}(\square X,\mbf{u}, \square L, \square P,W)\subseteq \mbf{J}$, then $\mbf{f}(X,\mbf{u},L,P,W) \subseteq Z_w \oplus \gzinclusion\left(\mbf{J},  Z - \bm{\gamma}_z \right)$.
\end{proposition}

\proof Choose any $(\mathbf{x},\bm{\lambda},\mbf{p},\mathbf{w})\in X\times L \times P \times W$. Lemma \ref{lem:joint_mve} (with $\bm{\alpha} \triangleq \mbf{f}$, $\mbf{r} \triangleq (\mbf{x},\bm{\lambda},\mbf{p})$, $\mbf{s} \triangleq \mbf{w}$, $R \triangleq X \times L \times P$, and $S \triangleq W$) ensures that there exists a real matrix $\hat{\mathbf{J}}\in \mathbf{J}$ such that $\mbf{f}(\mbf{x},\mbf{u},\bm{\lambda},\mbf{p},\mbf{w}) = \mbf{f}(\bm{\gamma}_x,\mbf{u},\bm{\gamma}_\lambda,\bm{\gamma}_p,\mbf{w}) + \hat{\mathbf{J}} (\mathbf{z} - \bm{\gamma}_z)$, with $\mbf{z} = (\mbf{x},\bm{\lambda},\mbf{p})$. By Theorem \ref{thm:ndyn_czinclusion} and the choice of $Z_w$, it follows that $\mbf{f}(\mbf{x},\mbf{u},\bm{\lambda},\mbf{p},\mbf{w}) \in Z_w \oplus \gzinclusion\left(\mathbf{J}, Z - \bm{\gamma}_z\right)$, as desired. \qed

\begin{corollary} (Joint prediction) \rm \label{thm:joint_mvejointprediction}
	Let $\mbf{f} : \realset^n \times \realset^{n_u} \times \realset^{n_\lambda} \times \realset^{n_p} \times \realset^{n_w} \to \realset^n$ be of class $\mathcal{C}^1$, and let $\nabla_x \mbf{f}$ denote the gradient of $\mbf{f}$ with respect to its first argument. For $k \geq 1$, let $\mbf{u}_{k-1} \in \realset^{n_u}$, and assume that $\mbf{z}_{k-1} = (\mbf{x}_{k-1},\bm{\lambda}_{k-1},\mbf{p}) \in \hat{Z}_{k-1} = \{\hat{\mbf{G}}_{k-1}, \hat{\mbf{c}}_{k-1}$, $\hat{\mbf{A}}_{k-1}, \hat{\mbf{b}}_{k-1}\}$. Moreover, let $\mbf{w}_{k-1} \in W = \{\mbf{G}_w, \mbf{c}_w, \mbf{A}_w, \mbf{b}_w\}$, $\bm{\lambda}_{k} \in L_k$, and $\mbf{p} \in P$. Choose any $\bm{\gamma}_z = (\bm{\gamma}_x, \bm{\gamma}_\lambda, \bm{\gamma}_p) \in \square \hat{Z}_{k-1}$. In addition, let $\mbf{E} = [\zeros{n_p}{n} \,\; \zeros{n_p}{n_\lambda} \,\; \eye{n_p}]$. Let $Z_w$ be a constrained zonotope such that $\mbf{f}(\bm{\gamma}_x,\mbf{u}_{k-1},\bm{\gamma}_\lambda,\bm{\gamma}_p,\mbf{w}_{k-1}) \subseteq Z_w$ for all $\mbf{w}_{k-1} \in W$, and let $\mbf{J}_z \in \intvalsetmat{n}{n}$ be an interval matrix satisfying $\nabla^T_z \mbf{f}(\mbf{x}_{k-1},\mbf{u}_{k-1}, \bm{\lambda}_{k-1}, \mbf{p}, \mbf{w}_{k-1} ) \subseteq \mbf{J}_z$ for all $(\mbf{x}_{k-1},\bm{\lambda}_{k-1},\mbf{p}) \in \square \hat{Z}_{k-1}$, and $\mbf{w}_{k-1} \in W$. If $\{\underline{\hat{\mbf{G}}},\underline{\hat{\mbf{c}}}\}$ is a zonotope with $\underline{n}_g$ generators satisfying $\hat{Z}_{k-1} - \bm{\gamma}_z \subseteq \{\underline{\hat{\mbf{G}}},\underline{\hat{\mbf{c}}}\}$, $\mbf{m} \triangleq (\mbf{J}_z - \text{mid}(\mbf{J}_z)) \underline{\hat{\mbf{c}}} \in \intvalset^n$, and $\hat{\mbf{P}} \in \realsetmat{n}{n}$ is a diagonal matrix with $\hat{P}_{ii} = \text{rad}(m_i) + \sum_{j=1}^{\underline{n}_g} \sum_{\ell=1}^{n+n_\lambda+n_p} \text{rad}(J_{z,i\ell})|\underline{\hat{G}}_{\ell j}|$, then $(\mbf{x}_k,\bm{\lambda}_k,\mbf{p}) = \mbf{z}_{k} \in \bar{Z}_k$, with
	\begin{equation} \label{eq:joint_mvejointprediction}
	\bar{Z}_k = \begin{bmatrix} \midpoint{\mbf{J}_z} \\ \mbf{0} \\ \mbf{E} \end{bmatrix} \hat{Z}_{k-1} \oplus \begin{bmatrix} \midpoint{\mbf{J}_z} \\ \mbf{0} \\ \mbf{0} \end{bmatrix} (-\bm{\gamma}_z) \oplus \begin{bmatrix} \hat{\mbf{P}} \\ \mbf{0} \\ \mbf{0} \end{bmatrix} B_\infty^n \oplus \begin{bmatrix} \mbf{I} \\ \mbf{0} \\ \mbf{0} \end{bmatrix} Z_w \oplus \begin{bmatrix} \mbf{0} \\ \mbf{I} \\ \mbf{0} \end{bmatrix} L_k.
	\end{equation}	
\end{corollary}
\proof Choose any $(\mbf{x}_{k-1},\bm{\lambda}_{k-1},\mbf{p}) = \mbf{z}_{k-1} \in \hat{Z}_{k-1}$, $\mbf{w}_{k-1} \in W$. From \eqref{eq:joint_systemf}, Proposition \ref{thm:joint_mvestatepred} and Theorem \ref{thm:ndyn_czinclusion}, there must exist $\bm{\delta} \in B_\infty^n$ such that
\begin{equation*}
\mbf{x}_k = \mbf{f}(\mbf{x}_{k-1},\mbf{u}_{k-1},\bm{\lambda}_{k-1},\mbf{p},\mbf{w}_{k-1}) = \mbf{f}(\bm{\gamma}_x,\mbf{u}_{k-1},\bm{\gamma}_\lambda,\bm{\gamma}_p,\mbf{w}_{k-1}) + \text{mid}(\mbf{J}_z)(\mbf{z}_{k-1} - \bm{\gamma}_z) + \mbf{P} \bm{\delta} 
\end{equation*}
with $\hat{\mbf{P}}$ defined as in the statement of the corollary. Then, since $\bm{\lambda}_k \in L_k$ and, by the definition of $\mbf{z}_{k-1}$ and $\mbf{E}$, $\mbf{p} = \mbf{E}\mbf{z}_{k-1}$ holds, we have that
\begin{align*}
(\mbf{x}_{k},\bm{\lambda}_k,\mbf{p}) & = (\mbf{f}(\bm{\gamma}_x,\mbf{u}_{k-1},\bm{\gamma}_\lambda,\bm{\gamma}_p,\mbf{w}_{k-1}) + \text{mid}(\mbf{J}_z)(\mbf{z}_{k-1} - \bm{\gamma}_z) + \hat{\mbf{P}} \bm{\delta}, \bm{\lambda}_k, \mbf{E} \mbf{z}_{k-1}) \\
& = (\text{mid}(\mbf{J}_z)\mbf{z}_{k-1} - \text{mid}(\mbf{J}_z)\bm{\gamma}_z + \hat{\mbf{P}} \bm{\delta} + \mbf{f}(\bm{\gamma}_x,\mbf{u}_{k-1},\bm{\gamma}_\lambda,\bm{\gamma}_p,\mbf{w}_{k-1}), \bm{\lambda}_k, \mbf{E} \mbf{z}_{k-1}) \\
& = \begin{bmatrix} \midpoint{\mbf{J}_z} \\ \mbf{0} \\ \mbf{E} \end{bmatrix} \mbf{z}_{k-1} + \begin{bmatrix} \midpoint{\mbf{J}_z} \\ \mbf{0} \\ \mbf{0} \end{bmatrix} (-\bm{\gamma}_z) + \begin{bmatrix} \hat{\mbf{P}} \\ \mbf{0} \\ \mbf{0} \end{bmatrix} \bm{\delta} + \begin{bmatrix} \mbf{I} \\ \mbf{0} \\ \mbf{0} \end{bmatrix} \mbf{f}(\bm{\gamma}_x,\mbf{u}_{k-1},\bm{\gamma}_\lambda,\bm{\gamma}_p,\mbf{w}_{k-1}) + \begin{bmatrix} \mbf{0} \\ \mbf{I} \\ \mbf{0} \end{bmatrix} \bm{\lambda}_k \\
& \in \begin{bmatrix} \midpoint{\mbf{J}_z} \\ \mbf{0} \\ \mbf{E} \end{bmatrix} \hat{Z}_{k-1} \oplus \begin{bmatrix} \midpoint{\mbf{J}_z} \\ \mbf{0} \\ \mbf{0} \end{bmatrix} (-\bm{\gamma}_z) \oplus \begin{bmatrix} \hat{\mbf{P}} \\ \mbf{0} \\ \mbf{0} \end{bmatrix} B_\infty^{n} \oplus \begin{bmatrix} \mbf{I} \\ \mbf{0} \\ \mbf{0} \end{bmatrix} Z_w \oplus \begin{bmatrix} \mbf{0} \\ \mbf{I} \\ \mbf{0} \end{bmatrix} L_k,
\end{align*}
which proves the corollary. \qed

The following two propositions provide enclosures $\tilde{Z}_k$ and $\hat{Z}_k$ satisfying the joint consistency and update steps defined by \eqref{eq:joint_consistency0} and \eqref{eq:joint_update0}, respectively. These results are both based on Lemma \ref{lem:joint_mve} and Theorem \ref{thm:ndyn_czinclusion}.

\begin{proposition} (Joint consistency) \rm \label{thm:joint_mvejointconsistency}
  	Let $\mbf{h}: \realset^n  \times \realset^{n_u} \times \realset^{n_\lambda} \times \realset^{n_p} \times \realset^{n_d} \to \realset^{n_h}$ be of class $\mathcal{C}^1$. For $k \geq 0$, let $\mbf{u}_{k} \in \realset^{n_u}$, assume that $\mbf{z}_{k} = (\mbf{x}_{k},\bm{\lambda}_{k}, \mbf{p}) \in \bar{Z}_{k} = \{\bar{\mbf{G}}_{k}, \bar{\mbf{c}}_{k}, \bar{\mbf{A}}_{k}, \bar{\mbf{b}}_{k}\}$, and let $\mbf{d}_{k} \in D = \{\mbf{G}_d, \mbf{c}_d, \mbf{A}_d, \mbf{b}_d\}$. Let $\mbf{h}(\mbf{x}_k,\mbf{u}_k,\bm{\lambda}_k,\mbf{p},\mbf{d}_k) = \mbf{0}$ for at least one $(\mbf{x}_k,\bm{\lambda}_k,\mbf{p}) \in \bar{Z}_k$, and $\mbf{d}_k \in D$. Choose any $\bm{\gamma}_z = (\bm{\gamma}_x, \bm{\gamma}_\lambda, \bm{\gamma}_p) \in \square \bar{Z}_k$. Let $Z_d = \{\mbf{G}_{Z_d}, \mbf{c}_{Z_d}, \mbf{A}_{Z_d}, \mbf{b}_{Z_d}\}$ be a constrained zonotope such that $-\mbf{h}(\bm{\gamma}_x,\mbf{u},\bm{\gamma}_\lambda, \bm{\gamma}_p, \mbf{d}_k) \in Z_d = \{\mbf{G}_{Z_d}, \mbf{c}_{Z_d}, \mbf{A}_{Z_d}, \mbf{b}_{Z_d}\}$ for all $\mbf{d}_k \in D$, and let $\mbf{J}_z \in \intvalsetmat{n_h}{n+n_\lambda+n_p}$ satisfy $\nabla^T_z \mbf{h}(\mbf{x}_k,\mbf{u}_k,\bm{\lambda}_k, \mbf{p}, \mbf{d}_k) \subseteq \mbf{J}_z$ for all $(\mbf{x}_k,\bm{\lambda}_k,\mbf{p}) \in \bar{Z}_k$ and $\mbf{d}_k \in D$. If $\{\underline{\bar{\mbf{G}}},\underline{\bar{\mbf{c}}}\}$ is a zonotope with $\underline{n}_g$ generators satisfying $\bar{Z}_{k} - \bm{\gamma}_z \subseteq \{\underline{\bar{\mbf{G}}},\underline{\bar{\mbf{c}}}\}$, $\mbf{m} \triangleq (\mbf{J}_z - \text{mid}(\mbf{J}_z)) \underline{\bar{\mbf{c}}} \in \intvalset^n$, and $\bar{\mbf{P}} \in \realsetmat{n_h}{n_h}$ is a diagonal matrix with $\bar{P}_{ii} = \text{rad}(m_i) + \sum_{j=1}^{\underline{n}_g} \sum_{\ell=1}^{n+n_\lambda+n_p} \text{rad}(J_{z,i\ell})|\underline{\bar{G}}_{\ell j}|$, then $\{ (\mbf{x}_{k}, \bm{\lambda}_k, \mbf{p}) \in \bar{Z}_k : \mbf{h}(\mbf{x}_k,\mbf{u}_k,\bm{\lambda}_k,\mbf{p},\mbf{d}_k) = \mbf{0}, \, \mbf{d}_k \in D \} \subseteq \tilde{Z}_k$, where
  	\begin{equation} \label{eq:joint_mveconsistencyjoint}
	\tilde{Z}_k = \left\{ \begin{bmatrix} \bar{\mbf{G}}_z & \mbf{0} & \mbf{0} \end{bmatrix}, \bar{\mbf{c}}_z, \begin{bmatrix} \bar{\mbf{A}}_z & \mbf{0} & \mbf{0} \\ \mbf{0} & \mbf{0} & \mbf{A}_{Z_d} \\ \midpoint{\mbf{J}_z} \bar{\mbf{G}}_z & -\bar{\mbf{P}} & -\mbf{G}_{Z_d} \end{bmatrix}, \begin{bmatrix} \bar{\mbf{b}}_z  \\ \mbf{b}_{Z_d} \\ \midpoint{\mbf{J}_z} \left(\bm{\gamma}_z - \bar{\mbf{c}}_z\right) + \mbf{c}_{Z_d} \end{bmatrix} \right\}. 
  	\end{equation}   		
\end{proposition}
   
\proof 
Choose any $(\mbf{x}_k,\bm{\lambda}_k,\mbf{p}) = \mbf{z}_{k} \in \bar{Z}_{k}$, $\mbf{d}_{k} \in D$, satisfying $\mbf{h}(\mbf{x}_k,\mbf{u}_k,\bm{\lambda}_k,\mbf{p},\mbf{d}_k) = \mbf{0}$. Lemma \ref{lem:joint_mve} (with $\bm{\alpha} \triangleq \mbf{h}$, $\mbf{r} \triangleq (\mbf{x}_k,\bm{\lambda}_k,\mbf{p})$, $\mbf{s} \triangleq \mbf{d}_k$, $R \triangleq \bar{Z}_k$, and $S \triangleq D$) ensures that there exists a real matrix $\hat{\mbf{J}}\in \mbf{J}_z$ such that $\mbf{h}(\mbf{x}_k,\mbf{u}_k,\bm{\lambda}_k,\mbf{p},\mbf{d}_k) = \mbf{h}(\bm{\gamma}_x,\mbf{u},\bm{\gamma}_\lambda,\bm{\gamma}_p,\mbf{d}_k) + \hat{\mbf{J}}_k (\mbf{z}_k - \bm{\gamma}_z)$, with $\mbf{z} = (\mbf{x},\bm{\lambda},\mbf{p})$. Therefore, from Theorem \ref{thm:ndyn_czinclusion}, there must exist $\bm{\delta} \in B_\infty^{n_h}$ such that
\begin{align*}
\mbf{0} = \mbf{h}(\mbf{x}_{k},\mbf{u}_{k},\bm{\lambda}_{k},\mbf{p},\mbf{d}_{k}) = \mbf{h}(\bm{\gamma}_x,\mbf{u}_{k},\bm{\gamma}_\lambda,\bm{\gamma}_p,\mbf{d}_{k}) + \text{mid}(\mbf{J}_z)(\mbf{z}_{k} - \bm{\gamma}_z) + \bar{\mbf{P}} \bm{\delta},
\end{align*}
with $\bar{\mbf{P}}$ defined as in the statement of this proposition. The equation above can be rewritten as $\text{mid}(\mbf{J}_z)\mbf{z}_{k} = - \mbf{h}(\bm{\gamma}_x,\mbf{u}_{k},\bm{\gamma}_\lambda,\bm{\gamma}_p,\mbf{d}_{k}) + \text{mid}(\mbf{J}_z)\bm{\gamma}_z - \bar{\mbf{P}} \bm{\delta} \in (\text{mid}(\mbf{J}_z)\bm{\gamma}_z) \oplus Z_d \oplus \bar{\mbf{P}} B_\infty^{n_h}$. Therefore, $\{(\mbf{x}_{k}, \bm{\lambda}_k, \mbf{p}) \in \bar{Z}_k : \mbf{h}(\mbf{x}_k,\mbf{u}_k,\bm{\lambda}_k,\mbf{p},\mbf{d}_k) = \mbf{0}, \, \mbf{d}_k \in D \} \subseteq \{ (\mbf{x}_{k}, \bm{\lambda}_k, \mbf{p}) = \mbf{z}_k \in \bar{Z}_k : \text{mid}(\mbf{J}_z)\mbf{z}_{k} \in  (\text{mid}(\mbf{J}_z)\bm{\gamma}_z) \oplus Z_d \oplus \bar{\mbf{P}} B_\infty^{n_h} \}$. By considering the generalized intersection \eqref{eq:pre_czintersection} and the definition of $\tilde{Z}_k$, the proposition is proven. \qed

\begin{proposition} (Joint update) \rm \label{thm:joint_mvejointupdate}
   	Let $\mbf{g}: \realset^n  \times \realset^{n_u} \times \realset^{n_\lambda} \times \realset^{n_p} \times \realset^{n_v} \to \realset^{n_y}$ be of class $\mathcal{C}^1$. For $k \geq 0$, let $\mbf{u}_{k} \in \realset^{n_u}$, assume that $\mbf{z}_{k} = (\mbf{x}_{k},\bm{\lambda}_{k}, \mbf{p}) \in \tilde{Z}_{k} = \{\tilde{\mbf{G}}_{k}, \tilde{\mbf{c}}_{k}, \tilde{\mbf{A}}_{k}, \tilde{\mbf{b}}_{k}\}$, and let $\mbf{v}_{k} \in V = \{\mbf{G}_v, \mbf{c}_v, \mbf{A}_v, \mbf{b}_v\}$. Let $\mbf{y}_k \in \realset^{n_y}$ such that $\mbf{y}_k=\mbf{g}(\mbf{x}_k,\mbf{u}_k,\bm{\lambda}_k,\mbf{p},\mbf{v}_k)$ for some $(\mbf{x}_k,\bm{\lambda}_k,\mbf{p}) \in \tilde{Z}_k$, and $\mbf{v}_k \in V$. Choose any $\bm{\gamma}_z = (\bm{\gamma}_x, \bm{\gamma}_\lambda, \bm{\gamma}_p) \in \square \tilde{Z}_k$. Let $Z_v = \{\mbf{G}_{Z_v}, \mbf{c}_{Z_v}, \mbf{A}_{Z_v}, \mbf{b}_{Z_v}\}$ be a constrained zonotope such that $-\mbf{g}(\bm{\gamma}_x,\mbf{u},\bm{\gamma}_\lambda, \bm{\gamma}_p, \mbf{v}_k) \in Z_v = \{\mbf{G}_{Z_v}, \mbf{c}_{Z_v}, \mbf{A}_{Z_v}, \mbf{b}_{Z_v}\}$ for all $\mbf{v}_k \in V$, and let $\mbf{J} \in \intvalsetmat{n_y}{n+n_\lambda+n_p}$ satisfy $\nabla^T_z \mbf{g}(\mbf{x}_k,\mbf{u}_k,\bm{\lambda}_k, \mbf{p}, \mbf{v}_k) \subseteq \mbf{J}_z$ for all $(\mbf{x}_k,\bm{\lambda}_k,\mbf{p}) \in \tilde{Z}_k$ and $\mbf{v}_k \in V$. If $\{\underline{\tilde{\mbf{G}}},\underline{\tilde{\mbf{c}}}\}$ is a zonotope with $\underline{n}_g$ generators satisfying $\tilde{Z}_{k} - \bm{\gamma}_z \subseteq \{\underline{\tilde{\mbf{G}}},\underline{\tilde{\mbf{c}}}\}$, $\mbf{m} \triangleq (\mbf{J}_z - \text{mid}(\mbf{J}_z)) \underline{\tilde{\mbf{c}}} \in \intvalset^n$, and $\tilde{\mbf{P}} \in \realsetmat{n}{n}$ is a diagonal matrix with $\tilde{P}_{ii} = \text{rad}(m_i) + \sum_{j=1}^{\underline{n}_g} \sum_{\ell=1}^{n+n_\lambda+n_p} \text{rad}(J_{z,i\ell})|\underline{\tilde{G}}_{\ell j}|$, then $\{ (\mbf{x}_{k}, \bm{\lambda}_k, \mbf{p}) \in \tilde{Z}_k : \mbf{g}(\mbf{x}_k,\mbf{u}_k,\bm{\lambda}_k,\mbf{p},\mbf{v}_k) = \mbf{y}_k, \, \mbf{v}_k \in V \} \subseteq \hat{Z}_k$, where
	\begin{equation} \label{eq:joint_mveupdatejoint}
	\hat{Z}_k = \left\{ \begin{bmatrix} \tilde{\mbf{G}}_z & \mbf{0} & \mbf{0} \end{bmatrix}, \tilde{\mbf{c}}_z, \begin{bmatrix} \tilde{\mbf{A}}_z & \mbf{0} & \mbf{0} \\ \mbf{0} & \mbf{0} & \mbf{A}_{Z_v} \\ \midpoint{\mbf{J}} \tilde{\mbf{G}}_z & -\tilde{\mbf{P}} & -\mbf{G}_{Z_v} \end{bmatrix}, \begin{bmatrix} \tilde{\mbf{b}}_z \\ \mbf{b}_{Z_v} \\ \midpoint{\mbf{J}_z} \left(\bm{\gamma}_z - \tilde{\mbf{c}}_z\right) + \mbf{y}_k + \mbf{c}_{Z_v}  \end{bmatrix} \right\}.
	\end{equation}   		
\end{proposition}

\proof 
Choose any $(\mbf{x}_k,\bm{\lambda}_k,\mbf{p}) = \mbf{z}_{k} \in \tilde{Z}_{k}$, $\mbf{v}_{k} \in V$, satisfying $\mbf{g}(\mbf{x}_k,\mbf{u}_k,\bm{\lambda}_k,\mbf{p},\mbf{v}_k) = \mbf{y}_k$. Lemma \ref{lem:joint_mve} (with $\bm{\alpha} \triangleq \mbf{g}$, $\mbf{r} \triangleq (\mbf{x}_k,\bm{\lambda}_k,\mbf{p})$, $\mbf{s} \triangleq \mbf{v}_k$, $R \triangleq \tilde{Z}_k$, and $S \triangleq V$) ensures that there exists a real matrix $\hat{\mbf{J}}\in \mbf{J}_z$ such that $\mbf{g}(\mbf{x}_k,\mbf{u}_k,\bm{\lambda}_k,\mbf{p},\mbf{v}_k) = \mbf{g}(\bm{\gamma}_x,\mbf{u},\bm{\gamma}_\lambda,\bm{\gamma}_p,\mbf{v}_k) + \hat{\mbf{J}}_k (\mbf{z}_k - \bm{\gamma}_z)$, with $\mbf{z} = (\mbf{x},\bm{\lambda},\mbf{p})$. Therefore, from Theorem \ref{thm:ndyn_czinclusion}, there must exist $\bm{\delta} \in B_\infty^{n_y}$ such that
\begin{align*}
\mbf{y}_k = \mbf{g}(\mbf{x}_{k},\mbf{u}_{k},\bm{\lambda}_{k},\mbf{p},\mbf{v}_{k}) = \mbf{g}(\bm{\gamma}_x,\mbf{u}_{k},\bm{\gamma}_\lambda,\bm{\gamma}_p,\mbf{v}_{k}) + \text{mid}(\mbf{J}_z)(\mbf{z}_{k} - \bm{\gamma}_z) + \tilde{\mbf{P}} \bm{\delta},
\end{align*}
with $\tilde{\mbf{P}}$ defined as in the statement of this proposition. The equation above can be rewritten as $\text{mid}(\mbf{J}_z)\mbf{z}_{k} = \mbf{y}_k - \mbf{g}(\bm{\gamma}_x,\mbf{u}_{k},\bm{\gamma}_\lambda,\bm{\gamma}_p,\mbf{v}_{k}) + \text{mid}(\mbf{J}_z)\bm{\gamma}_z - \tilde{\mbf{P}} \bm{\delta} \in (\mbf{y}_k + \text{mid}(\mbf{J}_z)\bm{\gamma}_z) \oplus Z_v \oplus \tilde{\mbf{P}} B_\infty^{n_y}$. Therefore, $\{(\mbf{x}_{k}, \bm{\lambda}_k, \mbf{p}) \in \tilde{Z}_k : \mbf{g}(\mbf{x}_k,\mbf{u}_k,\bm{\lambda}_k,\mbf{p},\mbf{v}_k) = \mbf{y}_k, \, \mbf{v}_k \in V \} \subseteq \{ (\mbf{x}_{k}, \bm{\lambda}_k, \mbf{p}) = \mbf{z}_k \in \tilde{Z}_k : \text{mid}(\mbf{J}_z)\mbf{z}_{k} \in  (\mbf{y}_k + \text{mid}(\mbf{J}_z)\bm{\gamma}_z) \oplus Z_v \oplus \tilde{\mbf{P}} B_\infty^{n_y} \}$. By considering the generalized intersection \eqref{eq:pre_czintersection} and the definition of $\hat{Z}_k$, the proposition is proven. \qed

\begin{remark} \rm \label{rem:joint_ZwZdZv}
	The constrained zonotopes $Z_w$, $Z_d$, and $Z_v$ in Corollary \ref{thm:joint_mvejointprediction}, and Propositions \ref{thm:joint_mvejointconsistency} and \ref{thm:joint_mvejointupdate}, can be obtained as $Z_w = \mbf{f}(\bm{\gamma}_x,\mbf{u}_{k-1},\bm{\gamma}_\lambda,\bm{\gamma}_p,\bm{\gamma}_w) \oplus \gzinclusion \left(\mbf{J}_w,  W - \bm{\gamma}_w \right) \supseteq \mbf{f}(\bm{\gamma}_x,\mbf{u},\bm{\gamma}_\lambda,\bm{\gamma}_p,W)$, $Z_d = -\mbf{h}(\bm{\gamma}_x,\mbf{u}_k,\bm{\gamma}_\lambda,\bm{\gamma}_p,\bm{\gamma}_d) \oplus \gzinclusion \left(-\mbf{J}_d,  D - \bm{\gamma}_d \right) \supseteq - \mbf{h}(\bm{\gamma}_x,\mbf{u},\bm{\gamma}_\lambda,\bm{\gamma}_p,D)$, and $Z_v = -\mbf{g}(\bm{\gamma}_x,\mbf{u},\bm{\gamma}_\lambda,\bm{\gamma}_p,\bm{\gamma}_v) \oplus \gzinclusion \left(-\mbf{J}_v,  V - \bm{\gamma}_v \right) \supseteq - \mbf{g}(\bm{\gamma}_x,\mbf{u},\bm{\gamma}_\lambda,\bm{\gamma}_p,V)$, respectively, for a chosen $(\bm{\gamma}_w,\bm{\gamma}_d,\bm{\gamma}_v) \in \square W \times \square D \times \square V$, and interval matrices $\mbf{J}_w \supseteq \nabla^T_w \mbf{f}(\bm{\gamma}_x,\mbf{u},\bm{\gamma}_\lambda,\bm{\gamma}_p, \square W), \mbf{J}_d \supseteq \nabla^T_d \mbf{h}(\bm{\gamma}_x,\mbf{u},\bm{\gamma}_\lambda,\bm{\gamma}_p, \square D)$, and $\mbf{J}_v \supseteq \nabla^T_v \mbf{g}(\bm{\gamma}_x,\mbf{u},\bm{\gamma}_\lambda,\bm{\gamma}_p, \square V)$. 
\end{remark}

\begin{remark} \rm \label{rem:joint_setcomplexity}
	If the constrained zonotopes $(\hat{Z}_{k-1}, \bar{Z}_k,\tilde{Z}_k, L_k, W, D, V)$ have $(\hat{n}_g,\bar{n}_g,\tilde{n}_g,n_{g_\lambda},$ $n_{g_w},n_{g_d},n_{g_v})$ generators, and $(\hat{n}_c,\bar{n}_c,\tilde{n}_c,n_{c_\lambda},n_{c_w},n_{c_d},n_{c_v})$ constraints, respectively, then the enclosures obtained by: (i) Corollary \ref{thm:joint_mvejointprediction} have $\hat{n}_g + 2n + n_{g_w} + n_{g_\lambda}$ generators and $\hat{n}_c + n_{c_w} + n_{g_\lambda}$ constraints; (ii) Proposition \ref{thm:joint_mvejointconsistency} have $\bar{n}_g + 2n_h + n_{g_d}$ generators and $\bar{n}_c + n_h + n_{c_d}$ constraints; and (iii) Proposition \ref{thm:joint_mvejointupdate} have $\tilde{n}_g + 2n_y + n_{g_v}$ generators and $\hat{n}_c + n_y + n_{c_v}$ constraints.
\end{remark}

The following Algorithm summarizes the set-based joint state and parameter estimation method proposed in this chapter, which is denoted by CZMV-J.

\begin{algorithm}[!htb]
	\caption{CZMV-J}
	\label{alg:joint_estimation}
	\begin{algorithmic}[1]	
		\State (Joint prediction step) Given the constrained zonotopes $\hat{Z}_{k-1} \times L_k \times W \subset \realset^{n+n_\lambda+n_p} \times \realset^{n_\lambda} \times \realset^{n_w}$, and input $\mbf{u}_{k-1} \in \realset^{n_u}$, compute the joint prediction enclosure $\bar{Z}_k$ given by \eqref{eq:joint_mvejointprediction}. This enclosure satisfies \eqref{eq:joint_prediction0}.
		\State (Joint consistency step) Given the constrained zonotopes $\bar{Z}_{k} \times D \subset \realset^{n+n_\lambda+n_p} \times \realset^{n_d}$, and input $\mbf{u}_{k} \in \realset^{n_u}$, compute the joint consistency enclosure $\tilde{Z}_k$ given by \eqref{eq:joint_mveconsistencyjoint}. This enclosure satisfies \eqref{eq:joint_consistency0}.
		\State (Joint update step) Given the constrained zonotopes $\tilde{Z}_{k} \times V \subset \realset^{n+n_\lambda+n_p} \times \realset^{n_v}$, and input $\mbf{u}_{k} \in \realset^{n_u}$, compute the joint update enclosure $\hat{Z}_k$ given by \eqref{eq:joint_mveupdatejoint}. This enclosure satisfies \eqref{eq:joint_update0}.		
	\end{algorithmic}
	\normalsize
\end{algorithm}

\begin{remark} \rm \label{rem:joint_choiceofgamma}
	In the set-based joint and state estimation method developed in this chapter, we use the approximation point $\bm{\gamma}_z$ given by \emph{C2} proposed in Chapter \ref{cha:nonlinearmeasinv} (see Section \ref{sec:nmeas_choiceofgamma}). 
\end{remark}

\section{Numerical example} \label{sec:joint_example}

This section presents numerical results for the set-based joint state and parameter estimation method proposed in this chapter. We also compare the results provided by CZMV-J with the method proposed in Chapter \ref{cha:nonlinearmeasinv} based on the mean value extension, denoted by CZMV (in which the model parameters $\mbf{p}$ are additional components to the uncertainties $\mbf{w}_k$), and with a `naive' methodology denoted by CZMV-N. In this latter, the set-based joint and state estimation in \eqref{eq:joint_prediction0}--\eqref{eq:joint_update0} is performed without the proposed unified framework, i.e., the sets $\bar{Z}_k$, $\tilde{Z}_k$, and $\hat{Z}_k$ are disjointed between each step of the algorithm into three lower-dimensional enclosures, given by $\bar{X}_k \times \bar{L}_k \times \bar{P}_k$, $\tilde{X}_k \times \tilde{L}_k \times \tilde{P}_k$, and $\hat{X}_k \times \hat{L}_k \times \hat{P}_k$, respectively. The disadvantage of CZMV-N with respect to CZMV-J is related to the fact that the mutual coupling between states, algebraic variables, and unknown parameters is ignored from one step of the algorithm to the next step.

\subsection{Example 1}

Consider a nonlinear discrete-time system with nonlinear dynamics and nonlinear measurement described by
\begin{equation} \label{eq:joint_example1}
\begin{aligned}
x_{1,k} & = 3 x_{1,k-1} - p x_{1,k-1}^2 - \frac{4 x_{1,k-1} x_{2,k-1}}{4 + x_{1,k-1}} + w_{1,k-1} \\
x_{2,k} & = -2 x_{2,k-1} + \frac{3 x_{1,k-1} x_{2,k-1}}{4 + x_{1,k-1}} + w_{2,k-1} \\
y_{1,k} & =  x_{1,k} - \sin\left(\frac{x_{2,k}}{2}\right) + v_{1,k}, \\
y_{2,k} & = -x_{1,k}x_{2,k} + x_{2,k} + (7p-1) + v_{2,k},
\end{aligned}
\end{equation}
with $\|\mbf{w}_k\|_\infty \leq 0.2$, $\|\mbf{v}_k\|_\infty \leq 0.1$, and $p \in P \subset \realset$ is an unknown model parameter. The initial state enclosure $\bar{X}_0$ and the known parameter bounds $P$ are zonotopes given by
\begin{equation} \label{eq:joint_example1_X0}
\bar{X}_0 = \left\{ \begin{bmatrix} 0.3 & 0.6 & -0.3 \\ 0.3 & 0.3 & 0 \end{bmatrix}, \begin{bmatrix} 10 \\ 0.5 \end{bmatrix} \right\}, P = \left\{ 5, 1/7 \right\}.
\end{equation}
To obtain process measurements, \eqref{eq:joint_example1} was simulated with $\mbf{x}_0 = (10.2,0.65) \in \bar{X}_0$, and $p = 1/7 \in P$, and the process and measurement uncertainties are generated from uniform random distributions. The numbers of generators and constraints of the estimated state and parameter enclosures were limited to 8 and 3, respectively.

\begin{figure*}[!tb]
	\centering{
		\def\svgwidth{0.6\textwidth}
		{\footnotesize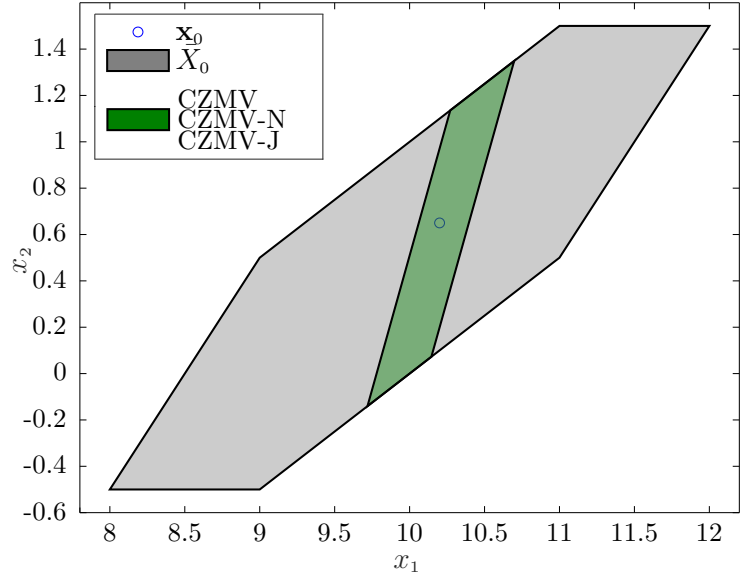}
		\caption{The initial state $\mbf{x}_0$, the set $\bar{X}_0$, and the projections onto $\mbf{x}$ of the enclosures obtained by CZMV, CZMV-N, and CZMV-J, in the update step using $\mbf{y}_0$ for the first example (which are exactly the same in this case).}\label{fig:joint_exampleX0}}
\end{figure*}

\begin{figure*}[!tb]
	\centering{
		\def\svgwidth{0.8\textwidth}
		{\footnotesize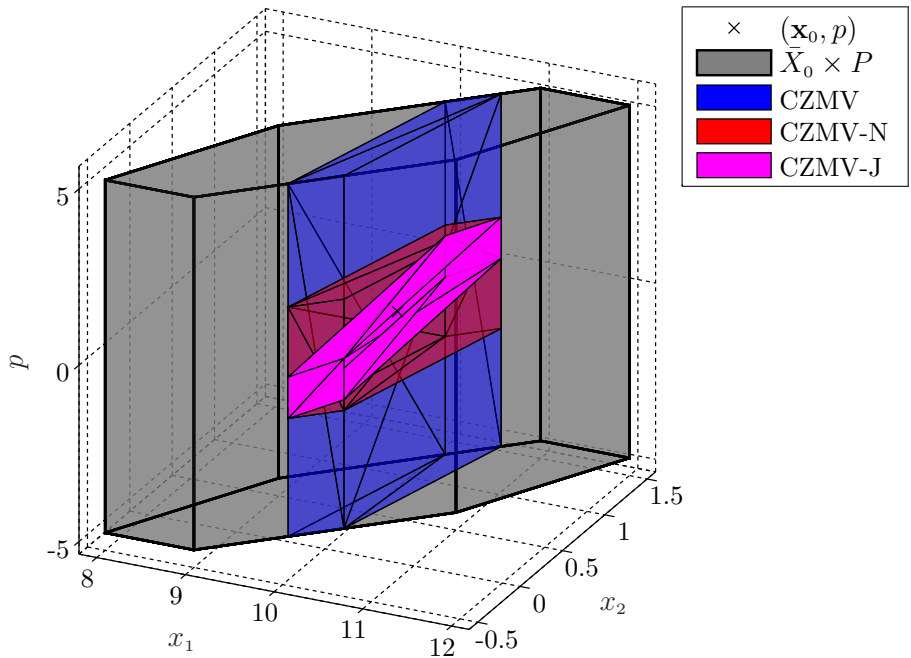}
		\caption{The variable $(\mbf{x}_0,p)$, the initial set $\bar{X}_0 \times P$, and the enclosures obtained by CZMV, CZMV-N, and CZMV-J, in the first update step for the first example.}\label{fig:joint_exampleZ0}}
\end{figure*}

Figure \ref{fig:joint_exampleX0} shows the initial set $\bar{X}_0$, together with the enclosures obtained by CZMV, CZMV-N, and CZMV-J, in the update step using $\mbf{y}_0$, projected into the state space. It can be noticed that the enclosure provided by all the estimators for the state $\mbf{x}_0$ are equivalent. However, in Figure \ref{fig:joint_exampleZ0}, which shows the set $\bar{X}_0 \times P$ together with the enclosures obtained by CZMV, CZMV-N, and CZMV-J in the first update step, it can be noticed that while the parameter uncertainty remains the same with CZMV (since the parameters are considered as disturbances, and therefore not estimated), while by using CZMV-N the enclosure $P$ is significantly refined. In addition, using the proposed unified approach given by CZMV-J, the obtained enclosure is still smaller, since it takes into account the mutual coupling between states and parameter, which is a result from the output equation in \eqref{eq:joint_example1}.

\begin{figure*}[!tb]
	\centering{
		\def\svgwidth{0.7\textwidth}
		{\scriptsize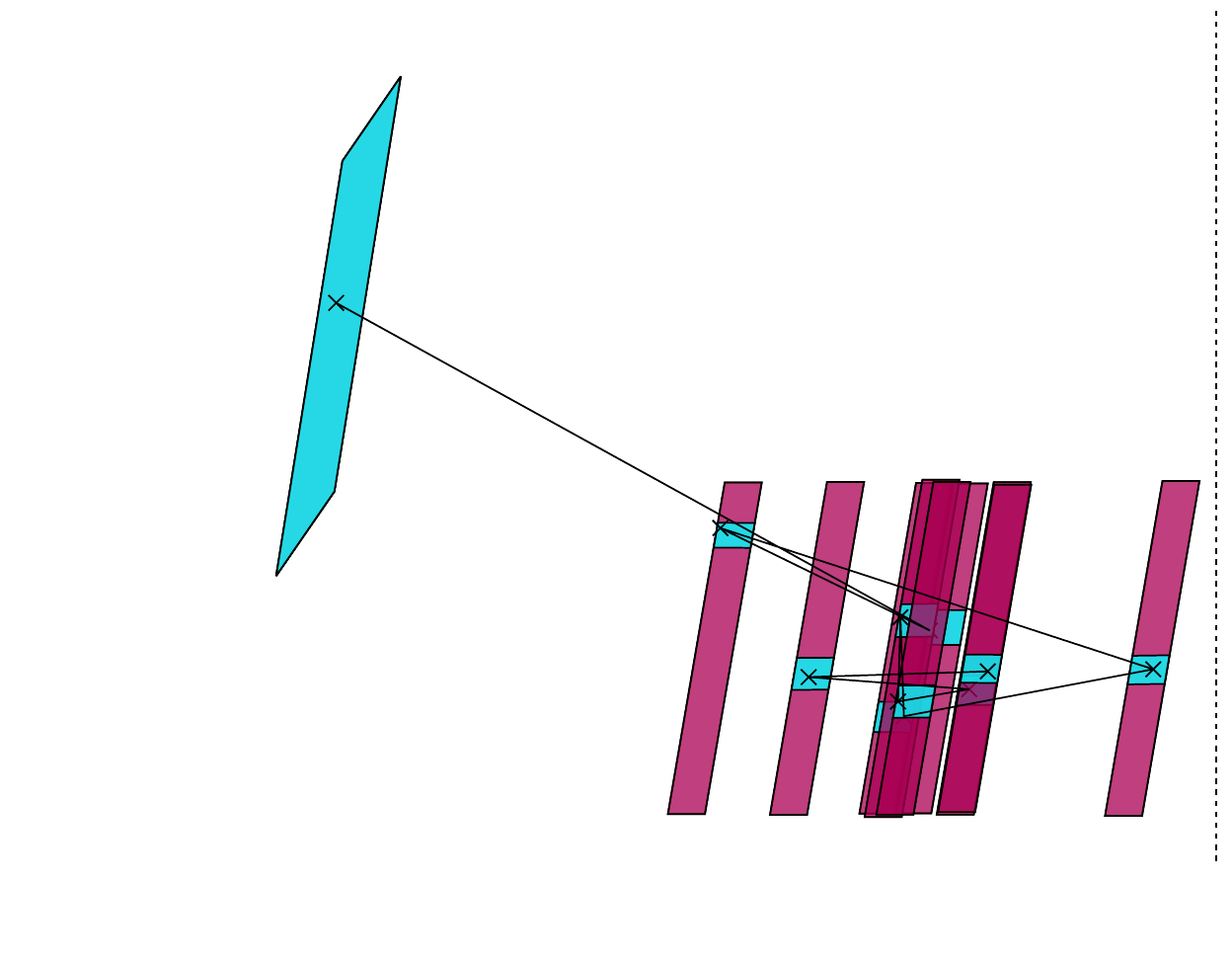}
		\caption{The states $\mbf{x}_k$, and the projections onto the state space of the enclosures provided by CZMV, CZMV-N, and CZMV-J, for the first example and $k \in [0,200]$.}\label{fig:joint_example2Dcomparison}}
\end{figure*}

\begin{figure*}[!tb]
	\centering{
		\def\svgwidth{0.7\textwidth}
		{\footnotesize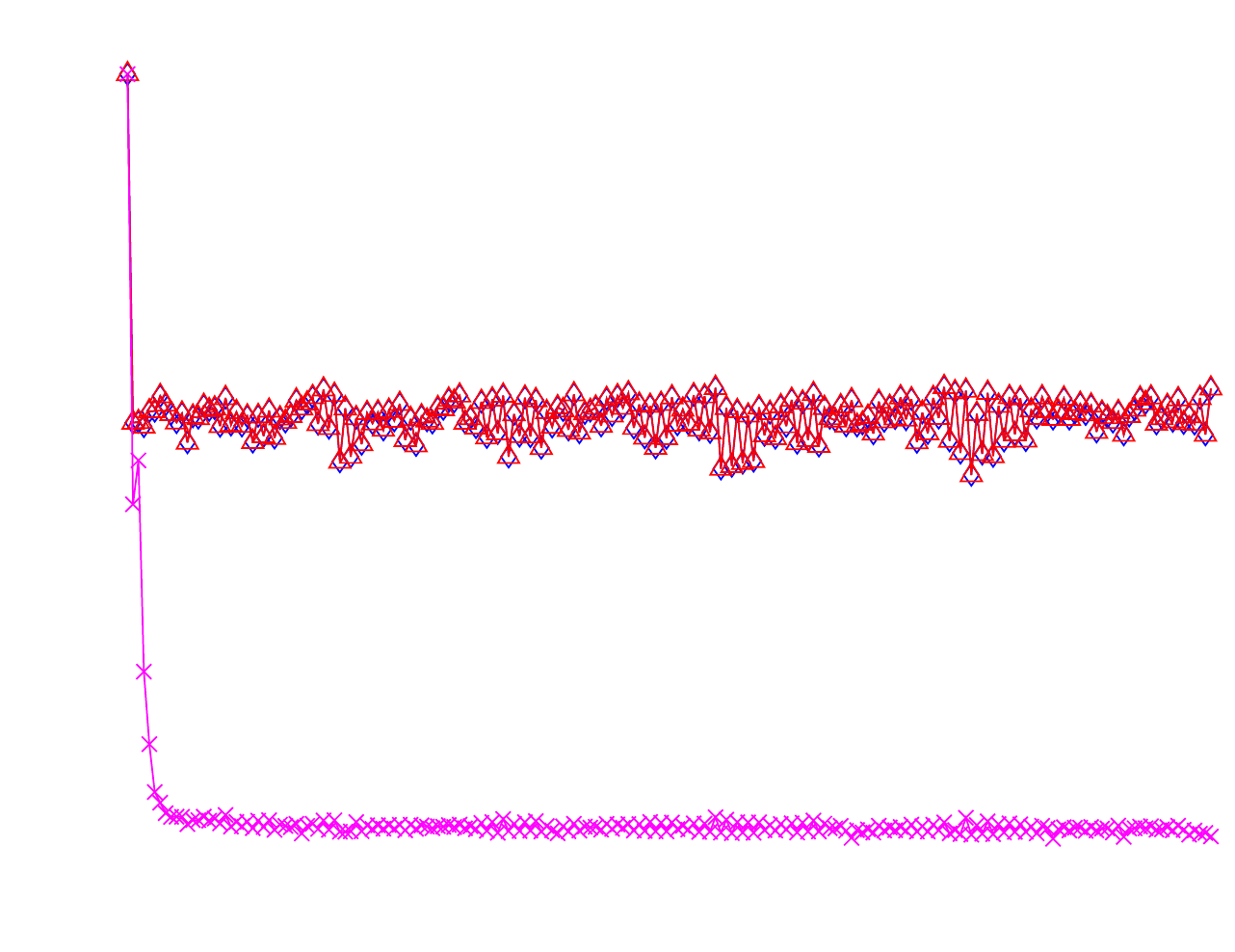}
		\caption{The radii of the projections onto the state space of the enclosures provided by CZMV, CZMV-N, and CZMV-J, for the first example and $k \in [0,200]$.}\label{fig:joint_exampleradiusX}}
\end{figure*}

\begin{figure*}[!tb]
	\centering{
		\def\svgwidth{0.7\textwidth}
		{\footnotesize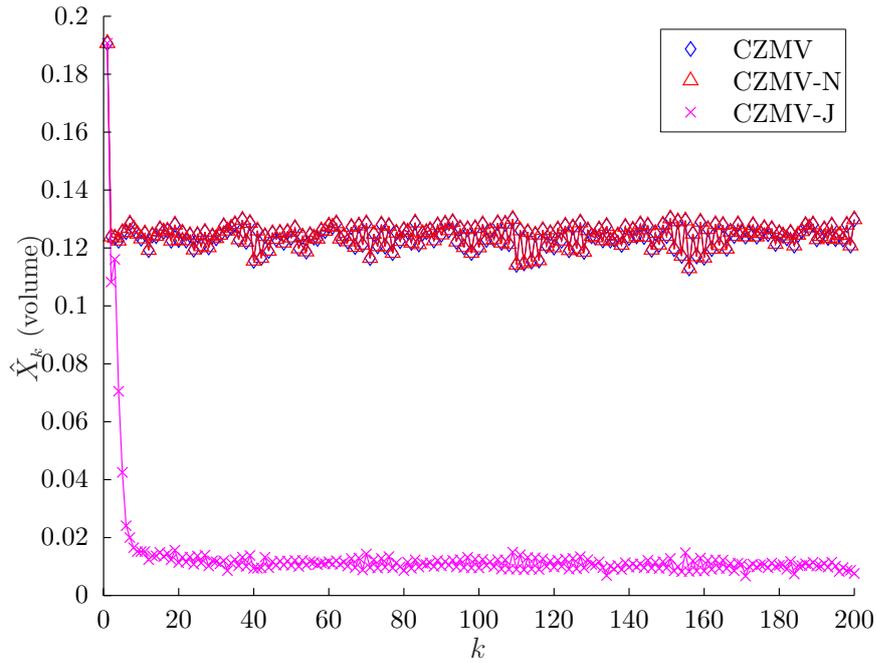}
		\caption{The volumes of the projections onto the state space of the enclosures provided by CZMV, CZMV-N, and CZMV-J, for the first example and $k \in [0,200]$.}\label{fig:joint_examplevolumeX}}
\end{figure*}

Figure \ref{fig:joint_example2Dcomparison} shows the projections $\hat{X}_K$ of the enclosures obtained using CZMV, CZMV-N, and CZMV-J, for $k \in [0,200]$, in intervals of 20 time steps each. It can be noticed that both the previous state estimation method CZMV and the naive strategy CZMV-N are not able to provide accurate enclosures of the states, while the proposed method CZMV-J is able to provide sharp bounds. This is corroborated by Figures \ref{fig:joint_exampleradiusX} and \ref{fig:joint_examplevolumeX}, which show the radii and volumes of the computed state enclosures.

\begin{figure*}[!tb]
	\centering{
		\def\svgwidth{0.7\textwidth}
		{\footnotesize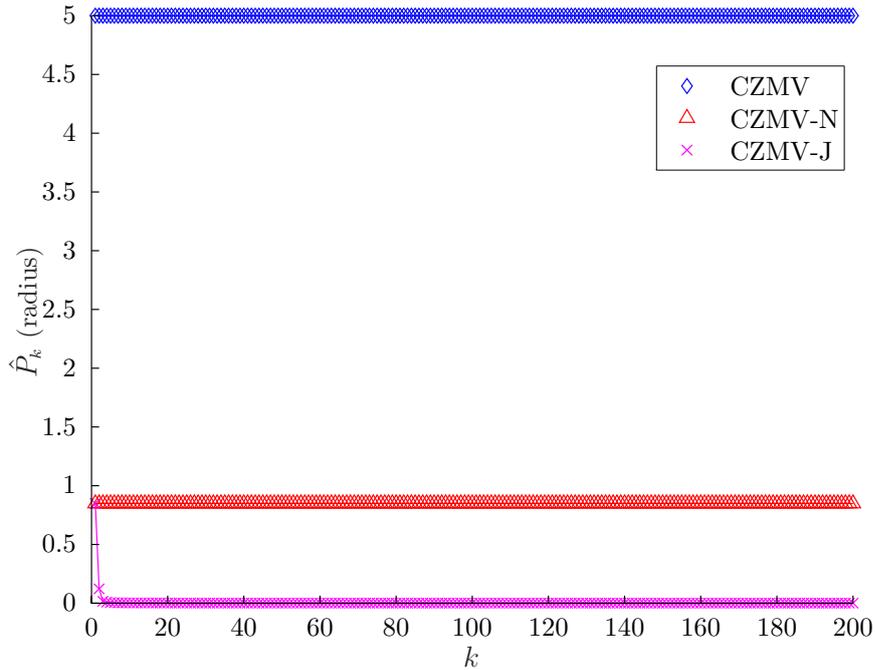}
		\caption{The enclosures of the unknown parameter $p$ provided by CZMV, CZMV-N, and CZMV-J, for the first example and $k \in [0,200]$.}\label{fig:joint_exampleradiusP}}
\end{figure*}

Figure \ref{fig:joint_exampleradiusP} illustrates the radii of the parameter enclosures provided by CZMV, CZMV-N, and CZMV-J. As it can be noticed, the parameter enclosures used in CZMV are not refined during the entire simulation. On the other hand, CZMV-N is able to refine the parameter bounds, but it is not sufficiently accurate to result in tighter state bounds than CZMV (as shown in Figure \ref{fig:joint_example2Dcomparison}). Nevertheless, the parameter bounds provided by CZMV-J are significantly refined over time. Consequently, this resulted in much more accurate state enclosures as expected, in comparison to the CZMV and CZMV-N, which do consider the unified framework proposed in this chapter.

\subsection{Example 2}

Consider a nonlinear discrete-time system with nonlinear dynamics and nonlinear algebraic constraints described by \citep{Sjoberg2005}
\begin{equation} \label{eq:joint_example2}
\begin{aligned}
x_{1,k} & = x_{1,k-1} + T_s(p x_{2,k-1}), \\
x_{2,k} & = x_{2,k-1} + T_s(e^{\lambda_{k-1}} - 1 + 0.5u_{k-1}), \\
      0 & = x_{1,k} - \arcsin(1 - e^{\lambda_{k}} + 0.5u_{k}), \\
y_{1,k} & =  p x_{1,k} + v_{1,k}, \\
y_{2,k} & =    x_{2,k} + v_{2,k},
\end{aligned}
\end{equation}
with $\|\mbf{v}_k\|_\infty \leq 0.01$, $T_s = 0.01$ s, and $p \in P \subset \realset$ is an unknown model parameter. The initial state enclosure $\bar{X}_0$, the known bounds $L_k$ on the algebraic variable $\lambda_k$, and the known parameter bounds $P$, are zonotopes given by
\begin{equation} \label{eq:joint_example2_X0}
\bar{X}_0 = \left\{ \begin{bmatrix} 0.1 & 0.2 & -0.1 \\ 0.1 & 0.1 & 0 \end{bmatrix}, \begin{bmatrix} -0.2 \\ -0.15 \end{bmatrix} \right\}, \quad L_k = \{ 0.1, \lambda_k\}, \quad P = \left\{ 0.1, 1 \right\}.
\end{equation}
The initial set $\bar{Z}_0$ is given by $\bar{Z}_0 = \bar{X}_0 \times L_0 \times P$. To obtain process measurements, \eqref{eq:joint_example2} was simulated with $\mbf{x}_0 = (0,0) \in \bar{X}_0$, and $p = 1 \in P$, and the process and measurement uncertainties are generated from uniform random distributions. For simulation purposes, the algebraic variable $\lambda_k$ is obtained by solving the nonlinear algebraic constraint \eqref{eq:joint_example2} at each time step $k$. The numbers of generators and constraints of the estimated state, algebraic variable, and parameter enclosures were limited to 8 and 3, respectively.

\begin{figure*}[!tb]
	\centering{
		\def\svgwidth{0.7\textwidth}
		{\scriptsize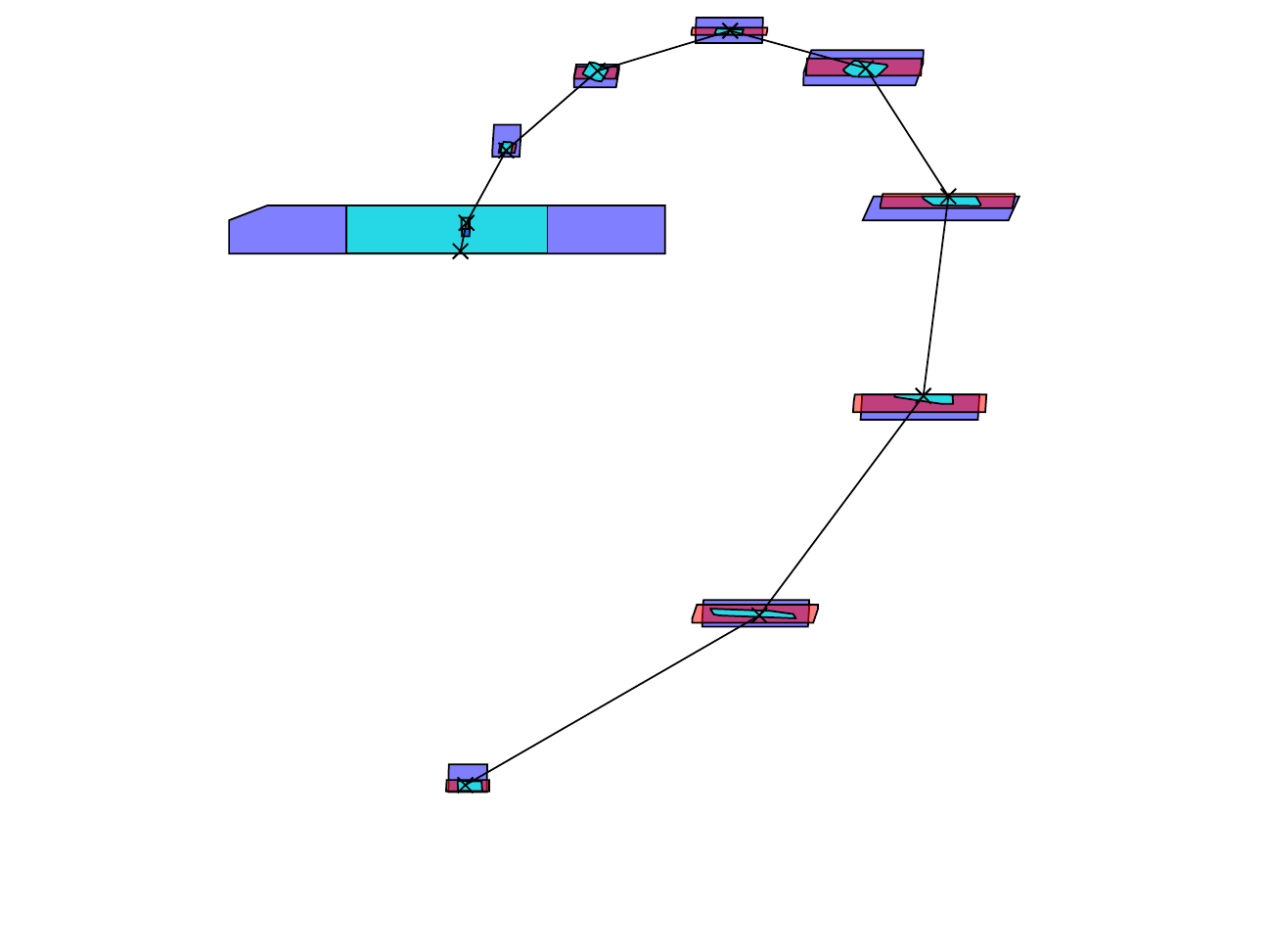}
		\caption{The states $\mbf{x}_k$, and the projections onto the state space of the enclosures provided by CZMV, CZMV-N, and CZMV-J, for the second example and $k \in [0,500]$.}\label{fig:joint_PLL2Dcomparison}}
\end{figure*}

\begin{figure*}[!tb]
	\centering{
		\def\svgwidth{0.7\textwidth}
		{\scriptsize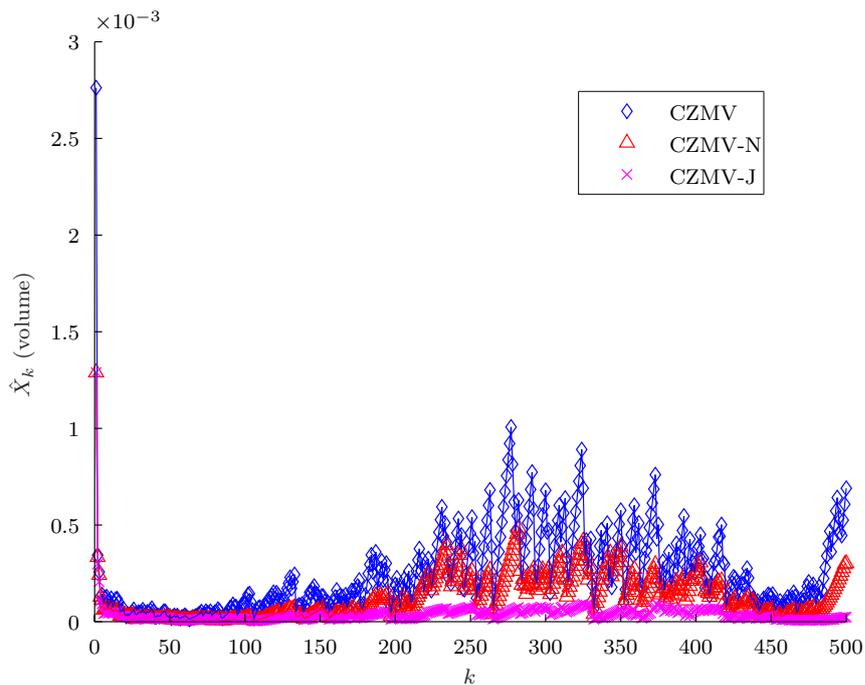}
		\caption{The volumes of the projections onto the state space of the enclosures provided by CZMV, CZMV-N, and CZMV-J, for the second example and $k \in [0,500]$.}\label{fig:joint_PLLvolumeX}}
\end{figure*}

Figure \ref{fig:joint_PLL2Dcomparison} shows the projections $\hat{X}_K$ of the enclosures obtained using CZMV, CZMV-N, and CZMV-J, for $k \in [0,500]$, in intervals of 50 time steps each. As it can be noticed, both CZMV and CZMV-N are able to provide reasonable enclosures of the states. However, CZMV-J is able to provide sharper bounds in comparison to the other methods. This is corroborated by Figure \ref{fig:joint_PLLvolumeX}, which show the volumes of the computed state enclosures.

\begin{figure*}[!tb]
	\centering{
		\def\svgwidth{0.7\textwidth}
		{\scriptsize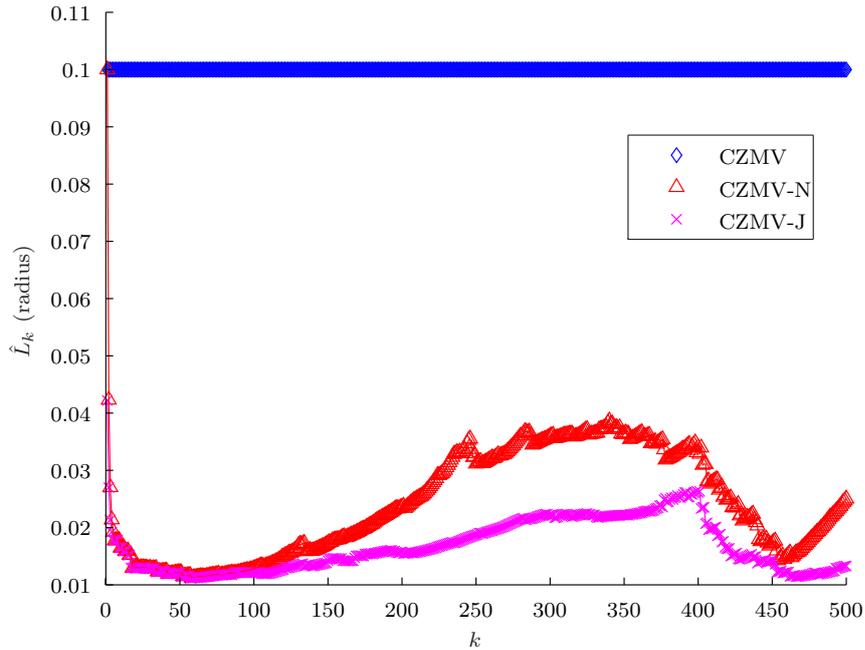}
		\caption{The enclosures of the algebraic variable $\lambda_k$ provided by CZMV, CZMV-N, and CZMV-J, for the second example and $k \in [0,500]$.}\label{fig:joint_PLLradiusL}}
\end{figure*}

\begin{figure*}[!tb]
	\centering{
		\def\svgwidth{0.7\textwidth}
		{\scriptsize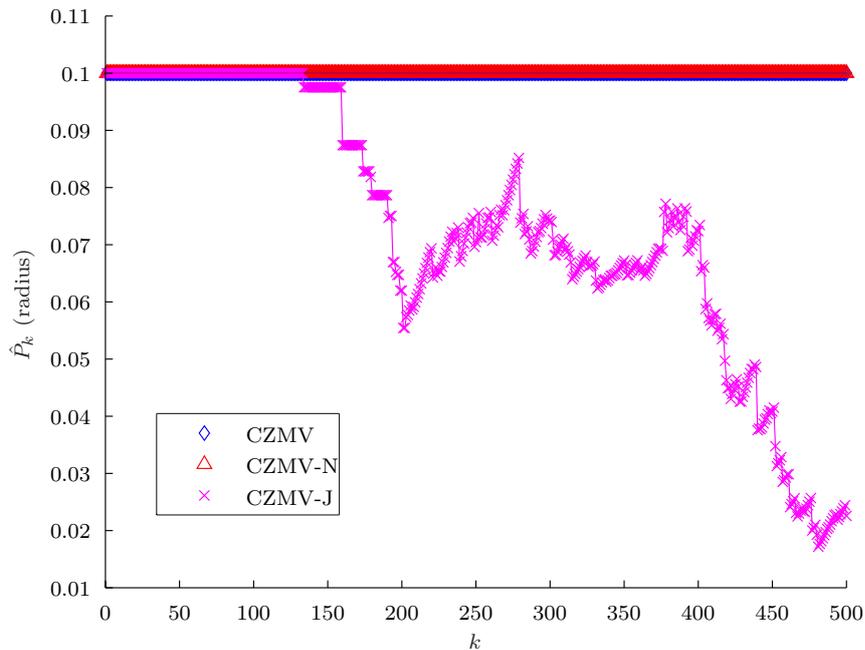}
		\caption{The enclosures of the unknown parameter $p$ provided by CZMV, CZMV-N, and CZMV-J, for the second example and $k \in [0,200]$.}\label{fig:joint_PLLradiusP}}
\end{figure*}

Figure \ref{fig:joint_PLLradiusL} illustrates the radii of the enclosures $\hat{L}_k$ of $\lambda_k$ provided by CZMV, CZMV-N, and CZMV-J. As expected, the enclosures $\hat{L}_k$ used in CZMV are not refined during the entire simulation. On the other hand, both CZMV-N and CZMV-J are able to refine the enclosures of the algebraic variable, but the latter is notably more accurate. In addition, Figure \ref{fig:joint_PLLradiusP} illustrates the radii of the parameter enclosures provided by CZMV, CZMV-N, and CZMV-J. As in the case of the algebraic variable $\lambda_k$, the parameter enclosures used in CZMV are not refined during the simulation. On the other side, CZMV-N is also not able to refine the parameter bounds in this case, while the parameter bounds provided by CZMV-J are significantly refined over time. This result corroborates the advantages of using the proposed unified framework for joint state and parameter estimation in comparison to other techniques.%

\section{Final remarks} \label{sec:joint_remarks}

This chapter developed a new method for set-based joint state and parameter estimation of nonlinear discrete-time systems with algebraic variables and unknown model parameters. By extending the nonlinear state estimation methods using constrained zonotopes to include parameter estimation in a unified framework, and therefore maintaining existing mutual dependencies between states, algebraic variables, and unknown parameters, the accuracy of both state and parameter estimation was significantly improved, as corroborated by two numerical examples. %
The next chapters present a few practical applications of the set-based state estimation and fault diagnosis methods developed in this thesis, specifically: unmanned aerial vehicles (Chapter \ref{cha:appuav}), water distribution networks (Chapter \ref{cha:appwdn}), and Lithium-ion batteries (Chapter \ref{cha:appbat}).

\chapter{Application: Unmanned aerial vehicles}\thispagestyle{headings} \label{cha:appuav}

This chapter presents the application of set-based state estimation and fault diagnosis methods using constrained zonotopes to unmanned aerial vehicles. It is organized according to the following topics: (i) linear state estimation applied to aerial load transportation; (ii) nonlinear state estimation applied to a quadrotor UAV considering sensors located at the UAV; and (iii) active fault diagnosis of a quadrotor UAV subject to actuator and sensor faults.

\section{Linear state estimation applied to aerial load transportation} \label{sec:uav_linearestimation}
\sectionmark{State estimation applied to aerial load transportation}

The problem of slung load transportation arises in a variety of essential tasks, such as transportation of containers in harbors \citep{Ngo2012}, aerial delivery of supplies in search-and-rescue missions \citep{Bernard2011}, and landmine detection \citep{BisgaardThesis}. The suspended load is usually connected to the mobile platform by means of a rope, considerably changing its dynamic behavior and adding unactuated degrees of freedom to the whole system. Moreover, the rope is a non-rigid body and is not always taut, which increases the task challenge. Several studies can be found in the literature, concerning different modeling approaches and control strategies for load transportation using overhead cranes \citep{Wu2015}, robotic manipulators \citep{Chen2007}, and aerial vehicles \citep{Bisgaard2009b}.%

Unmanned aerial vehicles have been extensively used for load transportation, thanks to their versatility and autonomous operation \citep{Sreenath2013b}. Control objectives include path tracking of the UAV with load swing attenuation~\citep{Raffo2016}, and path tracking of the suspended load~\citep{Sreenath2013b}. The latter is the appropriate goal for tasks requiring precise maneuvering of the load, for which the knowledge of the load position is of utmost importance.

As mentioned above, a recurrent issue in aerial load transportation is the necessity of knowing the load position to accomplish the task. Since available sensors are often embedded in the mobile platform, information on the load position may not be directly available. The problem of estimating the load position is commonly addressed through visual systems and Bayesian state estimators, such as the linear Kalman filter \citep{JainThesis}, and the unscented Kalman filter (UKF) \citep{Bisgaard2007a,Bisgaard2007b,Bisgaard2010}. In particular, in \cite{Bisgaard2007b}, algorithms based on the UKF are proposed for estimation of the full state vector of a helicopter with suspended load, with measurements provided by a Global Positioning System (GPS) equipment, a magnetometer, a camera, an IMU on the helicopter and another one on the load. However, such algorithms require knowledge on statistical properties of existing process and measurement disturbances, which may not be easily obtained. On the other hand, set-based estimation requires knowledge only on bounds of existing disturbances, and are guaranteed to include the system states consistent with available measurements.

Most of the unmanned aerial vehicles used in load transportation tasks are in helicopter and quadrotor configurations. These rotary-wing UAVs have vertical take-off and landing (VTOL) and hovering capabilities, and achieve high maneuverability in low velocities. However, due to their limited flight envelope, such UAVs are not appropriate for missions that require long distance traveling, such as deployment of supplies to risky zones. To overcome such constraint, recent researches are looking into the design of small-scale hybrid aircrafts, being the tilt-rotor configuration among the most popular ones \citep{Amiri2011,Park2013,Cardoso2016,Santos2017b,Cardoso2021}. Provided with both fixed and rotary wings, tilt-rotor UAVs achieve an enlarged flight envelope by switching between helicopter and airplane flight-modes through thrusters tilting. However, such advantages come with several design and control challenges, since these aircrafts are complex, underactuated mechanical systems with highly coupled dynamics. Additionally, when these UAVs are connected to a payload through a rope, the dynamic behavior of the system varies due to the load's swing, which can destabilize the whole system if it is not well attenuated. 

The use of tilt-rotor UAVs for load transportation is recent. A three-level cascade strategy composed of feedback linearization controllers is proposed in \cite{Almeida2015b}, which considers load swing attenuation, but unrealistically assumes that all the states are precisely known. On the other hand, a model predictive control (MPC) strategy is designed in \cite{Santos2017} for path tracking of a tilt-rotor UAV with suspended load, in which the aircraft followed a desired trajectory, while the load remained stable. The strategy takes into account time-varying load mass and rope length, and estimates the load position and orientation by means of an unscented Kalman filter. However, the state estimation is not guaranteed, and nothing can be said about the transient response of the closed-loop system. The use of set-based state estimation based on zonotopes for aerial load transportation with a tilt-rotor UAV was first proposed in \cite{Rego2016c,Rego2019}. However, the latter is considerably conservative, and requires iterative procedures to take into account multiple measurements.

This section solves the path tracking problem of a suspended load using a tilt-rotor UAV, with states obtained through set-based state estimation using constrained zonotopes. The realistic scenario in which the load position is not measured is taken into account, considering sensors with different sampling times, and unknown-but-bounded disturbances. In order to provide feedback to the controller based on the estimated state set, which is a constrained zonotope, a novel optimal state choice is proposed taking into consideration a \emph{constrained minimum-variance criteria}, resulting in the main contribution of this section. Path tracking control for the proposed aerial load transportation task is then designed by a robust discrete-time mixed $\mathcal{H}_2/\mathcal{H}_\infty$ state-feedback approach. The estimation and control strategies are validated through numerical experiments, performed in a platform based on the Gazebo simulator with a Computer Aided Design (CAD) model of the system\footnote{See https://github.com/Guiraffo/ProVANT-Simulator.} \citep{Lara2017}, to corroborate the performance of the proposed method. This section improves the work published in \cite{Rego2019} by: (i) the design of a far more accurate set-based state estimator; (ii) a novel criteria for improved feedback connection based on constrained minimum-variance; (iii) improved control design taking into account the time-varying nature of the trajectory; and (iv) rigorous validation of the proposed strategies considering the CAD model of the system. The content of this section was published in \cite{Rego2018b}, which is a continuity of the previous work published in \cite{Rego2019}.

\subsection{Tilt-rotor UAV with suspended load modeling} \label{sec:uav_tiltrotormodeling}

\subsubsection{Equations of motion}

\begin{figure}[!ht]
	\centering{
		\def\svgwidth{0.6\columnwidth}
		{\vspace{-3mm}\small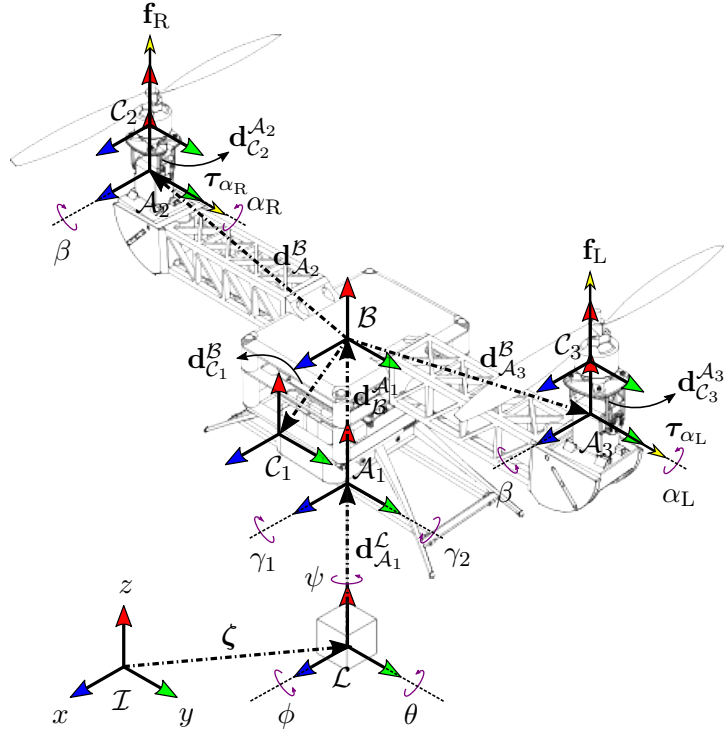}
		\caption{The tilt-rotor UAV with suspended load, kinematic definitions, input forces and torques.}\label{fig:uav_tiltrotorkinematics}}
\end{figure}

This section describes the tilt-rotor UAV with suspended load (Figure \ref{fig:uav_tiltrotorkinematics}) as a multi-body mechanical system composed of: (i) the aircraft's main structure (including batteries, landing skids, electronics and instrumentation); (ii) the right thruster group (right thruster and revolute joint); (iii) the left thruster group (left thruster and revolute joint); and (iv) the load group (load and rope). For control purposes, the rope is assumed to be rigid and massless.

The equations of motion of the entire system are formulated from the perspective of the load, by considering the load as a free rigid body with the tilt-rotor UAV rigidly attached to it. This choice of perspective allows state-feedback control strategies to directly steer the trajectory of the load \citep{Rego2019}. Nine reference frames are defined (Figure \ref{fig:uav_tiltrotorkinematics}): (i) the inertial reference frame, $\frI$; (ii) the load center of mass frame, $\frL$; (iii) the aircraft's geometric center frame, $\frB$; (iv) the main body center of mass frame, $\frC{1}$; (v) the right thruster center of mass frame, $\frC{2}$; (vi) the left thruster center of mass frame, $\frC{3}$; (vii) the rope's point of connection frame, $\frA{1}$; (viii) the right tilting axis frame, $\frA{2}$; and (ix) the left tilting axis frame, $\frA{3}$.

The position of the load with respect to $\frI$ is denoted by $\bm{\zeta} \triangleq [ x \;\, y \;\, z ]^T$. The displacement from $\frL$ to $\frA{1}$ is defined in $\frL$ as $\dLA{1} \triangleq [ 0 \;\, 0 \;\, l ]^T$, where $l$ corresponds to the rope length. The displacements from $\frA{1}$ to $\frB$, from $\frB$ to $\frC{1}$, from $\frB$ to $\frA{i}$, and from $\frA{i}$ to $\frC{i}$ are denoted by $\dAB{1}$, $\dBC{1}$, $\dBA{i}$, $\dAC{i}{i}$, respectively, in the respective preceding frames, with $i \in \{2,3\}$. The orientation of the load with respect to $\frI$ is parametrized by Euler angles, $\bm{\eta} \triangleq [ \phi \; \theta \; \psi ]^T$, using the $ZYX$ convention about local axes, and is described by
\begin{equation} \label{eq:uav_tiltrotorRIL}
\RIL {\triangleq} \mbf{R}_{z,\psi} \mbf{R}_{y,\theta} \mbf{R}_{x,\phi} {=}  \begin{bmatrix}
\text{c}_\psi \text{c}_\theta & \text{c}_\psi \text{s}_\theta \text{s}_\phi {-} \text{s}_\psi \text{c}_\phi & \text{c}_\psi \text{s}_\theta \text{c}_\phi {+} \text{s}_\psi \text{s}_\phi \\
\text{s}_\psi \text{c}_\theta & \text{s}_\psi \text{s}_\theta \text{s}_\phi {+} \text{c}_\psi \text{c}_\phi & \text{s}_\psi \text{s}_\theta \text{c}_\phi {-} \text{c}_\psi \text{s}_\phi \\
-\text{s}_\theta & \text{c}_\theta \text{s}_\phi & \text{c}_\theta \text{c}_\phi
\end{bmatrix} \text{.}
\end{equation}
with $\text{s}_{(\cdot)} {\triangleq} \sin(\cdot)$ and $\text{c}_{(\cdot)} {\triangleq} \cos(\cdot)$. The orientation of the UAV with respect to $\frL$ is described by $\bm{\gamma} \triangleq [ \gone \; \gtwo ]^T$, with $\RLA{1} \triangleq \mbf{R}_{x,-\gone} \mbf{R}_{y,-\gtwo}$. The orientation of the thrusters' groups with respect to $\frB$ are defined by $\RBA{2} \triangleq \mbf{R}_{x,-\beta} \mbf{R}_{y,\aR}$ and $\RBA{3} \triangleq \mbf{R}_{x,\beta} \mbf{R}_{y,\aL}$, where $\aR$ and $\aL$ are the right and left tilting angles, respectively, and $\beta$ is a fixed inclination angle of the thrusters towards $\frB$ \citep{Rego2019}. Moreover, $\frA{1}$, $\frB$, and $\frC{1}$ are parallel and attached to the same rigid body, thus $\RAB{1} = \RBC{1} = \RAC{i}{i} = \eye{3}$, $\RLB = \RLA{1}$, $\RBC{i} = \RBA{i}$, $i \in \{2,3\}$.

Let $\wLIL$ denote the angular velocity of frame $\frL$ with respect to $\frI$, expressed in $\frL$. Then, $\wLIL = \Weta \deta$, $\wALA{1} = \mbf{Q} \dgamma$, $\wABA{2} = \ay \daR$, $\wABA{3} = \ay \daL$, $\wBAB{1} = \wiBi{1} = \wiAi{i} = \zeros{3}{1}$\footnote{$\eye{n}$ is an identity matrix with dimension $n$, and $\zeros{n}{m}$ denotes an $n$ by $m$ matrix of zeros.}, $\wBLB = \wALA{1}$, and $\wiBi{i} = \wABA{i}$, with
\begin{equation*}
\Weta \triangleq \begin{bmatrix}
1 & 0 & -\text{s}_\theta \\
0 & \text{c}_\phi & \text{s}_\phi \text{c}_\theta \\
0 & -\text{s}_\phi & \text{c}_\phi \text{c}_\theta
\end{bmatrix}\text{,} \quad \mbf{Q} \triangleq \begin{bmatrix}
-\text{c}_\gtwo & 0 \\
0 & -1 \\
\text{s}_\gtwo & 0
\end{bmatrix}\text{,} \quad \ay \triangleq \begin{bmatrix} 0 \\ 1 \\ 0 \end{bmatrix}.
\end{equation*}

The generalized coordinates are defined according to $\mbf{q} \triangleq [ \bm{\zeta}^T \,\; \bm{\eta}^T \,\; \bm{\gamma}^T \,\; \aR \,\; \aL ]^T$, 
which include the load position and orientation given in the inertial frame and reflect the choice of perspective. Therefore, through the Euler-Lagrange formulation \citep{Rego2019}, the equations of motion are written as
\begin{equation} \label{eq:uav_eulerlagrangeCANONICAL}
\mbf{M}(\mbf{q})\ddot{\mbf{q}} + \mbf{C}(\mbf{q},\dot{\mbf{q}})\dot{\mbf{q}} + \mbf{g}(\mbf{q}) = \gforce(\mbf{q},\dot{\mbf{q}},\mbf{u},\mbf{d}) \text{,}
\end{equation}
where $\mbf{M}(\mbf{q})$ is the inertia matrix, $\mbf{C}(\mbf{q},\dot{\mbf{q}})$ is the Coriolis and centripetal forces matrix, $\mbf{g}(\mbf{q})$ is the gravitational forces vector, and $\gforce(\mbf{q},\dot{\mbf{q}},\mbf{u},\mbf{d}) = \mbf{L}_\text{in}(\mbf{q}) \mbf{u} - \mbf{L}_\text{fr} \dot{\mbf{q}} + \mbf{L}_\text{db} \mbf{d}$, with $\mbf{u} = [\infR \,\; \infL \,\; \intauaR \,\; \intauaL]^T \in \setreal^4$ denoting the system inputs (thrust forces and servomotor torques, see Figure \ref{fig:uav_tiltrotorkinematics}), $\mbf{d} \in \setreal^3$ denoting external disturbances affecting the load, $\mbf{L}_\text{in}(\mbf{q})$ is the input coupling matrix, $\mbf{L}_\text{fr}(\mbf{q})$ is the matrix of viscous friction coefficients, and $\mbf{L}_\text{d}(\mbf{q})$ is the disturbance mapping matrix  (see \cite{Rego2019} for details).

Finally, by defining the system states $\mbf{x} \triangleq [ \mbf{q}^T \,\; \dot{\mbf{q}}^T ]^T \in \setreal^{20}$, the equations \eqref{eq:uav_eulerlagrangeCANONICAL} are written as $\dot{\mbf{x}} = \mbf{\varphi}(\mbf{x}, \mbf{u}, \mbf{d})$, 
\begin{equation} \label{eq:uav_tiltrotorstatespacenonlinear}
\bm{\varphi}(\mbf{x}, \mbf{u}, \mbf{d}) {=} \begin{bmatrix}
\dot{\mbf{q}} \\
\mbf{M}^{-1} \left[ - (\mbf{C}  + \mbf{L}_\text{fr}) \dot{\mbf{q}} - \mbf{g}(\mbf{q}) + \mbf{L}_\text{in}(\mbf{q}) \mbf{u} + \mbf{L}_\text{db}\mbf{d} \right]
\end{bmatrix} \text{,}
\end{equation}
which describes the load behavior in the inertial frame explicitly, with the aircraft position and orientation being described only with respect to the load.

\subsubsection{Measurement equation} \label{sec:uav_tiltrotormeasurement}

This doctoral thesis assumes that the load position and orientation (and corresponding velocities) with respect to the inertial frame are not measured. Available measurements are provided by: (i) a Global Positioning System (GPS), providing the planar position of the tilt-rotor UAV with respect to $\frI$; (ii) a barometer, providing the altitude of the UAV with respect to $\frI$; (iii) an Inertial Measurement Unit (IMU), providing the orientation and angular velocities of the UAV with respect to $\frI$; (iv) a camera, providing the position of the load with respect to the $\frA{1}$; and (v) embedded sensors at the servos, providing the tilting angles and their time derivatives.

The GPS, barometer, and IMU are located at $\frB$, while the camera is placed at the origin of $\frA{1}$. Let $\zetaB \triangleq [x_\frB \,\; y_\frB \,\; z_\frB]^T$ denote the position of the UAV with respect to $\frI$. Then,
\begin{equation} \label{eq:uav_estimationgpsbarometer}
\zetaB(\bm{\zeta},\bm{\eta}) = \bm{\zeta} + \RIL \dLA{1} + \RIL \RLB \dAB{1} \text{.} 
\end{equation}

The orientation of the UAV with respect to $\frI$ is parametrized by $\etaB \triangleq [\phiB \; \theta_\frB \; \psi_\frB]^T$, using the local roll-pitch-yaw convention. Therefore, $\RIB \triangleq \mbf{R}_{z,\psiB} \mbf{R}_{y,\thetaB} \mbf{R}_{x,\phiB} = \RIL \RLB$, leading to %
\begin{gather}
\phiB(\bm{\eta},\bm{\gamma}) = \arctan({{(\RIL \RLB)}_{32}}/{{(\RIL \RLB)}_{33}}) \text{,} \label{eq:uav_estimationIMUphi}\\
\thetaB(\bm{\eta},\bm{\gamma}) = \arcsin(-{(\RIL \RLB)}_{31}) \text{,} \label{eq:uav_estimationIMUtheta}\\
\psiB(\bm{\eta},\bm{\gamma}) = \arctan({{(\RIL \RLB)}_{21}}/{{(\RIL \RLB)}_{11}}) \text{,} \label{eq:uav_estimationIMUpsi}
\end{gather}
for $\thetaB \neq \pm \pi/2$, where $(\cdot)_{ij}$ denotes the element from the $i$-th row and $j$-th column. Moreover, the angular velocity provided by the IMU is given by
\begin{equation}
\wBIB(\bm{\eta},\bm{\gamma},\deta,\dgamma) = \bm{\omega}^\frB_{\frI \frL} + \wBLB = (\RLB)^T \Weta \deta + \mbf{Q} \dgamma. \label{eq:uav_estimationIMUwBIB}
\end{equation}

On the other hand, let $\mbf{d}^\frA{1}_{\frA{1}\frL}$ denote the displacement from $\frA{1}$ to $\frL$, expressed in $\frA{1}$, provided by the camera. Then,
\begin{equation} \label{eq:uav_estimationcamera}
\mbf{d}^\frA{1}_{\frA{1}\frL}(\bm{\gamma}) = - (\RLA{1})^T \mbf{d}^\frL_{\frA{1}} = - (\RLB)^T \mbf{d}^\frL_{\frA{1}}.
\end{equation}

Putting together \eqref{eq:uav_estimationgpsbarometer} through \eqref{eq:uav_estimationcamera}, and considering $\aR$, $\aL$, $\daR$, and $\daL$, provided by sensors at the servos, lead to the nonlinear measurement equation $\mbf{y}_k = \bm{\pi}(\mbf{x}_k) + \mbf{v}_k$, where $\mbf{v}_k$ denotes measurement noise, and
\begin{equation*}
\bm{\pi}(\mbf{x}_k) = \left[\bm{\zeta}^T_\frB \,\; \bm{\eta}_\frB^T \,\; (\wBIB)^T \,\; (\mbf{d}^\frA{1}_{\frA{1}\frL})^T \,\; \aR \;\, \aL \;\, \daR \;\, \daL  \right]^T.
\end{equation*}

Finally, let $\setI_k$ denote the set of available measurements at time instant $k$, such that $\setI_k \subseteq \{1,2,\dots,16\}$, and $\iota_k$ denote the number of elements in $\setI_k$. Then, $\mbf{y}_k, \mbf{v}_k \in \setreal^{\iota_k}$, and %
\begin{equation} \label{eq:uav_estimationswitched}
\bm{\pi}^{[k]}(\mbf{x}_k) \triangleq \left[ \bm{\pi}(\mbf{x}_k)(i) \right]_{i \in \setI_k} \text{,}
\end{equation}
where $\bm{\pi}(\mbf{x}_k)(i)$ denotes the $i$-th row of $\bm{\pi}(\mbf{x}_k)$, and the brackets denote vertical concatenation.

\subsubsection{Linearized model for state estimation} \label{sec:uav_linearizedmodel}
Since the method proposed in this section is formulated for linear systems, the state-space equations \eqref{eq:uav_tiltrotorstatespacenonlinear} and measurement mapping \eqref{eq:uav_estimationswitched} are linearized around equilibrium values\footnote{Due to limited computational resources, analytical expressions for the Jacobian and Hessians of the nonlinear mapping $\bm{\varphi}(\mbf{x},\mbf{u},\mbf{d})$ could not be obtained. Therefore, we resort to linear state estimation for aerial load transportation using the tilt-rotor UAV. Moreover, equilibrium values are used since online evaluation of a time-varying model around trajectory values has demonstrated to be intractable.} $\mbf{x}^\text{eq}$, $\mbf{u}^\text{eq}$, $\mbf{d} = \bm{0}$, through first-order Taylor series expansion, and discretized by the zero-order-hold method, yielding to the disturbed switching linear discrete-time system
\begin{equation}
\begin{aligned} \label{eq:uav_tiltrotorlinearsystem0}
\Delta \mbf{x}_{k} & = \mbf{A} \Delta \mbf{x}_{k-1} + \mbf{B} \Delta \mbf{u}_{k-1} + \mbf{F} \mbf{d}_{k-1} + \mbf{w}_{k-1} \text{,} \\
\mbf{y}_{k} & = \mbf{H}^{[k]} \Delta \mbf{x}_{k} + \bm{\pi}^{[k]}(\mbf{x}^\text{eq}) + \mbf{v}_k,
\end{aligned}
\end{equation}
where the process noise $\mbf{w}_{k}$ corresponds to unmodeled dynamics due to linearization and discretization, and the measurement noise $\mbf{v}_{k}$ now includes the nonlinearities of \eqref{eq:uav_estimationswitched}. Moreover, $\Delta \mbf{x}_k \triangleq \mbf{x}_k - \mbf{x}^\text{eq}$ and $\Delta \mbf{u}_k \triangleq \mbf{u}_k - \mbf{u}^\text{eq}$, where $\mbf{A} \in \setrealmat{20}{20}$, $\mbf{B} \in \setrealmat{20}{4}$, $\mbf{F} \in \setrealmat{20}{3}$, and $\mbf{H}^{[k]} \in \setrealmat{\iota_k}{20}$ are the respective Jacobian matrices evaluated at the equilibrium values followed by the discretization procedure.

For robust state estimation, the state vector is augmented with the external disturbances $\mbf{d}_k$, yielding the augmented dynamics
\begin{equation} \label{eq:uav_tiltrotorlinearsystem}
\begin{aligned}
\begin{bmatrix} \Delta \mbf{x}_k \\ \mbf{d}_k \end{bmatrix} & {=} \begin{bmatrix} \mbf{A} & \mbf{F} \\ \zeros{3}{20} & \eye{3} \end{bmatrix} \begin{bmatrix} \Delta \mbf{x}_{k-1} \\ \mbf{d}_{k-1} \end{bmatrix} \!{+}\! \begin{bmatrix} \mbf{B} \\ \zeros{3}{4} \end{bmatrix} \Delta \mbf{u}_{k-1} {+}\! \begin{bmatrix} \mbf{w}_{k-1} \\ \tilde{\mbf{d}}_{k-1} \end{bmatrix}\!\!, \\
\mbf{y}_k & = \begin{bmatrix} \mbf{H}^{[k]} & \zeros{\iota_k}{3} \end{bmatrix} \begin{bmatrix} \Delta \mbf{x}_k \\ \mbf{d}_k \end{bmatrix} + \bm{\pi}^{[k]}(\mbf{x}^\text{eq}) + \mbf{v}_k,
\end{aligned}
\end{equation}
with $\tilde{\mbf{d}}_{k-1} \triangleq \mbf{d}_{k} - \mbf{d}_{k-1}$. 

Accordingly, define $\bm{\nu}_k \triangleq [ \Delta \mbf{x}_k^T \,\; \mbf{d}_k^T ]^T$, $\bar{\mbf{w}}_k \triangleq [ \mbf{w}_{k}^T \,\; \tilde{\mbf{d}}_{k}^T ]^T$, and $\bar{\mbf{v}}_k \triangleq \bm{\pi}^{[k]}(\mbf{x}^\text{eq}) + \mbf{v}_k$. %
Thus, the linear set-based state estimation method is formulated in the next subsection.

\subsection{Problem formulation} \label{sec:linearformulation}

As previously commented in Chapter \ref{cha:preliminaries}, constrained zonotopes were effectively used in \cite{Scott2016} for set-based state estimation of linear systems, using the method illustrated in Section \ref{sec:pre_linearestimation}. This chapter extends this estimation method for discrete-time switching systems. Consider the augmented linear discrete-time switching system described by \eqref{eq:uav_tiltrotorlinearsystem}, and let $\Anu$, $\Bnu$ and $\Hnu^{[k]}$ denote the associated matrices. Assuming bounded uncertainties $\bar{\mbf{w}}_{k} \in W_k$ and $\bar{\mbf{v}}_k \in V_k$, and $\bm{\nu}_0 \in X_0$\footnote{Note that $X_0 \subset \realset^{23}$, since this set is associated to the augmented state variables $\bm{\nu}_k$.}, where $X_0$, $W_k$ and $V_k$ are known compact sets, linear set-based state estimation consists in computing enclosures $\bar{X}_k$ and $\hat{X}_k$ such that%
\begin{align}
\bar{X}_k & \supseteq \{ \Anu \bm{\nu} + \Bnu \Delta \mbf{u}_{k-1} {+} \bar{\mbf{w}} : \bm{\nu} \in \hat{X}_{k-1}, \, \bar{\mbf{w}} \in W_{k-1} \}, \label{eq:UAVpredictionlinear0}\\
\hat{X}_k & \supseteq \{ \bm{\nu} \in \bar{X}_k : \Hnu^{[k]} \bm{\nu} + \bar{\mbf{v}} = \mbf{y}_k, \, \bar{\mbf{v}} \in V_k \}, \label{eq:UAVupdatelinear0}
\end{align}
with \eqref{eq:UAVpredictionlinear0} referred to as \emph{prediction step}, and \eqref{eq:UAVupdatelinear0} as \emph{update step}. These enclosures can be obtained exactly with constrained zonotopes through
\begin{align}
\bar{X}_k & = \Anu \hat{X}_{k-1} \oplus \Bnu \Delta\mbf{u}_{k-1} \oplus W_{k-1} , \label{eq:uav_tiltrotorprediction}\\
\hat{X}_k & = \bar{X}_k \cap_{\Hnu^{[k]}} (\mbf{y}_k \oplus (- V_k)), \label{eq:uav_tiltrotorupdate}
\end{align}
using \eqref{eq:pre_czlimage}--\eqref{eq:pre_czintersection}.

As discussed in Chapter \ref{cha:preliminaries}, the right-hand side of \eqref{eq:uav_tiltrotorprediction} cannot be computed using interval arithmetic without wrapping. On the other hand, the equality in \eqref{eq:uav_tiltrotorprediction} can be achieved using zonotopes and constrained zonotopes, but the intersection in \eqref{eq:uav_tiltrotorupdate} can not be computed exactly using zonotopes. As a consequence, the enclosures of the system states obtained after many iterations of prediction and update may be quite conservative using zonotopes. Nevertheless, with constrained zonotopes both \eqref{eq:uav_tiltrotorprediction} and \eqref{eq:uav_tiltrotorupdate} are easily computed using \eqref{eq:pre_czlimage}--\eqref{eq:pre_czintersection}. Moreover, iterated computations of \eqref{eq:uav_tiltrotorprediction} and \eqref{eq:uav_tiltrotorupdate} result in only a linear increase in the complexity of the CG-rep \eqref{eq:pre_cgrep}. This can be effectively addressed using the complexity reduction methods presented in Section \ref{sec:complexityreduction}.

\subsection{Minimum-variance feedback} \label{sec:uav_minimumvariance}

Despite the accuracy and low computational burden of the method described by \eqref{eq:uav_tiltrotorprediction}--\eqref{eq:uav_tiltrotorupdate}, to perform state-feedback control with estimated states one must choose a point satisfying $\hat{\bm{\nu}}_k \in \hat{X}_k$. Therefore, we propose an optimal choice according to a \emph{minimum-variance criterion}, which is one of the main contributions of this chapter, and can be straightforwardly adapted to more general classes of linear discrete-time switching dynamical systems.

The proposed criterion is based on the well-known Kalman filter algorithm \citep{Simon2006}. Let $\mbf{P}^{\nu}_k$, $\mbf{P}^{w}_k$, and $\mbf{P}^{v}_k$, denote the covariance matrices of the estimation error $\bm{\nu}_k - \hat{\bm{\nu}}_k$, process noise $\bar{\mbf{w}}_k$, and measurement noise $\bar{\mbf{v}}_k$, respectively. Consider the constrained optimization problem
\begin{gather}
\underset{\mbf{N}_k}{\min} ~ \trace{\mbf{P}^{\nu}_k} \quad \text{s.t.} \quad \hat{\bm{\nu}}_k \in \hat{X}_k, \label{eq:minCKF}
\end{gather}
where $\hat{\bm{\nu}}_k$ is given according to the correction step%
\begin{equation} \label{eq:CKF_correction}
\hat{\bm{\nu}}_{k} = \bar{\bm{\nu}}_{k} + \mbf{N}_k (\mbf{y}_k - (\Hnu^{[k]} \bar{\bm{\nu}}_{k} + \bm{\pi}^{[k]}(\mbf{x}^\text{eq}))),
\end{equation}
with $\bar{\bm{\nu}}_{k} = \mbf{A}_{\bm{\nu}} \hat{\bm{\nu}}_{k-1} + \mbf{B}_{\bm{\nu}} \Delta \mbf{u}_{k-1}$, and $\mbf{N}_k$ is the Kalman gain. Moreover,
\begin{equation} \label{eq:CKF_covariance}
\mbf{P}^{\bm{\nu}}_{k} = (\eye{23} - \mbf{N}_k \mbf{H}^{[k]}_{\bm{\nu}}) \bar{\mbf{P}}^{\bm{\nu}}_{k} (\eye{23} - \mbf{N}_k \mbf{H}^{[k]}_{\bm{\nu}})^T 
+ \mbf{N}_k \mbf{P}^{{\mbf{v}}}_k \mbf{N}_k^T,
\end{equation}
where $\bar{\mbf{P}}^{\bm{\nu}}_{k} = \mbf{A}_{\bm{\nu}} \mbf{P}^{\bm{\nu}}_{k-1} \mbf{A}_{\bm{\nu}}^T + \mbf{P}^{\mbf{w}}_k$. Considering Property \ref{prop:pre_czisemptyinside} and $\hat{X}_k = \{\mbf{G}_x, \mbf{c}_x, \mbf{A}_x, \mbf{b}_x \}$, the optimization problem \eqref{eq:minCKF} is rewritten as a quadratic program with linear constraints,
\begin{align}
\underset{\mbf{N}_k, \bm{\xi}}{\min} & ~ \trace{\mbf{P}^{\nu}_k} \label{eq:CKFQP}\\
\text{s.t.} & ~ \bar{\bm{\nu}}_{k} + \mbf{N}_k (\mbf{y}_k - (\Hnu^{[k]} \bar{\bm{\nu}}_{k} + \bm{\pi}^{[k]}(\mbf{x}^\text{eq}))) = \mbf{c}_x + \mbf{G}_x \bm{\xi}, \nonumber\\
& \ninf{\bm{\xi}} \leq 1, \quad \mbf{A}_x \bm{\xi} = \mbf{b}_x,\nonumber
\end{align}
with $\mbf{P}^{\nu}_k$ given by \eqref{eq:CKF_covariance}. The minimum-variance choice $\hat{\bm{\nu}}_k$ is then obtained by solving \eqref{eq:CKFQP}, and computing \eqref{eq:CKF_correction} afterwards. The optimization problem \eqref{eq:CKFQP} can be regarded as \emph{constrained Kalman filtering}, in which divergence issues are not present since $\hat{\bm{\nu}}_k \in \hat{X}_k$ is guaranteed by construction.\footnote{Despite this guarantee, the stochastic properties of Gaussian distribution associated to $\hat{\bm{\nu}}_{k}$ has no connection with the set $\hat{X}_k$. This issue is a subject of future investigation.}

\subsection{Path tracking control}

\subsubsection{Linearized parameter-varying error dynamics} \label{sec:lpvmodel}

The tilt-rotor UAV with suspended load is an underactuated mechanical system with four control inputs. For path tracking of the suspended load, the position $\bm{\zeta} = [x \; y \; z]^T$ and yaw angle $\psi$ of the load are chosen to be regulated. Let $\mbf{q}^\text{tr} = [\bm{\zeta}^\text{tr} \,\; \bm{\eta}^\text{tr} \,\; \bm{\gamma}^\text{eq} \,\; \alpha^\text{eq}_\text{R} \,\; \alpha^\text{eq}_\text{L}]$, with $\bm{\eta}^\text{tr} = [\phi^\text{eq} \,\; \theta^\text{eq} \,\; \psi^\text{tr}]$, where `tr' denotes trajectory values. Assuming $\psi^\text{tr}$ constant, the linearized error dynamics of \eqref{eq:uav_tiltrotorstatespacenonlinear} are given by $\delta \dot{\mbf{x}} = \mbf{A}_\text{c}(t) \delta \mbf{x} + \mbf{B}_\text{c} \delta \mbf{u} + \mbf{F}_\text{c} \mbf{d}$, where $\delta \mbf{x} \triangleq \mbf{x} - \mbf{x}^\text{tr}$, $\delta \mbf{u} \triangleq \mbf{u} - \mbf{u}^\text{tr}$, $\mbf{x}^\text{tr} \triangleq [(\mbf{q}^\text{tr})^T \;\, (\dot{\mbf{q}}^\text{tr})^T]^T$, and $\mbf{A}_\text{c}(t)$ is a function of $\mbf{u}^\text{tr}$ \citep{Rego2019}.

Considering a feasible reference trajectory, the desired input can be obtained from \eqref{eq:uav_tiltrotorstatespacenonlinear} as follows:
\begin{equation} \label{eq:feedforward}
\mbf{u}^\text{tr} = \mbf{L}_\text{in}(\mbf{q}^\text{tr})^+ \! \left[ \mbf{M}(\mbf{q}^\text{tr}) \ddot{\mbf{q}}^\text{tr} {+} (\mbf{C}(\mbf{q}^\text{tr}, \dot{\mbf{q}}^\text{tr}) {+} \mbf{L}_\text{fr}) \dot{\mbf{q}}^\text{tr} {+} \mbf{g}(\mbf{q}^\text{tr}) \right],
\end{equation}
where $\mbf{L}_\text{in}(\mbf{q}^\text{tr})^+$ denotes the pseudo-inverse of $\mbf{L}_\text{in}(\mbf{q}^\text{tr})$. It can be shown that \eqref{eq:feedforward} is a function only of the accelerations $\ddot{\bm{\zeta}}^\text{tr}$ \citep{Rego2019}. By defining the vector of parameters $\bm{\sigma} \triangleq \ddot{\bm{\zeta}}^\text{tr}$, we have that $\mbf{u}^\text{tr}(t) \triangleq \mbf{u}(\bm{\sigma})$, which is affine in $\bm{\sigma}$. Then, substituting \eqref{eq:feedforward} in $\mbf{A}_\text{c}(t)$ yields the parameter-varying matrix $\mbf{A}_\text{c}(\bm{\sigma})$, which is affine in $\bm{\sigma}$ since \eqref{eq:uav_tiltrotorstatespacenonlinear} is affine in $\mbf{u}$.

For an improved trajectory tracking and rejection of constant disturbances, the state vector $\delta\mbf{x}$ is augmented with integral actions over the regulated variables, as $\bm{\chi} \triangleq [\delta \mbf{x}^T \,\; \smallint (\bm{\zeta}-\bm{\zeta}^\text{tr})^T \,\; \smallint (\psi - \psi^\text{tr})]^T$. The resulting augmented dynamics are then discretized through Euler approximation\footnote{The Euler approximation is used here in order to obtain a discrete-time LPV system. On the other hand, the zero-order-hold method is used in \eqref{eq:uav_tiltrotorlinearsystem0} for reduced discretization error. }, yielding the discrete-time, augmented LPV error dynamics
\begin{equation} \label{eq:LPVsystem}
\bm{\chi}_{k+1} = \Achi (\bm{\sigma}) \bm{\chi}_{k} + \Bchi \delta \mbf{u}_{k} + \Fchi \mbf{d}_{k} \text{,}
\end{equation}
with $\Achi(\bm{\sigma}) \in \setrealmat{24}{24}$, $\Bchi \in \setrealmat{24}{4}$ and $\Fchi \in \setrealmat{24}{3}$. Finally, assuming bounded trajectory accelerations, the LPV system \eqref{eq:LPVsystem} can be rewritten in a convex polytopic representation, by defining $\Achi(\bm{\sigma}) \triangleq \Achi(\tilde{\bm{\sigma}}) = \sum_{i=1}^{8} \tilde{\sigma}_i\Achi^i$, with $\Achi^i$ denoting the $i$-th vertex of $\Achi(\tilde{\bm{\sigma}})$, and $\sum_{i=1}^{8} \tilde{\sigma}_i = 1$. Resulting errors from linearization and discretization are taken into account in this section as unmodeled dynamics. Note that the LPV dynamics \eqref{eq:LPVsystem} describe the behavior of the tracking error $\delta \mbf{x}$ augmented with integral actions, while the time-invariant linear dynamics \eqref{eq:uav_tiltrotorlinearsystem} describe the evolution of the system states $\mbf{x}$ around equilibrium values augmented with the external disturbances.

\subsubsection{Discrete-time mixed $\mathcal{H}_2/\mathcal{H}_\infty$ control with pole placement constraints}

The aim of the proposed controller is to perform robust trajectory tracking of the load with stabilization of the tilt-rotor UAV, with estimation provided by the novel method presented in Section \ref{sec:uav_minimumvariance}. The desired accelerations for the load are required to be smooth and bounded. The present method extends the approach in \cite{Gahinet1996} for discrete-time linear systems. Let
\begin{equation*}
\begin{aligned}
\mbf{z}_k^{(2)} & = \mbf{H}_{\mbf{z}} \bm{\chi}_k + \mbf{D}_{\mbf{zu}} \delta\mbf{u}_k \text{,} \\
\mbf{z}_k^{(\infty)} & = \mbf{H}_{\mbf{z}} \bm{\chi}_k + \mbf{D}_{\mbf{zu}} \delta\mbf{u}_k + \mbf{D}_{\mbf{zd}} \mbf{d}_k \text{,}
\end{aligned}
\end{equation*}
where $\mbf{H}_{\mbf{z}} \in \setrealmat{n_z}{24}$, $\mbf{D}_{\mbf{zu}} \in \setrealmat{n_z}{4}$, $\mbf{D}_{\mbf{zd}} \in \setrealmat{n_z}{3}$. %
Let $\bm{\Psi}^{(2)}_{\mbf{d}{\mbf{z}}}(\zdomain)$ and $\bm{\Psi}^{(\infty)}_{\mbf{d}{\mbf{z}}}(\zdomain)$ denote the discrete-time transfer matrices from $\mbf{d}_k$ to $\mbf{z}_k^{(2)}$ and $\mbf{d}_k$ to $\mbf{z}_k^{(\infty)}$, and $\| \bm{\Psi}^{(2)}_{{\mbf{d}}\mbf{z}} (\zdomain) \|_2$ and $\| \bm{\Psi}^{(\infty)}_{{\mbf{d}}{\mbf{z}}} (\zdomain) \|_\infty$ be the corresponding $\mathcal{H}_2$ and $\mathcal{H}_\infty$ norms, respectively. Given the state-feedback control law $\mbf{u}_k = - \mbf{K} \bm{\chi}$, the gain matrix $\mbf{K}$ that minimizes $\trace{\bm{\Omega}} > \| \bm{\Psi}^{(2)}_{{\mbf{d}}\mbf{z}} \|_2^2$ while guaranteeing an upper-bound $\tilde{\gamma} > \| \bm{\Psi}^{(\infty)}_{{\mbf{d}}{\mbf{z}}} \|_\infty^2$ is given by $\mbf{K} = -\mbf{Y}\mbf{X}^{-1}$, where $\mbf{Y}$ and $\mbf{X}$ are the solution of the optimization problem %
\begin{gather}
\underset{\mbf{P},\mbf{X},\mbf{Y},\bm{\Omega}}{\min} ~ \trace{\bm{\Omega}} \quad \text{s.t.} \label{eq:uav_h2hinf}\\
\begin{bmatrix}
\bm{\Omega} & \mbf{H}_{\mbf{z}} \mbf{X} + \mbf{D}_{\mbf{zu}} \mbf{Y} \\
* & \mbf{X} + \mbf{X}^T - \mbf{P}
\end{bmatrix} > 0, \nonumber\\
\begin{bmatrix}
\mbf{P} & \Achi^i \mbf{X} + \mbf{B}_{\bm{\chi}} \mbf{Y} & \mbf{F}_{\bm{\chi}} & \zeros{24}{n_z} \\
* & \mbf{X} + \mbf{X}^T - \mbf{P} & \zeros{24}{3} & \mbf{X}^T \mbf{H}_{\mbf{z}}^T + \mbf{Y}^T \mbf{D}_{\mbf{zu}}^T \\
* & * & \eye{3} & \mbf{D}_{\mbf{zd}}^T \\
* & * & * & \tilde{\gamma} \eye{n_z} 
\end{bmatrix} > 0, \nonumber
\end{gather}
with $i = 1,2,\dots,8$, $\mbf{P} = \mbf{P}^T > 0$, $\mbf{X} > 0$ and $\bm{\Omega} = \bm{\Omega}^T$.

To ensure transient properties for the closed-loop system, additional constraints are imposed to pole placement. Seeking minimum and maximum settling times, maximum percentage overshoot, and also to avoid ringing, three regions are of interest: $\mathbb{D}_1 \triangleq \{ \zdomain \in \mathbb{C} : \real{\zdomain} > \deps \geq 0\}$, $\mathbb{D}_2 \triangleq \{ \zdomain \in \mathbb{C} : 0 \leq |\zdomain| < \varpi \}$, and $\mathbb{D}_3 \triangleq \{ \zdomain \in \mathbb{C} : 0 \leq |\imag{\zdomain}| < \tau \}$. The eigenvalues of $\Achi(\bm{\sigma}) - \Bchi \mbf{K}$ belong to $\bigcap_{j=1}^{3} \mathbb{D}_j$ iff \citep{Rego2019}
\begin{gather*}
\mbf{X} (\Achi^i)^T + \Achi^i \mbf{X} + \mbf{Y}^T\mbf{B}_{\bm{\chi}}^T + \mbf{B}_{\bm{\chi}}\mbf{Y} - 2\varepsilon \mbf{X} > 0, \nonumber \\
\begin{bmatrix}
-\varpi \mbf{X} & \Achi^i \mbf{X} + \mbf{B}_{\bm{\chi}} \mbf{Y} \\
* & -\varpi \mbf{X}
\end{bmatrix} < 0, \nonumber \\
\begin{bmatrix} -2 \tau \mbf{X} ~& ~\mbf{X} (\Achi^i)^T \!{-} \Achi^i \mbf{X} + \mbf{Y}^T \Bchi^T {-} \Bchi \mbf{Y} \\ * & - 2 \tau \mbf{X} \end{bmatrix} < 0. \nonumber
\end{gather*}

Finally, the control signals applied to \eqref{eq:uav_tiltrotorstatespacenonlinear} are obtained from $\mbf{u}_k = \delta \mbf{u}_k + \mbf{u}_k^\text{tr} {=} -\mbf{K}\hat{\bm{\chi}}_k + \mbf{u}_k^\text{tr}$, with $\mbf{u}_k^\text{tr}$ given by \eqref{eq:feedforward}, where the estimated states $\hat{\bm{\chi}}_k$ are retrieved from $\hat{\bm{\nu}}_k$ according to the definitions of the state vectors $\bm{\nu}_k$ and $\bm{\chi}_k$.

\subsection{Numerical experiments} \label{sec:uav_tiltrotorexamples}

This section evaluates the performance of the proposed method through numerical experiments.

The experiment is performed in the ProVANT simulator platform, using a Computer Aided Design (CAD) model of the system. The model parameters are shown in Table\footnote{Parameter uncertainties can be considered immediately through $\bar{\mbf{w}}_k$, $\bar{\mbf{v}}_k$.} \ref{tab:uav_tiltrotormodelparameters}, where $m_j$ is the mass of each rigid body, $\Ii{j}$ is the inertia tensor, $j \in \{ \mathcal{L}, 1,2,3\}$, $\hat{\mbf{g}}$ is the gravity acceleration vector in $\frI$, $k_\tau$ and $b$ are drag torque parameters obtained experimentally, $\lambdaR,\lambdaL \in \{-1,1\}$ are given according to the direction of rotation of the propellers, $\mu_\gamma$ and $\mu_\alpha$ are constant parameters associated with viscous friction \citep{Rego2019}. These model parameters yield the equilibrium point
\begin{equation*}%
\begin{aligned}
\mbf{q}^{\text{eq}} = &~
[ \zeros{1}{6} \;\, 0.000132 \;\, 0.01396 \;\, 0.01401 \;\, 0.01381]^T \text{,} \\
\mbf{u}^{\text{eq}} = &~
[ 11.7323 \;\, 11.7676 \;\, 4.1389{\cdot}10^{-7} \;\, 1.0121{\cdot}10^{-5}]^T \text{.}
\end{aligned}
\end{equation*}

\begin{table}[!ht]
	\centering
	\caption{Model parameters of the system.} \vspace{-3mm}
	\begin{tabular}{c c}
		Parameter & Value \\ \hline
		$(\mL,m_1,m_2,m_3)$ & $(0.5,1.7068,0.08978,0.08978)$ kg \\
		$(\dLA{1},\dAB{1})$ & $([ 0 \; 0 \; 0.5 ]^T,[ 0 \; 0 \; 0.119 ]^T)$ m\\
		$\dBC{1}$ & $[ \text{-}0.004321 \; 0.000601 \; \text{-}0.045113]^T$ m\\
		$\dBA{2}$ & $[  0 \; \text{-}0.275433 \; 0.056262 ]^T$ m\\
		$\dBA{3}$ & $[  0 \; 0.275433 \; 0.056261 ]^T$ m\\
		$(\dAC{2}{2},\dAC{3}{3})$ & $([  0 \; 0 \; 0.05647 ]^T, [  0 \; 0 \; 0.05648 ]^T)$ m\\
		$\IL$ & $8.333 {\cdot} 10^{-6} {\cdot} \eye{3} $ kg$\cdot$m$^2$ \\
		$\Ii{1}$ & $\begin{bmatrix} 4047.04 & 0.8606 & 9.6577 \\ * &  881.62 & \text{-}0.8731 \\ * & * & 4173.18 \end{bmatrix} {\cdot} 10^{-6}$ kg${\cdot}$m$^2$ \\
		$\Ii{2},\Ii{3}$ & $\text{diag} (335.74, 335.74, 641.59) {\cdot} 10^{-6}$kg${\cdot}$m$^2$ \\
		$\hat{\mbf{g}}$ & $[ 0\;\, 0 \;\, \text{-}9.81]^T$ m/s$^2$ \\
		$(k_\tau, b)$ &  $(1.7 {\cdot} 10^{-7}$ N$\cdot$m$\cdot$s$^2, 9.5 {\cdot} 10^{-6}$ N$\cdot$s$^2)$\\
		$(\lambdaR, \lambdaL, \beta)$ & $(1, \text{-}1, 5^\text{o})$\\
		$\mu_\gamma$, $\mu_\alpha$ & $0.005$ N${\cdot}$m/(rad/s) \\
		\hline
	\end{tabular}
	\normalsize
	\label{tab:uav_tiltrotormodelparameters}
\end{table}

The desired trajectory is composed of several paths, with $\psi^\text{tr}$ = 0 rad. The sensor parameters are shown in Table \ref{tab:uav_tiltrotorsensorparameters}.

\begin{table}[!ht]
	\footnotesize
	\centering
	\caption{Parameters of the sensors.} \vspace{-3mm}
	\begin{tabular}{c c c c c}
		Sensor & $\mathbb{I}$ & Noise bound & Sampling & PDF \\ \hline
		GPS & $\{1,2\}$ & $\pm\!\!$ $0.15\!$ m & $120\!$ ms & Gauss.\\ 
		Barometer & $\{3\}$ & $\pm\!\!$ $0.51\!$ m & $12\!$ ms & Gauss.\\ 
		\multirow{2}{*}{IMU} & $\{4,5,6\}$ & $\pm\!\!$ $2.618 {\cdot} 10^{-3}\!$ rad & \multirow{2}{*}{$12\!$ ms} & \multirow{2}{*}{Gauss.} \\
		& $\{7,8,9\}$ & $\pm\!\!$ $16.558 {\cdot} 10^{-3}\!$ rad/s & \\ 
		\multirow{2}{*}{Camera} & $\{10,11\}$ & $\pm\!\!$ $0.005\!$ m & \multirow{2}{*}{$24\!$ ms} & \multirow{2}{*}{Unif.} \\
		& $\{12\}$ & $\pm\!\!$ $0.02\!$ m & \\		                           	                           
		\multirow{2}{*}{Servos} & $\{13,14\}$ & $\pm\!\!$ $5.67 {\cdot} 10^{-3}\!$ rad & \multirow{2}{*}{$12\!$ ms} & \multirow{2}{*}{Unif.} \\
		& $\{15,16\}$ & $\pm\!\!$ $0.50772\!$ rad/s & \\		                           
		\hline
	\end{tabular} \normalsize
	\label{tab:uav_tiltrotorsensorparameters}
\end{table}

The optimization problem \eqref{eq:uav_h2hinf} was solved using SDPT3, with $\varepsilon = 0.55$, $\varpi = 0.994$, $\tau = 0.3$, $\tilde{\gamma} = 81$, $\ddot{x}^\text{tr}(t) \in [-0.5,\; 0.5]$, $\ddot{y}^\text{tr}(t) \in [-0.5,\; 0.5]$, $\ddot{z}^\text{tr}(t) \in [-0.3,\; 0.3]$,
\begin{align*}%
\mbf{H}_{\mbf{z}} & = \text{diag} \Big(\tfrac{\sqrt{10}}{2} \ones{3}{1},\tfrac{\sqrt{0.5}}{\pi/2}\ones{2}{1},\tfrac{\sqrt{5}}{\pi},\tfrac{2}{\pi}\ones{2}{1},\tfrac{0.2}{\pi}\ones{2}{1}, 
\tfrac{1}{2}\ones{3}{1},\tfrac{3}{\pi}\ones{2}{1},\tfrac{4}{\pi},\tfrac{\sqrt{5}}{3\pi}\ones{2}{1},\tfrac{0.1}{3\pi}\ones{2}{1},\sqrt{5}{\cdot}\ones{3}{1},\sqrt{0.1}\Big) , \\
\mbf{D}_{\mbf{zu}} & = \begin{bmatrix} \text{diag}\left(
\frac{\sqrt{750}}{30-\infR^\text{eq}},\frac{\sqrt{750}}{30-\infL^\text{eq}},  \frac{\sqrt{5000}}{2-\intauaR^\text{eq}} \right)& \zeros{3}{1} \\
\zeros{2}{3} & \zeros{2}{1} \\
\zeros{1}{3} & \frac{\sqrt{5000}}{2-\intauaL^\text{eq}} \\
\zeros{18}{3} & \zeros{18}{1}
\end{bmatrix},\\
\mbf{D}_{\mbf{zd}} & = [
\zeros{3}{10} \;\,
\eye{3} \;\,
\mbf{V}^T \;\,
0.5{\cdot}\ones{3}{1}\;\,
\mbf{V}^T \;\,
\zeros{3}{2} \;\,
\eye{3} \;\,
0.5{\cdot}\ones{3}{1}
]^T,
\end{align*}
$\mbf{V} \triangleq \begin{bmatrix} 0 & 1 & 0 \\ 1 & 0 & 0 \end{bmatrix}$\footnote{$\ones{n}{m}$ is an $n$ by $m$ matrix of ones.}. These matrices were adjusted empirically using the Bryson's method \citep{Johnson1987} as starting point. For comparison with results obtained using the zonotope-based algorithm of \cite{Rego2019}, the sets $X_0$, $W_k$, and $V_k$, were chosen as the zonotopes $X_0 = \{\mbf{G}_{x_0}, [(\bm{\zeta}^\text{tr}_0)^T \; \zeros{1}{20}]^T\}$, $W_k = \{\mbf{G}_{w}, \zeros{23}{1}\}$ and $V_k = \{\mbf{G}_{v},\bm{\pi}(\mbf{x}^\text{eq})\}$, with $\mbf{G}_{x_0} = \text{diag} ( 0.5 {\cdot} \ones{3}{1},$ $ 0.02, $ $ 0.02, $ $\pi/180, $ $1.5 {\cdot} |\gamma_1^\text{eq}|, $ $1.5 {\cdot} |\gamma_2^\text{eq}|, $ $1.5 {\cdot} |\aR^\text{eq}|, $ $1.5 {\cdot} |\aL^\text{eq}|, $ $0.02 {\cdot} \ones{13}{1})$, $\mbf{G}_w = $ $\text{diag} (10^{-4} {\cdot} \ones{13}{1}, $ $0.01 {\cdot} \ones{3}{1}, $ $0.05 {\cdot} \ones{2}{1}, $ $10^{-4} {\cdot} \ones{2}{1}, $ $0.01 {\cdot} \ones{3}{1})$, $\mbf{G}_v = $ $\text{diag} (0.18 {\cdot} \ones{2}{1}, $ $0.612, $ $3.1416\ten{-3}{\cdot}\ones{2}{1}, $ $0.03, $ $19.872 {\cdot} 10^{-3} {\cdot} \ones{2}{1}, $ $0.24, $ $0.006 {\cdot} \ones{2}{1}, $ $0.06 , $  $6.8067 {\cdot} 10^{-3} {\cdot} \ones{2}{1}, $ $0.6093 {\cdot} \ones{2}{1})$. These bounds were adjusted by trial and error to accommodate the linearization errors. The number of generators and constraints of $\hat{X}_k$ were limited to 50 and 10, respectively, using Methods \ref{meth:genredB} and \ref{meth:czconelim}, while the number of generators of the zonotopes was limited to 1725 using Method \ref{meth:genredA}. The optimal estimate $\hat{\bm{\nu}}_k \in \hat{X}_k$ was obtained considering $\mbf{P}^\nu_0 = \eye{23}$, with $\hat{\bm{\nu}}_0$ as the center of the interval hull\footnote{The interval hull is computed as in Property \ref{prope:pre_czihull}.} of $\hat{X}_0 = X_0 \cap_{\Hnu^{[0]}}(\mbf{y}_0 \oplus (- V_0))$, $\mbf{P}^w_k = \frac{1}{3} \mbf{G}_w^2$, and $\mbf{P}^v_k = \frac{1}{3} \mbf{G}_v^2$. These choices of $\mbf{P}^w_k$ and $\mbf{P}^v_k$ assume $W_k$ and $V_k$ as being the respective 95\% confidence regions for the minimum-variance feedback. The quadratic program \eqref{eq:CKFQP} was solved using CPLEX \citep{cplex}. The initial state is $\mbf{x}_0 = [0\,\; 0\,\; 1 \,\; \zeros{1}{13}]^T$.

\begin{figure}[!ht]
	\centering{
		\def\svgwidth{0.6\columnwidth}
		{\scriptsize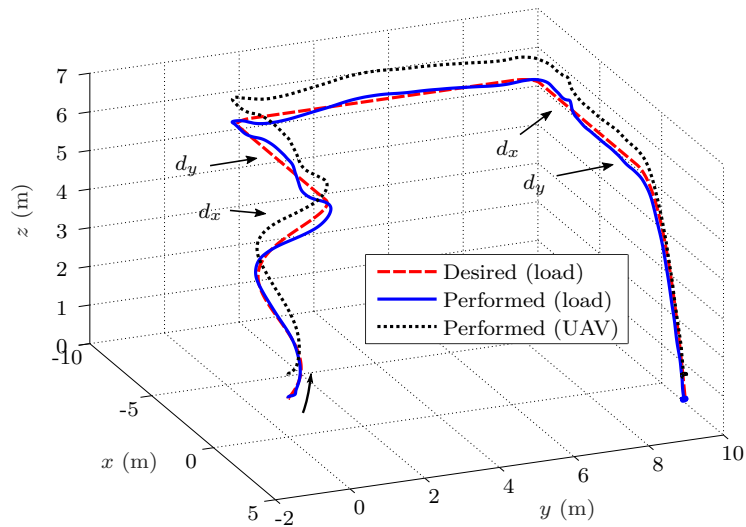}
		\caption{Trajectories performed by the UAV and the load.}\label{fig:uav_tiltrotortrajectory}}
\end{figure}

\begin{figure}[!ht]
	\centering{
		\def\svgwidth{0.7\columnwidth}
		{\scriptsize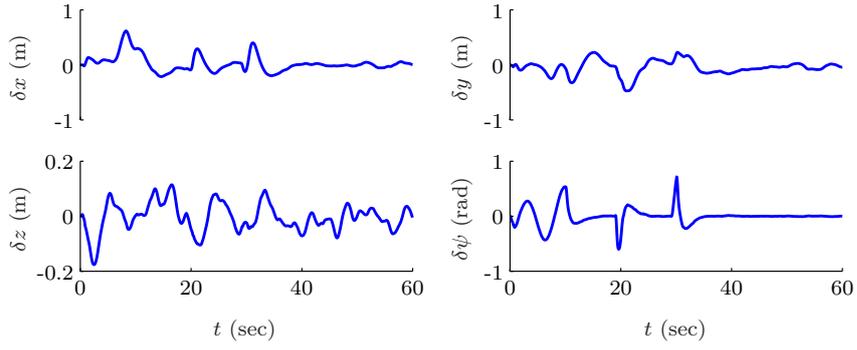}
		\caption{Tracking error of the regulated variables.}\label{fig:uav_error}}
\end{figure}

\begin{figure}[!ht]
	\centering{
		\def\svgwidth{0.7\columnwidth}
		{\scriptsize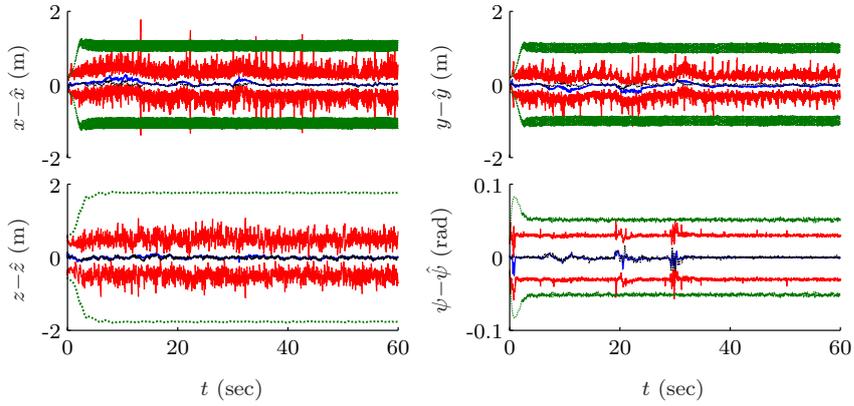}
		\caption{Estimation error of the regulated variables obtained using the CZs (blue solid line) and the zonotope-based algorithm from \cite{Rego2019} (black dotted line), along with confidence limits (red solid lines for CZs, and green dotted lines for zonotopes).}\label{fig:uav_tiltrotorestimation}}
\end{figure}

\begin{figure}[!ht]
	\centering{
		\def\svgwidth{0.7\columnwidth}
		{\scriptsize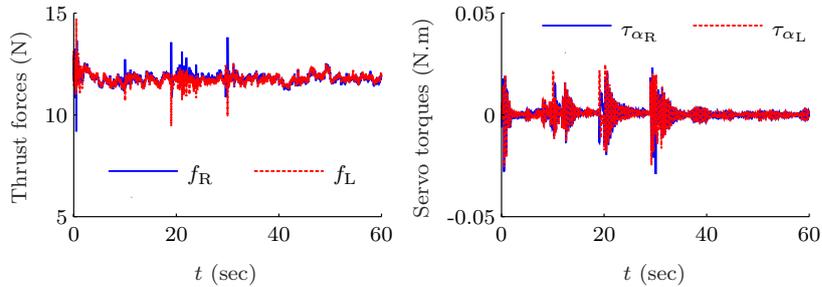}
		\caption{Control signals applied to the UAV.}\label{fig:uav_inputs}}
\end{figure}

The trajectories performed by the load and the UAV are shown in Figure \ref{fig:uav_tiltrotortrajectory}, and the tracking error is shown in Figure \ref{fig:uav_error}. External forces were added and removed from the load, as filtered steps with magnitude 0.5 N. The estimation error of the regulated variables is shown in Figure \ref{fig:uav_tiltrotorestimation}, along with confidence limits (interval hull), and results using the strategy described in \cite{Rego2019}. Sharper bounds are observed for the proposed method, which demonstrate the reduced conservatism of constrained zonotopes. This is a direct result of the exact intersection \eqref{eq:uav_tiltrotorupdate}, which was severely over-estimated by zonotopic enclosures even with a much higher number of generators. Note that the confidence limits are not symmetric, since the optimal choice $\bm{\nu}_k$ is not the geometric center of $\hat{X}_k$ in general. Moreover, since outliers can happen due to unmodeled dynamics, if $\hat{X}_k = \emptyset$ (verified through Property \ref{prop:pre_czisemptyinside}), $\bar{X}_k$ was used instead. The path tracking was performed successfully, as shown by the tracking error of the regulated variables, which remained bounded during the experiment\footnote{A video of the performed experiment is available in https://youtu.be/x-oJNQqdavs .}. The control signals applied to the aircraft are shown in Figure \ref{fig:uav_inputs}. To conclude, experiments were conducted with an alternative feedback $\hat{\bm{\nu}}_k \in \hat{X}_k$ given by the center of the interval hull of $\hat{X}_k$. Although less computationally demanding (since it requires the solution of LPs instead of a QP), such choice resulted to be substantially noisy, destabilizing the closed-loop system in the experiment.

\section{Nonlinear state estimation of a quadrotor UAV} \label{sec:uav_quadrotorestimation}

This section presents numerical results provided by the mean value and first-order Taylor extensions proposed in Chapter \ref{cha:nonlineardynamics} when applied to nonlinear set-based state estimation of a quadrotor UAV. The content of this section as published in \cite{Rego2020c}.

	Consider the quadrotor UAV described in \cite{Mistler2001} with state vector $\bm{\zeta} = [x\,\;y\,\;z\,\;u\,\;v\,\;w\,\;\phi\,\;\theta\,\;\psi\,\;p\,\;q\,\;r]^T$, where $[x\,\;y\,\;z]^T$ is the position of the UAV with respect to the inertial frame, $[u\,\;v\,\;w]^T$ is the velocity vector expressed in the inertial frame, $[\phi\,\;\theta\,\;\psi]^T$ are Euler angles describing the orientation of the UAV, and $[p\,\;q\,\;r]^T$ is the angular velocity vector expressed in the body frame. The equations of motion are \citep{Mistler2001}:
	\begin{equation} \label{eq:uav_quadrotor}
    \!\!\!\dot{\bm{\zeta}} = \!
	\left\{ \begin{aligned}
	\dot{x} & = u, & & \dot{u} = \frac{1}{m} (\cos\psi \sin\theta \cos\phi + \sin\psi \sin\phi)U_1 {+} \frac{1}{m}D_x, \\
	\dot{y} & = v, & & \dot{v} = \frac{1}{m} (\sin\psi \sin\theta \cos\phi - \cos\psi \sin\phi)U_1 {+} \frac{1}{m}D_y, \\
	\dot{z} & = w, & & \dot{w} = -g + \frac{1}{m} (\cos\theta \cos\phi)U_1 + \frac{1}{m}D_z, \\
	\dot{\phi} & = p + q \sin\phi \tan\theta + r \cos\phi \tan\theta,\!  & & \dot{p} = \frac{I_{yy} - I_{zz}}{I_{xx}} qr + \frac{l}{I_{xx}} U_2, \\
	\dot{\theta} & = q \cos\phi- r \sin\phi, & & \dot{q} = \frac{I_{zz} - I_{xx}}{I_{yy}} pr + \frac{l}{I_{yy}} U_3, \\
	\dot{\psi} & = q \sin\phi \sec\theta + r \cos\phi \sec\theta, & & \dot{r} = \frac{I_{xx} - I_{yy}}{I_{zz}} pq + \frac{1}{I_{zz}} U_4,
	\end{aligned} \right.
	\end{equation}
	where $m$, $I_{xx}$, $I_{yy}$, $I_{zz}$, and $l$ are physical parameters, $g$ is the gravitational acceleration, $U_1$ is the total thrust generated by the propellers, $U_2$ is the difference of thrusts between the left and right propellers, $U_3$ is the difference of thrusts between the front and back propellers, $U_4$ is the difference of torques between clockwise and counter-clockwise turning propellers, and $\mathbf{d} = [D_x \,\; D_y \,\; D_z]^T$ are disturbance forces applied to the UAV with $\ninf{\mathbf{d}} \leq 1$. The experiment consists in obtaining guaranteed bounds on the system states $\bm{\zeta}$ while the quadrotor UAV tracks a vertical helix trajectory defined by the reference values $x^\text{ref}(t) = \frac{1}{2} \cos\left(\frac{t}{2}\right)$, $y^\text{ref}(t) = \frac{1}{2} \sin\left(\frac{t}{2}\right)$, $z^\text{ref}(t) = 1 + \frac{t}{10}$, $\psi^\text{ref} = \frac{\pi}{3}$,
	subject to the disturbance forces described by $D_x = 1$ N for $t \in [5,15)$ s, $D_y = 1$ N for $t \in [8,15)$ s, and $D_z = 1$ N for $t \in [10,15)$ s. These forces are zero otherwise. The dynamic feedback controller in \cite{Mistler2001} is used to track the reference trajectory above\footnote{In this experiment, the control action is computed using the real states $\bm{\zeta}_k$. An extension of the minimum-variance approach proposed in Section \ref{sec:uav_minimumvariance} can be used for feedback connection using a point that belongs to $\hat{X}_k$.}. The simulation parameters are shown in Table \ref{tab:uav_quadrotormodelparametersestmation}. %

\begin{table}[!tb]
	\footnotesize
	\centering
	\caption{Model parameters of the quadrotor UAV considered in the state estimation experiment.}
	\begin{tabular}{c c c c c c c c c} \hline
		Parameter & $m$ & $l$ & $I_{xx},I_{yy},I_{zz}$ & $g$  \\ \hline
		Value & $0.7$ kg & $0.3$ m & $1.2416$ kg${\cdot}$m$^2$ & $9.81$ m/s$^2$ \\ 
		\hline
	\end{tabular} \normalsize
	\label{tab:uav_quadrotormodelparametersestmation}
\end{table}

	We consider a realistic scenario in which the available measurements are provided by sensors located at the quadrotor UAV, which include: (i) a Global Positioning System (GPS); (ii) a barometer; and (iii) an Inertial Measurement Unit (IMU). The measurements are affected by bounded uncertainties as described in Table \ref{tab:uav_quadrotorsensorparameters}. The velocity vector $[u\,\;v\,\;w]^T$ is not measured.

	\begin{table}[!htb]
		\footnotesize
		\centering
		\caption{Measured variables with error bounds.}
		\begin{tabular}{c c c} \hline
			Sensor & Variables & Noise bounds \\ \hline
			GPS & $\{x,y\}$ & $\pm\!\!$ $0.15\!$ m \\ 
			Barometer & $\{z\}$ & $\pm\!\!$ $0.51\!$ m \\ 
			\multirow{2}{*}{IMU} & $\{\phi,\theta,\psi\}$ & $\pm\!\!$ $2.618 {\cdot} 10^{-3}\!$ rad \\
			& $\{p,q,r\}$ & $\pm\!\!$ $16.558 {\cdot} 10^{-3}\!$ rad/s \\
			\hline
		\end{tabular} \normalsize
		\label{tab:uav_quadrotorsensorparameters}
	\end{table}

\begin{figure}[!tb]
		\centering{
			\def\svgwidth{0.67\columnwidth}
			{\scriptsize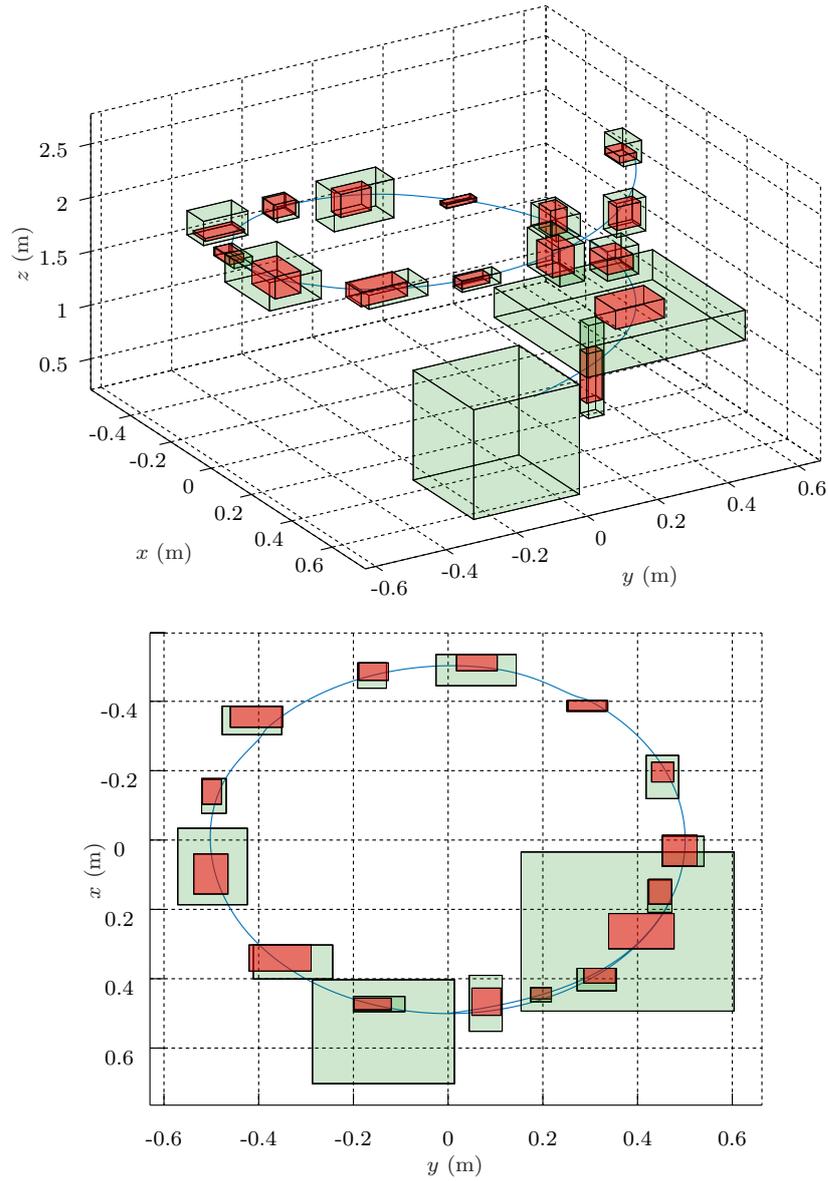}
			\caption{The trajectory performed by the quadrotor UAV (solid line), and the interval hulls of the constrained zonotopes (red boxes) and zonotopes (green boxes) estimated by CZMV and ZMV, respectively, projected onto $(x, y, z)$.}\label{fig:uav_quadrotor_trajectory_boxes}}
	\end{figure}

	The nonlinear equations \eqref{eq:uav_quadrotor} were discretized by Euler approximation with sampling time $0.01$ s. The initial states $\bm{\zeta}_0$ are bounded by $X_0 = \{\mathbf{G}_0,\bm{0}\}$, where $\mathbf{G}_0 = \text{diag}\left(2, 2, 2, 1, 1, 1, \frac{\pi}{6}, \frac{\pi}{6}, \frac{\pi}{2}, \frac{\pi}{12}, \frac{\pi}{12}, \frac{\pi}{12}\right).$ To generate process measurements, the discrete-time dynamics were simulated with $\bm{\zeta}_0  = [0.5\,\;0\,\;1\,\; \zeros{1}{5} \,\; \pi/3\,\; \zeros{1}{3} ]^T \in X_0$ and process and measurement noises drawn from uniform distributions. Figure \ref{fig:uav_quadrotor_trajectory_boxes} shows the trajectory performed by the quadrotor UAV along with the interval hulls\footnote{The conversion from CG-rep to H-rep (see Property \ref{prop:pre_czhreptocgrep}) for the purposes of exact drawing is intractable for the constrained zonotopes in this example.} of the enclosures computed by the methods CZMV and {\alamobravo}, projected onto $(x,y,z)$-axes. CZMV was implemented with $\bm{\gamma}_x$ given by \emph{C2}, and since \eqref{eq:uav_quadrotor} is also affine in $\mbf{w}_k \triangleq \mbf{d}_k$, Theorem \ref{thm:ndyn_meanvalue} was implemented with $Z_w \supseteq \bm{\mu}(\bm{\gamma}_w,W)$ computed as described at the end of Remark \ref{rem:ndyn_affine}. The number of constraints and generators of the computed constrained zonotopes was limited to 40 and 12, respectively, while the number of generators of the computed zonotopes was limited to 40.  
	
\begin{figure}[!tb]
		\centering{
			\def\svgwidth{0.7\columnwidth}
			{\scriptsize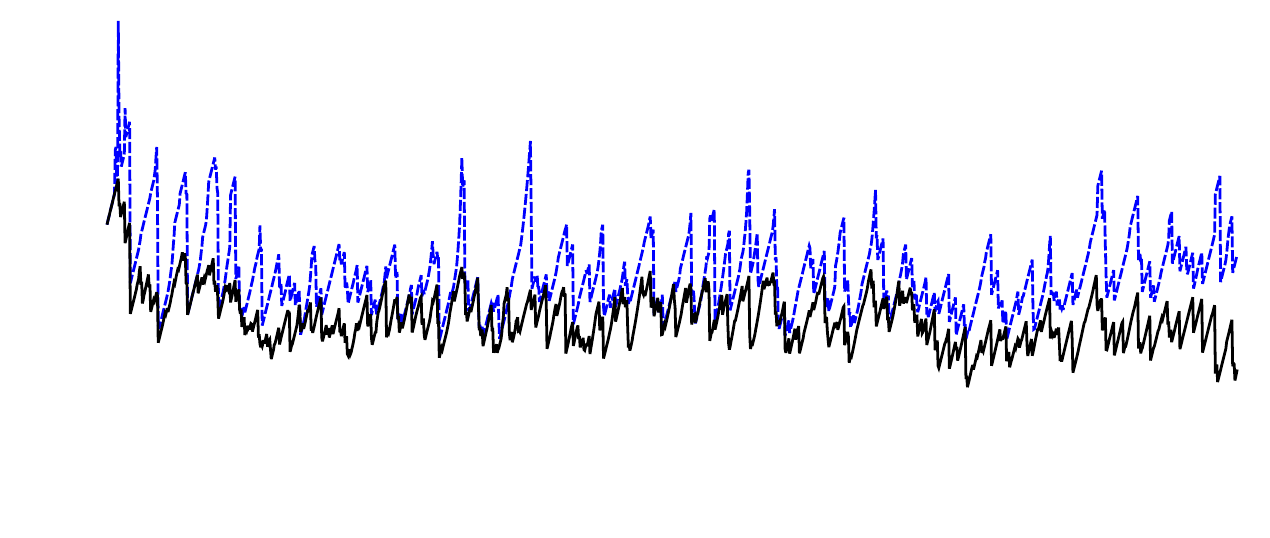}
			\caption{Radii of the update sets computed by CZMV (solid black line) and {\alamobravo} (dashed blue line) for the quadrotor UAV experiment.}\label{fig:uav_quadrotor_meanvalue_radius}}
	\end{figure}

\begin{figure}[!tb]
		\centering{
			\def\svgwidth{0.7\columnwidth}
			{\scriptsize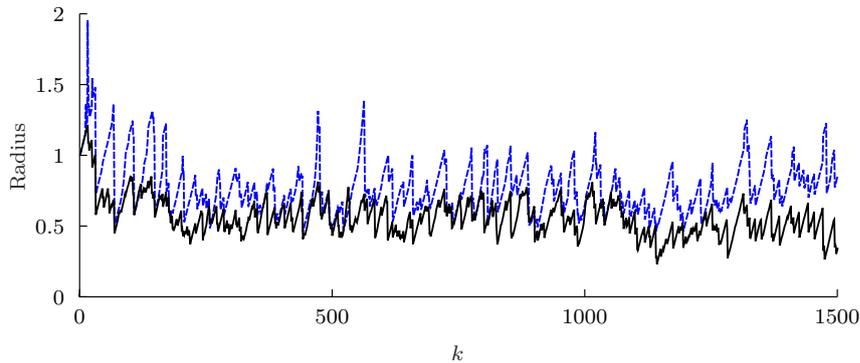}
			\caption{Radii of the update sets computed by CZFO (solid black line) and {\combastelbravo} (dashed blue line) for the quadrotor UAV experiment.}\label{fig:uav_quadrotor_firstorder_radius}}
	\end{figure}

	The interval hulls of the constrained zonotopes obtained by CZMV were smaller than those from {\alamobravo}, demonstrating the accuracy of the proposed method. Figure \ref{fig:uav_quadrotor_meanvalue_radius} shows the radii of the constrained zonotopes and zonotopes computed by CZMV and {\alamobravo}, respectively. Both algorithms were capable of providing tight bounds on the system states $\bm{\zeta}_k \in \realset^{12}$. Nevertheless, CZMV provided less conservative bounds than {\alamobravo}, even for a high-order nonlinear dynamical system such as \eqref{eq:uav_quadrotor} (the CZMV-to-{\alamobravo} ARR was $74.41\%$). %
	Finally, Figure \ref{fig:uav_quadrotor_firstorder_radius} compares the radii of the update sets computed by {\combastelbravo} and CZFO with $\bm{\gamma}_x$ given by \emph{C3}\footnote{The increased complexity of the constrained zonotopes provided by \emph{C4} proved to be intractable for this example.}. Once again, CZFO provided less conservative bounds than {\combastelbravo} (the CZFO-to-{\combastelbravo} ARR was $74.45\%$). The results from CZMV and CZFO were again very similar, with CZMV providing marginally better results (the CZMV-to-CZFO ARR was $99.93\%$). The ARR for different numbers of constraints are shown in Table \ref{tab:uav_quadrotorARR}, with the average computed considering in addition simulations with different numbers of generators. Execution times are shown in Table \ref{tab:uav_quadrotortimes}. Note that most of the CZFO execution times are smaller than the ones presented in Table \ref{tab:ndyn_example2times}. This might be counter-intuitive since the current example has more state variables. However, the use of \emph{C4} in Example 1 results in a relatively more complex enclosure and therefore requires a much higher execution time for generator reduction and constraint elimination. Note that in this example, the computational times of the state estimators were greater than the considered sampling time of $0.01$ s. Nevertheless, this fact does not invalidate the obtained results since these were run in a numerical simulation. Better times can be achieved by optimized implementation of the algorithms and using more powerful hardware, for instance. Besides, note that even though the current execution times of CZMV and CZFO would in principle prevent their use in fast applications, the improved accuracy can significantly reduce the number of time steps required for guaranteed fault diagnosis for systems in which execution time is not critical.%
	
	\begin{table}[!htb]
		\footnotesize
		\centering
		\caption{Average radius ratio of the estimators with varying numbers of constraints. Each average is taken over 3 separate simulations with $n_g \in \{ 25,30,40\}$.}
		\begin{tabular}{c c c c} \hline
			$n_c$ & CZMV/ZMV & CZFO/ZFO & CZFO/CZMV \\ \hline
			$3$ & $82.5\%$ & $82.2\%$ & $99.9\%$ \\
			$6$ & $76.7\%$ & $77.1\%$ & $100.7\%$ \\		
			$12$ & $75.0\%$ & $74.8\%$ & $99.7\%$ \\
			\hline
		\end{tabular} \normalsize
		\label{tab:uav_quadrotorARR}
	\end{table}
	
	\begin{table}[!htb]
		\footnotesize
		\centering
		\caption{Average total times per iteration of the estimators with varying numbers of constraints. Each average is taken over 15 separate simulations using $n_g \in \{25,30,40\}$. Times spent only on complexity reduction are shown in parenthesis.}
		\begin{tabular}{c c c c c} \hline
			$n_c$ & ZMV & ZFO & CZMV & CZFO \\ \hline
			$0$ & $53.3~(1.0)$ ms & $174.2~(54.2)$ ms & -- & -- \\
			$3$ & -- & -- & $104.7~(7.3)$ ms & $266.9~(128.6)$ ms \\
			$6$ & -- & -- & $109.5~(8.4)$ ms & $615.0~(471.6)$ ms \\			
			$12$ & -- & -- & $127.8~(12.3)$ ms & $2.62~(2.46)$ s \\
			\hline	
		\end{tabular} \normalsize
		\label{tab:uav_quadrotortimes}
	\end{table}
		

\section{Fault diagnosis of a quadrotor UAV} \label{sec:uav_quadrotorfaultdiagnosis}

This section presents an example consisting in active fault isolation of a quadrotor UAV subject to actuator and sensor fault using the methods developed in Chapter \ref{cha:faultdiagnosis}.

Let $\bm{\zeta} = [x \,\; y \,\; z \,\; u \,\; v \,\; w \,\; \phi \,\; \theta \,\; \psi\,\;p\,\;q\,\;r]^T$ be the system state, where $[x\,\;y\,\;z]^T$ is the position of the UAV with respect to the inertial frame, $[u\,\;v\,\;w]^T$ is the velocity vector expressed in the inertial frame, $[\phi\,\;\theta\,\;\psi]^T$ are Euler angles describing the orientation of the UAV, and $[p\,\;q\,\;r]^T$ is the angular velocity vector expressed in the body frame. Let $\mbf{u} \in \realset^4$ denote the thrust forces generated by the propellers. The equations of motion are\footnote{This is an adaptation of the model in Section \ref{sec:uav_quadrotorestimation} considering the thrust forces generated by the propellers as inputs, with inclination angle as described in Section 4.4.1 in \cite{RaffoThesis}.}%

\begin{equation} \label{eq:uav_quadrotorfaulty}
\dot{\bm{\zeta}} = 
\left\{ \begin{aligned}
\dot{x} & = u,~ & & \dot{u} = \frac{1}{m} F_1 + \frac{1}{m}D_x, \\
\dot{y} & = v,~ & & \dot{v} = \frac{1}{m} F_2 + \frac{1}{m}D_y, \\
\dot{z} & = w,~ & & \dot{w} = -g + \frac{1}{m} F_3 + \frac{1}{m}D_z, \\
\dot{\phi} & = p + q \sin\phi \tan\theta + r \cos\phi \tan\theta,~ & & \dot{p} = \frac{I_{yy} - I_{zz}}{I_{xx}} qr + \frac{1}{I_{xx}} T_1, \\
\dot{\theta} & = q \cos\phi- r \sin\phi,~ & & \dot{q} = \frac{I_{zz} - I_{xx}}{I_{yy}} pr + \frac{1}{I_{yy}} T_2, \\
\dot{\psi} & = q \sin\phi \sec\theta + r \cos\phi \sec\theta,~ & & \dot{r} = \frac{I_{xx} - I_{yy}}{I_{zz}} pq + \frac{1}{I_{zz}} T_3,
\end{aligned} \right.
\end{equation}
where
\begin{equation*}
\begin{bmatrix} F_1 \\ F_2 \\ F_3 \\ T_1 \\ T_2 \\ T_3 \end{bmatrix} = 
\begin{bmatrix} \eye{3} & \bm{0} \\ \bm{0} & \RIB \end{bmatrix} \begin{bmatrix} 0 & l \cos(\alpha) & 0 & -l \cos(\alpha) \\
-l \cos(\alpha) & 0 & l \cos(\alpha) & 0 \\
\ktaub \cos(\alpha) & -\ktaub \cos(\alpha) & \ktaub \cos(\alpha) & -\ktaub \cos(\alpha) \\
-\sin(\alpha) & 0 & \sin(\alpha) & 0 \\
0 & -\sin(\alpha) & 0 & \sin(\alpha) \\
\cos(\alpha) & \cos(\alpha) & \cos(\alpha) & \cos(\alpha)	
\end{bmatrix} \bm{\Omega}^{[i]} \mbf{u},
\end{equation*}
with $\RIB$ defined equivalently as in \eqref{eq:uav_tiltrotorRIL} (Section \ref{sec:uav_tiltrotormodeling}), $m$, $I_{xx}$, $I_{yy}$, $I_{zz}$, $l$, $g$, $\alpha$, $k_\tau$, $b$ are physical parameters given in Table \ref{tab:uav_quadrotormodelparameters}, and $\mbf{d} = [D_x \,\; D_y \,\; D_z]^T$ are disturbance forces applied to the UAV with $\ninf{\mbf{d}} \leq 1$. The available measurement is given by
\begin{equation} \label{eq:uav_quadrotormeasurement}
\mbf{y}_k = \mbf{s}^{[i]} + \mbf{C}^{[i]}\mbf{x}_k + \mbf{D}_v^{[i]} \mbf{v}_k,
\end{equation}
with bounds on the measurement noise $\mbf{v}_k$ as described in Table \ref{tab:uav_quadrotorsensorparameters}. The actuator faults are modeled by $\bm{\Omega}^{[i]} \neq \eye{4}$, in which the faulty scenario corresponds to partial loss of potency of some of the thrusters. The sensor faults are modeled by $\mbf{s}^{[i]} \neq \bm{0}$, which corresponds to measurement bias. This leads to a total of four possible scenarios, defined by
\begin{align*}
& \bm{\Omega}^{[1]} = \bm{\Omega}^{[3]} = \eye{4}, \quad \bm{\Omega}^{[2]} = \bm{\Omega}^{[4]} = \text{diag}(1, 0.8, 0.3, 1), \\
& \mbf{s}^{[1]} = \mbf{s}^{[2]} = \zeros{9}{1}, \quad \mbf{s}^{[3]} = \mbf{s}^{[4]} = [ \zeros{1}{5} \,\; \pi/6 \,\; \zeros{1}{3}]^T, \\
& \mbf{C}^{[1]} = \mbf{C}^{[2]} = \mbf{C}^{[3]} = \mbf{C}^{[4]} = \begin{bmatrix} \eye{3} & \zeros{3}{3} & \zeros{3}{6} \\ \zeros{6}{3} & \zeros{6}{3} & \eye{6} \end{bmatrix},\\
& \mbf{D}_v^{[1]} = \mbf{D}_v^{[2]} = \mbf{D}_v^{[3]} = \mbf{D}_v^{[4]} = \eye{9}.
\end{align*}

\begin{table}[!htb]
	\footnotesize
	\centering
	\caption{Model parameters of the quadrotor UAV considered in the fault diagnosis experiment.}
	\begin{tabular}{c c c c c} \hline
		Parameter & $m$ & $l$ & $I_{xx},I_{yy}$ & $I_{zz}$ \\ \hline
		Value & $2.24$ kg & $0.3332$ m & $0.0363$ kg${\cdot}$m$^2$ & $0.0615$ kg${\cdot}$m$^2$ \\ 
		\hline
		Parameter & $g$  & $\alpha$ & $k_\tau$ & $b$  \\ \hline
		Value & $9.81$ m/s$^2$ & $5^{\circ}$ & $1.7\ten{-7}$ N${\cdot}$m${\cdot}$s$^2$ & $9.5\ten{-6}$ N${\cdot}$s$^2$ \\
		\hline		
	\end{tabular} \normalsize
	\label{tab:uav_quadrotormodelparameters}
\end{table}

The nonlinear equations \eqref{eq:uav_quadrotorfaulty} were discretized by Euler approximation with sampling time $0.01$ s. The initial state belongs to the constrained zonotope
\begin{equation*}
X_0 = \Big\{[0.05{\cdot}\eye{12} \,\; 0.05{\cdot}\ones{12}{1}], \zeros{12}{1}, [1 \,\; 0.5{\cdot}\ones{1}{12}], 1 \Big\}.
\end{equation*}

For fault isolation, we consider an input sequence with length $N = 2$. The cost function is $J(\seq{\mbf{u}}) = \|\seq{\mbf{u}}\|_{\seq{\mbf{R}}}^2 = \sum_{j=0}^{N-1} \mbf{u}_j^T \mbf{R} \mbf{u}_j$, with $\mbf{R} = \eye{4}$. The complexity of the constrained zonotopes $\seq{Z}(i,j)$ is limited to 27 generators and 3 constraints using Methods \ref{meth:genredB} and \ref{meth:czconelim}, respectively. Moreover $\seq{\bm{\gamma}}_w = \bm{0}$, and $\seq{\bm{\gamma}}_u = (\mbf{u}^\text{eq}, \mbf{u}^\text{eq})$, with $\mbf{u}^\text{eq} = (1/4)mg \sec(\alpha) {\cdot}\ones{4}{1}$. The vector $\mbf{h}_0$ was chosen as $\bm{\gamma}_{x,0} = \bm{0}$ (in this case $\bm{0} \in X_0$), with $\bm{\gamma}_{x,k}$ obtained by propagating $(\bm{\gamma}_{x,k-1},\bm{\gamma}_{u,k-1},\bm{\gamma}_{w,k-1})$ through the discretized version of the dynamics \eqref{eq:uav_quadrotorfaulty} for each model, $k \in \{1, 2\}$. The minimum separation threshold is $\varepsilon = 1.0\ten{-8}$. The admissible input set is $U = \{  3{\cdot}\ones{4}{1}, \mbf{u}^\text{eq} \}$.

The MIQP \eqref{eq:fault_optimalseparatingczfinal} was solved using MATLAB 9.1 and GUROBI 8.0.1, which resulted in the optimal separating input
\begin{equation} \label{eq:uav_quadrotorseparatinginput}
\seq{\mbf{u}} = \left( \begin{bmatrix} 2.5146 \\ 2.5146 \\ 2.5146 \\ 2.5146 \end{bmatrix}, \begin{bmatrix} 2.5146 \\ 2.5146 \\ 2.5146 \\ 2.5146  \end{bmatrix} \right).
\end{equation}

This input sequence was found with execution time $3.58$ s. The intersection \eqref{eq:separatinginputdef} was verified to be empty for every $i,j \in \modelset = \{1,2,3,4\}$, $i \neq j$, using the separating input \eqref{eq:uav_quadrotorseparatinginput}. Moreover, the sequence \eqref{eq:uav_quadrotorseparatinginput} was injected into each model $i \in \modelset$, and $\seq{\mbf{y}}\zerospace^{[i]} \in \underline{\seq{\Psi}}\zerospace_{1:2}^{[i]}$ was satisfied for every $i \in \modelset$, with $\seq{\mbf{y}}\zerospace^{[i]} \notin \underline{\seq{\Psi}}\zerospace_{1:2}^{[j]}$ for every $i \neq j$. Therefore, the active model in \eqref{eq:uav_quadrotorfaulty}--\eqref{eq:uav_quadrotormeasurement} was guaranteed to be isolated by the injection of the separating input \eqref{eq:uav_quadrotorseparatinginput}, allowing robust fault diagnosis of the quadrotor UAV when subject to actuator and sensor faults using the proposed active fault isolation method.%

\section{Final remarks}

This chapter presented the application of set-based state estimation and fault diagnosis methods using constrained zonotopes to unmanned aerial vehicles. 

The first application involved linear state estimation and state-feedback control for trajectory tracking of a suspended load using a tilt-rotor UAV. A constrained minimum-variance criterion was proposed for feedback connection with estimated states. The linear state estimation method based on constrained zonotopes was validated through numerical experiments carried out in the ProVANT simulator, a platform based on the Gazebo simulator with a CAD model of the system, which demonstrated the reduced conservativeness of the proposed linear set-based state estimator, the performance of the optimal selection of estimated states, and the robustness of the designed controller.

The second application was a numerical experiment considering nonlinear state estimation of a quadrotor UAV, taking into account a realistic scenario in which measurements are provided by sensors located at the UAV. The third application considered a numerical experiment involving the optimal input design for nonlinear active fault diagnosis of a quadrotor UAV subject to actuator and sensor faults, which addressed scenarios of complete actuator failure and the existence of sensor bias.

\chapter{Application: Water distribution networks}\thispagestyle{headings} \label{cha:appwdn}

State estimation in water distribution networks (WDNs) is a challenging task due to the scarcity of measurements and the presence of several modeling uncertainties. 
In the literature, pseudo-measurements are often required to obtain an observable model for state estimation, which are highly uncertain and mostly represented by bounded sets.
This chapter proposes two new methods for set-based state estimation in WDNs, considering unknown-but-bounded measurement and model uncertainties to calculate bounds on the system states. 
A new interval method is proposed based on iterative computation of tight enclosures of the nonlinear head-loss functions, and bounding the solution of the algebraic equations using rescaling. 
This method, referred to as Iterative Slope Approximation and Rescaling (ISAR), is mildly more conservative than the Iterative Hydraulic Interval State Estimation (IHISE) method previously proposed in \cite{Vrachimis2019}, but significantly more computationally efficient.
However, since intervals are not capable of capturing the dependencies between state variables, we propose the use of constrained zonotopes as an additional step to both IHISE and ISAR. 
This yields two new algorithms for set-based state estimation of WDNs, capable of capturing the dependencies between hydraulic states, which result in sets with significantly smaller volumes than intervals. 
The benefits of constrained zonotopes are also highlighted when the new enclosures are used for leakage detection. These provide higher leakage detection rates compared to intervals, as demonstrated in two case studies using benchmark WDNs.
The content presented in this chapter has been submitted for publication in \cite{Rego2020d}.

This chapter is organized as follows. Section \ref{sec:wdn_introduction} introduces the main topic of this chapter. The topology of the water distribution networks considered in this chapter is described in Section \ref{sec:wdn_problemformulation} along with the corresponding mathematical formulation. Section \ref{sec:wdn_stateestimation} reviews the interval method proposed in \cite{Vrachimis2019}, introduces the required mathematical background on CZs, describes a new interval method with reduced computational cost, and present the final enclosures described by CZs. The use of these enclosures for leakage detection in WDNs is discussed in Section \ref{sec:wdn_leakdetection}. Numerical examples are presented in Section \ref{sec:wdn_numericalexperiments}, and Section \ref{sec:wdn_finalremarks} concludes the chapter.

\section{Introduction} \label{sec:wdn_introduction}
A Water Distribution Network (WDN) is a system responsible for delivering water to consumers from water sources. 
The water industry is being modernized with the installation of sensors for monitoring WDNs, the use of supervisory control, and data acquisition
(SCADA) systems. 
Integrated monitoring platforms can combine real-time sensor measurements with hydraulic models and geographical information systems. 
However, there is still the need for intelligent algorithms to process the acquired data, and use them for important tasks such as state estimation. 
State estimation algorithms in WDNs infer the internal system states, such as water flows in pipes and pressures at nodes, given the available measurements. 
A complete view of system states supports the operators in decision-making and enables the early detection of hydraulic faults, such as leakages. 
It is estimated that every day more than $45$ million $m^3$ of treated water is lost due to leakages in developing countries, which could have served $200$ million consumers \citep{Kingdom2006}.

State estimation in WDNs is a challenging task due to the scarcity of measurements compared to the system size. 
Indeed, observability in WDNs is only achievable in networks of large pipes which bring water from sources to water utilities, called the transport networks \citep{Vrachimis2018a}.
In these networks, all the inflows and outflows are measured in real-time, which is a sensor configuration that guarantees observability \citep{Diaz2017}. 
However, most pipe failures in WDNs occur in urban distribution networks which may have hundreds of outlets that are not measured in real time. 
Water utilities typically sectorize urban networks into District Metered Areas (DMAs) and measure the inlet flow and pressure of each sub-network.  
A number of pressure sensors may exist in a DMA, typically used for pressure monitoring. 
To obtain hydraulic observability, groups of consumer nodes are modeled as a single outlet and a high resolution estimate of their consumption is determined using knowledge about population distributions and historical data \citep{Avni2015}. 
However, these so-called pseudo-measurements of consumer demands are highly uncertain with no statistical characterization of their error to the true consumption.
The best way to characterize their uncertainty is with the use of error bounds \citep{Arsene2011}. 
Moreover, WDN hydraulic models are often uncertain in their parameters, which clearly affects the quality of state estimation \citep{Diaz2018}. 
The most notable uncertain parameter is pipe roughness, which can be described as a measure of the impedance each pipe exhibits to the water flowing through it. 
This parameter cannot be determined experimentally for every pipe, thus an estimate based on the pipe material and age is usually adopted \citep{Boulos2006}.
	
Standard state estimation techniques used in WDNs, such as Kalman filtering and weighted least squares, require a measurement set that makes the system observable and the statistical characterization of sensor measurement error to give more weight to measurements originating from more-accurate sensors \citep{Nagar2000,Diaz2016}.
Then, using a mathematical model of the system, they produce an estimated point and covariance \citep{Kang2009}.
Given the nature of uncertainties in WDNs, the approach of modeling uncertainties of input data as intervals has been proposed in \cite{Bargiela2003} and \cite{Vrachimis2019}.
This approach results in a set-based state estimate which is often more useful to an operator than providing point and covariance estimates, due to the lack of statistical characterization of all the uncertainties involved.
	
	Set-based state estimation methods are often called Confidence Limit Analysis in the WDNs literature. The use of bounds for the representation of measurement uncertainty in WDNs was first introduced in \cite{Bargiela1989}, where it was incorporated to estimate interval enclosures of the states. A common way to obtain set-based states in WDNs using a nonlinear model is through Monte-Carlo simulations (MCS) \citep{Eliades2015}.
		Nevertheless, this approach is computationally intensive, and the actual lower and upper bounds may not be reached even with a large number of simulations.
	On the other hand, bounds can be obtained without the need for extensive simulations using different methods such as the Error Maximization method \citep{Arsene2011}, the Ellipsoid method and Linear Programming \citep{Bargiela2003}.
		The bounds generated by these approaches are not guaranteed because of the linearization uncertainty introduced when using a linear approximation of the WDN hydraulic model, while they often neglect parameter uncertainty.
	An interesting approach when dealing with systems with bounded uncertainties is interval-based state estimation which has been studied extensively for linear systems \citep{Efimov2013c,Wang2015}, and also used for specific classes of nonlinear systems \citep{Raissi2012,Vrachimis2019}. In particular, a recent interval-based state estimation approach, specific for water systems, was proposed in \cite{Vrachimis2019}, in which an optimization-based method calculates interval enclosures of the states using a nonlinear hydraulic model and considering parameter, input, and measurement uncertainties.
	It is shown that the provided bounds on the states approximate closely those obtained by performing a large number of MCS. Such bounds can be used not only for state estimation, but also for leak detection \citep{Vrachimis2018c}.
	
	In contrast to intervals, zonotopes are capable of capturing potential dependencies between state variables, while mitigating the wrapping effect inherent to state estimation methods based entirely on interval analysis \citep{Kuhn1998}. Zonotopes have been used in parameter estimation and leak detection of WDNs in \cite{Blesa2010}, and leak localization in \cite{Escofet2015}. Specifically, the latter requires linearization of the output equations, and conversions between the generator representation of zonotopes to polytopes either in half-space representation or in vertex representation. These conversions have combinatorial and exponential complexities, respectively, and can lead to dramatic computational times in higher dimensions \citep{Althoff2010}. On the other hand, while maintaining the computational advantages of zonotopes and also being able to capture the dependency of state variables, constrained zonotopes incorporate equality constraints between the state variables into the mathematical description (see Chapters \ref{cha:nonlinearmeasinv}, \ref{cha:descriptor}, and \ref{cha:stateparameter}). 	
	Nevertheless, despite these advantages, to the best of our knowledge the properties of constrained zonotopes have not been yet explored in methods for set-based state estimation and leak detection in WDNs.

	This chapter proposes two new methods for set-based state estimation of WDNs based on constrained zonotopes, considering unknown-but-bounded water demands and model parameters. Specifically, we first formulate the mathematical model of a WDN from the laws of conservation of energy and mass, which is later approximated by linear algebraic constraints. Accordingly, we propose an interval method based on iterative computation of tight enclosures of the water flows by bounding the solution of the linear algebraic equations using rescaling \citep{Scott2016}, and subsequently re-approximating the nonlinear equalities for the new flow bounds. This method is referred to as Iterative Slope Approximation and Rescaling (ISAR), which is mildly more conservative but significantly faster than the previous approach in \cite{Vrachimis2019}. However, as mentioned before, intervals are not capable of capturing the dependencies between state variables. Therefore, in addition we propose the use of CZs to obtain tights bounds of the solution of the nonlinear algebraic equations, with linear approximation based on the final interval enclosures provided by both the approach described in \cite{Vrachimis2019}, and ISAR. These two new algorithms for set-based state estimation of WDNs, with a fair trade-off between efficiency and accuracy, are both capable of capturing the dependencies between hydraulic heads and water flows. This result in sets with significantly smaller volumes than intervals, and therefore with improved accuracy. The obtained enclosures are further used in passive leak detection of WDNs, in which, by being less conservative than intervals, allow a significant higher number of detections in several scenarios with different leak magnitudes. We investigate the key advantages of the proposed methods for set-based state estimation and leak detection in WDNs in numerical examples and also a realistic network, which highlight their improved efficiency and accuracy in comparison to intervals.

	\section{Problem formulation} \label{sec:wdn_problemformulation}
	The topology of a WDN is modeled by a directed graph denoted as $\mathcal{G}=(\mathcal{N},\mathcal{L})$. 
	Let $\mathcal{N} = \{ 1, \ldots ,{n_h} \}$ be the set of all nodes, where $| \mathcal{N} | = n_h$ is the total number of nodes.
	These represent junctions of pipes, consumer water demand locations, reservoirs and tanks. 
	The hydraulic state associated with the nodes is the \emph{hydraulic head},  indicated by $h_j,\,j\in \mathcal{N}$.
	The hydraulic head consists of the pressure at node $j$, whose minimum is the node elevation $ \eta_i $ with respect to a geodesic reference.
	Each node $j$ is also associated with a water consumer demand at the node location, denoted by $d_j$. 
	Demands drive the dynamics of a WDN and are an uncertain input to the system. 
	Let $\mathcal{L} = \{1, \cdots ,{n_q}\}$ be the set of links, where $| \mathcal{L} | = n_q$ is the total number of links. 
	These represent network pipes, water pumps and pipe valves.
	The hydraulic quantity associated with a link $i \in \mathcal{L}$ is the water flow, indicated by $q_i$. Moreover, $u_i$ denotes the contribution of the level of water reservoirs and tanks to the $i$-th link.
	
	The system of hydraulic algebraic equations which describe the behavior of the plant are derived from the laws of conservation of energy and mass for water networks.
	Consider a WDN with $n_q$ links and $n_h$ nodes, described by the nonlinear system of equations
	\begin{subequations} \label{eq:wdn_system}
		\begin{align}
		& \mbf{f}(\mbf{q}_k) + \mbf{B} \mbf{h}_k = \mbf{u}_k, \label{eq:wdn_systemA}\\
		& \mbf{B}^T \mbf{q}_k = \mbf{d}_k, \label{eq:wdn_systemB}
		\end{align}
	\end{subequations}
	where $k \geq 0$ denotes the time instant, $\mbf{q}_k \in \realset^{n_q}$ denote the water flows at the links, $\mbf{h}_k \in \realset^{n_h}$ denote the node heads, $\mbf{B} \in \realsetmat{n_q}{n_h}$ denotes the incidence matrix (whose elements are $-1$, $0$, or $1$ depending on the connectivity of nodes with links), $\mbf{u}_k \in \realset^{n_q}$ is the contribution of the level of water reservoirs and tanks, and $\mbf{d}_k \in \realset^{n_h}$ is the water demand by the consumers at the nodes. The nonlinear mapping $\mbf{f}(\mbf{q}_k): \realset^{n_q} \to \realset^{n_q}$ denotes the head loss at the links, with its $i$-th component $f_i(\mbf{q}_k) \triangleq f_i(q_{i,k})$ being given by
	\begin{equation} \label{eq:wdn_systemf}
	f_i(q_{i,k}) = \left\{ \begin{aligned} r_i |q_{i,k}|^{(\nu-1)} q_{i,k} & & & \text{ if link }i\text{ is a pipe},   \\ 
	-(w_1 - w_2 q_{i,k}^{w_3}) & & & \text{ if link }i\text{ is a pump}, \end{aligned} \right.
	\end{equation}
	with $\nu$, $w_1$, $w_2$, $w_3 \in \realset$ being known parameters derived from physical properties of their corresponding elements (i.e., pipes and pumps) \citep{Boulos2006}. On the other hand, the parameter $\mbf{r} \in \realset^{n_q}$ is assumed to be uncertain and to belong to a known interval $R$. 
	
	Given the known information on $\mbf{u}_k$ and $\mbf{d}_k$ at instant $k$, for $k \geq 0$, the objective of this chapter is to obtain tight bounds $Q_k$ and $H_k$ satisfying $\mbf{q}_k \in Q_k$ and $\mbf{h}_k \in H_k$. It is assumed that interval bounds $U_k$ and $D_k$ of $\mbf{u}_k$ and $\mbf{d}_k$, respectively, are known from historical data, such that $\mbf{u}_k \in U_k$, $\mbf{d}_k \in D_k$, for all $k \geq 0$. In addition, we assume that conservative interval bounds on the state $\mbf{x}_k \triangleq (\mbf{q}_k,\mbf{h}_k)$ can be derived from physical constraints or available historical data.
		
	\section{Set-based state estimation of WDNs}  \label{sec:wdn_stateestimation}
	
	\subsection{Iterative Hydraulic Interval State Estimation} \label{sec:wdn_IHISE}
	Iterative Hydraulic Interval State Estimation (IHISE) is a set-based state estimation methodology for WDNs first proposed in \cite{Vrachimis2018a} and further developed in \cite{Vrachimis2019}.
	The methodology calculates bounds on the system states satisfying the nonlinear algebraic equations \eqref{eq:wdn_system}, given bounds on water demands and the model parameter $\mbf{r}$.
	This is achieved using bounding linearization, a technique which restricts the nonlinearities within a convex set, thus converting the hydraulic equations in a form where state bounds can be calculated using convex optimization techniques. 
	An iterative procedure minimizes the distance between upper and lower state bounds, by reducing the feasible set defined by bounding linearization at each step and converging to the tightest possible interval bounds.
	The IHISE is summarized in the following four steps.
	\subsubsection{Initial bounds on the state vector} \label{sec:wdn_initialbounds}
	The initial enclosure of the state $\mbf{x}_k\in X_k$ is {obtained by considering the physical properties of the network}.
	The initial lower bounds $\mbf{h}_k^{\text{L}\,(0)}$ on the unknown head vector $\mbf{h}_k$ are set as equal to the elevation of each node, while the initial upper bounds $\mbf{h}_k^{\text{U}\,(0)}$ are equal to the sum of reservoir and pump heads, which is physically the maximum head that any node in the WDN can achieve. Given $[\mbf{h}_k^{\text{L}\,(0)}, \, \mbf{h}_k^{\text{U}\,(0)}]$, and the known bounds $U_k$ of the reservoirs and tanks levels $\mbf{u}_k$, by noting that each equation in \eqref{eq:wdn_systemA} is a function of only one $q_{i,k}$, and that the head-loss functions $ f_i(q_{i,k}) $ are inclusion isotonic \citep{Moore2009} (meaning that $A \subseteq B \implies {f}(A) \subseteq {f}(B)$), we have that the initial bounds on the unknown water flows can be obtained using interval arithmetic. This is done by rearranging \eqref{eq:wdn_systemA} with respect to $q_i$, $i \in \mathcal{L}$, and applying the inverse of $f_i$. The initial bounds of the flow and head states are indicated by $Q^{(0)}_k$, and $H^{(0)}_k$, respectively.
	
	\subsubsection{Bounding linearization of the interval nonlinear terms} \label{sec:wdn_boundinglinearization}
	Let $m \geq 1$ denote the iteration index.
	This step encloses the uncertain nonlinear terms $\mbf{f}(\mbf{q}_k)$ by a convex set, for all\footnote{For notational simplicity, we omit the dependence on the iteration index $m$ of some variables, such as most of the lower and upper bounds of intervals.} $\mbf{q}_{k}\in Q_k^{(m-1)} = [\mbf{q}_{k}^\text{L}, \mbf{q}_{k}^\text{U}]$ and all the uncertain parameters $\mbf{r} \in R$, using bounding linearization \citep{Vrachimis2019}. 
	One way to achieve this is to enclose the image of each component of the nonlinear function ${f}_i(q_{i,k})$, for all $q_{i,k}\in [ q_{i,k}^\text{L}, q_{i,k}^\text{U} ]$ and $\mbf{r} \in R$, between a lower line $ {\bar{f}_{i}^\text{L}(q_{i,k})}$ and an upper line ${\bar{f}_{i}^\text{U}(q_{i,k})}$. These lines are defined as
	\begin{subequations}\label{eq:wdn_lines}
		\begin{align}
		&{\bar{f}_{i}^\text{L}(q_{i,k})}=\lambda^\text{L}_i q_{i,k}+\beta^\text{L}_i, \ \forall i\in\mathcal{L},\\
		&{\bar{f}_{i}^\text{U}(q_{i,k})}=\lambda^\text{U}_i q_{i,k}+\beta^\text{U}_i, \ \forall i\in\mathcal{L},
		\end{align}
	\end{subequations}
	with $\lambda^\text{L}_i$, $\lambda^\text{U}_i$, $\beta^\text{L}_i$, and $\beta^\text{U}_i$ computed as explained in \cite{Vrachimis2019}. The corresponding convex enclosure is defined component-wise by the linear inequalities
	\begin{equation}\label{eq:wdn_BoundingLinearization}
	{\bar{f}_{i}^\text{L}(q_{i,k})} \leq f_i(q_{i,k}) \leq  {\bar{f}_{i}^\text{U}(q_{i,k})}, \forall {q_{i,k}}\in\left[{q_{i,k}^\text{L},q_{i,k}^\text{U}}\right], i\in\mathcal{L}
	\end{equation}
	
	\subsubsection{Solution of the linear interval system of equations using linear programming}
	{The procedure described in Step 2 yields linear inequality constraints that allow one to} replace $ {f}_i({q_{i,k}})$ in \eqref{eq:wdn_systemA} with new variables 
	$ \zeta_i $, thus allowing the transformation of all the algebraic equations in \eqref{eq:wdn_system} into a set of linear inequalities.
	Let $Q_k^{(m-1)} = [\mbf{q}^\text{L}_k, \mbf{q}^\text{U}_k]$, $H_k^{(m-1)} = [\mbf{h}^\text{L}_k, \mbf{h}^\text{U}_k]$, $U_k = [\mbf{u}^\text{L}_k, \mbf{u}^\text{U}_k]$, $D_k = [\mbf{d}^\text{L}_k, \mbf{d}^\text{U}_k]$, $\bm{\Lambda}^\text{L} = \text{diag}(\bm{\lambda}^\text{L})$, and $\bm{\Lambda}^\text{U} = \text{diag}(\bm{\lambda}^\text{U})$, with $\bm{\lambda}^\text{L} = [ \lambda_1^\text{L} \cdots \lambda_{n_q}^\text{L}]^T$, and $\bm{\lambda}^\text{U} = [ \lambda_1^\text{U} \cdots \lambda_{n_q}^\text{U}]^T$.
	An enclosure of the state set satisfying \eqref{eq:wdn_systemA}--\eqref{eq:wdn_systemB}, can be obtained using the following linear inequalities:
	\begin{subequations}\label{eq:wdn_constraint}
		\begin{align}
		& \mbf{u}^\text{L}_k \leq \bm{\zeta} + \mbf{B} \mbf{h} \leq \mbf{u}^\text{U}_k, \label{eq:wdn_IHISEconstraint1}\\ 
		& \mbf{d}^\text{L}_k \leq \mbf{B}^T \mbf{q} \leq \mbf{d}^\text{U}_k, \label{eq:wdn_IHISEconstraint2}\\
		& \bm{\Lambda}^\text{L} \mbf{q} + \bm{\beta}^\text{L} \leq \bm{\zeta} \leq \bm{\Lambda}^\text{U} \mbf{q} + \bm{\beta}^\text{U}, \label{eq:wdn_IHISEconstraint3}\\
		& \mbf{q}^\text{L}_k \leq \mbf{q} \leq \mbf{q}^\text{U}_k, \label{eq:wdn_IHISEconstraint4}\\
		& \mbf{h}^\text{L}_k \leq \mbf{h} \leq \mbf{h}^\text{U}_k, \label{eq:wdn_IHISEconstraint5}
		\end{align}
	\end{subequations}
	where $\bm{\beta}^\text{L} = [ \beta_1^\text{L} \cdots \beta_{n_q}^\text{L}]^T$ and $\bm{\beta}^\text{U} = [ \beta_1^\text{U} \cdots \beta_{n_q}^\text{U}]^T$. Note that all the interval sets have been replaced by their respective lower and upper  bounds. Then, a total of $2(n_q+n_h)$ linear programs (LPs) are then formulated using \eqref{eq:wdn_IHISEconstraint1}--\eqref{eq:wdn_IHISEconstraint5}, with decision variables given by $[\mbf{q}^T \; \mbf{h}^T \; \bm{\zeta}^T]^T = [\mbf{x}^T \; \bm{\zeta}^T]^T$. 
	Each pair of the following LPs provide the lower and upper bounds of the $j$-th state variable $x_j$, with $j = 1,\ldots,n_h+n_q$:
	\begin{equation} \label{eq:wdn_IHISELPs}
	\left\{
	\begin{aligned}
	{x}^\text{L}_{j} = \underset{\mbf{x},\bm{\zeta}}{\min}& ~ x_{j} \\
	\text{s.t.} & ~ \eqref{eq:wdn_IHISEconstraint1}\text{--}\eqref{eq:wdn_IHISEconstraint5},
	\end{aligned}\right. \quad   \left\{
	\begin{aligned}
	{x}^\text{U}_{j} = \underset{\mbf{x},\bm{\zeta}}{\max}& ~ x_{j} \\
	\text{s.t.} & ~ \eqref{eq:wdn_IHISEconstraint1}\text{--}\eqref{eq:wdn_IHISEconstraint5}.
	\end{aligned} \right.
	\end{equation}

	\subsubsection{Iterative solution of the nonlinear system of equations}
	The procedure in Step 3 leads to an interval solution of \eqref{eq:wdn_system}, denoted by $\tilde{X}_k^{(m)}$, which encloses the exact solution set denoted by $X_k^*$.
	An iterative method is then used to find a tight $X_k^{(m)}$ with respect to $X_k^*$, resulting from $X_k^{(m)} = X_k^{(m-1)} \cap \tilde{X}_k^{(m)}$, with $X^{(0)} = Q^{(0)} \times H^{(0)}$. At each iteration $m$, the previous refined set $X_k^{(m-1)}$ is used in the bounding linearization procedure of Step 2 to reduce the convex set defined by the inequalities \eqref{eq:wdn_constraint}.
	Step 3 is then repeated to produce a smaller set $\tilde{X}_k^{(m)}$.
	The iterations of the IHISE terminate when the relative change in the obtained enclosure $X_k^{(m)}$, given by the metric $e_k^{(m)}$, is smaller than a chosen tolerance value $\varepsilon$, i.e., $e_k^{(m)}<\varepsilon$.
	This metric $e_k^{(m)}$ is given by
	\begin{equation} \label{eq:wdn_convergencetolerance}
	e_k^{(m)} \triangleq {\big| {\big( {\mbf{x}_k^{\text{U}\,(m)} - \mbf{x}_k^{\text{L}\,(m)}} \big) - \big( {\mbf{x}_k^{\text{U}\,(m - 1)} - \mbf{x}_k^{\text{L}\,(m - 1)}} \big)} \big|_1}.
	\end{equation}
	The overall procedure of the IHISE methodology is described in Algorithm \ref{alg:wdn_IHISE}.
	
	\begin{algorithm}[!htb]
		\caption{IHISE \citep{Vrachimis2019}} 
		\label{alg:wdn_IHISE}
		\begin{algorithmic}[1]	
			\State Assign initial bounds $H^{(0)}_k$ using physical constraints.
			\State Compute initial bounds $Q^{(0)}_k$ according to Step 1.
			\State Assign $X^{(0)}_k = Q^{(0)}_k \times H^{(0)}_k$, and $m = 1$.
			\State Perform bounding linearization of $\mbf{f}(\mbf{q}_k)$ for $\mbf{q}_k \in Q^{(m-1)}_k$.
			\State Solve the pair of LPs \eqref{eq:wdn_IHISELPs} for $j=1,\ldots,n_q+n_h$, and obtain $\tilde{X}^{(m)}_k$.
			\State Assign $X^{(m)}_k = X^{(m-1)}_k \cap \tilde{X}^{(m)}_k$.
			\State If $e^{(m)} < \varepsilon$, return $X^{(m)}_k$ and terminate. Otherwise, assign $m = m + 1$ and go to 4.
		\end{algorithmic}
		\normalsize
	\end{algorithm}
	
	Note that each iteration in Algorithm \ref{alg:wdn_IHISE} requires the solutions of $2(n_h + n_q)$ LPs. Assuming that the solution of each LP has the average complexity reported in \cite{Kelner2006}, i.e., it has average complexity of $O(N_d N_c^3)$, in which $N_d$ and $N_c$ are the number of decision variables and constraints, respectively, then the computational complexity of Algorithm \ref{alg:wdn_IHISE} is $O(\kappa_mn^5)$, where $n = n_h + n_q$, and $\kappa_m$ corresponds to the number of performed iterations.

	\subsection{Slope approximation of the nonlinear equalities}
	
	Consider the initial bounds $Q_k^{(0)}$ and $H_k^{(0)}$ computed as in Section \ref{sec:wdn_initialbounds}. The exact solution set of \eqref{eq:wdn_system} for all $\mbf{u}_k \in U_k$, $\mbf{d}_k \in D_k$, is the set of all $\mbf{q}_k \in Q_k^{(0)}$ and $\mbf{h}_k \in H_k^{(0)}$ such that
	\begin{equation} \label{eq:wdn_systemsolution}
	\begin{aligned}
	& \mbf{f}(\mbf{q}_k) + \mbf{B} \mbf{h}_k \in U_k, \\
	& \mbf{B}^T \mbf{q}_k \in D_k. 
	\end{aligned}
	\end{equation}
	
	The following lemma presents an approximation of the nonlinear equalities \eqref{eq:wdn_system} based on the bounding linearization procedure described in Section \ref{sec:wdn_boundinglinearization}. This is referred to as \textit{slope approximation}. Instead of the inequalities \eqref{eq:wdn_constraint}, Lemma \ref{lem:wdn_slope} approximates \eqref{eq:wdn_system} by set of linear algebraic equations.
	
	\begin{lemma} \rm \label{lem:wdn_slope}
		Let $\mbf{q}_k \in [\mbf{q}^\text{L}_k, \mbf{q}^\text{U}_k] \subset \intvalset^{n_q}$, $\mbf{h}_k \in \realset^{n_h}$, and $(\mbf{d}_k, \mbf{u}_k) \in D_k \times U_k$, such that the nonlinear algebraic equations \eqref{eq:wdn_system} hold. Then, there exists $\bm{\delta} \in B_\infty^{n_q}$, $\bm{\Delta} = \text{diag}(\bm{\delta})$, $\bm{\Lambda}_+ \in \realsetmat{n_q}{n_q}$, $\bm{\Lambda}_- \in \realsetmat{n_q}{n_q}$, $\bm{\beta}_+ \in \realset^{n_q}$, and $\bm{\beta}_- \in \realset^{n_q}$ such that
		\begin{equation} \label{eq:wdn_slopeapprox}
		\left\{
		\begin{aligned}
		& \bm{\Lambda}_+ \mbf{q}_k + \mbf{B} \mbf{h}_k = \mbf{u}_k - \bm{\beta}_+ - \bm{\Delta} \left( \bm{\beta}_- + \bm{\Lambda}_- \mbf{q}_k\right), \\
		& \mbf{B}^T \mbf{q}_k = \mbf{d}_k.
		\end{aligned} \right.
		\end{equation}
	\end{lemma}
	
	\proof	
	For a given interval bound $[\mbf{q}_k^\text{L}, \mbf{q}_k^\text{U}]$, there exist real vectors $\bm{\lambda}^\text{L},\bm{\lambda}^\text{U},\bm{\beta}^\text{L},\bm{\beta}^\text{U} \in \realset^{n_q}$ such that $\lambda_i^\text{L}q_i + \beta_i^\text{L} \leq f_i(q_i) \leq \lambda_i^\text{U} q_i + \beta_i^\text{U}$ for all $i = 1,2,\ldots,n_q$, and $\mbf{q}_k \in [\mbf{q}^\text{L}_k, \mbf{q}^\text{U}_k]$ (see Section \ref{sec:wdn_IHISE}). Therefore,
	\begin{align*}
	& \lambda_i^\text{L}q_{i,k} + \beta_i^\text{L} \leq f_i(q_{i,k}) \leq \lambda_i^\text{U} q_{i,k} + \beta_i^\text{U} \\ & \iff f_i(q_{i,k}) \in [\lambda_i^\text{L}q_{i,k} + \beta_i^\text{L},~ \lambda_i^\text{U} q_{i,k} + \beta_i^\text{U}] \\ 
	& \iff f_i(q_{i,k}) \in \left(\frac{1}{2} \left(\beta_i^\text{U} + \beta_i^\text{L}\right) + \frac{1}{2} \left(\lambda_i^\text{U} + \lambda_i^\text{L}\right) q_{i,k}\right) \\ & \quad\quad\quad\quad\quad \oplus \left(\frac{1}{2} \left(\beta_i^\text{U} - \beta_i^\text{L}\right) + \frac{1}{2} \left(\lambda_i^\text{U} - \lambda_i^\text{L} \right) q_{i,k} \right) B_\infty.
	\end{align*}
	Let $\bm{\Lambda}_{+} = (1/2)\text{diag}(\bm{\lambda}^\text{U} + \bm{\lambda}^\text{L})$, $\bm{\Lambda}_{-} = (1/2)\text{diag}(\bm{\lambda}^\text{U} - \bm{\lambda}^\text{L})$, $\bm{\beta}_{+} = (1/2)(\bm{\beta}^\text{U} + \bm{\beta}^\text{L})$, $\bm{\beta}_{-} = (1/2)(\bm{\beta}^\text{U} - \bm{\beta}^\text{L})$. Then, there exists $\bm{\delta} \in B_\infty^{n_q}$ such that
	\begin{equation} \label{eq:wdn_slopef}
	\begin{aligned} 
	\mbf{f}(\mbf{q}_k) = \bm{\Lambda}_+ \mbf{q}_k + \bm{\beta}_+ + \bm{\Delta} \left( \bm{\beta}_- + \bm{\Lambda}_- \mbf{q}_k\right),
	\end{aligned}
	\end{equation}
	where $\bm{\Delta} = \text{diag}(\bm{\delta})$. Substituting \eqref{eq:wdn_slopef} in the nonlinear equations \eqref{eq:wdn_system} results in the system of equalities \eqref{eq:wdn_slopeapprox}. \qed
	
	\subsection{Iterative Slope Approximation and Rescaling}
	
	In this section, we propose an iterative method to find interval bounds of the solution set of \eqref{eq:wdn_slopeapprox} with modest computational complexity, without requiring to solve linear programs as in Algorithm \ref{alg:wdn_IHISE}. This method in based on iteratively enclosing the solution set of the slope approximation \eqref{eq:wdn_slopeapprox} by a constrained box in CG-rep, followed by the use of the rescaling procedure described in Chapter \ref{cha:preliminaries}, to obtain sharp interval bounds of the solution. The resulting method is referred to as Iterative Slope Approximation and Rescaling. %
	
	Let $s \geq 1$ denote the iteration index, and $\tilde{\mbf{u}}_k \triangleq \mbf{u}_k - \bm{\beta}_+ - \bm{\Delta} \left( \bm{\beta}_- + \bm{\Lambda}_- \mbf{q}_{k}\right)$. Moreover, let $\tilde{U}_k^{(s)} \triangleq [\tilde{\mbf{u}}_k^\text{L},\tilde{\mbf{u}}_k^\text{U}]$ be the interval obtained by evaluating the right-hand-side of $\tilde{\mbf{u}}_k$ for all $\mbf{u}_k \in [\mbf{u}_k^\text{L},\mbf{u}_k^\text{U}]$, $\mbf{q}_k \in [\mbf{q}^\text{L}_k,\mbf{q}_k^\text{U}]$, and $\Delta_{ii} \in B_\infty$, using interval arithmetic. By defining the augmented vector $\tilde{\mbf{p}}_k = (\mbf{q}_k, \mbf{h}_k, \mbf{\tilde{u}}_k, \mbf{d}_k) \in \tilde{P}_k^{(s)} \in \intvalset^{2n_q + 2n_h}$, with
	\begin{align} 
	\tilde{P}_k^{(s)} & = Q_k^{(s-1)} \times H_k^{(s-1)} \times \tilde{U}_k^{(s)} \times D_k, \label{eq:wdn_Ptilde} \\
	\tilde{\mbf{B}} & = \begin{bmatrix} \bm{\Lambda}_+ & \mbf{B} & - \eyenoarg & \bm{0} \\ \mbf{B}^T & \bm{0} & \bm{0} & -\eyenoarg \end{bmatrix}, \nonumber
	\end{align}
	the equalities \eqref{eq:wdn_slopeapprox} can be rewritten in the form $\tilde{\mbf{B}} \tilde{\mbf{p}}_k = \bm{0}$. In addition, the relation $\tilde{\mbf{p}}_k \in \tilde{P}_k^{(s)}$ implies that there must exist a vector $\bar{\mbf{p}} \in B_\infty^{2n_q + 2n_h}$ satisfying
	\begin{equation} \label{eq:wdn_ptilde}
	\tilde{\mbf{p}}_k = \text{mid}(\tilde{P}_k^{(s)}) + \text{diag}(\text{rad}(\tilde{P}_k^{(s)})) \bar{\mbf{p}}.
	\end{equation}
	Then, replacing \eqref{eq:wdn_ptilde} into the equalities $\tilde{\mbf{B}} \tilde{\mbf{p}}_k = \bm{0}$ results in $\tilde{\mbf{B}} (\text{mid}(\tilde{P}_k^{(s)}) + \text{diag}(\text{rad}(\tilde{P}_k^{(s)}) \bar{\mbf{p}})) = \bm{0}$, which, in turn, implies that $\tilde{\mbf{B}}(\text{diag}(\text{rad}(\tilde{P}_k^{(s)}))) \bar{\mbf{p}} = - \tilde{\mbf{B}} (\text{mid}(\tilde{P}_k^{(s)}))$. Therefore, by defining the variables $\tilde{\mbf{c}}_k = \text{mid}(\tilde{P}_k^{(s)})$, $\tilde{\mbf{G}}_k = \text{diag}(\text{rad}(\tilde{P}_k^{(s)}))$, $\tilde{\mbf{A}} = \tilde{\mbf{B}}(\text{diag}(\text{rad}(\tilde{P}_k^{(s)}))) \bar{\mbf{p}}$, and $\tilde{\mbf{b}}_k = - \tilde{\mbf{B}} (\text{mid}(\tilde{P}_k^{(s)}))$, we have from \eqref{eq:wdn_ptilde} that the following relation holds:
	\begin{equation} \label{eq:wdn_ptildeCZ}
	\tilde{\mbf{p}}_k \in \{ \tilde{\mbf{c}}_k + \tilde{\mbf{G}}_k \bar{\mbf{p}} : \|\bar{\mbf{p}}\|_\infty \leq 1, \tilde{\mbf{A}}_k \bar{\mbf{p}} = \tilde{\mbf{b}}_k \},
	\end{equation}
	in which the right-hand side is a constrained box in CG-rep (note that $\tilde{\mbf{G}}_k$ is diagonal). The rescaling procedure described by Method \ref{meth:rescaling} in Chapter \ref{cha:preliminaries} is then applied to the right-hand side of \eqref{eq:wdn_ptildeCZ} to obtain an approximated interval hull of the solution of $\tilde{\mbf{A}}_k \bar{\mbf{p}} = \tilde{\mbf{b}}_k$ and, consequently, obtain a tight interval bound of $\tilde{\mbf{p}}_k$. Sharp bounds can be obtained using the iterative procedure described in Algorithm \ref{alg:pre_rescaling} (Section \ref{sec:complexityreduction}), whose complexity, in the case of \eqref{eq:wdn_ptildeCZ}, is $O(n^3)$ (assuming that every variable is proportional to the state dimension $n = n_q + n_h$). In addition, further accurate results can be obtained if the constraints $\tilde{\mbf{A}}_k \bar{\mbf{p}} = \tilde{\mbf{b}}_k$ are first preconditioned using Gauss-Jordan elimination.

	The proposed ISAR algorithm involving the rescaling of  \eqref{eq:wdn_ptildeCZ} is illustrated in Algorithm \ref{alg:wdn_ISAR}. To obtain an interval $X^{(s)}_k$ to be used in the subsequent iteration, we neglect the equality constraints of the constrained box obtained in the rescaling procedure, which, in this case, is equivalent to computing its interval hull. Moreover, we use the metric $\varepsilon^{(s)}$ given by \eqref{eq:wdn_convergencetolerance}, with $m \triangleq s$, as stop criterion. Since Algorithm \ref{alg:pre_rescaling} has complexity $O(n^3)$, we have that ISAR has complexity $O(\kappa_s n^3)$, with $\kappa_s$ being the number of performed iterations by Algorithm \ref{alg:wdn_ISAR}, while IHISE has average complexity $O(\kappa_m n^5)$ (see Section \ref{sec:wdn_IHISE}).%

	\begin{algorithm}[!htb]
		\caption{ISAR} 
		\label{alg:wdn_ISAR}
		\begin{algorithmic}[1]
			\State Assign initial bounds $H^{(0)}_k$ using physical constraints.
			\State Compute initial bounds $Q^{(0)}_k$ according to Section \ref{sec:wdn_initialbounds}.
			\State Assign $X^{(0)}_k = Q^{(0)}_k \times H^{(0)}_k$, $s = 1$.
			\State Compute the slope approximation \eqref{eq:wdn_slopeapprox} for $\mbf{q}_k \in Q^{(s-1)}_k$.
			\State Compute $\tilde{P}_k^{(s)}$ as in \eqref{eq:wdn_Ptilde}.
			\State Rescale the constrained box in \eqref{eq:wdn_ptildeCZ} using Method \ref{meth:rescaling} and Algorithm \ref{alg:pre_rescaling}.
			\State Retrieve the interval $X^{(s)}_k$ from the resulting CG-rep by neglecting its equality constraints.
			\State If $e^{(s)} < \varepsilon$, return $X^{(s)}_k$ and terminate. Otherwise, assign $s = s + 1$ and go to 4.
		\end{algorithmic}
		\normalsize
	\end{algorithm}
	
	\begin{remark} \rm
		Step 7 in Algorithm \ref{alg:wdn_ISAR} is equivalent to computing the interval hull of the set obtained by Method \ref{meth:rescaling}.
	\end{remark}	
	
	\begin{remark} \rm
		The interval provided by Algorithm \ref{alg:wdn_ISAR} may not necessarily correspond to the interval hull of the solution of the linear algebraic equations \eqref{eq:wdn_slopeapprox}. However, these are still accurate enclosures of the solution set, as illustrated in Section \ref{sec:wdn_numericalexperiments}.
	\end{remark}

	\subsection{Final enclosure with constrained zonotopes} \label{sec:wdn_czenclosure}

	The interval bounds obtained by ISAR can be tight, but due to the wrapping effect inherent to intervals, these are not able to capture the dependencies between all the state variables. Therefore, we consider a final step in the state estimation method using constrained zonotopes to obtain a more accurate enclosure that takes into account these dependencies. The complete algorithm is denoted by ISAR-CZ. For the sake of completeness, we also consider applying this final step to the IHISE, with the corresponding algorithm denoted by IHISE-CZ.
	
	Consider the slope approximation \eqref{eq:wdn_slopeapprox}, the augmented vector $\mbf{x}_k = (\mbf{q}_k,\mbf{h}_k)$ and the sets satisfying $(\mbf{q}_k,\mbf{h}_k) \in Q_k^{(\kappa_l)} \times H_k^{(\kappa_l)} = X_k^{(\kappa_l)}$, where $\kappa_l \triangleq \kappa_m$ or $\kappa_l \triangleq \kappa_s$ if the enclosure was obtained by IHISE or by ISAR, respectively. Let $\tilde{\mbf{u}}_k = \mbf{u}_k - \bm{\beta}_+ - \bm{\Delta} \left( \bm{\beta}_-+\bm{\Lambda}_- \mbf{q}_k \right)$, with $\mbf{u}_k \in U_k$.  Then, it is true that $\tilde{\mbf{u}}_k \in \check{U}_k$, where $\check{U}_k$ is a CZ given by
	\begin{equation*}
	\check{U}_k \triangleq U_k \oplus \left( - \bm{\beta}_+ \right) \oplus \triangleleft\big(\mbf{D}, \bm{\beta}_- \oplus \bm{\Lambda}_- Q_k^{(\kappa_l)} \big),
	\end{equation*}
	with $\mbf{D}$ being a diagonal interval matrix satisfying $D_{ii} = B_\infty^{n_q}$ for $i=1,2,\ldots,n_q$, such that $\bm{\Delta} \in \mbf{D}$, and the operator $\triangleleft(\cdot, \cdot)$ is the CZ-inclusion operator defined in Theorem \ref{thm:ndyn_czinclusion}. The final enclosure used for state estimation is a CZ outer approximation of the solution set of \eqref{eq:wdn_slopeapprox}, and is denoted by $\hat{X}_k$. For $\mbf{u}_k \times \mbf{d}_k \in U_k \times D_k$, this CZ is obtained by the generalized intersection
	\begin{equation} \label{eq:wdn_finalcz}
	\hat{X}_k = \{\mbf{x} \in X_k^{(\kappa_l)} : \mbf{F} \mbf{x} \in \tilde{U}_k \times D_k\} =X_k^{(\kappa_l)} \cap_\mbf{F} (\check{U}_k \times D_k),
	\end{equation}
	where $\mbf{F} = \begin{bmatrix} \bm{\Lambda}_+ & \mbf{B} \\ \mbf{B}^T & \bm{0} \end{bmatrix}$.
	
	In this chapter, we take advantage of the state coupling resulting from the nonlinear equality constraints \eqref{eq:wdn_system}, which are captured by the constrained zonotope $\hat{X}_k$ given by \eqref{eq:wdn_finalcz}. Moreover, since $Q_k^{(\kappa_l)}$, $H_k^{(\kappa_l)}$, $D_k$, and $U_k$, are all intervals (hence have $n_q$, $n_h$, $n_h$, and $n_q$ generators, respectively), the resulting enclosure $\hat{X}_k$ has $3n_q + 2n_h$ generators and $n_q+n_h$ constraints. Finally, to avoid numerical issues in $\hat{X}_k$ when one or both the sets $U_k$ and $D_k$ are degenerate, we add tolerance intervals $\{\tau_u,0\}$ and $\{\tau_d,0\}$ to each of the corresponding degenerate dimensions in $U_k$ and $D_k$, respectively.

	\begin{remark} \rm \label{rem:wdn_zonotopes}
		Although all the sets $Q_k^{(\kappa_l)}$, $H_k^{(\kappa_l)}$, $U_k$ and $D_k$, are intervals (and therefore zonotopes), constrained zonotopes are required to compute \eqref{eq:wdn_finalcz} because the generalized intersection \eqref{eq:pre_czintersection} is conservative and difficult to enclose using zonotopes \citep{Bravo2006}.
	\end{remark}
	
	\begin{remark} \rm \label{rem:wdn_measurement}
		In this chapter, the state estimation is assumed to be performed offline, using only the information obtained from physical constraints and historical data of the WDN. However, if measurements of $\mbf{q}_k$ and $\mbf{h}_k$ are available, these can be used to reduce the conservativeness of the initial bounds $Q^{(0)}_k$ and $H^{(0)}_k$ by intersecting them with the measurement bounds. This procedure can be done prior to computing the IHISE-CZ and ISAR-CZ, and will be explored in future work.
	\end{remark}
	
	\section{Passive leak detection in WDNs} \label{sec:wdn_leakdetection}
	
	Consider a WDN in the presence of leakages, with $n_q$ links and $n_h$ nodes, described by the nonlinear algebraic equation \eqref{eq:wdn_systemA}, and
	\begin{equation}
	\mbf{B}^T \mbf{q}_k = \mbf{d}_k + \bm{\ell}_k, \label{eq:wdn_systemfaultB}
	\end{equation}
	where $\mbf{q}_k \in \realset^{n_q}$ are the water flows at the links, $\mbf{B} \in \realsetmat{n_q}{n_h}$ is the incidence matrix, and $\mbf{d}_k \in \realset^{n_h}$ is the water demand by the consumers at the nodes. In \eqref{eq:wdn_systemfaultB}, $\bm{\ell}_k$ denotes the presence of pressure-dependent leakages modeled in this chapter at the nodes \citep{Greyvenstein2007a}. The purpose of leak detection is to check whether $\ell_{i,k} \neq 0$ for at least one node $i$.

	Consider the output equation
	\begin{equation}\label{eq:wdn_measurements}
	\mbf{y}_k = \mbf{C} \mbf{x}_k + \mbf{v}_k,
	\end{equation}
	where $\mbf{v}_k \in \realset^{n_y}$ is the measurement uncertainty, and $Y_k \triangleq [\mbf{y}_k^\text{L}, \mbf{y}_k^\text{U}]$ is the obtained measurement interval, with $\text{mid}(Y_k) = \mbf{y}_k$. 
	In this chapter, this condition can be verified conservatively\footnote{Since $\hat{X}_k$ is computed conservatively, \eqref{eq:wdn_faultdetection} $\implies \ell_{i,k} \neq 0$ for at least one $i$, but the converse is not true in general.} by checking if the following relation holds:
	\begin{equation} \label{eq:wdn_faultdetection}
	\mbf{C} \hat{X}_k \cap Y_k = \emptyset,
	\end{equation}
	in which $\hat{X}_k$ is the set obtained by state estimation IHISE-CZ and ISAR-CZ without considering leakage. In the case of IHISE and ISAR, $\hat{X}_k$ is the interval provided by Algorithms \ref{alg:wdn_IHISE} and \ref{alg:wdn_ISAR}, respectively.
	
	Despite being conservative, the advantage in verifying if \eqref{eq:wdn_faultdetection} holds is that no false positives can occur during leakage detection, given that the bounds on the uncertain parameters are valid, i.e., $\mbf{r} \in R$, $\mbf{u}_k \in U_k$ and $\mbf{d}_k \in D_k$, for all $k \geq 0$. 
	The resulting constrained zonotope from the left-hand-side in \eqref{eq:wdn_faultdetection} can be computed using \eqref{eq:pre_czlimage} and \eqref{eq:pre_czintersection}. Property \ref{prop:pre_czisemptyinside} can be used to verify if this constrained zonotope is empty by solving one LP.

	\section{Numerical experiments} \label{sec:wdn_numericalexperiments}
	
	In this section, we show the efficiency and accuracy of the proposed CZ based methods (IHISE-CZ and ISAR-CZ) for state estimation of WDNs. In addition, we show how these methods can be employed for leak detection (we consider scenarios where the leakage is present for the entire simulation) and compare their performance with the interval-based methods IHISE and ISAR. For these latter, the convergence tolerance \eqref{eq:wdn_convergencetolerance} was set to $\varepsilon=1$.
	
	\subsection{Two-pipe example}
	\begin{figure}[!tb]	
		\centering
		\includegraphics[width=0.6\columnwidth]{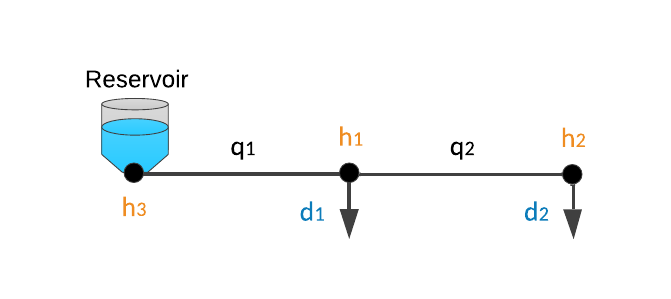}
		\caption{The two-pipe network example.}
		\label{fig:wdn_TwoPipeNet}
	\end{figure}
	
	In this first example we consider a network composed of 3 nodes and 2 pipes (see Figure \ref{fig:wdn_TwoPipeNet}). A 24-hour simulation was performed with EPANET \citep{Rossman2000epanet}, using conditions (water demands and reservoir levels) that vary every half of an hour.
	We consider that the physical parameters $\mbf{r} \in R$ have uncertainty of $5\%$, the demands $\mbf{d}_k \in D_k$ are measured with uncertainty of $5\%$, and the reservoir level $\mbf{u}_k$ is precisely known. The tolerance values adopted in Section \ref{sec:wdn_czenclosure} are set to $\tau_u = 1.0 {\times} 10^{-4}$ m and $\tau_d = 1.0 {\times} 10^{-4}$ m$^3$/h.
	
\begin{figure}[!tb]
		\centering{
			\def\svgwidth{0.6\columnwidth}
			{\scriptsize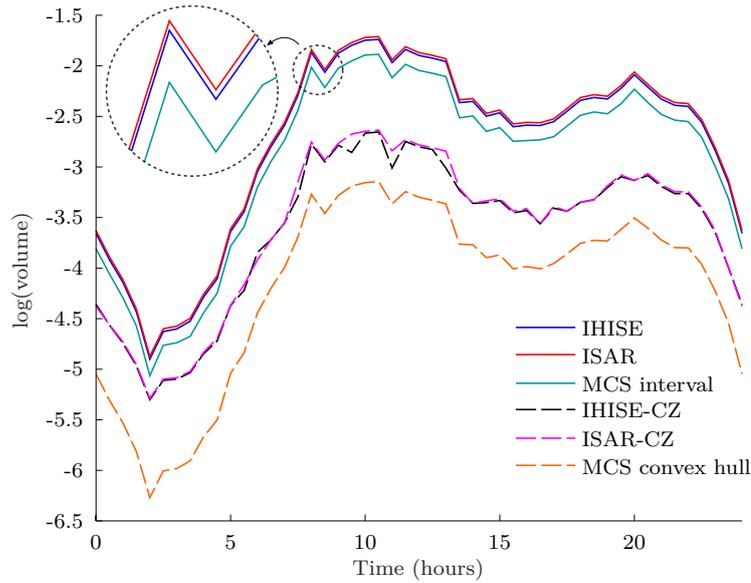}
			\caption{The volumes of the sets $\hat{X}_k$ obtained using IHISE, IHISE-CZ, ISAR, ISAR-CZ, and also by the interval hull and convex hull of the trajectories resulting from 1000 Monte-Carlo simulations (MCS) using EPANET for the two-pipe network.}\label{fig:wdn_volume}}
	\end{figure}

	Figure \ref{fig:wdn_volume} shows the volumes of the sets\footnote{For the sake of simplicity, in this section we use the slight abuse of notation $\hat{X}_k \triangleq X_k^{(\kappa_l)}$ for the sets obtained using IHISE and ISAR.} $\hat{X}_k$, obtained using IHISE, ISAR, IHISE-CZ, and ISAR-CZ, for time $k \in [0,24]$ in hours, in intervals of 30 minutes each. Figure \ref{fig:wdn_volume} also reports the volumes of `MCS interval' and `MCS convex hull'. The former provides an interval approximation of the solution of \eqref{eq:wdn_system} obtained performing 1000 MCS, considering parameters $\mbf{r}$ and demands $\mbf{d}_k$ drawn randomly from their corresponding uncertainty bounds. The latter is obtained by computing the convex hull of the same 1000 trajectories. Note that both results, which are based on a finite number of samples, represent only an inner approximation of the solution of \eqref{eq:wdn_system} but are useful as a reference for the set-based state estimation methods.
	
	Table \ref{tab:wdn_twopipetable} shows the execution times of IHISE, ISAR, IHISE-CZ, and ISAR-CZ, averaged over the time horizon $k \in [0,24]$, together with the number of iterations performed by the IHISE and ISAR for the given convergence tolerance $\varepsilon$. Note that the average execution time of ISAR is only $48.6\%$ of the IHISE, even if a higher number of iterations is required. 
	As expected, the volumes of the intervals/CZs provided by ISAR/ISAR-CZ are slightly bigger than the ones provided by IHISE/IHISE-CZ. Nevertheless, as shown in Figure \ref{fig:wdn_volume}, this increase is reasonable, demonstrating that the drawback of using ISAR/ISAR-CZ is mild while the advantage in terms of execution time is significant.
	
\begin{figure}[!tb]
		\centering{
			\def\svgwidth{0.6\columnwidth}
			{\scriptsize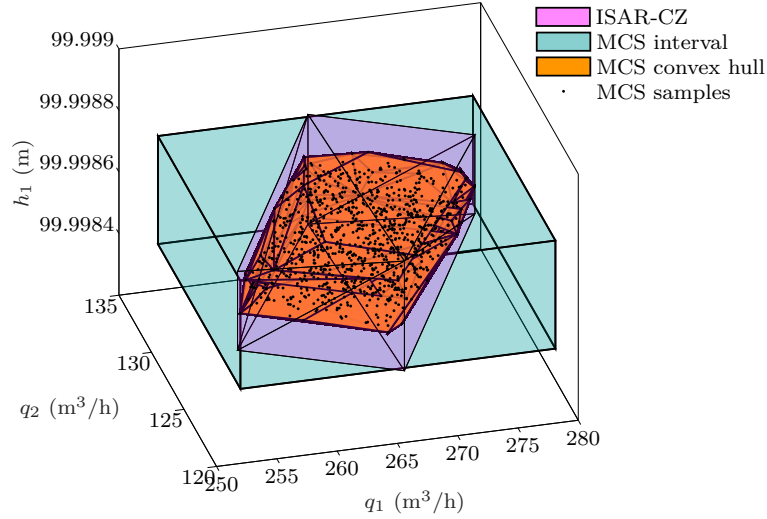}
			\caption{The projections of $\hat{X}_0$ onto $(q_1,q_2,h_1)$, obtained using ISAR-CZ, the intervals computed component-wise by maximum and minimum values obtained from 1000 MCS, as well as the convex hull and samples of the corresponding trajectories.}\label{fig:wdn_MCS}}
	\end{figure}

	In addition, the volumes of both the outer convex approximations {given} by the CZs in IHISE-CZ and ISAR-CZ are smaller than the interval approximations provided by IHISE and ISAR (Figure \ref{fig:wdn_volume}), illustrating that both IHISE-CZ and ISAR-CZ can be significantly less conservative.
	Note that the volumes of the CZs are smaller even than MCS interval, demonstrating that CZs are indeed capable of capturing the existing coupling between the state variables. 
	This effect is highlighted in Figure \ref{fig:wdn_MCS}, which shows the projections of $\hat{X}_0$ onto $(q_1,q_2,h_1)$ obtained using ISAR-CZ, the sets corresponding to MCS interval and MCS convex hull, and also the MCS trajectories.

	\begin{table}[!htb]
		\scriptsize
		\centering
		\caption{Average execution times and number of iterations per time step in the two-pipe example.}
		\begin{tabular}{c c c } \hline
			& Average execution times (seconds)  & Number of iterations (\#)  \\ \hline
			IHISE & $0.0243$ & $2$ \\
			ISAR & $0.0118$ & $3$ \\
			IHISE-CZ & $0.0286$ & $2$ \\
			ISAR-CZ & $0.0167$ & $3$  \\
			\hline
		\end{tabular} \normalsize
		\label{tab:wdn_twopipetable}
	\end{table}	
	
	In the following, we consider a scenario with the presence of a leakage at node 1.
	For illustration purposes, it is considered that all the water flows and hydraulic heads are measured, i.e., $\mbf{C}$ in \eqref{eq:wdn_measurements} is the identity matrix.
	Specifically, the flows at pipes 1 and 2 and the heads at nodes 1 and 2 are all measured with uncertainty of $2\%$, i.e., the measurement uncertainty in \eqref{eq:wdn_measurements} is bounded by $\mbf{v}_k \in V_k$, with $V_{k,i} = [-2\%|y_{k,i}|, 2\%|y_{k,i}|]$.
	
	To generate a leakage at node $i$, this leakage is modelled by $\ell_{i,k} = \theta_i (h_{i,k} - \eta_{i})^{\phi}$, 
		where $\theta_i$ and $\phi$ denote the leak emitter coefficient and exponent respectively \citep{Greyvenstein2007a}, which are completely unknown to the fault detection methodologies. The leakage emitter exponent for all the leakages is given by $\phi = 0.5$.
		The emitter coefficient $\theta_i$ for a leakage at node $i$ is given based on the average system inflow of the WDN (i.e., the sum of all nominal demands $d_{i,k}$ averaged over a given time horizon $k \in [0,N]$), denoted by $\rho$. Then, to generate $\ell_i \neq 0$, the emitter coefficient $\theta_i$ is set to
	\begin{equation*}
	\theta_i = \frac{\alpha \rho}{(\bar{h}_{i} - \eta_{i})^{\phi}}
	\end{equation*}
	where $\bar{h}_{i}$ is the nominal head of node $i$ averaged over $k \in [0,N]$, and $\alpha \in [0, 100]\%$ is a parameter defining the magnitude of the leakage in comparison to the average total inflow $\rho$, referred as the leak magnitude.
	
\begin{figure}[!tb]
		\centering{
			\def\svgwidth{0.6\columnwidth}
			{\scriptsize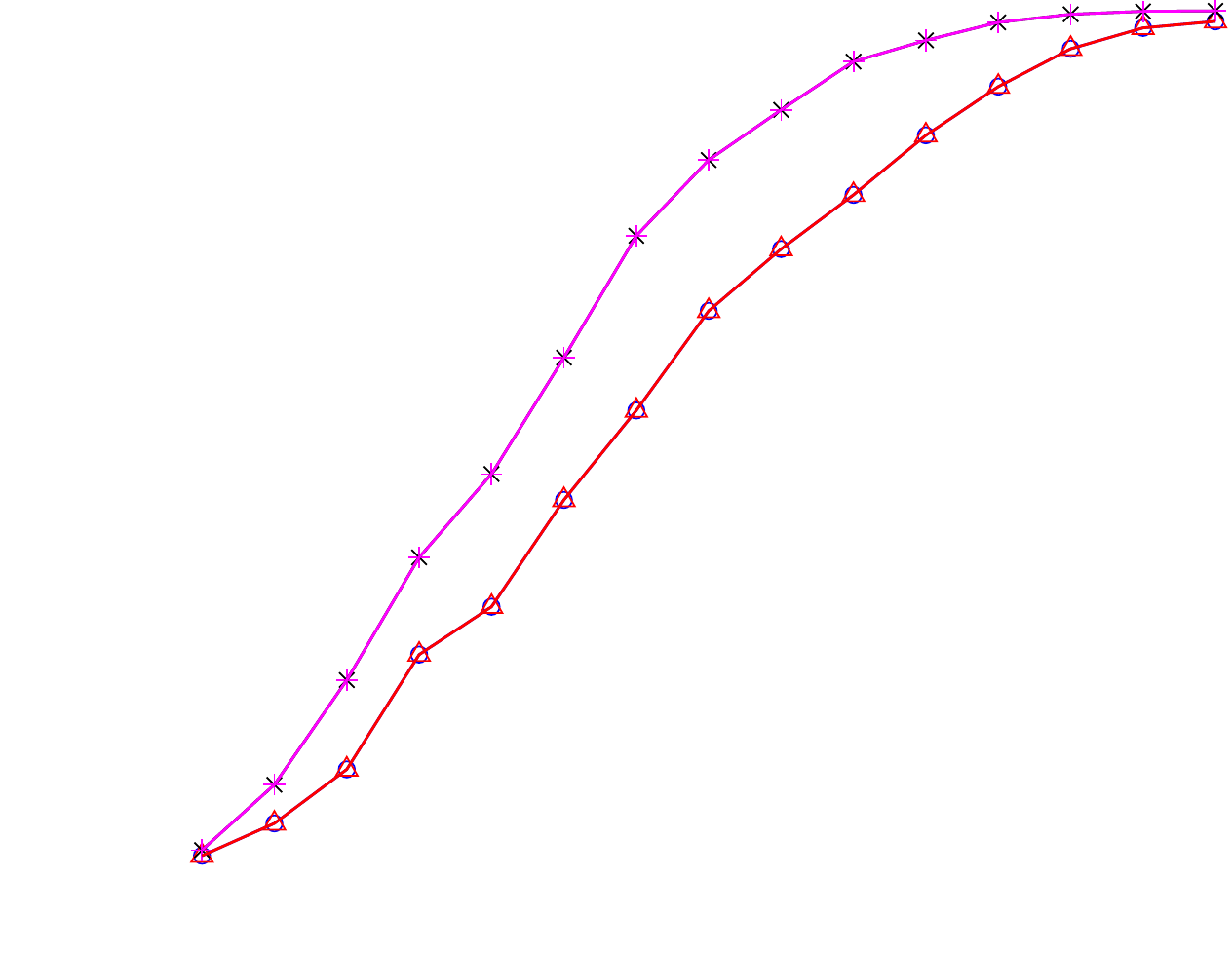}
			\caption{Leak detection rate of the IHISE, ISAR, IHISE-CZ, and ISAR-CZ for the two-pipe network (leak at node 1), considering different leak magnitudes, averaged over 100 random simulations and the time interval $k \in [0, 24]$.}\label{fig:wdn_leaktwopipemagnitude}}
	\end{figure}

\begin{figure}[!tb]
		\centering{
			\def\svgwidth{0.6\columnwidth}
			{\scriptsize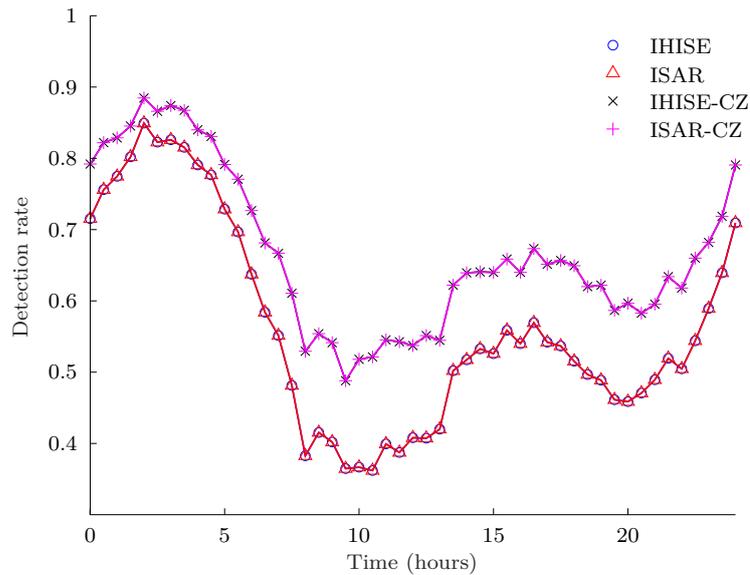}
			\caption{Leak detection rate of the IHISE, ISAR, IHISE-CZ, and ISAR-CZ for the two-pipe network (leak at node 1) over time, averaged over 100 random simulations and the leak magnitudes in the interval $[1, 15]\%$.}\label{fig:wdn_leaktwopipetime}}
	\end{figure}

\begin{figure}[!tb]
		\centering{
			\def\svgwidth{0.62\columnwidth}
			{\scriptsize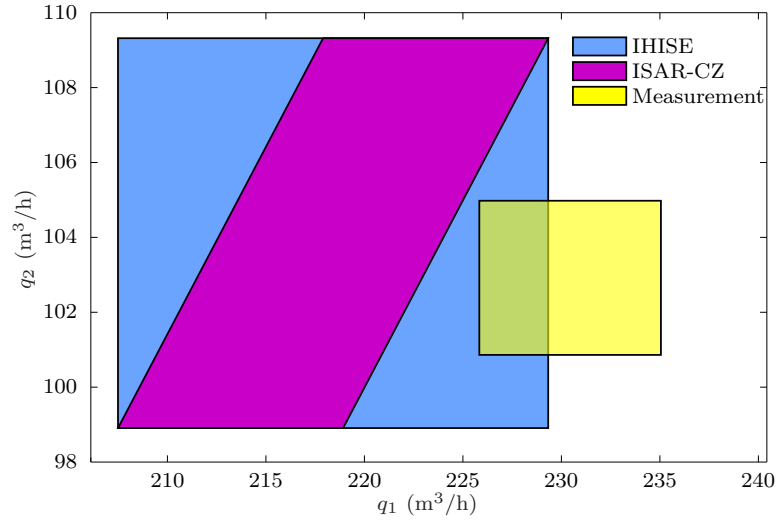}		        
			\caption{The projection onto $\mbf{q}_k \in \realset^2$ of the sets $\hat{X}_k$ obtained using IHISE and ISAR-CZ, together with the measured flow interval for $\mbf{q}_k$ at time $k = 2$ in one of the 100 random simulations considering leakage at node 1 of $4\%$ of the average nominal inflow in the two-pipe example.}\label{fig:wdn_leak2d}}
	\end{figure}

	Figures \ref{fig:wdn_leaktwopipemagnitude} and \ref{fig:wdn_leaktwopipetime} show the leak detection rate (a successful individual detection is denoted by 1, while 0 corresponds to the absence of detection), averaged over 100 MCS. The detection rates shown in Figures \ref{fig:wdn_leaktwopipemagnitude} and \ref{fig:wdn_leaktwopipetime} are further averaged over the leak magnitude range $\alpha = 1, 2, \ldots, 15\%$, and the time interval $k = 0,1,\ldots,24$ hours, respectively. In this example, IHISE-CZ and IHISE provided exactly the same detection rates of ISAR-CZ and ISAR, respectively. Moreover, IHISE-CZ and ISAR-CZ provide up to about $20\%$ more detections than IHISE and ISAR, demonstrating the improvements achieved in leak detection when using a set representation capable of taking into account the dependencies between water flows and hydraulic heads. This feature is highlighted in Figure \ref{fig:wdn_leak2d}, which shows the projection of the sets $\hat{X}_k$ provided by IHISE, ISAR-CZ, onto the water flows $(q_{1,k},q_{2,k})$, as well as the corresponding projection of the measurement set $Y_k$, for time $k = 2$, in one of the 100 MCS considering leakage magnitude of $4\%$. This illustrates a case in which the leak has been successfully detected by the ISAR-CZ but not by the IHISE, which can be verified by the fact that the intersection of $Y_k$ with the CZ provided by ISAR-CZ is empty, while the intersection with the interval provided by IHISE is not empty.
	
	\subsection{Transport network}
	
	\begin{figure}[!tb]	
		\centering
		\includegraphics[width=0.6\textwidth]{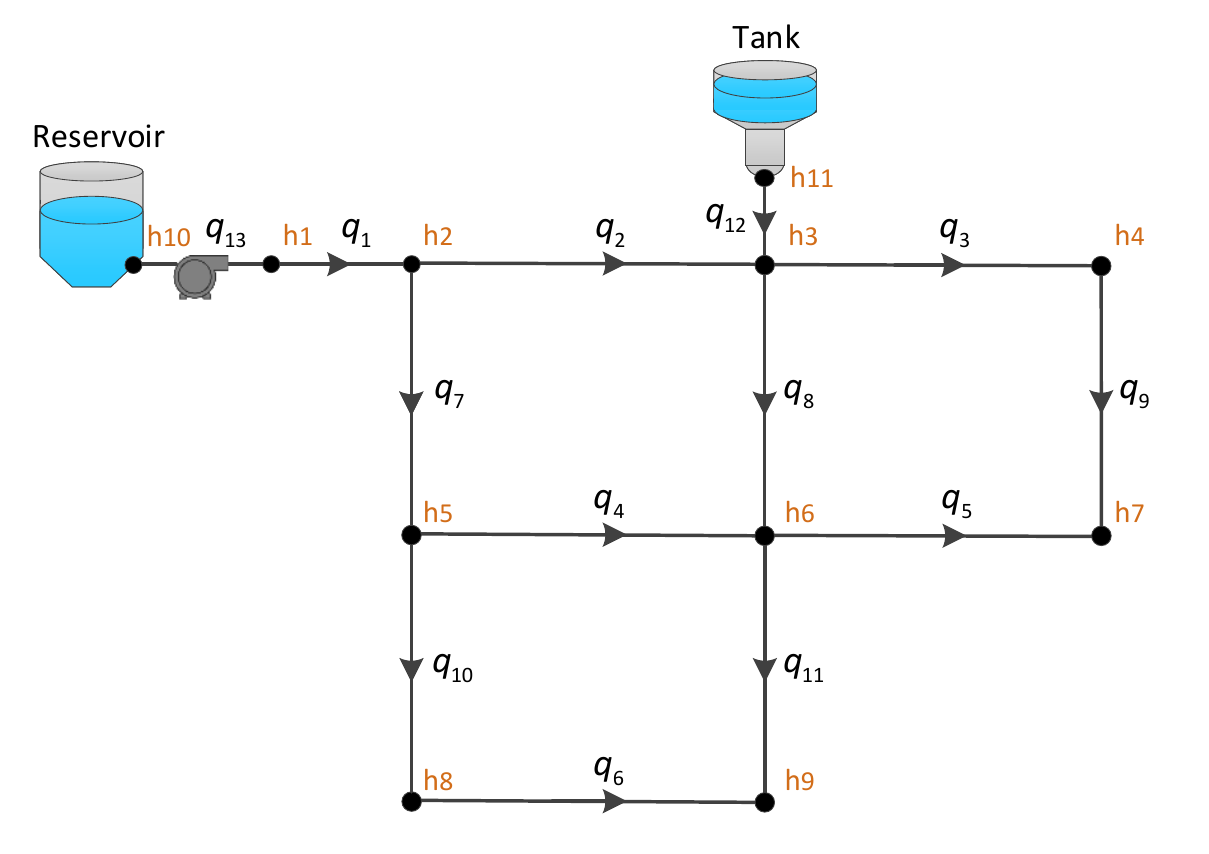}
		\caption{The benchmark network ``Net1'' used here as an example of a transport network.}
		\label{fig:wdn_Net1}
	\end{figure}
	The second example considers the benchmark network `Net1', which is provided by EPANET, and is illustrated in Figure \ref{fig:wdn_Net1}.
	This network is composed of 9 nodes and 13 pipes, and it is used here as an example of a real transport network. 
	In the following, we consider that pipe parameters $\mbf{r} \in R$ have uncertainty of $5\%$, the water demands at nodes $\mbf{d}_k \in D_k$, which are inflows to DMAs, are measured with uncertainty of $5\%$, and the reservoir and tank levels $\mbf{u}_k$ are precisely known. 
	The tolerance values adopted in Section \ref{sec:wdn_czenclosure} are set to $\tau_u = 0.1$ m and $\tau_d = 0.1$ m$^3$/h. 
	
\begin{figure}[!tb]
		\centering{
			\def\svgwidth{0.6\columnwidth}
			{\scriptsize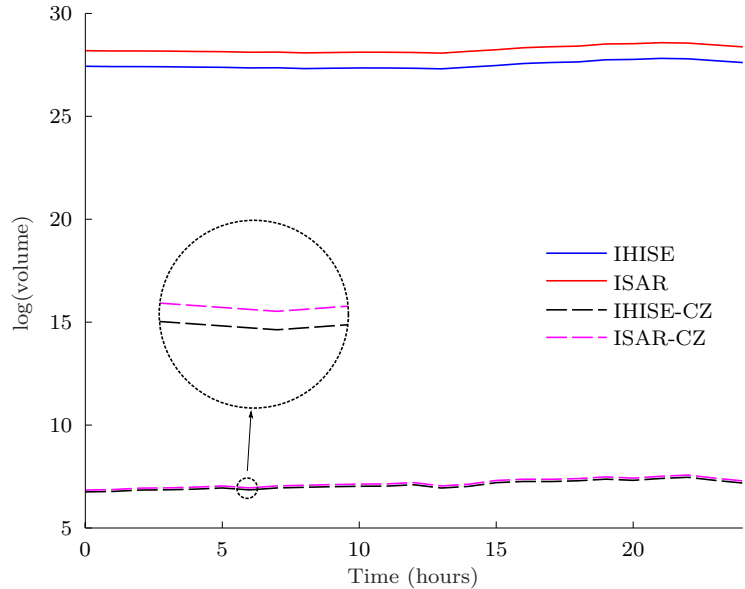}
			\caption{The volumes of the sets $\hat{X}_k$ obtained using IHISE, IHISE-CZ, ISAR, and ISAR-CZ for the transport network `Net1'.}\label{fig:wdn_volume13pipe}}
	\end{figure}

	Figure \ref{fig:wdn_volume13pipe} shows the volumes of the sets\footnote{In this example, the fact that the sets $\hat{X}_k$ are in $\realset^{22}$ results in intractable computational cost in obtaining the exact volumes of the CZs. Therefore, we show instead the volumes of the parallelotopes obtained by eliminating all the constraints in $\hat{X}_k$ followed by generator reduction, using the complexity reduction methods presented in Section \ref{sec:complexityreduction}. The volumes are computed using the formula in \cite{Chisci1996}.} $\hat{X}_k$ obtained using IHISE, ISAR, IHISE-CZ, ISAR-CZ, for time $k \in [0,24]$. As in the previous example, for this transport network the volumes of the CZs generated by both IHISE-CZ and ISAR-CZ are significantly smaller than the volumes of the intervals provided by IHISE and ISAR, demonstrating that IHISE-CZ and ISAR-CZ can be significantly less conservative than IHISE and ISAR. Note also that the volumes of enclosures provided by IHISE-CZ and ISAR-CZ are very similar, showing that, also in this case, there is almost no drawback in terms of accuracy when using ISAR instead of IHISE. The corresponding execution times averaged over the time horizon $k \in [0,24]$ are shown in Table \ref{tab:wdn_13pipetable}, together with the number of iterations performed by the IHISE and ISAR. In this case, the average execution time of ISAR, which requires a higher number of iterations than IHISE, is only $7.4\%$ of the time of IHISE.

	\begin{table}[!htb]
		\scriptsize
		\centering
		\caption{Average execution times and number of iterations per time step for the `Net1' transport network.}
		\begin{tabular}{c c c } \hline
			& Average execution times (seconds)  & Number of iterations (\#)  \\ \hline
			IHISE & $0.6364$ & $7$--$8$ \\
			ISAR & $0.0470$ & $9$ \\
			IHISE-CZ & $0.6410$ & $7$--$8$ \\
			ISAR-CZ & $0.0516$ & $9$  \\
			\hline
		\end{tabular} \normalsize
		\label{tab:wdn_13pipetable}
	\end{table}

	We now assume that the flows at pipes 12 and 13, and the pressure at node 6 are measured, i.e., $\mbf{y}_k = {(q_{k,12}, q_{k,13}, h_{k,6})} + \mbf{v}_k$, where the measurement uncertainty is bounded by $\mbf{v}_k \in V_k$, with $V_{k,i} = [-2\%|y_{k,i}|, 2\%|y_{k,i}|]$. Two leakage scenarios are considered, with the presence of leakage at node 8 and node 4, respectively. 

\begin{figure}[!tb]
		\centering{
			\def\svgwidth{0.6\columnwidth}
			{\scriptsize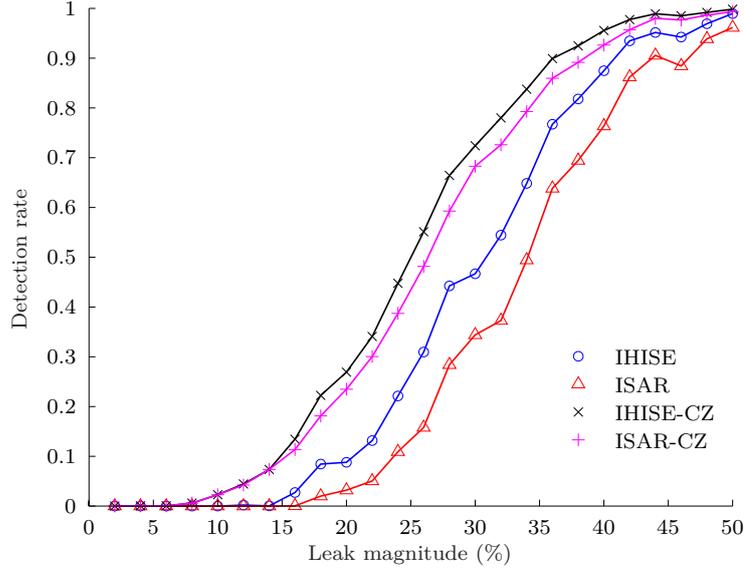}
			\caption{Leak detection rate of the IHISE, ISAR, IHISE-CZ, and ISAR-CZ for the transport network `Net1' (leak at node 8), considering different leak magnitudes, averaged over 100 random simulations and the time interval $k \in [0, 24]$.}\label{fig:wdn_leak13pipemagnitude}}
	\end{figure}

\begin{figure}[!tb]
		\centering{
			\def\svgwidth{0.6\columnwidth}
			{\scriptsize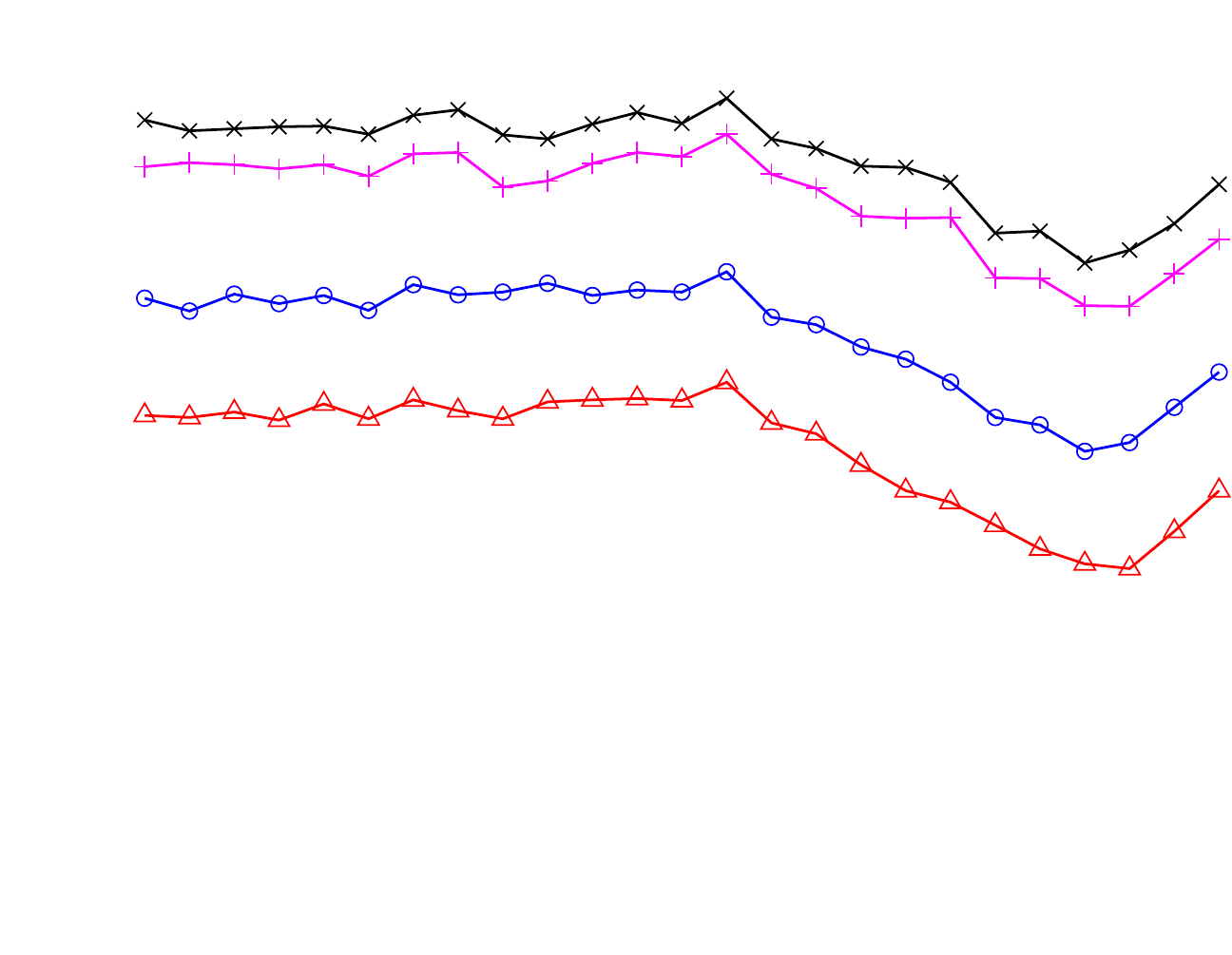}
			\caption{Leak detection rate of the IHISE, ISAR, IHISE-CZ, and ISAR-CZ for the transport network `Net1' (leak at node 8) over time, averaged over 100 random simulations and the leak magnitudes in the interval $[2, 50]\%$.}\label{fig:wdn_leak13pipetime}}
	\end{figure}

\begin{figure}[!tb]
		\centering{
			\def\svgwidth{0.6\columnwidth}
			{\scriptsize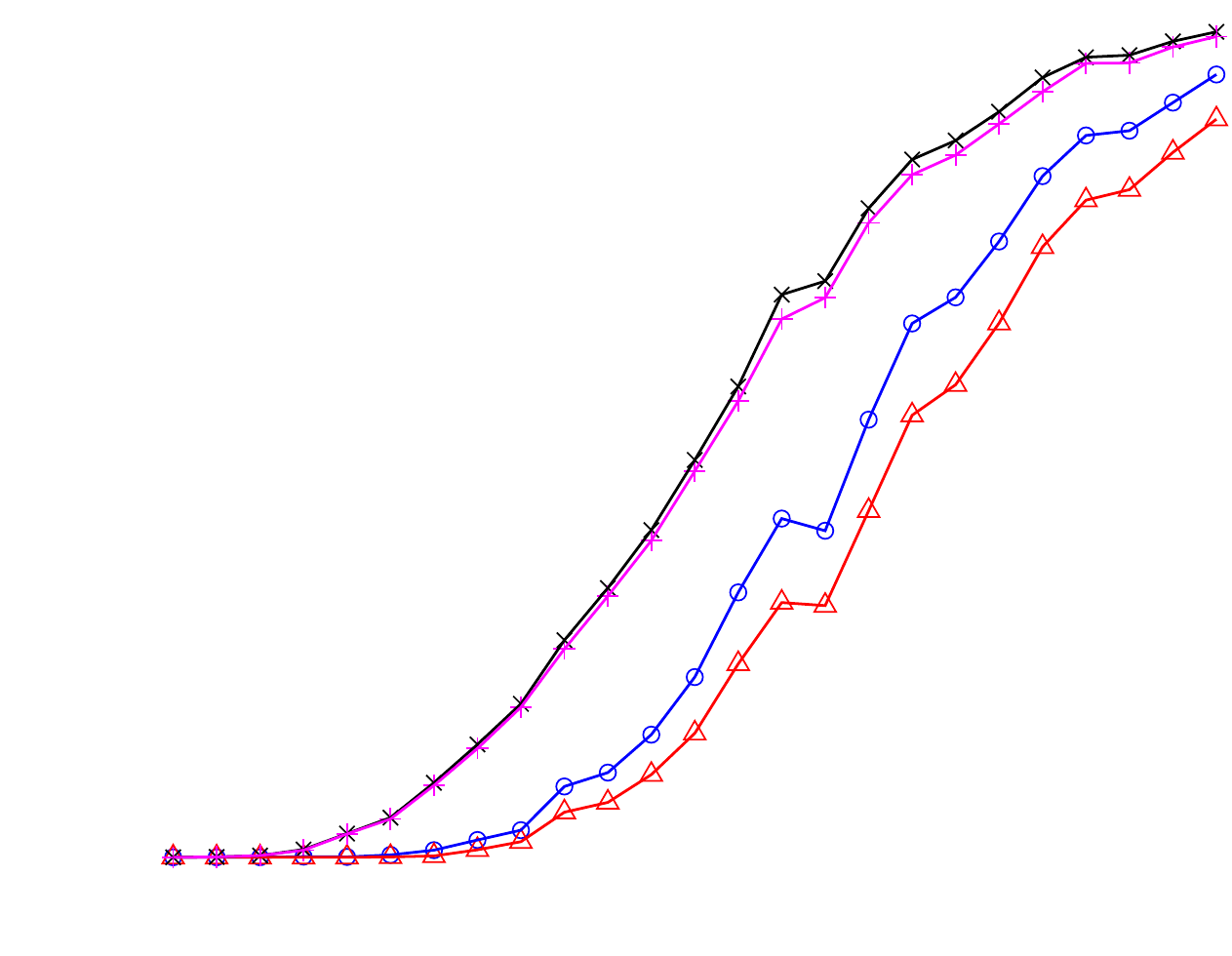}
			\caption{Leak detection rate of the IHISE, ISAR, IHISE-CZ, and ISAR-CZ for the transport network `Net1' (leak at node 4), considering different leak magnitudes, averaged over 100 random simulations and the time interval $k \in [0, 24]$.}\label{fig:wdn_leak13pipemagnitude2}}
	\end{figure}

\begin{figure}[!tb]
		\centering{
			\def\svgwidth{0.6\columnwidth}
			{\scriptsize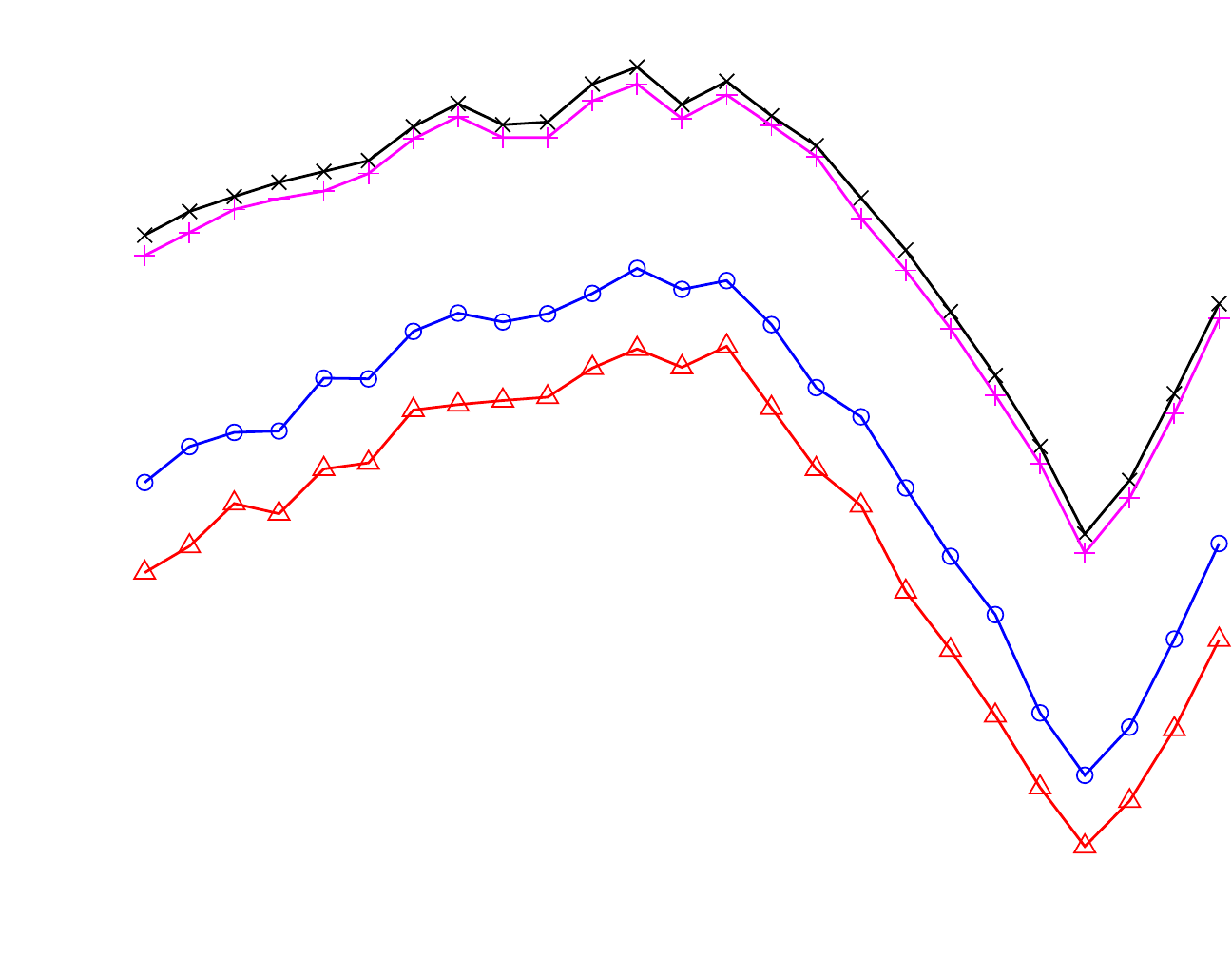}
			\caption{Leak detection rate of the IHISE, ISAR, IHISE-CZ, and ISAR-CZ for the transport network `Net1' (leak at node 4) over time, averaged over 100 random simulations and the leak magnitudes in the interval $[2, 50]\%$.}\label{fig:wdn_leak13pipetime2}}
	\end{figure}

	Figures \ref{fig:wdn_leak13pipemagnitude} and \ref{fig:wdn_leak13pipetime} show the leak detection rate for the first scenario, averaged over 100 MCS, and further over the leak magnitude range $\alpha = 2, 4, \ldots, 50\%$, and the time interval $k = 0,1,\ldots,24$ hours, respectively. In this example, IHISE-CZ provides up to about $20\%$ more detections than IHISE, demonstrating once again the advantages of using CZs instead of intervals for leak detection of WDNs. In addition, Figures \ref{fig:wdn_leak13pipemagnitude2} and \ref{fig:wdn_leak13pipetime2} show the leak detection rate for the second scenario, averaged over 100 MCS, and further over the leak magnitude range $\alpha = 2, 4, \ldots, 50\%$, and the time interval $k = 0,1,\ldots,24$ hours, respectively. In the second scenario, once again the IHISE-CZ provides up to about $20\%$ more detections than IHISE.

	\subsection{CY-DMA network}
	\begin{figure}[!tb]	
		\centering
		\includegraphics[width=0.6\columnwidth]{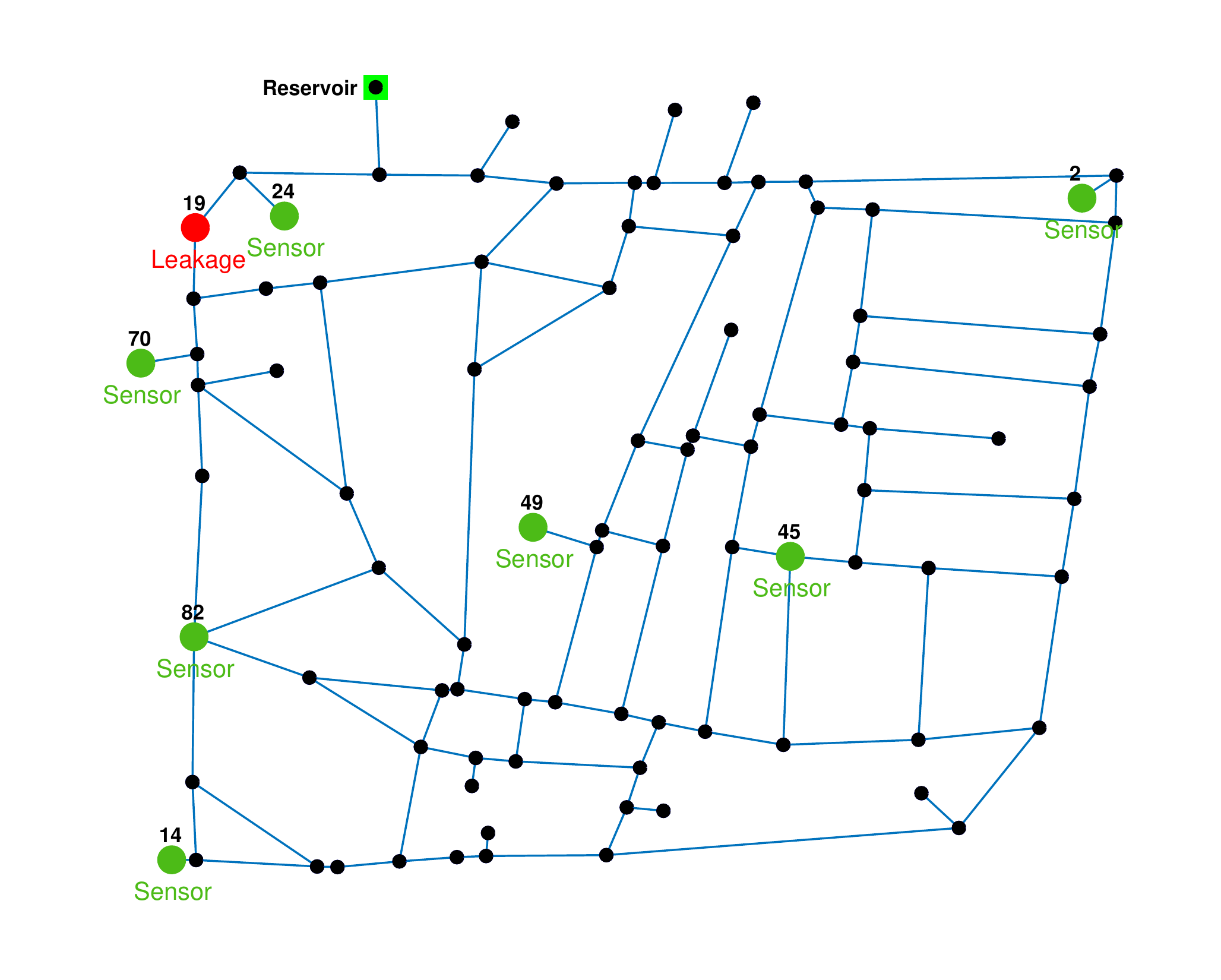}
		\caption{The `CY-DMA' network, a realistic example of a district metered area network.}
		\label{fig:wdn_cydma}
	\end{figure}
	
\begin{figure}[!tb]
		\centering{
			\def\svgwidth{0.6\columnwidth}
			{\scriptsize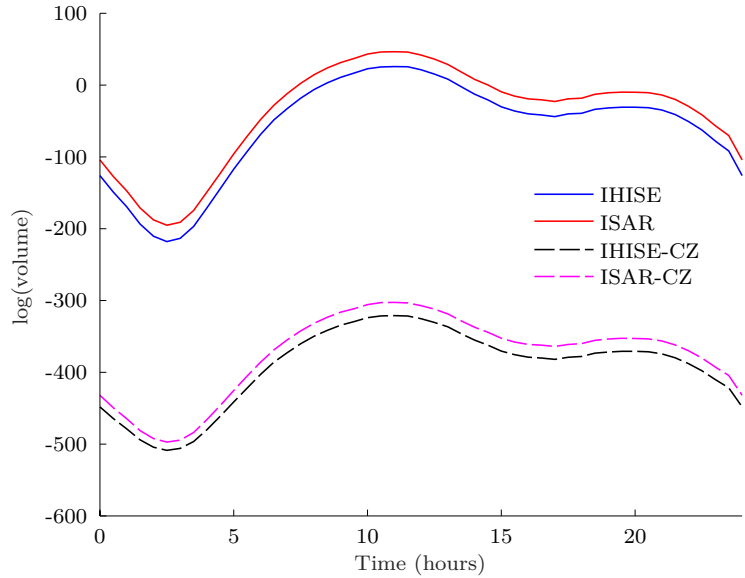}
			\caption{The volumes of the sets $\hat{X}_k$ obtained using IHISE, IHISE-CZ, ISAR, and ISAR-CZ for the CY-DMA network.}\label{fig:wdn_volumecydma}}
	\end{figure}

	The third example considers a realistic {DMA} network from a large water utility in Cyprus, denoted by `CY-DMA' \citep{Vrachimis2020}.
	The network is composed of 90 nodes and 130 pipes and is illustrated in Figure \ref{fig:wdn_cydma}. 
	In this example, we consider that the physical parameters $\mbf{r} \in R$ have uncertainty of $5\%$, the node demands $\mbf{d}_k \in D_k$, which represent groups of residential consumers, are measured with uncertainty of $10\%$, and the reservoir level $\mbf{u}_k$ is precisely known. The tolerance values adopted in Section \ref{sec:wdn_czenclosure} are set to $\tau_u = 1.0 {\times} 10^{-4}$ m and $\tau_d = 1.0 {\times} 10^{-4}$ m$^3$/h. %
	Figure \ref{fig:wdn_volumecydma} shows the volumes of the sets\footnote{As in the previous example, for the CZs we show the volumes of the parallelotopes obtained by eliminating all the constraints in $\hat{X}_k$ followed by generator reduction.} $\hat{X}_k$ obtained using IHISE, ISAR, IHISE-CZ, ISAR-CZ, for time $k \in [0,24]$. For this realistic network, the volumes of the constrained zonotopes generated by both IHISE-CZ and ISAR-CZ are, once again, significantly smaller than the bounds provided by IHISE and ISAR. Note that, due to the dimensionality of the CY-DMA network, considering, for instance, a rectangular set with uncertainty of $0.1$ would result in $\text{log(volume)} = -220$. Taking this into account, in particular the IHISE-CZ is capable of providing sets with about $\text{log(volume)} = -500$. The corresponding execution times averaged over $k \in [0,24]$ are shown in Table \ref{tab:wdn_cydmatable}, together with the number of iterations performed by the IHISE and ISAR. Note that the average execution time of ISAR, which requires a higher number of iterations than IHISE, is only $10.7\%$ of the time of IHISE.
	
	\begin{table}[!htb]
		\scriptsize
		\centering
		\caption{Average execution times and number of iterations per time step for the CY-DMA network.}
		\begin{tabular}{c c c } \hline
			& Average execution times (seconds)  & Number of iterations (\#)  \\ \hline
			IHISE & $54.5293$ & $15$--$16$ \\
			ISAR & $5.8429$ & $34$--$37$ \\
			IHISE-CZ & $54.5549$ & $15$--$16$ \\
			ISAR-CZ & $5.8670$ & $34$--$37$ \\
			\hline
		\end{tabular} \normalsize
		\label{tab:wdn_cydmatable}
	\end{table}
	
	We now consider a leakage scenario, with the presence of leakage at node 19. 
	It is considered that the inlet flow (pipe {14}) and pressure ({node `Reservoir'}) are measured.
	Moreover it is considered that the water utility has installed pressures sensors at nodes 2, 14, 24, 45, 49, 70, and 82, with a total of 7 pressure sensors.
	The sensors' positions were chosen using a sensor placement procedure which calculates the {sensitivity matrix of the WDN} and, using a heuristic optimization approach, computes the sensor locations that maximize the collective sensitivity of all sensors to all possible leakages \citep{Casillas2013}.
	The measured flows and pressures have uncertainties of $2\%$ and $0.35\%$, respectively. 
	
\begin{figure}[!tb]
		\centering{
			\def\svgwidth{0.6\columnwidth}
			{\scriptsize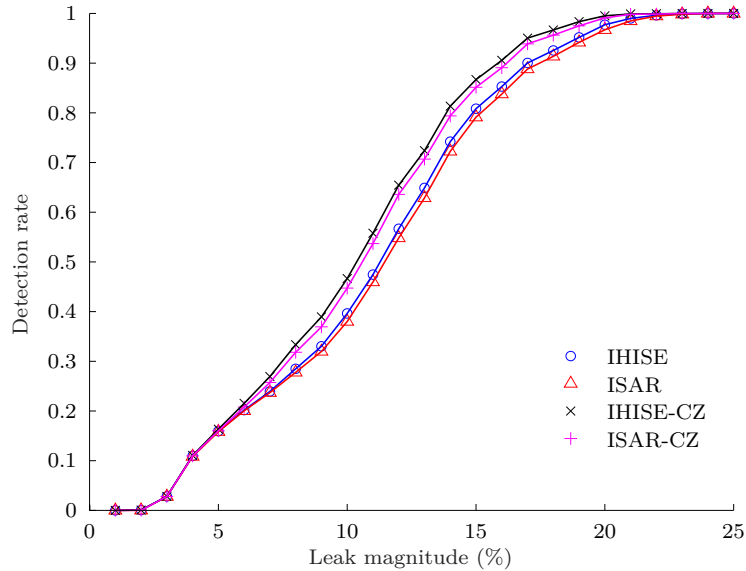}
			\caption{Leak detection rate of the IHISE, ISAR, IHISE-CZ, and ISAR-CZ for the CY-DMA network (leak at node 19), considering 7 pressure sensors, different leak magnitudes, averaged over 100 random simulations and the time interval $k \in [0, 24]$.}\label{fig:wdn_leakcydma9magnitude}}
	\end{figure}

\begin{figure}[!tb]
		\centering{
			\def\svgwidth{0.6\columnwidth}
			{\scriptsize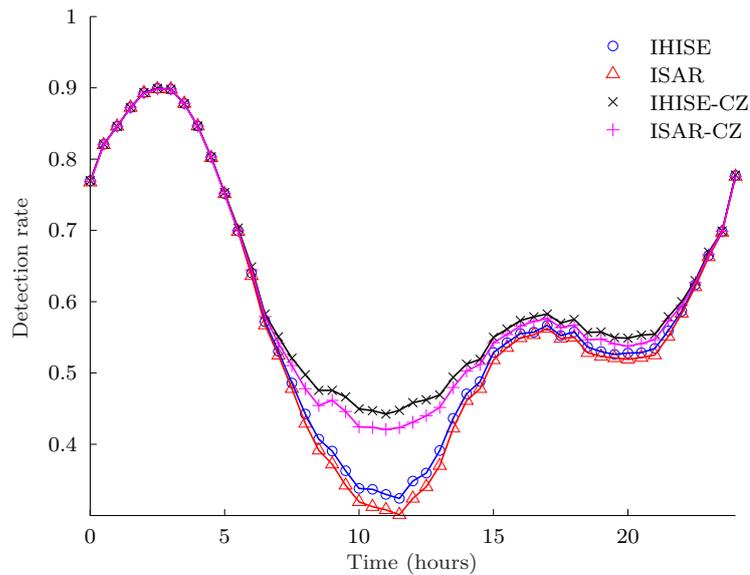}
			\caption{Leak detection rate of the IHISE, ISAR, IHISE-CZ, and ISAR-CZ for the CY-DMA network (leak at node 19) over time, considering 7 pressure sensors, averaged over 100 random simulations and the leak magnitudes in the interval $[1, {25}]\%$.}\label{fig:wdn_leakcydma9time}}
	\end{figure}

	Figures \ref{fig:wdn_leakcydma9magnitude} and \ref{fig:wdn_leakcydma9time} show the leak detection rate, averaged over 100 MCS, and further over the leak magnitude range $\alpha = 1, 2, \ldots, {25}\%$, and the time interval $k = 0,1,\ldots,24$ hours, respectively. In this scenario, IHISE-CZ could provide up to about $10\%$ more detections than IHISE.
	In this scenario, it is interesting to note that the advantage of using CZs instead of intervals is more expressive for leakage detection during midday (12:00-13:00) rather than night hours (3:00-4:00). 
	This realistic network uses typical residential demand patterns with peak during midday, and minimum flow demands during night.
	The high pressure conditions resulting from the low demands during night hours cause the leakage magnitude to increase, which makes the leakage more evident to the leakage detection methodologies.
	The ability of CZs to consider the dependencies between state variables results in the interesting capability of providing higher detection rates even at peak demand hours, which reduces the time of detection and water loss.
	Finally, a few different leakage scenarios have been simulated taking into account several additional pressure sensors in the CY-DMA. 
	However, due to the optimal placement of pressure sensors, the improvement in leak detection of all the methods was mild, which do not cope with the cost of placing additional sensors in a realistic network, and for this reason were not reported.
		
	\section{Final remarks} \label{sec:wdn_finalremarks}
	
	This chapter proposed two new methods for set-based state estimation of WDN based on CZs, namely IHISE-CZ and ISAR-CZ. With respect to interval-based methods, CZ-based ones are able to capture the dependencies between hydraulic states, resulting in sets with significantly smaller volumes. The benefits of IHISE-CZ and ISAR-CZ were also demonstrated for leakage detection in WDNs. In particular, the simulation results on a benchmark transport network and on a realistic DMA network show an increase in detection rates of up to $20\% $ when using the proposed methods, while also requiring significantly less computational resources to compute the enclosures. Moreover, it is observed that the proposed methods perform well during peak demand hours, thus reducing significantly the time-to-detection and the cost of water loss for water utilities.

\chapter{Application: Li-Ion batteries}\thispagestyle{headings} \label{cha:appbat}

In this chapter, we propose an interval state estimation of a lithium-ion cell when using an Equivalent Circuit Model (ECM), taking into account bounded parametric uncertainty and measurement noise. Battery management systems (BMSs) are responsible for controlling and monitoring battery operations. In order to be effective, BMSs rely on mathematical models. The parameters of these latter are usually obtained through an identification procedure which provides limited accuracy. In addition, only cell current and voltage are usually measurable, and therefore, the model states need to be estimated. Simulation results show the suitability of the proposed interval estimator to provide a tight enclosure of the states, which is essential for fault detection and model-based control of a lithium-ion cell. The results of this chapter were obtained in collaboration with the PhD candidate Diego Locatelli, supervised by Prof. Davide M. Raimondo. The content has been presented in \cite{Locatelli2021}. 

This chapter is organized as follows. In Section \ref{sec:batt_models}, the Single Particle Model with electrolyte dynamics (SPMe) and ECM models are described. The identification procedure and set-based state estimation are presented in Sections \ref{sec:batt_ID} and \ref{sec:batt_setbased}, respectively. The results are reported in Section \ref{sec:batt_results} and, finally, Section \ref{sec:batt_conclusions} concludes the chapter.

\section{Introduction}

Lithium-ion batteries  have become of particular interest due to their use in many commercial sectors, including portable electronic devices, hybrid and electric vehicles, and grid energy storage. 
High cell voltage, cycling durability, high energy, and power density, are the main key factors that make Lithium-ion the best trade-off between costs, performance and efficiency, when compared to other battery types on the market.

Battery management systems aim to make the battery safer, reliable and efficient. 
As shown in \cite{lu2013review}, violation of temperature, current and voltage restrictions, may result in lower performance as well as safety issues. Advanced BMSs rely on mathematical models in order to improve efficiency and monitor the battery cell, by estimating fundamental quantities such as the State of Charge (SOC). 
Mathematical models for Lithium-ion battery dynamics fall within two main categories: Electrical Circuit Models and Electrochemical Models (EMs).  
EM are very accurate and, therefore, very useful when high fidelity is required, such as in simulations. They are generally very complex models, and this affects the computational cost. 
On the other hand, ECMs are simpler models, and they are used in real-time control system applications especially for online SOC and state estimation.

Look-up table methods are used for SOC estimation exploiting its relations with other measurable parameters as the Open Circuit Potential (OCP), but they are not suitable for run-time operations \citep{xiong2017critical}. This is due to the fact that the OCP can be measured only in stationary conditions.
Coulomb counting is also one of the main tools used for SOC estimation \citep{piller2001methods}, but it is not completely accurate because of drift errors due to disturbances on the current measurement and uncertainty on the cell capacity. Moreover, while SOC estimation is important, in order to employ a model-based controller, full state estimation is required.

Several stochastic approaches on state estimation have been applied in the battery field, including Extended Kalman Filtering  \citep{di2010lithium,bizeray2015lithium}, sliding-mode observer \citep{kim2009technique}, Sigma-point Kalman Filter \citep{plett2006sigma}, Particle filtering \citep{tulsyan2016state}, and Moving Horizon Estimator \citep{7867062}. 
The stochastic estimation approaches assume that probabilistic distributions of the uncertainties are known. 
On the other side, set-based estimation considers unknown-but-bounded uncertainties.  
When physical bounds are available (as for the case of ECM parameters), this second option is more reliable since knowing the exact distribution of the uncertainties is rarely the case in practice. %
To the best of our knowledge, set-based estimation was used within the battery context only in \cite{zhang2020interval}. This latter proposes a continuous-time ECM-based interval observer for SOC estimation in a parallel connected Li-ion battery pack. The proposed observer assumes that continuous-time measurements are available, which is not true in practice since sensors have finite sampling rates. Moreover, the coupling between parameters is not considered, which is usually present when an identification procedure is performed. 

In this chapter, we develop a discrete-time interval observer for a single Li-ion cell based on the forward-backward method described in \cite{jaulin2001nonlinear}. In particular, parametric uncertainties are considered and obtained by performing an identification procedure on data collected on a well known EM, i.e. the SPMe. The contributions of this chapter are: (i) the identification of ECM parameter bounds based on the Fisher Information Matrix, (ii) a discrete-time interval state estimation method based on inclusion functions and constraint propagation, which handles discrete-time measurement. Numerical experiments show that the proposed methodology is efficient and  provides accurate enclosures for both the SOC and the electric state variables of the ECM.

\section{Lithium-ion cell modeling}\label{sec:batt_models}

This section introduces the two models considered in this chapter. The SPMe, which is a well known EM, is used as the ``real plant''. Its complete parameterization is
obtained according to \cite{ecker2015parameterization} and represents a Kokam SLPB 75106100 cell. Since the SPMe is computational expensive,  for estimation purposes we adopt the Thévenin ECM instead. Its parametrization and  uncertainty quantification are obtained  relying on input-output data (current-voltage) collected from the SPMe.

\subsection{Single Particle Model With Electrolyte dynamics}\label{sec:SPMe}

The SPMe is an approximation of the Doyle-Fuller-Newman (DFN) model proposed in \cite{Doyle_1993}, which predicts the charge/discharge behaviour of dual insertion cells. The model is general for cells utilizing two composite electrodes with active insertion material, electrolyte and inert conducting material. Cathode, separator and anode are the three sections composing the battery cell and are here denoted by $\{ p,s,n \}$.  In this chapter, the model described by \cite{pozzi2020balancing} is taken as a reference, but we assume the cell is kept in  a climatic chamber at constant temperature (isothermal process). The single cell dynamics is described by a system of Ordinary Differential Equations (ODEs). A polynomial approximation of the solid ions concentration along the particle radius is used to reduce the Fick's law for each electrode to an ODE (see \cite{subramanian2005efficient}). Moreover, the Finite Volume Method (FVM) is used to spatially discretize the Partial Differential Equations (PDEs) describing the electrolyte ions diffusion.  

The dynamics of the SPMe is described by $3 + 3N$ electric state variables $[\overline{\theta}_p(t) \,\; \overline{q}_p(t)$ $\,\; \overline{q}_n(t) \,\; c_{e,p}^{[1]}(t) \,\; \dots \,\; c_{e,n}^{[N]}(t)]$, where $N$ is the number of discretization volumes  for each section of the cell,  $\overline{\theta}_p(t)$ is the cathode average stoichiometry,  $\overline{q}_i(t)$ is the volume-averaged concentration flux for the $i$-th electrode, and $c_{e,j}^{[k]}(t)$ is the electrolyte concentration in the $k$-th volume of the $j$-th cell section. Note that indices $j$ and $i$ will be used to refer to the three cell sections $ \{p, s, n\}$ and to the two electrodes $ \{p, n\}$, respectively.

The dynamics of the cathodic average stoichiometry can be expressed as $$\dot{\overline{\theta}}_p(t) = -\frac{\Delta \theta_p}{C_\text{batt}^\text{S}}I(t),$$ where $C_\text{batt}^\text{S}$ is the nominal cell capacity,  $\Delta \theta_i = \theta_{i}^{100\%} - \theta_{i}^{0\%}$ is the difference between the values of the stoichiometry  at battery fully charged and fully discharged, and $I$ is the current, input of the system\footnote{In this chapter, we adopt the convention that the battery is being charged when a negative current is applied.}.
In order to avoid a further differential equation, the anodic average stoichiometry is computed according to the algebraic condition (see \cite{di2010lithium}) $$\overline{\theta}_n(t)= \theta_n^{0\%} + \frac{\overline{\theta}_p(t)-\theta^{0\%}_p}{\Delta \theta_p}\Delta \theta_n.$$ For each electrode, the volume-averaged concentration flux dynamics is described by $$\dot{\overline{q}}_i = -30\frac{D_{s,i}}{R_{p,i}^2} {\overline{q}}_i(t) - \frac{45 \Delta \theta_i c_{s,i}^\text{max}}{6 R_{p,i}{C_\text{batt}^\text{S}}}I(t),$$ where $D_{s,i}$ is the solid diffusion coefficient, $c_{s,i}^\text{max}$ is the maximum solid concentration, $R_{p,i}$ is the solid particle radius, and $\theta_i$ is the surface stoichiometry given by $$\theta_i(t) = \overline{\theta}_i (t)+ \frac{8 R_{p,i} \overline{q}_i(t)}{35 c_{s,i}^\text{max}}-\frac{R_{p,i}^2\Delta \theta_i}{105 D_{s,i}{C_\text{batt}^\text{S}}}I(t).$$ 

The ODEs describing the diffusion of the electrolyte concentration derive from the finite volume discretization of the original PDEs proposed in \cite{moura2016battery}. Each section of the cell is divided into $N$ finite volumes. The $k$-th volume in the $j$-th section is centered in the spatial coordinate $x_{j,k}$ and ranges within the interval $[ x_{j,k-\frac{1}{2}} \, , x_{j,k+\frac{1}{2}}]$, with width $\Delta x_j = L_j/N$, where $L_j $ is the section thickness. For each volume, the
electrolyte concentration dynamics is given by
\begin{equation} \label{eq:batt_ce}
\epsilon_j \frac{\partial c_{e,j}^{[k]}(t)}{\partial t}=
\begin{bmatrix} 
\frac{\tilde{D}_e(x)}{\Delta x_j}\frac{\partial c_{e,j}(x,t)}{\partial x}
\end{bmatrix}
^{x_{j,k+\frac{1}{2}}}_{x_{j,k-\frac{1}{2}}}
+ K(j),
\end{equation}
with $\epsilon_j$ the material porosity, $A$ the contact area between solid and electrolyte phases, $F$ the Faraday constant, $t_+$ the transference number, and $\tilde{D}_e(x)$ the electrolyte diffusion coefficient, whose computation is explained more in detail in \cite{pozzi2020balancing}. The term $K(j)$ is null for the separator section and $\mp (1-t_+)\,I(t)/FAL_n$ for cathode and anode respectively. The state of charge $z^{\text{S}}(t)$ can be computed according to the cathodic or anodic convention, as shown by the following relation 
\begin{equation} \label{eq:batt_SOC}
z^{\text{S}}(t)=\frac{\overline{\theta}_p(t) - \theta_p^{0\%}}{\Delta \theta_p}=\frac{\overline{\theta}_n(t) - \theta_n^{0\%}}{\Delta \theta_n}.
\end{equation}
Finally, the output terminal voltage is given by
\begin{equation} \label{eq:batt_Vspmet}
V^{\text{S}}(t) \triangleq V_\text{oc}(t) + \overline{\eta}_p(t) - \overline{\eta}_n(t) + \Delta \Phi_e(t),
\end{equation}
where $V_\text{oc}(t)$ is the overall open circuit potential defined as
\begin{equation} \label{eq:U}
V_\text{oc}(t) \triangleq U_p(t) - U_n(t),
\end{equation}
with $U_p(t) $ and $U_n(t)$ the cathodic and anodic open circuit potentials, respectively. The expression of these latter is obtained using a fitting procedure on the experimental data provided in \cite{ecker2015parameterization}, as
\begin{gather}
U_p = 4.571 + 0.02414\theta_p -7.8370\theta_p^2 +8.07\theta_p^3 + 20.94\theta_p^4 -40.7\theta_p^5 +18.45\theta_p^6,  \label{eq:Up} \\
U_n = \frac{0.00694+0.1261\theta_n}{\theta_n^2+0.6995\theta_n+0.00405}. \label{eq:Un}
\end{gather}

The overpotentials $\overline{\eta}_i(t) $ are defined, for each electrode, as $\overline{\eta}_i(t) {=} \frac{2RT}{F}\text{sinh}^{-1} \!\Big(\!\frac{\Delta \theta_i R_{p,i}}{6\overline{i}_{0,i}(t)C_\text{batt}^{\text{S}}}I(t)\!\Big)$, in which the quantity $\overline{i}_{0,i}(t)$ is given by $\overline{i}_{0,i}(t) = Fk_i \sqrt{\overline{c}_{e,i}(t)\theta_i(t)(1-\theta_i(t))}$, where $k_i$ is the rate reaction constant, and $\overline{c}_{e,i}(t)$ is the average electrolyte concentration in the $i$-th section. Returning to \eqref{eq:batt_Vspmet}, $\Delta \Phi_e(t)$ is evaluated by
\begin{equation} \label{eq:batt_delta_phi}
\Delta \Phi_e(t)= \Phi_{e}^{\text{drop}}(t) + \frac{2RT(t)}{F}(1-t_+)~\text{log}\biggl( \frac{c_{e,p}^{[1]}(t)}{c_{e,n}^{[N]}(t)}\biggl),
\end{equation}
whose terms are already defined except for $\Phi_e^\text{drop}(t)$, which is approximated by $\Phi_{e}^\text{drop}(t) \simeq - (I(t)/2N) (\phi_p(t) + 2\phi_s(t) + \phi_n(t))$. The quantities $\phi_j(t)$, $j \in \{p,s,n\}$ depend on the electrolyte conductivity and are computed assuming a trapezoidal shape  of the ionic current $i_e(x,t)$ along the spatial domain. We refer to \cite{pozzi2020balancing} for further details. 
The SPMe parameters used in this chapter are taken from \cite{ecker2015parameterization} and are summarized in Table \ref{tab:batt_spmet_param}.\footnote{
	Activation energies are used in Arrhenius's law for temperature-dependent parameters as $\psi (T) = \psi_0 \, e{\frac{-E_{\alpha,\psi}}{R\,T}}$. }
\begin{table}[htb]
	\begin{center}
		\caption{SPMe cell section varying parameters.}\label{tab:batt_spmet_param}
		\begin{tabular}{cccc} \hline
			Parameter & Value & Parameter & Value   \\\hline
			$\theta^{0\%}_p$ & $0.86$ & $L_{p}$ (m) & $54.5\ten{-6}$  \\
			$\theta^{0\%}_n$ & $0.04$ &  $L_{s}$ (m) &  $19\ten{-6}$  \\
			$\theta^{100\%}_p$ & $0.26$ & $L_{n}$ (m) & $73.7\ten{-6}$  \\
			$\theta^{100\%}_n$ & $0.75$ & $\epsilon_{p}$ & $1.5443$  \\
			$R_{p,p}$ (m) & $6.49 {\cdot}10^{-6}$ & $\epsilon_{s}$ &  $1.7572$ \\
			$R_{p,n}$ (m) & $8.7\ten{-6}$  & $\epsilon_{n}$ &  $1.6369$ \\
			$c_{s,p}^\text{max}$ $\big(\frac{\text{mol}}{\text{m}^3}\big)$ & $48580$ &  $k_{p}$ $\big(\frac{\text{m}^{5/2}}{\text{mol}^{1/2}}\big)$ & $3.0496\ten{-11}$  \\
			$c_{s,n}^\text{max}$ $\big(\frac{\text{mol}}{\text{m}^3}\big)$ &  $31920$ & $k_{n}$ $\big(\frac{\text{m}^{5/2}}{\text{mol}^{1/2}}\big)$ &  $1.4656\ten{-11}$ \\
			$D_{s,i}$ $\big(\frac{\text{m}^2}{\text{s}}\big)$ & $10^{-14}$  & $E_{\alpha_{k},p}$ $\big(\frac{\text{J}}{\text{mol}}\big)$ & $43600$ \\ 
			$E_{\alpha_{D_s},p}$ $\big(\frac{\text{J}}{\text{mol}}\big)$ & $80600$  & $E_{\alpha_{k},n}$ $\big(\frac{\text{J}}{\text{mol}}\big)$ & $53400$ \\ 
			$E_{\alpha_{D_s},n}$ $\big(\frac{\text{J}}{\text{mol}}\big)$ &  $40800$ & $C_\text{batt}^\text{S}$ (A${\cdot}$h) &  $7.5$ \\
			$D_{e,j}$ $\big(\frac{\text{m}^2}{\text{s}}\big)$ & $2.4 \ten{-10}$  & $t_+$ &  $0.26$ \\
			$E_{\alpha_{D_e},j}$ $\big(\frac{\text{J}}{\text{mol}}\big)$ & $17100$  & $A$ (m$^2$) &  $0.4121$ \\
			$R$ $\big(\frac{\text{J}}{\text{mol} \cdot \text{K}}\big)$ & $8.314472$  & $F$ $\big(\frac{\text{s}\cdot \text{A}}{\text{mol} }\big)$ &  $96485$ \\
			\hline
		\end{tabular}
	\end{center}
\end{table}

\subsection{Equivalent Circuit Model}\label{sec:batt_ECM}
\begin{figure}[!tb]
	\begin{center}
		\includegraphics[width=0.55\columnwidth]{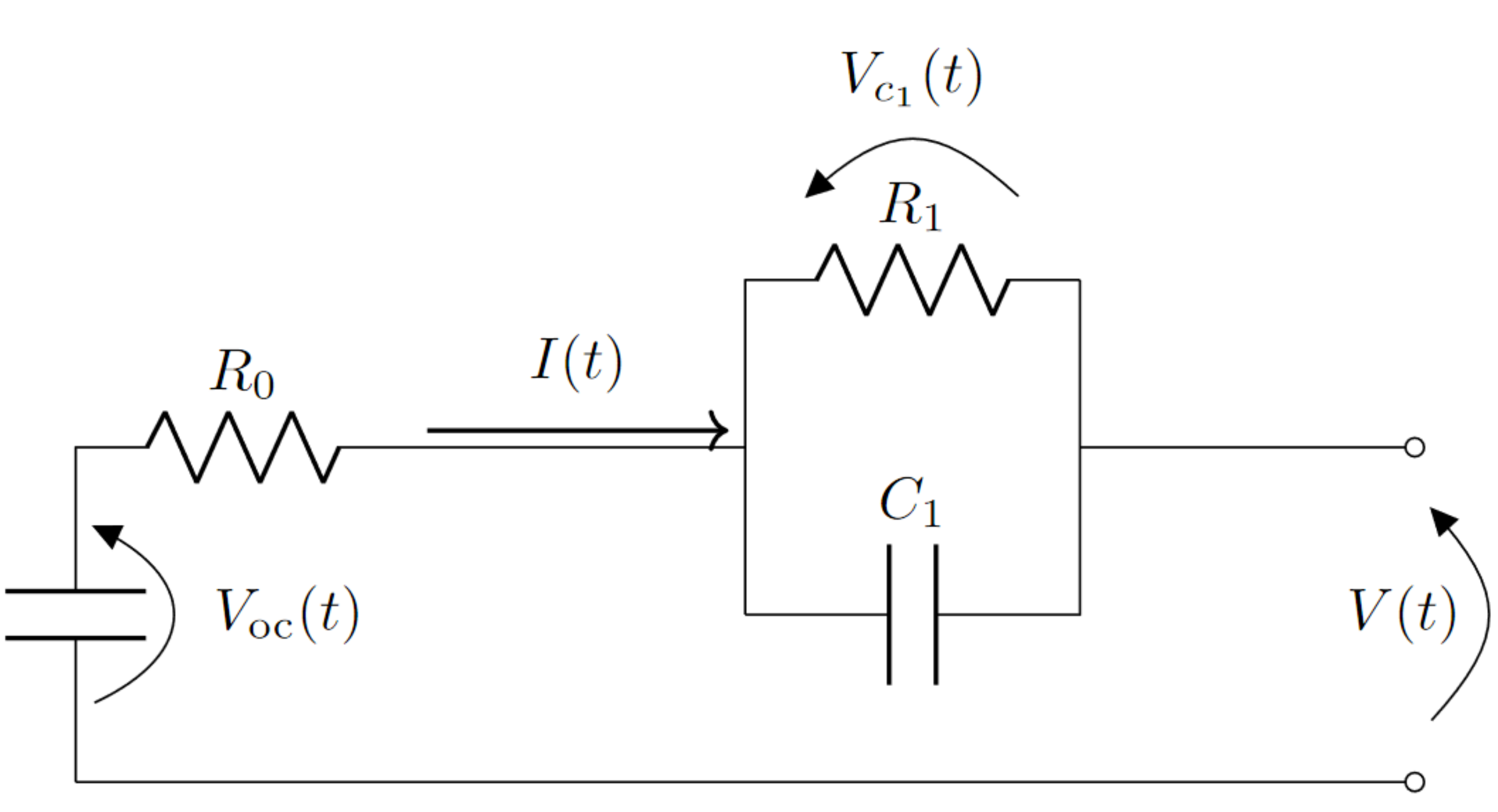}    
		\caption{Thévenin equivalent circuit model of the Lithium-ion cell.}
		\label{fig:ECM}
	\end{center}
\end{figure}

The ECM considered in this chapter for state estimation purposes is the Thévenin model depicted in Figure \ref{fig:ECM}.
It consists of three parts: (i) the open circuit potential $V_\text{oc}(t)$; (ii) two internal resistors $R_0$ and $R_1$; and (iii) the capacitor $C_1$. This latter is useful to describe the charge and discharge transient of the cell.
The electrical behavior of the ECM is derived from Kirchhoff and Ohm's laws as described by
\begin{equation} \label{eq:ECM_dyn}
\left \{ \begin{aligned}
\dot{z}^{\text{E}}(t)& = -\frac{I(t)}{C_\text{batt}^\text{E}},\\
\dot{V}_{c_1}(t)&=-\frac{V_{c_1}(t)}{R_1(t)C_1(t)}+\frac{I(t)}{C_1(t)} ,
\end{aligned}
\right.
\end{equation}
where $z^\text{E}(t)$ is the cell SOC, $I(t)$ is the input current\footnote{Note that, also in this case, we adopt the convention that the battery is charged by a negative current.}, $V_{c_1}(t)$ is the voltage drop across the parallel ${R_1}-{C_1}$, and $C_\text{batt}^\text{E}$ is the nominal cell capacity. 

\begin{remark} \rm
	Since the ECM will only approximate the SPMe dynamics and its parameters will  be estimated, there is no guarantee that battery capacity, SOC and OCP will have the exact same  value/time behavior. For this reason, we distinguish them by using the superscripts E and S, respectively.
\end{remark}

\begin{figure}
	\begin{center}
		\includegraphics[width=0.5\columnwidth]{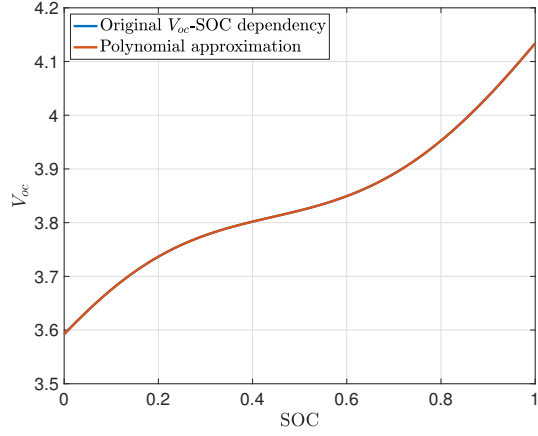}    
		\caption{Polynomial approximation of OCP-SOC dependency.}
		\label{fig:batt_OCP_fit}
	\end{center}
\end{figure}

In accordance with \cite{perez2017optimal}, in order the ECM to better approximate the SPMe, $R_1$ and $C_1$ can be expressed as  nonlinear functions of the SOC, given by
\begin{align} 
R_1(t) & = R_{1,0} +  R_{1,1}z^{\text{E}}(t) + R_{1,2}z^{\text{E}}(t)^2, \label{eq:R1}\\
C_1(t) & = C_{1,0} +  C_{1,1}z^{\text{E}}(t) + C_{1,2}z^{\text{E}}(t)^2. \label{eq:C1}
\end{align}
Finally, the output voltage equation can be computed as
\begin{equation} \label{eq:batt_ECM_V}
V^{\text{E}}(t) = V_\text{oc}^{\text{E}}(t) - R_0\,I(t) - V_{c_1}(t),
\end{equation}
where $V_\text{oc}^{\text{E}}(t)$ is the overall open circuit potential. Unlike \eqref{eq:U}, this latter is here expressed as a function of the SOC, i.e. $z^{\text{E}}(t)$. This is   necessary since $\theta_p$ and $\theta_n$ are not states of \eqref{eq:ECM_dyn}. However, since the OCP represents
the potential difference between the negative and positive electrodes when no current is applied,
then,  once the transient is over, $\overline{\theta}_i=\theta_i$. This, together with equation \eqref{eq:batt_SOC}, allows to express $\theta_p$ and $\theta_n$, and consequently \eqref{eq:Up}, \eqref{eq:Un} and the OCP,  in terms of SOC. Nevertheless, rather than using such an expression, we fit on it a polynomial function of the SOC
\begin{equation} \label{eq:batt_OCP}
\begin{aligned}
V_\text{oc}^{\text{E}}(t) & = 3.592 + 0.9082\, z^{\text{E}}(t) - 0.57 \,z^{\text{E}}(t)^2 -2.979\,z^{\text{E}}(t)^3 + 6.56\, z^{\text{E}}(t)^4 -4.238 \,z^{\text{E}}(t)^5 \\& +0.8608
\,z^{\text{E}}(t)^6 -1.676\ten{-10}\, z^{\text{E}}(t)^7 + 1.143\ten{-10} \,z^{\text{E}}(t)^8 -2.982\ten{-11}\,z^{\text{E}}(t)^9,
\end{aligned}
\end{equation}
which is simpler (and therefore may lead to less conservatism when evaluated through interval arithmetic), but still very accurate (see Figure \ref{fig:batt_OCP_fit}).

The identification of the ECM parameters, besides the $V_{\text{oc}}$ which is assumed known a priori, relies on the procedure described in Section \ref{sec:batt_ID}.

\section{Identification of the ECM parameters}\label{sec:batt_ID}
The ECM described in Section \ref{sec:batt_ECM} is formally written as
\begin{equation} \label{eq:formal_cont}
\left\{\begin{aligned}
\dot{\mathbf{x}}(t)& = \mbf{f}(\mathbf{x}(t),\mathbf{u}(t),\bm{\phi}),\\
\mathbf{y}(t)&=\mbf{g}(\mathbf{x}(t),\mathbf{u}(t),\bm{\phi}),\\
\mathbf{x}(t_0)&=\mathbf{x}_0,
\end{aligned}
\right.
\end{equation}
with
\begin{equation} \label{eq:batt_phi_param}
\bm{\phi} \triangleq
\begin{bmatrix} 
R_0   & R_{1,0} & R_{1,1} & R_{1,2}&  C_{1,0} & C_{1,1} & C_{1,2}& C_\text{batt}^\text{E}  
\end{bmatrix}.
\end{equation}

In order to apply the interval state estimation method proposed in Section \ref{sec:batt_setbased}, the ECM is discretized according to the Euler's method, thus obtaining
\begin{equation} \label{eq:batt_formal_disc}
\left \{ \begin{aligned}
\mathbf{x}_{k+1}& = \mathbf{x}_{k}+T_s \mbf{f}(\mathbf{x}_k,u_k,\bm{\phi}),\\
\mathbf{y}_k&=\mbf{g}(\mathbf{x}_k,u_k,\bm{\phi}),\\
\mathbf{x}_{k_0}&=\mathbf{x}_0,
\end{aligned}
\right.
\end{equation}
where $T_s$ is the sampling time, and $k\geq k_0$ is the time step.

\begin{figure}[tb]
	\begin{center}
		\includegraphics[width=0.5\columnwidth]{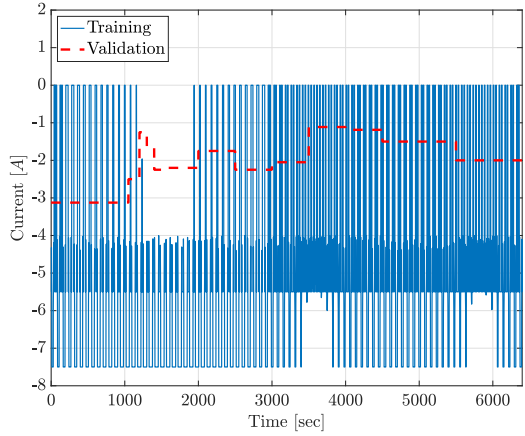}    
		\caption{Training and validation current input profiles.} 
		\label{fig:input_training_val}
	\end{center}
\end{figure}

According to \cite{perez2017optimal}, $R_1$ and $C_1$ exhibit a different SOC relation for the charging and discharging phase. In this chapter, we consider the charging phase only. The parameter estimation is performed starting from data collected on the ``real plant'', here assumed to be the SPMe.
We consider as initial condition an almost completely discharged cell at rest, i.e.  $\mbf{x_0}= (0.017,0)$, and as input signal, defined over a time horizon $\bar{k}$, the one reported in Figure \ref{fig:input_training_val}.  
Note that a sufficiently exciting input profile helps the identifiability and reduces parameters uncertainties.

\begin{remark} \rm \label{rem:batt_initialstate}
	Even though the ECM parameters still need to be estimated, $\mathbf{x_0}$ can be easily obtained when a cell is at rest. In this case, the voltage across the parallel $R_1-C_1$ is zero, $V^{\text{E}}$ coincides with 
	$V_\text{oc}^{\text{E}}$, and $z^{\text{E}}(0)$ can be computed from  equation \eqref{eq:batt_OCP}\footnote{The OCP is assumed known a priori.}.
\end{remark}

The system is affected by measurement noise that ranges within the interval
$[-3mV,3mV]$. In order to use the following identification procedure, in this section we approximate it as i.i.d. Gaussian noise $\mathbf{d}_k\sim \mathcal{N}(0,\sigma_y^{2})$ with $\sigma_y=1mV$. In particular, for the true value of the parameter vector $\bm{\phi}^*$, one has $\hat{\mbf{y}}_k(\bm{\phi}^*)=g(\mathbf{x}_k,\mathbf{u}_k,\bm{\phi}^*)+\mathbf{d}_k$. 
Let $\mathbf{\hat{Y}(\bm{\phi}^*)}\in \mathbb{R}^{\overline{k}+1}$ denote the vector of observed output data over the time horizon $k=k_0,\ldots, k_0+\bar{k}$. Then, it holds that  $\hat{\mathbf{Y}}(\bm{\phi}^*) \sim \mathcal{N}({\mathbf{Y}}(\bm{\phi}^*),\mbf{C}_y)$, where ${\mathbf{Y}}(\bm{\phi}^*)$ stands for the output vector in the absence of measurement noise and  $\mbf{C}_y\in  \mathbb{R}^{(\overline{k}+1)\times (\overline{k}+1)}$ for the diagonal measurement covariance matrix given by $\mbf{C}_y = \sigma_y^2\mbf{I}_{\overline{k}+1}$. 

Once the training data is collected, the parameters $\hat{\bm{\phi}}$ can be estimated by solving the following maximum likelihood optimization problem:
\begin{subequations}\label{opt}
	\begin{align}
	\min_{\bm{\phi}} ~ &
	(\hat{\mathbf{Y}}(\bm{\phi}^*)-\mathbf{Y}(\bm{\phi}))^\text{T}\,\mbf{C}_y^{-1}(\hat{\mathbf{Y}}(\bm{\phi}^*)-\mathbf{Y}(\bm{\phi})) \nonumber \\
	\textrm{s.t.} ~~ & \text{model dynamics \eqref{eq:batt_formal_disc}}, \nonumber \\
	& \mathbf{h}(\mathbf{x}_k,\mathbf{u}_k,\bm{\phi}) \geq \mbf{0}, \label{eq:procine} \\
	& \bm{\phi}^\text{min}\leq  \bm{\phi}  \leq \bm{\phi}^\text{max}, \label{eq:boundsOPT}
	\end{align}
\end{subequations}
where $\mathbf{Y}(\bm{\phi})$
stands for the output vector obtained by solving, for a given $\bm{\phi}$, system \eqref{eq:batt_formal_disc} 
over the time horizon $k=k_0,\ldots, k_0+\bar{k}$. Equation \eqref{eq:boundsOPT} allows to account for physical bounds on the parameters, e.g. $R_0\geq 0$, $C^{\text{E}}_\text{batt} \geq 0$. Similarly, equation (\ref{eq:procine}) is required to bound SOC-dependent variables such as $R_1$ and $C_1$. In practice, assuming functions \eqref{eq:R1} and \eqref{eq:C1} have the same concavity as in \cite{perez2017optimal}, one can force the positivity of these quantities by setting constraints on their values for   $z^\text{E}=1$ and $z^\text{E}=0$ (concave case), or for the $z^\text{E}$ corresponding to the minimum value of the function (convex case). This latter can be found analytically. Further constraints can be added when prior knowledge on the parametric bounds is available. 
The value of the estimated parameters, obtained by solving problem \eqref{opt}, is reported in Table \ref{tb:phi_est}.

\begin{table}[htb]
	\begin{center}
		\caption{Estimated ECM parameters.}\label{tb:phi_est}
		\begin{tabular}{cccc} \hline
			$R_0(\Omega)$ & $R_{1,0}(\Omega)$ & $R_{1,1}(\Omega)$ & $R_{1,2}(\Omega)$ \\\hline
			$0.093$ & $0.0221$ & $-0.07$  & $0.0672$\\\hline
			$C_{1,0}(\text{As})$ & $C_{1,1}(\text{As})$ & $C_{1,2}(\text{As})$ & $C_\text{batt}^{\text{E}}(\text{As})$ \\\hline
			$235.52$ &$7.7613\,10^{4}$ & $-7.0974\,10^{4}$ & $2.6963\,10^{4}$ \\\hline
		\end{tabular}
	\end{center}
\end{table}

\begin{figure}[tb]
	\begin{center}
		\includegraphics[width=0.7\columnwidth]{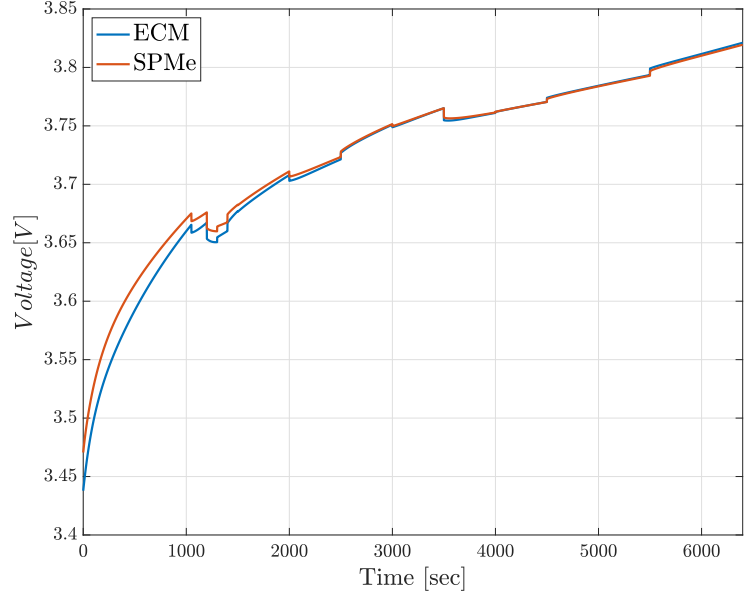}    
		\caption{Validation profile of the terminal voltage.} \label{fig_val}
	\end{center}
\end{figure}

The ECM model with the estimated parameters has been tested in validation against the SPMe, using the validation profile in Figure \ref{fig:input_training_val} and the same initial condition of the training phase. As shown in Figure \ref{fig_val}, the results are good but some discrepancy is present.
For state-estimation and control purposes, it is important to quantify how accurate the obtained parameter values are. Since $\mathbf{\hat{Y}(\bm{\phi}^*)}$ is a random variable, one has that also the estimated  $\hat{\bm{\phi}}$ is a random variable with covariance matrix $\mbf{C}_{\hat{\bm{\phi}}}\in \mathbb{R}^{8\times 8}$. Similarly to \cite{pozzi2018optimal}, we rely on the Fisher Information Matrix (FIM) to obtain an estimate of $\mbf{C}_{\hat{\bm{\phi}}}$. According to the Cramer-Rao bound, the FIM provides a lower bound on the parameters covariance matrix as follows:
\begin{equation} \label{eq:FIM}
\mbf{C}_{\hat{\phi}} \geq (\mbf{F}(\bm{\hat{\phi}}))^{-1}=((\mbf{S}(\bm{\hat{\phi}}))^\text{T}\,\mbf{C}_y^{-1}\,\mbf{S}(\bm{\hat{\phi}}))^{-1},
\end{equation}
where $\mbf{F}(\bm{\hat{\phi}}) \in \mathbb{R}^{8\times 8}$ is the FIM, and $\mbf{S}(\bm{\hat{\phi}})$ is the sensitivity matrix computed as $\mbf{S}(\bm{\hat{\phi}}) \triangleq \nabla_{\hat{\bm{\phi}}}\mbf{Y}(\bm{\hat{\phi}})$.

\begin{figure}
	\begin{center}
		\includegraphics[width=0.5\columnwidth]{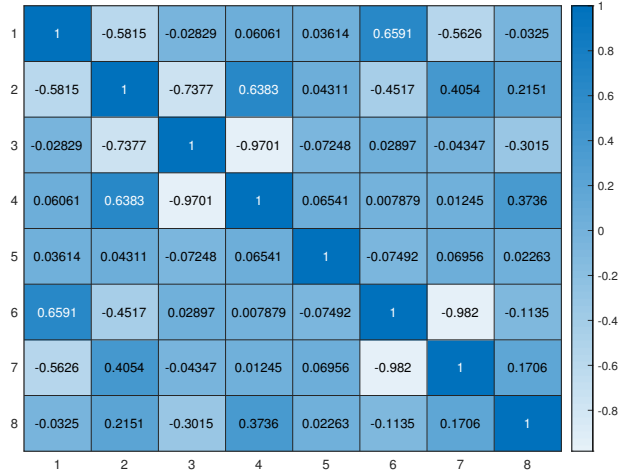}   
		\caption{Parametric correlation matrix.} 
		\label{fig:CORR}
	\end{center}
\end{figure}

\begin{remark} \rm
	Despite the FIM inverse being commonly used in the literature to quantify the parametric uncertainty, this method can be inaccurate for several reasons: (i) it provides only a lower bound of $\mbf{C}_{\hat{\bm{\phi}}}$; (ii) it relies on local parameter sensitivity (which assumes a linear relationship between model parameter variations and simulation results); (iii) it is not able to capture the non-Gaussian 
	case since it estimates only the covariance matrix; (iv) it assumes that the model structure used in the identification phase is correct. Future work will consider better estimates by using 
	global sensitivity analysis and take into account the error committed when approximating a high-fidelity model with a reduced one (see e.g.  \cite{weber2019process}).
\end{remark}

The estimated correlation matrix $\mbf{R_{\hat{\bm{\phi}}}}$\footnote{Given $\mbf{D}\triangleq\sqrt{\text{diag}(\mbf{C_{\hat{\phi}}})}$, the correlation matrix can been derived by $\mbf{R_{\hat{\phi}}}=\mbf{D}^{-1}\mbf{C_{\hat{\phi}}}\mbf{D}^{-1}$.}, obtained using the FIM, is reported in Figure \ref{fig:CORR}.
Non-zero off-diagonal values testify a correlation between the parameters in (\ref{eq:batt_phi_param}). In order to take into account this feature when using the interval state estimation of the next section, a linear change of variable is made $\hat{\bm{\psi}}=\mbf{T}\hat{\bm{\phi}}$,
where $\mbf{T}$ is chosen as the inverse  of the  eigenvector matrix of $\mathbf{C}_{\hat{\bm{\phi}}}$\footnote{Since $\mathbf{C}_{\hat{\bm{\phi}}}$ is symmetric, then it has a complete set of orthogonal eigenvectors, and therefore, $\mbf{T}$ is invertible.}. By doing so, one obtains that the covariance matrix $\mathbf{C}_{\hat{\bm{\psi}}}$ of the new parameter set $\hat{\bm{\psi}}$ is diagonal. 

The use of an unbounded distribution to describe the parametric uncertainty is overly conservative since physical electric parameters cannot be negative nor exceed 
meaningful values (it does not make any sense for the Kokam SLPB 75106100 cell we consider to have a resistance of M$\Omega$).
For this reason, in the following section, we restrict each element of the parametric uncertainty $\hat{\bm{\psi}}_i$ to the following domain
$[\hat{\psi}_i-3\,\sqrt{C_{\hat{\bm{\psi}}}(i,i)},\hat{\psi}_i+3\,\sqrt{C_{\hat{\bm{\psi}}}(i,i)}]$, thus truncating to $\pm$ 3 times the standard deviation of the parameter. These sets are further reduced by considering the physical limits of the original parameters $\hat{\bm{\phi}}$ (e.g. positivity for each SOC value). Note that, considering a reference system with a diagonal covariance matrix is very important since  intervals are not able to capture any coupling between variables.

The identification phase has been implemented using CasADi \citep{Andersson2019}. The optimization problem has been solved with IPOPT, a primal-dual interior point method interfaced with CasADi.

\section{Set-based state estimation method}\label{sec:batt_setbased}

The set-based state estimation method proposed in this chapter relies on intervals. See Section \ref{sec:pre_iarithmetic} for basic operations and notations of interval analysis.

\subsection{Problem formulation}

Consider the discrete-time nonlinear model written below
\begin{equation} \label{eq:batt_SB_MODEL}
\left \{ \begin{aligned}
\mbf{x}_k& = \mbf{f}(\mbf{x}_{k-1},\mbf{u}_{k-1},\mbf{w}_{k-1}),\\ 
\mbf{y}_k&= \mbf{g}(\mbf{x}_k,\mbf{u}_k,\mbf{v}_k), 
\end{aligned} \right.
\end{equation}
for $k \geq 1$, with $\mbf{y}_0 =\mbf{g}(\mbf{x}_0,\mbf{u}_0,\mbf{v}_0)$, $\mbf{f}: \realset^{n} \times \realset^{n_u} \times \realset^{n_w} \to \realset^{n} $, $\mbf{g}: \realset^{n} \times \realset^{n_u} \times \realset^{n_v} \to \realset^{n_y}$, where $\mbf{w} \in W$ and $\mbf{v} \in V$ are the unknown-but-bounded process and measurement disturbances, respectively, with $W$ and $V$ being intervals.

The objective is to obtain accurate interval enclosures $\hat{X}_k$ of the state variables $\mbf{x}_k$ which are consistent with the nonlinear system \eqref{eq:batt_SB_MODEL} and the measurement $\mbf{y}_k$. Given an initial set $\bar{X}_0$ such that $\mbf{x}_0 \in \bar{X}_0$, in this chapter we proceed through the well-known prediction-update algorithm, which is based on computing intervals $\bar{X}_k$ and $\hat{X}_k$ such that
\begin{align}
\bar{X}_k \supseteq & \{ \mbf{f}(\mbf{x}_{k-1}, \mbf{u}_{k-1}, \mbf{w}_{k-1}) : \mbf{x}_{k-1} \in \bar{X}_{k-1}, \, \mbf{w}_{k-1} \in W \}, \label{eq:batt_prediction0}\\
\hat{X}_k \supseteq & \{ \mbf{x}_k \in \bar{X}_k : \mbf{g}(\mbf{x}_k, \mbf{u}_k, \mbf{v}_k) = \mbf{y}_k , \, \mbf{v}_k \in V \}, \label{eq:batt_update0}
\end{align}
where \eqref{eq:batt_prediction0} is referred to as the \emph{prediction step}, and \eqref{eq:batt_update0} as the \emph{update step}. We assume that an interval enclosure $\bar{X}_0$ of the initial state $\mbf{x}_0$ and the current measurement $\mbf{y}_k$ for $k \geq 0$ are known.

In this chapter, the guaranteed state estimation is performed in an efficient way by combining inclusion functions with \emph{forward-backward constraint propagation} (FBCP) \citep{jaulin2001nonlinear}. For each time $k$, given the previous state set $\hat{X}_{k-1}$, the prediction step \eqref{eq:batt_prediction0} is performed using inclusion functions \citep{Moore2009}, resulting in the predicted interval $\bar{X}_k$ such that $\mbf{x}_k \in \bar{X}_k$. On the other hand, the update step \eqref{eq:batt_update0} is computed by solving a Constraint Satisfaction Problem (CSP).

To obtain $\hat{X}(k)$, the general problem is to refine $\bar{X}_k$ and $\bar{Y}_k$ by removing values in the respective domains that are not consistent with each other.
This corresponds to a CSP $\mathcal{H}$, which is formulated as
\begin{equation*} 
\begin{aligned}
\mathcal{H}:\mbf{y}_k= & \mbf{g}(\mbf{x}_k,\mbf{u}_k,\mbf{v}_k),\; (\mbf{x}_k,\mbf{y}_k,\mbf{v}_k) \in \bar{X}_k \times \bar{Y}_k \times V,
\end{aligned}
\end{equation*}
whose solution set is defined as $\mathcal{S}\triangleq\{ (\mbf{y}_k, \mbf{x}_k) \in \bar{Y}_k \times \bar{X}_k: \mbf{y}_k=\mbf{g}(\mbf{x}_k,\mbf{u}_k,\mbf{v}_k), \mbf{v}_k \in V\}$. To solve $\mathcal{H}$, during the forward constraint propagation phase, the state set $\bar{X}_k$ is first propagated through $\mbf{g}$ using inclusion functions, considering intermediate variables, yielding an output interval $\bar{Y}_k$ which is refined by intersecting it with the measurement $\mbf{y}_k$. During the backward propagation, the nonlinear mapping $\mbf{g}$ is swept backwards and the interval $\bar{Y}_k$ obtained in forward propagation phase is used to refine $\bar{X}_k$. The proposed FBCP algorithm for the ECM is detailed in the next section.

\subsection{State estimation of the Lithium-ion cell using FBCP}

The FBCP algorithm proposed in \cite{jaulin2001nonlinear} is composed of three intermediate steps (\emph{contractor decomposition}, \emph{forward update}, and \emph{backward update}), here applied to the ECM example. 

We consider \eqref{eq:batt_formal_disc} describing the ECM with output function $\mbf{g}$ given by \eqref{eq:batt_ECM_V}. The process disturbance $\mbf{w}_k = \mbf{T}\hat{\bm{\phi}}$ is determined by the parameters uncertainties in the ECM by means of the transformation matrix in Section \ref{sec:batt_ID}, bounded by $\mbf{w}_k \in W \triangleq \mbf{T} [\hat{\bm{\phi}}]$. 
On the other hand, the output disturbance $\mbf{v}_k$ comprises both the parameters uncertainties and output additive measurement noise $\mbf{d}_k \in D \subset \realset$, with $\mbf{d}$ defined in Section \ref{sec:batt_ID}, as $\mbf{v}(k) \triangleq [\mbf{w}(k)^T \,\; \mbf{d}(k)]^T$. Note that, since $R_0$ is correlated to the other parameters appearing in \eqref{eq:batt_phi_param} (see Figure \ref{fig:CORR}), $\mbf{v}_k$ depends also on $\mbf{w}_k$.

In order to mitigate the dependency effect that arises when considering the polynomial function \eqref{eq:batt_OCP}, $V_\text{oc}^\text{E}(k)$ is rewritten using the centered form suggested in \cite{hansen2003global}, as
\begin{equation} \label{eq:Taylor}
V_\text{oc}^\text{E}(x_{1,k}) = p_{0,k} + \sum_{i=1}^{9} p_{i,k}(x_{1,k}-c)^i,
\end{equation}
where $c \triangleq \text{mid}(\bar{X}_{1,k})$, with $x_{1,k} \triangleq z^\text{E}(k) \in \bar{X}_{1,k}$ being the first component of $\bar{X}_k$, $p_i$ being auxiliary variables given by $p_{0,k} \triangleq V_\text{oc}^\text{E}(c)$, $p_{i,k} \triangleq (1/i!)V_\text{oc}^{\text{E},[i]}(c)$, $i \in \{1,\ldots,9\}$, and $V_\text{oc}^{\text{E},[i]}(c)$ denoting the $i$-th derivative of \eqref{eq:batt_OCP} with respect to $z^\text{E}(k)$, evaluated at $c$.

\emph{1) Contractor decomposition.} The function $\mbf{g}$ derived from \eqref{eq:batt_ECM_V} is first decomposed into a ``primitive'' form comprised of simplified expressions in which only one function or one elementary operation is present. This is achieved by introducing intermediate variables $h_{i,k} \in H_{i,k}$ given by
\begin{equation} \label{eq:batt_inter}
h_{i,k} \triangleq (x_{1,k}-c)^i, \quad i \in \{1,\ldots,9\},
\end{equation}
which allow to rewrite the polynomial function \eqref{eq:batt_OCP} as
\begin{equation} \label{eq:batt_ocp_constract}
V_\text{oc}^\text{E}(k) = p_{0,k}+\sum_{i=1}^9p_{i,k} h_{i,k}.
\end{equation}
Therefore, intervals $H_{i,k}$ and $\mathcal{V}^\text{E}_k \triangleq \{V_\text{oc}^\text{E}(k) \text{ given by } \eqref{eq:batt_ocp_constract} :  h_{i,k} \in H_{i,k}\}$ are computed by evaluating \eqref{eq:batt_inter} for all $x_{1,k} \in \bar{X}_{1,k}$, and \eqref{eq:batt_ocp_constract} for all $h_{i,k} \in H_{i,k}$, respectively, using interval arithmetic.

\emph{2) Forward update.} Given the measurement $\mbf{y}(k)$, we first subtract the noise interval $[\mbf{d}]$, obtaining $\bar{Y} \triangleq \mbf{y}(k) - \mbf{d}$, for all $\mbf{d} \in D$. Then, we compute the intersection 
\begin{equation} \label{eq:batt_[Y]}
\bar{Y}_k \gets \bar{Y}_k \cap (\mathcal{V}^\text{E}_k \oplus \bar{X}_{2,k} \oplus R_0 u_k),
\end{equation}
in order to refine $\bar{Y}_k$ (Minkowski sums are obtained using interval addition). Note that if no measurement is available at time $k$, this set is initialized as $\bar{Y}_k \gets (-\infty,+\infty)$. 

\emph{3) Backward update.} The decomposed function $\mbf{g}$ is swept backwards. Every variable appearing on the right-hand-side of $\mbf{g}$ is made a explicit function of the other variables appearing in the forward update, as
\begin{subequations}\label{eq:step3}
	\begin{align} 
	\mathcal{V}^\text{E}_k & \gets \mathcal{V}^\text{E}_k \cap (\bar{Y}_k \oplus (-\bar{X}_{2,k}) \oplus (-R_0u_k)), \label{eq:back1} \\
	\hat{X}_{2,k} & \gets \bar{X}_{2,k} \cap (\bar{Y}_k \oplus (-\mathcal{V}^\text{E}_k) \oplus (-R_0u_k)). \label{eq:back2}
	\end{align}
\end{subequations}
For assessing the state of charge $\hat{X}_{1,k}$, equation \eqref{eq:batt_ocp_constract} is swept backwards and decomposed into multiple expressions, passing first through the intermediate variables $H_{i,k}$, for $i=1,\ldots,9$, as
\begin{equation} \label{eq:batt_backOCP}
\begin{aligned}[]
H_{i,k} \gets H_{i,k} \cap \Big[\frac{1}{p_{i,k}} & \Big(\mathcal{V}^\text{E}_k \oplus (-p_{0,k}) \bigoplus_{j \neq i} (- p_{j,k} H_{j,k})\Big)\Big].
\end{aligned}
\end{equation}

The interval $H_{1,k}$ is further refined as follows. Since $H_{1,k}$ is zero centered by definition (see \eqref{eq:batt_inter}), the following logic is necessary for retrieving both positive and negative even roots to $H_{1,k}$:
\begin{align*}
H_{1,k} \gets & H_{1,k} \cap_{i \in\{2,4,6,8\}} H_{i,k}^{\frac{1}{i}}  = H_{1,k} \cap_{i \in\{2,4,6,8\}}  [-|\overline{h}_{i,k}|^{\frac{1}{i}},|\overline{h}_{i,k}|^{\frac{1}{i}}].
\end{align*}
For odd roots, $$H_{1,k} \gets H_{1,k} \cap_{i \in\{3,5,7,9\}} ([\underline{\alpha}_{i,k},\overline{\alpha}_{i,k}]),$$ where (i) $\underline{\alpha}_{i,k} \triangleq \underline{h}_{i,k}\!\,^{\frac{1}{i}}$ if $h_{i,k} > 0$ for all $h_{i,k} \in H_{i,k}$, $\underline{\alpha}_{i,k} \triangleq -(|\underline{h}_{i}|^{\frac{1}{i}})$ otherwise, and (ii) $\overline{\alpha}_{i,k} \triangleq -(|\overline{h}_{i}|^{\frac{1}{i}})$ if $h_{i,k} < 0$ for all $h_{i,k} \in H_{i,k}$, $\overline{\alpha}_{i} \triangleq \overline{h}_{i}\!\,^{\frac{1}{i}}$ otherwise, with $\underline{h}_{i,k}$ and $\overline{h}_{i,k}$ being the lower and upper bounds of $H_{i,k}$, respectively. Finally, the refinement of the SOC interval is done through the intersection $\hat{X}_{1,k} \gets \bar{X}_{1,k} \cap (H_{1,k} \oplus c)$. 
The interval enclosure $\hat{X}_k$ for the update step \eqref{eq:batt_update0} is then given by $\hat{X}_k \gets \hat{X}_{1,k} \times \hat{X}_{2,k}$.

\begin{remark} \rm
	Steps 2 and 3 can be repeated iteratively to obtain a more refined enclosure $\hat{X}_k$ \citep{jaulin2001nonlinear}. Nevertheless, for the ECM, one iteration demonstrated to provide sufficiently accurate enclosures, as illustrated in Section \ref{sec:batt_results}.
\end{remark}

\begin{remark} \rm
	Since intervals cannot capture the fact that $\mbf{w}_k$ is constant, as well as the dependencies between $\mbf{v}_k$ and $\mbf{w}_k$, the application of the FBCP algorithm to the ECM is then conservative. Future work will explore other set representations such as zonotopes, which allow to effectively manage this dependency and to reduce conservatism. 
\end{remark}

\section{Numerical results}\label{sec:batt_results}

\begin{figure}
	\begin{center}
		\includegraphics[width=0.6\columnwidth]{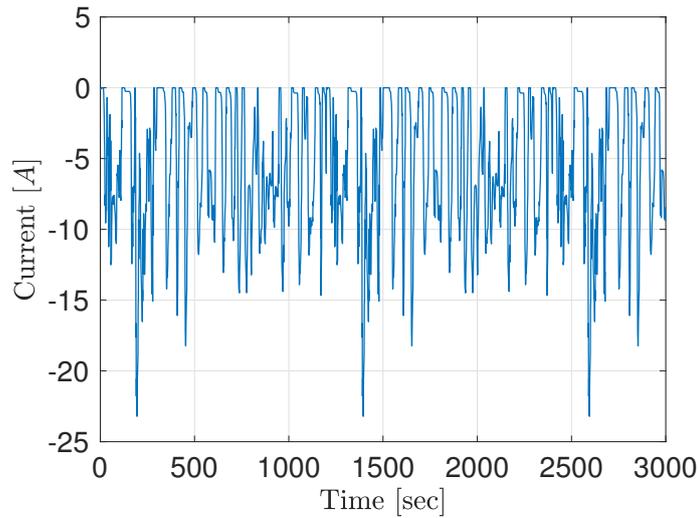}  
		\caption{Current input profile $I(t)$ applied to the ECM.} 
		\label{fig:batt_INPUT_PROF}
	\end{center}
\end{figure}

This section presents the results obtained by applying the interval state estimation method to the considered Li-ion cell during charging phase with input profile in Figure \ref{fig:batt_INPUT_PROF}.  

\begin{figure*}
	\begin{center}
		\includegraphics[width=0.9\textwidth]{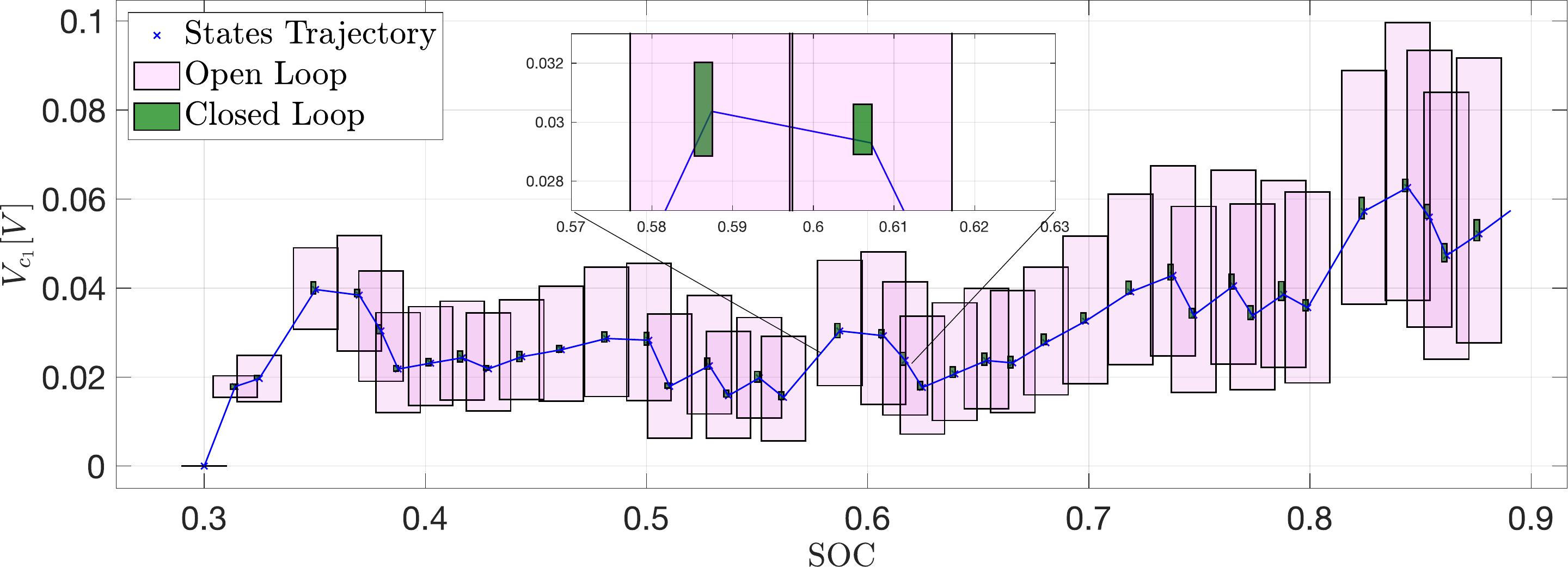} 
		\caption{Enclosures obtained using interval state estimation.} 
		\label{fig:batt_St_est}
	\end{center}
\end{figure*}

The initial state set $\bar{X}_0$ is centered in the initial state condition $(z^\text{E}(0),V_{c1}(0))=(0.3,0)$. We consider an initial state uncertainty of $1\%$ for the SOC and a much lower uncertainty on the voltage $V_{c1}$ of $0.0001\% V$. This is justified by assuming that the battery is in stationary
condition at $t=0$ after a rest period (see Remark \ref{rem:batt_initialstate}). 
The open loop state estimation is computed using only the prediction step \eqref{eq:batt_prediction0} with natural inclusion function \citep{Moore2009}, at each time step $k$. 
Figure \ref{fig:batt_St_est} depicts a comparison between states set computed in open loop (pink) and closed loop (green) respectively, highlighting the result obtained after the refinement procedure related to FBCP.

\begin{figure}
	\begin{center}
		\includegraphics[width=0.7\columnwidth]{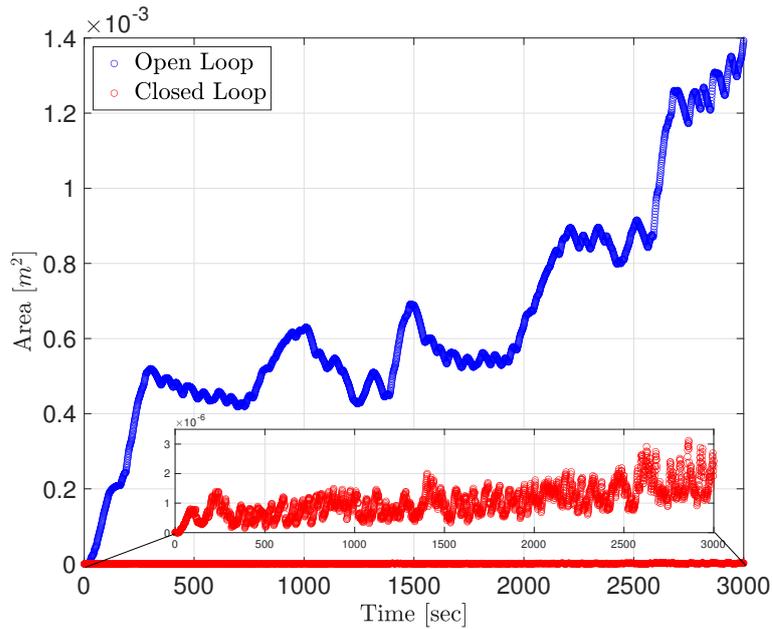}    
		\caption{Area of the estimated enclosures over time.} 
		\label{fig:batt_SURF}
	\end{center}
\end{figure}

A further comparison between open and closed loop estimation is also shown in Figure \ref{fig:batt_SURF}, in which the areas of the enclosures are calculated at each time $k$. Note that, despite the improved accuracy, the sets mildly increase over time. This is due to the increase of the SOC with time due to charging, which results in more conservative interval enclosure of the polynomial \eqref{eq:batt_ocp_constract}. As it can be noticed, the use of the FBCP approach (closed-loop) significantly reduces the size of the obtained intervals and still guarantees that the real state trajectory $\mbf{x}_k$ (depicted in blue in Figure \ref{fig:batt_St_est}) belongs to $\hat{X}_k$. 
The simulation has been done with MATLAB 2020a, while all interval operations were performed using INTLAB 9 \citep{Rump1999}. The plots in Figure \ref{fig:batt_St_est} has been done using MPT \citep{kvasnica2004multi}.

\section{Final remarks}\label{sec:batt_conclusions}
This chapter developed a discrete-time interval observer for a single cell Li-ion battery based on \cite{Jaulin2001}. Parameters have been identified by mean of least squares method using voltage data collected on an SPMe. Parametric uncertainties have been derived exploiting FIM, whose inverse approximates the covariance of the parameters. A variable transformation of the parameters have been performed to take their correlation into account. Simulation results demonstrated tight enclosures on the states over time, which is corroborated by comparing open-loop and closed-loop state estimation. %

\chapter{Conclusions}\thispagestyle{headings} \label{cha:conclusions}

This chapter summarizes the main contributions obtained in this doctoral thesis, and concludes the text. Future work proposals are also presented and detailed at the end of this chapter. 

\section{Overview and contributions}

This doctoral thesis developed new methods for set-based state estimation and active fault diagnosis of nonlinear discrete-time systems with unknown-but-bounded uncertainties, discrete-time nonlinear systems whose trajectories are known to satisfy nonlinear equality constraints (called invariants), set-based state estimation and active fault diagnosis of linear descriptor systems, and joint state and parameter estimation of nonlinear descriptor systems. Set-based estimation aims to compute tight enclosures of the possible system states in each time step subject to unknown-but-bounded uncertainties. Most existing methods employ a standard prediction-update framework with set-based prediction and update steps based on various set representations and techniques. However, achieving accurate enclosures for nonlinear systems remains a significant challenge. When these enclosures are represented by simple sets such as intervals, ellipsoids, parallelotopes, and zonotopes, certain set operations can be very conservative. Yet, using general convex polytopes is much more computationally demanding. 

To address this challenge, this thesis presented new methods for efficiently propagating constrained zonotopes through nonlinear mappings:
\begin{itemize}
	\item Chapter \ref{cha:nonlineardynamics} developed three new different approaches for computing the prediction step \eqref{eq:ndyn_prediction0} based on constrained zonotopes. The first method (Section \ref{sec:ndyn_czib}) combined properties of interval arithmetic and constrained zonotopes, leading to a highly tunable and accurate state estimation algorithm. The other methods (Sections \ref{sec:ndyn_meanvalue} and \ref{sec:ndyn_firstorder}) expanded the tools proposed in \cite{Scott2016} to the class of nonlinear discrete-time systems \eqref{eq:ndyn_system}. 
	\item These methods extended, in a consistent way, two existing approaches for propagating zonotopes through nonlinear mappings. The key advantage of these new extensions is that they allowed the entire state estimation procedure to be done using constrained zonotopes in CG-rep. Therefore, the update step could be done by exact intersection (with linear measurements), which is known to generate highly asymmetrical sets that cannot be accurately enclosed by ellipsoids, intervals, parallelotopes, and zonotopes. 
	\item Using the methods proposed in Chapter \ref{cha:nonlineardynamics}, such sets could be directly propagated to the next time step without prior simplification to a symmetric set. This overcomes a major source of conservatism in existing methods based on the aforementioned enclosures, while largely retaining the efficiency of computations with zonotopes. 
\end{itemize}

In addition, this thesis improved the standard prediction-update framework for systems with invariants by adding a consistency step. This new step used invariants to reduce conservatism and was enabled by new algorithms for refining constrained zonotopes based on nonlinear constraints. Also, new update algorithms were developed allowing nonlinear measurement equations in CZ-based state estimation for the first time:
\begin{itemize}
	\item Chapter \ref{cha:nonlinearmeasinv} presented new methods for set-valued state estimation of discrete-time nonlinear systems whose trajectories are known to satisfy nonlinear equality constraints, called invariants. 
	\item The new consistency step used invariants to reduce conservatism and is enabled by new algorithms for refining constrained zonotopes based on nonlinear constraints. 
	\item Chapter \ref{cha:nonlinearmeasinv} also presented significant improvements to prediction and update steps proposed in Chapter \ref{cha:nonlineardynamics}. Specifically, new update algorithms were developed that allow nonlinear measurement equations, and existing prediction methods based on conservative approximation techniques were modified to allow a more flexible choice of the approximation point, which led to tighter enclosures. 
	\item Numerical results demonstrated that the proposed methods can provide significantly tighter enclosures than existing zonotope-based methods while maintaining comparable efficiency.
\end{itemize}

This thesis also introduced a new approach for set-based active fault diagnosis of a class of nonlinear discrete-time systems. The new enclosures based on constrained zonotopes are employed for passive fault detection of a class of nonlinear discrete-time systems:

\begin{itemize}
	\item Chapter \ref{cha:faultdiagnosis} proposed an adaptation of the first-order Taylor extension (Section \ref{sec:ndyn_firstorder}) to obtain an affine parametrization of the reachable sets of systems with nonlinear dynamics. Based on this affine parametrization of reachable sets, a new method for input design aiming active fault diagnosis of nonlinear discrete-time systems has been proposed. The method was based on the open-loop fault diagnosis approach proposed in \cite{Scott2014} for linear systems using zonotopes. \item Chapter \ref{cha:faultdiagnosis} also extended the active fault diagnosis method to constrained zonotopes, based on the approach proposed in \cite{Raimondo2016}. 
	
	\item The resulting problem consisted in solving an MIQP to find the optimal input sequence that allows guaranteed fault diagnosis in a given time horizon. The performance of the proposed active fault diagnosis method was demonstrated in a numerical experiment, which consisted in fault diagnosis of a robotic manipulator subject to actuator faults. The diagnosis of the current possible faults was guaranteed in the experiment by injecting the designed optimal input sequence into the nonlinear system.
\end{itemize}

In addition, this thesis presented new methods based on constrained zonotopes for set-valued state estimation and active fault diagnosis of linear descriptor systems, besides a new set representation, called mixed zonotope:

\begin{itemize}
	\item Chapter \ref{cha:descriptor} proposed novel algorithms for set-based state estimation and AFD of linear descriptor systems with unknown-but-bounded uncertainties. The first methods used constrained zonotopes. In contrast to other sets representations, linear static constraints on the state variables, typical of descriptor systems, can be directly incorporated in the mathematical description of constrained zonotopes. Thanks to this feature, set-based methods based on CZs could provide less conservative enclosures for the trajectories of linear descriptor systems.
	
	\item Chapter \ref{cha:descriptor} proposed also a new representation for unbounded sets based on zonotopes, called mixed zonotopes. These are a generalization of zonotopes capable of describing unbounded sets, which allowed to develop methods for state estimation and active fault diagnosis of linear descriptor systems without assuming the knowledge of an admissible set that encloses all the possible trajectories of the system. 	
	
	\item The new methods based on constrained zonotopes and mixed zonotopes led to significantly tighter results than zonotope approaches. In addition, active fault diagnosis was enabled without assuming rank properties on the structure of the system. The effectiveness of the new methods was corroborated by numerical examples.
	
\end{itemize}

This thesis also presented a new method for set-based joint state and parameter estimation of nonlinear discrete-time systems:

\begin{itemize}
	\item Chapter \ref{cha:stateparameter} developed a new method for set-based joint state and parameter estimation of nonlinear discrete-time systems with algebraic variables and unknown model parameters. This method extended the nonlinear state estimation methods using constrained zonotopes proposed in Chapter \ref{cha:nonlinearmeasinv} to include the estimation of both algebraic variables and unknown parameters in a unified framework.
	
	\item Unlike other methods in the literature, the proposed method maintained existing dependencies between states, algebraic variables, and unknown parameters, which resulted in the accuracy of both state and parameter estimation being significantly improved. The effectiveness of the new methodology was highlighted in two numerical examples, with the second example being a nonlinear descriptor system.
	
\end{itemize}
 
Lastly, this thesis presented the application of the proposed set-based state estimation and fault diagnosis methods using constrained zonotopes to unmanned aerial vehicles, water distribution networks, and a lithium-ion cell:

\begin{itemize}
	\item Chapter \ref{cha:appuav}, initially, dealt with the problem of path tracking of a suspended load using a tilt-rotor UAV. Such task requires the knowledge of the load position, which may not be provided by available sensors. Therefore, to accomplish the path tracking, a linear set-based state estimator based on constrained zonotopes was proposed to solve the problem of estimating the load position and orientation, considering sensors with different sampling times, and unknown-but-bounded disturbances. In order to provide feedback to the controller using the estimated state set, an optimal choice of the estimated point was given according to a proposed constrained minimum-variance criterion. The path tracking was then solved by a discrete-time mixed $\mathcal{H}_2/\mathcal{H}_\infty$ controller with pole-placement constraints. Results from numerical experiments, performed in a platform based on the Gazebo simulator with a CAD model of the system were presented to corroborate the performance of the linear set-based state estimator along with the designed control strategy. The second application was a numerical experiment considering nonlinear state estimation of a quadrotor UAV, taking into account a realistic scenario in which measurements are provided by sensors located at the UAV. The third application considered a numerical experiment involving the optimal input design for nonlinear active fault diagnosis of a quadrotor UAV subject to actuator and sensor faults, which addressed scenarios of complete actuator failure and the existence of sensor bias.
	
	\item Chapter \ref{cha:appwdn} proposes two new methods for set-based state estimation in WDNs, considering unknown-but-bounded measurement and model uncertainties to calculate bounds on the system states. A new interval method was proposed based on iterative computation of tight enclosures of the nonlinear head-loss functions, and bounding the solution of the algebraic equations using rescaling. This method, denoted by ISAR, was mildly more conservative than the IHISE method previously proposed in \cite{Vrachimis2019}, but significantly more computationally efficient. However, since intervals are not capable of capturing the dependencies between state variables, Chapter \ref{cha:appwdn} also proposed the use of constrained zonotopes as an additional step to both IHISE and ISAR.
	This led to two new algorithms for set-based state estimation of WDNs, capable of capturing the dependencies between hydraulic states, which result in sets with significantly smaller volumes than intervals. The benefits of constrained zonotopes were also highlighted when the new enclosures were used for leakage detection. These provided higher leakage detection rates compared to intervals, as demonstrated in two case studies using benchmark WDNs.	
	
	\item Chapter \ref{cha:appbat} developed a discrete-time interval state estimator for a single Li-ion cell, based on the forward-backward method described in \cite{jaulin2001nonlinear}. In particular, parametric uncertainties were considered and obtained by performing an identification procedure on data collected on a well known EM, i.e. the SPMe. Among the contributions of the chapter, were: (i) the identification of ECM parameter bounds based on the Fisher Information Matrix, which was performed by the PhD candidate Diego Locatelli, being supervised by Prof. Davide M. Raimondo; and (ii) a discrete-time interval state estimation method based on inclusion functions and constraint propagation, which handles discrete-time measurement, and is the contribution concerning this doctoral thesis. Numerical experiments showed that the proposed methodology is efficient and provided accurate enclosures for both the SOC and the electric state variables of the ECM.
\end{itemize}

\section{Future work}

This section describes some possible directions for future work derived from the contributions presented in this doctoral thesis.

\begin{itemize}
	\item \emph{Efficient implementation of the nonlinear state estimation methods via parallelization}. The methods developed in Chapters \ref{cha:nonlineardynamics} and \ref{cha:nonlinearmeasinv} were able to provide more accurate enclosures than existing zonotopic methods in the literature. However, the proposed methods have an increased computational burden, which may prevent their application in some practical systems. To circumvent this issue, an immediate future work is the implementation of the proposed algorithms using parallelization procedures in Graphics Processing Units (GPUs). This would allow for reduced computation times, and therefore the application of the proposed methods in an experimental setup for unmanned aerial vehicles, for instance.
	
	\item \emph{Active fault diagnosis of discrete-time systems with nonlinear dynamics with reduced conservatism}. Chapter \ref{cha:faultdiagnosis} developed an active fault diagnosis method for discrete-time systems with nonlinear dynamics based on the affine parameterization of reachable sets in the inputs, using the first-order Taylor extension proposed in Chapter \ref{cha:nonlineardynamics}. However, the affine parameterization obtained requires the evaluation of the reachable sets over the entire admissible input set, which may result in severe conservatism. Future work would seek different parameterizations of the reachable sets such that this conservatism is reduced.
	
	\item \emph{Active fault diagnosis of discrete-time systems with nonlinear dynamics, nonlinear measurement and nonlinear invariants}. The active fault diagnosis method proposed in Chapter \ref{cha:faultdiagnosis} was developed for discrete-time systems assuming nonlinear dynamics and linear measurement. In this way, the reachable sets could be approximated using the first-order Taylor extension proposed in Chapter \ref{cha:nonlineardynamics}. A direct continuation of this research is to rather consider the first-order Taylor extension developed in Chapter \ref{cha:nonlinearmeasinv}, which allows also for nonlinear measurement and takes into account the existence of nonlinear invariants. In addition, the choice of approximation point would be improved with respect to the current active fault diagnosis method, since it would be allowed to belong to the interval hull of the previous enclosure, as proposed in Chapter \ref{cha:nonlinearmeasinv}.
	
	\item \emph{Closed-loop active fault diagnosis of nonlinear discrete-time systems.} The active fault diagnosis method proposed in Chapter \ref{cha:faultdiagnosis} is an open-loop structure, i.e., it does not make use of measurements obtained by injecting the optimal input into the system to improve the diagnosis procedure. After achieving the results suggested in the two future work proposals above, a continuation of this research is to extend the developed fault diagnosis method to a closed-loop structure as proposed in \cite{Raimondo2016}, using a moving horizon scheme and nonlinear observers to shorten the input sequence required to perform the fault diagnosis.
	
	\item \emph{Closed-loop fault diagnosis of linear descriptor systems.} Similarly to the proposal above, the active fault diagnosis methods developed in Chapter \ref{cha:descriptor} using constrained zonotopes and mixed zonotopes are open-loop structures. A direct continuation of this research is to develop closed-loop methods for linear descriptor systems by using a moving-horizon scheme and set-based observers to take into account the measurements obtained during the fault diagnosis procedure.
	\item \emph{State estimation and fault diagnosis of nonlinear systems using mixed zonotopes.} A new representation for unbounded sets has been proposed in Chapter \ref{cha:descriptor}, referred to as mixed zonotopes. This new set representation allowed for state estimation and fault diagnosis of linear descriptor systems without assuming the knowledge of an admissible state set. Moreover, in the state estimation case, it allowed also for obtaining accurate enclosures of the system states without assuming the knowledge of a bounded initial set. A future work proposal is to investigate the development of state estimation and fault diagnosis methods for nonlinear systems using mixed zonotopes. The main challenge in the development of such methods is the evaluation of the remainder set in the enclosure approximations, since the previous set may be unbounded.
	\item \emph{Investigate the conditions in which mixed zonotopes are able to provide bounded enclosures for linear descriptor systems}. The results obtained in Chapter \ref{cha:descriptor} showed that it was possible to obtain bounded enclosures for the trajectories of linear descriptor systems using mixed zonotopes. However, this boundedness property is not guaranteed by the proposed methods. A direct continuation of this research is to investigate which are the necessary conditions and properties of the linear descriptor system which result in bounded enclosures for the trajectories of the states.	
\end{itemize}

\def\thispagestyle#1{}
\bibliographystyle{apalike2}

\bibliography{\bibfolder/masterthesis_bib,\bibfolder/appendices_bib,\bibfolder/UAVControl_bib,\bibfolder/BackgroundHist_bib,\bibfolder/Surveys_bib,\bibfolder/PassiveFTC_bib,\bibfolder/ActiveFTC_bib,\bibfolder/UAVFTC,\bibfolder/SetTheoretic_bib,\bibfolder/SetTheoreticFTCFDI_bib,\bibfolder/Davide_bib,\bibfolder/paperAutomatica_bib,\bibfolder/paperCDC_bib,\bibfolder/paperECC_bib,\bibfolder/paperIFAC_bib,\bibfolder/paperNonlinearMeas_bib,\bibfolder/Robotic_bib,\bibfolder/stelios_bibliography,\bibfolder/phdthesis_bib,\bibfolder/Diego_bib,\bibfolder/paperParameter_bib,\bibfolder/BackgroundHist_bib}

\appendix

\chapter{Derivation of computational complexities}\thispagestyle{headings} \label{app:computationalcomplexity}

This appendix details the derivation of the computational complexities of basic operations on zonotopes and constrained zonotopes, and the set-based state estimators presented in this doctoral thesis.

\section{Zonotopes}

\subsection{Basic operations}

\subsubsection{Linear image}

Let $Z = \{\mbf{G}_z,\mbf{c}_z\} \subset \realset^n$, $\mbf{R} \in \realsetmat{n_r}{n}$, with $\mbf{c}_z \in \realset^n$, $\mbf{G}_z \in \realsetmat{n}{n_g}$, and consider the linear image $\mbf{R}Z = \{\mbf{R}\mbf{G}_z, \mbf{R} \mbf{c}_z\}$. Then,
\begin{equation*}
O(\mbf{R}Z) = O(\mbf{R}\mbf{c}_z) + O(\mbf{R}\mbf{G}_z) = O(nn_r) + O(nn_gn_r) = O(nn_gn_r).
\end{equation*}

\subsubsection{Minkowski sum}

Let $Z = \{\mbf{G}_z,\mbf{c}_z\} \subset \realset^n$, $W = \{\mbf{G}_w,\mbf{c}_w\} \subset \realset^n$, with $\mbf{c}_z \in \realset^n$, $\mbf{G}_z \in \realsetmat{n}{n_g}$, $\mbf{c}_w \in \realset^n$, $\mbf{G}_z \in \realsetmat{n}{n_{g_w}}$, and consider the Minkowski sum $$Z \oplus W =\left\{ \begin{bmatrix} \mathbf{G}_z \,\; \mathbf{G}_w \end{bmatrix}, \mathbf{c}_z + \mathbf{c}_w \right\}.$$ Then,
\begin{equation*}
O(Z \oplus W) = O(\mbf{c}_z + \mbf{c}_w) = O(n).
\end{equation*}

\subsubsection{Generator reduction}

Let $Z = \{\mbf{G},\mbf{c}\} \subset \realset^n$, with $\mbf{c} \in \realset^n$, $\mbf{G} \in \realsetmat{n}{n_g}$. Let $k_g \leq (n_g-n)$ denote the number of generators to be eliminated. The computational complexity of generator reduction of zonotopes is the same presented in \cite{Scott2016} and \cite{Scott2018}:
\begin{equation*}
O(\text{eliminate }k_g\text{ generators from }Z) = O(n^2n_g + k_g n_g n).
\end{equation*}

\subsubsection{Interval hull}

Let $Z = \{\mbf{G},\mbf{c}\} \subset \realset^n$, with $\mbf{c} \in \realset^n$, $\mbf{G} \in \realsetmat{n}{n_g}$. Consider the interval hull of $Z$ given by $\left[\mbf{c} - \bm{\zeta}, \mbf{c} + \bm{\zeta}\right]$, where \citep{Kuhn1998}
\begin{equation*}
\zeta_i = \sum_{j=1}^{n_g} |\mbf{G}_{ij}|.
\end{equation*}
Then,
\begin{align*}
O(\text{interval hull of } Z) & = O(\mbf{c} - \bm{\zeta}) + O(\mbf{c}+ \bm{\zeta}) + nO\big(\sum_{j=1}^{n_g} |\mbf{G}_{ij}|\big) = O(\mbf{c} + \bm{\zeta}) + nO\big(\sum_{j=1}^{n_g} |\mbf{G}_{ij}|\big)\\
& = O(n) +  nO(n_g) = O(nn_g).
\end{align*}

\subsubsection{Zonotope inclusion}

Let $Z = \mbf{p} \oplus \mbf{M} B_\infty^{n_g}$, with $\mbf{p} \in \realset^n$, $\mbf{M} \in \intvalsetmat{n}{n_g}$. Consider the zonotope inclusion $\diamond(Z) = \mbf{p} \oplus [\midpoint{\mbf{M}} \,\; \mbf{P}] B_\infty^{n_g+n}$, where $P_{ii} = \sum_{j=1}^{n_g} \frac{1}{2} \text{diam}(M_{ij})$, $P_{ij} = 0$ for $i \neq j$ \citep{Alamo2005a}. Then,
\begin{align*}
O(\diamond(Z)) & = O(\midpoint{\mbf{M}}) + O(\mbf{P}) = O(\midpoint{\mbf{M}}) + nO(P_{ii}) \\
& = O(nn_g) + nO(n_g) = O(nn_g).
\end{align*}

\subsection{ZMV}

\subsubsection{Prediction step}

Let $\bm{\mu} : \realset^n \times \realset^{n_w} \to \realset^n$ be continuously differentiable, and let $\nabla_x \bm{\mu}$ denote the gradient of $\bm{\mu}$ with respect to its first argument. Let $\hat{X}_{k-1} = \{\mbf{G}_x,\mbf{c}_x\} \subset \realset^n$, and $W = \{\mbf{G}_w,\mbf{c}_w\} \subset \realset^{n_w}$ with $\mbf{c}_x \in \realset^n$, $\mbf{G}_x \in \realsetmat{n}{n_g}$, $\mbf{c}_w \in \realset^{n_w}$, and $\mbf{G}_w \in \realsetmat{n_w}{n_{g_w}}$. Let $Z$ be a zonotope such that $\bm{\mu}(\mbf{c}_x,W) \subseteq Z$, and let $\mathbf{M} \in \intvalsetmat{n}{n_g}$ be an interval matrix satisfying $\nabla^T_x \bm{\mu}(\hat{X}_{k-1}, W) \mbf{G_x} \subseteq \mathbf{M}$. According to Theorem 4 in \cite{Alamo2005a}, $$\bm{\mu}(\hat{X}_{k-1},W) \subseteq \bar{X}_k = Z \oplus \diamond(\mbf{M} B_\infty^{n_g}),$$ in which $\bar{X}_{k}$ has $n_g + n_{g_w} + 2n$ generators if $Z$ is also computed by the mean value extension. Therefore,
\begin{align*}
O(\text{ZMV}_\text{prediction}) & = O(Z \oplus \diamond(\mbf{M} B_\infty^{n_g})) + O(\diamond(\mbf{M} B_\infty^{n_g})) + O(Z) + O(\text{interval hull of }\hat{X}_{k-1},W) \\
& = O(n) + O(nn_g + n^2n_g) + O(n_wn_{g_w} + nn_wn_{g_w}) + O(nn_g + n_wn_{g_w}) \\
& = O(n^2n_g + nn_wn_{g_w}).
\end{align*}

\subsubsection{Update step}

Let $\mbf{y}_k = \mbf{C} \mbf{x}_k + \mbf{D}_u \mbf{u}_k + \mbf{D}_v \mbf{v}_k$, with $\mbf{y}_k \in \realset^{n_y}$, $\mbf{x}_k \in \realset^{n}$, $\mbf{u}_k \in \realset^{n_u}$, $\mbf{v}_k \in \realset^{n_v}$. Consider the zonotopes $\bar{X}_k = \{\mbf{G}_x, \mbf{c}_x\} \subset \realset^n$, $V = \{\mbf{G}_v, \mbf{c}_v\} \subset \realset^{n_v}$ with $\mbf{c}_x \in \realset^n$, $\mbf{G}_x \in \realsetmat{n}{\bar{n}_g}$, $\mbf{c}_v \in \realset^{n_v}$, and $\mbf{G}_v \in \realsetmat{n_v}{n_{g_v}}$. For simplicity, define $m_g = n_g + n_{g_w}$. Following the intersection method proposed in \cite{Bravo2006}, the update step is computed iteratively using one row of $\mbf{y}_k$ at a time, as described below.

\paragraph{Algorithm: Update step}
\begin{enumerate}[(1)]
	\item Assign $\tilde{X}_k = \{\tilde{\mbf{G}}_x, \tilde{\mbf{c}}_x\} \gets \bar{X}_k$, $i \gets 1$.
	\item Assign $\tilde{y} \gets i$-th row of $\mbf{y}_k$, $\mbf{p}^T \gets i$-th row of $\mbf{C}$, $\mbf{d}_u^T \gets i$-th row of $\mbf{D}_u$, $\mbf{d}_v^T \gets i$-th row of $\mbf{D}_v$.
	\item Compute the strip $S \gets \{\mbf{x} : |\mbf{p}^T\mbf{x} - d| \leq \sigma\}$, where $d = \tilde{y} - \mbf{d}_u^T\mbf{u}_k - \mbf{d}_v \mbf{c}_v$, $\sigma = \sum_{\ell=1}^{n_{g_v}}\sum_{j=1}^{n_v} |(\mbf{D}_v)_{ij} (\mbf{G}_v)_{j\ell}|$.
	\item Compute the tight strip $\tilde{S} \gets \{\mbf{x} : |\mbf{p}^T\mbf{x} - \tilde{d}| \leq \tilde{\sigma}\}$, where
	\begin{align*}
	\tilde{\sigma} = \frac{1}{2}(\bar{\sigma}^\text{U}- \bar{\sigma}^\text{L}), \quad \tilde{d} = d + \frac{1}{2}(\bar{\sigma}^\text{U} + \bar{\sigma}^\text{L}),
	\end{align*}
	with $\bar{\sigma}^\text{L} = \max\{-\sigma, \mbf{p}^T\tilde{\mbf{c}}_x - \|\tilde{\mbf{G}}_x^T\mbf{p}\|_1 - d\} $, $\bar{\sigma}^\text{U} = \min\{\sigma, \mbf{p}^T\tilde{\mbf{c}}_x + \|\tilde{\mbf{G}}_x^T\mbf{p}\|_1 - d\}$.%
	\item Assign $\tilde{X}_k \gets \tilde{X}_k \cap \tilde{S} = \{\bm{G}(j^*), \mbf{c}(j^*)\}$, where
	\begin{equation*}
	j^* = \text{arg } \underset{j}{\text{min}} \det(\mbf{G}(j)\mbf{G}(j)^T),	
	\end{equation*}
	with $j \in \{0,1,\ldots,\bar{n}_g\}$, %
	\begin{align*}
	& \mbf{c}(j) = \left\{ \begin{aligned} & \tilde{\mbf{c}}_x + \left(\frac{\tilde{d} - \mbf{p}^T\tilde{\mbf{c}}_x}{\mbf{p}^T(\tilde{\mbf{G}}_x)_{:,j}}\right) (\tilde{\mbf{G}}_x)_{:,j} & & \quad \text{ if }1\leq j \leq \bar{n}_g \text{ and } \mbf{p}^T(\tilde{\mbf{G}}_x)_{:,j} \neq 0 \\
	& \tilde{\mbf{c}}_x & & \quad \text{ otherwise} \end{aligned} \right. \\
	& 	\mbf{G}(j) = \left\{ \begin{aligned} & [\mbf{g}^j_1 \,\; \mbf{g}^j_2 \,\; \ldots \mbf{g}^j_{\bar{n}_g}] & & \quad \text{ if }1\leq j \leq \bar{n}_g \text{ and } \mbf{p}^T(\tilde{\mbf{G}}_x)_{:,j} \neq 0 \\
	& \tilde{\mbf{G}}_x & & \quad \text{ otherwise} \end{aligned} \right. \\
	& 	\mbf{g}^j_\ell = \left\{ \begin{aligned} & (\tilde{\mbf{G}}_x)_{:,\ell} - \left(\frac{\mbf{p}^T(\tilde{\mbf{G}}_x)_{:,\ell}}{\mbf{p}^T(\tilde{\mbf{G}}_x)_{:,j}}\right) (\tilde{\mbf{G}}_x)_{:,j} & & \quad \text{ if } \ell \neq j \\
	& \left(\frac{\tilde{\sigma}}{\mbf{p}^T(\tilde{\mbf{G}}_x)_{:,j}}\right) (\tilde{\mbf{G}}_x)_{:,j} & & \quad \text{ if }\ell = j \end{aligned} \right.
	\end{align*}
	\item If $i < n_y$, go to Step 2 and assign $i \gets i + 1$. Otherwise, assign $\hat{X}_k \gets \tilde{X}_k$.
\end{enumerate}

The computational complexity is
\begin{align*}
O(\text{ZMV}_\text{update}) & = n_y O(\text{update with the }i\text{-th row of }\mbf{y}_k) \\
& = n_y\left(O(\text{Step (3)}) + O(\text{Step (4)}) + O(\text{Step (5)}) \right).
\end{align*}

But,
\begin{align*}
O(\text{Step (3)}) & = O(1) + O(n_u) + O(n_v) + O(n_vn_{g_v}) = O(n_u + n_vn_{g_v}),
\end{align*}
\begin{align*}
O(\text{Step (4)}) & = O(1) + O(n) + O(n\bar{n}_g) = O(n \bar{n}_g),
\end{align*}
\begin{align*}
O(\text{Step (5)}) & = \bar{n}_gO(n^3) + \bar{n}_gO(n^2\bar{n}_g) + \bar{n}_gO(n) + \bar{n}_gO(n\bar{n}_g) = O(n^3\bar{n}_g + n^2 \bar{n}_g^2).
\end{align*}

Since $\bar{n}_g = m_g + 2n$, hence
\begin{align*}
O(\text{ZMV}_\text{update}) & = n_y\left( O(n_u + n_vn_{g_v}) + O(n \bar{n}_g) + O(n^3\bar{n}_g + n^2 \bar{n}_g^2 + n \bar{n}_g^2) \right) \\
& = n_y\left( O(n^3\bar{n}_g + n^2 \bar{n}_g^2 + n_u + n_vn_{g_v}) \right) \\
& = O(n_y(n^3(m_g + n) + n^2 (m_g+n)^2 + n_u + n_vn_{g_v})).
\end{align*}

\subsubsection{Order reduction}

Let $n_g$ denote the desired number of generators of $\hat{X}_k$ after order reduction, i.e., $\hat{X}_{k}$ must have the same complexity as $\hat{X}_{k-1}$. Order reduction is performed by eliminating $k_g$ generators from $\hat{X}_k$. For simplicity, define $m_g = n_g + n_{g_w}$. The prediction and update steps of the ZMV lead to $\hat{X}_{k}$ with $m_g + 2n$ generators. Therefore, $k_g = n_{g_w} + 2n$.  Consequently,
\begin{align*}
O(\text{ZMV}_\text{reduction}) & = O(\text{eliminate }k_g\text{ generators from }\hat{X}_k) \\
& = O(n^2(m_g + n) + k_g (m_g + n)n) \\ 
& = O(n^2(m_g + n) + (n_{g_w} + n) (m_g + n)n) = O(n^2(m_g + n) + n(n_{g_w} + n) (m_g + n)). 
\end{align*}

\subsection{ZFO}

\subsubsection{Prediction step}

Let $\bm{\eta}: \realset^{m} \to \realset^{n}$ be of class $\mathcal{C}^2$, and let $\mathbf{z} = (\mbf{x}_{k-1}, \mbf{w}_{k-1}) \in \realset^{m}$ denote its argument, where $m = n + n_w$. Let $Z = \hat{X}_{k-1} \times W = \{\mathbf{G}, \mathbf{c}\} \subset \realset^{m}$ be a zonotope with $m_g = n_g + n_{g_w}$ generators. For each $q = 1,2,\dots,n$, let $\mathbf{Q}^{[q]}\in\mathbb{IR}^{m\times m}$ and $\tilde{\mathbf{Q}}^{[q]}\in\mathbb{IR}^{m_g\times m_g}$ be interval matrices satisfying $\mathbf{Q}^{[q]} \supseteq \mathbf{H} \eta_q (Z)$ and $\tilde{\mathbf{Q}}^{[q]} \supseteq \mathbf{G}^T \mathbf{Q}^{[q]} \mathbf{G}$. Moreover, define $\tilde{\mbf{c}}$, $\tilde{\mbf{G}}$, $\tilde{\mbf{G}}_\mbf{d}$ as in Theorem 3. Then, the method in \cite{Combastel2005} ensures that
\begin{equation*}%
\bm{\eta}(Z) \subseteq \bar{X}_k = \bm{\eta}(\mbf{c}) \oplus \nabla^T \bm{\eta}(\mathbf{c})(\mbf{G}B_\infty^{m_g}) \oplus \tilde{\mathbf{c}} \oplus \left[ \tilde{\mathbf{G}} \,\; \tilde{\mathbf{G}}_{\mathbf{d}} \right] B_\infty^{\tilde{m}_g},
\end{equation*}
where $\tilde{m}_g = (1/2)m_g^2 + (3/2)m_g + n$, and therefore $\bar{X}_k$ has $(1/2)m_g^2 + (5/2)m_g + n$ generators. Thus,
\begin{align*}
O(\text{CZFO}_\text{prediction}) & = O(\bm{\eta}(\mbf{c}) \oplus \nabla^T \bm{\eta}(\mathbf{c})(\mbf{G}B_\infty^{m_g}) \oplus \tilde{\mathbf{c}} \oplus \left[ \tilde{\mathbf{G}} \,\; \tilde{\mathbf{G}}_{\mathbf{d}} \right] B_\infty^{\tilde{m}_g}) \\
& = O(n) + O(\tilde{\mbf{c}}) + O(\tilde{\mbf{G}}) + O(\tilde{\mbf{G}}_\mbf{d}) + nO(\tilde{\mbf{Q}}^{[q]}) +  O(\text{interval hull of }Z) \\
& = O(n) + O(nm_g) + O(nm_g^2) + O(nm_g^2) + nO(m^2m_g + mm_g^2) + O(mm_g) \\
& = O(n(m^2m_g + mm_g^2)).
\end{align*}

\subsubsection{Update step}

Let $\mbf{y}_k = \mbf{C} \mbf{x}_k + \mbf{D}_u \mbf{u}_k + \mbf{D}_v \mbf{v}_k$, with $\mbf{y}_k \in \realset^{n_y}$, $\mbf{x}_k \in \realset^{n}$, $\mbf{u}_k \in \realset^{n_u}$, and $\mbf{v}_k \in \realset^{n_v}$. Consider the zonotopes $\bar{X}_k = \{\mbf{G}_x, \mbf{c}_x\} \subset \realset^n$ and $V = \{\mbf{G}_v, \mbf{c}_v\} \subset \realset^{n_v}$, with $\mbf{c}_x \in \realset^n$, $\mbf{G}_x \in \realsetmat{n}{\bar{n}_g}$, $\mbf{c}_v \in \realset^{n_v}$, and $\mbf{G}_v \in \realsetmat{n_v}{n_{g_v}}$. For simplicity, define $m_g = n_g + n_{g_w}$. The ZFO also follows the intersection method proposed in \cite{Bravo2006}. Therefore, the update step is equivalent to the ZMV update, but with $\bar{n}_g = (1/2)m_g^2 + (5/2)m_g + n$. Hence
\begin{align*}
O(\text{ZFO}_\text{update}) & = n_y\left( O(n^3\bar{n}_g + n^2 \bar{n}_g^2 + n_u + n_vn_{g_v}) \right) \\
& = O(n_y(n^3(m_g^2 + n) + n^2 (m_g^2+n)^2 + n_u + n_vn_{g_v})).
\end{align*}

\subsubsection{Order reduction}

Let $n_g$ denote the desired number of generators of $\hat{X}_k$ after order reduction, i.e., $\hat{X}_{k}$ must have the same complexity as $\hat{X}_{k-1}$. Order reduction is performed by eliminating $k_g$ generators from $\hat{X}_k$. For simplicity, define $m_g = n_g + n_{g_w}$. The prediction and update steps of ZFO lead to $\hat{X}_{k}$ with $(1/2)m_g^2 + (5/2)m_g + n$ generators. Therefore, $k_g = (1/2)m_g^2 + (5/2)m_g + n - n_g$.  Consequently,
\begin{align*}
O(\text{ZFO}_\text{reduction}) & = O(\text{eliminate }k_g\text{ generators from }\hat{X}_k) \\
& = O(n^2(m_g^2 + m_g + n) + k_g (m_g^2 + m_g + n)n) \\ 
& = O(n^2(m_g^2 + n) + (m_g^2 + n) (m_g^2 + n)n) \\
& = O(n^2(m_g^2 + n) + n(m_g^2 + n)^2). 
\end{align*}

\section{Constrained zonotopes}

\subsection{Basic operations}

\subsubsection{Linear image}

Let $Z = \{\mbf{G}_z,\mbf{c}_z,\mbf{A}_z,\mbf{b}_z\} \subset \realset^n$, $\mbf{R} \in \realsetmat{n_r}{n}$, with $\mbf{c}_z \in \realset^n$, $\mbf{G}_z \in \realsetmat{n}{n_g}$, $\mbf{A}_z \in \realsetmat{n_c}{n_g}$, $\mbf{b}_z \in \realset^{n_c}$, and consider the linear image $\mbf{R}Z = \{\mbf{R}\mbf{G}_z, \mbf{R} \mbf{c}_z, \mbf{A}_z, \mbf{b}_z\}$. Then,
\begin{equation*}
O(\mbf{R}Z) = O(\mbf{R}\mbf{c}_z) + O(\mbf{R}\mbf{G}_z) = O(nn_r) + O(nn_gn_r) = O(nn_gn_r).
\end{equation*}

\subsubsection{Minkowski sum}

Let $Z = \{\mbf{G}_z,\mbf{c}_z,\mbf{A}_z,\mbf{b}_z\} \subset \realset^n$, $W = \{\mbf{G}_w,\mbf{c}_w,\mbf{A}_w,\mbf{b}_w\} \subset \realset^n$, with $\mbf{c}_z \in \realset^n$, $\mbf{G}_z \in \realsetmat{n}{n_g}$, $\mbf{A}_z \in \realsetmat{n_c}{n_g}$, $\mbf{b}_z \in \realset^{n_c}$, $\mbf{c}_w \in \realset^n$, $\mbf{G}_w \in \realsetmat{n}{n_{g_w}}$, $\mbf{A}_w \in \realsetmat{n_{c_w}}{n_{g_w}}$, $\mbf{b}_w \in \realset^{n_{c_w}}$, and consider the Minkowski sum $$Z \oplus W =\left\{ \begin{bmatrix} \mathbf{G}_z \,\; \mathbf{G}_w \end{bmatrix}, \mathbf{c}_z + \mathbf{c}_w, \begin{bmatrix} \mathbf{A}_z & \bm{0} \\ \bm{0} & \mathbf{A}_w \end{bmatrix}, \begin{bmatrix} \mathbf{b}_z \\ \mathbf{b}_w \end{bmatrix} \right\}.$$ Then,
\begin{equation*}
O(Z \oplus W) = O(\mbf{c}_z + \mbf{c}_w) = O(n).
\end{equation*}

\subsubsection{Generalized intersection}

Let $Z = \{\mbf{G}_z,\mbf{c}_z,\mbf{A}_z,\mbf{b}_z\} \subset \realset^n$, $Y =\{\mbf{G}_y, \mbf{c}_y, \mbf{A}_y, \mbf{b}_y\}$, $\mbf{R} \in \realsetmat{n_r}{n}$, with $\mbf{G}_z \in \realsetmat{n}{n_g}$, $\mbf{c}_z \in \realset^n$, $\mbf{A}_z \in \realsetmat{n_c}{n_g}$, $\mbf{b}_z \in \realset^{n_c}$, $\mbf{G}_y \in \realsetmat{n_r}{n_{g_r}}$, $\mbf{c}_y \in \realset^{n_r}$, $\mbf{A}_y \in \realsetmat{n_{c_r}}{n_{g_r}}$, $\mbf{b}_y \in \realset^{n_{c_r}}$. Consider the generalized intersection $$Z \cap_{\mathbf{R}} Y = \left\{ \begin{bmatrix} \mathbf{G}_z \,\; \bm{0} \end{bmatrix}, \mathbf{c}_z, \begin{bmatrix} \mathbf{A}_z & \bm{0} \\ \bm{0} & \mathbf{A}_y \\ \mathbf{R} \mathbf{G}_z & -\mathbf{G}_y \end{bmatrix}, \begin{bmatrix} \mathbf{b}_z \\ \mathbf{b}_y \\ \mathbf{c}_y - \mathbf{R} \mathbf{c}_z \end{bmatrix} \right\}.$$ Then,
\begin{align*}
O(Z \cap_\mbf{R} Y) = O(\mbf{R}\mbf{G}_z) + O(-\mbf{G}_y) + O(\mbf{c}_y - \mbf{R} \mbf{c}_z) & = O(nn_gn_r) + O(n_rn_{g_r}) + O(n_r + nn_r) \\
& = O(nn_gn_r + n_rn_{g_r}).
\end{align*}

\subsubsection{Generator reduction}

Let $Z = \{\mbf{G},\mbf{c},\mbf{A},\mbf{b}\} \subset \realset^n$, with $\mbf{c} \in \realset^n$, $\mbf{G} \in \realsetmat{n}{n_g}$, $\mbf{A} \in \realsetmat{n_c}{n_g}$, $\mbf{b} \in \realset^{n_c}$. Let $k_g \leq n_g-(n+n_c)$ denote the number of generators to be eliminated. The computational complexity of generator reduction of constrained zonotopes is the same presented in \cite{Scott2016} and \cite{Scott2018}, with reduction operated over the lifted zonotope $$Z^+ = \left\{ \begin{bmatrix} \mbf{G} \\ \mbf{A} \end{bmatrix}, \begin{bmatrix} \mbf{c} \\ - \mbf{b} \end{bmatrix} \right\}.$$ Therefore,
\begin{equation*}
O(\text{eliminate }k_g\text{ generators from }Z) = O((n+n_c)^2n_g + k_g n_g (n+n_c)).
\end{equation*}

\subsubsection{Constraint elimination}

Let $Z = \{\mbf{G},\mbf{c},\mbf{A},\mbf{b}\} \subset \realset^n$, with $\mbf{c} \in \realset^n$, $\mbf{G} \in \realsetmat{n}{n_g}$, $\mbf{A} \in \realsetmat{n_c}{n_g}$, $\mbf{b} \in \realset^{n_c}$.  Let $k_c \leq n_c$ denote the number of constraints to be eliminated from $Z$. Following the constraint elimination algorithm in \cite{Scott2016}, for each eliminated constraint the remaining constraints are first preconditioned through Gauss-Jordan elimination with full pivoting and then subjected to a rescaling procedure before the next constraint is eliminated. Therefore,
\begin{align*}
O(\text{eliminate }1\text{ constraint from }Z) & = O(\text{pre-conditioning}) + O(\text{rescaling}) \\
& + O(\text{eliminate the constraint}).
\end{align*}

From the Appendix in \cite{Scott2016}, we have that
\begin{equation*}
O(\text{pre-conditioning}) + O(\text{rescaling}) = O(n_c^2 n_g + n_cn_g^2).
\end{equation*}

To eliminate the constraint, one of the generators of $Z$ is first chosen by minimizing an approximation of the Hausdorff distance between $Z$ and the set obtained by constraint elimination. This requires solving \eqref{eq:pre_hausdorff} for each $j \in \{1,2,\ldots,n_g\}$, which has the reported complexity $O((n_g+n_c)^3)$, and computing the Hausdorff distance $H_j^* = \|\mbf{G}\mbf{d}_j^*\|_2^2 + \|\mbf{d}_j^*\|_2^2$, with $\mbf{d}_j^*$ the solution of \eqref{eq:pre_hausdorff} for each $j$. This is done with the following complexity:
\begin{align*}
O(\text{eliminate the constraint}) & = O(\text{Hausdorff distance minimization}) + n_gO(H_j^*)\\ & = O((n_g+n_c)^3) +n_gO(nn_g) =  O((n_g+n_c)^3 + nn_g^2).
\end{align*}

Then,
\begin{equation*}
O(\text{eliminate }1\text{ constraint from }Z) = O(n_c^2 n_g + n_cn_g^2) + O((n_g+n_c)^3 + nn_g^2) = O((n_g+n_c)^3 + nn_g^2).
\end{equation*}

Finally, the complexity of eliminating $k_c$ constraints from $Z$ can be bounded by
\begin{align*}
O(\text{eliminate }k_c\text{ constraints from }Z) & = k_cO(\text{eliminate }1\text{ constraint from }Z) \\
& = O(k_c(n_g+n_c)^3 + k_cnn_g^2).
\end{align*}

\subsubsection{Interval hull (Property 1)}

Let $Z = \{\mbf{G},\mbf{c},\mbf{A},\mbf{b}\} \subset \realset^n$, with $\mbf{c} \in \realset^n$, $\mbf{G} \in \realsetmat{n}{n_g}$, $\mbf{A} \in \realsetmat{n_c}{n_c}$, $\mbf{b} \in \realset^{n_c}$. Consider the interval hull of $Z$ computed as in Property 1. In the standard form, each LP in Property 1 has $N_d = n_g+1$ decision variables and $N_c = 2n_g+n_c+1$ constraints. Then,
\begin{equation*}
O(\text{interval hull of }Z) = 2n O(N_d N_c^3) {=} O(n N_d N_c^3) {=} O(n(1{+}n_g)(1{+}2n_g{+}n_c)^3) = O(nn_g(n_g+n_c)^3).
\end{equation*}

\subsubsection{CZ-inclusion (Theorem 1)}

Let $X = \mbf{p} \oplus \mbf{M} B_\infty(\mbf{A},\mbf{b}) \subset \realset^m$, $\mbf{J} \in \intvalsetmat{n}{m}$, $\bar{X} = \bar{\mbf{p}} \oplus \bar{\mbf{M}}B_\infty^{\bar{n}_g} \supseteq X$, $\mbf{m} \supseteq (\mbf{J} - \midpoint{\mbf{J}}) \bar{\mbf{p}}$, $\triangleleft(\mbf{J},X) = \midpoint{\mbf{J}} X \oplus \mbf{P} B_\infty^n$, with $\mbf{p} \in \realset^m$, $\mbf{M} \in \realsetmat{m}{n_g}$, $\mbf{A} \in \realsetmat{n_c}{n_g}$, $\mbf{b} \in \realset^{n_c}$, $P_{ii} = \frac{1}{2} \diam{m_i} + \frac{1}{2} \sum_{j=1}^{\bar{n}_g} \sum_{k=1}^{m} \text{diam}(J_{ik})|\bar{M}_{kj}|$, $P_{ij} = 0$ for $i \neq j$. Assume that $\bar{X}$ is obtained by eliminating $n_c$ constraints from $X$, and $\mbf{m}$ is computed using interval arithmetic. Then,
\begin{align*}
O(\triangleleft(\mbf{J},X)) & = O(\midpoint{\mbf{J}} X \oplus \mbf{P} B_\infty^n) + O((\mbf{J} - \midpoint{\mbf{J}}) \bar{\mbf{p}}) + nO\Big((1/2) \diam{m_i} \\
& \quad + \frac{1}{2} \sum_{j=1}^{\bar{n}_g} \sum_{k=1}^{m} \text{diam}(J_{ik})|\bar{M}_{kj}|\Big)+ O(\text{eliminate }n_c\text{ constraints from }X) \\
& = O(nm + nmn_g + n) + O(nm + nm) + nO(\bar{n}_gm) + O(n_c(n_g+n_c)^3 + n_cmn_g^2) \\
& = O(nmn_g) + O(nm) + O(nmn_g) + O(n_c(n_g+n_c)^3 + mn_cn_g^2) \\
& = O(nmn_g + n_c(n_g+n_c)^3 + mn_g^2n_c).
\end{align*}

\subsubsection{Closest point (Proposition 1)}

Let $Z = \{\mbf{G},\mbf{c},\mbf{A},\mbf{b}\} \subset \realset^n$, $\mbf{h} \in \realset^{n}$, with $\mbf{c} \in \realset^n$, $\mbf{G} \in \realsetmat{n}{n_g}$, $\mbf{A} \in \realsetmat{n_c}{n_g}$, $\mbf{b} \in \realset^{n_c}$. In the standard form, the LP in Proposition 1 has $N_d = n+n_g+1$ decision variables and $N_c = 2n+2n_g+n_c+1$ constraints. Then,
\begin{align*}
O(\text{Proposition 1}) = O(\mbf{c} + \mbf{G} \bm{\xi}^*) {+} O(N_dN_c^3) & = O(n + nn_g) {+} O((n+n_g+1)(2n+2n_g+n_c+1)^3) \\ & = O((n+n_g)(n+n_g+n_c)^3).
\end{align*}

\subsubsection{Change center (Proposition 2)}

Let $$Z = \{\mbf{G},\mbf{c},\mbf{A},\mbf{b}\} = \left\{ \begin{bmatrix} \mbf{G}\mbf{E}_r & \bm{0} \end{bmatrix},\mbf{h},\begin{bmatrix} \mbf{A} \mbf{E}_r & \bm{0} \\ \bm{0} & \mbf{A} \\ \mbf{G} \mbf{E}_r & - \mbf{G} \end{bmatrix}, \begin{bmatrix} \mbf{b} - \mbf{A} \bm{\xi}_\text{m} \\ \mbf{b} \\ \mbf{c} - \mbf{h} \end{bmatrix} \right\} \subset \realset^n,$$ with $\mbf{c} \in \realset^n$, $\mbf{G} \in \realsetmat{n}{n_g}$, $\mbf{A} \in \realsetmat{n_c}{n_g}$, $\mbf{b} \in \realset^{n_c}$, $\mbf{h} \in \realset^{n}$, $\mbf{E}_r = \frac{1}{2} \text{diag}(\bm{\xi}^\text{U} - \bm{\xi}^\text{L})$, $\bm{\xi}_\text{m} = \half(\bm{\xi}^\text{U} + \bm{\xi}^\text{L})$. Consider $(\bm{\xi}^\text{L},\bm{\xi}^\text{U})$ computed by solving the LP in Proposition 2, in which $(\tilde{\bm{\xi}}^\text{L},\tilde{\bm{\xi}}^\text{U})$ is obtained by Algorithm 1 in the Appendix in \cite{Scott2016}. In the standard form, the LP in Proposition 2 has $N_d = 3n_g$ decision variables and $N_c = n+4n_g$ constraints. Then,
\begin{align*}
O( \text{Proposition 2}) & = O(\mbf{G}\mbf{E}_r) + O(\mbf{A}\mbf{E}_r) + O(-\mbf{G}) + O(\mbf{b} - \mbf{A} \bm{\xi}_\text{m}) + O(\mbf{c}-\mbf{h}) \\
& \quad + O((1/2) \text{diag}(\bm{\xi}^\text{U} - \bm{\xi}^\text{L})) + O((1/2)(\bm{\xi}^\text{U} + \bm{\xi}^\text{L})) + O(N_dN_c^3) \\ &\quad + O(\text{Algorithm 1})\\
& = O(nn_g^2) + O(n_cn_g^2) + O(nn_g) + O(n_c) + O(n_cn_g) + O(n) + O(n_g) + O(n_g) \\
& \quad + O(N_dN_c^3) + O(\text{Algorithm 1}) \\
& = O(nn_g^2) + O(n_cn_g^2) + O(N_dN_c^3) + O(\text{Algorithm 1}) \\
& = O(nn_g^2) + O(n_cn_g^2) + O((3n_g)(n+4n_g)^3) + O(\text{Algorithm 1}) \\
& = O(nn_g^2) + O(n_cn_g^2) + O(n_g(n+n_g)^3) + O(\text{Algorithm 1}) \\
& = O(n_g(n+n_g)^3 + n_g^2n_c) + O(n_g^2 n_c) \\
& = O(n_g(n+n_g)^3 + n_g^2n_c).
\end{align*}

\subsection{CZMV}

\subsubsection{Prediction step (Theorem 2)}

Let $\bm{\mu} : \realset^n \times \realset^{n_w} \to \realset^n$ be continuously differentiable, and let $\nabla_x \bm{\mu}$ denote the gradient of $\bm{\mu}$ with respect to its first argument. Let $\hat{X}_{k-1} = \{\mbf{G}_x,\mbf{c}_x,\mbf{A}_x,\mbf{b}_x\} \subset \realset^n$ and $W = \{\mbf{G}_w,\mbf{c}_w,\mbf{A}_w,\mbf{b}_w\} \subset \realset^{n_w}$, with $\mbf{c}_x \in \realset^n$, $\mbf{G}_x \in \realsetmat{n}{n_g}$, $\mbf{A}_x \in \realsetmat{n_c}{n_g}$, $\mbf{b}_x \in \realset^{n_c}$, $\mbf{c}_w \in \realset^{n_w}$, $\mbf{G}_w \in \realsetmat{n_w}{n_{g_w}}$, $\mbf{A}_w \in \realsetmat{n_{c_w}}{n_{g_w}}$, and $\mbf{b}_w \in \realset^{n_{c_w}}$, and choose any $\mathbf{h} \in \hat{X}_{k-1}$. Let $Z$ be a constrained zonotope such that $\bm{\mu}(\mathbf{h},W) \subseteq Z$, and let $\mathbf{J} \in \intvalsetmat{n}{n}$ be an interval matrix satisfying $\nabla^T_x \bm{\mu}(\hat{X}_{k-1}, W)\subseteq \mathbf{J}$. Theorem 2 ensures that $$\bm{\mu}(\hat{X}_{k-1},W) \subseteq \bar{X}_k = Z \oplus \gzinclusion(\mathbf{J},  \hat{X}_{k-1} - \mathbf{h}),$$ where $\bar{X}_{k}$ has $n_g + n_{g_w} + 2n$ generators and $n_c + n_{c_w}$ constraints if $Z$ is also computed by the mean value extension. Therefore,
\begin{align*}
O(\text{CZMV}_\text{prediction}) & = O(Z \oplus \gzinclusion(\mathbf{J},  \hat{X}_{k-1} - \mathbf{h})) + O(\gzinclusion(\mathbf{J},  \hat{X}_{k-1} - \mathbf{h})) + O(Z) \\
& \quad + O(\text{interval hull of }\hat{X}_{k-1},W) \\
& = O(n) + O(n^2n_g + n_c(n_g+n_c)^3 + nn_g^2n_c) + O(nn_wn_{g_w} + n_{c_w}(n_{g_w}+n_{c_w})^3 \\ & \quad + n_wn_{g_w}^2n_{c_w}) + O(nn_g(n_g+n_c)^3 + n_wn_{g_w}(n_{g_w}+n_{c_w})^3) \\
& =O(n^2n_g + nn_wn_{g_w} + (nn_g+n_c)(n_g+n_c)^3 + (n_wn_{g_w}+n_{c_w})(n_{g_w}+n_{c_w})^3).
\end{align*}

\subsubsection{Update step}

Let $\bar{X}_{k} = \{\mbf{G}_x,\mbf{c}_x,\mbf{A}_x,\mbf{b}_x\} \subset \realset^n$, $V = \{\mbf{G}_v,\mbf{c}_v,\mbf{A}_v,\mbf{b}_v\} \subset \realset^{n_w}$, $\mbf{u}_k \in \realset^{n_u}$, and $\mbf{y}_k \in \realset^{n_y}$ with $\mbf{c}_x \in \realset^n$, $\mbf{G}_x \in \realsetmat{n}{\bar{n}_g}$, $\mbf{A}_x \in \realsetmat{\bar{n}_c}{\bar{n}_g}$, $\mbf{b}_x \in \realset^{\bar{n}_c}$, $\mbf{c}_v \in \realset^{n_v}$, $\mbf{G}_v \in \realsetmat{n_v}{n_{g_v}}$, $\mbf{A}_v \in \realsetmat{n_{c_v}}{n_{g_v}}$, and $\mbf{b}_w \in \realset^{n_{c_v}}$. The update step is performed according to$$\hat{X}_k = \bar{X}_k \cap_{\mathbf{C}} ((\mathbf{y}_k - \mathbf{D}_u \mathbf{u}_k) \oplus (-\mathbf{D}_v V)),$$ where $\mbf{C} \in \realsetmat{n_y}{n}$, $\mbf{D}_u \in \realsetmat{n_y}{n_u}$, and $\mbf{D}_v \in \realsetmat{n_y}{n_v}$. As a result, from the prediction step of CZMV, $\bar{X}_k$ has $\bar{n}_g = n_g + n_{g_w} + 2n$ generators and $\bar{n}_c = n_c + n_{c_w}$ constraints. Therefore,
\begin{align*}
O(\text{CZMV}_\text{update}) & = O(\bar{X}_k \cap_{\mathbf{C}} ((\mathbf{y}_k - \mathbf{D}_u \mathbf{u}_k) \oplus (-\mathbf{D}_v V))) \\ & = O(n(n_g+n_{g_w} + 2n)n_y + n_yn_{g_v}) + O(n_y) + O(n_y + n_yn_u) + O(n_yn_vn_{g_v}) \\ & = O(n_yn(m_g+n) + n_yn_u + n_yn_vn_{g_v}), 
\end{align*}
where $m_g = n_g + n_{g_w}$.

\subsubsection{Order reduction}

Let $n_g$ and $n_c$ denote the desired number of generators and constraints of $\hat{X}_k$ after order reduction, i.e., $\hat{X}_{k}$ must have the same complexity as $\hat{X}_{k-1}$. Order reduction is performed by eliminating $k_c$ constraints and then $k_g$ generators from $\hat{X}_k$. For simplicity, define $m_g = n_g + n_{g_w}$, $m_c = n_c + n_{c_w}$, $\delta_n = n - n_y$, $\delta_w = n_{g_w} - n_{c_w}$, and $\delta_v = n_{g_v} - n_{c_v}$. The prediction and update steps of CZMV lead to $\hat{X}_{k}$ with $n_g + n_{g_w} + 2n + n_{g_v}$ generators and $n_c + n_{c_w} + n_{c_v} + n_y$ constraints. Therefore, $k_c = n_{c_w} + n_{c_v} + n_y$, and $k_g = n_{g_w} + 2n + n_{g_v} - k_c = n_{g_w} + 2n + n_{g_v} - n_{c_w} - n_{c_v} - n_y$.  Consequently,
\begin{align*}
O(\text{eliminate }k_c\text{ constraints from }\hat{X}_k) & = O(k_c(m_g + 2n + n_{g_v} + m_c + n_{c_v} + n_y)^3 \\ & \quad + k_cn(m_g + 2n + n_{g_v})^2) \\
& =  O((n_{c_w} + n_{c_v} + n_y)(m_g+m_c+n_{g_v} +n_{c_v}+n+n_y)^3 \\ & \quad + (n_{c_w} + n_{c_v} + n_y)n(m_g + n_{g_v} + n)^2) \\
& = O((n_{c_w} + n_{c_v} + n_y)(m_g+m_c+n_{g_v} +n_{c_v}+n+n_y)^3.
\end{align*}

Note that by eliminating $k_c$ constraints from $\hat{X}_k$, the same number of generators (i.e., $k_c$) is also removed \citep{Scott2016}. Hence,
\begin{align*}
O(\text{elim. }k_g\text{ gen. from }\hat{X}_k\text{ after const. elim.}) & = O((n+n_c)^2(m_g + 2n + n_{g_v} - k_c) \\& \quad + k_g (m_g + 2n + n_{g_v} - k_c)(n+n_c)) \\ 
& = O((n+n_c)^2(\delta_n + \delta_w + \delta_v + n_g) \\& \quad + (\delta_n + \delta_w + \delta_v) (\delta_n + \delta_w + \delta_v + n_g)(n+n_c)).
\end{align*}

Finally,
\begin{align*}
O(\text{CZMV}_\text{reduction}) & = O(\text{eliminate }k_c\text{ constraints from }\hat{X}_k) \\
& \quad + O(\text{elim. }k_g\text{ gen. from }\hat{X}_k\text{ after const. elim.}) \\ & = O((n_{c_w} + n_{c_v} + n_y)(m_g+m_c+n_{g_v} +n_{c_v}+n+n_y)^3 \\& \quad + (n+n_c)^2(n_g + \delta_n + \delta_w + \delta_v)  + (n+n_c)(\delta_n + \delta_w + \delta_v) (n_g + \delta_n + \delta_w + \delta_v)).
\end{align*}

\subsection{CZFO}

\subsubsection{Prediction step}

Let $\bm{\eta}: \realset^{m} \to \realset^{n}$ be of class $\mathcal{C}^2$, and let $\mathbf{z} = (\mbf{x}_{k-1}, \mbf{w}_{k-1}) \in \realset^{m}$ denote its argument, where $m = n + n_w$. Let $Z = \hat{X}_{k-1} \times W = \{\mathbf{G}, \mathbf{c}, \mathbf{A}, \mathbf{b}\} \subset \realset^{m}$ be a constrained zonotope	 with $m_g = n_g + n_{g_w}$ generators and $m_c = n_c + n_{c_w}$ constraints. For each $q = 1,2,\dots,n$, let $\mathbf{Q}^{[q]}\in\mathbb{IR}^{m\times m}$ and $\tilde{\mathbf{Q}}^{[q]}\in\mathbb{IR}^{m_g\times m_g}$ be interval matrices satisfying $\mathbf{Q}^{[q]} \supseteq \mathbf{H} \eta_q (Z)$ and $\tilde{\mathbf{Q}}^{[q]} \supseteq \mathbf{G}^T \mathbf{Q}^{[q]} \mathbf{G}$. Moreover, define $\tilde{\mbf{c}}$, $\tilde{\mbf{G}}$, $\tilde{\mbf{G}}_\mbf{d}$, $\tilde{\mbf{A}}$, $\tilde{\mbf{b}}$, and $\mbf{L}$, as in Theorem 3. Choosing any $\mbf{h} \in Z$, Theorem 3 ensures that
\begin{equation*}
\bm{\eta}(Z) \subseteq \bar{X}_k = \bm{\eta}(\mathbf{h}) \oplus \nabla^T \bm{\eta}(\mathbf{h})(Z - \mathbf{h}) \oplus R,
\end{equation*}
where $R = \tilde{\mathbf{c}} \oplus \left[ \tilde{\mathbf{G}} \,\; \tilde{\mathbf{G}}_{\mathbf{d}} \right] B_\infty(\tilde{\mathbf{A}}, \tilde{\mathbf{b}}) \oplus \gzinclusion (\mathbf{L}, (\mathbf{c} - \mathbf{h}) \oplus 2\mathbf{G} B_\infty(\mathbf{A},\mathbf{b}) )$ and $\bar{X}_k$ has $(1/2)m_g^2 + (5/2)m_g + 2n$ generators and $(1/2)m_c^2 + (5/2)m_c$ constraints. Therefore,
\begin{align*}
O(\text{CZFO}_\text{prediction}) & = O(\bm{\eta}(\mathbf{h}) \oplus \nabla^T \bm{\eta}(\mathbf{h})(Z - \mathbf{h}) \oplus R) + O(R) \\
& =O(n) + O(nmm_g) + O(R) = O(nmm_g) + O(R).
\end{align*}

But
\begin{align*}
O(R) & = O(\tilde{\mathbf{c}} \oplus \left[ \tilde{\mathbf{G}} \,\; \tilde{\mathbf{G}}_{\mathbf{d}} \right] B_\infty(\tilde{\mathbf{A}}, \tilde{\mathbf{b}}) \oplus \gzinclusion (\mathbf{L}, (\mathbf{c} - \mathbf{h}) \oplus 2\mathbf{G} B_\infty(\mathbf{A},\mathbf{b}) )) \\
& = O(n) + O(\tilde{\mbf{c}}) + O(\tilde{\mbf{G}}) + O(\tilde{\mbf{G}}_\mbf{d}) + O(\tilde{\mbf{A}}) + O(\tilde{\mbf{b}}) + O(\mbf{L}) \\
& \quad + O(\gzinclusion (\mathbf{L}, (\mathbf{c} - \mathbf{h}) \oplus 2\mathbf{G} B_\infty(\mathbf{A},\mathbf{b}))) + nO(\tilde{\mbf{Q}}^{[q]}) +  O(\text{interval hull of }Z) \\
& = O(n) + O(nm_g) + O(nm_g^2) + O(nm_g^2) + O(m_gm_c^2 + m_g^2m_c^2) + O(m_gm_c^2) + O(nm^2) \\ & \quad + O(n + nmm_g + m_c(m_g+m_c)^3 + mm_g^2m_c) + nO(m^2m_g + mm_g^2) + O(mm_g(m_g+m_c)^3) \\
& = O(n(m^2m_g + mm_g^2) + (mm_g+m_c)(m_g+m_c)^3).
\end{align*}

Therefore,
\begin{align*}
O(\text{CZFO}_\text{prediction}) = O(nmm_g) + O(R) = O(n(m^2m_g + mm_g^2) + (mm_g+m_c)(m_g+m_c)^3).
\end{align*}

\subsubsection{Update step}

Let $\bar{X}_{k} = \{\mbf{G}_x,\mbf{c}_x,\mbf{A}_x,\mbf{b}_x\} \subset \realset^n$, $V = \{\mbf{G}_v,\mbf{c}_v,\mbf{A}_v,\mbf{b}_v\} \subset \realset^{n_w}$, $\mbf{u}_k \in \realset^{n_u}$, and $\mbf{y}_k \in \realset^{n_y}$ with $\mbf{c}_x \in \realset^n$, $\mbf{G}_x \in \realsetmat{n}{\bar{n}_g}$, $\mbf{A}_x \in \realsetmat{\bar{n}_c}{\bar{n}_g}$, $\mbf{b}_x \in \realset^{\bar{n}_c}$, $\mbf{c}_v \in \realset^{n_v}$, $\mbf{G}_v \in \realsetmat{n_v}{n_{g_v}}$, $\mbf{A}_v \in \realsetmat{n_{c_v}}{n_{g_v}}$, and $\mbf{b}_w \in \realset^{n_{c_v}}$. The update step is performed according to$$\hat{X}_k = \bar{X}_k \cap_{\mathbf{C}} ((\mathbf{y}_k - \mathbf{D}_u \mathbf{u}_k) \oplus (-\mathbf{D}_v V)),$$ where $\mbf{C} \in \realsetmat{n_y}{n}$, $\mbf{D}_u \in \realsetmat{n_y}{n_u}$, and $\mbf{D}_v \in \realsetmat{n_y}{n_v}$. As a result, from the prediction step of CZFO, $\bar{X}_{k}$ has $\bar{n}_g = (1/2)m_g^2 + (5/2)m_g + 2n$ generators and $\bar{n}_c = (1/2)m_c^2 + (5/2)m_c$ constraints, where $m_g = n_g + n_{g_w}$, and $m_c = n_c + n_{c_w}$. Therefore,
\begin{align*}
O(\text{CZFO}_\text{update}) & = O(\bar{X}_k \cap_{\mathbf{C}} ((\mathbf{y}_k - \mathbf{D}_u \mathbf{u}_k) \oplus (-\mathbf{D}_v V))) \\ & = O(n((1/2)m_g^2 {+} (5/2)m_g + 2n)n_y + n_yn_{g_v}) {+} O(n_y) {+} O(n_y + n_yn_u) {+} O(n_yn_vn_{g_v}) \\ & = O(n_yn(m_g^2 + n) + n_yn_u + n_yn_vn_{g_v}).
\end{align*}

\subsubsection{Order reduction}

Let $n_g$ and $n_c$ denote the desired number of generators and constraints of $\hat{X}_k$ after order reduction, i.e., $\hat{X}_{k}$ must have the same complexity as $\hat{X}_{k-1}$. Order reduction is performed by eliminating $k_c$ constraints and then $k_g$ generators from $\hat{X}_k$. For simplicity, define $m_g = n_g + n_{g_w}$, $m_c = n_c + n_{c_w}$, $\delta_n = n - n_y$, $\delta_v = n_{g_v} - n_{c_v}$, and $\tilde{\delta} = m_g^2 - m_c^2$. The prediction and update steps of CZFO lead to $\hat{X}_{k}$ with $(1/2)m_g^2 + (5/2)m_g +  2n + n_{g_v}$ generators and $(1/2)m_c^2 + (5/2)m_c + n_{c_v} + n_y$ constraints. Therefore, $k_c = (1/2)m_c^2 + (5/2)m_c + n_{c_v} + n_y - n_c$ and $k_g = (1/2)m_g^2 + (5/2)m_g +  2n + n_{g_v} - n_g - k_c = (1/2)m_g^2 + (5/2)m_g +  2n + n_{g_v} - n_g - (1/2)m_c^2 - (5/2)m_c - n_{c_v} - n_y + n_c$.  Consequently,
\begin{align*}
O(\text{eliminate }k_c\text{ constraints from }\hat{X}_k) & = O(k_c(m_g^2 + m_g + n + n_{g_v} + m_c^2 + m_c + n_{c_v} + n_y)^3 \\ & \quad + k_cn(m_g^2 + m_g + n)^2) \\
& = O((m_c^2 + n_{c_v} + n_y)(m_g^2 + n + n_{g_v} + m_c^2 + n_{c_v} + n_y)^3 \\ & \quad + (m_c^2 + n_{c_v} + n_y)n(m_g^2 + n)^2) \\
& = O((m_c^2 + n_{c_v} + n_y)(m_g^2 + m_c^2 + n_{g_v} + n_{c_v} + n + n_y)^3.
\end{align*}

Once again, by eliminating $k_c$ constraints from $\hat{X}_k$, the same number of generators (i.e., $k_c$) is also removed \citep{Scott2016}. Hence,
\begin{align*}
O(\text{elim. }k_g\text{ gen. from }\hat{X}_k\text{ after const. elim.}) & = O((n+n_c)^2(m_g^2 + n + n_{g_v} - m_c^2 - n_{c_v} - n_y) \\& \quad + k_g (m_g^2 + n + n_{g_v} - m_c^2 - n_{c_v} - n_y)(n+n_c)) \\ 
& = O((n+n_c)^2(\tilde{\delta} + \delta_n + \delta_v) \\& \quad + (\tilde{\delta}+ \delta_n + \delta_v) (\tilde{\delta} + \delta_n + \delta_v)(n+n_c)) \\ 
& = O((n+n_c)^2(\tilde{\delta} {+} \delta_n {+} \delta_v) {+} (\tilde{\delta} {+} \delta_n {+} \delta_v)^2 (n{+}n_c)).
\end{align*}

Finally,
\begin{align*}
O(\text{CZFO}_\text{reduction}) & = O(\text{eliminate }k_c\text{ constraints from }\hat{X}_k) \\
& \quad + O(\text{elim. }k_g\text{ gen. from }\hat{X}_k\text{ after const. elim.}) \\ & = O((m_c^2 + n_{c_v} + n_y)(m_g^2 + m_c^2 + n_{g_v} + n_{c_v} + n + n_y)^3 \\& \quad + (n+n_c)^2(\tilde{\delta} + \delta_n + \delta_v) + (\tilde{\delta}+ \delta_n + \delta_v)^2 (n+n_c)).
\end{align*}

\end{document}